\documentclass[a4paper,10pt]{article}
\usepackage{ifluatex}

\ifluatex 
    \usepackage{fontspec}
    \defaultfontfeatures{Ligatures={TeX}}
\else
    \usepackage[latin1]{inputenc}
    \usepackage[T2A,T1]{fontenc}
\fi

\usepackage{etoolbox}
\robustify{\underline}
\usepackage{setspace}
\usepackage{enumerate}
\usepackage[english]{babel}
\usepackage{enumerate}
\usepackage{amsmath,amssymb}
\usepackage{array,booktabs}
\usepackage{tcolorbox}
\usepackage{graphicx}
\usepackage{amsthm}
\usepackage{bbm}
\usepackage{empheq}
\usepackage{bm}
\usepackage{mathtools}
\usepackage[nottoc,numbib]{tocbibind}
\usepackage{tikz}
\usepackage[artemisia]{textgreek}
\usepackage{selinput}
\usepackage{
nameref,
cleveref,
}

\usetikzlibrary{positioning ,shadows,matrix,calc,decorations.pathmorphing,shapes,decorations.pathreplacing,calligraphy}
\usepackage{upgreek}
\usepackage{cmbright}
\usepackage{thmtools}
\usepackage{wasysym}
\usepackage{slashed}
\usepackage{concmath}
\usepackage{bigints}
\usepackage{setspace}
\usepackage{scalerel,stackengine}
\stackMath
\newcommand\reallywidehat[1]{%
\savestack{\tmpbox}{\stretchto{%
  \scaleto{%
    \scalerel*[\widthof{\ensuremath{#1}}]{\kern-.6pt\bigwedge\kern-.6pt}%
    {\rule[-\textheight/2]{1ex}{\textheight}}
  }{\textheight}%
}{0.5ex}}%
\stackon[1pt]{#1}{\tmpbox}%
}

\SelectInputMappings{  
  cacute={ć},          
  lslash={ł}           
}

\AtBeginDocument{\let\latexlabel\label}

\newcommand*{\doubleequation}[3][]{%
    \par\vskip\abovedisplayskip\noindent
    \if\relax\detokenize{#1}\relax
       \let\@dblLabelI\@empty
       \let\@dblLabelII\@empty
    \else 
       \@dblequationAux #1,%
    \fi
    \makebox[0.5\linewidth-1.5em]{%
     \hspace{\stretch2}%
     \makebox[0pt]{$\displaystyle #2$}%
     \hspace{\stretch1}%
    }%
    \makebox[0.5\linewidth-1.5em]{%
     \hspace{\stretch1}%
     \makebox[0pt]{$\displaystyle #3$}%
     \hspace{\stretch2}%
    }%
    \makebox[3em][r]{(%
  \refstepcounter{equation}\theequation\@dblLabelI, 
  \refstepcounter{equation}\theequation\@dblLabelII)}%
  \par\vskip\belowdisplayskip
}

\makeatletter
\newcommand{\AlignFootnote}[1]{%
  \ifmeasuring@
    \chardef\@tempfn=\value{footnote}%
    \footnotemark
    \setcounter{footnote}{\@tempfn}%
  \else
    \iffirstchoice@
      \footnote{#1}%
    \fi
  \fi}
\makeatother

\newtheorem{thm}{Theorem}
\newtheorem{theorem*}{Theorem}
\newtheorem{lemma}[thm]{Lemma}
\newtheorem{proposition}[thm]{Proposition}
\newtheorem{corollary}{Corollary}
\newtheorem{defin}{Definition}
\theoremstyle{definition}

\newtheorem{remark}{Remark}
\numberwithin{thm}{section}
\numberwithin{lemma}{section}
\numberwithin{proposition}{section}
\numberwithin{corollary}{section}
\numberwithin{remark}{section}
\numberwithin{defin}{section}
\newtheorem*{notation*}{Notation}
\numberwithin{equation}{section}

\makeatletter
\def\namedlabel#1#2{\begingroup
   \def\@currentlabel{#2}%
   \label{#1}\endgroup
}
\makeatother

\usepackage{lmodern}
\usepackage[margin=3cm]{geometry}
\usepackage{verbatim}
\usepackage{fullpage}
\usepackage{amsthm}
\usepackage[affil-it]{authblk}
\usepackage{mathrsfs}
\usepackage{graphicx}
\newcommand{\ubar}{\overline{u}}

\newcommand{\divo}{\mathring{\slashed{\mathrm{div}}}\,}
\newcommand{\divr}{\slashed{\mathrm{div}}}

\newcommand{\curlo}{\mathring{\slashed{\text{curl}}}}
\newcommand{\curlr}{\slashed{\text{curl}}}

\newcommand{\nablau}{\Omega\slashed{\nabla}_3}
\newcommand{\nablav}{\Omega\slashed{\nabla}_4}
\newcommand{\nablagml}{\slashed{\nabla}_3}
\newcommand{\nabladlt}{\slashed{\nabla}_4}
\newcommand{\fancyd}{\slashed{\mathcal{D}}}
\newcommand{\fancydring}{\mathring{\slashed{\mathcal{D}}}}
\newcommand{\fancydstar}{\slashed{\mathcal{D}}^\star}
\newcommand{\fancydstarring}{\mathring{\slashed{\mathcal{D}}}^\star}

\newcommand{\fbar}{\underline{f}}

\newcommand{\overone}{\stackrel{\mbox{\scalebox{0.4}{(1)}}}}
\newcommand{\slashednabla}{\slashed{\nabla}}

\newcommand{\trx}{\overone{\Omega tr\chi}}
\newcommand{\trxbar}{\overone{\Omega tr\underline\chi}}
\newcommand{\invertedalpha}{\raisebox{\depth}{\scalebox{1}[-1]{$\alpha$}}}
\newcommand{\invertedpsi}{\raisebox{\depth}{\scalebox{1}[-1]{$\Psi$}}}

\def\J{\ensuremath\mathcal{J}}

\DeclareMathOperator{\tr}{tr}

\DeclareFontFamily{U} {cmmi}{}

\DeclareFontShape{U}{cmmi}{m}{n}{
  <-6> cmmi5
  <6-7> cmmi6
  <7-8> cmmi7
  <8-9> cmmi8
  <9-10> cmmi9
  <10-12> cmmi10
  <12-> cmmi12}{}

\DeclareSymbolFont{Xcmmi} {U} {cmmi}{m}{n}

\DeclareMathSymbol{\Sigma}{\mathord}{Xcmmi}{24}

\usepackage{multirow}
\usepackage{titling}
\usepackage{breqn}
\usepackage{tensor}
\graphicspath{ {/home/hm532} }
\usepackage[backend=biber]{biblatex}
\usepackage{accents}
\newcommand\bref[1]{(\ref{#1})}
\makeatletter
\patchcmd{\@maketitle}{\LARGE \@title}{\fontsize{16}{19.2}\selectfont\@title}{}{}
\makeatother

\newtheoremstyle{named}{}{}{\itshape}{}{\bfseries}{.}{.5em}{\thmnote{#3}}
\theoremstyle{named}
\newtheorem*{namedtheorem}{Theorem}
\newtheorem*{namedcorollary}{Corollary}

\newcommand{\dw}{\sin\theta d\theta d\phi}

\DeclareSymbolFont{lettersA}{U}{txmia}{m}{it}
\DeclareMathSymbol{\rhoup}{\mathord}{lettersA}{26}
\DeclareMathSymbol{\phiup}{\mathord}{lettersA}{30}
\DeclareMathSymbol{\alphaup}{\mathord}{lettersA}{11}
\DeclareMathSymbol{\psiup}{\mathord}{lettersA}{32}
\DeclareMathSymbol{\varrhoup}{\mathord}{lettersA}{37}
\DeclareMathSymbol{\etaup}{\mathord}{lettersA}{17}

\newcommand{\fr}{\mathrm{f}}

\newcommand{\Pscri}{\accentset{\scalebox{.6}{\mbox{\tiny (1)}}}{{\mathrm{P}}}}

\newcommand{\Qscri}{\accentset{\scalebox{.6}{\mbox{\tiny (1)}}}{{\mathrm{Q}}}}

\newcommand{\Olin}{\Omega^{-1}\accentset{\scalebox{.6}{\mbox{\tiny (1)}}}{\Omega}}
\newcommand{\Olino}{\accentset{\scalebox{.6}{\mbox{\tiny (1)}}}{\Omega}}
\newcommand{\glinh}{\accentset{\scalebox{.6}{\mbox{\tiny (1)}}}{\hat{\slashed{g}}}}
\newcommand{\glin}{\accentset{\scalebox{.6}{\mbox{\tiny (1)}}}{\slashed{g}}}
  \newcommand{\glinto}{\accentset{\scalebox{.6}{\mbox{\tiny (1)}}}{\sqrt{\slashed{g}}}}
\newcommand{\bmlin}{\accentset{\scalebox{.6}{\mbox{\tiny (1)}}}{b}}

\newcommand{\Phii}[1]{\Phi^{(#1)}}

\newcommand{\OlinK}{\Omega^{-1}\accentset{\scalebox{.6}{\mbox{\tiny (1)}}}{\widetilde{\Omega}}}
\newcommand{\OlinoK}{\accentset{\scalebox{.6}{\mbox{\tiny (1)}}}{\Omega}}
\newcommand{\glinhK}{\accentset{\scalebox{.6}{\mbox{\tiny (1)}}}{\widetilde{\hat{\slashed{g}}}}}
\newcommand{\glinK}{\accentset{\scalebox{.6}{\mbox{\tiny (1)}}}{\widetilde{\slashed{g}}}}

\newcommand{\bmlinK}{\accentset{\scalebox{.6}{\mbox{\tiny (1)}}}{\widetilde{b}}}

\newcommand{\Olins}{\Omega^{-1}\accentset{\scalebox{.6}{\mbox{\tiny (1)}}}{\mathtt{\Omega}}}
\newcommand{\Olinos}{\accentset{\scalebox{.6}{\mbox{\tiny (1)}}}{\mathtt{\Omega}}}
\newcommand{\glinhs}{\accentset{\scalebox{.6}{\mbox{\tiny (1)}}}{\hat{\slashed{\mathtt{g}}}}}
\newcommand{\glins}{\accentset{\scalebox{.6}{\mbox{\tiny (1)}}}{\slashed{\mathtt{g}}}}
  
\newcommand{\bmlins}{\accentset{\scalebox{.6}{\mbox{\tiny (1)}}}{\mathtt{b}}}

\newcommand{\xblin}{\accentset{\scalebox{.6}{\mbox{\tiny (1)}}}{\underline{\hat{\chi}}}}
\newcommand{\xlin}{\accentset{\scalebox{.6}{\mbox{\tiny (1)}}}{{\hat{\chi}}}}

\newcommand{\eblin}{\accentset{\scalebox{.6}{\mbox{\tiny (1)}}}{\underline{\eta}}}
\newcommand{\elin}{\accentset{\scalebox{.6}{\mbox{\tiny (1)}}}{{\eta}}}
\newcommand{\otx}{\accentset{\scalebox{.6}{\mbox{\tiny (1)}}}{\left(\Omega\tr\chi\right)}}
\newcommand{\otxb}{\accentset{\scalebox{.6}{\mbox{\tiny (1)}}}{\left(\Omega\tr\underline{\chi}\right)}}
\newcommand{\olin}{\accentset{\scalebox{.6}{\mbox{\tiny (1)}}}{\omega}}
\newcommand{\olinb}{\accentset{\scalebox{.6}{\mbox{\tiny (1)}}}{\underline{\omega}}}

\newcommand{\xblinK}{\accentset{\scalebox{.6}{\mbox{\tiny (1)}}}{\widetilde{\underline{\hat{\chi}}}}}
\newcommand{\xlinK}{\accentset{\scalebox{.6}{\mbox{\tiny (1)}}}{{\widetilde{\hat{\chi}}}}}

\newcommand{\eblinK}{\accentset{\scalebox{.6}{\mbox{\tiny (1)}}}{\widetilde{\underline{\eta}}}}
\newcommand{\elinK}{\accentset{\scalebox{.6}{\mbox{\tiny (1)}}}{{\widetilde{\eta}}}}
\newcommand{\otxK}{\accentset{\scalebox{.6}{\mbox{\tiny (1)}}}{\widetilde{\left(\Omega \tr \chi\right)}}}
\newcommand{\otxbK}{\accentset{\scalebox{.6}{\mbox{\tiny (1)}}}{\widetilde{\left(\Omega \tr \underline{\chi}\right)}}}
\newcommand{\olinK}{\accentset{\scalebox{.6}{\mbox{\tiny (1)}}}{\widetilde{\omega}}}
\newcommand{\olinbK}{\accentset{\scalebox{.6}{\mbox{\tiny (1)}}}{\widetilde{\underline{\omega}}}}

\newcommand{\xblins}{\accentset{\scalebox{.6}{\mbox{\tiny (1)}}}{\underline{\hat{\upchi}}}}
\newcommand{\xlins}{\accentset{\scalebox{.6}{\mbox{\tiny (1)}}}{{\hat{{\upchi}}}}}

\newcommand{\eblins}{\accentset{\scalebox{.6}{\mbox{\tiny (1)}}}{\underline{{\upeta}}}}
\newcommand{\elins}{\accentset{\scalebox{.6}{\mbox{\tiny (1)}}}{{{\upeta}}}}
\newcommand{\otxs}{\accentset{\scalebox{.6}{\mbox{\tiny (1)}}}{\left(\mathtt{\Omega}\mathtt{tr} {\upchi}\right)}}
\newcommand{\otxbs}{\accentset{\scalebox{.6}{\mbox{\tiny (1)}}}{\left(\mathtt{\Omega tr} \underline{{\upchi}}\right)}}
\newcommand{\olins}{\accentset{\scalebox{.6}{\mbox{\tiny (1)}}}{{\upomega}}}
\newcommand{\olinbs}{\accentset{\scalebox{.6}{\mbox{\tiny (1)}}}{\underline{{\upomega}}}}

\newcommand{\ablins}{\accentset{\scalebox{.6}{\mbox{\tiny (1)}}}{\underline{\upalpha}}}
\newcommand{\alins}{\accentset{\scalebox{.6}{\mbox{\tiny (1)}}}{{\upalpha}}}

\newcommand{\ablin}{\accentset{\scalebox{.6}{\mbox{\tiny (1)}}}{\underline{\alpha}}}
\newcommand{\alin}{\accentset{\scalebox{.6}{\mbox{\tiny (1)}}}{{\alpha}}}
\newcommand{\pblin}{\accentset{\scalebox{.6}{\mbox{\tiny (1)}}}{\underline{\psi}}}
\newcommand{\plin}{\accentset{\scalebox{.6}{\mbox{\tiny (1)}}}{{\psi}}}

\newcommand{\bblin}{\accentset{\scalebox{.6}{\mbox{\tiny (1)}}}{\underline{\beta}}}
\newcommand{\blin}{\accentset{\scalebox{.6}{\mbox{\tiny (1)}}}{{\beta}}}
\newcommand{\rlin}{\accentset{\scalebox{.6}{\mbox{\tiny (1)}}}{\rho}}
\newcommand{\slin}{\accentset{\scalebox{.6}{\mbox{\tiny (1)}}}{{\sigma}}}
\newcommand{\Klin}{\accentset{\scalebox{.6}{\mbox{\tiny (1)}}}{K}}

\newcommand{\ablinK}{\accentset{\scalebox{.6}{\mbox{\tiny (1)}}}{\widetilde{\underline{\alpha}}}}
\newcommand{\alinK}{\accentset{\scalebox{.6}{\mbox{\tiny (1)}}}{\widetilde{\alpha}}}
\newcommand{\pblinK}{\accentset{\scalebox{.6}{\mbox{\tiny (1)}}}{\widetilde{\underline{\psi}}}}

\newcommand{\bblinK}{\accentset{\scalebox{.6}{\mbox{\tiny (1)}}}{\widetilde{\underline{\beta}}}}
\newcommand{\blinK}{\accentset{\scalebox{.6}{\mbox{\tiny (1)}}}{{\widetilde{\beta}}}}
\newcommand{\rlinK}{\accentset{\scalebox{.6}{\mbox{\tiny (1)}}}{\widetilde{\rho}}}
\newcommand{\slinK}{\accentset{\scalebox{.6}{\mbox{\tiny (1)}}}{{\widetilde{\sigma}}}}
\newcommand{\KlinK}{\accentset{\scalebox{.6}{\mbox{\tiny (1)}}}{\widetilde{K}}}

\newcommand{\pblins}{\accentset{\scalebox{.6}{\mbox{\tiny (1)}}}{\underline{{\uppsi}}}}
\newcommand{\plins}{\accentset{\scalebox{.6}{\mbox{\tiny (1)}}}{{{\uppsi}}}}

\newcommand{\bblins}{\accentset{\scalebox{.6}{\mbox{\tiny (1)}}}{\underline{{\upbeta}}}}
\newcommand{\blins}{\accentset{\scalebox{.6}{\mbox{\tiny (1)}}}{{{\upbeta}}}}
\newcommand{\rlins}{\accentset{\scalebox{.6}{\mbox{\tiny (1)}}}{{\uprho}}}
\newcommand{\slins}{\accentset{\scalebox{.6}{\mbox{\tiny (1)}}}{{{\upsigma}}}}

\newcommand{\ustar}{{u^*}}
\newcommand{\g}{\mathtt{g}}

\newcommand{\pastmemoryPsilin}{\mathscr{P}_{-}}

\newcommand{\glinhn}[1]{\,\hat{\mathrm{g}}^{(#1)}}
\newcommand{\glinhnscr}[1]{\,\hat{\slashed{\mathtt{g}}}^{(#1)}}

\newcommand{\Ylin}{\accentset{\scalebox{.6}{\mbox{\tiny (1)}}}{Y}}
\newcommand{\Psilin}{\accentset{\scalebox{.6}{\mbox{\tiny (1)}}}{\Psi}}
\newcommand{\Psilinb}{\accentset{\scalebox{.6}{\mbox{\tiny (1)}}}{\underline{\Psi}}}

\newcommand{\upPsilin}{\accentset{\scalebox{.6}{\mbox{\tiny (1)}}}{\bm{\uppsi}}}
\newcommand{\upPsilinb}{\accentset{\scalebox{.6}{\mbox{\tiny (1)}}}{\underline{\bm{\uppsi}}}}

\newcommand{\fullsystem}{\bref{metric transport in 4 direction trace}-\bref{Gauss}\;}

\newcommand{\fullsystemK}{\bref{metric transport in 4 direction Kruskal}-\bref{Gauss Kruskal}\;}

\newcommand{\nablaubar}{\slashednabla_{{u^*}}}

\newcommand{\outwardsgaugefunction}{\underline{\mathfrak{f}}_{\mathscr{I}^+}}

\newcommand{\inwardsgaugefunction}{{\mathfrak{f}}_{\mathscr{H}^+}}

\newcommand{\vsigmap}[1]{v_{{\Sigma^*_+},#1}}
\newcommand{\usigmap}[1]{u_{{\Sigma^*_+},#1}}

\newcommand{\usigmapt}[1]{u_{{\widetilde{\Sigma}_+},#1}}

\newcommand{\vsigmam}[1]{v_{{\Sigma^*_-},#1}}
\newcommand{\usigmam}[1]{u_{{\Sigma^*_-},#1}}

\newcommand{\flin}{\accentset{\scalebox{.6}{\mbox{\tiny (1)}}}{f}}
\newcommand{\glinn}{\accentset{\scalebox{.6}{\mbox{\tiny (1)}}}{g}}

\newcommand{\flinl}{\accentset{\scalebox{.6}{\mbox{\tiny (1)}}}{f}_{(\ell,m)}}
\newcommand{\glinnl}{\accentset{\scalebox{.6}{\mbox{\tiny (1)}}}{g}_{(\ell,m)}}

\newcommand{\inhom}{\widetilde{M}}

\newcommand{\flinslp}{{\accentset{\scalebox{.6}{\mbox{\tiny (1)}}}{\mathpzc{f}}}_{{\mathscr{I}^-},(\ell,m)}}
\newcommand{\flinslf}{{\accentset{\scalebox{.6}{\mbox{\tiny (1)}}}{\mathpzc{f}}}_{{\mathscr{I}^+},(\ell,m)}}
\newcommand{\flins}{{\accentset{\scalebox{.6}{\mbox{\tiny (1)}}}{\mathpzc{f}}}_{(\ell,m)}}

\newcommand{\glinnslp}{{\accentset{\scalebox{.6}{\mbox{\tiny (1)}}}{\mathpzc{g}}}_{{\mathscr{I}^-},(\ell,m)}}
\newcommand{\glinnslf}{{\accentset{\scalebox{.6}{\mbox{\tiny (1)}}}{\mathpzc{g}}}_{{\mathscr{I}^+},(\ell,m)}}

\newcommand{\Flins}[2]{{\accentset{\scalebox{.6}{\mbox{\tiny (1)}}}{\mathpzc{F}}}_{(#1,m)}^{(#2)}}
\newcommand{\Flinslp}[2]{{\accentset{\scalebox{.6}{\mbox{\tiny (1)}}}{\mathpzc{F}}}_{\mathscr{I}^-,(#1,m)}^{(#2)}}

\DeclareMathAlphabet{\mathpzc}{OT1}{pzc}{m}{it}

\author{Hamed Masaood}

\affil{ Imperial College London, Department of Mathematics,\\
South Kensington Campus, London~SW7~2AZ, United Kingdom}

\title{A Scattering Theory for Linearised Gravity\\ on the Exterior of the Schwarzschild Black Hole II:\\ The Full System}

\newcommand{\Sscriscri}{ \mathfrak{S}_{\mathscr{I}^-\cup\overline{\mathscr{H}^-}\rightarrow\mathscr{I}^+\cup\overline{\mathscr{H}^+}}}

\newcommand{\Sscriscriscri}{ \mathfrak{S}_{\mathscr{I}^-\rightarrow\mathscr{I}^+}}

\newcommand{\SscriHp}{\mathfrak{S}_{\mathscr{I}^-\cup\overline{\mathscr{H}^-}}}
\newcommand{\SscriHf}{\mathfrak{S}_{\mathscr{I}^+\cup\overline{\mathscr{H}^+}}}

\newcommand{\Gscriscri}{ \mathfrak{G}_{\mathscr{I}^-\cup\overline{\mathscr{H}^-}\rightarrow\mathscr{I}^+\cup\overline{\mathscr{H}^+}}}

\DeclareMathVersion{normal2}

\begin{document}

\maketitle

\begin{abstract}
    We construct a scattering theory for the linearised Einstein equations on a Schwarzschild background in a double null gauge. We build on the results of Part I \cite{Mas20}, where we used the energy conservation enjoyed by the Regge--Wheeler equation associated with the stationarity of the Schwarzschild background to construct a scattering theory for the Teukolsky equations of spin $\pm2$. We now extend the scattering theory of Part I to the full system of linearised Einstein equations by treating it as a system of transport equations which is sourced by solutions to the Teukolsky equations, leading to Hilbert space-isomorphisms between spaces of finite energy initial data and corresponding spaces of scattering states under suitably chosen gauge conditions on initial and scattering data. As a corollary, we show that for a solution which is Bondi-normalised at both past and future null infinity, past and future linear memories are related by an antipodal map.
\end{abstract}

\tableofcontents

\section{Introduction and overview}

In this paper we complete the series started in Part I \cite{Mas20} on the construction of a scattering theory for the (vacuum) Einstein equations,
\begin{align}\label{EVE}
    \text{Ric}[g]=0,
\end{align}
when linearised in a double null gauge against a Schwarzschild background,
\begin{align}\label{Schwarzschild}
    g=-\left(1-\frac{2M}{r}\right)dt^2+\left(1-\frac{2M}{r}\right)^{-1}dr^2+r^2(d\theta^2+\sin^2\theta d\phi^2).
\end{align}

In Part I \cite{Mas20}, a scattering theory was constructed for the \textbf{Teukolsky equations of spin} $\bm{\pm2}$ which govern the gauge invariant components of the linearised curvature. We also showed in \cite{Mas20} how the \textbf{Teukolsky--Starobinsky identities} can be propagated in solutions arising from scattering data for the linearised system, giving rise to a scattering theory completely governing the radiating degrees of freedom of solutions to the linearised Einstein equations. A crucial element of that construction is the $\partial_t$ isometry of \bref{Schwarzschild} and the corresponding energy conservation law enjoyed by the \textbf{Regge--Wheeler equation}, itself a consequence of the Teukolsky equations on a Schwarzschild spacetime via the Chandrasekhar transformations introduced in \cite{Chandrasekhar} and used in physical-space form to show the linear stability of the Schwarzschild exterior in \cite{DHR16}.\\

\indent As we pass here from the scattering theory constructed in \cite{Mas20} for the Teukolsky equations to one encompassing the full system of linearised Einstein equations, the main new feature of the analysis is the study of the role of the gauge ambiguity in defining spaces of scattering states and in evolving solutions from data at infinity. Namely, we must identify the essential asymptotic information needed to construct solutions to the full linearised system, and we must specify how gauge freedom is restricted by the demands of evolving solutions from scattering data. We show that, starting from solutions to the initial value problem with asymptotically flat Cauchy data, it is possible to pass to a gauge where the radiation fields representing scattering states satisfy the conditions defining a \textbf{Bondi-normalised double null gauge}. We show that in such a gauge, the Hilbert spaces of scattering states defined for the Teukolsky equations in \cite{Mas20} carry over to corresponding spaces of scattering states for the full linearised Einstein system. We show how to construct solutions to the linearised Einstein equations out of scattering data in these spaces, leading to a Hilbert-space isomorphism between spaces of scattering data and the space of initial data defined using the $\partial_t$-energy conservation of the Regge--Wheeler equation. Thus the  scattering theory we construct here achieves the three requirements for a satisfactory solution of the scattering problem:

\begin{enumerate}[I]
\item \textit{Existence of scattering states}: That a given class of solutions can be associated with past/future scattering states.\label{QI}
\item \textit{Uniqueness of scattering states}: That the above association be injective; solutions that give rise to the same scattering state must coincide.\label{QII}
\item \textit{Asymptotic completeness}: That this association must exhaust the class of initial data of interest.\label{QIII}
\end{enumerate}

As part of our construction, we classify the collection of pure gauge transformations consistent with a Bondi-normalised double null gauge and we find that the effect of these gauge transformations on the coordinates at null infinity is identical to that of the \textbf{Bondi--Metzner--Sachs (BMS)} group.\\

This overview is organised as follows: \Cref{Section 1.1 overview of scattering theory} provides a recap of relevant results on asymptotic analysis and scattering theory in general relativity, highlighting the foundational work of Bondi, Penrose and Christodoulou--\mbox{Klainerman}. In  \Cref{Section 1.2} we recap the results of Part I, sketch the unique challenges associated with the study of scattering in the presence of gauge ambiguity, and present the scheme used in this work to show that the scattering problem is well-posed in a double null gauge, followed by a preliminary statement of the main results of this paper in \Cref{Section 1.3 summary of results}.



\subsection{Milestones in the study of gravitational radiation}\label{Section 1.1 overview of scattering theory}

We begin with a brief overview of the history of gravitational scattering theory, highlighting the contributions of Bondi, Metzner and van der Burg  \cite{BMV1962}, Sachs \cite{Sachs1962}, Penrose \cite{Penrose:1964ge}, \cite{PenroseZeroRestMass},  \cite{PenroseAsymptoticPropertiesOfFieldsAndSpacetimes} and Christodoulou \cite{ChristodoulouMemoryEffect}.


\subsubsection{The Bondi--Sachs formalism}

\indent The idea that small-data solutions to the equations \bref{EVE} disperse to asymptotically freely propagating waves was a widely believed prediction since the earliest days of the theory of general relativity, as it was realised that general relativity implies a wave-like character in weak gravitational fields \cite{1916SPAW.......688E}. However, the sense in which weak gravitational fields from isolated sources asymptote to a nearly Minkowskian state had not been systematically treated until the work of Bondi et al.~\cite{BMV1962} and Sachs \cite{Sachs1962}, who aimed to clear up the confusion surrounding this subject. \\

In \cite{BMV1962}, Bondi et~al. propose the identification of asymptotically flat spacetimes with the asymptotics of the metric components when expressed in what are now known as \textit{Bondi coordinates}: a coordinate system $(u,r,\theta^A)$ with $u$ being a retarded optical function, and $r$ the area-radius function. Bondi et~al. define asymptotically flat spacetimes by demanding that the metric components admit an analytic expansion in terms of $r^{-1}$ near the values attained by the Minkowski metric. The authors of \cite{BMV1962} attempt to derive the Poincar\'e group as an approximate isometry in the large-$r$ limit to reconcile their definition with the isometry group of Minkowski space, only to discover that their definition of asymptotic flatness admits an additional  collection of asymptotic isometries, now known as ``supertranslations", which furthermore defines an \textit{infinite dimensional} extension of the Poincar\'e group. Up to the point of the publication of~\cite{BMV1962}, the extent to which the spacetimes identified by Bondi represent a typical state of weak gravitational fields was far from clear, even with the caveat that the assumption of analyticity can be weakened provided the metric components converge in a sufficiently regular fashion to the Minwkowski metric in Bondi coordinates. With these developments, the notion of asymptotic flatness was born and the systematic mathematical study of the scattering problem came to be regarded as essential to clarifying the issues raised by the work of Bondi and his group.\\

\subsubsection{The conformal picture and Penrose diagrams}

Within a year of the publication of the work of Bondi et al.~\cite{BMV1962}, Roger Penrose proposed the use of conformal methods to give the asymptotic structure of a spacetime a concrete geometrical representation, via a conformal extension of the metric towards ``null infinity" and the related concept of the Penrose diagram  \cite{Penrose:1964ge}. Crucially, Penrose proposed an alternative definition of an asymptotically flat spacetime as one which admits a null infinity which is null in a suitable conformal extension, in analogy with the Penrose diagram of Minkowski space. The definition of asymptotic flatness made by Penrose reproduces the notions and results obtained by Bondi et~al. and does so using a covariant geometric language. In particular, Penrose derives the BMS group by restricting the gauge redundancy associated to the choice of a conformal factor,  so that null infinity defines in the conformally extended picture a complete null hypersurface, which furthermore can be foliated by unit round spheres. See \cite{PenroseRindler} for the full exposition. The terms and notions introduced by Penrose were subsequently widely adopted, independently of the details of any conformal extension.
\subsubsection{Christodoulou--\mbox{Klainerman} and the laws of gravitational radiation}

In spite of the impressive revelations afforded by the methods of Penrose, there was still lacking a concrete connection of the results found by Bondi and Penrose to the dynamics of the Einstein equations \bref{EVE}, and it was unclear that the notions used to define asymptotic flatness were representative of any physics at all. A confirmation of the relevance of an important proportion of the ideas of Bondi and Penrose was finally achieved in the case of small perturbations to the Minkowski spacetime in the work of Christodoulou and \mbox{Klainerman} \cite{Ch-K}. As a corollary to the stability results of \cite{Ch-K}, it was shown that the asymptotics defined by Bondi in \cite{BMV1962} are realised in evolution to leading order. While it was not stated as an explicit theorem in \cite{Ch-K}, it can be deduced from the results of \cite{Ch-K} that the BMS group is indeed the residual group of gauge transformations which are compatible with the existence of null infinity and the laws of gravitational radiation. However, in the full generality of the data considered in \cite{Ch-K}, the higher order regularity assumptions of \cite{BMV1962}, \cite{Sachs1962} do not hold. For a discussion of the physical significance of the lack of  higher order regularity, see \cite{KehrbergerI}, \cite{KehrbergerII}, \cite{KehrbergerIII}.\\

\indent Beyond putting the insights of Bondi and Penrose on solid mathematical foundation, Christodoulou revealed a yet-unknown feature of gravitational radiation. The work \cite{Ch-K} establishes a set of laws that govern gravitational radiation at null infinity, which were further elaborated upon in the work of Christodoulou \cite{ChristodoulouPRL}. The laws of gravitational radiation laid out in \cite{Ch-K} and in \cite{ChristodoulouPRL} not only provide a framework that codifies the dynamical emergence of the \textit{memory effect} and subsumes earlier known heuristics about the memory effect in linearised gravity (see \cite{BRAGINSKY}, \cite{Zeldovich}), but they also reveal the presence of 
a \textit{nonlinear} \textit{memory effect}, where the passage of a pulse of gravitational radiation leaves behind an imprint in the form of a permanent spatial displacement in affected test bodies \cite{ChristodoulouMemoryEffect}. The experimental confirmation of the nonlinear memory effect has become a highly sought-after goal with the arrival of gravitational wave detectors, and the theoretical study of the effect has sparked exciting new avenues of research on other connections of black holes and fundamental physics, see for instance \cite{Hawking:2016msc}.


\subsubsection{Recent advances in gravitational scattering}

The mathematical study of scattering in general relativity has seen a substantial growth in recent years with works that address the problem taking into account requirements \bref{QI}, \bref{QII}, \bref{QIII}. The work of Dafermos, Rodnianski and Shlapentokh-Rothman \cite{DRSR14} resolves the scattering problem for the scalar wave equation on the exterior of the Kerr black hole, while the work of Alford \cite{Fred} studies the same problem on the Oppenheimer--Snyder collapsing star model. Other works address the scattering problem for certain components of gravitational radiation, such as \cite{Mas20} and the work of Teixeira da Costa \cite{Rita2019}. A scattering construction, under the fully nonlinear equations \bref{EVE}, of spacetimes that settle down to a member of the Kerr family has been done in \cite{DHR18}, providing an affirmative answer to \bref{QI} for exponentially decaying gravitational radiation.

\subsection{From Teukolsky to linearised gravity: the gauge ambiguity and the laws of gravitational radiation}\label{Section 1.2}

We now outline the results proven in this paper and the key mathematical points as they relate to the results of this paper or to the literature. We will give a brief summary of the main results of this paper, namely Theorem II, Theorem III and Corollary II, in \Cref{Section 1.3 summary of results}

\subsubsection{Gauge freedom and long-time dynamics of the Einstein equations}\label{Section 1.2 Gauge freedom}

The laws of gravitational radiation conceived by Bondi, Penrose and Christodoulou provide a convenient starting point in posing the scattering problem for the Einstein equations \bref{EVE}: are the quantities featuring in the laws of gravitational radiation, as defined by Christodoulou and Klainerman in \cite{Ch-K}, sufficient to construct solutions to \bref{EVE}, and do these solutions asymptotically realise the given gravitational radiation data in a manner consistent with the picture provided by forwards evolution from Cauchy data? Posing the scattering problem in this language, and taking into view the requirements \bref{QI}, \bref{QII}, \bref{QIII} of a satisfactory answer, it is clear that a satisfactory scattering theory amounts to a well-posedness result for \bref{EVE} with data ``at infinity". And indeed, many of the challenges encountered in studying scattering for \bref{EVE} arise also in the usual Cauchy problem.\\

For example, while the Einstein equations \bref{EVE} are of evolutionary character, it is far from obvious how this property could be exhibited and studied using the geometric formulation given in \bref{EVE}. To show that the system \bref{EVE} is of hyperbolic type and to prove its well-posedness, Choquet-Bruhat resorted in \cite{Choquet-Bruhat} to a well-chosen coordinate system, namely a system of wave coordinates $(x^\mu)$ with $\Box_g x^\mu=0$, which reduces \bref{EVE} to a system of quasilinear wave equations. Indeed, finding a suitable coordinate system which makes manifest the hyperbolicity of the equations \bref{EVE} was the main obstacle in the way of solving the initial value problem for \bref{EVE}.\\

When it comes to the scattering problem, making a judicious choice of coordinates is of even greater importance than it is for the well-posedness of the initial value problem. For a coordinate system to allow for a scattering theory for \bref{EVE} to exist, it must support global existence and enable control of solutions for late times, thus global existence results are a natural starting point in the search for a candidate gauge for scattering theory. Global existence for the nonlinear equations \bref{EVE} has only recently seen breakthroughs in the study of small data perturbations to the Minkowski space and the Schwarzschild family of black holes (see also the recent \cite{LukOh21} establishing global existence for a large class of large data perturbations to the Minkowski spacetime). For example, the proof of stability of the Minkowski space by Christodoulou and \mbox{Klainerman} \cite{Ch-K} uses a combination of coordiante systems, including a system $(u,t,\theta^A)$, where $u$ is a retarded optical function and $t$ is a time function which generates a maximal foliation of spacetime. The proof of the linear stability of the Schwarzschild solution given in \cite{DHR16}, and its recent sequel \cite{DHRT21} establishing nonlinear stability, make crucial use of the structure afforded by casting \bref{EVE} in a double null gauge. These gauge in essence exhibit the natural null structure in  the nonlinearity in \eqref{EVE}. Another example is the use of wave coordinates to provide an alternative proof of the nonlinear stability of the Minkowski space. See also \cite{Johnson}, where a generalisation of the wave gauge was used to provide another proof of the linear stability of the Schwarzschild black hole. Although the aforementioned wave-like coordinate systems do not fully exhibit the null structure of the Einstein equations \eqref{EVE}, they still reveal a weak form of the null structure, sometimes dubbed as the ``weak null condition" in the sense of still capturing the essential decay hierarchy at late times that enables a stability argument. See the introduction of \cite{Lin-Rod} for a greater discussion. In addition to making way for the study of long time behaviour of solutions, the significance of a well-chosen coordinate system to the problem of global existence is to allow for the proper identification of the final state of evolution.\\

Here we make use of the method of \cite{DHR16}, utilising a double null gauge in linearising the system \bref{EVE} and studying scattering for the resulting system. It is of great interest to study the scattering properties of \bref{EVE} when cast in other coordinate systems, such as the wave coordinates used by Choquet-Bruhat and the variation thereof used in \cite{Johnson}.

\subsubsection{The double null gauge and the Teukolsky equations}\label{Section 1.3 double null gauge intro}

By a double null gauge, we mean a coordinate system $(\bm{u},\bm{v},\bm{\theta}^A)$ where $\bm{u}, \bm{v}$ are the optical functions of the metric $\bm{g}$ and $\bm{\theta}^A$ define a coordinate system on $\mathcal{S}_{\bm{u},\bm{v}}$, the 2-dimensional manifolds formed by the intersection loci of $\bm{u}, \bm{v}$. In such a coordinate system, the metric takes the form
\begin{align}\label{DNGmetric}
    \bm{g}=-4\bm{\Omega}^2\bm{dudv}+\bm{\slashed{g}}_{AB}(\bm{d\theta}^A-\bm{b}^A\bm{dv})(\bm{d\theta}^B-\bm{b}^B\bm{dv}).
\end{align}

Here, $\bm{b}_A$ and $\bm{\slashed{g}}_{AB}$ are tensor fields which are everywhere tangential to the surfaces $\mathcal{S}_{\bm{u},\bm{v}}$. $\bm{\slashed{g}}$ is the metric induced on $\mathcal{S}_{\bm{u},\bm{v}}$. For example, the Eddington--Finkelstein null coordinates $u$, $v$ define such a double null gauge on the Schwarzschild exterior, and the metric \bref{Schwarzschild} can then be cast in the form \bref{DNGmetric}:
\begin{align}
    g=-4\Omega^2dudv+r^2(d\theta^2+\sin^2\theta\, d\phi^2),\qquad\qquad \Omega^2:=1-\frac{2M}{r},
\end{align}
where $M$ is the mass parameter of the background Schwarzschild spacetime. The equations \bref{EVE} are linearised by introducing a linear perturbation,  parametrised by a smallness parameter $\epsilon$, to each component of the metric, its Levi-Civita connection and curvature:
\begin{align}
    \bm{g}=g\;+\stackrel{\;\;\mbox{\scalebox{0.4}{(1)}}}{\epsilon g}, \qquad \bm{\Gamma}=\Gamma+\stackrel{\,\,\;\;\mbox{\scalebox{0.4}{(1)}}}{\epsilon\; \Gamma}, \qquad \bm{R}=R+\stackrel{\,\;\;\mbox{\scalebox{0.4}{(1)}}}{\epsilon R}.
\end{align}
In particular,
\begin{align}
 \bm{\slashed{g}}=r^2\gamma_{S^2}+\epsilon\glin=r^2\gamma_{S^2}+\epsilon(\glinh+\frac{1}{2}\tr\glin\cdot r^2\gamma_{S^2}),\qquad\bm{\Omega}^2=\Omega^2(1+2\,\epsilon\,\Olin),\qquad\bm{b}=0+\epsilon\,\bmlin,
\end{align}
where $\glin$ was decomposed into its trace and traceless components (under $r^2\gamma_{S^2}$). We keep only the terms in \bref{EVE} which are linear in $\epsilon$.
The linearisation of the structure equations relating the metric, connection and curvature components leads to the equations defining their linearised counterparts. For example, the metric perturbations $\glinh$, $\tr\glin$, $\bmlin$ define the linearised outgoing shear and expansion, $\xlin$, $\otx$ respectively,  via
\begin{align}\label{outgoing shear and expansion intro}
    \nablav \glinh=2\Omega\xlin+{\slashednabla}_A\bmlin_B+{\slashednabla}_B\bmlin_A-r^2{\gamma_{S^2}}_{AB}\slashednabla^C\bmlin_C,\qquad \partial_v \tr\glinh=2\otx+2\slashednabla_A\bmlin^A,
\end{align}
the operator $\slashednabla$ is the Levi-Civita covariant derivative associated to $g$, projected onto the cotangent space of $\mathcal{S}_{u,v}=S^2_{u,v}$. The ingoing shear and expansion, $\xblin$, $\otxb$ are defined by
\begin{align}\label{ingoing shear and expansion intro}
     \nablau \glinh=2\Omega\xblin,\qquad \partial_u \tr\glinh=2\otxb.
\end{align}
The outgoing and ingoing shear components, $\xlin$ and $\xblin$ respectively, define the linearised curvature components $\overone{\alpha}_{AB}=\overone{W}_{A4B4}$, $\overone{\underline\alpha}_{AB}=\overone{W}_{A3B3}$ via
\begin{align}\label{example2}
    \nablav\; \frac{r^2}{\Omega}\xlin=-r^2\alin,\qquad\qquad  \nablau\; \frac{r^2}{\Omega}\xblin=-r^2\ablin.
\end{align}
Aside from the structure equations defining the linearised Levi-Civita connection and curvature components, the linearised Bianchi identities lead to equations of the form
\begin{align}\label{coupling}
   \stackrel{\,\;\;\;\;\mbox{\scalebox{0.4}{(1)}}}{\nabla \; W}+\stackrel{\mbox{\scalebox{0.4}{(1)}}\;\;\;\;\;\;}{\Gamma\; W}=0.
\end{align}
 For example,
\begin{align}\label{example1}
    \frac{1}{\Omega}\nablagml r\Omega^2\overone{\alpha}\;=-2r\fancydstar_2 \Omega\overone{\beta} +\frac{6M}{r^2}\Omega\overone{\hat{\chi}}, \qquad\qquad \nablav r^4\Omega\overone{\beta}-2M r^2\Omega\overone{\beta}\;= r\slashed{div}\;r^3\Omega^2\overone{\alpha},
\end{align}
where $\overone{\beta}_A=\overone{W}_{A434}$. A similar set of equations governs $\ablin$. Note already that choosing a double null gauge leaves a residual ambiguity in the choice of the coordinates that define the double null foliation.

Notice that nontrivial curvature components of the Schwarzschild background couple to the linearised connection in \bref{coupling}, as exhibited by the example  \bref{example1}. Were it not for this coupling, the method of energy currents could have been applied to \bref{coupling} using the Bel--Robinson tensor as an energy-momentum tensor to define the relevant currents (see the introduction of \cite{Mas20} for a discussion of this method). An alternative route towards control of the linearised equations may be followed by noticing that quantities $\alin$, $\ablin$, which are in fact invariant under residual gauge transformations, can be shown via equations \bref{example2} and their ingoing counterparts to obey \textit{decoupled} wave equations. These are the famous Teukolsky equations of spin $\bm{\pm2}$ \cite{TeukP74}, \cite{Teu73}:
\begin{align}\label{wave equation +}
     \Box_g \Omega^2\alin +\frac{4}{r\Omega^2}\left(1-\frac{3M}{r}\right)\partial_u \Omega^2\alin=V(r) \Omega^2\alin,
\end{align}
\begin{align}\label{wave equation -}
     \Box_g \Omega^2\ablin -\frac{4}{r\Omega^2}\left(1-\frac{3M}{r}\right)\partial_v \Omega^2\ablin =V(r) \Omega^2\ablin.
\end{align}
Obtaining boundedness and decay results for the Teukolsky equations \bref{wave equation +}, \bref{wave equation -} remained an open problem for a long time until they were finally obtained in \cite{DHR16}, where the key insight was to make use of a time domain version of a trick conceived by Chandrasekhar \cite{Chandrasekhar} to map solutions of \bref{wave equation +}, \bref{wave equation -} to solutions of the {Regge--Wheeler equation},
\begin{align}\label{RWintro}
    \nablau\nablav\Psi-\Omega^2\slashed{\Delta}\Psi+V(r)\Psi=0,
\end{align}
where $V(r)=\frac{\Omega^2(3\Omega^2+1)}{r^2}$, by considering $r$-weighted 2\textsuperscript{nd} order null derivatives of $\alin$, $\ablin$ according to
\begin{align}\label{transport}
    \overone\Psi=\left(\frac{r^2}{\Omega^2}\nablau\right)^2r\Omega^2\overone\alpha,\qquad\qquad \overone{\underline\Psi}=\left(\frac{r^2}{\Omega^2}\nablav\right)^2r\Omega^2\overone{\underline\alpha}.
\end{align}
The Regge--Wheeler equation \bref{RWintro} above has the same structure as the scalar wave equation on Schwarzschild, and is thus malleable to the method of energy currents leading to boundedness and decay results for solutions arising from initial data where a suitable $L^2$ energy is finite. The presence of the Teukolsky equations \bref{wave equation +}, \bref{wave equation -} and the Regge--Wheeler equation \bref{RWintro} is the main factor behind the success of the double null gauge in controlling the long-time behaviour of solutions to \bref{EVE}, in the cases of both the linear and nonlinear regimes.

Another set of relations decoupling $\alin$, $\ablin$ from the remaining components of the linearised system derive from the linearised Einstein equations. These are the well-known Teukolsky--Starobinsky identities,

\begin{align}
\frac{\Omega^2}{r^2}\Omega\slashed{\nabla}_3 \left(\frac{r^2}{\Omega^2}\nablau\right)^3\alpha=2r^4\slashed{\mathcal{D}}^*_2\slashed{\mathcal{D}}^*_1\overline{\slashed{\mathcal{D}}}_1\slashed{\mathcal{D}}_2 r\Omega^2{\underline\alpha}+12M\partial_t\hspace{.5mm}r\Omega^2{\underline\alpha}, \label{eq:227intro1}\\
\frac{\Omega^2}{r^2}\Omega\slashed{\nabla}_4 \left(\frac{r^2}{\Omega^2}\Omega\slashed{\nabla}_4\right)^3{\underline\alpha}=2r^4\slashed{\mathcal{D}}^*_2\slashed{\mathcal{D}}^*_1\overline{\slashed{\mathcal{D}}}_1\slashed{\mathcal{D}}_2 r\Omega^2\alpha-12M\partial_t\hspace{.5mm}r\Omega^2\alpha.\label{eq:228intro1}
\end{align}

Note that one may utilise coordinates where the shift vector multiplies the differential in $\bm{u}$:
\begin{align}\label{DNGmetric other orientation}
    \bm{g}=-4\bm{\Omega}^2\bm{dudv}+\bm{\slashed{g}}_{AB}(\bm{d\theta}^A+\bm{b}^A\bm{du})(\bm{d\theta}^B+\bm{b}^B\bm{du}).
\end{align}
Linearising the Einstein equations using the form of the metric given in \bref{DNGmetric other orientation} produces a different set of equations than that obtained through \bref{DNGmetric}, and the two systems can be transformed into one another through a time reversal $t\longrightarrow-t$, or using the antipodal map $(\theta,\varphi)\longrightarrow(\pi-\theta,2\pi-\varphi)$.

\subsubsection{The results of Part I: a scattering theory for the Teukolsky equations}\label{section 1.4 summary of part I}
\indent The similarity of the Regge--Wheeler equation \bref{RWintro} to the scalar wave equation implies that the equation \bref{RWintro} obeys a conservation law of the energy associated to the isometry generated by $\partial_t$ of the Schwarzschild background. The $\partial_t$-energy conservation law was shown to lead to a scattering theory for the scalar wave equation satisfying the requirements \bref{QI}, \bref{QII}, \bref{QIII}, as shown in \cite{DRSR14} for the whole of the subextremal Kerr family. In \cite{Mas20}, we showed that this is indeed equally true for the Regge--Wheeler equation \bref{RWintro}, constructing such a scattering theory for \bref{RWintro} in the process. In contrast, the Teukolsky equations are not known to enjoy a similar conservation law, except to the extent that the energy conservation associated to \bref{RWintro} implies one for \bref{wave equation +}, \bref{wave equation -} via \bref{transport}.\\

The existence of a satisfactory scattering theory for \bref{RWintro} begs the question: is it possible, viewing the relations \bref{transport} as transport equations for $\alin$, $\ablin$, to pass from the scattering theory available to \bref{RWintro} to a scattering statement for the Teukolsky equations \bref{wave equation +}, \bref{wave equation -}? Our work in Part I \cite{Mas20} shows that this is indeed the case; we identify a collection of spaces of scattering states at $\mathscr{H}^\pm$, $\mathscr{I}^\pm$, for which the following statement applies: Denote by $\overline{\Sigma}$ the surface $\{t=0\}$ extended to include the bifurcation sphere. Then we have

\begin{namedtheorem}[Theorem I \textnormal{\cite{Mas20}}]\label{Theorem 1}
For the Teukolsky equations \bref{wave equation +}, \bref{wave equation -} of spins $\pm2$, evolution from smooth, compactly supported data on a Cauchy surface extends to unitary Hilbert space isomorphisms:
\begin{align}
    {}^{(+2)}\mathscr{F}^{+}:\mathcal{E}^{T,+2}_{\overline\Sigma}\longrightarrow \mathcal{E}^{T,+2}_{\mathscr{I}^+}\oplus\mathcal{E}^{T,+2}_{\overline{\mathscr{H}^+}},\qquad\qquad{}^{(-2)}\mathscr{F}^{+}:\mathcal{E}^{T,-2}_{\overline\Sigma}\longrightarrow \mathcal{E}^{T,-2}_{\mathscr{I}^+}\oplus\mathcal{E}^{T,-2}_{\overline{\mathscr{H}^+}},\\
    {}^{(+2)}\mathscr{F}^{-}:\mathcal{E}^{T,+2}_{\overline\Sigma}\longrightarrow \mathcal{E}^{T,+2}_{\mathscr{I}^-}\oplus\mathcal{E}^{T,+2}_{\overline{\mathscr{H}^-}},\qquad\qquad{}^{(-2)}\mathscr{F}^{-}:\mathcal{E}^{T,-2}_{\overline\Sigma}\longrightarrow \mathcal{E}^{T,-2}_{\mathscr{I}^-}\oplus\mathcal{E}^{T,-2}_{\overline{\mathscr{H}^-}}.
\end{align}
The spaces $\mathcal{E}^{T,\pm2}_{\overline\Sigma}$ are defined to be spaces of Cauchy data for \bref{wave equation +}, \bref{wave equation -}, respectively, such that the $\partial_t$-energy of the quantities $\Psi$, $\underline\Psi$ defined from $\alpha$, $\underline\alpha$ respectively via \bref{transport} is finite. The spaces of past/future scattering states  $\mathcal{E}^{T,\pm2}_{\mathscr{I}^\pm},\mathcal{E}^{T,\pm2}_{\mathscr{H}^\pm},$ are the Hilbert spaces obtained by completing suitable smooth, compactly supported data on $\mathscr{I}^\pm, \mathscr{H}^\pm$ under the corresponding norms in the following:

\begin{changemargin}{-1cm}{2cm}
\begin{center}
\setstretch{1.5}
\begin{tikzpicture}[scale=0.6,on grid]
\node (I)    at ( 0,0)   {};

\path 
   (I) +(90:4)  coordinate (Itop) coordinate[label=90:$i^+$]
       +(-90:4) coordinate (Ibot) coordinate[label=-90:$i^-$]
       +(180:4) coordinate (Ileft)
       +(0:4)   coordinate (Iright) coordinate[label=0:$i^0$]
       ;
\draw  (Ileft) --  node[align=center,yshift=15,xshift=15]{$\Big\|(\mathring{\slashed{\Delta}}-2)(\mathring{\slashed{\Delta}}-4)\left(2M\int^{\infty}_v d\bar{v}e^{\frac{1}{2M}({v}-\bar{v})}\Omega^2\alpha\right)\Big\|^2_{L^2(\overline{\mathscr{H}^+})}\qquad\qquad\qquad\qquad\qquad\qquad\qquad\qquad\qquad\qquad\qquad\qquad$\\$+\Big\|6M\partial_v\left(2M\int^{\infty}_v d\bar{v}e^{\frac{1}{2M}({v}-\bar{v})}\Omega^2\alpha\right)\Big\|_{L^2(\overline{\mathscr{H}^+})}^2\qquad\qquad\qquad\qquad\qquad\qquad\qquad\qquad\qquad\qquad\qquad\;\qquad$} node[rotate=45,below]{$\overline{\mathscr{H}^+}$} (Itop) ;
\draw  (Ileft) -- node[yshift=-15,xshift=15]{$\Big\|2M\left(-2(2M\partial_u)+3(2M\partial_u)^2-(2M\partial_u)^3\right)2M\Omega^{-2}\alpha\Big\|^2_{L^2(\overline{\mathscr{H}^-})}\qquad\qquad\qquad\qquad\qquad\qquad\qquad\qquad\qquad\qquad\qquad\qquad\qquad$} node[rotate=-45,above]{$\overline{\mathscr{H}^-}$} (Ibot) ;
\draw[dash dot dot] (Ibot) -- node[align=center][yshift=-10,xshift=-15]{$\qquad\qquad\qquad\qquad\qquad\qquad\qquad\qquad\qquad\Big\|6M\upalpha_{\mathscr{I}^-}\Big\|^2_{L^2(\mathscr{I}^-)}$\\[1mm] $\qquad\qquad\qquad\qquad\qquad\qquad\qquad\qquad\qquad+\left\|(\mathring{\slashed{\Delta}}-2)(\mathring{\slashed{\Delta}}-4)\left(\int^{v}_{-\infty}\upalpha_{\mathscr{I}^-} d\bar{v}\right)\right\|^2_{L^2(\mathscr{I}^-)}$} node[rotate=45,above]{$\mathscr{I}^-$}(Iright) ;
\draw[dash dot dot] (Iright) -- node[yshift=10,xshift=-10]{$\qquad\qquad\;\;\;\;\;\;\;\;\;\;\;\;\;\;\;\;\;\;\;\;\qquad\left\|(\partial_u)^3\upalpha_{\mathscr{I}^+}\right\|^2_{L^2(\mathscr{I}^+)}$} node[rotate=-45,below]{$\mathscr{I}^+$}(Itop) ;

\filldraw[white] (Itop) circle (3pt);
\draw[black] (Itop) circle (3pt);

\filldraw[white] (Ibot) circle (3pt);
\draw[black] (Ibot) circle (3pt);
\draw[black] (Ileft) circle (3pt);
\filldraw[black] (Ileft) circle (3pt);
\filldraw[white] (Iright) circle (3pt);
\draw[black] (Iright) circle (3pt);
\end{tikzpicture}

\end{center}

\end{changemargin}
\begin{changemargin}{-1.4cm}{2cm}

\begin{center}
\begin{tikzpicture}[scale=0.6]
\node (I)   at ( 0,0)   {};

\path 
   (I) +(90:4)  coordinate (Itop) coordinate[label=90:$i^+$]
       +(-90:4) coordinate (Ibot) coordinate[label=-90:$i^-$]
       +(180:4) coordinate (Ileft)
       +(0:4)   coordinate (Iright) coordinate[label=0:$i^0$]
       ;
\draw  (Ileft) --  node[yshift=20,xshift=25]{$\Big\|2M\left(2(2M\partial_v)+3(2M\partial_v)^2+(2M\partial_v)^3\right)2M\Omega^{-2}\underline\alpha\Big\|^2_{L^2(\overline{\mathscr{H}^+})}\qquad\qquad\qquad\qquad\qquad\qquad\qquad\qquad\qquad\qquad\qquad\qquad\qquad\qquad$} node[rotate=45,below]{$\overline{\mathscr{H}^+}$} (Itop) ;
\draw  (Ileft) -- node[align=center][yshift=-10,xshift=10]{$\Big\|6M\partial_u\left(2M\int^{u}_{-\infty}d\bar{u}e^{\frac{1}{2M}(u-\bar{u})}\Omega^2\underline\alpha\right)\Big\|^2_{L^2(\overline{\mathscr{H}^-})}\qquad\qquad\qquad\qquad\qquad\qquad\qquad\qquad\qquad\qquad\qquad$\\[1mm] $+\left\|(\mathring{\slashed{\Delta}}-2)(\mathring{\slashed{\Delta}}-4)\left(2M\int^{u}_{-\infty}d\bar{u}e^{\frac{1}{2M}(u-\bar{u})}\Omega^2\underline\alpha\right)\right\|^2_{L^2(\overline{\mathscr{H}^-})}\qquad\qquad\qquad\qquad\qquad\qquad\qquad\qquad\qquad\qquad\qquad$} node[rotate=-45,above]{$\overline{\mathscr{H}^-}$} (Ibot) ;
\draw[dash dot dot] (Ibot) -- node[yshift=-12,xshift=-12]{$\qquad\qquad\;\;\;\;\;\;\;\;\;\;\;\;\;\;\qquad\qquad\left\|(\partial_v)^3\underline\upalpha_{\mathscr{I}^-}\right\|_{L^2(\mathscr{I}^-)}^2$} node[rotate=45,above]{$\mathscr{I}^-$}(Iright) ;
\draw[dash dot dot] (Iright) -- node[align=center][yshift=10,xshift=-20]{ $\qquad\qquad\qquad\qquad\qquad\qquad\qquad\qquad\qquad\qquad\left\|(\mathring{\slashed{\Delta}}-2)(\mathring{\slashed{\Delta}}-4)\left(\int^{u}_{-\infty}\underline\upalpha_{\mathscr{I}^+} d\bar{u}\right)\right\|^2_{L^2(\mathscr{I}^+)}$ \\$\qquad\qquad\qquad\qquad\qquad\qquad\qquad\qquad\qquad+\Big\|6M\underline\upalpha_{\mathscr{I}^+}\Big\|^2_{L^2(\mathscr{I}^+)}$} node[rotate=-45,below]{$\mathscr{I}^+$}(Itop) ;

\filldraw[white] (Itop) circle (3pt);
\draw[black] (Itop) circle (3pt);

\filldraw[white] (Ibot) circle (3pt);
\draw[black] (Ibot) circle (3pt);

\filldraw[white] (Iright) circle (3pt);
\draw[black] (Iright) circle (3pt);
\filldraw[black] (Ileft) circle (3pt);
\draw[black] (Iright) circle (3pt);
\end{tikzpicture}

\end{center}
\end{changemargin}
The maps ${}^{(\pm2)}\mathscr{F}^{\pm}$ lead to the unitary Hilbert space isomorphisms
\begin{align}
\begin{split}
    &\mathscr{S}^{+2}: \mathcal{E}^{T,+2}_{\mathscr{I}^+}\oplus\mathcal{E}^{T,+2}_{\overline{\mathscr{H}^+}}\longrightarrow \mathcal{E}^{T,+2}_{\mathscr{I}^-}\oplus\mathcal{E}^{T,+2}_{\overline{\mathscr{H}^-}},\\
    &\mathscr{S}^{-2}: \mathcal{E}^{T,-2}_{\mathscr{I}^+}\oplus\mathcal{E}^{T,-2}_{\overline{\mathscr{H}^+}}\longrightarrow \mathcal{E}^{T,-2}_{\mathscr{I}^-}\oplus\mathcal{E}^{T,-2}_{\overline{\mathscr{H}^-}}.
\end{split}
\end{align}
\end{namedtheorem}

Importantly, one may incorporate the Teukolsky--Starobinsky identities \bref{eq:227intro1}, \bref{eq:228intro1} into the statement of Theorem I to obtain the following corollary: 

\begin{namedcorollary}[Corollary I \textnormal{\cite{Mas20}}]\label{Corollary 1}
Given a smooth, compactly supported $\upalpha_{\mathscr{I}^-}$ on $\mathscr{I}^-$ such that $\int_{-\infty}^\infty d\bar{v} \; \upalpha_{\mathscr{I}^-}=0$, and an $\underline\upalpha_{\mathscr{H}^-}$ such that $U^{-2}\underline\upalpha_{\mathscr{H}^-}$ is smooth, compactly supported on $\overline{\mathscr{H}^-}$, there exists a unique smooth pair $(\alpha, \underline\alpha)$ on the exterior region of Schwarzschild, satisfying equations \bref{wave equation +}, \bref{wave equation -} respectively, where $\alpha$ realises $\upalpha_{\mathscr{H}^+}$ as its radiation field on $\overline{\mathscr{H}^+}$, $\underline\alpha$ realises $\underline\upalpha_{\mathscr{I}^+}$ as its radiation field on ${\mathscr{I}^+}$, such that constraints \bref{eq:227intro1} and \bref{eq:228intro1} are satisfied. Moreover, $\alpha, \underline\alpha$ induce smooth radiation fields $\underline{\upalpha}_{\mathscr{I}^+}, \upalpha_{\mathscr{H}^+}$ in $\mathcal{E}^{T,-2}_{{\mathscr{I}^+}}, \mathcal{E}^{T,+2}_{\overline{\mathscr{H}^+}}$ respectively. This extends to a unitary Hilbert space isomorphism:
\begin{align}
    \mathscr{S}^{-2,+2}:\mathcal{E}^{T,+2}_{\mathscr{I}^-}\oplus\mathcal{E}^{T,-2}_{\overline{\mathscr{H}^-}}\longrightarrow \mathcal{E}^{T,-2}_{\mathscr{I}^+}\oplus\mathcal{E}^{T,+2}_{\overline{\mathscr{H}^+}}.
\end{align}
\end{namedcorollary}
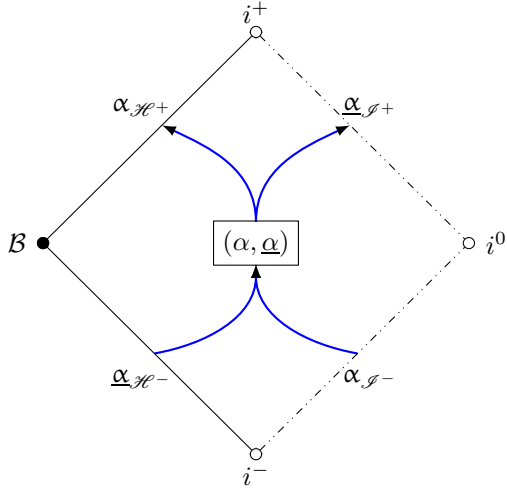
\begin{center}
\begin{tikzpicture}[->,scale=0.7, arrow/.style={
            color=black,
            draw=blue,thick,
            -latex,
                font=\fontsize{8}{8}\selectfont},
        ]
\node (I)    at ( 0,0)   {};

\path 
   (I) +(90:4)  coordinate (Itop) coordinate [label={$i^+$}]
       +(180:4) coordinate (Ileft) coordinate [label=180:{$\mathcal{B}\;$}]
       +(0:4)   coordinate (Iright) coordinate [label=0:{$\;i^0$}]
       +(270:4) coordinate (Ibot) coordinate [label=-90:{$i^-$}]
       ;
\draw[arrow] ($(Itop)+(-90:3.6cm)$) to [in=-25,out=90] ($(Itop)+(-135:2.5cm)$);

\draw[arrow] ($(Itop)+(-90:3.6cm)$) to [in=205,out=90]($(Itop)+(-45:2.5cm)$);

\draw[arrow] ($(Ibot)+(135:2.7cm)$) to [out=10,in=-90] ($(Ibot)+(90:3.6cm)$);

\draw[arrow] ($(Ibot)+(45:2.7cm)$) to [out=170,in=-90] ($(Ibot)+(90:3.6cm)$);


\draw  (Ileft) -- node[yshift=4mm,xshift=-1mm]{$\upalpha_{\mathscr{H}^+}$} (Itop) ;
\draw[dash dot dot] (Iright) --  node[yshift=4mm,xshift=1.mm]{$\underline\upalpha_{\mathscr{I}^+}$}(Itop) ;
\node[draw] at ($(Itop)+(-90:4cm)$) {$(\alpha,\underline\alpha)$};
\draw[dash dot dot] (Iright) --  node[yshift=-4mm,xshift=1.mm]{$\upalpha_{\mathscr{I}^-}$}(Ibot) ;
\draw  (Ileft) -- node[yshift=-4mm,xshift=-1mm]{$\underline\upalpha_{\mathscr{H}^-}$} (Ibot) ;


\filldraw[white] (Itop) circle (3pt);
\draw[black] (Itop) circle (3pt);

\filldraw[white] (Ibot) circle (3pt);
\draw[black] (Ibot) circle (3pt);

\filldraw[white] (Iright) circle (3pt);
\draw[black] (Iright) circle (3pt);
\filldraw[black] (Ileft) circle (3pt);
\draw[black] (Ileft) circle (3pt);
\end{tikzpicture}
\end{center}

\subsubsection{The gauge ambiguity and radiation fields}

Following on from Theorem I, our goal is to find functional spaces of data at $\mathscr{I}^\pm$, $\overline{\mathscr{H}^\pm}$, 
and at $\overline{\Sigma}=\{t=0\}$ for which an analogous statement to Theorem I applies for the case of the linearised Einstein equations. Considering the spaces of scattering data given by Corollary I and the second equation in \bref{example2}, we can already consider as a candidate for the radiation field at $\mathscr{I}^+$ the quantity
\begin{align}\label{radiation field at scri+ intro}
    \xblins_{\mathscr{I}^+}(u,\theta^A):=\lim_{v\longrightarrow\infty}r\xblin(u,v,\theta^A),
\end{align}
for solutions arising from suitably decaying initial data. This is in fact nothing but the double null gauge expression for the \textbf{Bondi news function} defined in \cite{BMV1962}. Similarly, one may consider defining the radiation field $\xlins_{\mathscr{H}^+}$ at $\overline{\mathscr{H}^+}$ in terms of $\xlin$, and vice versa for $\mathscr{I}^-$, $\overline{\mathscr{H}^-}$. At this point, we must confront the problem of the gauge ambiguity inherited by the linearised system, particularly as it relates to the asymptotics of solution towards null infinity. Indeed, the candidate radiation fields defined above, though themselves gauge invariant, arise from quantities which are gauge-dependent in the bulk and, more importantly, it is not possible to source all of the linearised quantities via the full linearised system knowing only $\xblins_{\mathscr{I}^+}$ at $\mathscr{I}^+$ and $\xlins_{\mathscr{H}^+}$ at $\overline{\mathscr{H}^+}$; note that in order to solve the characteristic value problem for \bref{EVE} when linearised in the form \bref{DNGmetric}, the components for which data is required include the linearised lapse $\Olino$ and the traceless part of the linearised metric on both null cones, as well as the linearised shift $\bmlin$ on the outgoing null cone \cite{DHR16}. One expects that the scattering problem will require data of a similar number of degrees of freedom. However, in view of how radiation energy is defined via the norms $\|\;\|_{\mathcal{E}^{T,+2}_{\mathscr{I}^+}}$, $\|\;\|_{\mathcal{E}^{T,+2}_{\overline{\mathscr{H}^+}}}$ we see that
$\|\;\|_{\mathcal{E}^{T,+2}_{\mathscr{I}^+}}$, $\|\;\|_{\mathcal{E}^{T,+2}_{\overline{\mathscr{H}^+}}}$ do not couple to the linearised metric components $\Olin$, $\bmlin$, thus we may regard them as ``pure gauge" degrees of freedom in the sense that they do not carry any radiation energy. We aim to solve the scattering problem, say from data on $\mathscr{I}^+$, $\overline{\mathscr{H}^+}$, given only radiation fields that appear in the expression for radiation energy fluxes given by Theorem I, so we will prescribe gauge conditions that ensures a unique, full set of scattering data can be constructed out of any given set of radiation fields of finite energy, i.e.~of finite norm in the spaces $\mathcal{E}^{T,+2}_{\mathscr{I}^+}$, $\mathcal{E}^{T,+2}_{\overline{\mathscr{H}^+}}$ for future scattering data. Similar considerations apply to past scattering data.\\

\subsubsection{The initial value problem for the linearised Einstein equations in a double null gauge}

Note that once the system \bref{EVE} is linearised in a double null gauge, the infinitesimal gauge ambiguity inherited by the system preserves the double null structure of the system. Thus solving the initial value problem with a Cauchy hypersurface for the system \bref{EVE} when linearised in a double null gauge is necessary if we are to attempt to study the asymptotics of solutions near $i^0$. In contrast to the initial characteristic value problem, the initial value problem for \bref{EVE} for Cauchy data on a spacelike surface has not received much attention when the Einstein equations are posed in a double null gauge. In the present work we solve the spacelike initial value problem for the linearised Einstein equation in a double null gauge in \Cref{Section 5: well-posedness for IVP}. In particular, we identify those components of the linearised metric and their time derivatives which can be prescribed as free seed data on $\overline{\Sigma}$, and we show how the constraint equations can be solved to construct out of the given seed data a full initial data set consisting of data for all of the metric components and their normal derivatives. On $\overline{\Sigma}$, a seed data set consists of linearised Kerr parameters $(\mathfrak{m},\mathfrak{a})\in \mathbb{R}\times \mathbb{R}^3$ and the following components:
\begin{align}\label{seed data on Sigmabar}
    \glinh,\; V^{-1}\bmlin_{\ell\geq2},\;  \curlr \slashednabla_TV^{-1}\bmlin_{\ell\geq2}, \;   \tr\glin_{\ell\geq2},\;   \slashednabla_T \tr\glin_{\ell\geq2},\;   \Olin_{\ell\geq2},
\end{align}
where $T$ is the vector field $\partial_U+\partial_V$ defined via Kruskal coordinates.\\

Having resolved the initial value problem for the given linearised system, we may define a space of initial states on $\overline{\Sigma}$ via the spaces  $\mathcal{E}^{T,\pm2}_{\overline{\Sigma}}$. (see already \Cref{remark on choice of energy for full system}). The problem with this prescription is that $\|\;\|_{\mathcal{E}^{T,\pm2}_{\overline{\Sigma}}}$ do not distinguish between data corresponding to any two solutions differing by a gauge transformation, and thus they cannot define a norm on initial data in the presence of pure gauge solutions. In order for a smooth element of the space of initial states to correspond to a unique initial seed data set, we must define a space of initial states on $\overline{\Sigma}$ specifying a gauge-fixing condition on seed data.\\

The role of gauge conditions on initial data on the Cauchy surface goes beyond making the space of initial states formally more tractable. Already in the linear theory one encounters the fact that solutions may not even be bounded unless specific gauge conditions are chosen, as was demonstrated in the linear stability result for Schwarzschild in \cite{DHR16}. In line with the prescription given in \cite{DHR16}, we choose our initial data to satisfy
\begin{align}\label{bifurcation gauge conditions intro}
\divr(\elin-\eblin)\Big|_{\mathcal{B}}=0,\qquad\qquad \left[\rlin-\rlin_{\ell=0}+{\slashed{\Delta}}\Olin\right]\Big|_{\mathcal{B}}=0.
\end{align}
We complement the gauge conditions \bref{bifurcation gauge conditions intro} with additional conditions in order to fix the gauge freedom completely. The conditions we use are
\begin{align}\label{Sigmabar + gauge intro}
    \left[(2M\partial_r)^2+6M\partial_r+2\right]\divr\,\divr\, V\Omega^{-1}\xlin|_{\overline{\Sigma}}=0,\qquad\qquad \divr\,\divr\, V^{-1}\Omega\xblin|_{\overline{\Sigma}}=0,\qquad\qquad \glinh|_{\overline{\Sigma}}=0.
\end{align}

Initial data satisfying the conditions \bref{bifurcation gauge conditions intro} and \bref{Sigmabar + gauge intro} above will be said to be in the {$\bm{\overline{\Sigma}}_{\bm{+}}$}\textbf{-gauge} and we denote by $\mathcal{E}^T_{\overline{\Sigma},+}$ the space of initial states given by the completion of smooth, compactly supported seed data in the $\overline{\Sigma}_+$ gauge. The ``$+$" here refers to future scattering, as the conditions \bref{Sigmabar + gauge intro} are chosen to enable decay of solutions evolved towards $\overline{\mathscr{H}^+}$, $\mathscr{I}^+$. Analogous conditions hold for evolution towards $\overline{\mathscr{H}^-}$, $\mathscr{I}^-$, retaining \bref{bifurcation gauge conditions intro},
exchanging $\xlin$ and $\xblin$ in \bref{Sigmabar + gauge intro} and taking into account the appropriate weights in $V$, giving rise to the \textbf{{$\bm{\overline{\Sigma}}_{\bm{-}}$}-gauge}. Similarly to $\mathcal{E}^T_{\overline{\Sigma},+}$, one may define a space of initial states $\mathcal{E}^T_{\overline{\Sigma},-}$ via the $\overline{\Sigma}_-$ gauge conditions. The Hilbert spaces $\mathcal{E}^T_{\overline{\Sigma},-}$, $\mathcal{E}^T_{\overline{\Sigma},+}$ are isomorphic and, moreover, the transformation from one gauge to the other can be bounded in terms of the starting data.

\subsubsection{Asymptotic regularity and Bondi-normalised double null gauges}

Having established the parameters of constructing initial data for the linearised Einstein equations in a double null gauge in \Cref{Section 5: well-posedness for IVP}, the next step is to study the question of \textbf{existence}, i.e.~question \bref{QI} in our discussion of the scattering problem at the beginning of this introduction. As mentioned earlier, the question of existence has so far remained a point of confusion, not least because of the gauge ambiguity discussed above, but also because before the publication of \cite{Ch-K}, there was no clear dynamical picture supporting any particular hypothesis regarding the asymptotics of gravitational radiation. In particular, the work \cite{ChristodoulouMarcelGrossman} suggested that physically relevant spacetimes do not admit peeling estimates or conformal smoothness. A recent series of papers has given mathematical evidence towards this view, see \cite{KehrbergerGajic}, \cite{KehrbergerII}, \cite{KehrbergerIII}, \cite{KehrbergerI}. (See also the upcoming \cite{KehrbergerM} for the case of the linearised system considered here).\\ 

In \cite{Ch-K} (and subsequently also in \cite{DHRT21}), it was crucial to work in a gauge which asymptotically satisfies the prescription given by Bondi {et~al.} in \cite{BMV1962} in order to obtain boundedness and decay for perturbations evolved under \bref{EVE}. In \cite{Ch-K}, the laws of gravitational radiation first devised by Bondi were shown to be satisfied by the asymptotics of generic perturbations to the Minkowski space. We follow suit in this paper, and in evolving solutions towards $\mathscr{I}^+$ we pass to a \textbf{future Bondi-normalised double null gauge}, a gauge that satisfies
\begin{align}
    &\lim_{v\longrightarrow\infty}\Olino(u,v,\theta^A)=-\frac{1}{2}\mathfrak{m},\label{future Bondi gauge intro 1}\\ 
    &\lim_{v\longrightarrow\infty} r^2\Klin_{\ell\geq2}(u,v,\theta^A)=0.\label{future Bondi gauge intro 2}
\end{align}
where $\mathfrak{m}$ is the linearised ADM mass. For the sake of prescribing a gauge fixing scheme that fully fixes the gauge freedom, we will also demand
\begin{align}
    &\lim_{v\longrightarrow\infty}\glinh(u,v,\theta^A)=0,\label{fixing residual freedom Bondi intro}\\
    &\lim_{v\longrightarrow\infty}\tr\glin(u,v,\theta^A)=0.\label{fixing residual freedom Bondi trace intro}
\end{align}
We will refer to the combination of conditions \eqref{future Bondi gauge intro 1}, \eqref{future Bondi gauge intro 2}, \eqref{fixing residual freedom Bondi intro}, \eqref{fixing residual freedom Bondi trace intro} as the $\bm{\mathscr{I}^+}$ \textbf{gauge conditions}.\\ 

In studying forwards scattering, we work with initial data that have polynomial decay towards $i^0$,
\begin{align}\label{rate of decay intro}
    \glin\sim O_{\infty}(r^{-n}),
\end{align}
where $f\sim O_k(r^{-n})$ here means $\partial_x^i f\sim O(r^{-n-i})$ for $i=0\dots k$. We find that the laws of gravitational radiation encoded in Chapter 17 of \cite{Ch-K} are attainable if $n>1$ in \bref{rate of decay intro}. For $n=1$, the conditions \bref{future Bondi gauge intro 1}, \bref{future Bondi gauge intro 2} can be attained, but not all the laws of gravitational radiation are satisfied, e.g. the convergence of $r^2\xlin$ is not assured, and neither is the boundedness of the limit towards the past end of $\mathscr{I}^+$. For the full set of gravitational radiation laws to be satisfied, additional conditions are needed beyond the boundedness of $r^i \partial^i\glin$, 
However, one may still claim the existence of $\xblins_{\mathscr{I}^+}$ defined in \bref{radiation field at scri+ intro} if $n\geq\frac{1}{2}$ in \bref{rate of decay intro}.\\

Note that though we fix the value of the linearised lapse at $\mathscr{I}^+$, we do not require it to vanish. This is because when the equations \bref{EVE} are linearised in the gauge \bref{DNGmetric} it can be shown that any spherically symmetric solution to the linearised equations which is bounded everywhere at $\overline{\mathscr{H}^+}$ and has vanishing linearised lapse at $\mathscr{I}^+$ must be pure gauge. See already \Cref{Bondi normalisation ell=0 1 is not regular at H+}. 

\subsubsection*{The Bondi--Metzner--Sachs group}\label{Intro: Statement of the Bondi gauge}

In our study of the scattering problem towards $\overline{\mathscr{H}^+}$, $\mathscr{I}^+$ we show in \Cref{Section 7 BMS} that the conditions \bref{future Bondi gauge intro 1}, \bref{future Bondi gauge intro 2} restrict the residual gauge freedom so that the coordinate $u$ can only transform according to 
\begin{align}\label{u part of BMS intro}
    u\longrightarrow u+\epsilon\left[L_{\ell=1}(\theta^A)+T_u(\theta^A)\right],
\end{align}
where $L_{\ell=1}, T_u$ are smooth scalar functions on the unit sphere such that $L_{\ell=1}$ is in the span of the $\ell=1$ spherical harmonics. The null coordinate $v$ is restricted to transformations of the form
\begin{align}\label{v part of BMS intro}
    v\longrightarrow v+\epsilon f(v,\theta^A)    \text{ with }\partial_vf(v,\theta^A)\longrightarrow -L_{\ell=1} \text{ as } v\longrightarrow\infty.
\end{align}
The conditions \bref{future Bondi gauge intro 1}, \bref{future Bondi gauge intro 2} above do not restrict the freedom to perturb a double null gauge in the angular direction. Thus transformations to the coordinates $\theta^A$ of the form 
\begin{align}\label{diffeos of sphere at infinity intro}
    \theta^A\longrightarrow \theta^A+\epsilon\,j^A(v,\theta^A)
\end{align}
are allowed for $j$ having a smooth limit as $v\longrightarrow\infty$. The action of the above transformation induces a diffeomorphism on the $S^2$ cross sections of $\mathscr{I}^+$ (sometimes referred to as the ``sphere at infinity"). We may restrict this gauge freedom by hand so that only infinitesimal isometries in the angular directions are allowed:
\begin{align}\label{isometries of sphere at infinity intro}
    j^A\longrightarrow \slashed{\epsilon}_A{}^B\slashednabla_B q,\qquad\qquad q=\sum_{m=-1,0,1} c_mY^{\ell=1}_m.
\end{align}
where $\slashed{\epsilon}$ is the volume form on $S^2_{u,v}$.\\

The transformations \bref{u part of BMS intro} and \bref{isometries of sphere at infinity intro} constitute what is known as the \textbf{Bondi--Metzner--Sachs} group at future null infinity, which we will denote by BMS\textsuperscript{$+$}. Asymptotic Lorentz boosts are generated by $L_{\ell=1}$. Rotations are generated by $q$ in \bref{isometries of sphere at infinity intro}. Translations are induced by the $\ell=0, 1$ modes of $T_u$ in \bref{u part of BMS intro}. The transformations generated by the $\ell\geq2$ modes of $G_u$ are known as \textit{supertranslations}. Some authors consider an extension of BMS\textsuperscript{$+$} by allowing the full range of transformations given by \bref{diffeos of sphere at infinity intro}. See for instance \cite{Campiglia_2015} and \cite{Flanagan_2020}.\\

\indent The collection of residual gauge transformations described above by \bref{u part of BMS intro}, \bref{v part of BMS intro}, and \bref{diffeos of sphere at infinity intro} properly contains BMS\textsuperscript{$+$}. Note that translations in the $v$-directions can have a nontrivial action on the radiation fields at $\mathscr{I}^+$, though the quantities affected by the action induced by \bref{v part of BMS intro} do not carry any gravitational energy, and should thus be regarded as ``pure gauge" degrees of freedom.\\

\indent An analogous treatment applies to the asymptotics near  $\mathscr{I}^-$, giving rise to a BMS\textsuperscript{$-$} group defined via \bref{u part of BMS intro}, \bref{v part of BMS intro} exchanging $u$ with $v$ and $\fbar(u,\theta^A)$ with $f(v,\theta^A)$ and restricting $q_1,q_2$ to be constant in $v$. The scattering problem from $\overline{\Sigma}$ towards $\mathscr{I}^+$, $\overline{\mathscr{H}^+}$ is studied in \Cref{Section 8.7: Scattering from Sigmabar to H- and I-}. Bondi-normalised double null gauges are studied in \Cref{Section 7 BMS}. 

\subsubsection{Gauge conditions at the event horizon and backwards scattering}

It is possible to eliminate the considerable residual gauge freedom left over by the $\mathscr{I}^+$ gauge conditions by imposing the following conditions at $\overline{\mathscr{H}^+}$: denoting by $\mathfrak{m}\in\mathbb{R}$ the linearised ADM mass and by $\mathfrak{a}\in\mathbb{R}^3$ the linearised ADM angular momentum, we demand
\begin{equation}\label{future horizon gauge conditions intro}
\begin{alignedat}{2}
\begin{split}
    &\Olin|_{\overline{\mathscr{H}^+}}=-\frac{1}{2}\mathfrak{m},\qquad && \bmlin|_{\overline{\mathscr{H}^+}}=2\slashed{\epsilon}^{AB}\sum_{m\in\{-1,0,1\}}\mathfrak{a}_m\partial_A Y_m^{\ell=1}, \\ & \divo\divo V^{-1}\Omega\xblin|_{\mathcal{B}}=0,\qquad && V^{-1}\otxb_{\ell=0,1}|_{\mathcal{B}}=0.
\end{split}
\end{alignedat}
\end{equation}
in addition to the conditions \bref{bifurcation gauge conditions intro}. When the conditions \bref{future horizon gauge conditions intro} and \bref{bifurcation gauge conditions intro} are satisfied we say that the solution is \textbf{future horizon-normalised} or that it satisfies the  \textbf{$\bm{\overline{\mathscr{H}^+}}$-gauge} conditions. As discussed earlier, in order to construct solutions out of scattering data at $\mathscr{I}^+$, $\overline{\mathscr{H}^+}$ it is necessary to prescribe gauge conditions on data given on $\mathscr{I}^+$, $\overline{\mathscr{H}^+}$ in order to solve for the full scattering data set for the full system given only the radiation fields that carry the energy fluxes of Theorem I, i.e.~$\xblins_{\mathscr{I}^+}$, $\xlins_{\mathscr{H}^+}$. The conditions \bref{bifurcation gauge conditions}, the vanishing of $\Olin$ at $\overline{\mathscr{H}^+}$ and $\divr\,\divr\, V\Omega^{-1}\xblin$ at the bifurcation sphere $\mathcal{B}$ are a prerequisite for the linearised system to be bounded at late times, as shown in \cite{DHR16}.\\

In the backwards scattering problem, studied here in \Cref{Section 6 Backwards scattering}, we take $\xblins_{\mathscr{I}^+}$, $\xlins_{\mathscr{H}^+}$ to be compactly supported. We further use the conditions \bref{future horizon gauge conditions intro} to solve for the remaining components at $\overline{\mathscr{H}^+}$, but we take $\divr\,\divr\, V\Omega^{-1}\xblin$ to vanish where the future support of $\xlins_{\mathscr{H}^+}$ ends rather than at $\mathcal{B}$. This is to ensure that prior to the support in time of both $\xblins_{\mathscr{I}^+}$, $\xlins_{\mathscr{H}^+}$, spacetime is exactly Schwarzschildean and is represented in the Eddington--Finkelstein coordinates, i.e.~free of infinitesimal gauge transformations there. One may then perform a gauge transformation so that \bref{future horizon gauge conditions intro} are satisfied. The horizon gauge conditions are defined via \bref{future horizon gauge conditions intro} because \bref{future horizon gauge conditions intro} do not refer to the asymptotics of fields at $\overline{\mathscr{H}^+}$, thus having $\divr\,\divr\,V\Omega^{-1}\xblin$ vanish at $\mathcal{B}$ is more suitable for the forwards scattering problem.\\

\indent When evolving to $\mathscr{I}^-\cup\overline{\mathscr{H}^-}$, analogous conditions can be imposed at $\overline{\mathscr{H}^-}$: the \textbf{$\bm{\mathscr{H}^-}$-gauge} is defined by \bref{bifurcation gauge conditions} and
\begin{align}\label{past horizon gauge conditions intro}
    \Olin|_{\overline{\mathscr{H}^-}}=-\frac{1}{2}\mathfrak{m},\qquad \divr\,\divr \,U^{-1}\Omega\xlin|_{\mathcal{B}}=0,\qquad U^{-1}\otx_{\ell=0,1}|_{\mathcal{B}}=0.
\end{align}
Horizon-normalised double null gauges are studied in \Cref{Section 7: horizon normalised gauges}.\\


\subsubsection{A summary of the argument leading to Theorem II}\label{summary of Thm II argument intro}

With the $\mathscr{I}^\pm$ and $\overline{\mathscr{H}^\pm}$ gauge conditions we are now able to fully fix the gauge on solutions evolved from scattering data. Thus far we have devised four gauge-fixing schemes: the \textbf{future scattering gauge}, consisting of the $\overline{\mathscr{H}^+}$, $\mathscr{I}^+$ gauge conditions, the \textbf{past scattering gauge} consisting of $\overline{\mathscr{H}^-}$, $\mathscr{I}^-$ gauge conditions, and the $\bm{\overline{\Sigma}}_\pm$-gauges. We can now summarise our construction of the scattering theory constructed in this paper: in evolving from $\overline{\Sigma}$ towards $\mathscr{I}^+$,  $\overline{\mathscr{H}^+}$ we start from smooth, compactly supported seed data on $\overline{\Sigma}$, construct a full initial data set via the constraint equations and the $\overline{\Sigma}_+$ gauge conditions and find that the resulting data set decays exponentially towards $i^0$. We evolve forwards to obtain a solution $\mathfrak{S}$ and construct the radiation fields at $\mathscr{I}^+$, $\overline{\mathscr{H}^+}$, keeping track of the minimum polynomial decay rates required from initial data in order for the radiation fields to exist. We then add pure gauge solutions $\mathfrak{G}_{\mathscr{I}^+}$, $\underline{\mathfrak{G}}_{\overline{\mathscr{H}^+}}$ so that $\mathfrak{S}+\mathfrak{G}_{\mathscr{I}^+}+\underline{\mathfrak{G}}_{\overline{\mathscr{H}^+}}$ satisfies both the $\mathscr{I}^+$ and $\overline{\mathscr{H}^+}$ gauge conditions, and we show that the solutions $\mathfrak{G}_{\mathscr{I}^+}$, $\underline{\mathfrak{G}}_{\overline{\mathscr{H}^+}}$ can be bounded in terms of the data giving rise to $\mathfrak{S}$. This shows that forwards evolution under the linearised equations defines an injective, map $\mathscr{F}^+$ from  $\mathcal{E}^T_{\overline{\Sigma},+}$ to  $\mathcal{E}^T_{\overline{\mathscr{H}^+}}\oplus \mathcal{E}^T_{\mathscr{I}^+}$. We construct the inverse map $\mathscr{B}^-$ starting from data in the dense subset of $\mathcal{E}^T_{\overline{\mathscr{H}^+}}\oplus \mathcal{E}^T_{\mathscr{I}^+}$ consisting of smooth, compactly supported data at $\overline{\mathscr{H}^+}$, $\mathscr{I}^+$, we use the gauge conditions at $\mathscr{I}^+$, $\overline{\mathscr{H}^+}$ to construct scattering data for all the components of the linearised system, we use Theorem I to construct solutions to the Teukolsky equations and construct a solution $\mathfrak{S}$ to the full system treating it as a transport system which is sourced by the Teukolsky variables $\alin$, $\ablin$ and the scattering data constructed via the gauge conditions at $\mathscr{I}^+$, $\overline{\mathscr{H}^+}$. We find a pure gauge solution $\mathfrak{G}_{\overline{\Sigma}_+}$ such that the data induced by $\mathfrak{S}+\mathfrak{G}_{\overline{\Sigma}_+}$ satisfies the $\overline{\Sigma}_+$ gauge, and we show that the induced data decay like $O_\infty(r^{-1})$. The unitarity of the maps $\mathscr{B}^-$ implies that the extensions of $\mathscr{F}^+$, $\mathscr{B}^-$ invert one another on the whole of the spaces $\mathcal{E}^T_{\overline{\Sigma},+}$,  $\mathcal{E}^T_{\overline{\mathscr{H}^+}}\oplus \mathcal{E}^T_{\mathscr{I}^+}$ defining Hilbert space isomorphisms. We do the same evolving between $\overline{\Sigma}$ and $\mathscr{I}^-$,  $\overline{\mathscr{H}^-}$ (now using the $\overline{\Sigma}_-$ gauge) to obtain Hilbert space isomorphisms $\mathscr{F}^-:\mathcal{E}^T_{\overline{\Sigma},-}\longrightarrow\mathcal{E}^T_{\mathscr{I}^-}\oplus\mathcal{E}^T_{\overline{\mathscr{H}^-}}$, $\mathscr{B}^+:\mathcal{E}^T_{\mathscr{I}^-}\oplus\mathcal{E}^T_{\overline{\mathscr{H}^-}}\longrightarrow\mathcal{E}^T_{\overline{\Sigma}_-}$. This leads to the Hilbert space isomorphism $\mathscr{S}=\mathscr{B}^+\circ\mathscr{F}^+:\mathcal{E}^T_{\mathscr{I}^-}\oplus\mathcal{E}^T_{\overline{\mathscr{H}^-}}\longrightarrow \mathcal{E}^T_{\mathscr{I}^+}\oplus\mathcal{E}^T_{\overline{\mathscr{H}^+}}$ with inverse $\mathscr{F}^-\circ\mathscr{B}^-$. The strategy outlined here leads to \textbf{Theorem II} which we state in \Cref{Section 1.6 statement of results intro} below (and again in more detail in \ref{forward scattering full system thm} of \Cref{Theorem IIA}, \ref{backwards scattering full system thm} of \Cref{Theorem IIB}, and \ref{thm for past scattering} of \Cref{Theorem IIC}). The boundedness of $\mathfrak{G}_{\mathscr{I}^+}$ is studied in \Cref{Section 10 Passing to Bondi gauge}. The asymptotics of $\underline{\mathfrak{G}}_{\overline{\mathscr{H}^+}}$ and its boundedness in terms of initial data is studied in \Cref{Section 11 Horizon gauge}. In backwards scattering, the boundedness of the transformation to the $\overline{\Sigma^+}$ gauge in terms of future scattering data, and the boundedness of the transformation to the $\overline{\Sigma^-}$ gauge in terms of past scattering data, is studied in \Cref{an estimate on initial gauge from backwards scattering}. The quantities bounding the gauge transformations listed here are also stated in \Cref{Theorem IIA}, \Cref{Theorem IIB} and \Cref{Theorem IIC}.\\


\subsubsection{The global scattering problem}\label{Section 1.2.9 Intro to memory effect}

Having established Theorem II, we then consider the scattering problem when solutions are evolved out of scattering data at $\overline{\mathscr{H}^-}$, $\mathscr{I}^-$ all the way to $\overline{\mathscr{H}^+}$, $\mathscr{I}^+$. Given that the radiation fields $\xblins_{\mathscr{I}^+}$, $\xlins_{\mathscr{H}^+}$, $\xlins_{\mathscr{I}^-}$, $\xblins_{\mathscr{H}^+}$ constructed in the course of establishing Theorem II arise as the traces of solutions of the linearised Einstein equations on the respective boundary components in a gauge-invariant manner, we may immediately compose the scattering maps defined in Theorem II to yield the global scattering map we are after. However, we would also like to exhibit explicitly the relation between past scattering states in the past scattering gauge (i.e.~in the gauge that is normalised at $\overline{\mathscr{H}^-}$ and $\mathscr{I}^-$) and future scattering states in the future scattering gauge (i.e.~in the gauge that is normalised at $\overline{\mathscr{H}^+}$ and $\mathscr{I}^+$). We also devise another gauge fixing scheme for the purpose of studying the linear memory effect, as the laws of gravitational radiation are best exhibited in a gauge that is Bondi-normalised at $\mathscr{I}^+$ and $\mathscr{I}^-$ simultaneously. Thus we denote by the \textbf{global scattering gauge} the gauge that is Bondi-normalised at both $\mathscr{I}^+$, $\mathscr{I}^-$, that satisfies the conditions \eqref{bifurcation gauge conditions intro}, and which moreover satisfies 
\begin{align}\label{auxiliary gauge conditions global BMS intro}
   U^{-1}\otxb_{\ell=0,1}\Big|_{\mathcal{B}}=0,\qquad \lim_{v\longrightarrow\infty}\glinh(u,v,\theta^A)=0,\qquad  \lim_{v\longrightarrow\infty}\tr\glin_{\ell=1}(u,v,\theta^A)=0.
\end{align}

In the global scattering problem, we start from compactly supported past scattering data $(\xlins_{\mathscr{I}^-},\xblins_{\mathscr{H}^-})$ and solve for a solution $\mathfrak{S}_{\mathscr{I}^-\cup \overline{\mathscr{H}^-}\longrightarrow \mathscr{I}^+\cup \overline{\mathscr{H}^+}}$ which satisfies both the $\mathscr{I}^-$ and $\overline{\mathscr{H}^-}$ gauge conditions (with the gauge at $\overline{\mathscr{H}^-}$ chosen to make the solution exactly Schwarzschildean near $i^-$). This solutions extends to define smooth limits for $r\xblin$, $\Olino$ and $r^2\Klin$ at $\mathscr{I}^+$ such that $\xblins_{\mathscr{I}^+}:=\lim_{v\longrightarrow\infty}r\xblin$ belongs to $\mathcal{E}^T_{\mathscr{I}^+}$. We may then define a pure gauge solution $\mathfrak{G}$, which can be estimated in terms of the data $\xlins_{\mathscr{I}^-}$ and $\xblins_{\mathscr{H}^-}$, such that $\mathfrak{S}_{\mathscr{I}^-\cup \overline{\mathscr{H}^-}\longrightarrow \mathscr{I}^+\cup \overline{\mathscr{H}^+}}+\mathfrak{G}$ is both $\mathscr{I}^+$ and $\mathscr{I}^-$-normalised.\\

For a solution arising from past scattering data $\xlins_{\mathscr{I}^-}$, $\xblins_{\mathscr{I}^-}$ and which is $\overline{\mathscr{H}^-}$ and $\mathscr{I}^-$ normalised, we have at $\mathscr{I}^-$,
\begin{align}
    \partial_v \xblins_{\mathscr{I}^-}=\xlins_{\mathscr{I}^-},
\end{align}
where
\begin{align}
    \lim_{u\longrightarrow-\infty}r^2\xblin(u,v,\theta^A)=:\xblins_{\mathscr{I}^-}(u,\theta^A)
\end{align}
produces a well-defined field, which we denote here by $\xblins_{\mathscr{I}^-}$. As mentioned in our discussion of the backwards scattering problem in \Cref{summary of Thm II argument intro}, when the mean of $\xlins_{\mathscr{I}^-}$ in $v$ is nontrivial, the resulting solution decays towards $i^0$ to order $O_\infty(r^{-1})$. While, at face value, this decay rate is not sufficient to obtain smooth pointwise limits at $\mathscr{I}^+$ for $r^2\xlin, \Psilin$ and $\Psilinb$ by applying $r^p$-estimates with $p>1$, we find that by breaking up the solution into its spherical harmonic modes, we can apply the approximate conservation laws used in \cite{KehrbergerIII} to the Regge--Wheeler equation \eqref{RWintro} to prove that $r^2\xlin$ $\Psilin$ and $\Psilinb$ attain smooth pointwise limits at $\mathscr{I}^+$. The trick of breaking down the solution to individual spherical harmonic modes overcomes the problem posed by the slow decay of the $r^{-2}$ weight of the $S^2$ Laplacian term in \eqref{RWintro}, which comes into play precisely when $\xblins_{\mathscr{I}^+}$ has a nontrivial mean in $v$. The improved control over $\Psilin$ and $\Psilinb$ at $\mathscr{I}^+$ allows us to ensure that the gauge transformation implemented via $\mathfrak{G}$ preserves the conditions \eqref{bifurcation gauge conditions intro}. We are then able to perform another gauge transformation that leads to a solution to the linearised Einstein equations which is in the future scattering gauge. Thus, we are able to construct the Hilbert space isomorphism
\begin{align}\label{Domain of S}
    \mathscr{S}:\mathcal{E}^T_{\mathscr{I}^-}\oplus \mathcal{E}^T_{\overline{\mathscr{H}^-}}\longrightarrow \mathcal{E}^T_{\mathscr{I}^+}\oplus \mathcal{E}^T_{\overline{\mathscr{H}^+}}.
\end{align} 

The results above are the subject of \textbf{Theorem III}, which we state in \Cref{statement of theorem iii intro} and again in more detail in \Cref{statement of corollary ii}. Section \Cref{Section 14: global scattering problem} is devoted to the proof of Theorem III.\\

It is worth noting at this point that an explicit construction of $\mathscr{S}$ is also possible by starting with the subspace of $\mathcal{E}^T_{\mathscr{I}^-}$ consisting of smooth, compactly supported radiation fields $\xlins_{\mathscr{I}^-}$ with a vanishing mean in the null direction:
\begin{align}\label{zero mean past intro}
    \int_{-\infty}^\infty d\bar{v}\,\xlins_{\mathscr{I}^-}=0.
\end{align}
Even with the restriction \eqref{zero mean past intro}, this set is a dense subspace of $\mathcal{E}^T_{\mathscr{I}^-}$. Moreover, the argument of Theorem II shows that the resulting solution decays towards $i^0$ to order $O_\infty(r^{-2})$, and the improved decay near $i^0$ allows us to immediately use Theorem II to extend the solution to a future scattering state in the future scattering gauge, without the need for a spherical harmonic analysis of the solution near $i^0$, and the fact that the restriction \eqref{zero mean past intro} still leaves us with a dense subspace of $\mathcal{E}^T_{\mathscr{I}^-}$ allows us to extend the scattering map to the full domain and range \eqref{Domain of S}. Of course, the drawback of such treatment is that it leaves open the question of scattering theory for data that possesses a nontrivial linear memory.

\subsubsection{The linear memory effect and matching near spacelike infinity}

To show that, for solutions constructed via Theorem II from smooth, compactly supported scattering data $\xlins_{\mathscr{I}^-}$ and $\xblins_{\mathscr{H}^-}$, the quantities $r^2\xlin$, $\Psilin$, $\Psilinb$, $r^3\rlin$, $r^3\slin$ have smooth pointwise limits at $\mathscr{I}^+$, we show that $\Psilin$, $\Psilinb$ are uniformly bounded throughout $\overline{\mathscr{M}}$, and then we use the approximate conservation laws of \cite{KehrbergerIII} to show that each individual $(\ell,m)$ coefficient in the spherical harmonic expansion of $\Psilin$, $\Psilinb$ attains a smooth pointwise limit at $\mathscr{I}^+$. In particular, we show that 
$\mathscr{I}^-$ and $\mathscr{I}^+$ limits of $\Psilin$, $\Psilinb$ are related via 
\begin{align}
    &\lim_{u\longrightarrow-\infty}{\upPsilin_{\mathscr{I}^+}}_{,\ell m}=(-1)^\ell\lim_{v\longrightarrow\infty}{\upPsilin_{\mathscr{I}^-}}_{,\ell m},\label{antipodal 1 intro}\\
    &\lim_{u\longrightarrow-\infty}{\upPsilinb_{\mathscr{I}^+}}_{,\ell m}=(-1)^\ell\lim_{v\longrightarrow\infty}{\upPsilinb_{\mathscr{I}^-}}_{,\ell m}.\label{antipodal 2 intro}
\end{align}
This implies that $\upPsilin_{\mathscr{I}^+}$ and $\upPsilinb_{\mathscr{I}^+}$ are related near $i^0$ by an \textbf{antipodal map}: in the standard atlas on $S^2$ we get
\begin{align}
    &\lim_{u\longrightarrow-\infty}{\upPsilin_{\mathscr{I}^+}}(\theta,\varphi)=\lim_{v\longrightarrow\infty}{\upPsilin_{\mathscr{I}^-}}(\pi-\theta,\pi+\varphi),\label{antipodal 3 intro}\\
    &\lim_{u\longrightarrow-\infty}{\upPsilinb_{\mathscr{I}^+}}(\theta,\varphi)=\lim_{v\longrightarrow\infty}{\upPsilinb_{\mathscr{I}^-}}(\pi-\theta,\pi+\varphi).\label{antipodal 4 intro}
\end{align}

Note that the asymptotic analysis near $i^0$ of solutions to the linear wave equation on Schwarzschild performed in \cite{KehrbergerIII} already includes the analogous results to \bref{antipodal 1 intro}--\bref{antipodal 4 intro} in the case of the wave equation as special cases (see Eqns.~1.24), (10.3), (10.48) and (10.50) in \cite{KehrbergerIII}), although the particular relevance for linear gravitational memory was not mentioned.
Here, we apply the methods of \cite{KehrbergerIII} in the context of linearised gravity to exhibit the relationships between past and future memory \bref{antipodal 1 intro}--\bref{antipodal 4 intro} in a self-contained way.\\

The behaviour of $\upPsilinb_{\mathscr{I}^+}$, $\upPsilin_{\mathscr{I}^-}$ is related to the {linear memory effect} as follows: in a double null gauge, the memory effect is exhibited via the quantities 
\begin{align}
    \xlins_{\mathscr{I}^+}(u,\theta^A):=\lim_{v\longrightarrow\infty}r^2\xlin(u,v,\theta^A),\qquad\qquad \xblins_{\mathscr{I}^-}(v,\theta^A):=\lim_{u\longrightarrow-\infty}r^2\xblin(u,v,\theta^A).
\end{align}
When $\xlins_{\mathscr{I}^+}$, $\xblins_{\mathscr{I}^-}$ are well-defined, the future linear memory effect is defined as the quantity $\Sigma^-_{\mathscr{I}^+}-\Sigma^+_{\mathscr{I}^+}$, where 
\begin{align}
     \Sigma_{\mathscr{I}^+}^+(\theta^A):=\lim_{u\longrightarrow\infty}&\xlins_{\mathscr{I}^+}(u,\theta^A),\qquad \Sigma_{\mathscr{I}^+}^-(\theta^A):=\lim_{u\longrightarrow-\infty}\xlins_{\mathscr{I}^+}(u,\theta^A).
\end{align}
Similarly, past memory is given by 
\begin{align}
     \Sigma_{\mathscr{I}^-}^+(\theta^A)- \Sigma_{\mathscr{I}^-}^-(\theta^A):=\lim_{v\longrightarrow\infty}\xblins_{\mathscr{I}^-}(v,\theta^A)-\lim_{v\longrightarrow-\infty}\xblins_{\mathscr{I}^-}(v,\theta^A),
\end{align}
provided the above limits exist. Define the quantities $\Pscri$, $\Qscri$ via
\begin{align}\label{def of rlins intro}
   \rlins_{\mathscr{I}^+}  (u,\theta^A)=\lim_{v\longrightarrow\infty}r^3\rlin_{\ell\geq2},(u,v,\theta^A),\qquad \rlins_{\mathscr{I}^-}(v,\theta^A)=\lim_{u\longrightarrow-\infty}r^3\rlin_{\ell\geq2}\,(u,v,\theta^A),
\end{align}
\begin{align}\label{def of slins intro}
   \slins_{\mathscr{I}^+}(u,\theta^A)=\lim_{v\longrightarrow\infty}r^3\slin_{\ell\geq2}\,(u,v,\theta^A),\qquad \slins_{\mathscr{I}^-}(u,\theta^A)=\lim_{u\longrightarrow-\infty}r^3\slin_{\ell\geq2}\,(u,v,\theta^A).
\end{align}
Let
\begin{align}\label{def of Pscri}
    \Pscri_{\mathscr{I}^+}^\pm=\lim_{u\longrightarrow\pm\infty} \rlins_{\mathscr{I}^+},\qquad \Pscri_{\mathscr{I}^-}^\pm=\lim_{v\longrightarrow\pm\infty} \rlins_{\mathscr{I}^-},
\end{align}
\begin{align}\label{def of Qscri}
    \Qscri_{\mathscr{I}^+}^\pm=\lim_{u\longrightarrow\pm\infty} \slins_{\mathscr{I}^+},\qquad \Qscri_{\mathscr{I}^-}^\pm=\lim_{v\longrightarrow\pm\infty} \slins_{\mathscr{I}^-}.
\end{align}
We show that, for a solution which is Bondi-normalised at both $\mathscr{I}^+$ and $\mathscr{I}^-$ (e.g.~the global scattering gauge), the limits of $\upPsilin_{\mathscr{I}^-}$, $\upPsilinb_{\mathscr{I}^+}$, $\Pscri_{\mathscr{I}^+}$, $\Qscri_{\mathscr{I}^+}$, $\Pscri_{\mathscr{I}^-}$, $\Qscri_{\mathscr{I}^-}$ are related by 
\begin{align}
    \lim_{u\longrightarrow-\infty}\upPsilinb_{\mathscr{I}^+}=2\fancydstarring_2\fancydstarring_1\left(\Pscri_{\mathscr{I}^+}^-,\Qscri_{\mathscr{I}^+}^-\right),\qquad \lim_{v\longrightarrow\infty}\upPsilin_{\mathscr{I}^-}=2\fancydstarring_2\fancydstarring_1\left(\Pscri_{\mathscr{I}^-}^+,-\Qscri_{\mathscr{I}^-}^+\right).
\end{align}
Given \bref{antipodal 3 intro} and \bref{antipodal 4 intro}, we obtain
\begin{align}
    \Pscri_{\mathscr{I}^+}^-(\theta,\varphi)= \Pscri_{\mathscr{I}^-}^+(\pi-\theta,\pi+\varphi),\qquad \Qscri_{\mathscr{I}^+}^-(\theta,\varphi)= \Qscri_{\mathscr{I}^-}^+(\pi-\theta,\pi+\varphi).
\end{align}
In the global scattering gauge, $\upPsilinb_{\mathscr{I}^+}$ is related to $\xlins_{\mathscr{I}^+}$ via
\begin{align}
    \upPsilinb_{\mathscr{I}^+}=-(\mathring{\slashed{\Delta}}-2)(\mathring{\slashed{\Delta}}-4)\xlins_{\mathscr{I}^+}-6M\xblins_{\mathscr{I}^+}.
\end{align}
Similarly, we have on $\mathscr{I}^-$
\begin{align}
    \upPsilin_{\mathscr{I}^-}=-(\mathring{\slashed{\Delta}}-2)(\mathring{\slashed{\Delta}}-4)\xblins_{\mathscr{I}^-}+6M\xlins_{\mathscr{I}^-},
\end{align}
where $\mathring{\slashednabla}$ in the above denotes the Levi-Civita covariant derivative on the unit sphere and $\mathring{\slashed{\Delta}}$ denotes the associated Laplacian on $S^2$.

As we have that $\lim_{v\longrightarrow-\infty}{\upPsilin_{\mathscr{I}^-}(v,\theta^A)}=0$, $\lim_{u\longrightarrow\infty}{\upPsilinb_{\mathscr{I}^+}(u,\theta^A)}=0$, the same extends to $\xblins_{\mathscr{I}^-}$, $\xlins_{\mathscr{I}^+}$ respectively in the global scattering gauge since $\xblins_{\mathscr{I}^+}, \xlins_{\mathscr{I}^-}\in H^1(\mathbb{R}\times S^2)$ by Theorem II below. Thus we also have that 
\begin{align}
    \Pscri_{\mathscr{I}^+}^+= \Pscri_{\mathscr{I}^-}^-=\Qscri_{\mathscr{I}^+}^+= \Qscri_{\mathscr{I}^-}^-=0.
\end{align}
Given that in a Bondi-normalised double null gauge we have 
\begin{align}
    &\divo\divo \xlins_{\mathscr{I}^+}=-\rlins_{\mathscr{I}^+}, \qquad \curlo\divo\xlins_{\mathscr{I}^+}=-\slins_{\mathscr{I}^+},\\
    &\divo\divo \xblins_{\mathscr{I}^-}=-\rlins_{\mathscr{I}^-}, \qquad \curlo\divo\xblins_{\mathscr{I}^-}=\slins_{\mathscr{I}^-},
\end{align}
we deduce the relations
\begin{align}\label{BMS invariant statement}
  &\divo\divo\left(\Sigma^-_{\mathscr{I}^+}-\Sigma^+_{\mathscr{I}^+}\right)=\left(-\Pscri_{\mathscr{I}^+}^-+\Pscri_{\mathscr{I}^+}^+\right),\qquad \curlo\divo\left(\Sigma^-_{\mathscr{I}^+}-\Sigma^+_{\mathscr{I}^+}\right)=\left(-\Qscri_{\mathscr{I}^+}^-+\Qscri_{\mathscr{I}^+}^+\right),\\
  &\divo\divo\left(\Sigma^+_{\mathscr{I}^-}-\Sigma^-_{\mathscr{I}^-}\right)=\left(-\Pscri_{\mathscr{I}^-}^-+\Pscri_{\mathscr{I}^-}^+\right),\qquad \curlo\divo\left(\Sigma^+_{\mathscr{I}^-}-\Sigma^-_{\mathscr{I}^-}\right)=\left(\Qscri_{\mathscr{I}^-}^+-\Qscri_{\mathscr{I}^+}^-\right).
\end{align}

Note that by including $\Sigma^+_{\mathscr{I}^+}, \Sigma^-_{\mathscr{I}^-}, \Pscri_{\mathscr{I}^+}^+, \Pscri_{\mathscr{I}^-}^-, \Qscri_{\mathscr{I}^+}^+, \Qscri_{\mathscr{I}^-}^-$ in the formulae above we render it invariant under the actions of either the \text{BMS}\textsuperscript{+} or the \text{BMS}\textsuperscript{-} groups.\\

The formulae \bref{BMS invariant statement} reproduce the results of \cite{ChristodoulouPRL} at $\mathscr{I}^+$ in the linear regime: The laws that determine the memory effect at $\mathscr{I}^+$ for the full system read
\begin{align}\label{curl Z+-Z-}
    \curlo\, \divo \left(\bm{\Sigma}^+_{\mathscr{I}^+}-\bm{\Sigma}^-_{\mathscr{I}^+}\right)=\left(\bm{\mathrm{Q}}_{\mathscr{I}^+}-\overline{\bm{\mathrm{Q}}}_{\mathscr{I}^+}\right)^+-\left(\bm{\mathrm{Q}}_{\mathscr{I}^+}-\overline{\bm{\mathrm{Q}}}_{\mathscr{I}^+}\right)^-
\end{align}
\begin{align}\label{div Z+-Z-}
   \divo\divo \left(\bm{\Sigma}^+_{\mathscr{I}^+}-\bm{\Sigma}^-_{\mathscr{I}^+}\right)=\left(\bm{\mathrm{P}}_{\mathscr{I}^+}-\overline{\bm{\mathrm{P}}}_{\mathscr{I}^+}\right)^+-\left(\bm{\mathrm{P}}_{\mathscr{I}^+}-\overline{\bm{\mathrm{P}}}_{\mathscr{I}^+}\right)^--\frac{1}{4}\int_{-\infty}^\infty d\bar{u} |\bm{\widehat{\underline{\chi}}}_{\mathscr{I}^+}|^2 +\frac{1}{4}\int_{S^2}\int_{-\infty}^\infty d\bar{u}d\gamma_{S^2}\,|\bm{\widehat{\underline{\chi}}}_{\mathscr{I}^+}|^2,
\end{align}
where $\left(\bm{\Sigma}^+_{\mathscr{I}^+}-\bm{\Sigma}^-_{\mathscr{I}^+}\right), \left(\bm{\mathrm{P}}_{\mathscr{I}^+}-\overline{\bm{\mathrm{P}}}_{\mathscr{I}^+}\right)^\pm, \left(\bm{\mathrm{Q}}_{\mathscr{I}^+}-\overline{\bm{\mathrm{Q}}}_{\mathscr{I}^\pm}\right)^+$ are defined for the full system in an analogous manner to the definitions \eqref{def of rlins intro}, \bref{def of slins intro}, \eqref{def of Pscri}, \eqref{def of Qscri}, and $\overline{\bm{\mathrm{P}}}, \overline{\bm{\mathrm{Q}}}$ refer to the means of ${\bm{\mathrm{P}}}, {\bm{\mathrm{Q}}}$ over $S^2$.\\

For the example of radiation arising from small perturbations to the Minknowski spacetime for which the stability theorem of \cite{Ch-K} applies, the Cauchy data for the solutions considered is asymptotically flat to order $\left(\frac{3}{2},4\right)$, and this leads to vanishing  $\left(\bm{\mathrm{Q}}_{\mathscr{I}^+}-\overline{\bm{\mathrm{Q}}}_{\mathscr{I}^+}\right)^+-\left(\bm{\mathrm{Q}}_{\mathscr{I}^+}-\overline{\bm{\mathrm{Q}}}_{\mathscr{I}^+}\right)^-$. This is consistent with our results, where we show in \Cref{Section 7 forward scattering} that for Cauchy data that is asymptotically flat to order $(1+\delta,5)$ for any $\delta>0$, the quantities $\Pscri_{\mathscr{I}^+}^\pm$, $\Qscri_{\mathscr{I}^+}^\pm$ vanish. On the other hand, for past scattering data for the linearised Einstein equations for which  $\Qscri_{\mathscr{I}^-}^+-\Qscri_{\mathscr{I}^-}^-$ is nontrivial, the quantity $\Qscri_{\mathscr{I}^+}^--\Qscri_{\mathscr{I}^+}^+$ is also nontrivial and it appear in the right hand side of the linearised version of \eqref{curl Z+-Z-}. The linearisation of \eqref{div Z+-Z-} gets rid of the flux terms of $\bm{\widehat{\underline{\chi}}}$ since the background quantity $\widehat{\underline{\chi}}_{Schw}$ vanishes. It is worth keeping in mind that  In \cite{ChristodoulouPRL}, the laws \eqref{curl Z+-Z-}, \eqref{div Z+-Z-} go beyond the setting of the Minkwoski stability theorem and are stated in a general nonlinear setting. They are in particular expected to apply to the physics of gravitational radiation arising from the dynamics of isolated systems, see for example \cite{ChristodoulouMarcelGrossman}. For such setups, it is expected that $\left(\bm{\mathrm{Q}}_{\mathscr{I}^+}-\overline{\bm{\mathrm{Q}}}_{\mathscr{I}^+}\right)^+-\left(\bm{\mathrm{Q}}_{\mathscr{I}^+}-\overline{\bm{\mathrm{Q}}}_{\mathscr{I}^+}\right)^-$. On the other hand, for solutions to \eqref{EVE} arising from scattering data with   $\left(\bm{\mathrm{Q}}_{\mathscr{I}^-}-\overline{\bm{\mathrm{Q}}}_{\mathscr{I}^-}\right)^--\left(\bm{\mathrm{Q}}_{\mathscr{I}^-}-\overline{\bm{\mathrm{Q}}}_{\mathscr{I}^-}\right)^+$ is already nontrivial, we expect that $\left(\bm{\mathrm{Q}}_{\mathscr{I}^+}-\overline{\bm{\mathrm{Q}}}_{\mathscr{I}^+}\right)^+-\left(\bm{\mathrm{Q}}_{\mathscr{I}^+}-\overline{\bm{\mathrm{Q}}}_{\mathscr{I}^+}\right)^-$ will not vanish. Indeed, this is what we show here in the linear setting. \\

Given that $\left(\Sigma^-_{\mathscr{I}^+}-\Sigma^+_{\mathscr{I}^+}\right)$, $\left(\Sigma^+_{\mathscr{I}^-}-\Sigma^+_{\mathscr{I}^-}\right)$ are symmetric traceless 2-contravariant tensor fields on $S^2$, the above shows that there exists two scalar functions $f,g$ on $S^2$ such that 
\begin{align}\label{future memory intro 1}
    &\left(\Sigma^+_{\mathscr{I}^-}-\Sigma^-_{\mathscr{I}^-}\right)=2\fancydstarring_2\fancydstarring_1(f,g)=-\frac{1}{2}\left[\mathring{\slashednabla}\widehat{\otimes}\mathring{\slashednabla}f+\mathring{\slashednabla}\widehat{\otimes}\star\mathring{\slashednabla}g\right],
\end{align}
\begin{align}\label{past memory intro 1}
    &\left(\Sigma^-_{\mathscr{I}^+}-\Sigma^-_{\mathscr{I}^+}\right)=2\fancydstarring_2\fancydstarring_1(f\circ\mathscr{A},-g\circ\mathscr{A})=-\frac{1}{2}\left[\mathring{\slashednabla}\widehat{\otimes}\mathring{\slashednabla}f\circ\mathscr{A}-\mathring{\slashednabla}\widehat{\otimes}\star\mathring{\slashednabla}g\circ\mathscr{A}\right],
\end{align}
where $\mathring{\slashednabla}$ in the above denotes the Levi-Civita covariant derivative on the unit sphere, $\star\mathring{\slashednabla}_A=\slashed{\epsilon}_{AB}\mathring{\slashednabla}^B$ with $\slashed{\epsilon}$ denoting the volume form on the unit sphere, the notation $\mathring{\slashednabla}\widehat{\otimes}\mathring{\slashednabla}f$ denoting the symmetric traceless component of $\mathring{\slashednabla}_A\mathring{\slashednabla}_Bf$ and similarly $\mathring{\slashednabla}\widehat{\otimes}\star\mathring{\slashednabla}g$ denoting the symmetric traceless component of $\mathring{\slashednabla}_A\star\mathring{\slashednabla}_Bg$, and $\mathscr{A}$ is the antipodal map on $S^2$, which acts on the standard atlas on $S^2$ via
\begin{align}
    \mathscr{A}(\theta,\varphi)=(\pi-\theta,\pi+\varphi).
\end{align}

The results outlined here lead to \textbf{Corollary II}, which we state in  \Cref{statement of corollary ii intro 2} below (and again in more detail in \Cref{statement of corollary ii 2}). The proof of Corollary II is the subject of \Cref{Section 15 Proof of Theorem III}. As of the moment of writing, the antipodal matching between the past and future memory effects in the context of scattering has only been considered in setups where a priori assumptions are made for the asymptotics of perturbations of the Einstein equations near $i^0$, see for example \cite{Prabhu}, \cite{Capone}. \\

Finally, note that for the scattering problem for the full Einstein equations \eqref{EVE} on a black hole spacetime, the energy flux contribution to the right hand side of \eqref{curl Z+-Z-} indicates that one expects that any relation between past and future memory will also include a contribution from the energy flux on the future event horizon.

\subsection{Summary of main results}\label{Section 1.3 summary of results}

We now present the gauge conditions used to represent initial and scattering data, and give a preliminary statement of the main results of this paper.



\subsubsection{Theorem II: the scattering theory of the full linearised system}\label{Section 1.6 statement of results intro}

We define the space of scattering states $\mathcal{E}^{T}_{\overline{\Sigma},+}$ on $\overline{\Sigma}$ as the completion of smooth, asymptotically flat initial data sets on $\overline{\Sigma}$ which are in the $\overline{\Sigma}_+$-gauge, under the norm defining $\mathcal{E}^{T,-2}_{\overline{\Sigma}}$ of Theorem I of \cite{Mas20} (see already \Cref{remark on choice of energy for full system}). An analogous definition describes  $\mathcal{E}^{T}_{\overline{\Sigma},-}$,  and it is easy to see that it is isometric to $\mathcal{E}^{T}_{\overline{\Sigma},+}$. The spaces of scattering states at $\mathscr{I}^\pm$, $\overline{\mathscr{H}^\pm}$ are defined as the completion of smooth, compactly supported data on the respective regions under the norms defined in the following diagram:
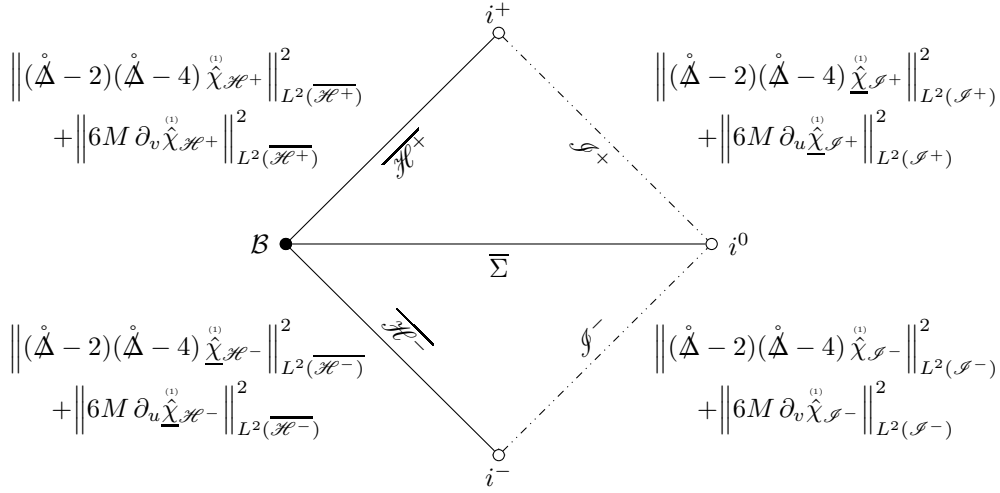
\begin{center}
\begin{tikzpicture}[->,scale=0.7, arrow/.style={
            color=black,
            draw=blue,thick,
            -latex,
                font=\fontsize{8}{8}\selectfont},
        ]
\node (I)    at ( 0,0)   {};

\path 
   (I) +(90:4)  coordinate (Itop) coordinate [label={$i^+$}]
       +(180:4) coordinate (Ileft) coordinate [label=180:{$\mathcal{B}\;$}]
       +(0:4)   coordinate (Iright) coordinate [label=0:{$\;i^0$}]
       +(270:4) coordinate (Ibot) coordinate [label=-90:{$i^-$}]
       ;






\draw (Ileft)--node[below]{$\overline{\Sigma}$}(Iright);

\draw  (Ileft) -- node[align=center][yshift=4mm,xshift=-20mm]{$ \left\|(\mathring{\slashed{\Delta}}-2)(\mathring{\slashed{\Delta}}-4)\,\xlin_{\mathscr{H}^+}\right\|^2_{L^2(\overline{\mathscr{H}^+})}\qquad\qquad$\\$+ \Big\|6M\,\partial_v\xlin_{\mathscr{H}^+}\Big\|^2_{L^2(\overline{\mathscr{H}^+})}\qquad\qquad$} node[rotate=45,below]{$\overline{\mathscr{H}^+}$} (Itop) ;
\draw[dash dot dot] (Iright) --  node[align=center][yshift=4mm,xshift=22.mm]{$\qquad\qquad \left\|(\mathring{\slashed{\Delta}}-2)(\mathring{\slashed{\Delta}}-4)\,\xblin_{\mathscr{I}^+}\right\|^2_{L^2(\mathscr{I}^+)}$\\$\qquad\qquad+ \Big\|6M\,\partial_u\xblin_{\mathscr{I}^+}\Big\|^2_{L^2({\mathscr{I}^+})}$}node[rotate=-45,below]{${\mathscr{I}^+}$}(Itop) ;
\draw[dash dot dot] (Iright) --  node[align=center][yshift=-4mm,xshift=22.mm]{$\qquad\qquad \left\|(\mathring{\slashed{\Delta}}-2)(\mathring{\slashed{\Delta}}-4)\,\xlin_{\mathscr{I}^-}\right\|^2_{L^2(\mathscr{I}^-)}$\\$\qquad\qquad+ \Big\|6M\,\partial_v\xlin_{\mathscr{I}^-}\Big\|^2_{L^2({\mathscr{I}^-})}$}node[rotate=45,above]{${\mathscr{I}^-}$}(Ibot) ;
\draw  (Ileft) -- node[align=center][yshift=-4mm,xshift=-20mm]{$ \left\|(\mathring{\slashed{\Delta}}-2)(\mathring{\slashed{\Delta}}-4)\,\xblin_{\mathscr{H}^-}\right\|^2_{L^2(\overline{\mathscr{H}^-)}}\qquad\qquad$\\$+ \Big\|6M\,\partial_u\xblin_{\mathscr{H}^-}\Big\|^2_{L^2(\overline{\mathscr{H}^-})}\qquad\qquad$} node[rotate=-45,above]{$\overline{\mathscr{H}^-}$}(Ibot) ;



\filldraw[white] (Itop) circle (3pt);
\draw[black] (Itop) circle (3pt);

\filldraw[white] (Ibot) circle (3pt);
\draw[black] (Ibot) circle (3pt);

\filldraw[white] (Iright) circle (3pt);
\draw[black] (Iright) circle (3pt);
\filldraw[black] (Ileft) circle (3pt);
\draw[black] (Ileft) circle (3pt);
\end{tikzpicture}
\end{center}

\begin{namedtheorem}[Theorem II]\label{Theorem 2}
For any Cauchy data set $\mathfrak{D}_{\overline{\Sigma}}$ for the linearised Einstein equations which is smooth and suitable asymptotically flat and which is in the $\overline{\Sigma}_{+}$-gauge, there exists a unique solution $\mathfrak{S}$ restricting to $\mathfrak{D}_{\overline{\Sigma}}$ on $\overline{\Sigma}$. The solution $\mathfrak{S}$ is the sum of a member of the 
linearised Kerr family defined in \Cref{reference future linearised Kerr solution} and a component which induces smooth radiation fields at $\overline{\mathscr{H}^+}$, $\mathscr{I}^+$ via
\begin{align}\label{def of radiation on scri+ intro}
    \xblins_{\mathscr{I}^+}(u,\theta^A)=\lim_{v\longrightarrow\infty} r\xblin(u,v,\theta^A),\qquad\qquad \xlins_{\mathscr{H}^+}=\Omega\xlin|_{\overline{\mathscr{H}^+}},
\end{align}

Furthermore, there exist unique pure gauge solutions $\underline{\mathfrak{G}}_{\mathscr{I}^+}$ $\underline{\mathfrak{G}}_{\overline{\mathscr{H}^+}}$ which are supported on $\ell\geq2$ spherical harmonics, and which can be estimated from the initial data for $\mathfrak{S}$, such that the $\ell\geq2$ component of  $\mathfrak{S}+\mathfrak{G}_{\mathscr{I}^+}+\underline{\mathfrak{G}}_{\overline{\mathscr{H}^+}}$ satisfies the conditions of the $\overline{\mathscr{H}^+}$ and the $\mathscr{I}^+$ gauges. We have
\begin{align}\label{future energy conservation intro}
    \|\mathfrak{D}_{\overline{\Sigma}}\|^2_{\mathcal{E}^{T}_{\overline{\Sigma},+}}=\|\xblins_{\mathscr{I}^+}\|^2_{\mathcal{E}^T_{\mathscr{I}^+}}+\|\xlins_{\mathscr{H}^+}\|^2_{\mathcal{E}^{T}_{\overline{\mathscr{H}^+}}}.
\end{align}

In turn, given real parameters $(\mathfrak{m},\mathfrak{a})\in\mathbb{R}\times \mathbb{R}^3$ and smooth, compactly supported data $\xblins_{\mathscr{I}^+}$ on $\mathscr{I}^+$ and $\xlins_{\mathscr{H}^+}$ at $\overline{\mathscr{H}^+}$, there exists a unique solution $\mathfrak{S}_{\mathscr{I}^+\cap\overline{\mathscr{H}^+}}$ which is the sum of a linearised Kerr reference solution of parameters $(\mathfrak{m},\mathfrak{a})$, and a component which is supported on $\ell\geq2$ and which satisfies both the $\overline{\mathscr{H}^+}$ and the $\mathscr{I}^+$ gauge conditions, with \bref{def of radiation on scri+ intro} satisfied. There exists a unique pure gauge solution $\mathfrak{G}_{\overline{\Sigma},+}$ which is supported on $\ell\geq2$  spherical harmonics and which can be estimated in terms of the scattering data $\xblins_{\mathscr{I}^+}$, $\xlins_{\mathscr{H}^+}$, such that $\mathfrak{S}_{\mathscr{I}^+\cup\overline{\mathscr{H}^+}}+\mathfrak{G}_{\overline{\Sigma},+}$ induces asymptotically flat data on $\overline{\Sigma}$ which is in the $\overline{\Sigma}_+$-gauge. The induced data from $\mathfrak{S}_{\mathscr{I}^+\cup\overline{\mathscr{H}^+}}+\mathfrak{G}_{\overline{\Sigma},+}$ satisfy \bref{future energy conservation intro}.

Completely analogous statements apply to scattering between $\overline{\Sigma}$ and $\mathscr{I}^-$, $\overline{\mathscr{H}^-}$. The above correspondence extends to Hilbert space isomorphisms
\begin{align}
\begin{split}
    &\mathscr{F}^+:\mathcal{E}^{T}_{\overline{\Sigma},+}\longrightarrow \mathcal{E}^T_{\mathscr{I}^+}\oplus \mathcal{E}^{T}_{\overline{\mathscr{H}^+}},\qquad\qquad \mathscr{B}^-: \mathcal{E}^T_{\mathscr{I}^+}\oplus \mathcal{E}^{T}_{\overline{\mathscr{H}^+}}\longrightarrow\mathcal{E}^{T}_{\overline{\Sigma},+},\\
    &\mathscr{F}^-:\mathcal{E}^{T}_{\overline{\Sigma},-}\longrightarrow \mathcal{E}^T_{\mathscr{I}^-}\oplus \mathcal{E}^{T}_{\overline{\mathscr{H}^-}},\qquad\qquad \mathscr{B}^+: \mathcal{E}^T_{\mathscr{I}^-}\oplus \mathcal{E}^{T}_{\overline{\mathscr{H}^-}}\longrightarrow\mathcal{E}^{T}_{\overline{\Sigma},-},
\end{split}
\end{align}
such that 
\begin{align}
\begin{split}
    &\mathscr{F}^+\circ\mathscr{B}^-=Id_{\mathcal{E}^T_{\overline{\Sigma},+}},\qquad\qquad \mathscr{B}^-\circ\mathscr{F}^+=Id_{\mathcal{E}^T_{\mathscr{I}^+}\oplus \mathcal{E}^{T}_{\overline{\mathscr{H}^+}}},\\
    &\mathscr{F}^-\circ\mathscr{B}^+=Id_{\mathcal{E}^T_{\overline{\Sigma},-}},\qquad\qquad \mathscr{B}^+\circ\mathscr{F}^-=Id_{\mathcal{E}^T_{\mathscr{I}^-}\oplus \mathcal{E}^{T}_{\overline{\mathscr{H}^-}}}.
\end{split}
\end{align}
\end{namedtheorem}
For the estimate on the pure gauge transformation required to pass to a Bondi-normalised gauge at both $\mathscr{I}^+$ and $\mathscr{I}^-$ in terms of past scattering data on $\mathscr{I}^-$, $\mathscr{H}^-$, see \Cref{an estimate on initial gauge from backwards scattering}.

\subsubsection{Theorem III: scattering from $\mathscr{I}^-$, $\overline{\mathscr{H}^-}$ to $\mathscr{I}^+$, $\overline{\mathscr{H}^+}$}\label{statement of theorem iii intro}

\begin{namedtheorem}[Theorem III]
Given smooth, compactly supported scattering data $\xlins_{\mathscr{I}^-}$, $\xblins_{\mathscr{H}^-}$ and real parameters $\mathfrak{m}\in\mathbb{R}$ and $\mathfrak{a}\in \mathbb{R}^3$, there exists a unique solution $\mathfrak{S}_{\overline{\mathscr{H}^-}\cup\mathscr{I}^-}$ to the linearised Einstein equations which is the sum of a member of the linearised Kerr family with mass parameter $\mathfrak{m}$ and rotation parameter $\mathfrak{a}$, and a dispersive component which realises $\xlins_{\mathscr{I}^-}$, $\xblins_{\mathscr{H}^-}$ via
\begin{align}\label{def of radiation on scri- intro}
    \xlins_{\mathscr{I}^-}(v,\theta^A)=\lim_{u\longrightarrow-\infty} r\xlin(u,v,\theta^A),\qquad\qquad \xblins_{\mathscr{H}^-}=\Omega\xblin|_{\overline{\mathscr{H}^-}}.
\end{align}
The solution $\mathfrak{S}_{\overline{\mathscr{H}^-}\cup\mathscr{I}^-}$ is $\mathscr{I}^-$ and $\overline{\mathscr{H}^-}$-normalised, and it induces smooth radiation data on $\mathscr{I}^+$, $\overline{\mathscr{H}^+}$ via \bref{def of radiation on scri+ intro}. Furthermore, there exists a unique pure gauge solution $\mathfrak{G}_{\mathscr{I}^+}$, which can be estimated in terms of the initial data of $\mathfrak{S}_{\overline{\mathscr{H}^-}\cup\mathscr{I}^-}$, such that $\mathfrak{S}_{\overline{\mathscr{H}^-}\cup\mathscr{I}^-}+\mathfrak{G}_{\mathscr{I}^+}$ is in the global scattering gauge, and there exists a unique pure gauge solution $\mathfrak{G}_{\overline{\mathscr{H}^+}}$ such that $\mathfrak{S}_{\overline{\mathscr{H}^-}\cup\mathscr{I}^-}+\mathfrak{G}_{\mathscr{I}^+}+\mathfrak{G}_{\overline{\mathscr{H}^+}}$ is in the future scattering gauge. The solution $\mathfrak{G}_{\mathscr{I}^+}+\mathfrak{G}_{\overline{\mathscr{H}^+}}$ is the unique pure gauge solution such that $\mathfrak{S}_{\overline{\mathscr{H}^-}\cup\mathscr{I}^-}+\mathfrak{G}_{\mathscr{I}^+}+\mathfrak{G}_{\overline{\mathscr{H}^+}}$ is in the future scattering gauge.

Finally, we have
\begin{align}\label{future past energy conservation intro}
  \|\xlins_{\mathscr{I}^-}\|^2_{\mathcal{E}^T_{\mathscr{I}^-}}+\|\xblins_{\mathscr{H}^-}\|^2_{\mathcal{E}^{T}_{\overline{\mathscr{H}^-}}}=\|\xblins_{\mathscr{I}^+}\|^2_{\mathcal{E}^T_{\mathscr{I}^+}}+\|\xlins_{\mathscr{H}^+}\|^2_{\mathcal{E}^{T}_{\overline{\mathscr{H}^+}}}.
\end{align}
An analogous statement applies for evolution from scattering data on $\overline{\mathscr{H}^+}$, $\mathscr{I}^+$. The above extends to a Hilbert space isomorphism
\begin{align}
    \mathscr{S}:\mathcal{E}^T_{\mathscr{I}^-}\oplus \mathcal{E}^{T}_{\overline{\mathscr{H}^-}}\longrightarrow \mathcal{E}^T_{\mathscr{I}^+}\oplus \mathcal{E}^{T}_{\overline{\mathscr{H}^+}}.
\end{align}
\end{namedtheorem}
 

\begin{remark} The fact that time translation and angular momentum operators commute with $\Box_g$ means that we can project scattering data on individual spherical harmonic modes and consider solutions in frequency space. Since $\overone{\hat{\chi}}, \overone{\hat{\underline\chi}}$ are supported on $\ell\geq2$, we can translate the unitarity clause of Theorem II in terms of fixed frequency, fixed azimuthal mode solutions to the following statement:
\begin{align}
    \Big\|\overone{\hat{\upchi}}_{\mathscr{H}^+,\;\omega,m,\ell}\Big\|_{L^2_\omega }^2+\Big\|\overone{\hat{\underline\upchi}}_{\mathscr{I}^+,\;\omega,m,\ell}\Big\|^2_{L^2_\omega }\;=\;\Big\|\overone{\hat{\underline\upchi}}_{\mathscr{H}^-,\;\omega,m,\ell}\Big\|_{L^2_\omega }^2+\Big\|\overone{\hat{\upchi}}_{\mathscr{I}^-,\;\omega,m,\ell}\Big\|^2_{L^2_\omega }.
\end{align}
Resumming in $\ell_{m,\ell}^2$ and using Plancherel, we obtain the identity 
\begin{align}\label{conservation law}
     \Big\|\overone{\hat{\upchi}}_{\mathscr{H}^+}\Big\|_{L^2(\overline{\mathscr{H}^+})}^2+\Big\|\overone{\hat{\underline\upchi}}_{\mathscr{I}^+}\Big\|_{L^2(\mathscr{I}^+)}^2\;=\; \Big\|\overone{\hat{\underline\upchi}}_{\mathscr{H}^-}\Big\|^2_{L^2(\overline{\mathscr{H}^-})}+\Big\|\overone{\hat{\upchi}}_{\mathscr{I}^-}\Big\|^2_{L^2(\mathscr{I}^-)}.
\end{align}
The statement \bref{conservation law} above ties up with the work by Holzegel \cite{Holzegel_2016}, where a set of conservation laws are derived for the full system of linearised Einstein equations on the Schwarzschild exterior \bref{Schwarzschild} (using purely physical-space methods).

\end{remark}

\indent Note that in particular, for past scattering data that is vanishing on $\overline{\mathscr{H}^-}$, the identity  \bref{conservation law} has the interpretation that the energy of the gravitational energy radiated to $\mathscr{I}^+$ is bounded \underline{with constant 1} by the incoming gravitational energy radiated from $\mathscr{I}^-$, i.e.~there is no superradiant amplification of reflected gravitational radiation on the Schwarzschild exterior.

\subsubsection{Corollary II: matching the past and future linear memory effects}\label{statement of corollary ii intro 2}

\begin{namedtheorem}[Corollary II]
    Let $\mathfrak{S}$ be a solution to the linearised Einstein equations arising from smooth, compactly supported past scattering data and which is normalised to the global scattering gauge according to Theorem III. Let
     \begin{align}
            \Sigma_{\mathscr{I}^+}^\pm:=\lim_{u\longrightarrow\pm\infty}\xlins_{\mathscr{I}^+},\qquad \Sigma_{\mathscr{I}^-}^\pm:=\lim_{v\longrightarrow\pm\infty}\xblins_{\mathscr{I}^-}.
        \end{align}
     Then $\left(\Sigma^-_{\mathscr{I}^+}-\Sigma^+_{\mathscr{I}^+}\right)$, $\left(\Sigma^+_{\mathscr{I}^-}-\Sigma^-_{\mathscr{I}^-}\right)$ are related as follows: there exist smooth scalar functions $f,g$ on $S^2$ such that
\begin{align}\label{future memory intro 2}
    &\left(\Sigma^+_{\mathscr{I}^-}-\Sigma^-_{\mathscr{I}^-}\right)=2\fancydstarring_2\fancydstarring_1(f,g)=-\frac{1}{2}\left[\mathring{\slashednabla}\widehat{\otimes}\mathring{\slashednabla}f+\mathring{\slashednabla}\widehat{\otimes}\star\mathring{\slashednabla}g\right],
\end{align}
\begin{align}\label{past memory intro 2}
    &\left(\Sigma^-_{\mathscr{I}^+}-\Sigma^-_{\mathscr{I}^+}\right)=2\fancydstarring_2\fancydstarring_1(f\circ\mathscr{A},-g\circ\mathscr{A})=-\frac{1}{2}\left[\mathring{\slashednabla}\widehat{\otimes}\mathring{\slashednabla}f\circ\mathscr{A}-\mathring{\slashednabla}\widehat{\otimes}\star\mathring{\slashednabla}g\circ\mathscr{A}\right],
\end{align}
where $\mathscr{A}$ is the antipodal map on $S^2$.
\end{namedtheorem}

\subsection{Outline of the paper}

In \Cref{subsection 2.1 schwarzschild in dng} we define the background Eddington--Finkelstein double null foliation and other $(3+1)$ coordinate systems that we will use throughout the paper, followed by listing certain elliptic estimates for tensor fields on $S^2$ in addition to basic Sobolev, Poincar\'e and Sobolev estimates. The linearised Einstein equations in a double null gauge, derived in \cite{DHR16}, are listed in \Cref{subsection 2.2 equations in double null gauge}, and the Teukolsky and Regge--Wheeler equations (as well as related identities) are derived in \Cref{TRW}.\\

In \Cref{Section 5: well-posedness for IVP} we study the well-posedness of the linearised Einstein equations in a double null gauge, classify the pure gauge solutions arising from infinitesimal diffeomorphisms preserving the double null structure, define the linearised Kerr family as solutions to the linearised system, and provide a detailed treatment of the gauge fixing scheme defined above for initial data. \Cref{Section 7 BMS} is devoted to the gauge conditions defined on future/past null infinity, while \Cref{Section 7: horizon normalised gauges} is devoted to the gauge conditions defined on the future/past event horizons. \Cref{Section 7: statement of results} contains a detailed statement of the theorems described above.\\ 

Forwards scattering theory towards the future null infinity and event horizon are studied in full detail in \Cref{Section 7 forward scattering}. In particular, the transition to a Bondi-normalised frame in forward scattering is described in \Cref{Section 10 Passing to Bondi gauge}, deriving in particular boundedness estimates on the associated gauge solution, and in \Cref{Section 11 Horizon gauge} the solution obtained in \Cref{Section 7 forward scattering} is made future horizon normalised, after which is a brief description of the analogous construction towards past null infinity and event horizon in \Cref{Section 8.7: Scattering from Sigmabar to H- and I-}. In \Cref{Section 6 Backwards scattering}, scattering from future/past null infinity and event horizon is studied. Finally, \Cref{Section 14: global scattering problem} contains a proof of Theorem III, and the linear memory effect is studied in \Cref{Section 15 Proof of Theorem III}, where Corollary II is proven.

\section{Preliminaries}
\subsection{The Schwarzschild exterior in a double null gauge}\label{subsection 2.1 schwarzschild in dng}

\indent Denote by $\mathscr{M}$ the exterior of the maximally extended Schwarzschild spacetime. Denoting by $S^2$ the standard unit sphere, we define the Schwarzschild exterior $\mathscr{M}$, using Kruskal coordinates, as the manifold with corners 
\begin{align}
    \mathscr{M}=\{(U,V,\theta^A)\in(-\infty,0]\times [0,\infty)\times S^2\}
\end{align}
The function $r(U,V)$ is determined by $-UV=\left(\frac{r}{2M}-1\right)e^{\frac{r}{2M}}$, $(\theta^A)$ is a coordinate system on $S^2$ and $\gamma_{AB}$ is the standard metric on the unit sphere $S^2$. The time-orientation of $\mathscr{M}$ is defined by the vector field $\partial_U+\partial_V$. 

The boundary of $\mathscr{M}$ consists of the two null hypersurfaces
\begin{align}
    \mathscr{H}^+&=\{0\}\times(0,\infty)\times S^2,\\
    \mathscr{H}^-&=(-\infty,0)\times \{0\}\times S^2,
\end{align}
and the 2-sphere $\mathcal{B}$ where $\mathscr{H}^+$ and $\mathscr{H}^-$ bifurcate:
\begin{align}
    \mathcal{B}=\{U,V=0\}\cong S^2 .
\end{align}
We define $\overline{\mathscr{H}^+}=\mathscr{H}^+\cup \mathcal{B}$, $\overline{\mathscr{H}^-}=\mathscr{H}^-\cup \mathcal{B}$. \\
\indent The interior of $\mathscr{M}$ can be covered with the familiar Schwarzschild coordinates $(t,r,\theta^A)$ and the metric takes the form \bref{Schwarzschild}, i.e.
\begin{align}
    ds^2=-\left(1-\frac{2M}{r}\right)dt^2+\left(1-\frac{2M}{r}\right)^{-1}dr^2+r^2\gamma_{AB}d\theta^Ad\theta^B.
\end{align}
Let $\Omega^2=\left(1-\frac{2M}{r}\right)$. It will be convenient to work instead in Eddington--Finkelstein coordinates
\begin{align}\label{EF null coordinates}
    u=\frac{1}{2}(t-r_*),\qquad\qquad\qquad v=\frac{1}{2}(t+r_*),
\end{align}
where $r_*$ is defined by
\begin{align}
    r_*=r+2M\log\left(\frac{r}{2M}-1\right)-2M
\end{align}
The coordinates $(u,v,\theta^A)$ also define a double null foliation (see Appendix B) of the interior of $\mathscr{M}$ since the metric takes the form
\begin{align}\label{DNGSchw}
    ds^2=-4\left(1-\frac{2M}{r}\right)dudv+r(u,v)^2(d\theta^2+\sin^2\theta d\phi^2).
\end{align}
In particular the null frame defined by the coordinates \bref{EF null coordinates} is given by (see Appendix B):
\begin{align}
    e_3=\frac{1}{\Omega}\partial_u,\qquad\qquad e_4=\frac{1}{\Omega}\partial_v.
\end{align}
We may relate $U,V$ to $u,v$ after fixing the residual freedom in defining $t,r_*$ by
\begin{align}\label{Kruskal}
    U=-e^{-\frac{u}{2M}},\qquad\qquad V=e^{\frac{v}{2M}},
\end{align}
Note that the intersections of null hypersurfaces of constant $u,v$ are spheres with metric $\slashed{g}_{AB}:=r^2\gamma_{AB}$. We denote these spheres by $S^2_{u,v}$. \\

\indent The $(u,v)$-coordinate system degenerates on $\overline{\mathscr{H}^+}$ and $\overline{\mathscr{H}^-}$ where $u=\infty,v=-\infty$ respectively. To compensate for this we can use the Kruskal coordinates to introduce weighted quantities in the coordinates $(u,v,\theta^A)$ that are regular on $\mathscr{H}^\pm$.
The Kruskal frame and the Eddington--Finkelstein frame are related by 
\begin{align}
    &\frac{UV}{\Omega^2}=-\fr(r),\qquad\qquad \fr(r)=\frac{r}{2M}e^{\frac{r\Omega^2}{2M}}\\
    &\partial_v=\frac{V}{2M}\partial_V,\qquad\qquad \partial_u=-\frac{U}{2M}\partial_U.
\end{align}
We also define the parameters $T,R$ by
\begin{align}\label{T R Kruskal coordinates}
    T=V+U,\qquad\qquad R=V-U.
\end{align}

Thus
\begin{align}
\partial_T=\frac{1}{2}\left(\partial_U+\partial_V\right)=2M\left(\frac{1}{V}\partial_v-\frac{1}{U}\partial_u\right),\qquad \partial_R=\frac{1}{2}\left(\partial_V-\partial_U\right).
\end{align}

We note already at this stage that the regularity of $\partial_U,\partial_V$ on the event horizons implies that $\frac{1}{\Omega}e_3, \Omega e_4$ are regular on $\mathscr{H}^+$ and $\frac{1}{\Omega}e_4, \Omega e_3$ are regular on $\mathscr{H}^-$ (but not $\overline{\mathscr{H}^\pm}$, which include $\mathcal{B}$). In the Kruskal frame, the metric has the form
\begin{align}\label{metric Kruskal}
    g=-4\Omega^2_{K}(U,V)dUdV+r^2(U,V)(d\theta^2+\sin^2\theta d\phi^2),
\end{align}
where
\begin{align}
    \Omega^2_{K}(U,V)=\frac{8M^3}{r(U,V)}e^{-\frac{r(U,V)}{2M}}.
\end{align}

\indent We denote by $\mathscr{C}_{u^*}$ the ingoing null hypersurface of constant $u=u^*$, and similarly $\underline{\mathscr{C}}_{v^*}$ denotes the outgoing null hypersurface $v=v^*$; define $\mathscr{C}_{u^*}\cap[v_1,v_2]$ to be the subset of $\mathscr{C}_{u^*}$ for which $v\in[v_1,v_2]$, $\underline{\mathscr{C}}_v\cap[u_1,u_2]$ denotes the subset of  $\underline{\mathscr{C}}_v$ for which $u\in[u_-,u_+]$. We denote the spacetime region bounded by $\mathscr{C}_{u_0}\cap[v_0,v_1], \mathscr{C}_{u_1}\cap[v_0,v_1], \underline{\mathscr{C}}_{v_0}\cap[u_0,u_1], \underline{\mathscr{C}}_{v_1}\cap[u_0,u_1]$ by $\mathscr{D}^{u_1,v_1}_{u_0,v_0}$. We also denote the spacetime region bounded by $\mathscr{C}_u,\underline{\mathscr{C}}_v, {\Sigma^*_+}$ by $\mathscr{D}^{u,v}_{{\Sigma^*_+}}$.

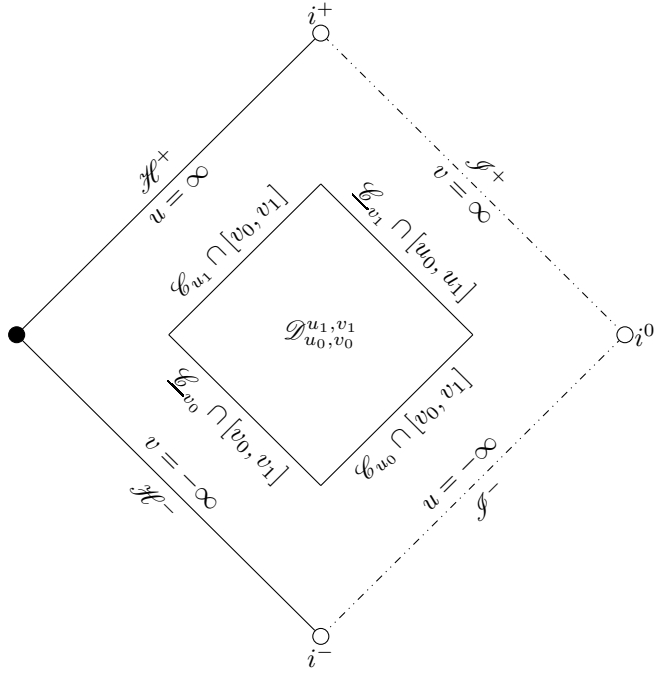
\begin{center}
\begin{tikzpicture}[scale=1]
\node (I)    at ( 0,0)   {$\mathscr{D}^{u_1,v_1}_{u_0,v_0}$};

\path 
   (I) +(90:4)  coordinate (Itop) coordinate[label=90:$i^+$]
       +(-90:4) coordinate (Ibot) coordinate[label=-90:$i^-$]
       +(180:4) coordinate (Ileft)
       +(0:4)   coordinate (Iright) coordinate[label=0:$i^0$]
       ;

\path 
   (I) +(90:2)  coordinate (Ictop)
       +(-90:2) coordinate (Icbot)
       +(180:2) coordinate (Icleft)
       +(0:2)   coordinate (Icright)
       ;

\draw  (Ileft) -- node[rotate=45,below] {$u=\infty$} node[rotate=45,above]{$\mathscr{H}^+$} (Itop) ;
\draw  (Ileft) -- node[rotate=-45,above] {$v=-\infty$} node[rotate=-45,below]{$\mathscr{H}^-$}(Ibot) ;
\draw[dash dot dot] (Ibot) -- node[rotate=45,above] {$u=-\infty$} node[rotate=45,below]{$\mathscr{I}^-$}(Iright) ;
\draw[dash dot dot] (Iright) -- node[rotate=-45,below] {$v=\infty$} node[rotate=-45,above]{$\mathscr{I}^+$}(Itop) ;


\draw(Icleft) --node[rotate=45,above] {$\mathscr{C}_{u_1}\cap[v_0,v_1]$} (Ictop);
\draw(Ictop) -- node[rotate=-45,above] {$\underline{\mathscr{C}}_{v_1}\cap[u_0,u_1]$}(Icright);
\draw(Icright) -- node[rotate=45,below] {$\mathscr{C}_{u_0}\cap[v_0,v_1]$}(Icbot);
\draw(Icbot) -- node[rotate=-45,below] {$\underline{\mathscr{C}}_{v_0}\cap[v_0,v_1]$}(Icleft);


\filldraw[white] (Itop) circle (3pt);
\draw[black] (Itop) circle (3pt);

\filldraw[white] (Ibot) circle (3pt);
\draw[black] (Ibot) circle (3pt);

\filldraw[white] (Iright) circle (3pt);
\draw[black] (Iright) circle (3pt);
\filldraw[black] (Ileft) circle (3pt);

\end{tikzpicture}
\end{center}

Let ${\Sigma}$ be the spacelike surface $\{t=0\}$ and let $\overline{\Sigma}=\Sigma\cup\mathcal{B}$ be the topological closure of $\Sigma$ in $\mathscr{M}$. $\overline{\Sigma}$ is a smooth Cauchy surface for $\mathscr{M}$ which connects $\mathcal{B}$ with ``spacelike infinity"; in Kruskal coordinates it is given by $\{U+V=0\}$. We also work with a spacelike hypersurface ${\Sigma^*_+}$
intersecting $\mathscr{H}^+$ to the future of $\mathcal{B}$, defined as follows: let 
\begin{align}\label{t star}
    t^*_+=t+2M\log\left(\frac{r}{2M}-1\right)=v-r+2M.
\end{align}
The function $t^*_+$ can be extended to $\overline{\mathscr{H}^\pm}$ to define a smooth function on all of $\mathscr{M}$. We denote by ${\Sigma^*_+}$ the hypersurface
\begin{align}\label{def of Sigmastar future}
    {\Sigma^*_+}=\{t^*_+=0\}.
\end{align}
\noindent Note that ${\Sigma^*_+}$ intersects $\mathscr{H}^+$ at $v=0$ and asymptotes to spacelike infinity. Define $\mathscr{H}^+_{\geq 0}:=\mathscr{H}^+\cap  J^+({\Sigma^*_+})$.

Analogously, we may define $t^*_-$ by
\begin{align}
    t^*_-=-t+2M\log\left(\frac{r}{2M}-1\right)=-u-r+2M.
\end{align}
Analogously to \bref{def of Sigmastar future}, we define
\begin{align}\label{def of Sigmastar future}
    {\Sigma^*_-}=\{t^*_-=0\}.
\end{align}
Note that $\Sigma^*_-$ intersects $\mathscr{H}^-$ at $u=0$. We define $\mathscr{H}^-_{\leq0}:=\mathscr{H}^-\cap J^-(\Sigma^*_-)$. For any $u$ we define $\vsigmap{u}$ by demanding $S^2_{u,\vsigmap{u}}\in \Sigma^*_+$. For any $v$ we define $\usigmap{u}$ by $S^2_{\usigmap{u},v}\in \Sigma^*_+$. We similarly define $\vsigmam{u}$, $\usigmam{v}$ for any $u$, $v$ respectively. The functions 

In addition to $\Sigma^*_\pm$, $\Sigma$ and $\overline{\Sigma}$, we define
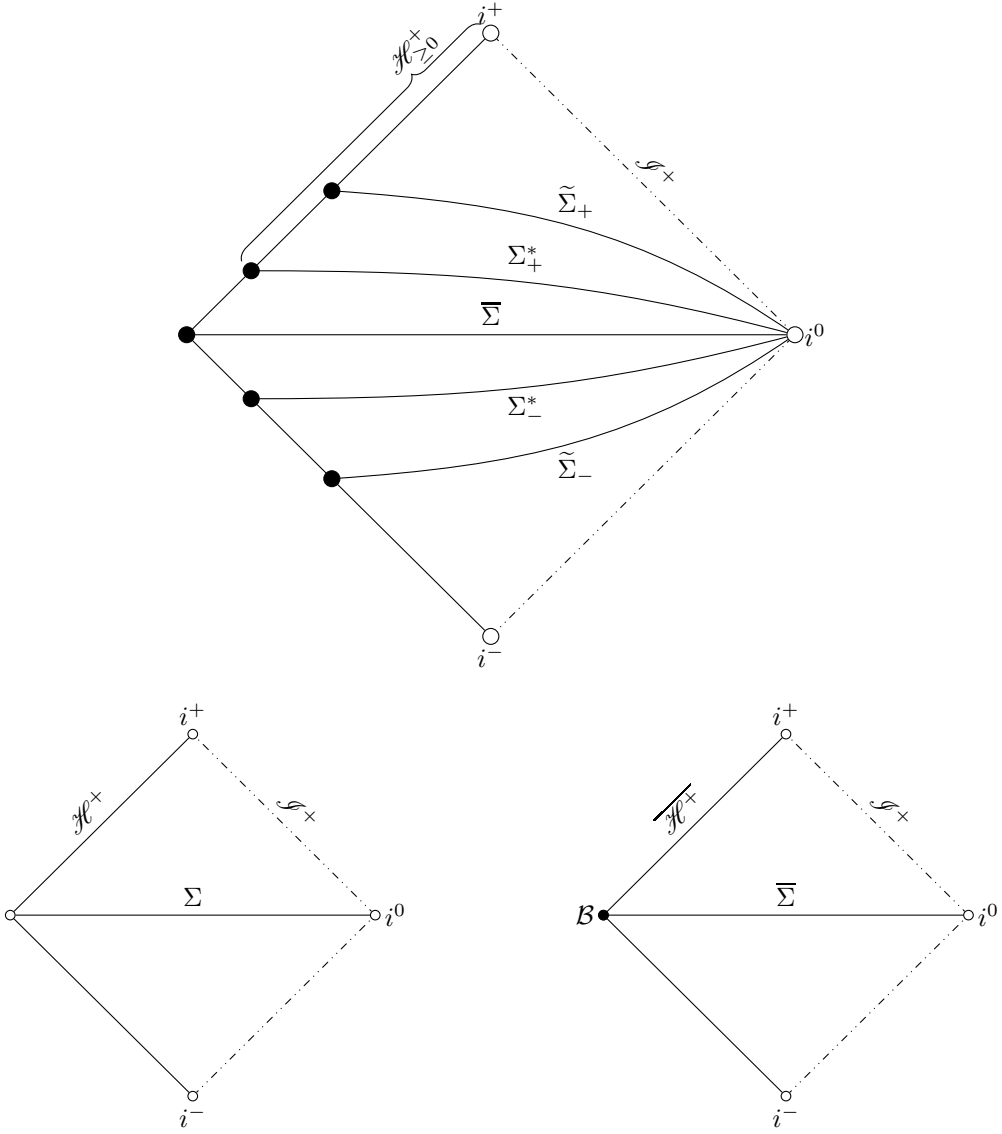
\begin{align}
    \widetilde{\Sigma}_+=\{r=v\},\qquad\qquad \widetilde{\Sigma}_-=\{r=|u|\}.
\end{align}

\begin{center}
\begin{tikzpicture}[scale=1]
\node (I)    at ( 0,0)   {};

\path 
   (I) +(90:4)  coordinate (Itop) coordinate[label=90:$i^+$]
       +(-90:4) coordinate (Ibot) coordinate[label=-90:$i^-$]
       +(180:4) coordinate (Ileft)
       +(0:4)   coordinate (Iright) coordinate[label=0:$i^0$]
       ;

\draw  (Ileft) --  (Itop) ;

\draw [black,
    decorate, 
    decoration = {brace,
        raise=5pt,
        amplitude=5pt,
        aspect=0.75}] ($(Ileft)+(45:1.2)$) --  ($(Itop)$)
node[pos=0.75,above=8pt,black,rotate=45]{$\mathscr{H}^+_{\geq0}$};

\draw  (Ileft) -- (Ibot) ;
\draw[dash dot dot] (Ibot) -- (Iright) ;
\draw[dash dot dot] (Iright) --  node[rotate=-45,above]{$\mathscr{I}^+$}(Itop) ;


\draw ($(Ileft)+(45:1.2)$)  to[out=-0, in=165, edge node={node [above] {${\Sigma^*_+}$}}] ($(Iright)$);

\draw ($(Ileft)+(45:2.7)$)  to[out=-4, in=145, edge node={node [above] {${\widetilde{\Sigma}_+}$}}] ($(Iright)$);

\draw ($(Ileft)$)  to[out=0, in=180, edge node={node [above] {$\overline{\Sigma}$}}] ($(Iright)$);

\draw ($(Ileft)+(-45:2.7)$)  to[out=4, in=215, edge node={node [below] {${\widetilde{\Sigma}_-}$}}] ($(Iright)$);

\draw ($(Ileft)+(-45:1.2)$) to[out=-0, in=195, edge node={node [below] {${\Sigma^*_-}$}}] ($(Iright)$);




\filldraw[white] (Itop) circle (3pt);
\draw[black] (Itop) circle (3pt);

\filldraw[white] (Ibot) circle (3pt);
\draw[black] (Ibot) circle (3pt);

\filldraw[white] (Iright) circle (3pt);
\draw[black] (Iright) circle (3pt);
\filldraw[black]($(Ileft)+(45:1.2)$) circle (3pt);
\filldraw[black]($(Ileft)+(45:2.7)$) circle (3pt);

\filldraw[black]($(Ileft)$) circle (3pt);

\filldraw[black]($(Ileft)+(-45:1.2)$) circle (3pt);
\filldraw[black]($(Ileft)+(-45:2.7)$) circle (3pt);
\end{tikzpicture}
\end{center}

\begin{center}
\begin{tikzpicture}[scale=0.6]
\node (I)    at ( 0,0)   {};

\path 
   (I) +(90:4)  coordinate (Itop) coordinate[label=90:$i^+$]
       +(-90:4) coordinate (Ibot) coordinate[label=-90:$i^-$]
       +(180:4) coordinate (Ileft)
       +(0:4)   coordinate (Iright) coordinate[label=0:$i^0$]
       ;

\draw  (Ileft) --  node[rotate=45,above]{$\mathscr{H}^+$} (Itop) ;
\draw  (Ileft) --(Ibot) ;
\draw[dash dot dot] (Ibot) -- (Iright) ;
\draw[dash dot dot] (Iright) --  node[rotate=-45,above]{$\mathscr{I}^+$}(Itop) ;

\draw ($(Ileft)$)  to[out=0, in=180, edge node={node [above] {$\Sigma$}}] ($(Iright)$);

\filldraw[white] (Itop) circle (3pt);
\draw[black] (Itop) circle (3pt);

\filldraw[white] (Ibot) circle (3pt);
\draw[black] (Ibot) circle (3pt);

\filldraw[white] (Iright) circle (3pt);
\draw[black] (Iright) circle (3pt);
\filldraw[white] (Ileft) circle (3pt);
\draw[black] (Ileft) circle (3pt);
\end{tikzpicture}\hspace{2cm}\begin{tikzpicture}[scale=0.6]
\node (I)    at ( 0,0)   {};

\path 
   (I) +(90:4)  coordinate (Itop) coordinate[label=90:$i^+$]
       +(-90:4) coordinate (Ibot) coordinate[label=-90:$i^-$]
       +(180:4) coordinate (Ileft) coordinate[label=180:$\mathcal{B}$]
       +(0:4)   coordinate (Iright) coordinate[label=0:$i^0$]
       ;

\draw  (Ileft) --  node[rotate=45,above]{$\overline{\mathscr{H}^+}$} (Itop) ;
\draw  (Ileft) --(Ibot) ;
\draw[dash dot dot] (Ibot) -- (Iright) ;
\draw[dash dot dot] (Iright) --  node[rotate=-45,above]{$\mathscr{I}^+$}(Itop) ;

\draw ($(Ileft)$)  to[out=0, in=180, edge node={node [above] {$\overline{\Sigma}$}}] ($(Iright)$);

\filldraw[white] (Itop) circle (3pt);
\draw[black] (Itop) circle (3pt);

\filldraw[white] (Ibot) circle (3pt);
\draw[black] (Ibot) circle (3pt);

\filldraw[white] (Iright) circle (3pt);
\draw[black] (Iright) circle (3pt);
\filldraw[black] (Ileft) circle (3pt);
\draw[black] (Ileft) circle (3pt);
\end{tikzpicture}
\end{center}

\subsubsection*{The coordinates $(t^*_+,\overline{u})$}

To work with initial data on the surface ${\Sigma^*_+}$, we will use the coordinate system $(t^*_+,\overline{u}^*,\theta^A)$, where $t^*_+$ is as in \bref{t star} and $u^*=u$. 
We then have
\begin{align}
    dt^*_+=dt+\frac{2M}{r\Omega^2}dr=dv+du+\left(1-\Omega^2\right)(dv-du)=(2-\Omega^2)dv+\Omega^2du.
\end{align}
The Jacobian of the transformation $(u,v)\longrightarrow(\ustar,t^*_+)$ and its inverse are
\begin{align}
    J=\begin{pmatrix}
    \frac{\partial t^*_+}{\partial v} & \frac{\partial t^*_+}{\partial u}\\[6pt]
    \frac{\partial \ustar}{\partial v} & \frac{\partial \ustar}{\partial u}
    \end{pmatrix}=\begin{pmatrix}
    2-\Omega^2 & \Omega^2\\[6pt]
    0 & 1
    \end{pmatrix}
    \qquad\qquad
    J^{-1}=\begin{pmatrix}
    \frac{\partial v}{\partial t^*_+} & \frac{\partial v}{\partial \ustar}\\[6pt]
    \frac{\partial u}{\partial t^*_+} & \frac{\partial u}{\partial \ustar}
    \end{pmatrix}
   = \frac{1}{2-\Omega^2}\begin{pmatrix}
    1 & -{\Omega^{2}}\\[6pt]
    0 & \;2-\Omega^2
    \end{pmatrix}
\end{align}
The background Schwarzschild metric now takes the form
\begin{align}\label{Schw metric in t* u* coordinates}
    g=-4\Omega^2dudv+\slashed{g}_{AB}d\theta^A d\theta^B&=-\frac{4\Omega^2}{2-\Omega^2}d\ustar dt^*_++\frac{4\Omega^4}{2-\Omega^2}d{\ustar}^2+\slashed{g}_{AB}d\theta^A d\theta^B,
\end{align}
and the inverse metric has components
\begin{align}
    g^{{t^*_+}{t^*_+}}=-(2-\Omega^2),\qquad\qquad g^{\ustar{t^*_+}}=-\frac{2-\Omega^2}{2\Omega^2},\qquad\qquad g^{AB}=\slashed{g}^{AB}.
\end{align}

The unit normal vector to ${\Sigma^*_+}$ is given by
\begin{align}
    n_\mu=\frac{1}{\sqrt{2-\Omega^2}}\delta_{\mu t^*_+},\qquad n^\mu=-\frac{1}{\sqrt{2-\Omega^2}}g^{\mu t^*_+}=\frac{1}{\sqrt{2-\Omega^2}}\begin{pmatrix} 2-\Omega^2 \\ \frac{2-\Omega^2}{2\Omega^2} \\ 0 \\ 0 \end{pmatrix}
\end{align}
We also define the vector field
\begin{align}
    N:=\frac{1}{2}\left(\frac{1}{\Omega^2}\partial_u+\frac{1}{2-\Omega^2}\partial_v\right).
\end{align}
We then have
\begin{align}
    \partial_v=(2-\Omega^2)\left(N-\frac{1}{2\Omega^2}\partial_\ustar\right),\quad\qquad \partial_u=\Omega^2\left(N+\frac{1}{2\Omega^2}\partial_\ustar\right),\qquad\quad \frac{1}{\Omega^2}\partial_\ustar=\frac{1}{\Omega^2}\partial_u-\frac{1}{2-\Omega^2}\partial_v.
\end{align}
\subsubsection*{Null infinity $\mathscr{I}^\pm$}

We define the notion of null infinity by directly attaching it as a boundary to $\mathscr{M}$. Define $\mathscr{I}^+,\mathscr{I}^-$ to be the manifolds
\begin{align}
\mathscr{I}^+,\mathscr{I}^-:=\mathbb{R}\times S^2
\end{align}
and define $\overline{\mathscr{M}}$ to be the extension 
\begin{align}
\overline{\mathscr{M}}=\mathscr{M}\cup\mathscr{I}^+\cup\mathscr{I}^-.
\end{align}
For sufficiently large $R$ and any open set $\mathcal{O}\subset\mathbb{R}\times S^2$, declare the sets $\mathcal{O}^+_R=(R,\infty]\times\mathcal{O}$ to be open in $\overline{\mathscr{M}}$, identifying $\mathscr{I}^+$ with the points $(u,\infty,\theta,\phi)$. To the set $\mathcal{O}_R^+$ we assign the coordinate chart $(u,s,\theta,\phi)\in \mathbb{R}\times[0,1)\times S^2$ via the map
\begin{align}
(u,v,\theta,\phi)\longrightarrow(u,\frac{R}{v},\theta,\phi),
\end{align}
where $(u,v,\theta,\phi)$ are the Eddington--Finkelstein coordinates we defined earlier. The limit $\lim_{v\longrightarrow\infty} (u,v,\theta,\phi)$ exists and is unique, and we use it via the above charts to fix a coordinate system $(u,\theta,\phi)$ on $\mathscr{I}^+$. The same can be repeated to define an atlas attaching $\mathscr{I}^-$ as a boundary to $\overline{\mathscr{M}}$.
\subsubsection{$S^2_{u,v}$-projected connection and angular derivatives}\label{D1D2}
We will be working primarily with tensor fields that are everywhere tangential to the $S^2_{u,v}$ spheres foliating $\mathscr{M}$. By this we mean any tensor fields of type $(k,l)$, $\digamma\in \mathcal{T}^{(k,l)}\mathscr{M}$ on $\mathscr{M}$ such that for any point $q=(u,v,\theta^A)\in\mathscr{M}$ we have $\digamma|_q\in \mathcal{T}^{(k,l)}_{(\theta^A)}S^2_{u,v}$. (Note that a vector $X^A\in \mathcal{T}_{(\theta^A)}S^2_{u,v}$ is canonically identified with a vector $X^a\in\mathcal{T}_q\mathscr{M}$ via the inclusion map, whereas we make the identification of a 1-form $\eta_A\in\mathcal{T}^*_{(\theta^A)}\mathscr{M}$ as an element in the cotangent bundle of $\mathscr{M}$ by declaring that $\eta(X)=0$ if $X$ is in the orthogonal complement of $\mathcal{T}S^2_{u,v}$ under the spacetime metric \bref{metric Kruskal}.) We will refer to such tensor fields as ``$S^2_{u,v}$-tangent" tensor fields in the following. It will also be convenient to work with an ``$S^2_{u,v}$ projected" version of the covariant derivative belonging to the Levi-Civita connection of the metric \bref{Schwarzschild}. We define these notions as follows:\\ 
\indent We denote by $\slashed{\nabla}_A$ (or sometimes simply $\slashed{\nabla}$) the covariant derivative on $S^2_{u,v}$ with the metric $\slashed{g}_{AB}$. Note that $r\slashed{\nabla}=\slashed{\nabla}_{\mathbb{S}^2}$ which we also denote by $\mathring{\slashed{\nabla}}$.\\
\indent For an $S^2_{u,v}$-tangent 1-form $\xi$, define $\fancyd_1 \xi$ to be the pair of functions
\begin{align}
    \fancyd_1{\xi}=(\slashed{\text{div}}\xi,\slashed{\text{curl}}\xi),
\end{align}
where $\slashed{\text{div}}\xi=\slashed{\nabla}^A \xi_A$ and $\slashed{\text{curl}}\xi=\slashed{\epsilon}^{AB} \slashednabla_A\xi_B$. Similarly, define
\begin{align}
    \overline{\fancyd}_1\xi:=(\slashed{\text{div}}\xi,-\slashed{\text{curl}}\xi).
\end{align}
For an $S^2_{u,v}$-tangent symmetric 2-tensor $\Xi_{AB}$ we define $\fancyd_2 \Xi$ to be the 1-form given by
\begin{align}
    (\fancyd_2 \Xi)_A:=(\slashed{\text{div}}\,\Xi)_A:=\slashednabla^B \Xi_{BA}.
\end{align}
\indent We define the operator $\fancydstar_1 $ to be the $L^2({S^2_{u,v}})$-dual to $\fancyd_1$. For scalars $(f,g)$ the 1-form $\fancydstar_1(f,g)$ is given by
\begin{align}\label{def of 0 to 1 spin}
    \fancydstar_1(f,g)=-\slashednabla_A f +\epsilon_{AB}\slashednabla^B g.
\end{align}
\indent Similarly we denote by $\fancydstar_2$ the $L^2({S^2_{u,v}})$-dual to $\fancyd_2$. For an $S^2_{u,v}$-tangent 1-form $\xi$ this is given by 
\begin{align}\label{def of 1 to 2 spin}
    (\fancydstar_2\xi)_{AB}=-\frac{1}{2}\left(\slashednabla_A\xi_B+\slashednabla_B\xi_A-\slashed{g}_{AB}\divr\xi\right).
\end{align}
We also use the notation
\begin{align}
\begin{split}
    \fancydring_1:=r\fancyd_1,\qquad\qquad\fancydstarring_1:=r\fancydstar_1,\\
    \fancydring_2:=r\fancyd_2,\qquad\qquad\fancydstarring_2:=r\fancydstar_2.
\end{split}
\end{align}
For example, if $\xi$ is a 1-form on $S^2_{u,v}$ then
\begin{align}
    ({\fancydstarring_2\xi})_{AB}=-\frac{1}{2}\left(\mathring{\slashednabla}_A\xi_B+\mathring{\slashednabla}_B\xi_A-\slashed{g}_{AB}\divo\xi\right).
\end{align}
and so on. We also define $\divo:=r\,\divr$, $\curlo:=r\,\curlr$.\\

We denote by $D\xi$ and $\underline{D}{\xi}$ the projected Lie derivative of $\xi$ in the $3$- and $4$-directions respectively. In Eddington--Finkelstein coordinates we have
\begin{align}
    (D\xi)_{A_1 A_2...A_n}=\partial_u(\xi_{A_1 A_2 ... A_n})\qquad (\underline{D}\xi)_{A_1 A_2...A_n}=\partial_v(\xi_{A_1 A_2 ... A_n}).
\end{align}
Similarly, we define $\nablagml\xi$ and $\nabladlt \xi$ to be the projections of the covariant derivatives $\nabla_3 \xi$ and $\nabla_4 \xi$ to $S^2_{u,v}$. 
Following the definitions above, we will need the  functional spaces defined below throughout:

\begin{defin}
Let $\mathcal{S}$ be a smooth hypersurface in $\mathcal{M}$. We say that a tensor field $T$ on $\mathcal{S}$ is an $S^2_{u,v}$-tangent tensor if for all $p=(u_p,v_p,\theta^A)\in\mathcal{S}$, $T|_p$ is equal to its projection onto the tangent bundle of $S^2_{u_p,v_p}$. Define
\begin{itemize}
\item $\Gamma^{(2)}(\mathcal{S})$ to be the space of smooth symmetric traceless $S^2_{u,v}$-2-tensors,
\item $\Gamma^{(1)}(\mathcal{S})$ to be the space of smooth $S^2_{u,v}$-1-forms,
\item $\Gamma^{(0)}(\mathcal{S})$ to be the space $C^\infty(\mathcal{S})$ of smooth scalar fields on $\mathcal{S}$.
\end{itemize}
We denote by $\Gamma^{(i)}_{c}(\mathcal{S})$ the versions of the spaces defined above consisting of smooth, compactly supported fields of each respective type for $i=0,1,2$.
\end{defin}

\subsubsection{Elliptic estimates on $S^2_{u,v}$}\label{subsubsection 2.1.2 Elliptic estimates on S2}

For a $k$-covariant $S^2_{u,v}$-tangent tensor field $\Xi$ on $\mathscr{M}$, define 
\begin{align}
    |\Xi|=\sqrt{\gamma^{A_1B_1}\gamma^{A_2B_2}\cdot\cdot\cdot\gamma^{A_pB_p}\Xi_{A_1...A_p}\Xi_{B_1...B_p}},\qquad |\Xi|_{S^2_{u,v}}=r(u,v)^{-k}|\Xi|,
\end{align}

We will use the notation $|\Xi_{A_1\dots A_p}|$ to denote the absolute value of the component $\Xi_{A_1\dots A_p}$ of $\Xi$ in a given coordinate system on $S^2$.

The operators $\fancyd_1,\fancyd_2,\fancydstar_1,\fancydstar_2$ defined in \Cref{D1D2} can be combined to give
\begin{align}\label{D1 D2 into Laplacians}
\begin{split}
    &-2r^2\fancydstar_2\fancyd_2=\mathring{\slashed{\Delta}}-2\qquad\qquad-r^2\fancydstar_1\fancyd_1=\mathring{\slashed{\Delta}}-1\\
    &-2r^2\fancyd_2\fancydstar_2=\mathring{\slashed{\Delta}}+1\qquad\qquad -r^2\fancyd_1\fancydstar_1=\mathring{\slashed{\Delta}}.
\end{split}
\end{align}
See also Lemma 2.2.1 of \cite{Ch-K}. The operator $\mathring{\slashed\Delta}$ is the Laplacian on the unit 2-sphere $S^2$. Note in particular:
\begin{align}
\begin{split}
    2r^4\fancydstar_2\fancydstar_1\fancyd_1\fancyd_2&=-2r^2\fancydstar_2\left(\mathring{\slashed{\Delta}}-1\right)\fancyd_2=2r^2\fancydstar_2\left(2+2r^2\fancyd_2\fancydstar_2\right)\fancyd_2\\&=4r^2\fancyd_2\fancydstar_2+\left(-2r^2\fancyd_2\fancydstar_2\right)^2=\mathcal{A}_2\left(\mathcal{A}_2-2\right),
\end{split}
\end{align}
using the notation
\begin{align}
    \mathcal{A}_2:=-2r^2\fancydstar_2\fancyd_2=\mathring{\slashed{\Delta}}-2.
\end{align}
The following is extracted from Section 4.4 of \cite{DHR16}:
\begin{proposition}
Let $\xi$ be a smooth symmetric traceless $S^2_{u,v}$ 1-form. We have the following identities:
\begin{align}\label{D1 controls 1 derivative}
    \int_{S^2_{u,v}}\dw\,\left[|\slashednabla\xi|^2+K|\xi|^2\right]=\int_{S^2_{u,v}}\dw\, |\fancyd_1\xi|^2,
\end{align}
\begin{align}\label{D1D1 controls 2 derivatives}
    \int_{S^2_{u,v}}\dw\,\left[|\slashed{\Delta}\xi|^2+K^2|\xi|^2+K|\slashednabla\xi|^2\right]=\int_{S^2_{u,v}}\dw\,|\fancydstar_1\fancyd_1\xi|^2.
\end{align}
\end{proposition}
\begin{proposition}
Let $\Xi$ be a smooth symmetric traceless $S^2_{u,v}$ 2-tensor. We have the following identities:
\begin{align}\label{D2 controls 1 derivative}
    \int_{S^2_{u,v}}\sin\theta d\theta d\phi\left[ |\slashednabla\Xi|^2+2K|\Xi|^2\right]=2\int_{S^2_{u,v}} \sin\theta d\theta d\phi |\fancyd_2 \Xi|^2,
\end{align}
\begin{align}\label{A2 controls 2 derivatives}
    \int_{S^2_{u,v}}\sin\theta d\theta d\phi\left[\frac{1}{4}|\slashed{\Delta}\Xi|^2+K|\Xi|^2 +K^2|\slashednabla\Xi|^2\right]=\int_{S^2_{u,v}} \sin\theta d\theta d\phi |\fancydstar_2\fancyd_2 \Xi|^2,
\end{align}
where $K=\frac{1}{r^2}$ is the Gaussian curvature of $S^2_{u,v}$. 
\end{proposition}
\begin{remark}\label{Kernel of D1 D2 is trivial}
Note that \bref{D1 controls 1 derivative} implies that the kernel of $\fancyd_1$ is trivial in the space of smooth 1-form fields on $S^2_{u,v}$. Similarly, \bref{D2 controls 1 derivative} implies that the kernel of $\fancyd_2$ is trivial in the space of smooth symmetric traceless 2-tensor fields on $S^2_{u,v}$. Using these facts and \bref{D1 D2 into Laplacians}, it is easy to prove the following
\end{remark}

\begin{proposition}\label{2-forms on S^2}
    Let $\Xi$ be a smooth symmetric traceless $S^2_{u,v}$-tangent 2-tensor field. Then there exist a unique pair $(f,g)$ of scalar functions supported on $\ell\geq2$ such that
    \begin{align}
        \Xi=r^2\fancydstar_2\fancydstar_1(f,g).
    \end{align}
\end{proposition}

\begin{proposition}\label{1-forms on S^2}
    Let $\xi$ be a smooth $S^2_{u,v}$-tangent 1-form. Then $\xi$ can be decomposed into
    \begin{align}
        \xi=\xi_{\ell=1}+\xi_{\ell\geq2},
    \end{align}
    where $\xi_{\ell\geq2}$ can be written as
    \begin{align}
        \xi_{\ell\geq2}=r^2\fancydstar_2\fancydstar_1(f,g)
    \end{align}
    for a unique pair $(f,g)$ of scalar functions supported on $\ell\geq2$, and $\xi_{\ell=1}$ is such that the scalars
    \begin{align}
        \slashed{div}\xi_{\ell=1}, \slashed{curl}\xi_{\ell=1}
    \end{align}
    are supported on $\ell=1$.
\end{proposition}

\begin{defin}\label{definition of conjugates}
For a smooth symmetric traceless $\mathcal{S}^2_{u,v}$ 2-tensor field $\Xi$, we define the \textbf{conjugate} of $\Xi$, denoted by $\overline{\Xi}$ as follows: let $f,g$ be the scalar functions generating $\Xi$ as in \Cref{2-forms on S^2}. Then $\overline{\Xi}$ is given by
\begin{align}
    \overline{\Xi}=2r^2\fancydstar_2\fancydstar_1(f,-g).
\end{align}
We make a similar definition for smooth $\mathcal{S}^2_{u,v}$ 1-tensor fields.
\end{defin}

\begin{lemma}
Let 
\begin{align}
    \mathcal{T}=r^2\fancydstar_2\fancydstar_1.
\end{align}
The kernel of $\mathcal{T}$ on $C^\infty(S^2)\times C^\infty(S^2)$ consists of pairs $(f,g)$ supported on $\ell=0,1$.
\end{lemma}
As a consequence, we have
\begin{lemma}
Let $\Xi$ be a smooth $S^2_{u,v}$-tangent 2-tensor field. There exists a unique smooth $S^2_{u,v}$-tangent 1-form $\xi$ that is supported on $\ell\geq2$ such that
\begin{align}
    \fancydstar_2\xi=\Xi.
\end{align}
\end{lemma}

The above reductions of $S^2_{u,v}$-tangent 1-forms and symmetric traceless 2-tensors into scalar functions on $S^2$ can be used to obtain the following Poincar\'e type inequalities:
\begin{lemma}\label{poincaresection}
Let $\Xi$ be a smooth symmetric traceless $S^2_{u,v}$ 2-tensor, then we have
\begin{align}\label{poincare}
    2K\int_{S^2_{u,v}}\sin\theta d\theta d\phi|\Xi|^2\leq \int_{S^2_{u,v}} \sin\theta d\theta d\phi|\slashednabla \Xi|^2
\end{align}
\end{lemma}
\begin{lemma}\label{scalar poincare}
    Let $(f_1,f_2)$ be a pair of scalar functions supported on $\ell\geq2$. Then we have
    \begin{align}\label{scalar poincare}
        \sum_{i=0}^2 \int_{S^2_{u,v}}\sin\theta d\theta d\phi\,\left(|\mathring{\slashednabla}^if_1|^2+|\mathring{\slashednabla}^if_2|^2\right)\lesssim \int_{S^2_{u,v}}\sin\theta d\theta d\phi\,|\fancydstarring_2\fancydstarring_1(f_1,f_2)|^2
    \end{align}
\end{lemma}

\begin{remark}
When integrating over $S^2_{u,v}$, we may occasionally abbreviate $\dw$ by $d\omega$.
\end{remark}

\subsubsection{Asymptotics of $S^2_{u,v}$-tensor fields}\label{subsubsection 2.1.3 Asymptotics of S2 tensor fields}
\begin{defin}\label{def of convergence}
Let $\digamma$ be a $k$-covariant $S^2_{u,v}$-tangent tensor field on $\mathscr{M}$. We say that $\digamma$ converges to $F=F_{A_1A_2...A_p}(u)$ as $v\longrightarrow\infty$ if $r^{-k}\digamma\longrightarrow F$ in the norm $|\;\;|_{S^2}$.
\end{defin}
Note that
\begin{align}
\begin{split}
    \left|\frac{1}{r^k}\digamma(u,v,\theta^A)-F(u,\theta^A)\right|_{S^2}&=\left|\int_{v}^\infty d\bar{v} \frac{\partial}{\partial v}\frac{1}{r^k}\digamma \right|_{S^2}\leq \int_{v}^\infty d\bar{v} \left|\frac{\partial}{\partial v}\frac{1}{r^k}\digamma\right|_{S^2}
    \\&=\int_{v}^{\infty}d\bar{v}\left|r^k\frac{\partial}{\partial v}\frac{1}{r^k}\digamma\right|=\int_{v}^\infty d\bar{v}|\nablav \digamma|.
\end{split}
\end{align}
Therefore, if $\nablav\digamma$ is integrable in $L^1(\underline{\mathscr{C}}_v)$ then $\digamma$ has a limit towards $\mathscr{I}^+$. It is easy to see that if $\{\digamma_n\}_n^\infty$ is a Cauchy sequence in $|\;\;|$ then $\digamma_n$ converges in the sense of \Cref{def of convergence}. The above extends to tensors of rank $(k,\tilde{k})$, where $r^{-k}$ is replaced by $r^{-k+\tilde{k}}$. Similar considerations apply when taking the limit towards $\mathscr{I}^-$.
For a symmetric tensor $\Psi$ of rank $(2,0)$, it will be simpler to work with $\Psi^{A}{}_B$. Note that $\nablav\Psi^A{}_B=\partial_v\Psi^A{}_B$, $\nablau\Psi^A{}_B=\partial_u \Psi^A{}_B$. Unless otherwise indicated, in taking the limits of symmetric traceless 2-tensor fields we will work with their  $S^2_{u,v}$-tangent $(1,1)$-tensor representatives instead. For the quantities $\bmlin$, $\elin$, $\eblin$, $\blin$, $\bblin$ we will always consider the limiting behaviour of the $1$-form representative.

\subsection{Hardy and Gr\"onwall inequalities}

We will need the following form of Gr\"onwall's inequality:

\begin{proposition}\label{Gronwall inequality}
Let $f,g,\lambda: \mathbb{R}\longrightarrow [0,\infty)$ be in $L^1_{c}(\mathbb{R})$ and assume $\lambda$ is non-decreasing. Assume that for a given $u_+$ and all $u\leq u_+$,
\begin{align}
    f(u)\leq \int_u^{u_+} f(\bar{u})g(\bar{g})d\bar{u} +\lambda(u),
\end{align}
then $f$ satisfies 
\begin{align}
    f(u)\leq \lambda(u) e^{\int_u^{u_+}g(\bar{u})d\bar{u}}\;,\qquad\qquad \int_{u}^{u_+}d\bar{u}\, g(\bar{u})f(\bar{u})\leq \lambda(u)\left[e^{\int_u^{u_+}d\bar{u}g(\bar{u})}-1\right]
\end{align}
\end{proposition}

Let us finally note the following technical lemma, which we will need to pointwise obtain decay from certain integrated decay estimates:

\begin{lemma}\label{technical lemma}
    Let $\mathtt{F}$ be a smooth tensor field on $J^+({\Sigma^*_+})$. Assume 
    \begin{align}
        \int_{J^+({\Sigma^*_+})\cap\{r\leq R\}} \Omega^2 du dv \sin\theta d\theta d\phi\; \left[|\mathtt{F}|^2+|\slashednabla_{t^*_+}\mathtt{F}|^2\right]<\infty,
    \end{align}
    then we have
    \begin{align}\label{result Morrey}
        \lim_{\lambda\longrightarrow\infty}\int_{\{t^*_+=\lambda, r\leq R\}} du \sin\theta d\theta d\phi \;|\mathtt{F}|^2=0
    \end{align}
\end{lemma}

\begin{proof}
Let $\mathtt{f}(\lambda)=\sqrt{\int_{\{t^*_+=\lambda, r\leq R\}} \Omega^2du \sin\theta d\theta d\phi \;|\mathtt{F}|^2}$ for $\lambda\in[0,\infty)$. Assume $v$ is such that $\mathtt{f}(v)=0$, then it is easy to show that $\frac{d}{d{\lambda}}\mathtt{f}|_{\lambda}=\sqrt{\int_{\{t^*_+=\lambda, r\leq R\}}\Omega^2 du \sin\theta d\theta d\phi \;|\slashednabla_{T^*_+}\mathtt{F}|^2}$. When $\mathtt{f}(\lambda)\neq0$, Cauchy--Schwarz implies
\begin{align}
\begin{split}
    \left|\frac{d}{d\lambda} \,\mathtt{f}\right|^2&=\left[\frac{1}{f(\lambda)}\int_{\{t^*_+=\lambda, r\leq R\}} \Omega^2 du \sin\theta d\theta d\phi \;\mathtt{F}\cdot\slashednabla_{t^*_+} \mathtt{F}\right]^2\\
    &\leq\int_{\{t^*_+=\lambda, r\leq R\}} \Omega^2du \sin\theta d\theta d\phi \;|\slashednabla_{T^*_+}\mathtt{F}|^2.
\end{split}
\end{align}
Thus $\int_{0}^\infty d\lambda \left[|\mathtt{f}|^2+\left|\frac{d}{d\lambda} \mathtt{f}\right|^2\right]<\infty$, which leads to \bref{result Morrey} by Morrey's inequality. 
\end{proof}

\begin{remark}
Note that $\partial_{t^*_+}$ generates an isometry of the Schwarzschild metric, as can be seen considering \bref{Schw metric in t* u* coordinates}.
\end{remark}

We will need the following application of Lebesgue's dominated convergence theorem, which we state here for definiteness:
\begin{proposition}\label{Lebesgue's BCT}
Assume $f(u,v,\theta^A)$ converges in $L^\infty([u_-,u_+]\times S^2)$ as $v\longrightarrow\infty$ to $f_{\mathscr{I^+}}(u,\theta^A)$. Then
\begin{align}
    \int_{u}^{u_+}d\bar{u} f(\bar{u},v,\theta^A)\longrightarrow \int_u^{u_+}d\bar{u} f_{\mathscr{I}^+}(u,\theta^A).
\end{align}
as $v\longrightarrow\infty$.
\end{proposition}

\section{The linearised Einstein equations in a double null gauge}\label{subsection 2.2 equations in double null gauge}

The linearisation of the Einstein equations \bref{EVE} in a double null gauge on a Schwarzschild background was performed in detail in \cite{DHR16}. Below we list the resulting linearised system.

\begin{defin}\label{def of linearised einstein equations and solutions}
The linearised Einstein equations in a double null gauge on a Schwarzschild exterior background are the following:

\begin{itemize}
    \item The transport equations governing the linearised metric components:
    \begin{alignat*}{2}
        \refstepcounter{equation}\latexlabel{metric transport in 4 direction trace}
        \refstepcounter{equation}\latexlabel{metric transport in 4 direction traceless}
        \refstepcounter{equation}\latexlabel{metric transport in 3 direction trace}
        \refstepcounter{equation}\latexlabel{metric transport in 3 direction traceless}
        \refstepcounter{equation}\latexlabel{partial_u b}
        \partial_v\tr\glin\,=\,2\otx-2\divr\bmlin\,,&\qquad &&\nablav\, \glinh\,=\,2\Omega\xlin+2\fancydstar_2\bmlin\,,\;\;\;
        \tag{\ref{metric transport in 4 direction trace},\;\ref{metric transport in 4 direction traceless}} 
        \\[9pt]
        \partial_u\tr\glin\,=\,2\otxb\,,&\qquad &&\nablau\,\glinh\,=\,2\Omega\xblin\,,\;\;\;
        \tag{\ref{metric transport in 3 direction trace},\;\ref{metric transport in 3 direction traceless}}
        \\[9pt]
        \partial_u\,\bmlin\,&=\,2\Omega^2\Big(\elin&&-\eblin\Big)\,,
        \tag{\ref{partial_u b}}
    \end{alignat*}
    \begin{equation*}
        \refstepcounter{equation}\latexlabel{partial v Olin}
        \refstepcounter{equation}\latexlabel{partial u Olin}
        \refstepcounter{equation}\latexlabel{elin eblin Olin}
        \qquad\qquad\partial_v\left(\frac{\Olino}{\Omega}\right)\,=\,\olin,\qquad\qquad \partial_u\left(\frac{\Olino}{\Omega}\right)\,=\,\olinb,\qquad\qquad \elin_A+\eblin_A=2\slashednabla_A\left(\frac{\Olino}{\Omega}\right),
        \tag{\ref{partial v Olin},\;\ref{partial u Olin},\;\ref{elin eblin Olin}}
    \end{equation*}
    \\[1pt]
    \item The equations governing the linearised connection coefficients:\\
    \begin{equation} \label{D4TrChiBar}
\partial_v\; r\otxb=2\Omega^2\left(\divr\; r\,\eblin+r\,\rlin -\frac{4M}{r^2}\,\frac{\Olino}{\Omega}\right)+\Omega^2\,\otx,
\end{equation}

\begin{equation}\label{D3TrChi}
\partial_u\; r\otx=2\Omega^2\left(\divr\; r\,\elin+r\,\rlin-\frac{4M}{r^2}\,\frac{\Olino}{\Omega}\right) -\Omega^2\, \otxb,
\end{equation}

\begin{alignat*}{2}
    \refstepcounter{equation}\latexlabel{D4TrChi}
    \refstepcounter{equation}\latexlabel{D3TrChibar}
    \refstepcounter{equation}\latexlabel{D4Chihat}
    \refstepcounter{equation}\latexlabel{D3Chihatbar}
    \qquad\qquad\partial_v\left[\frac{r^2}{\Omega^2}\otx\right]=4r\olin\,,&\qquad\qquad &&\partial_u\left[\frac{r^2}{\Omega^2}\otxb\right]=-4r\olinb\,,
    \tag{\ref{D4TrChi},\;\ref{D3TrChibar}}
    \\[10pt]
    \qquad\qquad\nablav\,\frac{r^2\xlin}{\Omega}\,=\,-r^2\alin\,,\;\;&\qquad\qquad &&\;\;\;\nablau\,\frac{r^2\xblin}{\Omega}\,=\,-r^2\ablin\,,
    \tag{\ref{D4Chihat},\;\ref{D3Chihatbar}}
\end{alignat*}

\begin{equation}\label{D3Chihat}
\nablau\; r\Omega\,\xlin\,=\,-2r\fancydstar_2 \,\Omega^2 \,\elin\,-\,\Omega^2 \left(\Omega\, \xblin\right),
\end{equation}

\begin{equation}\label{D4Chihatbar}
\nablav\;r\Omega\,\xblin\,=\,-2r\fancydstar_2\,\Omega^2\,\eblin\,+\,\Omega^2\left(\Omega\,\xlin\right),
\end{equation}

\begin{equation*}
    \refstepcounter{equation}\latexlabel{D3etabar}
    \refstepcounter{equation}\latexlabel{D4eta}
    \qquad\qquad\nablau\,r\,\eblin\,=\,r\Omega\,\bblin-\,\Omega^2\elin\,,\qquad\qquad \nablav\,r\,\elin\,=\,-r\Omega\,\blin+\Omega^2\,\eblin\,,
    \tag{\ref{D3etabar},\;\ref{D4eta}}
\end{equation*}

\begin{equation*}
    \refstepcounter{equation}\latexlabel{D4etabar}
    \refstepcounter{equation}\latexlabel{D3eta}
    \qquad\qquad\nablav\,r^2\,\eblin\,=\,2r^2\slashednabla \olin\,+\,r^2\Omega\,\blin\,,\qquad\qquad \nablau\,r^2\,\elin\,=\,2r^2\slashednabla\olinb\,-\,r^2\Omega\,\bblin\,,
    \tag{\ref{D4etabar},\;\ref{D3eta}}
\end{equation*}

\begin{equation}\label{transport equation of little omega}
\partial_v \overone{\underline\omega}=-\Omega^2\left(\overone{\rho}-\frac{4M}{r^3}\frac{\overone{\Omega}}{\Omega}\right)=\partial_u \overone{\omega}
\end{equation}\\[1pt]

\item The equations governing the curvature components:

\begin{alignat*}{2}
 \refstepcounter{equation}\latexlabel{Bianchi+2}
    \refstepcounter{equation}\latexlabel{Bianchi-2}
    \refstepcounter{equation}\latexlabel{Bianchi+1a}
    \refstepcounter{equation}\latexlabel{Bianchi-1a}
    \refstepcounter{equation}\latexlabel{Bianchi+1b}
    \refstepcounter{equation}\latexlabel{Bianchi-1b}
    \refstepcounter{equation}\latexlabel{Bianchi+0}
    \refstepcounter{equation}\latexlabel{Bianchi-0}
    \refstepcounter{equation}\latexlabel{Bianchi+0*}
    \refstepcounter{equation}\latexlabel{Bianchi-0*}
    \nablau \,r\Omega^2\,\alin\,=\,-2r\fancydstar_2\Omega^2\Omega\,\blin+\frac{6M\Omega^2}{r^2}\Omega\,\xlin\,,&\qquad\qquad &&\nablav \,r\Omega^2\,\ablin\,=\,2r\fancydstar_2\Omega^2\Omega\,\bblin+\frac{6M\Omega^2}{r^2}\Omega\,\xblin\,,
    \tag{\ref{Bianchi+2},\;\ref{Bianchi-2}}
    \\[6pt]\qquad\qquad\qquad\nablav\,\frac{r^4\blin}{\Omega}\,=\,r\divr\,r^3\alin\,,&\qquad\qquad\quad &&\nablau\,\frac{r^4\bblin}{\Omega}\,=\,-r\divr\,r^3\ablin\,,
    \tag{\ref{Bianchi+1a},\;\ref{Bianchi-1a}}
    \\[18pt]
    \nablav\,r^2\Omega\,\bblin=r\fancydstar_1(r\Omega^2\,\rlin,r\Omega^2\,\slin)\,+\,\frac{6M\Omega^2}{r}\,\eblin\,,&\qquad\qquad &&\nablau\,r^2\Omega\,\blin=r\fancydstar_1(-r\Omega^2\,\rlin,r\Omega^2\,\slin)\,-\,\frac{6M\Omega^2}{r}\,\elin\,,
     \tag{\ref{Bianchi+1b},\;\ref{Bianchi-1b}}
     \\[8pt]
     \partial_v\,r^3\rlin\,=\,r\divr\,r^2\Omega\,\blin\,+\,3M\,\otx\,,&\qquad\qquad&&\partial_u\,r^3\rlin\,=\,r\divr\,r^2\Omega\,\bblin\,+\,3M\,\otxb\,,
     \tag{\ref{Bianchi+0},\;\ref{Bianchi-0}}
     \\[6pt]
     \partial_v\,r^3\slin\,=\,-r\curlr\,r^2\Omega\,\blin\,,&\qquad\qquad&&\partial_u\,r^3\slin\,=\,r\curlr\,r^2\Omega\,\bblin\,,
     \tag{\ref{Bianchi+0*},\;\ref{Bianchi-0*}}
\end{alignat*}

\item The linearised Codazzi equations

\begin{alignat*}{2}
    \refstepcounter{equation}\latexlabel{elliptic equation 1}
    \refstepcounter{equation}\latexlabel{elliptic equation 2}
    \refstepcounter{equation}\latexlabel{elliptic equation 3 and 4}
    &\divr\,\xblin\,=\,\frac{\Omega}{r}\,\elin\,+\,\bblin\,+\,\frac{1}{2\Omega}\,\slashednabla\,\otxb\,,&&\tag{\ref{elliptic equation 1}}
    \\[10pt]
    &\divr\,\xlin\,=\,-\frac{\Omega}{r}\,\eblin\,+\,\blin\,+\,\frac{1}{2\Omega}\,\slashednabla\,\otx\,,&&\tag{\ref{elliptic equation 2}}
    \\[12pt]
    \curlr&\,\elin\,=\,\slin\qquad\qquad\qquad\qquad \curlr\,\eblin\,=\,-&&\slin,\tag{\ref{elliptic equation 3 and 4}}
\end{alignat*}

and the linearised Gauss equation
\begin{align}\label{Gauss}
\Klin&\,=\,-\,\rlin\,-\frac{1}{2r}\left(\overone{\Omega\tr\underline\chi}-\overone{\Omega\tr\chi}\right)-\frac{2\Omega}{r^2}\;\Olino.
\end{align}
\end{itemize}
Here, $\glinh$, $\xlin$, $\xblin$, $\alin$, $\ablin$ are symmetric traceless $S^2_{u,v}$-tangent tensor fields, $\bmlin$, $\elin$, $\eblin$, $\blin$, $\bblin$ are $S^2_{u,v}$-tangent $1$-form fields, and $\glinto$, $\Olino$, $\olin$, $\olinb$, $\otx$, $\otxb$, $\rlin$, $\slin$ are scalar fields. The linearised Gaussian curvature $\overone{K}$ is given by
    \begin{align}\label{Formula for K}
        \Klin[\,\glin\,]:=-\left(\frac{1}{4}{\slashed\Delta}+\frac{1}{2r^2}\right)\tr_{\slashed{g}}\glin+\frac{1}{2}\slashed{div}\slashed{div}\overone{\hat{\slashed{g}}}.
    \end{align}

Let $\mathcal{S}\subset \mathscr{M}$ be open. If the tuple
\begin{align}\label{tuple}
    \mathfrak{S}=\left(\;\glinh\;,\;\tr\glin\;,\;\bmlin\;,\;\Olino\;,\;\xlin\;,\;\xblin\;,\;\trx\;,\;\trxbar\;,\;\elin\;,\;\eblin\;,\;\olin\;,\;\olinb\;,\;\alin\;,\;\ablin\;,\;\blin\;,\;\bblin\;,\;\rlin\;,\;\slin\;\right),
\end{align}
satisfies the equations \fullsystem above on $\mathcal{S}$, we say that $\mathfrak{S}$ is a solution to the linearised Einstein equations, or alternatively, a solution to linearised gravity, on $\mathcal{S}$. 
\end{defin}

\begin{remark}\label{don't know}
    We may refer to a component $\upsilon$ of a solution $\mathfrak{S}$ by $\upsilon[\mathfrak{S}]$. For example, $\xlin[\mathfrak{S}]$ is the component $\xlin$ of the tuple \bref{don't know} defining the solution $\mathfrak{S}$. We will often drop the reference to $\mathfrak{S}$ when no ambiguity arises.
\end{remark}

\begin{remark}
In defining a solution to the system \fullsystem above, we treat the metric, connection and curvature components as independent quantities, related by transport equations \fullsystem above. 
\end{remark}

\subsection{The linearised equations in Kruskal coordiantes}\label{EF degeneration remark}

It is necessary to cast the equations \fullsystem in the Kruskal frame to compensate for the degeneration of the Eddington--Finkelstein frame near the bifurcation sphere $\mathcal{B}$. When re-scaling the quantities defining a solution to \fullsystem as to be consistent with linearisation in the Kruskal frame, 
\begin{align}\label{Regular linearised quantities}
\begin{split}
     &\qquad\glinhK:=\glinh,\qquad \tr\glinK:=\tr\glin,\qquad \bmlinK:=V^{-1}\bmlin,\qquad\OlinK:=\Olin,\\[3pt]
    &\xlinK:=V^{-1}\Omega\xlin,\qquad\elinK:=\elin,\qquad\olinK:=V^{-1}\olin,\qquad\otxK:=V^{-1}\otx,\\[4pt]
    &\xblinK:=U^{-1}\Omega\xblin,\qquad\eblinK:=\eblin,\qquad\olinbK:=U^{-1}\olinb,\qquad\otxbK:=U^{-1}\otxb,\\[5pt]
    &\qquad\qquad\qquad\alinK:=V^{-2}\Omega^2\alin,\qquad\qquad\qquad \ablinK:=U^{-2}\Omega^2\ablin, \\[5pt]
    &\;\qquad\qquad\qquad\blinK:=V^{-1}\Omega\blin,\qquad\qquad\qquad\; \bblinK:=U^{-1}\Omega\bblin, \\[6pt]
    &\qquad\qquad\qquad\qquad\rlinK:=\rlin,\qquad\qquad\qquad\qquad\quad \slinK:=\slin,
\end{split}
\end{align}
we obtain the following system:
\begin{align}
    (2M)^{-1}\partial_V \tr\glin\;=\;2\otxK-2\;\slashed{div}\,\bmlinK,\qquad\qquad&(2M)^{-1}\slashednabla_V \glinhK\;=\;2\xlinK+2\fancydstar_2\bmlinK,\label{metric transport in 4 direction Kruskal}\\
    (2M)^{-1}\partial_U \tr\glinK\;=\;-2\otxbK,\qquad\qquad &(2M)^{-1}\slashednabla_U\glinhK\;=\;-2\xblinK.\label{metric transport in 3 direction Kruskal}\\
    \frac{\fr}{2M}\partial_U\bmlinK^A\;=\;2(\elinK-&\eblinK)^A.\label{partial_u b Kruskal}\\
    (2M)^{-1}\partial_V\left(\OlinK\right)\;=\;\olinK,\qquad\qquad(2M)^{-1}\partial_U\left(\OlinK\right)&\;=\;-\olinbK,\qquad\qquad\elinK+\eblinK\;=\;2\slashednabla_A \left(\OlinK\right),\label{omega omegabar eta etabar Kruskal}
\end{align}
\begin{equation} \label{D4TrChiBar Kruskal}
    \frac{\fr}{2M}\partial_V \otxbK=2\left(\slashed{div}\; \elinK+\rlinK-\frac{4M}{r^3}\OlinK\right) -\frac{1}{r}\left(V\otxK-U\otxbK\right)=\frac{\fr}{2M}\partial_U \otxK,
\end{equation}
\begin{equation}\label{D4TrChi Kruskal}
(2M)^{-1}\partial_V\otxK-\frac{U}{2M\fr}\left(1+\frac{6M}{r}\right)\otxK=-\frac{4U}{r\fr}\olinK,
\end{equation}
\begin{equation}\label{D3TrChiBar Kruskal}
(2M)^{-1}\partial_U\otxbK-\frac{V}{2M\fr}\left(1+\frac{6M}{r}\right)\otxbK=\frac{4U}{r\fr}\olinbK,
\end{equation}
\begin{equation}\label{D4Chihat Kruskal}
(2M)^{-1}\slashednabla_V \xlinK-\frac{U}{2M\fr}\left(1+\frac{6M}{r}\right)\xlinK=-\alinK,
\end{equation}
\begin{align}\label{D3Chiharbar Kruskal}
  (2M)^{-1}\slashednabla_U \xblinK-\frac{V}{2M\fr}\left(1+\frac{6M}{r}\right)\xblinK=-\ablinK  ,
\end{align}
\begin{equation}\label{D3Chihat Kruskal}
\frac{\fr}{2M}\slashednabla_U\; r\xlinK=-2r\slashed{\mathcal{D}}^*_2 \elinK+{U}\xblinK,
\end{equation}
\begin{equation}\label{D4Chihatbar Kruskal}
\frac{\fr}{2M}\slashednabla_V\; r\xblinK=-2r\slashed{\mathcal{D}}^*_2 \eblinK-V\xlinK,
\end{equation}
\begin{equation}\label{D3etabar Kruskal}
(2M)^{-1}\slashednabla_Ur\eblinK=r\bblinK+\frac{U}{\fr}\eblinK,\qquad\qquad (2M)^{-1}\slashednabla_Vr\elinK=-r\blinK-\frac{U}{\fr}\eblinK,
\end{equation}
\begin{equation}\label{D4etabar Kruskal}
(2M)^{-1}\slashednabla_V r^2\eblinK=-2\mathring{\slashednabla}r\olinK-{r^2}\blinK,\qquad\qquad (2M)^{-1}\slashednabla_U r^2\elinK=-2\mathring{\slashednabla}r\olinbK-{r^2}\bblinK,
\end{equation}
\begin{equation}\label{transport equation of little omega Kruskal}
(2M)^{-1}\partial_V\olinbK\,=\,\frac{1}{\fr}\left(\rlinK-\frac{4M}{r^3}\OlinK\right)=-(2M)^{-1}\partial_U \olinK,
\end{equation}
\begin{equation}\label{Bianchi +2 Kruskal}
\frac{\fr}{2M}\slashednabla_U \;r\alinK=-2\mathring{\fancydstar_2}\,\blinK+\frac{6M}{r^2}\xlinK,\qquad\qquad \frac{\fr}{2M}\slashednabla_V \;r\ablinK=2\mathring{\fancydstar_2}\,\bblinK+\frac{6M}{r^2}\xblinK,
\end{equation}
\begin{equation}\label{Bianchi +1a Kruskal}
(2M)^{-1}\slashednabla_V r^2\blinK-\frac{U}{2M\fr}\left(1+\frac{6M}{r}\right)r^2\blinK=\divo\;r\alinK,
\end{equation}
\begin{align}\label{Bianchi -1a Kruskal}
(2M)^{-1}\slashednabla_U r^2\bblinK-\frac{V}{2M\fr}\left(1+\frac{6M}{r}\right)r^2\bblinK=\divo\;r\ablinK,
\end{align}
\begin{align}
   \frac{\fr}{2M}\slashednabla_U r^2\blinK=\mathring{\fancydstar_1}(-r\rlinK,r\slinK)-\frac{6M}{r}\elinK,&\qquad\qquad\frac{\fr}{2M}\slashednabla_V r^2\bblinK=-\mathring{\fancydstar_1}(r\rlinK,r\slinK)-\frac{6M}{r}\eblinK,\label{Bianchi +1b Kruskal}\\[4pt]
(2M)^{-1}\slashednabla_U\; r^3\rlinK=\divo\;r^2\bblinK-3M \otxbK,&\qquad \qquad(2M)^{-1}\slashednabla_V\; r^3\rlinK=\divo\;r^2\blinK+3M \otxK,\label{Bianchi 0 Kruskal}\\[6pt]
(2M)^{-1}\slashednabla_U\; r^3\slinK=\curlo \;r^2\bblinK,&\qquad\qquad (2M)^{-1}\slashednabla_V \; r^3\slinK=-\curlo\;r^2\blinK\label{Bianchi 0* Kruskal}
\end{align}
\begin{align}
&\slashed{{div}}\xblinK\;=\;-\frac{V}{r\fr}\elinK+\bblinK+\frac{1}{2}\slashednabla \otxbK,\label{elliptic equation 1 Kruskal}\\
&\slashed{{div}}\xlinK\;=\;\frac{U}{r\fr}\eblinK-\blinK+\frac{1}{2}\slashednabla \otxK,\label{elliptic equation 2 Kruskal}\\
\slashed{curl}& \elinK \;=\;\slinK,\qquad\qquad\qquad \slashed{curl} \eblinK \;=\;-\slinK,\label{elliptic equation 3 and 4 Kruskal}
\end{align}
\begin{align}\label{Gauss Kruskal}
\KlinK&=-\rlinK+\frac{1}{2r}\left(V\otxK-U\otxbK\right)-\frac{2\Omega^2}{r^2}\OlinK.
\end{align}

\section{The Teukolsky equations, the Teukolsky--Starobinsky identities and the Regge--Wheeler equation}\label{TRW}

\subsection{The spin $\pm2$ Teukolsky equations}

Let $\alin$, $\ablin$ belong to a solution to the linearised Einstein equations \fullsystem. It is easy to show that the equation \fullsystem implies that $\alin$, $\ablin$ obey 2\textsuperscript{nd} order hyperbolic equations that decouple from the rest of the system \fullsystem. These are the well-known Teukolsky equations:

\begin{align}\label{T+2}
\frac{\Omega^2}{r^2}\nablav \frac{r^4}{\Omega^4}\nablau r\Omega^2\alin=-2r^2\fancydstar_2\fancyd_2 r\Omega^2\alin-\frac{6M}{r}r\Omega^2\alin,
\end{align}
\begin{align}\label{T-2}
\frac{\Omega^2}{r^2}\nablau \frac{r^4}{\Omega^4}\nablav r\Omega^2\ablin=-2r^2\fancydstar_2\fancyd_2 r\Omega^2\ablin-\frac{6M}{r}r\Omega^2\ablin.
\end{align}
\Cref{T+2} is known as the \textbf{Spin $+2$ Teukolsky equation}, and \cref{T-2} is known as the \textbf{Spin $-2$ Teukolsky equation}.  For a derivation of \bref{T+2}, \bref{T-2} see \cite{Mas20}.
We now state some well-posedness results concerning the Teukolsky equations \bref{T+2}, \bref{T-2}:
\begin{proposition}\label{WP+2Sigma*}
Prescribe on ${\Sigma^*_+}$ a pair of smooth symmetric traceless $S^2_{u,v}$ 2-tensor fields $(\upalpha,\upalpha')$. Then there exists a unique smooth symmetric traceless $S^2_{u,v}$ 2-tensor field $\Omega^2\alpha$ that satisfies \bref{T+2} on $D^+({\Sigma^*_+})$, with $\Omega^2\alpha|_{{\Sigma^*_+}}=\upalpha, \slashednabla_{n_{{\Sigma^*_+}}}\Omega^2\alpha|_{{\Sigma^*_+}}=\upalpha'$.
\end{proposition}
\begin{proposition}\label{WP-2Sigma*}
Prescribe on ${\Sigma^*_+}$ a pair of smooth symmetric traceless $S^2_{u,v}$ 2-tensor fields $(\underline\upalpha,\underline\upalpha')$. Then there exists a unique smooth symmetric traceless $S^2_{u,v}$ 2-tensor field $\Omega^{-2}\underline\alpha$ that satisfies \bref{T-2} on $D^+({\Sigma^*_+})$, with $\Omega^{-2}\underline\alpha|_{{\Sigma^*_+}}=\underline\upalpha, \slashednabla_{n_{{\Sigma^*_+}}}\Omega^2\underline\alpha|_{{\Sigma^*_+}}=\underline\upalpha'$.
\end{proposition}
\noindent The same applies replacing ${\Sigma^*_+}$ with any other $\mathscr{H}^+$-penetrating spacelike surface ending at $i^0$.\\
\indent The degeneration of the EF frame discussed in \Cref{EF degeneration remark} is inherited by \bref{T+2}, \bref{T-2}, and we must work with $\widetilde{\alpha}=V^{-2}\Omega^2\alpha, \widetilde{\underline\alpha}=U^2\Omega^{-2}\underline\alpha$ in order to study the Teukolsky equations with data on $\overline{\Sigma}$. The weighted quantities $\widetilde{\alpha}, \widetilde{\underline\alpha}$ satisfy the following equations:

\begin{align}\label{T+2B}
    \frac{1}{\Omega^2}\nablau\nablav r\widetilde{\alpha}+\frac{1}{M}(4-3\Omega^2)\nablau r\widetilde{\alpha}-\frac{1}{r}(3\Omega^2-5)\widetilde{\alpha}-\slashed{\Delta}r\widetilde{\alpha}=0,
\end{align}
\begin{align}\label{T-2B}
    \frac{1}{\Omega^2}\nablau\nablav r\widetilde{\underline\alpha}-\frac{1}{M}(4-3\Omega^2)\nablav r\widetilde{\underline\alpha}-\frac{1}{r}(3\Omega^2-5)\widetilde{\underline\alpha}-\slashed{\Delta}r\widetilde{\underline\alpha}=0.
\end{align}
Equations (\ref{T+2B}) and (\ref{T-2B}) do not degenerate near $\mathcal{B}$ and we can make the following well-posedness statement:
\begin{proposition}\label{WP+2Sigmabar}
Prescribe a pair of smooth symmetric traceless $S^2_{U,V}$ 2-tensor fields $(\widetilde{\upalpha},\widetilde{\upalpha}')$ on $\overline{\Sigma}$. Then there exists a unique smooth symmetric traceless $S^2_{u,v}$ 2-tensor field $\Omega^2{\alpha}$ that satisfies (\ref{T+2}) on $ D^+(\overline{\Sigma})$ with $V^{-2}\Omega^2\alpha|_{\overline{\Sigma}}=\widetilde{\upalpha}$ and $\slashednabla_{n_{\overline{\Sigma}}}V^{-2}\Omega^2\alpha|_{\overline{\Sigma}}=\widetilde{\upalpha}'$.
\end{proposition}
\begin{proposition}\label{WP-2Sigmabar}
Prescribe a pair of smooth symmetric traceless $S^2_{U,V}$ 2-tensor fields $(\widetilde{\underline\upalpha},\widetilde{\underline\upalpha}')$ on $\overline{\Sigma}$. Then there exists a unique smooth symmetric traceless $S^2_{u,v}$ 2-tensor field $\Omega^{-2}{\underline\alpha}$ that satisfies (\ref{T-2}) on $ D^+(\overline{\Sigma})$ with $V^{2}\Omega^{-2}\underline\alpha|_{\overline{\Sigma}}=\widetilde{\underline\upalpha}$ and $\slashednabla_{n_{\overline{\Sigma}}}V^{2}\Omega^{-2}\underline\alpha|_{\overline{\Sigma}}=\widetilde{\underline\upalpha}'$.
\end{proposition}
\noindent Analogous statements to the above apply to past development from $\overline{\Sigma}$ with $U,\Omega^2$ switching places with $V,\Omega^{-2}$ respectively.\\

\begin{remark}[\textbf{The effect of time inversion on the Teukolsky equations}]\label{time inversion}
Under the transformation $t\longrightarrow-t$, $u\longrightarrow -v$ and $v\longrightarrow -u$ and thus $\alpha(u,v,\theta^A)\longrightarrow\alpha(-v,-u,\theta^A)=:\invertedalpha(u,v,\theta^A)$ and $\underline\alpha(u,v,\theta^A)\longrightarrow\underline\alpha(-v,-u,\theta^A)=:\underline\invertedalpha(u,v,\theta^A)$.\\
\indent It is clear $\invertedalpha(u,v,\theta^A)$ satisfies the $-2$ Teukolsky equation, i.e.~the equation satisfied by $\underline\alpha$. Similarly,  $\underline\invertedalpha(u,v,\theta^A)$ satisfies the $+2$ Teukolsky equation, i.e.~the equation satisfied by $\alpha$. This observation means that the asymptotics of $\alpha$ towards the future are identical to those of $\underline\alpha$ towards the past.  Therefore, determining the asymptotics of both $\underline\alpha$ and $\alpha$ towards the future is enough to determine the asymptotics of either $\alpha$ or $\underline\alpha$ in both the past and future directions. We will use this fact to obtain bijective scattering maps from studying the forward evolution of the fields $\alpha,\underline\alpha$. 
\end{remark}

\subsection{Physical-space Chandrasekhar transformations and the Regge--Wheeler equation}\label{Chandra}
The \textbf{Regge--Wheeler equation} for a symmetric traceless $S^2_{u,v}$ 2-tensor $\Psi$ is given by
\begin{align}\label{RW}
    \nablav\nablau\Psi-\Omega^2\slashed{\Delta}\Psi+\frac{\Omega^2}{r^2}(3\Omega^2+1)\Psi=0.
\end{align}
\indent In the Schwarzschild spacetime, the Teukolsky equations lead to the Regge--Wheeler equations in the following sense: define the following hierarchy of fields
\begin{alignat}{2}
     r^3\Omega \psi&:=\frac{r^2}{\Omega^2}\nablau r\Omega^2\alpha,\qquad\qquad\Psi&&:=\frac{r^2}{\Omega^2}\nablau r^3\Omega \psi=\left(\frac{r^2}{\Omega^2}\nablau\right)^2 r\Omega^2\alpha,\label{hier}\\
     r^3\Omega \underline\psi&:=\frac{r^2}{\Omega^2}\nablav r\Omega^2\underline\alpha,\qquad\qquad\underline\Psi&&:=\frac{r^2}{\Omega^2}\nablav r^3\Omega \underline\psi=\left(\frac{r^2}{\Omega^2}\nablav\right)^2 r\Omega^2\underline\alpha.\label{hier'}
\end{alignat}
Then we have
\begin{lemma}\label{+-2 implies RW}
Assume $\alpha$ satisfies the $+2$ Teukolsky equation \bref{T+2}, $\underline\alpha$ satisfies the $-2$ Teukolsky equation \bref{T-2}. Then $\Psi, \underline\Psi$ of \cref{hier} satisfy the Regge--Wheeler equation \bref{RW}.
\end{lemma}
\begin{proof}
   Direct computation. See \cite{Mas20} for the details.
\end{proof}

In light of the appearance of the Regge--Wheeler equation \bref{RW} as a consequence of the linearised system \fullsystem, we make the following definitions:
\begin{defin}\label{linearised hier}
Let $\alin$, $\ablin$ belong to a solution to the system \fullsystem. We define
\begin{align}\label{definition of plin and pblin}
    r^3\Omega\plin:=\frac{r^2}{\Omega^2}\nablau r\Omega\alin,\qquad\qquad r^3\Omega\pblin:=\frac{r^2}{\Omega^2}\nablav r\Omega\ablin,
\end{align}
\begin{align}\label{definition of Psilin and Psilinb}
    \Psilin:=\frac{r^2}{\Omega^2}\nablau r^3\Omega\plin=\left(\frac{r^2}{\Omega^2}\nablau\right)^2r\Omega\alin,\qquad\qquad\Psilin:=\frac{r^2}{\Omega^2}\nablav r^3\Omega\pblin=\left(\frac{r^2}{\Omega^2}\nablav\right)^2r\Omega\ablin.
\end{align}
\end{defin}
\begin{corollary}\label{Regge--Wheeler applies to linearised gravity around Schw}
Let $\alin, \ablin$ belong to a solution to the linearised Einstein equations \fullsystem. 
Define $\Psilin:=\left(\frac{r^2}{\Omega^2}\nablau\right)^2r\Omega^2\alin$, $\Psilinb:=\left(\frac{r^2}{\Omega^2}\nablav\right)^2r\Omega^2\ablin$. Then $\Psilin$, $\Psilinb$ both satisfy the Regge--Wheeler equation \bref{RW}.
\end{corollary}

\begin{corollary}\label{formula for Psilin Psilinb in terms of linearised system}
Let $\alin$, $\ablin$, $\Psilin$, $\Psilinb$ be as in \Cref{Regge--Wheeler applies to linearised gravity around Schw}. Tracking the definitions \bref{definition of Psilin and Psilinb} through the system \fullsystem, we get
\begin{align}
     \Psilin&=2r^2\fancydstar_2\fancydstar_1\left(r^3\rlin,\,-r^3\slin\right)+6M\left(r\Omega\xlin-r\Omega\xblin\right),\label{expression for Psilin}\\
    \Psilinb&=2r^2\fancydstar_2\fancydstar_1\left(r^3\rlin,\,r^3\slin\right)+6M\left(r\Omega\xlin-r\Omega\xblin\right).\label{expression for Psilinb}
\end{align}
\end{corollary}


\indent We now state a standard well-posedness result for (\ref{RW}) for data on ${\Sigma^*_+}$:
\begin{proposition}\label{RWwpCauchy}
For any pair $(\uppsi,\uppsi')$ of smooth symmetric traceless $S^2_r$ 2-tensor fields on ${\Sigma^*_+}$, there exists a unique smooth symmetric traceless $S^2_{u,v}$ 2-tensor field $\Psi$ which solves \cref{RW} in $ D^+({\Sigma^*_+})$ such that $\Psi|_{{\Sigma^*_+}}=\uppsi$ and $\slashednabla_{n_{{\Sigma^*_+}}} \Psi|_{{\Sigma^*_+}}=\uppsi'$. The same applies when data are posed on $\Sigma$ or $\overline{\Sigma}$.
\end{proposition}
In contrast to the Teukolsky equations \bref{T+2}, \bref{T-2}, the Regge--Wheeler equation \bref{RW} does not suffer from additional regularity issues near $\mathcal{B}$, as can be seen by rewriting \cref{RW} in Kruskal coordinates:
\begin{align}\label{RW in Kruskal}
    \slashednabla_U\slashednabla_V\Psi-{\slashed{\Delta}}\Psi+\frac{3\Omega^2+1}{r^2}\Psi=0.
\end{align}
Note that the relation between $\Psi$ and $\alpha$ in \bref{hier} takes the following form in the Kruskal frame:
\begin{align}\label{Kruskal hier+}
    \Psi=\left(\frac{r^2}{\Omega^2}\nablau\right)^2r\Omega^2\alpha=\left(2Mr^2\fr(r)\slashednabla_U\right)^2r\tilde{\alpha}.
\end{align}

\begin{proposition}\label{RWwpSigmabar}
\Cref{RWwpCauchy} is valid replacing ${\Sigma^*_+}$ with $\overline{\Sigma}$ everywhere.
\end{proposition}
For backwards scattering we will need the following well-posedness statement:
\begin{proposition}\label{RWwpBackwards}
Let $u_+<\infty, v_+<v_*<\infty$. Let $\widetilde{\Sigma}$ be a spacelike hypersurface connecting $\mathscr{H}^+$ at $v=v_+$ to $\mathscr{I}^+$ at $u=u_+$ and let $\underline{\mathscr{C}}=\underline{\mathscr{C}}_{v_*}\cap J^+(\widetilde{\Sigma})\cap\{t\geq0\}$. Prescribe a pair of smooth symmetric traceless $S^2_{u,v}$ 2-tensor fields:
\begin{itemize}
\item $\Psi_{{\mathscr{H}^+}}$ on ${\overline{\mathscr{H}^+}}\cap\{v<v_+\}$ vanishing in a neighborhood of $\widetilde{\Sigma}$,
\item $\Psi_{0,in}$ on $\underline{\mathscr{C}}$ vanishing in a neighborhood of $\widetilde{\Sigma}$.
\end{itemize}
Then there exists a unique smooth symmetric traceless $S^2_{u,v}$ 2-tensor $\Psi$ on $D^-\left(\overline{\mathscr{H}^+}\cup\widetilde{\Sigma}\cup\underline{\mathscr{C}}\right)\cap J^+(\overline{\Sigma})$ such that $\Psi|_{\overline{\mathscr{H}^+}}=\Psi_{{\mathscr{H}^+}}$, $\Psi|_{\underline{\mathscr{C}}}=\Psi_{0,in}$ and $\left(\Psi|_{\widetilde{\Sigma}},\slashednabla_{n_{\widetilde{\Sigma}}}\Psi|_{\widetilde{\Sigma}}\right)=(0,0)$.
\end{proposition}
We will also need
\begin{proposition}\label{RWwp local statement near B}
Let $(\uppsi,\uppsi')$ be smooth symmetric traceless $S^2_{u,v}$ 2-tensor fields on ${\Sigma^*_+}$, $\uppsi_{\mathscr{H}^+}$ be a smooth symmetric traceless $S^2_{\infty,v}$ 2-tensor field on $\overline{\mathscr{H}^+}\cap\{t^*_+\leq0\}$. Then there exists a unique smooth symmetric traceless $S^2_{u,v}$ 2-tensor field $\Psi$ on $J^-({\Sigma^*_+})$ such that $\Psi|_{\overline{\mathscr{H}^+}\cap\{t^*_+\leq0\}}=\uppsi_{\mathscr{H}^+}, \left(\Psi|_{{\Sigma^*_+}},\slashednabla_{n_{{\Sigma^*_+}}}\Psi|_{{\Sigma^*_+}}\right)=(\uppsi,\uppsi')$.
\end{proposition}

\begin{remark}\label{time inversion of RW}
Unlike the Teukolsky equations \bref{T+2}, \bref{T-2}, the Regge--Wheeler equation \bref{RW} is invariant under time inversion. If $\Psi(u,v)$ satisfies \bref{RW}, then $\invertedpsi(u,v):=\Psi(-v,-u)$ also satisfies \bref{RW}.
\end{remark}

\subsubsection[Further constraints among $\protect\Psilin$, $\protect\alin$ and $\protect\Psilinb$, $\protect\ablin$]{Further constraints among $\Psilin$, $\alin$ and $\Psilinb$, $\ablin$}.

Using that $\alin$ satisfies \bref{T+2}, $\Psilin$ satisfies \bref{RW} and $\Psilin$, $\alin$ are related by \bref{hier}, a computation shows
\begin{align}\label{further constraint --}
\Omega\slashed{\nabla}_4\Psilin=-(3\Omega^2-1)\frac{r^2}{\Omega^2}\Omega\slashed{\nabla}_3r\Omega^2\alin-6Mr\Omega^2\alin+\mathcal{A}_2\frac{r^2}{\Omega^2}\Omega\slashed{\nabla}_3r\Omega^2\alin,
\end{align}
\begin{align}\label{further constraint -}
    \frac{\Omega^2}{r^2}\nablav\frac{r^2}{\Omega^2}\nablav\Psilin=\mathcal{A}_2\left(\mathcal{A}_2-2\right)r\Omega^2\alin-6M(\nablau+\nablav)r\Omega^2\alin.
\end{align}
Similar constraints relates $\ablin$, $\Psilinb$:
\begin{align}\label{further constraint ++}
\Omega\slashed{\nabla}_3\Psilinb=-(3\Omega^2-1)\frac{r^2}{\Omega^2}\Omega\slashed{\nabla}_4r\Omega^2\ablin+6Mr\Omega^2\ablin+\mathcal{A}_2\frac{r^2}{\Omega^2}\Omega\slashed{\nabla}_4r\Omega^2\ablin,
\end{align}
\begin{align}\label{further constraint +}
    \frac{\Omega^2}{r^2}\nablau\frac{r^2}{\Omega^2}\nablau\Psilinb=\mathcal{A}_2\left(\mathcal{A}_2-2\right)r\Omega^2\ablin+6M(\nablau+\nablav)r\Omega^2\ablin.
\end{align}

\subsubsection{Aside: the Teukolsky--Starobinsky identities}\label{derivation of the Teukolsky--Starobinsky identities}
Assume $\alin, \ablin$ belong to a solution to the linearised Einstein equations. In addition to satisfying the Teukolsky equations \bref{T+2}, \bref{T-2}, the quantities $\alin, \ablin$ satisfy the \textbf{Teukolsky--Starobinsky identities}:
\begin{align}\label{eq:TS-}
\frac{\Omega^2}{r^2}\Omega\slashed{\nabla}_3 \left(\frac{r^2}{\Omega^2}\nablau\right)^3r\Omega^2\overone\alpha=2r^4\slashed{\mathcal{D}}^*_2\slashed{\mathcal{D}}^*_1\overline{\slashed{\mathcal{D}}}_1\slashed{\mathcal{D}}_2 r\Omega^2\overone{\underline\alpha}+6M\left[\Omega\slashed{\nabla}_4+\Omega\slashed{\nabla}_3\right]r\Omega^2\overone{\underline\alpha}.
\end{align}
\begin{align}\label{eq:TS+}
\frac{\Omega^2}{r^2}\Omega\slashed{\nabla}_4 \left(\frac{r^2}{\Omega^2}\nablav\right)^3r\Omega^2\overone{\underline\alpha}=2r^4\slashed{\mathcal{D}}^*_2\slashed{\mathcal{D}}^*_1\overline{\slashed{\mathcal{D}}}_1\slashed{\mathcal{D}}_2 r\Omega^2\overone{\alpha}-6M\left[\Omega\slashed{\nabla}_4+\Omega\slashed{\nabla}_3\right]r\Omega^2\overone{\alpha}.
\end{align}
\noindent The equations above were derived from the full system \fullsystem in Section 3.2 of Part I \cite{Mas20}. We used the Teukolsky--Starobinsky identities in \cite{Mas20} to couple solutions to the Teukolsky equations and construct a combined scattering theory which governs the radiating degrees of freedom of the linearised Einstein equations \fullsystem. The results of this paper will indeed confirm that the radiation energy of the full system can be completely characterised via the system composed of the Teukolsky equations \bref{T+2}, \bref{T-2} and the Teukolsky--Starobinsky identities \bref{eq:TS-}, \bref{eq:TS+}. 

\section{Well-posedness of the linearised Einstein equations in a double null gauge}\label{Section 5: well-posedness for IVP}

In this section we study the well-posedness of the linearised Einstein equations with smooth initial data on ${\Sigma^*_+}$ and $\overline{\Sigma}$. First, we look at the constraint equations on $\Sigma^*_+$ and show how we can prescribe a set of free seed data and solve the constraint equations to obtain a full Cauchy data set. We then prove local existence and uniqueness of solutions to the linearised Einstein system starting from smooth seed data that satisfy the constraint equations.

\subsection{The constraint equations}

We first list the constraints implied by the system \fullsystem on each of ${\Sigma^*_+}, \overline{\Sigma}$. We then describe a scheme to solve the equations and obtain compactly supported Cauchy data out of compactly supported seed data.

\subsubsection{The constraints on ${\Sigma^*_+}$}

A solution $\mathfrak{S}$ to the linearised Einstein system satisfies the following equations on the surface ${\Sigma^*_+}$:

\begin{defin}
The linearised Einstein system \fullsystem induces the following constraints on ${\Sigma^*_+}$:
\begin{align}\label{Spacelike Codazzis at Sigma star}
\begin{split}
    \slashednabla_N\divo{\glinh}\,-&\frac{1}{r(2-\Omega^2)}\mathring{\fancyd_2}\mathring{\fancydstar_2}\,\bmlin=\frac{8M}{r^2(2-\Omega^2)}\mathring{\slashed{d}}\left({\Olin}\right)+2N\left[\mathring{\slashed{d}}\left({\Olin}\right)\right]\\&+\frac{1}{\Omega^2(2-\Omega^2)}\partial_u \bmlin-\frac{1}{4\Omega^2}\slashednabla_{\ustar}\frac{r}{\Omega^2}\partial_u\bmlin+\frac{1}{2r}\mathring{\slashednabla}\divo\frac{\bmlin}{2-\Omega^2}+\frac{1}{2}\mathring{\slashednabla}\,\,N\tr\glin,
\end{split}
\end{align}

\begin{align}\label{Spacelike Gauss at Sigma star}
    \begin{split}
       &-\frac{3}{8r\Omega^2}\partial_{\ustar}\mathring{\slashed{div}}\bmlin+\frac{1}{4r}\mathring{\slashed{div}}\,N\, \bmlin+\frac{\Omega^2+3}{2r(2-\Omega^2)}\divr{\bmlin}+\frac{1}{r^2}\left(\mathring{\slashed{\Delta}}-2\right)\Olin+\frac{\Omega^4-3\Omega^2+4}{4r(2-\Omega^2)}\,N\,\tr\glin\\&+\frac{3\Omega^4-5\Omega^2-4}{4r(2-\Omega^2)}\frac{1}{2\Omega^2}\partial_{\ustar}\tr\glin-\frac{2-\Omega^2}{4\Omega^2}\partial_{\ustar}\left(N-\frac{1}{2\Omega^2}\partial_{\ustar}\right)\tr\glin-\frac{2\Omega^2}{r}\left(N-\frac{1}{2\Omega^2}\partial_{\ustar}\right)\Olin-\Klin=0
    \end{split}
\end{align}

\begin{align}\label{Spacelike connection constraint at Sigma star}
    \begin{split}
        -\frac{2}{\Omega^2}\partial_{\ustar}&N\tr\glin+\frac{2(3\Omega^4-10\Omega^2+7)}{r(2-\Omega^2)^2}\frac{1}{2\Omega^2}\partial_{\ustar}\tr\glin+\frac{2(3-\Omega^2)}{r(2-\Omega^2)^2}N\tr\glin-\frac{2\Omega^4-5\Omega^2+2}{r(2-\Omega^2)^3}\slashed{div}\bmlin\\&+\frac{2}{r(2-\Omega^2)}\left(N-\frac{1}{2\Omega^2}\partial_{\ustar}\right)\mathring{\slashed{div}}\bmlin-\frac{16}{r(2-\Omega^2)}\left(N+\frac{2M}{r}\frac{1}{2\Omega^2}\partial_{\ustar}\right)\Olin=0.
    \end{split}
\end{align}
\end{defin}

Note that \bref{Spacelike Codazzis at Sigma star} derives from the Codazzi equations \bref{elliptic equation 1}, \bref{elliptic equation 2} and the equations \bref{D3etabar}, \bref{D4eta}. Equation \bref{Spacelike Gauss at Sigma star} follows from the Gauss equation \bref{Gauss} and the equation \bref{D3TrChi} (or equivalently \bref{D4TrChiBar}), and either one of \bref{D4TrChi} and  \bref{D3TrChibar}. The equation \bref{Spacelike connection constraint at Sigma star} is derived from the equations \bref{D4TrChi} and \bref{D3TrChibar}.\\

We can derive from \bref{Spacelike Gauss at Sigma star} and \bref{Spacelike connection constraint at Sigma star}


\begin{align}\label{constraint equation for Olin on Sigma star}
    \begin{split}
        \frac{32M}{r^2}N(\Olin)+&\frac{8M^2}{r^3(2-\Omega^2)}N\tr\glin=-(2-\Omega^2)\left(\frac{1}{\Omega^2}\partial_{\ustar}\right)^2\tr\glin+\frac{11-5\Omega^2}{r(2-\Omega^2)}\frac{1}{\Omega^2}\partial_{\ustar}\tr\glin+8\Klin\\&-\frac{8}{r}\frac{1}{\Omega^2}\partial_\ustar\,\Olin-\frac{8}{r}\left(\mathring{\slashed{\Delta}}-2\right)\Olin+\frac{2}{r\Omega^2}\divo\partial_\ustar\bmlin+\frac{2\Omega^4+9\Omega^2-26}{r(2-\Omega^2)^2}\divo\bmlin,
    \end{split}
\end{align}

\subsubsection{The constraints on $\overline{\Sigma}$}
We study the constraint equations on $\overline{\Sigma}$ in the Kruskal coordinates and the coordinates $(R,\theta^A)$ induced by it on $\overline{\Sigma}$ ($R:=V-U$ was defined in \bref{T R Kruskal coordinates}):

\begin{defin}
The linearised Einstein system \fullsystem induces the following constraints on $\overline{\Sigma}$:

\begin{align}\label{Spacelike Codazzis at Sigma bar}
\begin{split}
    \slashednabla_T\divr\glinh-2M\fancyd_2\fancydstar_2\frac{\bmlin}{V}=&-\frac{2}{r}\slashednabla_T{\slashednabla}\Olin+\frac{R}{r}\left(3+\frac{r}{2M}\right)(\partial_T-\partial_R)\frac{\bmlin}{V}-\frac{f}{4M}\slashednabla_R\left(\partial_T-\partial_R\right)\frac{\bmlin}{V}+\frac{1}{2}{\slashednabla}\partial_T\tr\glin\\&+\frac{1}{2}{\slashednabla}\divr\frac{\bmlin}{V},
\end{split}
\end{align}

\begin{align}\label{Spacelike Gauss at Sigma bar}
\begin{split}
    \Klin=&\frac{1}{r^2}\left(\mathring{\slashed{\Delta}}-2\right)\Olin+\frac{f}{8M}\divr\left(\partial_T-\partial_R\right)\frac{\bmlin}{V}+\frac{f}{8M^2}\partial_R\left(\partial_T-\partial_R\right)\tr\glin+\frac{R}{8M^2}\left(1+\frac{10M}{r}\right)\partial_R\tr\glin\\&-\frac{R}{8M^2}\left(1+\frac{6M}{r}\right)\partial_T\tr\glin+\frac{R}{Mr}\left(\partial_T-\partial_R\right)\Olin+\frac{1}{r}\divr\bmlin,
\end{split}
\end{align}
\begin{align}\label{Spacelike connection constraint at Sigma bar}
    \begin{split}
        2f\partial_R\partial_T\tr\glin+\frac{R}{2}\left(1+\frac{6M}{r}\right)\partial_T\tr\glin+{2Mf}(\partial_T+\partial_R) \divr\frac{\bmlin}{V}+MR\left(1+\frac{6M}{r}\right)\divr\frac{\bmlin}{V}=\frac{8M}{r}R\partial_T\Olin.
    \end{split} 
\end{align}

\end{defin}

Subtracting \bref{Spacelike connection constraint at Sigma bar} from \bref{Spacelike Gauss at Sigma bar} we get
\begin{align}
\begin{split}
    &2U\left(1+\frac{6M}{r}\right)\partial_T \tr\glin+\frac{R}{2}\left(1+\frac{10M}{r}\right)\partial_R \tr\glin-2f\partial_R^2\tr\glin+\frac{4MR}{r}\left(-4\partial_T+\partial_R\right)\Olin\\&-r\Omega^2(2\Omega^2-1)\divr\bmlin-2Mf\divr\,\partial_R\frac{\bmlin}{V}+16M^2\left(\frac{1}{r^2}(\mathring{\slashed{\Delta}}-2)\Olin-\Klin\right)=0
\end{split}
\end{align}

\subsection{Existence and uniqueness of solutions to the linearised Einstein equations}

\begin{defin}\label{def of initial data on spacelike surface}
Let $\mathcal{S}$ be a spherically symmetric subdomain of ${\Sigma^*_+}$. A \textit{smooth initial data set} on $\mathcal{S}$ is a tuple ($\glinhs, \tr\glins$, $\Olinos$, $\bmlins$, $\glinhs'$, $\tr\glins'$, $\Olinos'$, $\bmlins'$) where $\glinhs,\,\glinhs'\in\Gamma^{(2)}(\mathcal{S}),\; \bmlins,\,\bmlins'\in\Gamma^{(1)}(\mathcal{S}),$ and $\Olinos,\tr\glins,\,\Olinos',\tr\glins'\in\Gamma^{(0)}(\mathcal{S})$, such that the constraint equations \bref{Spacelike Codazzis at Sigma star}, \bref{Spacelike Gauss at Sigma star}, \bref{Spacelike connection constraint at Sigma star} are satisfied.
\end{defin}

We now prove that the linearised Einstein system is well-posed for smooth initial data on either ${\Sigma^*_+}$ or $\overline{\Sigma}$. In the following proposition, we use the well-posedness of the Teukolsky equations to construct $\alin, \ablin$, which are then used to source the remaining components of the solution.

\begin{proposition}\label{EinsteinWP}
Let $\mathcal{S}$ be a spherically symmetric subdomain in ${\Sigma^*_+}$ and let (\,$\glinhs, \tr\glins$, $\Olinos$, $\bmlins$, $\glinhs'$, $\tr\glins'$, $\Olinos'$, $\bmlins'$\,) be an initial data set on $\mathcal{S}$ in the sense of \Cref{def of initial data on spacelike surface}. Then there exists a unique solution 
\begin{align}
    \mathfrak{S}=\left(\;\glinh\;,\;\tr\glin\;,\;\bmlin\;,\;\Olino\;,\;\xlin\;,\;\xblin\;,\;\trx\;,\;\trxbar\;,\;\elin\;,\;\eblin\;,\;\olin\;,\;\olinb\;,\;\alin\;,\;\ablin\;,\;\blin\;,\;\bblin\;,\;\rlin\;,\;\slin\;\right),
\end{align}
to the linearised Einstein equations \fullsystem on $D^+(\mathcal{S})$ in the sense of \Cref{def of linearised einstein equations and solutions}, such that the restriction of $(\,\glinh\;,\;\tr\glin\;,\;\Olin\;,\;\bmlin\,)$ to $\mathcal{S}$ coincides with (\,$\glinhs, \tr\glins$, $\Olinos$, $\bmlins$), and the restriction of $(\,\slashednabla_{n_{{\Sigma^*_+}}}\glinh\;,\;\slashednabla_{n_{{\Sigma^*_+}}}\tr\glin\;$, $\;\slashednabla_{n_{{\Sigma^*_+}}}\Olin\;,\;\slashednabla_{n_{{\Sigma^*_+}}}\bmlin\,)$ to $\mathcal{S}$ coincides with (\,$\glinhs'$, $\tr\glins'$, $\Olinos'$, $\bmlins'$).
\end{proposition}

\begin{proof}
We will assume that the given initial data is supported on $\ell\geq2$ and we postpone the treatment of the $\ell=0,1$ modes until \Cref{linearised Kerr section}.
Define the following quantities on $\mathcal{S}$:
\begin{align}
     &\elins:=\slashednabla\Olinos+\frac{1}{4\sqrt{2-\Omega^2}}\bmlins'+\frac{1}{8\Omega^2}\partial_\ustar\bmlins,\qquad\eblins:=2\slashednabla\Olinos-\elins,\label{elin and eblin data on Sigma*}\\
     &\xlins:=\frac{\sqrt{2-\Omega^2}}{2}\glinhs'-\frac{2-\Omega^2}{4\Omega^2}\slashednabla_\ustar\glinhs-\fancydstar_2\bmlins,\qquad\qquad\xblins:=\frac{1}{\sqrt{2\Omega^2}}\glinhs'+\frac{1}{2\Omega^2}\slashednabla_\ustar \glinhs,\label{xlin and xblin data on Sigma*}\\
     &\alins:=2(2-\Omega^2)\fancydstar_2\elins-\frac{2\Omega^2+1}{r}\xlins+\frac{2-\Omega^2}{\Omega^2}\slashednabla_\ustar \xlins+\frac{\Omega^2(2-\Omega^2)}{r}\xblins,\label{alpha data on Sigma*}\\
     &\ablins:=\frac{2}{2-\Omega^2}\fancydstar_2\eblins+\frac{5-2\Omega^2}{r(2-\Omega^2)}\xblins-\frac{1}{\Omega^2}\slashednabla_\ustar\xblins-\frac{1}{r(2-\Omega^2)}\xlins,\label{alphabar data on Sigma^*}\\
     &\otxs=\frac{\sqrt{2-\Omega^2}}{2}\tr\glins'-\frac{2-\Omega^2}{4\Omega^2}\partial_\ustar\glins-\divr\bmlins,\qquad\qquad \otxbs=\frac{1}{2\sqrt{2-\Omega^2}}\tr\glins'-\frac{1}{4\Omega^2}\partial_\ustar\tr\glins,\label{otx data on Sigma*}\\
     &\blins=-\divr\,\xlins-\frac{\Omega^2}{r}\eblins+\frac{1}{2}\slashednabla\otxs,\qquad\qquad \bblins=\divr\,\xblins-\frac{1}{r}\elins-\frac{1}{2}\slashednabla\otxbs,\label{beta and betabar data on Sigma*}\\
     &\slins=\slashed{curl}\,\elins=-\slashed{curl}\,\eblins,\qquad\qquad \rlins=-\Klin[\glins]+\frac{1}{2r}\left(\otxs-\otxbs\right)-\frac{2\Omega^2}{r^2}\Olinos.\label{sigma and rho data on Sigma*}
\end{align}
Moreover, define
\begin{align}
&\alins'=\sqrt{2-\Omega^2}\left[-\frac{1}{2\Omega^2}\slashednabla_\ustar\alins+\frac{1}{r}\alins-2\fancydstar_2\blins+\frac{6M}{r^3}\xlins\right],\label{data for time derivative of alpha on Sigma star}\\ &\ablins'=\frac{\sqrt{2-\Omega^2}}{2\Omega^2}\slashednabla_{\ustar}\ablins-\frac{\sqrt{2-\Omega^2}}{r}\ablins+\frac{2}{\sqrt{2-\Omega^2}}\fancydstar_2\,\bblins+\frac{6M}{r^3\sqrt{2-\Omega^2}}\xblins.\label{data for time derivative of alphabar on Sigma star}
\end{align}
With data $(\alins,\alins')$ for \bref{T+2} and $(\ablins,\ablins')$ for \bref{T-2},  \Cref{WP+2Sigma*,,WP-2Sigma*} allow us to construct smooth solutions $\alin,\ablin$ to the Teukolsky equations \bref{T+2}, \bref{T-2}, respectively, on $D^+(\mathcal{S})$.  
We can immediately solve the transport equations \bref{D4Chihat}, \bref{D3Chihatbar} for $\xlin, \xblin$ using $\xlins, \xblins$ as data, respectively. Similarly, we can find $\blin_{\ell\geq2}, \bblin_{\ell\geq2}$ using \bref{Bianchi+1a}, \bref{Bianchi-1a} with \bref{beta and betabar data on Sigma*} as data. The $+2$ Teukolsky equation \bref{T+2} together with equations in \bref{D4Chihat}, \bref{Bianchi+1a} now imply
\begin{align}
    \nablav\left(-2\fancydstar_2 \frac{r^4\blin}{\Omega}+6M\frac{r^2\xlin}{\Omega}\right)=\left(\mathcal{A}_2-\frac{6M}{r}\right)r^3\alin=\nablav\left(\frac{r^4}{\Omega^4}\nablau r\Omega^2\alin\right),
\end{align}
so equation \bref{Bianchi+2} is satisfied, and a similar argument shows that equation \bref{Bianchi-2} is also satisfied. \\
We now define $\elin,\eblin$ up to $\ell=0,1$ via \bref{D3Chihat}, \bref{D4Chihatbar}:
\begin{align}
\begin{split}
    \frac{1}{\Omega^2}\nablau r\Omega\xlin&=-2\fancydstar_2 \elin_{\ell\geq2}-\Omega\xblin,\\
    \nablav\Omega^{-1}\xblin+\frac{1}{r}\Omega^{-1}\xblin&=-2\fancydstar_2\eblin_{\ell\geq2}+\Omega\xlin.  
\end{split}
\end{align}

Define $\glinh$ via \bref{metric transport in 3 direction traceless} with $\glinhs$ as data, and define $\bmlins$ up to $\ell=0,1$ via \bref{metric transport in 4 direction traceless} (note that by \bref{xlin and xblin data on Sigma*}, $\bmlin_{\ell\geq2}|_{\mathcal{S}}=\bmlins_{\ell\geq2}$). Note that
\begin{align}
\begin{split}
    \nablau r\nablav \glinh&=-\Omega^2\nablav\glinh+r\nablav\left[2\Omega\xblin\right]=-2\Omega^2\left[\Omega\xlin-\fancydstar_2\bmlin\right]+2\nablav r\Omega\xblin-2\Omega^2\Omega\xblin.\\
    &=-2\Omega^2\left[\fancydstar_2\bmlin+2\mathring{\fancydstar_2}\elin+\Omega\xblin\right].
\end{split}
\end{align}
On the other hand, using equation \bref{metric transport in 4 direction traceless} we have
\begin{align}
    \nablau r\nablav \glinh=2\nablau r\Omega\xlin+2\mathring{\fancydstar_2}\nablau \bmlin.
\end{align}
Thus \bref{D3Chihat} implies
\begin{align}
    \fancydstar_2\left[\nablau\bmlin_{\ell\geq2}-2\Omega^2(\elin_{\ell\geq2}-\eblin_{\ell\geq2})-\frac{\Omega^2}{r}\bmlin_{\ell\geq2}\right]=0,
\end{align}
which implies that \bref{partial_u b} holds up to $\ell=0,1$. An identical argument shows that equations \bref{D3etabar}, \bref{D4eta} are satisfied up to $\ell=0,1$ modes, using \bref{D4Chihat}, \bref{D3Chihat}, and \bref{D4Chihatbar}. Note that now 
\begin{align}
    \Omega^2\,\alin&=\frac{2-\Omega^2}{\Omega^2}\nablaubar \Omega\xlin+\frac{(2-\Omega^2)}{r}\Omega\xblin+2(2-\Omega^2)\fancydstar_2\elin_{\ell\geq2}-\frac{2\Omega^2+1}{r}\Omega\xlin,\\
    \Omega^{-2}\ablin&=-\frac{1}{\Omega^2}\nablaubar \Omega^{-1}\xblin-\frac{1}{r(2-\Omega^2)}\Omega\xlin+\frac{2}{2-\Omega^2}\fancydstar_2\eblin_{\ell\geq2}+\frac{5-2\Omega^2}{r(2-\Omega^2)}\Omega^{-1}\xblin,
\end{align}
so in particular $\elin_{\ell\geq2}|_{\mathcal{S}}=\elins_{\ell\geq2}$ and $\eblin_{\ell\geq2}|_{\mathcal{S}}=\eblins_{\ell\geq2}$, and thus we have $\frac{1}{\Omega^2}\nablau\bmlin_{\ell\geq2}^A\Big|_{\mathcal{S}}=\bmlins'+\frac{1}{2\Omega^2}\nablaubar \bmlins$. \\

With the quantities $\alin, \ablin, \Psilin, \Psilinb, \xlin, \xblin, \blin, \bblin, \elin, \eblin$ at hand, define
\begin{align}
    &\text{C}:=\curlr\left(\elin_{\ell\geq2}+\eblin_{\ell\geq2}\right),\label{C}\\
    &\text{Codazzi}^-:=-\curlr\divr\xblin+\frac{\Omega}{r}\curlr\elin_{\ell\geq2}+\curlr\bblin_{\ell\geq2},\label{Codazzi-}\\
    &\text{Codazzi}^+:=-\curlr\divr\xlin-\frac{\Omega}{r}\curlr\eblin_{\ell\geq2}-\curlr\blin_{\ell\geq2}.\label{Codazzi+}    
\end{align}

A computation using equations \bref{D4Chihat}, \bref{D3Chihatbar}, \bref{D3Chihat}, \bref{D4Chihatbar}, \bref{D3etabar}, \bref{D4eta} shows that the above quantities satisfy

\begin{align}\label{equation of C}
    \begin{split}
        \left(\nablau\nablav-\frac{\Omega^2}{r^2}\mathring{\slashed{\Delta}}\right) r^2\text{C}=\frac{1}{r^2}\left[\nablau (r^4\Omega \,\text{Codazzi}^+)+\nablav(r^4\Omega\,\text{Codazzi}^-)\right],
    \end{split}
\end{align}

\begin{align}
    \refstepcounter{equation}\latexlabel{equation of Codazzi+}
    \refstepcounter{equation}\latexlabel{equation of Codazzi-}
        \nablav\left[\frac{r^4}{\Omega}\text{Codazzi}^+\right]=-r\nablav(r^2\text{C}),\qquad\qquad \nablau\left[\frac{r^4}{\Omega}\text{Codazzi}^-\right]=r\nablau(r^2\text{C}).\tag{\ref{equation of Codazzi+},\;\ref{equation of Codazzi-}}
\end{align}

A standard local well-posedness argument using energy estimates and Gr\"onwall's inequality shows that the system \bref{equation of C}, \bref{equation of Codazzi+}, \bref{equation of Codazzi-} admits only the trivial solution for trivial data on $\mathcal{S}$.\\ 

We may then define $\Olino$ up to $\ell=0,1$ by $2\slashednabla_A\Olin_{\ell\geq2}=\elin_{\ell\geq2}+\eblin_{\ell\geq2}$ and we immediately have that $\Olin_{\ell\geq2}|_{\mathcal{S}}=\Olinos_{\ell\geq2}$ by \bref{elin and eblin data on Sigma*}.
We now define $\otx_{\ell\geq2}, \otxb_{\ell\geq2}$ via \bref{D4TrChi} using $\otxs, \otxbs$ as data. We can now establish that the Codazzi equations are satisfied: we have
\begin{align}
    \divo \nablau \frac{r^2\xblin}{\Omega}=\divo(-r^2\ablin),
\end{align}
\begin{align}
\begin{split}
    \nablau\frac{r^3}{\Omega}\left(\frac{\Omega}{r}\elin_{\ell\geq2}+\bblin_{\ell\geq2}+\frac{1}{2\Omega}\slashednabla\otxb_{\ell\geq2}\right)&=2r\mathring{\slashednabla}\olinb_{\ell\geq2}-r^2\Omega\bblin_{\ell\geq2}+\frac{\Omega^2}{r^2}\frac{r^4\bblin_{\ell\geq2}}{\Omega}-\frac{1}{r}\divo r^3\ablin-2\mathring{\slashednabla}r\olinb_{\ell\geq2}\\&=-\divo r^2\ablin,
\end{split}
\end{align}
and since $\bblins$ was chosen consistently with the Codazzi equation \bref{elliptic equation 2} via \bref{beta and betabar data on Sigma*}, we see that \bref{elliptic equation 2} is propagated. A similar argument shows that \bref{elliptic equation 1} is also propagated. \\

We now define $\rlin_{\ell\geq2}, \slin_{\ell\geq2}$ by solving \bref{Bianchi-0}, \bref{Bianchi-0*}:
\begin{align}
    \nablau r^3\rlin_{\ell\geq2}=-\divo r^2\Omega\bblin_{\ell\geq2}+3M\otxb_{\ell\geq2},\qquad\nablau r^3\slin_{\ell\geq2}=-\curlo\, r^2\Omega\bblin_{\ell\geq2},
\end{align}
with data $\rlins, \slins$. This defines $\rlin$ and $\slin$ up to $\ell=0,1$. Applying $\nablau \frac{r^2}{\Omega^2}$ to both sides \bref{Bianchi+1b} and using \bref{Bianchi-2}, \bref{Bianchi-1a}, \bref{Bianchi-0}, \bref{Bianchi-0*}, \bref{D3etabar} and the Codazzi equation \bref{elliptic equation 1}, we get
\begin{align}
    \nablau \frac{r^2}{\Omega^2}\nablav r^2\Omega\bblin_{\ell\geq2}=\nablau\left[\fancydstarring_1(r^3\rlin_{\ell\geq2},r^3\slin_{\ell\geq2})+6Mr\eblin_{\ell\geq2}\right].
\end{align}
We may check using the definitions of $\ablins, \bblins, \rlins, \slins$ and $\eblins$ on $\mathcal{S}$ and the constraint equations \bref{Spacelike Codazzis at Sigma star}--\bref{Spacelike connection constraint at Sigma star} that equation \bref{Bianchi+1b} is satisfied by the initial data on $\mathcal{S}$, thus \bref{Bianchi+1b} propagates on $D^+(\mathcal{S})$. In turn, applying $\nablav$ to \bref{elliptic equation 1} and using \bref{Bianchi+1b}, \bref{D4eta}, \bref{D4Chihatbar}, and \bref{elliptic equation 2}, we get that \bref{D4TrChiBar} holds.\\

Note that using \bref{Bianchi+2}, \bref{Bianchi+1a} and \bref{D4Chihat}, we have
\begin{align}\label{d4 d3 beta}
    &\nablav\frac{r^2}{\Omega^2}\nablau  r^2\Omega\blin_{\ell\geq2}=-2\mathring{\fancyd_2}\mathring{\fancydstar_2}r^2\Omega\blin-2(3\Omega^2-2)r^2\Omega\blin_{\ell\geq2}+6M\divo \Omega\xlin,
\end{align}
We can also deduce from \bref{D3etabar} and \bref{Bianchi+0*} that
\begin{align}
    \partial_u \left(\curlo\, r^2\eblin_{\ell\geq2}\right)=\curlo\, r^2\Omega\bblin_{\ell\geq2}
\end{align}
which when combined with the definition of $\slins$ in \bref{sigma and rho data on Sigma*} implies \bref{elliptic equation 3 and 4}, which then implies \bref{Bianchi+0*}. Since \bref{Bianchi-2}, \bref{Bianchi+1b} and \bref{D4Chihatbar} are satisfied, we have that the expression \bref{expression for Psilinb} for $\Psilinb$ applies:
\begin{align}
    \Psilinb=2r^2\fancydstar_2\fancydstar_1\left(r^3\rlin_{\ell\geq2},\,r^3\slin_{\ell\geq2}\right)+6M\left(r\Omega\xlin-r\Omega\xblin\right).
\end{align}
Given that $\text{A}=\text{C}=0$, this immediately implies $\curlr\elin_{\ell\geq2}=-\curlr\eblin_{\ell\geq2}=\slin_{\ell\geq2}$. Therefore, the vanishing of $\underline{\text{A}}$ implies

\begin{align}\label{Expression for Psi WP}
    \Psilin=2r^2\fancydstar_2\fancydstar_1\left(r^3\rlin_{\ell\geq2},-r^3\slin_{\ell\geq2}\right)+6M\left(r\Omega\xlin-r\Omega\xblin\right).
\end{align}

The above implies \bref{Bianchi-1b} is satisfied. Using \bref{d4 d3 beta}, we then get \bref{Bianchi+0}. Equation \bref{transport equation of little omega} follows from applying $\nablau$ to \bref{D4etabar} and following through using \bref{Bianchi+1b}, \bref{Bianchi-1b}. Equation \bref{D3TrChi} follows by applying $\nablau$ to \bref{D4TrChi} and using \bref{D4TrChiBar} and \bref{transport equation of little omega}. Now  equations \bref{metric transport in 4 direction trace}, \bref{metric transport in 3 direction trace} and the Gauss equation \bref{Gauss} follow immediately.
\end{proof}

A similar proof goes through to show the well-posedness of the system \fullsystemK with data on $\overline{\Sigma}$ defined as follows:

\begin{defin}\label{def of initial data on Sigmabar}
Let $\mathcal{S}$ be an spherically symmetric subdomain of $\overline{\Sigma}$. A \textit{smooth initial data set} on $\mathcal{S}$ is a tuple ($\glinhs, \tr\glins$, $\Olinos$, $\bmlins$, $\glinhs'$, $\tr\glins'$, $\Olinos'$, $\bmlins'$) where $\glinhs,\,\glinhs'\in\Gamma^{(2)}(\mathcal{S}),\; \bmlins,\,\bmlins'\in\Gamma^{(1)}(\mathcal{S}),$ and $\Olinos,\tr\glins,\,\Olinos',\tr\glins'\in\Gamma^{(0)}(\mathcal{S})$, such that the constraint equations \bref{Spacelike Codazzis at Sigma bar}, \bref{Spacelike Gauss at Sigma bar}, \bref{Spacelike connection constraint at Sigma bar} are satisfied.
\end{defin}

In the following, $n^+_{\overline{\Sigma}}$ is the future directed unit normal to $\overline{\Sigma}$:

\begin{proposition}\label{EinsteinWP Sigmabar}
Let $\mathcal{S}$ be an spherically symmetric subdomain of $\overline{\Sigma}$ and let (\,$\glinhs, \tr\glins$, $\Olinos$, $\bmlins$, $\glinhs'$, $\tr\glins'$, $\Olinos'$, $\bmlins'$\,) be an initial data set on $\mathcal{S}$ in the sense of \Cref{def of initial data on Sigmabar}. Then there exists a unique solution $\mathfrak{S}$
to the linearised Einstein equations \fullsystem on $D^+(\mathcal{S})$ in the sense of \Cref{def of linearised einstein equations and solutions}, such that the restriction of $(\,\glinh\;,\;\tr\glin\;,\;\Olin\;,\;V^{-1}\bmlin\,)$ to $\mathcal{S}$ coincides with (\,$\glinhs, \tr\glins$, $\Olinos$, $\bmlins$), and the restriction of $(\,\slashednabla_{n^+_{\overline{\Sigma}}}\glinh\;$, $\;\slashednabla_{n^+_{\overline{\Sigma}}}\tr\glin\;,\;\slashednabla_{n^+_{\overline{\Sigma}}}\Olin\;,\;\slashednabla_{n^+_{\overline{\Sigma}}}V^{-1}\bmlin\,)$ to $\mathcal{S}$ coincides with (\,$\glinhs'$, $\tr\glins'$, $\Olinos'$, $\bmlins'$).
\end{proposition}

An identical argument goes for the initial value problem with solutions to \fullsystem evolved from data on $\Sigma^*_-$ to $J^-(\Sigma^*_-)$, or solutions to \fullsystemK evolved on $J^-(\overline{\Sigma})$ from initial data $\overline{\Sigma}$. In the following, $n^-_{\overline{\Sigma}}=-n_{\overline{\Sigma}}$, the past directed unit normal to $\overline{\Sigma}$.

\begin{defin}\label{def of initial data on spacelike surface past}
Let $\mathcal{S}$ be a spherically symmetric subdomain of ${\Sigma^*_-}$. A \textit{smooth initial data set} on $\mathcal{S}$ is a tuple ($\glinhs, \tr\glins$, $\Olinos$, $\bmlins$, $\glinhs'$, $\tr\glins'$, $\Olinos'$, $\bmlins'$) where $\glinhs,\,\glinhs'\in\Gamma^{(2)}(\mathcal{S}),\; \bmlins,\,\bmlins'\in\Gamma^{(1)}(\mathcal{S}),$ and $\Olinos,\tr\glins,\,\Olinos',\tr\glins'\in\Gamma^{(0)}(\mathcal{S})$, such that the constraint equations induced by the system \fullsystem on $\Sigma^*_-$ are satisfied.
\end{defin}


\begin{proposition}\label{EinsteinWP past}
Let $\mathcal{S}$ be a spherically symmetric subdomain in ${\Sigma^*_-}$ and let (\,$\glinhs, \tr\glins$, $\Olinos$, $\bmlins$, $\glinhs'$, $\tr\glins'$, $\Olinos'$, $\bmlins'$\,) be an initial data set on $\mathcal{S}$ in the sense of \Cref{def of initial data on spacelike surface past}. Then there exists a unique solution $\mathfrak{S}$
to the linearised Einstein equations \fullsystem on $J^-(\mathcal{S})$ in the sense of \Cref{def of linearised einstein equations and solutions}, such that the restriction of $(\,\glinh\;,\;\tr\glin\;,\;\Olin\;,\;\Omega^{-2}\bmlin\,)$ to $\mathcal{S}$ coincides with (\,$\glinhs, \tr\glins$, $\Olinos$, $\bmlins$), and the restriction of $(\,\slashednabla_{n_{{\Sigma^*_-}}}\glinh\;,\;\slashednabla_{n_{{\Sigma^*_-}}}\tr\glin\;$, $\;\slashednabla_{n_{{\Sigma^*_-}}}\Olin\;,\;\slashednabla_{n_{{\Sigma^*_-}}}\Omega^{-2}\bmlin\,)$ to $\mathcal{S}$ coincides with (\,$\glinhs'$, $\tr\glins'$, $\Olinos'$, $\bmlins'$).
\end{proposition}

\begin{proposition}\label{EinsteinWP Sigmabar past}
Let $\mathcal{S}$ be a spherically symmetric subdomain of $\overline{\Sigma}$ and let (\,$\glinhs, \tr\glins$, $\Olinos$, $\bmlins$, $\glinhs'$, $\tr\glins'$, $\Olinos'$, $\bmlins'$\,) be an initial data set on $\mathcal{S}$ in the sense of \Cref{def of initial data on Sigmabar}. Then there exists a unique solution $\mathfrak{S}$
to the linearised Einstein equations \fullsystem on $J^-(\mathcal{S})$ in the sense of \Cref{def of linearised einstein equations and solutions}, such that the restriction of $(\,\glinh\;,\;\tr\glin\;,\;\Olin\;,\;V^{-1}\bmlin\,)$ to $\mathcal{S}$ coincides with (\,$\glinhs, \tr\glins$, $\Olinos$, $\bmlins$), and the restriction of $(\,\slashednabla_{n^-_{\overline{\Sigma}}}\glinh\;,\;\slashednabla_{n^-_{\overline{\Sigma}}}\tr\glin\;,\;\slashednabla_{n^-_{\overline{\Sigma}}}\Olin\;,\;\slashednabla_{n^-_{\overline{\Sigma}}}V^{-1}\bmlin\,)$ to $\mathcal{S}$ coincides with (\,$\glinhs'$, $\tr\glins'$, $\Olinos'$, $\bmlins'$).
\end{proposition}

For convenience, we state and prove a well-posedness theorem for the mixed Cauchy-characteristic value problem with data on $\Sigma^*_+$ and $\mathscr{H}^+\cap\{v\leq0\}$. 

\begin{proposition}\label{EinsteinWP Sigmastar H+ past}
    Let (\,$\glinhs, \tr\glins$, $\Olinos$, $\bmlins$, $\glinhs'$, $\tr\glins'$, $\Olinos'$, $\bmlins'$\,) be an initial data set on $\Sigma^*_+$ in the sense of \Cref{def of initial data on spacelike surface} and let $\Olinos_{\mathscr{H}^+}$ be a smooth scalar field on $\overline{\mathscr{H}^+}\cap\{v\leq 0\}$, $\glinhs_{\mathscr{H}^+}$ be a smooth symmetric traceless $S^2$ 2-tensor field on $\overline{\mathscr{H}^+}\cap\{v\leq 0\}$. Then there exists a unique solution 
\begin{align}
    \mathfrak{S}=\left(\;\glinh\;,\;\tr\glin\;,\;\bmlin\;,\;\Olino\;,\;\xlin\;,\;\xblin\;,\;\trx\;,\;\trxbar\;,\;\elin\;,\;\eblin\;,\;\olin\;,\;\olinb\;,\;\alin\;,\;\ablin\;,\;\blin\;,\;\bblin\;,\;\rlin\;,\;\slin\;\right),
\end{align}
to the linearised Einstein equations \fullsystemK on $D^+(\overline{\Sigma})\cap J^-(\Sigma^*_+)$ in the sense of \Cref{def of linearised einstein equations and solutions}, such that the restriction of $(\,\glinh\;,\;\tr\glin\;,\;\Olin\;,\;\bmlin\,)$ to $\Sigma^*_+$ coincides with (\,$\glinhs, \tr\glins$, $\Olinos$, $\bmlins$), the restriction of $(\,\slashednabla_{n_{{\Sigma^*_+}}}\glinh\;,\;\slashednabla_{n_{{\Sigma^*_+}}}\tr\glin\;$, $\;\slashednabla_{n_{{\Sigma^*_+}}}\Olin\;,\;\slashednabla_{n_{{\Sigma^*_+}}}\bmlin\,)$ to $\Sigma^*_+$ coincides with (\,$\glinhs'$, $\tr\glins'$, $\Olinos'$, $\bmlins'$), and the restriction of $\Olino$, $\glinh$ to $\overline{\mathscr{H}^+}\cap\{v\leq 0\}$ coincides with $\Olinos_{\mathscr{H}^+}$ and $\glinhs_{\mathscr{H}^+}$ respectively.
\end{proposition}
\begin{proof}
    With the given data on $\Sigma^*_+$, we may use an identical argument to that of the proof of \Cref{EinsteinWP} to construct $\mathfrak{S}$ on $D^-(\Sigma^*_+)$ restricting to the given Cauchy data on $\Sigma^*_+$. Combining this solution with the given data on $\overline{\mathscr{H}^+}\cap\{v\leq0\}$, we may adapt Theorem 8.1 of \cite{DHR16} to construct the solution $\mathfrak{S}$ in the region $J^+(\overline{\Sigma})\cap\{V\leq 1, U\leq0\}$.
\end{proof}

\subsection{Pure gauge solutions}

In \cite{DHR16}, it was shown that linearising infinitesimal diffeomorphisms which preserve the double null gauge generates solutions to the linearised Einstein equations. These solutions represent the residual gauge ambiguity in choosing a double null gauge. As shown in \cite{DHR16}, these solutions can be classified into three families, which we present in this section and refer the reader to Section 5 of \cite{DHR16} for the details of their derivation.

\subsubsection{Family $\mathfrak{G}_{out}$: Pure gauge solutions generated by variations along the outgoing null direction}

\begin{lemma}\label{outwards gauge solutions}
Let $f$ be a smooth function of $v,\theta^A$. The following is a solution of the system \fullsystem:
\begin{align}
2\Olin &= \frac{1}{\Omega^2} \partial_v \left(f \Omega^2\right)  , & \glinh&=  \frac{4}{r} r^2 \slashed{\mathcal{D}}_2^\star\fancydstar_1(f,0)  \ , &\tr\glin &= \frac{4}{r}\left(\mathring{\slashed{\Delta}}+\Omega^2\right)f  , \nonumber \\
\bmlin &= 2r\fancydstar_1 \left[ r\partial_v \left(\frac{f}{r}\right),0\right] , & \elin &=- \frac{\Omega^2}{r^2} r \fancydstar_1(f,0)  , & \eblin &= -r\fancydstar_1\left[\frac{1}{\Omega^2}\partial_v \left(\frac{\Omega^2}{r}f\right),0 \right]   \nonumber \, ,
\nonumber \\
\Omega\xblin &= 2\frac{\Omega^2}{r^2} r^2 \slashed{\mathcal{D}}_2^\star \fancydstar_1 (f,0)  , & \otx &= 2 \partial_v \left(\frac{f \Omega^2}{r}\right)  , & \otxb &=  2\frac{\Omega^2}{r^2} \left[\mathring{\slashed{\Delta}} + 2\Omega^2-1 \right]f  , \nonumber \\
\rlin &= \frac{6M \Omega^2}{r^4} f  , & \bblin&= -\frac{6M\Omega}{r^4}  r \fancydstar_1 (f,0) , & \Klin &= -\frac{\Omega^2}{r^3}\left(\mathring{\slashed{\Delta}} + 2\right)f \nonumber
\end{align}
and
\[
\xlin = \alin = \ablin = 0 \ \ \ , \ \ \ \blin = 0 \ \ \ , \ \ \ \slin = 0 \nonumber \, .
\]
\end{lemma}

\begin{defin}
Given $f=f(v,\theta^A)$, define $\mathfrak{G}_{out}[f]$ to be the pure gauge solution generated by $f$ in the sense of \Cref{outwards gauge solutions}. 
\end{defin}

\subsubsection{Family $\mathfrak{G}_{in}$ : Pure gauge solutions generated by variations in the outgoing null direction}

\begin{lemma} \label{inwards gauge solutions} 
Let $\fbar$ be a function of $(u,\theta^A)$ such that $\Omega^2\fbar$ is smooth on $J^+({\Sigma^*_+})$ including $\mathscr{H}^+_{\geq0}$. The following is a solution of the system \fullsystem:
\begin{align}
 2\Olin &= \frac{1}{\Omega^2} \partial_u \left(\fbar \Omega^2\right), \, & \tr\glin &= -\frac{4\Omega^2 }{r}\fbar , & \bmlin &= -2\frac{\Omega^2}{r} r\fancydstar_1(\fbar,0), \nonumber \\
\Omega\xlin &= 2\frac{\Omega^2}{r^2} r^2 \slashed{\mathcal{D}}_2^\star \fancydstar_1 (\fbar,0), & \elin &= -\frac{1}{\Omega^2} r\fancydstar_1  \left[\partial_u \left(\frac{\Omega^2}{r}\fbar\right),0 \right] , & \eblin &= \frac{\Omega^2}{r^2} r \fancydstar_1( \fbar ,0), \nonumber \, \\
\otxb &= -2 \partial_u \left(\frac{ \Omega^2}{r}\fbar\right) , \ & \otx &= 2\frac{\Omega^2}{r^2} \left[\mathring{\slashed{\Delta}}+ 2\Omega^2-1\right]\fbar , 
\nonumber  \\
\blin &= \frac{6M\Omega}{r^4}  r \fancydstar_1( \fbar,0) , &  \rlin &= -\frac{6M \Omega^2}{r^4} \fbar  \, , & \Klin &= +\frac{\Omega^2}{r^3} \left(\mathring{\slashed{\Delta}} + 2 \right)\fbar \nonumber
\end{align}
and
\[
0 = \glinh= \xblin = \alin = \ablin = 0 \ \ \ , \ \ \ \bblin = 0 \ \ \ , \ \ \ \slin = 0 \nonumber \, .
\]

\end{lemma}
\begin{defin}
Given $\underline{f}=\underline{f}(v,\theta^A)$, define $\mathfrak{G}_{in}[\underline{f}]$ to be the pure gauge solution generated by $\underline{f}$ in the sense of \Cref{inwards gauge solutions}. 
\end{defin}

\subsubsection{Family $\mathfrak{G}_{res}$: Residual pure gauge solutions}\label{Section 7.3: Residual gauge solutions}

\begin{lemma} \label{residual gauge solutions}
For any smooth functions $q_1\left(v,\theta,\phi\right)$ and $q_2\left(v,\theta,\phi\right)$ the following is a pure gauge solution of the system of gravitational perturbations:
\begin{align}
\glinh  &=2r^2 \slashed{\mathcal{D}}_2^\star \slashed{\mathcal{D}}_1^\star \left(q_1,q_2\right) , &\tr\glin &= 2r^2 \slashed{\Delta} q_1, & \bmlin &= r^2 \slashed{\mathcal{D}}_1^\star \left(\partial_v q_1, \partial_v q_2\right) \, , \nonumber 
\end{align}
while the linearised
metric coefficient $\Olino$ as well as all linearised connection coefficients and  curvature components vanish.
\end{lemma}

\begin{defin}
Given $q_1=q_1(v,\theta^A), q_2=q_2(v,\theta^A)$, define $\mathfrak{G}_{res}[q_1,q_2]$ to be the pure gauge solution generated by $q_1,q_2$ in the sense of \Cref{residual gauge solutions}. 
\end{defin}

\begin{defin}\label{def of pure gauge}
A pure gauge solution is a solution to the linearised equations \fullsystem that can be expressed as a linear combination of the pure gauge solutions given in \Cref{inwards gauge solutions,,outwards gauge solutions,,residual gauge solutions}
\end{defin}

\begin{remark}\label{globally regular gauge solutions}
Note that if $\fbar$ of \Cref{inwards gauge solutions} is such that $\Omega^2\fbar$ is smooth on $J^+({\Sigma^*_+})$ including $\mathscr{H}^+_{\geq0}$, then $U\fbar$ is smooth on the whole of $\mathscr{M}$, thus by regarding $U\fbar$ instead of $\fbar$ as the generator of the family of solutions to \fullsystem defined in \Cref{inwards gauge solutions}, we can extend this family to $\mathscr{M}$ as solutions to \fullsystemK. A similar remark goes for \Cref{outwards gauge solutions} by regarding $Vf$ instead of $f$ as the generator of outgoing gauge solutions. 
\end{remark}

\subsection{Constructing initial data on ${\Sigma^*_\pm}, \overline{\Sigma}$}

We now show how to construct an initial data set on ${\Sigma^*_+}$ which satisfies the constraints \bref{Spacelike Codazzis at Sigma star}--\bref{Spacelike Gauss at Sigma bar}. The $\ell=0,1$ modes of the solution to the system \fullsystem are determined by the linearised Kerr parameters, so we confine our attention here to the $\ell\geq2$ part and treat the $\ell=0,1$ modes in the next section. 

In the following we see that having seed data for certain components of the initial data for the metric and connection components on either $\Sigma^*_+$ or $\overline{\Sigma}$  can be extended to a full initial data set by solving 2-dimensional elliptic equations on each of the spheres foliating $\Sigma^*_+$ or $\overline{\Sigma}$:
\begin{proposition}\label{constructing compactly supported data}
Given smooth data on ${\Sigma^*_+}$ for 
\begin{align}\label{12 10 2021}
    \glinh,\; \bmlin_{\ell\geq2},\;  \curlr \slashednabla_N\bmlin_{\ell\geq2}, \;   \tr\glin_{\ell\geq2},\;   \slashednabla_N \tr\glin_{\ell\geq2},\;   \Olin_{\ell\geq2},
\end{align}
there exists a unique set of smooth, compactly supported data for
\begin{align}
\slashednabla_N \glinh,\;   \slashednabla_N \Olin_{\ell\geq2},\;   \slashednabla_N \bmlin_{\ell\geq2},
\end{align}
which complete \bref{12 10 2021} into an initial data set for the system \fullsystem which is supported on $\ell\geq2$ spherical harmonic modes.
\end{proposition}

\begin{proof}
    With the given data, $\slashednabla_N \Olin_{\ell\geq2}$ is given by \bref{constraint equation for Olin on Sigma star}. We can then solve for $\divr\slashednabla_N \bmlin$ using \bref{Spacelike connection constraint at Sigma star}. Assuming $f$, $g$ are scalar functions with vanishing $\ell=0,1$ spherical harmonic modes such that $\slashednabla_N\bmlin_{\ell\geq2}=\fancydstar_1(f,g)$ (see \Cref{1-forms on S^2}), we may find $f$, $g$ via
    \begin{align}
        {\slashed{\Delta}}f=\divr\slashednabla_N\bmlin_{\ell\geq2},\qquad  {\slashed{\Delta}}g=\curlr\slashednabla_N\bmlin_{\ell\geq2}.
    \end{align}
    We can finally solve for $\fancyd_2\slashednabla_N \glinh$ using \bref{Spacelike Codazzis at Sigma star}. Assuming $\slashednabla_N\glinh=2\fancydstar_2\fancydstar_1(f',g')$, for $f'$, $g'$ scalar functions with vanishing $\ell=0,1$ spherical harmonic modes, we can find $f'$, $g'$ via
    \begin{align}
        \fancydring_1\fancydring_2\slashednabla_N\glinh=\mathring{\slashed{\Delta}}(\mathring{\slashed{\Delta}}+2)(f',g').
    \end{align}
\end{proof}
An identical argument applied to data on $\overline{\Sigma}$ gives
\begin{proposition}\label{constructing compactly supported data Sigma bar}
Given smooth, compactly supported data on $\overline{\Sigma}$ for 
\begin{align}\label{12 10 2021 2}
    \glinh,\; V^{-1}\bmlin_{\ell\geq2},\;  \curlr \slashednabla_TV^{-1}\bmlin_{\ell\geq2}, \;   \tr\glin_{\ell\geq2},\;   \slashednabla_T \tr\glin_{\ell\geq2},\;   \Olin_{\ell\geq2},
\end{align}
there exists a unique set of smooth, compactly supported data for
\begin{align}
\slashednabla_T \glinh,\;   \slashednabla_T \Olin_{\ell\geq2},\;  \divr \slashednabla_T V^{-1}\bmlin_{\ell\geq2},
\end{align}
which complete \bref{12 10 2021 2} into an initial data set for the system \fullsystemK which is supported on $\ell\geq2$ spherical harmonic modes.
\end{proposition}

In order to use the estimates of \cite{DHR16}, it is necessary to work in a gauge where
the initial data satisfy the following on ${\Sigma^*_+}\cap\mathscr{H}^+$:
\begin{align}\label{initial horizon gauge condition}
    \otx|_{{\Sigma^*_+}\cap\mathscr{H}^+}=0,\qquad\qquad\left[\rlin-\rlin_{\ell=0}+\divr\elin\right]_{{\Sigma^*_+}\cap\mathscr{H}^+}=0.
\end{align}
Similarly we will need the following conditions on $\Sigma^*_-\cap\mathscr{H}^-$ to estimate solutions evolved from data on $\Sigma^*_-$ towards $\mathscr{I}^-$, $\mathscr{H}^-_{\leq0}$:
\begin{align}\label{initial horizon gauge condition past}
    \otxb|_{{\Sigma^*_-}\cap\mathscr{H}^-}=0,\qquad\qquad\left[\rlin-\rlin_{\ell=0}+\divr\eblin\right]_{{\Sigma^*_-}\cap\mathscr{H}^-}=0.
\end{align}
We will also study solutions from data on $\overline{\Sigma}$ on  $J^+(\overline{\Sigma})$ as well as $J^-(\overline{\Sigma})$. In order for data on $\overline{\Sigma}$ to allow for  \bref{initial horizon gauge condition} to be satisfied on $\mathscr{H}^+$ and \bref{initial horizon gauge condition past} to be satisfied on $\mathscr{H}^-$ we will need the following gauge conditions on $\mathcal{B}$:
\begin{align}\label{bifurcation gauge conditions}
\divo(\elin-\eblin)\Big|_{\mathcal{B}}=0,\qquad\qquad \left[\rlin-\rlin_{\ell=0}+{\slashed{\Delta}}\Olin\right]\Big|_{\mathcal{B}}=0.
\end{align}

We can always transform an initial data set constructed via \Cref{constructing compactly supported data} to a gauge where \bref{initial horizon gauge condition} is satisfied by the $\ell\geq2$ component:
\begin{proposition}
Given an initial data set $\mathfrak{D}$ constructed in the sense of \Cref{constructing compactly supported data}, there exists a compactly supported function $\fbar(u,\theta^A)$ generating a pure gauge solution according to \Cref{inwards gauge solutions} which, when subtracted from the solution generated by $\mathfrak{D}$, gives a solution whose initial data set $\mathfrak{D}'$ is smooth, compactly supported and satisfies the conditions \bref{initial horizon gauge condition}.
\end{proposition}

\begin{proof}
Let $\xi$ be a smooth cutoff function on the interval $[0,1]$ with $\xi(0)=1$, $\xi(1)=0$ and $\xi'|_{[0,\frac{1}{4}]}=\xi'|_{[\frac{3}{4},1]}=0$. Define $\fbar_1(\theta^A)$, $\fbar_2(\theta^A)$ to be the unique pair satisfying
\begin{align}
    &\frac{2}{(2M)^2}\left[\mathring{\slashed{\Delta}}-1\right]\fbar_1=-\otx|_{\mathscr{H}^+\cap\Sigma^*_+},\\
    &\frac{1}{(2M)^3}\mathring{\slashed{\Delta}}\fbar_2=-\left(\rlin-\rlin_{\ell=0}+\divr\elin\right)|_{\mathscr{H}^+\cap\Sigma^*_+}+\frac{1}{(2M)^3}\left[3\left(\fbar_1-{(\fbar_1)}_{\ell=0}\right)-2\mathring{\slashed{\Delta}}\fbar_1\right].
\end{align}
Define $\fbar$ by $U\fbar=\xi(U)\times\left(\fbar_1+U\fbar_2\right)$.
\end{proof}

\subsubsection{The $\Sigma^*_\pm$, $\overline{\Sigma}^{\pm}
$ gauges}

We now prescribe a scheme to \textit{fix} the gauge freedom on the $\ell\geq2$ component of initial data on ${\Sigma^*_+}$ which is compatible with the conditions \bref{initial horizon gauge condition} on $\mathscr{H}^+_{\geq0}\cap{\Sigma^*_+}$.
\begin{defin}\label{Sigma* gauge}
Let $\mathfrak{D}$ be an initial data set on ${\Sigma^*_+}$ for the system \fullsystem. We say that $\mathfrak{D}$ is in the ``${\Sigma^*_+}$-gauge" if, using the coordinate system $(r,\theta^A)$ on $\Sigma^*_+$, $\mathfrak{D}$ on ${\Sigma^*_+}$ satisfies
\begin{align}\label{Sigma*+ gauge conditions}
    \left[(2M\partial_r)^2+3(2M\partial_{r})+2\right]\divo^2\Omega\xlin|_{\Sigma^*_+}=0,\qquad \divo^2\Omega^{-1}\xblin|_{\Sigma^*_+}=0,\qquad \glinh|_{\Sigma^*_+}=0.
\end{align}
in addition to the horizon gauge conditions \bref{initial  horizon gauge condition}.
\end{defin}

\begin{corollary}\label{realising Sigmastar gauge future}
Let $\mathfrak{D}$ be a smooth initial data set on ${\Sigma^*_+}$ for the system \fullsystem whose components are supported on the spherical harmonics with $\ell\geq2$. There exists a unique initial data set $\mathfrak{D}'$, related to $\mathfrak{D}$ by a gauge transformation, such that $\mathfrak{D}'$ is in the ${\Sigma^*_+}$-gauge of \Cref{Sigma* gauge}.
\end{corollary}

\begin{proof}
We find the unique $\fbar(u,\theta^A)$ satisfying
\begin{align}\label{instinctive dismissal}
    \left[(2M\partial_r)^2+6M\partial_{r}+2\right]\mathring{\slashed{\Delta}}(\mathring{\slashed{\Delta}}+2)\Omega^2\fbar=\left[(2M\partial_r)^2+3(2M\partial_{r})+2\right]\divo^2\Omega\xlin|_{\Sigma^*_+},
\end{align}
with initial conditions $\Omega^2\fbar|_{\mathscr{H}^+}=\fbar_1$, $\partial_r\Omega^2\fbar|_{\mathscr{H}^+}=\fbar_2$ where
\begin{align}
    &(\mathring{\slashed{\Delta}}-1)\fbar_1=2M^2\otx|_{\mathscr{H}^+\cap\Sigma^*_+},\qquad\mathring{\slashed{\Delta}}\fbar_2=-2(2M)^2\left(\rlin_{\ell\geq2}+\divr\elin_{\ell\geq2}\right)|_{\mathscr{H}^+\cap\Sigma^*_+}-\frac{1}{2M}(\mathring{\slashed{\Delta}}-3)\fbar_1.
\end{align}
We take $f(v,\theta^A)$ such that
\begin{align}
    \Omega^{-1}\xblin|_{\Sigma^*_+}=2\fancydstar_2\fancydstar_1(f,0).
\end{align}
We now take $q_1, q_2$ such that
\begin{align}
    \glinh=2\fancydstarring_2\fancydstarring_1\left(\frac{2f}{r}+q_1,q_2\right).
\end{align}

We modify $\mathfrak{D}$ by the gauge transformation generated by $\fbar$ according to \Cref{inwards gauge solutions}, $(q_1, q_2)$ according to
\Cref{residual gauge solutions} and $f$ according to \Cref{outwards gauge solutions} to arrive at $\mathfrak{D}'$
\end{proof}

\begin{defin}\label{Sigma* gauge past}
Let $\mathfrak{D}$ be an initial data set on ${\Sigma^*_-}$ for the system \fullsystem. We say that $\mathfrak{D}$ is in the ``${\Sigma^*_-}$-gauge" if, using the coordinate system $(r,\theta^A)$ on $\Sigma^*_-$, $\mathfrak{D}$ on ${\Sigma^*_-}$ satisfies
\begin{align}
    \left[(2M\partial_r)^2+6M\partial_{r}+2\right]\divo^2\Omega\xblin|_{\Sigma^*_-}=0,\qquad \divo^2\Omega^{-1}\xlin|_{\Sigma^*_-}=0,\qquad \glinh|_{\Sigma^*_-}=0.
\end{align}
in addition to the horizon gauge conditions \bref{initial horizon gauge condition past}.
\end{defin}

A similar scheme may be devised to fix the gauge freedom in data on $\overline{\Sigma}$. Here we replace the horizon gauge conditions \bref{initial horizon gauge condition} by \bref{bifurcation gauge conditions}

\begin{defin}\label{Sigma bar gauge}
Let $\mathfrak{D}$ be an initial data set on ${\overline{\Sigma}}$ for the system \fullsystem. We say that $\mathfrak{D}$ is in the ``$\overline{\Sigma}_+$-gauge" if $\mathfrak{D}$ on $\overline{\Sigma}$ satisfies
\begin{align}
    \left[\left(2M\partial_r\right)^2+6M\partial_r+2\right]\divo^2V\Omega^{-1}\xlin|_{\overline{\Sigma}\,}=0,\qquad \divo^2V^{-1}\Omega\xblin|_{\overline{\Sigma}\,}=0,\qquad \glinh|_{\overline{\Sigma}\,}=0,
\end{align}
in addition to the horizon gauge conditions \bref{bifurcation gauge conditions}. It is said to be in the $\overline{\Sigma}_-$ gauge if it satisfies
\begin{align}
    \left[\left(2M\partial_r\right)^2+6M\partial_r+2\right]\divo^2V^{-1}\Omega\xblin|_{\overline{\Sigma}\,}=0,\qquad \divo^2V\Omega^{-1}\xlin|_{\overline{\Sigma}\,}=0,\qquad \glinh|_{\overline{\Sigma}\,}=0,
\end{align}
in addition to the horizon gauge conditions \bref{bifurcation gauge conditions}.
\end{defin}

\begin{corollary}\label{realising initial gauge at Sigmabar}
Let $\mathfrak{D}$ be a smooth, compactly supported initial data set on $\overline{\Sigma}$ for the system \fullsystem whose components are supported on the spherical harmonics with $\ell\geq2$. There exists a unique initial data set $\mathfrak{D}'$, related to $\mathfrak{D}$ by a gauge transformation, such that $\mathfrak{D}'$ is in the $\overline{\Sigma}_+$-gauge of \Cref{Sigma bar gauge}. A similar statement applies to the $\overline{\Sigma}_-$.
\end{corollary}

\subsection{The $\ell=0,1$ component: a Linearised Kerr family}\label{linearised Kerr section}

In Section 6.2 of \cite{DHR16}, a family of linearised Kerr solutions was defined by linearising the Kerr metric in the parameters $(M,a)$ against a Schwarzschild background $(M,a=0)$. The resulting linearised metric defines a solution to the linearised Einstein equations \fullsystem. In the following we define reference linearised Kerr solutions and show that any smooth solution to the system \fullsystem supported only on $\ell=0,1$ modes is a sum of the reference Kerr family and a pure gauge solution. We treat the cases $\ell=0$ and $\ell=1$ separately, yielding reference Schwarzschild solutions in the first case and slowly rotating Kerr solutions in the second case. 

\subsubsection{The $\ell=0$ mode and reference linearised Schwarzschild solutions}

\begin{defin}\label{reference future linearised Schwarzschild solution}
    Given $\mathfrak{m}\in\mathbb{R}$, define the solution $\mathfrak{K}_{\ell=0}(\mathfrak{m})$ by
    \begin{align}
        2\Olin=-\mathfrak{m},\qquad \tr \glin=-2\mathfrak{m},\qquad \rlin=-\frac{2M}{r^3}\mathfrak{m},\qquad \Klin=\frac{\mathfrak{m}}{r^2},
    \end{align}
    with all remaining components vanishing.
\end{defin}

We now show the following:
\begin{proposition}\label{ell=0 classification proposition}
Any solution to the linearised Einstein equations \fullsystem which is supported on $\ell=0$ spherical harmonics is a sum of the reference linearised Schwarzschild solution defined above in \Cref{reference future linearised Schwarzschild solution} and a pure gauge solution. 
\end{proposition}

For $\ell=0$ modes, only the linearised metric components $\Olin_{\ell=0}, \tr\glin_{\ell=0}$ are nontrivial, thus $\otx_{\ell=0}, \otxb_{\ell=0}, \olin_{\ell=0}, \olinb_{\ell=0}$ are the only nontrivial connection components and $\rlin_{\ell=0}$ is the only nontrivial curvature component. We may then derive the following equations among the nontrivial components:
\begin{lemma}\label{ell=0 important equations}
The $\ell=0$ components of a solution $\mathfrak{S}$ to \fullsystem satisfy
\begin{align}\label{ell=0 diff between rlin and trglin}
    \partial_v\left(r^3\rlin_{\ell=0}-\frac{3M}{2}\tr\glin_{\ell=0}\right)=\partial_u\left(r^3\rlin_{\ell=0}-\frac{3M}{2}\tr\glin_{\ell=0}\right)=0,
\end{align}
\begin{align}\label{ell=0 partial v}
    \partial_v\left[\frac{r}{\Omega^2}\otx+\frac{1}{2}\tr\glin-4\Olin\right]=0,
\end{align}
\begin{align}\label{ell=0 partial u}
    \partial_u\left[\frac{r}{\Omega^2}\otxb-\frac{1}{2}\tr\glin+4\Olin\right]=0,
\end{align}
\begin{align}\label{ell=0 Gauss}
    \frac{1}{2r^2}\tr\glin=\rlin-\frac{1}{2r}\left[\otx-\otxb\right]+\frac{2\Omega^2}{r^2}\Olin.
\end{align}
\end{lemma}

Note that \bref{ell=0 diff between rlin and trglin} implies that $r^3\rlin_{\ell=0}$ and $\frac{3M}{2}\tr\glin_{\ell=0}$ differ by a constant. Looking at the gauge solutions defined in \Cref{outwards gauge solutions}, \Cref{inwards gauge solutions}, \Cref{residual gauge solutions} we see that the difference between $r^3\rlin_{\ell=0}$ and $\frac{3M}{2}\tr\glin_{\ell=0}$ is in fact gauge invariant. Inspired by the reference linearised Schwarzschild family given by \Cref{reference future linearised Schwarzschild solution}, we denote this difference by
\begin{align}\label{ell=0 diff between rlin and trglin defined}
    r^3\rlin_{\ell=0}-\frac{3M}{2}\tr\glin_{\ell=0}=:M\mathfrak{m}
\end{align}
for $\mathfrak{m}$ a real constant. We will now use the equations \bref{ell=0 diff between rlin and trglin}--\bref{ell=0 Gauss} of \Cref{ell=0 important equations} to find an explicit parametrisation of all solutions to the $\ell=0$ linearised Einstein equations which exhibits the classification given by \Cref{ell=0 classification proposition}. It suffices to deal with the case of data on $\Sigma^*_+$ as data on $\Sigma^*_-$ or $\overline{\Sigma}$ are dealt with similarly.

\begin{proposition}\label{classification of ell=0 proposition}
For any solution $\mathfrak{S}$ to the linearised Einstein equations \fullsystem which is supported on $\ell=0$ spherical harmonic modes in an open set $\mathcal{S}\in J^+(\Sigma^*_+)$ and which is $C^2$ away from $\mathscr{H}^+_{\geq0}$, there exists scalar functions $f_{\ell=0}(v)$, $\fbar_{\ell=0}(u)$ and a real parameter $\mathfrak{m}$ such that $\mathfrak{S}$ is given by
\begin{align}
    \mathfrak{S}=\mathfrak{G}_{in}[\fbar_{\ell=0}]+\mathfrak{G}_{out}[f_{\ell=0}]+\mathfrak{K}_{\ell=0}[\mathfrak{m}].
\end{align}
\end{proposition}

\begin{proof}
Equations \bref{ell=0 partial v}, \bref{ell=0 partial u} imply
\begin{align}\label{ell=0 classification first step}
    \frac{r}{\Omega^2}\otx+\frac{1}{2}\tr\glin-4\Olin=-2\partial_u\fbar_{\ell=0},\qquad  \frac{r}{\Omega^2}\otxb-\frac{1}{2}\tr\glin+4\Olin=2\partial_v f_{\ell=0}
\end{align}
for some scalar functions $f_{\ell=0}(v)$, $\fbar_{\ell=0}(u)$. Therefore,
\begin{align}
    \partial_t\tr\glin_{\ell=0}=4\partial_t\left(\frac{\Omega^2}{r}(f-\fbar)\right),
\end{align}
which in turn implies
\begin{align}
    \tr\glin_{\ell=0}=\frac{4\Omega^2}{r}(f-\fbar)+\inhom(r).
\end{align}
For some scalar function $\inhom(r)$. Using \bref{metric transport in 3 direction trace}, \bref{metric transport in 4 direction trace}, \bref{ell=0 diff between rlin and trglin defined} and \bref{ell=0 classification first step} we derive
\begin{align}
    2\Olin_{\ell=0}=\frac{1}{\Omega^2}\partial_u(\Omega^2\fbar_{\ell=0})+\frac{1}{\Omega^2}\partial_v(\Omega^2f_{\ell=0})+\frac{1}{4}(r\inhom)',
\end{align}
\begin{align}
    r^3\rlin_{\ell=0}=\frac{6M\Omega^2}{r}(f_{\ell=0}-\fbar_{\ell=0})+\frac{3M}{2}\inhom+M\mathfrak{m},
\end{align}
\begin{align}
    \otx_{\ell=0}=-\frac{2\Omega^2}{r^2}(2\Omega^2-1)(f_{\ell=0}-\fbar_{\ell=0})+\frac{2\Omega^2}{r}\partial_vf_{\ell=0}+\frac{\Omega^2}{2}(\inhom)',
\end{align}
\begin{align}
    \otxb_{\ell=0}=\frac{2\Omega^2}{r^2}(2\Omega^2-1)(f_{\ell=0}-\fbar_{\ell=0})-\frac{2\Omega^2}{r}\partial_u\fbar_{\ell=0}-\frac{\Omega^2}{2}(\inhom)'.
\end{align}
where $(\inhom)'=d(\inhom)/dr$. Gauss' equation now implies
\begin{align}
    \Omega^2(\inhom)'=\frac{\inhom}{r}\left(-1+\frac{4M}{r}\right)+\frac{4M\mathfrak{m}}{r^2}.
\end{align}
The general solution to the above equation is given by 
\begin{align}
    \inhom=\lambda\frac{\Omega^2}{r}+\frac{\Omega^2}{r}\int dr \left[\frac{4M\mathfrak{m}}{r^2\Omega^2}\times\frac{r}{\Omega^2}\right]=(\lambda+4M\mathfrak{m})\frac{\Omega^2}{r}+\frac{2\mathfrak{m}\Omega^2}{r}r_*-2\mathfrak{m}.
\end{align}
Note that the first term on the right hand side can be obtained according to \Cref{outwards gauge solutions} by a pure gauge solution generated by $f(v)=\lambda+4M\mathfrak{m}$, $\fbar(u)=0$, while the second term can similarly be generated according to \Cref{outwards gauge solutions}, \Cref{inwards gauge solutions} by $f(v)=\frac{\mathfrak{m}}{2}v$, $\fbar(u)=\frac{\mathfrak{m}}{2}u$.
\end{proof}

We now indicate how to explicitly find $\mathfrak{m}$, $f_{\ell=0}$, $\fbar_{\ell=0}$ of \Cref{classification of ell=0 proposition}. Given initial data for $\Olin_{\ell=0}$, $N\Olin_{\ell=0}$, $\tr\glin_{\ell=0}$, $N\tr\glin_{\ell=0}$, we may solve for $\rlin$ via the Gauss equation \bref{ell=0 Gauss}, thus $\mathfrak{m}$ can then be found via \bref{ell=0 diff between rlin and trglin defined}. Subtracting a reference Schwarzschild solution with mass parameter $\mathfrak{m}$ leads to a pure gauge solution as shown in \Cref{classification of ell=0 proposition}. We may now determine $f_{\ell=0}|_{\Sigma^*_+}$ via the first relation in \bref{ell=0 classification first step}: Let $f_{\ell=0, \Sigma^*_+}(r)$ be given by
\begin{align}\label{ell=0 inverse gauge transform 1}
    f_{\ell=0, \Sigma^*_+}(r)=C-\int_{2M}^rd\bar{r}\left(\frac{r}{\Omega^2}{\otxb}_{\ell=0}-\frac{1}{2}\tr\glin_{\ell=0}+4\Olin_{\ell=0}\right)\Big|_{\Sigma^*}(\bar{r}),
\end{align}
where $C$ is given by
\begin{align}\label{ell=0 inverse gauge transform constant}
    C=-\frac{M}{6}\left[M\partial_r\tr\glin_{\ell=0}+\frac{4M}{\Omega^2}\otxb_{\ell=0}-\tr\glin_{\ell=0}\right]\Big|_{\Sigma^*_+\cap\mathscr{H}^+_{\geq0}}.
\end{align}
Now take $\fbar_{\ell=0, \Sigma^*_+}(r)=-\frac{r}{4}\tr\glin_{\ell=0}+\Omega^2f_{\ell=0, \Sigma^*_+}$. Then 
\begin{align}\label{ell=0 inverse gauge transform 2}
    &f_{\ell=0}(v)=f_{\ell=0, \Sigma^*_+}(r-2M),\qquad\qquad
    \fbar_{\ell=0}(u)=\fbar_{\ell=0, \Sigma^*_+}(r(u,\vsigmap{u}))
\end{align}
give the generators of gauge solutions that comprise the remainder of $\mathfrak{S}$.

Finally, it suffices to have as initial data the values of $\Olin_{\ell=0}|_{\Sigma^*_+}$, $\tr\glin_{\ell=0}|_{\Sigma^*_+}$, as equation \bref{constraint equation for Olin on Sigma star} gives an algebraic constraint between $N\Olino_{\ell=0}|_{\Sigma^*_+}$ and   $N\tr\glin_{\ell=0}|_{\Sigma^*_+}$, which allows us to use \bref{Spacelike connection constraint at Sigma star} to solve an ODE for $N\tr\glin_{\ell=0}|_{\Sigma^*_+}$ with the condition that $\otx|_{\Sigma^*_+\cap\mathscr{H}^+_{\geq0}}=0$. Given uniformly bounded data $\Olin_{\ell=0}|_{\Sigma^*_+}$, $\tr\glin_{\ell=0}|_{\Sigma^*_+}$, this process leads to uniformly bounded initial data for the $\ell=0$ modes, which then can be gauge-transformed to the reference linearised Schwarzschild solution given by \Cref{reference future linearised Schwarzschild solution}. We may estimate this gauge transformation via \bref{ell=0 inverse gauge transform 1}--\bref{ell=0 inverse gauge transform 2}.

\subsubsection{The $\ell=1$ modes and rotating reference linearised Kerr solutions}

We now turn to the $\ell=1$ modes

\begin{defin}\label{reference future linearised Kerr solution}
    Let $Y^{\ell=1}_m$ be the spherical harmonics of eigenvalue $\ell=1$ and azimuthal numbers $m=-1,0,1$. For any $\bm{s}=(s_1,s_0,s_1)\in \mathbb{R}^3$, denote $Y_{\bm{s}}=\sum_{i=-1,0,1}s_{i}Y^{\ell=1}_{m=s_i}$. 
    The following is a family of solutions to the system \fullsystem: The nonvanishing metric components are
    \begin{align}\label{Kerr shift}
        \bmlin_A=\frac{4M}{r}\slashed{\epsilon}_A{}^B\partial_B Y_{\bm{s}}.
    \end{align}
    The nonvanishing connection components are
    \begin{align}
        \elin=\frac{3M}{r^2}\slashed{\epsilon}_A{}^B\partial_B Y_{\bm{s}},\qquad \eblin=-\elin.
    \end{align}
    The nonvanishing curvature components are
    \begin{align}
        \blin=\frac{\Omega}{r}\elin,\qquad\bblin=-\blin,\qquad \slin=\frac{6}{r^4}MY_{\bm{s}}.
    \end{align}
    We denote the solution above by $\mathfrak{K}_{\ell=1}(\bm{s})$.
\end{defin}

In analogy with \Cref{ell=0 classification proposition}, we will now show that if a solution to \fullsystem has $\slin_{\ell=1}=0$ then its $\ell=1$ component is a pure gauge solution.

\begin{proposition}\label{ell=1 classification proposition}
Let $\mathfrak{S}$ be a $C^2$ solution to the linearised Einstein equations \fullsystem on an open subset $\mathcal{S}$ of $J^+(\Sigma^*_+)$ which is supported on $\ell=1$ only spherical harmonic modes, such that $\slin=0$. Then there exist scalar functions $f_{\ell=1}(v,\theta^A)$, ${q_1}_{\ell=1}(v,\theta^A)$, ${q_2}_{\ell=1}(v,\theta^A)$, $\fbar_{\ell=1}(u,\theta^A)$ supported on $\ell=1$ spherical harmonic modes such that
\begin{align}\label{ell=1 classification statement}
    \mathfrak{S}=\mathfrak{G}_{in}[\fbar_{\ell=1}]+\mathfrak{G}_{out}[f_{\ell=1}]+\mathfrak{G}_{res}[({q_1}_{\ell=1},{q_2}_{\ell=1})].
\end{align}
\end{proposition}

\begin{proof}
We define $f_{\ell=1}$, $\fbar_{\ell=1}$ of the proposition using the equations \bref{Bianchi+1a}, \bref{Bianchi-1a}, that $\alin$, $\ablin$ have vanishing projections on $\ell=1$ modes, and the vanishing of $\slin_{\ell=1}$,
\begin{align}
    r^2\Omega\blin=6M\fancydstarring_1\left(\frac{\Omega^2}{r^2}\fbar_{\ell=1},0\right),\qquad\qquad r^2\Omega\bblin=-6M\fancydstarring_1\left(\frac{\Omega^2}{r^2}f_{\ell=1},0\right).
\end{align}
The equations \bref{Bianchi+1b}, \bref{Bianchi-1b} imply
\begin{align}
    \frac{\Omega^2}{r}\left(\eblin_{\ell=1}-\elin_{\ell=1}\right)=\fancydstarring_1\left[\partial_u\left(\frac{\Omega^2}{r^2}\fbar_{\ell=1}\right)-\partial_v\left(\frac{\Omega^2}{r^2}f_{\ell=1}\right),\;0\right].
\end{align}
Integrating \bref{partial_u b}, there exists ${q_1}_{\ell=1}(v,\theta^A)$, ${q_2}_{\ell=1}(v,\theta^A)$ such that
\begin{align}
    \bmlin_{\ell=1}=\fancydstarring_1\left[-2\frac{\Omega^2}{r}\fbar_{\ell=1}+2r\partial_v\left(\frac{f_{\ell=1}}{r}\right)+r\partial_v{q_1}_{\ell=1},\; r\partial_v{q_2}_{\ell=1}\right].
\end{align}
The Gauss equation \bref{Gauss} implies
\begin{align}
    \otx_{\ell=1}-\otxb_{\ell=1}=2r\rlin_{\ell=1}+\frac{4\Omega^2}{r}\Olin,
\end{align}

whereas the Codazzi equations \bref{elliptic equation 1}, \bref{elliptic equation 2} imply
\begin{align}
    \frac{1}{2\Omega}\slashednabla\left(\otx_{\ell=1}-\otxb_{\ell=1}\right)=\blin+\bblin+\frac{\Omega}{r}\left(\elin_{\ell=1}+\eblin_{\ell=1}\right).
\end{align}
Taking the divergence gives
\begin{align}
    \otx_{\ell=1}-\otxb_{\ell=1}=-\divo\left(r\Omega\blin_{\ell=1}+r\Omega\bblin_{\ell=1}\right),
\end{align}
Thus we find 
\begin{align}
    \rlin_{\ell=1}=-\frac{6M\Omega^2}{r^4}\left(f_{\ell=1}-\fbar_{\ell=1}\right).
\end{align}
Continuing this procedure we show that in fact $\mathfrak{S}$ is given by \bref{ell=1 classification statement}.
\end{proof}

\begin{remark}
The proof of \Cref{ell=1 classification proposition} gives a recipe for finding the gauge transformation that brings the solution back to the reference form given in \Cref{reference future linearised Kerr solution}.
\end{remark}

\begin{defin}\label{def of full linearised family}
    Let $\mathfrak{m}\in\mathbb{R}$, $\bm{s}=(s_{-1},s_0,s_1)\in\mathbb{R}^3$. We denote by $\mathfrak{K}(\mathfrak{m},\bm{s})$ the sum of the solutions $\mathfrak{K}_{\ell=0}(\mathfrak{m})$ and $\mathfrak{K}_{\ell=1}(\bm{s})$, where $\mathfrak{K}_{\ell=0}$ is parametrised by $\mathfrak{m}$ as in \Cref{reference future linearised Schwarzschild solution} and $\mathfrak{K}_{\ell=1}$ is parametrised by $\bm{s}$ as in \Cref{reference future linearised Kerr solution}. 
\end{defin}

\subsection{Asymptotic flatness at spacelike infinity}

When working with non-compactly supported initial data sets, we will use the following definition of asymptotic flatness at spacelike infinity:

\begin{defin}\label{Basic Big O notation}
    Let $\digamma$ be a scalar function on $\mathbb{R}$. We say that $\digamma$ is $O_n(y^{-k})$ as $r\longrightarrow\infty$ if 
    \begin{align}
    \begin{split}
        \partial_y^i \digamma =O\left(\frac{1}{y^{k+i}}\right) \text{ for } i=0,\dots n.
    \end{split}
    \end{align}
    We say that $\digamma$ is $O_\infty(y^{-k})$ if it is $O_n(y^{-k})$ for any $n\in\mathbb{N}$.
\end{defin}
\begin{defin}
If $(\uppsi,\uppsi')$ is an initial data set for the Regge--Wheeler equation \bref{RW} on ${\Sigma^*_+}$, $\Sigma^*_-$ or $\overline{\Sigma}$, we say that $(\uppsi,\uppsi')$ is $O_n(r^{-k})$ if $\uppsi=O_n(r^{-k})$ and $\uppsi'=O_n(r^{-(k+1)})$ in the sense of \Cref{Basic Big O notation}. The same defintion applies to initial data on ${\Sigma^*_+}$, $\Sigma^*_-$ or $\overline{\Sigma}$ for the Teukolsky equations \bref{T+2}, \bref{T-2}.
\end{defin}

Given the results of \Cref{linearised Kerr section}, we will always evolve from data for which the $\ell=0,1$ modes are induced by the reference linearised Kerr. Thus we define the notion of asymptotic flatness at spacelike infinity as follows:

\begin{defin}\label{def of asymptotic flatness at spacelike infinity}
Let $\mathfrak{D}$ be an initial data set on ${\Sigma^*_+}$ for the system \fullsystem in the sense of \Cref{def of initial data on spacelike surface}. We say that $\mathfrak{D}$ is asymptotically flat to order $(\gamma,N)$ at spacelike infinity if the $\ell=0,1$ components of $\mathfrak{D}$ arise from the restriction of $\mathfrak{K}(\mathfrak{m},\bm{s})$ given in for some real parameters $\mathfrak{m}\in\mathbb{R}$,  $\bm{s}=(s_1,s_0,s_1)\in\mathbb{R}^3$ as in \Cref{def of full linearised family}, while the $\ell\geq2$ component of $\mathfrak{D}$ satisfies
\begin{align}
    &\partial_r^n \Olinos, \partial_r^n \tr\glin=O\left(\frac{1}{r^{\gamma+n}}\right),\\
    &\partial_r^n \bmlins_A=O\left(\frac{1}{r^{\gamma+n-1}}\right),\\
    &\partial_r^n \glinhs_{AB}=O\left(\frac{1}{r^{\gamma+n-2}}\right),
\end{align}
for all integers $ n\leq N$, with respect to any coordinate system on the unit sphere $S^2$ (see \Cref{subsubsection 2.1.3 Asymptotics of S2 tensor fields}).

 A similar definition applies for initial data on $\overline{\Sigma}$.
We will refer to initial data for the linearised system \fullsystem simply as ``asymptotically flat initial data" if it is asymptotically flat to order $(\gamma,\infty)$ with $\gamma> 1$. If $\gamma=1$, we say the initial data is ``weakly asymptotically flat".
\end{defin}

\begin{remark}
An initial data set on $\Sigma^*_+$ satisfying the conclusions of \Cref{realising Sigmastar gauge future} satisfies
\begin{align}
    \left|\overone{R}_{\ell\geq2}|_{\Sigma^*_+}\right|,\;\left|\overone{\Gamma}_{\ell\geq2}|_{\Sigma^*_+}\right|,\,\left|\overone{g}_{\ell\geq2}|_{\Sigma^*_+}\right|\sim O_{\infty}\left(e^{-\frac{r}{2M}}\right),
\end{align}
as $r\longrightarrow\infty$, where $\overone{R}$ stands for any linearised curvature component, $\overone{\Gamma}$ stands for any linearised connection component, and $\overone{g}$ stands for any component of the linearised metric. The same applies to initial data on $\overline{\Sigma}$ satisfying the $\overline{\Sigma}_+$-gauge and to initial data on $\Sigma^*_-$ satisfying the $\Sigma^*_-$-gauge. An initial data set on $\overline{\Sigma}$ satisfying the conditions of the $\overline{\Sigma}_-$ gauge of \Cref{realising initial gauge at Sigmabar} has
\begin{align}
    \left|\overone{R}_{\ell\geq2}|_{\overline{\Sigma}\,}\right|,\;\left|\overone{\Gamma}_{\ell\geq2}|_{\overline{\Sigma}\,}\right|,\,\left|\overone{g}_{\ell\geq2}|_{\overline{\Sigma}\,}\right|\sim O_\infty\left(\frac{r}{2M}e^{-\frac{r}{2M}}\right)
\end{align}
as $r\longrightarrow\infty$.
\end{remark}

\section{Asymptotic flatness at null infinity and Bondi-normalised coordinates}\label{Section 7 BMS}

In studying the asymptotics of solutions to \fullsystem towards future (and past) null infinity, we will adopt the notion of asymptotic flatness at null infinity enabled by the stability result of \cite{DHR16}\footnote{See Section 8.3 of \cite{DHR16}}. In the course of proving Theorems 1, 2 and Corollary 1 we will develop a theory of radiation fields which allows for a stronger notion of asymptotic flatness at null infinity as required for the purposes of the scattering theory to be constructed here. We now state the definition of asymptotic flatness at null infinity to be used in this paper:

\begin{defin}\label{defin of DHR flatness at scri+}
We say a solution $\mathfrak{S}$ to \fullsystem is \textup{asymptotically flat towards $\mathscr{I}^+$ with weight $s$} \textup{to order $n$} for some $0<s\leq1$ and any integer $k$ and index $\gamma$ with $k+|\gamma|\leq n$ the following holds along $\mathscr{C}\cap J^+(\Sigma)$ for any $u$:
\begin{align}\label{asym flatness boundedness of lapse at scri+}
    |(\mathring{\slashednabla})^\gamma\Olin|\,+\,|(\mathring{\slashednabla})^\gamma (r\nablav)^k r^{2} \olin|\,\leq C(k,\gamma),
\end{align}
\begin{align}\label{asym flatness boundedness of shift at scri+}
    |(\mathring{\slashednabla})^\gamma (r\nablav)^k r^{-1} \bmlin_A|\,\leq C(k,\gamma)
\end{align}
\begin{align}\label{asym flatness boundedness of glinh at scri+}
    |(\mathring{\slashednabla})^\gamma r^{-2}\glinh_{AB}|\,+\,|(\mathring{\slashednabla})^\gamma (r\nablav)^k  \xlin_{AB}|+|(\mathring{\slashednabla})^\gamma (r\nablav)^k r^{1+s} \alin_{AB}|\,\leq C(k,\gamma).
\end{align}
We say that $\mathfrak{S}$ is \textup{strongly asymptotically flat towards $\mathscr{I}^+$} with weight $s$ to order $n$ at $\mathscr{I}^+$ if, in addition to the above, the pointwise limits of the following weighted components of $\mathfrak{S}$ as $v\longrightarrow\infty$ for any finite fixed $u$ define smooth fields on $\mathscr{I}^+$ (in the sense of \Cref{def of convergence}):
\begin{align}\label{ansatz of radiation fields scri+}
    \glinh,\, \Olino,\, r^{-1}\bmlin,\, r^2\xlin,\, r\xblin,\, r\ablin,\, r^2\bblin,\, r^3\rlin,\, r^3\slin.
\end{align}
\end{defin}

\begin{defin}\label{defin of DHR flatness at scri-}
We say a solution $\mathfrak{S}$ to \fullsystem is \textup{asymptotically flat towards $\mathscr{I}^-$ with weight $s$} \textup{to order $n$} for some $0<s\leq1$ and any integer $k$ and index $\gamma$ with $k+|\gamma|\leq n$ the following holds along $\mathscr{C}\cap J^-(\Sigma)$ for any $v$:
\begin{align}\label{asym flatness boundedness of lapse at scri-}
    |(\mathring{\slashednabla})^\gamma\Olin|\,+\,|(\mathring{\slashednabla})^\gamma (r\nablau)^k r^{2} \olinb|\,\leq C(k,\gamma),
\end{align}
\begin{align}\label{asym flatness boundedness of shift at scri-}
    |(\mathring{\slashednabla})^\gamma (r\nablau)^k r^{-1} \bmlin_A|\,\leq C(k,\gamma)
\end{align}
\begin{align}\label{asym flatness boundedness of glinh at scri-}
    |(\mathring{\slashednabla})^\gamma r^{-2}\glinh_{AB}|\,+\,|(\mathring{\slashednabla})^\gamma (r\nablau)^k  \xblin_{AB}|+|(\mathring{\slashednabla})^\gamma (r\nablau)^k r^{1+s} \ablin_{AB}|\,\leq C(k,\gamma).
\end{align}
We say that $\mathfrak{S}$ is \textup{strongly asymptotically flat towards $\mathscr{I}^-$} with weight $s$ to order $n$ at $\mathscr{I}^-$ if, in addition to the above, the pointwise limits of the following weighted components of $\mathfrak{S}$ as $u\longrightarrow-\infty$ for any finite fixed $v$ define smooth fields on $\mathscr{I}^-$ (in the sense of \Cref{def of convergence}):
\begin{align}\label{ansatz of radiation at scri-}
    \glinh,\, \Olino,\, r^{-1}\bmlin,\, r^2\xblin,\, r\xlin,\, r\alin,\, r^2\blin,\, r^3\rlin,\, r^3\slin.
\end{align}
\end{defin}

\begin{remark}\label{labels of radiation fields at scri}
For a spacetime which is strongly asymptotically flat at $\mathscr{I}^+$ we denote the limits of the quantities listed in \bref{ansatz of radiation fields scri+} by
\begin{align}
    \Olinos_{\mathscr{I}^+},\, \bmlins_{\mathscr{I}^+},\, \xlins_{\mathscr{I}^+},\, \xblins_{\mathscr{I}^+},\, \ablins_{\mathscr{I}^+},\, \bblins_{\mathscr{I}^+},\, \rlins_{\mathscr{I}^+},\, \slins_{\mathscr{I}^+}.
\end{align}
When the limits of either of $r\tr\glin$, $r\glinh$ is well-defined at $\mathscr{I}^+$, we denote it by
\begin{align}
    \tr\glins_{\mathscr{I}^+},\; \glinhs_{\mathscr{I}^+}
\end{align}
respectively. An analogous notational prescription applies at $\mathscr{I}^-$.
\end{remark}

\begin{remark}
The reference linearised Kerr family defined in \Cref{linearised Kerr section} is strongly asymptotically flat in the sense of \Cref{defin of DHR flatness at scri+} above.
\end{remark}

\begin{remark}\label{other radiation fields at infinity}
In addition to the radiation fields defined at $\mathscr{I}^+$ via \Cref{defin of DHR flatness at scri+} and $\mathscr{I}^-$ via \Cref{defin of DHR flatness at scri-}, we will sometimes refer to the following limits when they are well-defined:
\begin{align}
    &\mathtt{A}^{(i)}_{\mathscr{I}^+}(u,\theta^A):=\lim_{v\longrightarrow\infty}\left(\frac{r^2}{\Omega^2}\nablav\right)^{i-2}r^5\Omega^{-2}\alin(u,v,\theta^A),
    \\&\Phii{i}_{\mathscr{I}^+}:=\lim_{v\longrightarrow\infty}\left(\frac{r^2}{\Omega^2}\nablav\right)^{i}\Psilin, \\&\underline{\Phi}^{(i)}_{\mathscr{I}^+}:=\lim_{v\longrightarrow\infty}\left(\frac{r^2}{\Omega^2}\nablav\right)^{i}\Psilinb, \\&\mathcal{X}^{(i)}_{\mathscr{I}^+}:=\lim_{v\longrightarrow\infty}\left(\frac{r^2}{\Omega^2}\nablav\right)^{i}r^2\xlin(u,v,\theta^A),
    \\&\underline{\mathcal{X}}^{(i)}_{\mathscr{I}^+}:=\lim_{v\longrightarrow\infty}\left(\frac{r^2}{\Omega^2}\nablav\right)^{i}r\xblin(u,v,\theta^A),
    \\&\hat{\mathtt{g}}^{(n)}:=\lim_{v\longrightarrow\infty}\left(\frac{r^2}{\Omega^2}\nablav\right)^{i}r\glinh(u,v,\theta^A).
\end{align}
An analogous prescription applies at $\mathscr{I}^-$.
\end{remark}

\subsection{Future Bondi-normalised double null gauges and the {BMS}\textsuperscript{$+$} group}

We have already defined the (future) Bondi normalisation in \Cref{Intro: Statement of the Bondi gauge} of the introduction. We now study it in more detail and derive the restrictions it imposes on the gauge freedom given by the pure gauge solution of \Cref{inwards gauge solutions}, \Cref{outwards gauge solutions} and \Cref{residual gauge solutions}.

\begin{defin}\label{future Bondi gauge}
Let $\mathfrak{S}$ be a solution to the system \fullsystem with linearised mass parameter $\mathfrak{m}$. We say that $\mathfrak{S}$ is in \textit{future Bondi-normalised} (or that $\mathfrak{S}$ is in the $\mathscr{I}^+$ gauge) if the solution is  asymptotically flat at $\mathscr{I}^+$ in the sense of \Cref{defin of DHR flatness at scri+} and if in addition the following pointwise limits are attained along any $\mathscr{C}_{u},\;|u|<\infty$:
\begin{align}
    &\lim_{v\longrightarrow\infty} \Olino(u,v,\theta^A)=-\frac{1}{2}\mathfrak{m},\label{Olino gauge condition at scri+}\\
    &\lim_{v\longrightarrow\infty}r^2\Klin_{\ell\geq2}(u,v,\theta^A)=0,\label{round sphere at scri+}
\end{align}
and in addition we have
\begin{align}\label{fixing residual freedom Bondi}
    \lim_{v\longrightarrow\infty}\glinh(u,v,\theta^A)=0,
\end{align}
\begin{align}\label{fixing residual freedom Bondi trace}
    \lim_{v\longrightarrow\infty}\tr\glin_{\ell=1}(u,v,\theta^A)=0.
\end{align}
\end{defin}

\begin{remark}
While the linearised Kerr solutions $\mathfrak{K}(\mathfrak{m},\bm{s})$ are not future Bondi-normalised in the sense defined above, their renormalised versions $\mathfrak{K}^+(\mathfrak{m},\bm{s})$ satisfy the conditions \bref{Olino gauge condition at scri+}--\bref{round sphere at scri+}.
\end{remark}

\begin{proposition}\label{round sphere}
Assume that $\mathfrak{G}$ is a pure gauge solution which is asymptotically flat at $\mathscr{I}^+$ in the sense of \Cref{defin of DHR flatness at scri+} to infinite order, and assume in addition that the conditions \bref{round sphere at scri+} and
\begin{align}
    \lim_{v\longrightarrow\infty} \Olino(u,v,\theta^A)=0,\label{Olino gauge condition at scri+ gauge}
\end{align}
are satisfied. Then the generator of ingoing gauge solutions $\fbar$ is of the form
\begin{align}\label{kirahvi nimelta tuike}
    \fbar(u,\theta^A)=\underline{L}(\theta^A)u+\underline{T}(\theta^A),
\end{align}
where $\underline{L}$ is a linear combination of the $\ell=1$ spherical harmonics and $\underline{T}$ is any scalar function on $S^2$.
Moreover, the generator of outgoing gauge solutions $f(v,\theta^A)$ grows at most linearly in $v$. Finally, the generators of residual gauge solutions $q_1, q_2$ have smooth limits as $v\longrightarrow\infty$.
\end{proposition}
\begin{proof}
Imposing \bref{Olino gauge condition at scri+} together with \bref{asym flatness boundedness of lapse at scri+}, we have
\begin{align}\label{BMS intermediate thing in Olin}
    \left|\partial_v(f\Omega^2)+\Omega^2\partial_u\fbar-\frac{2M\Omega^2}{r^2}\fbar\right|\lesssim\frac{B(u)\Omega^2}{r},
\end{align}
for some constant $B$ that may depend on $u$, thus for any fixed $u$, $\partial_v(\Omega^2f)$ is bounded. As $f$ does not depend on $u$,  there exists constants $C,C'$ such that 
\begin{align}
    |f(v,\theta^A)|\leq Cv+C'.
\end{align}
In particular, $v^{-1}f$ is bounded. Condition \bref{Olino gauge condition at scri+} now implies that $\partial_vf(v,\theta^A)$ converges as $v\longrightarrow\infty$ and we have
\begin{align}
    \partial_u\fbar=-\lim_{v\longrightarrow\infty}\partial_vf,
\end{align}
so $\partial_u\fbar$ is independent of $u$ and $\fbar$ must be affine in $u$: there exist smooth functions $L,T_u$ on $S^2$ such that
\begin{align}
    \fbar(u,\theta^A)=\underline{L}(\theta^A)u+\underline{T}(\theta^A).
\end{align}
Finally, note that the boundedness of $r^2\olin$ (condition \bref{asym flatness boundedness of lapse at scri+}) implies that
\begin{align}
    r\partial_v f-f+r\partial_u(\fbar)\lesssim \log(2M)^{-1}r   
\end{align}
as $v\longrightarrow\infty$. 


Iterating the above, we obtain
\begin{align}
    |(\mathring{\slashednabla})^\gamma(v\partial_v)^k v^2\partial_v^{2} f_{\ell\neq1}|+|(\mathring{\slashednabla})^\gamma(v\partial_v)^kv^2\partial_v q_1|+|(\mathring{\slashednabla})^\gamma(v\partial_v)^kv^2\partial_v q_2|\,\leq B_{k,\gamma}.
\end{align}
For $\ell\neq1$, the condition \bref{round sphere at scri+} implies $v^{-1}f_{\ell\neq1}\longrightarrow0$ as $v\longrightarrow\infty$. Since $\partial_v f_{\ell\neq1}\longrightarrow -L_{\ell\neq1}$, we must have that $L_{\ell\neq1}=0$. 
\end{proof}

\begin{remark}
    The restriction \bref{round sphere at scri+} imposes the geometric condition that the induced metric on the $S^2$ cross-sections of $\mathscr{I}^+$ must be that of the round sphere. It has been proposed in the literature that the residual gauge freedom to choose coordinates on these cross-sections should be used to define an extensions of the BMS group. See for example \cite{Campiglia_2015}. 
\end{remark}

\begin{remark}
The gauge conditions \eqref{Olino gauge condition at scri+} and \eqref{round sphere at scri+} comprise the Bondi gauge in the sense defined in e.g.~\cite{PenroseRindler}. Once \eqref{Olino gauge condition at scri+}, \eqref{round sphere at scri+} are imposed, the residual gauge freedom allows to freely specify coordinates on the ``sphere at infinity" while maintaining a standard round metric at $\mathscr{I}^+$. We fix this residual gauge freedom by requiring the condition \eqref{fixing residual freedom Bondi} be satisfied in addition to \eqref{Olino gauge condition at scri+}, \eqref{round sphere at scri+} in order to obtain a scheme to fully fix the gauge in terms of conditions on scattering data at $\mathscr{I}^+$, $\mathscr{H}^+_{\geq0}$.  
\end{remark}
\begin{remark}
The function $L$ in \Cref{round sphere} above encodes the action of asymptotic Lorentz boosts. Asymptotic rotations are induced by residual gauge solutions generated by $(q_1=0,q_2)$ in the sense of \Cref{residual gauge solutions}, where $q_2$ is restricted to $\ell=1$ modes. Asymptotic translations are encoded in the $\ell=0,1$ modes of $T_u$.
\end{remark}

\begin{defin}\label{BMS+ definition}
Let $\underline{L}(\theta^A)$, $\underline{T}(\theta^A)$ be smooth scalar functions on the unit sphere with $\underline{L}$ in the span of the spherical harmonics $Y^{\ell=1}_{m}$ of azimuthal number $\ell=1$. A BMS\textsuperscript{$+$} gauge solution is a pure gauge solution made of the sum of solutions $\underline{\mathfrak{G}}$, $\mathfrak{G}$ and $\mathfrak{G}_{res}$, where $\underline{\mathfrak{G}}$ is generated by $\fbar$ given by
\begin{align}
    \fbar(u,\theta^A)=\underline{L}(\theta^A)u+\underline{T}(\theta^A),
\end{align}
$\mathfrak{G}$ is generated by $f(v,\theta^A)$ grows at most linearly in $v$, and $\mathfrak{G}_{res}$ generated by the pair $(0,q_2)$ with $q_2$ in the span of the $\ell=1$ spherical harmonics. We refer to the addition of a BMS\textsuperscript{$+$} gauge solution as a BMS\textsuperscript{$+$} transformation. 
\end{defin}

\subsection{Past Bondi-normalised double null gauges and the BMS\textsuperscript{$-$} group}


In light of \Cref{defin of DHR flatness at scri-} we define the analogous past-Bondi normalisation on $\mathscr{I}^-$ as follows:

\begin{defin}\label{future Bondi gauge}
Let $\mathfrak{S}$ be a solution to the system \fullsystem with linearised mass parameter $\mathfrak{m}$. We say that $\mathfrak{S}$ is in \textit{future Bondi-normalised} (or that $\mathfrak{S}$ is in the $\mathscr{I}^+$ gauge) if the solution is  asymptotically flat at $\mathscr{I}^+$ in the sense of \Cref{defin of DHR flatness at scri+} and if in addition the following pointwise limits are attained along any $\mathscr{C}_{u},\;|u|<\infty$:
\begin{align}
    &\lim_{u\longrightarrow-\infty} \Olino(u,v,\theta^A)=-\frac{1}{2}\mathfrak{m},\label{Olino gauge condition at scri-}\\
    &\lim_{u\longrightarrow-\infty}r^2\Klin_{\ell\geq2}(u,v,\theta^A)=0,\label{round sphere at scri-}
\end{align}
and in addition we have
\begin{align}\label{fixing residual freedom Bondi past}
    \lim_{u\longrightarrow-\infty}\glinh(u,v,\theta^A)=0.
\end{align}
\begin{align}\label{fixing residual freedom Bondi past bmlin}
    \lim_{u\longrightarrow-\infty}\tr\glin_{\ell=1}(u,v,\theta^A)=0,
\end{align}
\end{defin}


Following similar steps to the proof of \Cref{round sphere}, we may obtain the following:

\begin{proposition}\label{Mihalis's round sphere past}
Assume that $\mathfrak{G}$ is a pure gauge solution which is asymptotically flat at $\mathscr{I}^-$ in the sense of \Cref{defin of DHR flatness at scri-}  to infinite order, and assume in addition that the condition \bref{Olino gauge condition at scri-} is satisfied. Then the generator of outgoing gauge solutions $f$ is of the form
\begin{align}\label{kirahvi nimelta tuike}
    f(v,\theta^A)=L(\theta^A)v+T(\theta^A),
\end{align}
where $L$ is a linear combination of the $\ell=1$ spherical harmonics and $T$ is any scalar function on $S^2$.
Moreover, the generator of outgoing gauge solutions $\fbar(u,\theta^A)$ grows at most linearly in $u$. Finally, the generators of residual gauge solutions $q_1, q_2$ are smooth functions which are independent of $v$.
\end{proposition}

\begin{defin}\label{BMS- definition}
Let $L(\theta^A)$, $T(\theta^A)$ be smooth scalar functions on the unit sphere with $L$ in the span of the spherical harmonics $Y^{\ell=1}_{m}$ of azimuthal number $\ell=1$. A BMS\textsuperscript{$-$} gauge solution is a pure gauge solution made of the sum of solutions $\underline{\mathfrak{G}}$, $\mathfrak{G}$ and $\mathfrak{G}_{res}$, where ${\mathfrak{G}}$ is generated by $f$ given by
\begin{align}
    f(v,\theta^A)={L}(\theta^A)u+{T}(\theta^A),
\end{align}
$\underline{\mathfrak{G}}$ is generated by $\fbar(u,\theta^A)$ grows at most linearly in $u$, and $\mathfrak{G}_{res}$ generated by the pair $(0,q_2)$ with $q_2$ in the span of the $\ell=1$ spherical harmonics. We refer to the addition of a BMS\textsuperscript{$-$} gauge solution as a BMS\textsuperscript{$-$} transformation. 
\end{defin}



\section{The horizon-normalised gauges}\label{Section 7: horizon normalised gauges}

In this section we introduce gauge conditions for radiation fields on $\mathscr{H}^+_{\geq0}$ or $\overline{\mathscr{H}^+}$ that will be used to fix the residual gauge freedom left by imposing the future Bondi gauge conditions \bref{Olino gauge condition at scri+}--\bref{round sphere at scri+} at $\mathscr{I}^+$.

\subsection{The $\mathscr{H}^+_{\geq0}$-normalised gauge}

\begin{defin}\label{def of H+ geq 0 gauge}
Let $\mathfrak{S}$ be a solution to the linearised Einstein equations \fullsystem on $D^+(\Sigma^*_+)$ with linearised Kerr parameters $(\mathfrak{m},\mathfrak{a})$. We say that $\mathfrak{S}$ is in the $\mathscr{H}^+_{\geq0}$ gauge if the conditions \bref{initial horizon gauge condition} are satisfied and
\begin{alignat}{2}
    &\Omega^{-2}\otxb_{\ell=0,1}\Big|_{\mathscr{H}^+_{\geq0}\cap\Sigma^*_+}=0,\qquad &&\divo^2\Omega^{-1}\xblin\Big|_{\mathscr{H}^+_{\geq0}\cap \Sigma^*_+}=0,\\
    &\Olin\Big|_{\mathscr{H}^+_{\geq0}}=-\frac{1}{2}\mathfrak{m},\qquad &&\bmlin^A|_{\mathscr{H}^+_{\geq0}}=\frac{4M}{r}\slashed{\epsilon}^{AB}\sum_{m\in\{-1,0,1\}}\mathfrak{a}_m\partial_A Y_m^{\ell=1}.
\end{alignat}
\end{defin}

We now show that combining the Bondi gauge conditions \bref{Olino gauge condition at scri+}--\bref{round sphere at scri+} with the conditions of \Cref{def of H+ geq 0 gauge} fixes the gauge freedom completely on $D^+(\Sigma^*_+)$.

\begin{lemma}\label{Horizon gauge fixes gauge}
    Assume that $\mathfrak{G}$ is a pure gauge solution to \fullsystem that is both $\mathscr{I}^+$ and $\mathscr{H}^+_{\geq0}$ normalised. Then $\mathfrak{G}$ is trivial.
\end{lemma}

\begin{proof}
    Let $f(v,\theta^A)$, $\fbar(u,\theta^A)$, $q_1(v,\theta^A)$, $q_2(v,\theta^A)$ be the generators of $\mathfrak{G}$ according to \Cref{outwards gauge solutions,,inwards gauge solutions,,residual gauge solutions} respectively. We handle the $\ell\geq2$ component first. The Bondi gauge conditions \bref{Olino gauge condition at scri+}--\bref{round sphere at scri+} give the restriction $\fbar_{\ell\geq2}=T_{u,\ell\geq2}(\theta^A)$ according to \Cref{round sphere}. The second condition in \bref{initial horizon gauge condition} forces $T_{u,\ell\geq2}=0$, thus $\fbar_{u,\ell\geq2}=0$. \\
    The condition $\bmlin_{\ell\geq2}[\mathfrak{G}]|_{\mathscr{H}^+_{\geq0}}=0$ means that ${q_2}_{\ell\geq2}(v,\theta^A)=C_2(\theta^A)$ is constant in $v$ and 
    \begin{align}
        \partial_v\left(f_{\ell\geq2}+M{q_1}_{\ell\geq2}\right)=0,
    \end{align}
    thus ${q_1}_{\ell\geq2}=-\frac{1}{M}f_{\ell\geq2}+C_1$ for some smooth scalar functions $C_1$, $C_2$ on $S^2$. The condition that $\Olin_{\ell\geq2}|_{\mathscr{H}^+_{\geq0}}=0$ implies
    \begin{align}\label{27 12 2021}
        f_{\ell\geq2}=S(\theta^A)e^{-\frac{1}{2M}v}.
    \end{align}
    The condition that $\divo^2\Omega^{-1}\xblin$ vanishes on $\mathscr{H}^+_{\geq0}\cap \Sigma^*_+$ implies that $S=0$. Thus $f_{\ell\geq2}=0$ and ${q_1}_{\ell\geq2}=C_1$. The condition \eqref{fixing residual freedom Bondi} implies $C_1=C_2=0$.\\
    
    Turning to the $\ell=1$ mode, the Bondi gauge conditions give the restriction $\fbar_{\ell=1}=L_{\ell=1}u+T_{u,\ell=1}$. Demanding that $\Olin$ is regular at $\mathscr{H}^+$ forces $L_{\ell=1}=0$, and the vanishing of $\rlin_{\ell=1}+\divr\elin_{\ell=1}$ on $\mathscr{H}^+_{\geq0}$ leads to $T_{u,\ell=1}=0$. Having $\Olin[\mathfrak{G}]$ vanish on $\mathscr{H}^+_{\geq0}$ restricts $f_{\ell=1}$ to the form given in \bref{27 12 2021}, and the condition $\Omega^{-2}\otx_{\ell=1}|_{\mathscr{H}^+_{\geq0}\cap\Sigma^*_+}=0$ leads to $S=0$. 
    Having condition \bref{Kerr shift} forces ${q_1}_{\ell=1}, {q_2}_{\ell=1}$ to be constant in $v$. The condition \eqref{fixing residual freedom Bondi trace} leads to ${q_1}_{\ell=1}=0$. Note that $(0,q_2)$ generates a trivial gauge solution via \Cref{residual gauge solutions} if $q_2$ is only supported on $\ell=1$ modes.
    
    We finally treat the $\ell=0$ mode. The vanishing of $\Olin$ at $\mathscr{H}^+_{\geq0}$ implies 
    \begin{align}
        f_{\ell=0}(v)=T_{u,\ell=0}+Ce^{-\frac{1}{2M}v},
    \end{align}
    where $C$ is a constant. The vanishing of $\Omega^{-2}\otxb$ on $\mathscr{H}^+_{\geq0}\cap\Sigma^*_+$ implies that $C=0$. Thus we can only have  $f_{\ell=0}=\fbar_{\ell=0}=T_{u,\ell=0}$, and it is easy to check that $\mathfrak{G}_{in}[T_{u,\ell=0}]+\mathfrak{G}_{out}[T_{u,\ell=0}]$ is trivial for any constant $T_{u,\ell=0}$.
\end{proof}

\begin{remark}
Let
\begin{align}
     \mathfrak{G}_{+,\;\ell=0}=\mathfrak{G}_{in}\left[\frac{1}{2}\mathfrak{m}v\right]+\mathfrak{G}_{out}\left[\frac{1}{2}\mathfrak{m}u\right].
    \end{align}
Note that $\mathfrak{K}_{\ell=0}+\mathfrak{G}_{+,\ell=0}$ is not regular near $\mathscr{H}^+_{\geq0}$, since
\begin{align}
    &2\Olin=\frac{M\mathfrak{m}\Omega^2}{r}+\frac{2M^2\mathfrak{m}}{r^2}\log\left|\frac{r}{2M}-1\right|\\
    &\Omega^{-2}\otxb=\frac{\mathfrak{m}}{r}\left\{(2\Omega^2-1)\left[\Omega^2+\frac{2M}{r}\log\left|\frac{r}{2M}-1\right|\right]-1\right\}.
\end{align}
In fact, we can show that a solution to \fullsystem supported on $\ell=0$ with $\Olin|_{\mathscr{H}^+_{\geq0}}=0$ can not be regular at $\mathscr{H}^+_{\geq0}$ unless $\mathfrak{m}$, defined via \bref{ell=0 diff between rlin and trglin defined}, is vanishing:
\end{remark}
\begin{lemma}\label{Bondi normalisation ell=0 1 is not regular at H+}
    If $\mathfrak{S}$ is a solution to the linearised Einstein equations \fullsystem which is spherically symmetric and future Bondi-normalised, such that $\Olin$ is finite in a neighbourhood of $\mathscr{H}^+_{\geq0}$. Then $\mathfrak{S}$ is a pure gauge solution.
\end{lemma}
\begin{proof}
    By \Cref{ell=0 classification proposition} it suffices to consider solutions of the form $\mathfrak{S}=\mathfrak{K}_{\ell=0}+\mathfrak{G}$ for a pure gauge solution $\mathfrak{G}$. With the conditions \bref{Olino gauge condition at scri+}--\bref{round sphere at scri+}, we can adapt the argument leading to \Cref{round sphere} to show that the scalar functions $f(v)$, $\fbar(u)$ generating $\mathfrak{G}$ are restricted to 
    \begin{align}
        \fbar(u)=Au+B,\qquad \lim_{v\longrightarrow\infty}\frac{f}{v}=\lim_{v\longrightarrow\infty}\partial_v f=\mathfrak{m}+A.
    \end{align}
    for real constant $A,B$. The solution $\mathfrak{G}$ is regular on $\mathscr{H}^+_{\geq0}$ iff $A=0$. We then have
    \begin{align}
        r^2\Klin=\mathfrak{m}-2\Omega^2\left(\mathfrak{m}-\frac{B}{r}\right)\longrightarrow-\mathfrak{m}
    \end{align}
    as $v\longrightarrow\infty$. Thus we must have $\mathfrak{m}=0$, and \Cref{ell=0 classification proposition} tells us that $\mathfrak{G}$ in this case is pure gauge.
\end{proof}
\begin{remark}
A similar statement to \Cref{Bondi normalisation ell=0 1 is not regular at H+} applies at $\overline{{\mathscr{H}^-}}$.
\end{remark}
\begin{remark} 
the reference linearised Kerr family defined in \Cref{reference future linearised Kerr solution} is already both future and past Bondi-normalised.
\end{remark}

\subsection{The $\overline{\mathscr{H}^+}$-normalised gauge}

\begin{defin}\label{def of H+ bar gauge}
Let $\mathfrak{S}$ be a solution to the linearised Einstein equations \fullsystemK on $D^+(\overline{\Sigma})$ with linearised Kerr parameters $(\mathfrak{m},\mathfrak{a})$. We say that $\mathfrak{S}$ satisfies the $\overline{\mathscr{H}^+}$ gauge conditions if  \bref{bifurcation gauge conditions} are satisfied and
\begin{alignat}{2}
    &\otxbK_{\ell=0,1}\Big|_{\mathcal{B}}=0,\qquad&& {\divo}^2\xblinK\Big|_{\mathcal{B}}=0,\\
    &\Olin\Big|_{\overline{\mathscr{H}^+}}=-\frac{1}{2}\mathfrak{m},\qquad&& \bmlin|_{\overline{\mathscr{H}^+}}=2\slashed{\epsilon}^{AB}\sum_{m\in\{-1,0,1\}}\mathfrak{a}_m\partial_A Y_m^{\ell=1}.
\end{alignat}
\end{defin}

\begin{remark}\label{labels of radiation fields at H}
When discussing radiation fields at either of $\mathscr{H}^+_{\geq0}$ or $\mathscr{H}^-_{\leq0}$, we will use an analogous notation to that given in \Cref{labels of radiation fields at scri} to denote the combinations that induce regular fields at $\mathscr{H}^\pm_{\geq0}$. For instance,
\begin{align}
    \xlins_{\mathscr{H}^+}:=\Omega\xlin|_{\mathscr{H}^+_{\geq0}},\qquad \xblins_{\mathscr{H}^+}:=\Omega^{-1}\xlin|_{\mathscr{H}^+_{\geq0}},
\end{align}
and so on. When discussing the radiation fields near $\mathcal{B}$ we will explicitly attach the weights in Kruskal variables that are necessary to obtain a regular quantity therein.
\end{remark}

\subsection{The $\overline{\mathscr{H}^-}$-normalised gauge}

\begin{defin}\label{def of H- bar gauge}
    Let $\mathfrak{S}$ be a solution to the linearised Einstien equations \fullsystemK on $D^-(\overline{\Sigma})$ with linearised Kerr parameters $(\mathfrak{m},\mathfrak{a})$. We say that $\mathfrak{S}$ satisfies the $\overline{\mathscr{H}^-}$ gauge conditions if \eqref{bifurcation gauge conditions} are satisfied and
    \begin{align}
        \otxK_{\ell=0,1}\Big|_{\mathcal{B}}=0,\qquad\divo^2\xlinK\Big|_{\mathcal{B}}=0,\qquad
        \Olin\Big|_{\overline{\mathscr{H}^-}}=-\frac{1}{2}\mathfrak{m}.
    \end{align}
\end{defin}

\subsection{The global scattering gauge}

We now define the gauge we will use to study scattering over the whole of $\overline{\mathscr{M}}$. 

\begin{defin}\label{defin of global scattering gauge}
Let $\mathfrak{S}$ be a solution to \fullsystem defined over the whole of $\overline{\mathscr{M}}$ (i.e. $\mathfrak{S}$ defines a solution to \fullsystemK over $\overline{\mathscr{M}}$) with linearised Kerr parameters $(\mathfrak{m},\mathfrak{a})$. We say $\mathfrak{S}$ is in the global scattering gauge if the conditions \eqref{Olino gauge condition at scri+}, \eqref{round sphere at scri+}, \eqref{fixing residual freedom Bondi} at $\mathscr{I}^+$ are satisfied, the conditions at $\mathscr{I}^-$ \eqref{Olino gauge condition at scri-}, \eqref{round sphere at scri-} are satisfied, the conditions \bref{bifurcation gauge conditions} at $\mathcal{B}$ are satisfied, and the condition
\begin{align}\label{auxiliary gauge conditions global BMS main body}
   U^{-1}\otxb_{\ell=0,1}\Big|_{\mathcal{B}}=0.
\end{align}
is satisfied.
\end{defin}


\begin{remark}
Note that if a solution $\mathfrak{S}$ is both future and past Bondi-normalised, the residual gauge freedom in $u,v$ has the form
\begin{align}
    u\longrightarrow u+\epsilon(\underline{L}u+\underline{T}),\qquad u\longrightarrow u+\epsilon({L}v+T).
\end{align}
The condition that $\Olino$ must vanish at both $\mathscr{I}^+$ and $\mathscr{I}^-$ forces that
\begin{align}
    \underline{L}=-{L}.
\end{align}
\end{remark}

The argument leading to \Cref{Horizon gauge fixes gauge} can be applied to the global scattering gauge and we can conclude the following: 

\begin{lemma}\label{global gauge fixes gauge}
    Assume that $\mathfrak{S}$ is a solution to \fullsystem, $\mathfrak{G}$ is a pure gauge solution to \fullsystem and that  both $\mathfrak{S}$ and $\mathfrak{S}+\mathfrak{G}$ are in the global scattering gauge. Then $\mathfrak{G}$ is trivial.
\end{lemma}

\section{Statement of results}\label{Section 7: statement of results}

We first define the spaces of scattering states in \Cref{section 7.1: spaces of scattering states}. The precise statement of the main results of this paper follows in \Cref{Section 6.2: statement of results}.

\subsection{Spaces of scattering states}\label{section 7.1: spaces of scattering states}

\begin{defin}
Let $\mathfrak{D}=(\,\glinhs, \tr\glins$, $\Olinos$, $\bmlins$, $\glinhs'$, $\tr\glins'$, $\Olinos'$, $\bmlins'\,)$ be a smooth initial data set on ${\Sigma^*_+}$ for the system \fullsystem in the sense of \Cref{def of initial data on spacelike surface}. Define
\begin{align}
    \left\|\mathfrak{D}\right\|_{\mathcal{E}^{T}_{{\Sigma^*_+}}}=\left\|(\Psilin|_{{\Sigma^*_+}},\slashednabla_{n_{{\Sigma^*_+}}}\Psilin|_{{\Sigma^*_+}})\right\|_{\mathcal{E}^{T,RW}_{{\Sigma^*_+}}}.
\end{align}
Define the space $\mathcal{E}^{T}_{{\Sigma^*_+}}$ to be the completion of data $\mathfrak{D}$ which is asymptotically flat to order $(2,\infty)$, and which satisfies the conditions of the $\Sigma^*_+$ gauge, under the norm $\|\;\|_{\mathcal{E}^{T}_{{\Sigma^*_+}}}$. The space $\mathcal{E}^{T}_{\overline{\Sigma},+}$ is defined similarly (with respect to the $\overline{\Sigma}_+$-gauge).
\end{defin}

\begin{proposition}
Assume $\mathfrak{D}$ is an initial data set on ${\Sigma^*_+}$ for \fullsystem which is in the ${\Sigma^*_+}$-gauge and with $\|\mathfrak{D}\|_{\mathcal{E}^{T,RW}_{{\Sigma^*_+}}}=0$. Then $\mathfrak{D}$ is trivial.
\end{proposition}

\begin{proof}
    The vanishing of $\|\mathfrak{D}\|_{\mathcal{E}^{T,RW}_{{\Sigma^*_+}}}$ implies that $\alin=\ablin=0$ on ${\Sigma^*_+}$ (see Remark 4.1.1 and Proposition 4.2.1 in \cite{Mas20}). Thus $\mathfrak{D}$ generates a pure gauge solution. The conditions defining the ${\Sigma^*_+}$-gauge imply that the generators of this gauge solution must all vanish. 
\end{proof}

\begin{remark}\label{remark on choice of energy for full system}
Note that $\left\|(\Psilin|_{{\Sigma^*_+}},\slashednabla_{n_{{\Sigma^*_+}}}\Psilin|_{{\Sigma^*_+}})\right\|_{\mathcal{E}^{T,RW}_{{\Sigma^*_+}}}$ and $\left\|(\Psilinb|_{{\Sigma^*_+}},\slashednabla_{n_{{\Sigma^*_+}}}\Psilinb|_{{\Sigma^*_+}})\right\|_{\mathcal{E}^{T,RW}_{{\Sigma^*_+}}}$ are not equal, the difference between them being related to the $-$ sign discrepancy in the contribution of $\slin$ between the expressions   \bref{expression for Psilin} and \bref{expression for Psilinb} for $\Psilin$ and $\Psilinb$. However, the norms defined on $\Sigma$, $\overline{\Sigma}$ satisfy
\begin{align}
    &\left\|(\Psilin|_{\Sigma},\slashednabla_{n_{\Sigma}}\Psilin|_{\Sigma})\right\|_{\mathcal{E}^{T,RW}_{\Sigma}}=\left\|(\Psilinb|_{\Sigma},\slashednabla_{n_{\Sigma}}\Psilinb|_{\Sigma})\right\|_{\mathcal{E}^{T,RW}_{\Sigma}},\\ &\left\|(\Psilin|_{\overline{\Sigma}},\slashednabla_{n_{\overline{\Sigma}}}\Psilin|_{\overline{\Sigma}})\right\|_{\mathcal{E}^{T,RW}_{\overline{\Sigma}}}=\left\|(\Psilinb|_{\overline{\Sigma}},\slashednabla_{n_{\overline{\Sigma}}}\Psilinb|_{\overline{\Sigma}})\right\|_{\mathcal{E}^{T,RW}_{\overline{\Sigma}}}.
\end{align}
Indeed, since
\begin{align}
\begin{split}
    \int_{S^2}\dw|\Psilin|^2&=\|2\mathring{\fancydstar_2}\mathring{\fancydstar_1}(\rlin,\slin)\|_{L^2(S^2)}^2+\|6M(r\Omega\xlin-r\Omega\xblin)\|_{L^2(S^2)}^2\\&+\int_{S^2}\dw\,24M\mathring{\fancydstar_2}\mathring{\fancydstar_1}(r^3\rlin,r^3\slin)\times(r\Omega\xlin-r\Omega\xblin)\\&
    =\|2\mathring{\fancydstar_2}\mathring{\fancydstar_1}(\rlin,\slin)\|_{L^2(S^2)}^2+\|6M(r\Omega\xlin-r\Omega\xblin)\|_{L^2(S^2)}^2\\&+\int_{S^2}\dw 24M\left[\rlin\times\divo^2(r\Omega\xlin-r\Omega\xblin)+\slin\times \curlo\divo(r\Omega\xlin-r\Omega\xblin)\right],
\end{split}
\end{align}
\begin{align}
\begin{split}
    \int_{S^2}\dw|\Psilinb|^2&=\|2\mathring{\fancydstar_2}\mathring{\fancydstar_1}(\rlin,\slin)\|_{L^2(S^2)}^2+\|6M(r\Omega\xlin-r\Omega\xblin)\|_{L^2(S^2)}^2\\&+\int_{S^2}\dw 12M\left[\rlin\times\divo^2(r\Omega\xlin-r\Omega\xblin)-\slin\times \curlo\divo(r\Omega\xlin-r\Omega\xblin)\right],
\end{split}
\end{align}
The Codazzi equations \bref{elliptic equation 1}, \bref{elliptic equation 2}, \bref{Bianchi-0*} and \bref{Bianchi+0*} imply
\begin{align}
    \curlo\,\divo(r\Omega\xlin-r\Omega\xblin)=2\partial_t r^3\slin.
\end{align}
Similarly, 
\begin{align}
    \begin{split}
        \int_{S^2}\dw|\mathring{\slashednabla}\Psilin|^2&=\|2\mathring{\slashednabla}\mathring{\fancydstar_2}\mathring{\fancydstar_1}(\rlin,\slin)\|_{L^2(S^2)}^2+\|6M\mathring{\slashednabla}(r\Omega\xlin-r\Omega\xblin)\|_{L^2(S^2)}^2\\&
        -\int_{S^2}\dw\,24M\mathring{\slashed{\Delta}}\mathring{\fancydstar_2}\mathring{\fancydstar_1}(r^3\rlin,r^3\slin)\times(r\Omega\xlin-r\Omega\xblin)\\&
        =\|2\mathring{\slashednabla}\mathring{\fancydstar_2}\mathring{\fancydstar_1}(\rlin,\slin)\|_{L^2(S^2)}^2+\|6M\mathring{\slashednabla}(r\Omega\xlin-r\Omega\xblin)\|_{L^2(S^2)}^2\\&
        -\int_{S^2}\dw\,24M\mathring{\fancydstar_2}\mathring{\fancydstar_1}\left((\mathring{\slashed{\Delta}}+4)r^3\rlin,(\mathring{\slashed{\Delta}}+4)r^3\slin\right)\times(r\Omega\xlin-r\Omega\xblin).
    \end{split}
\end{align}
The cross term evaluates to
\begin{align}
\begin{split}
    &-\int_{S^2}\dw 24M\left[(\mathring{\slashed{\Delta}}+4)\rlin\times\divo^2(r\Omega\xlin-r\Omega\xblin)+(\mathring{\slashed{\Delta}}+4)\slin\times \curlo\divo(r\Omega\xlin-r\Omega\xblin)\right],
\end{split}
\end{align}
and the term containing $\slin$ above becomes
\begin{align}
    -24M\int_{S^2}\dw(\mathring{\slashed{\Delta}}+4)\slin\cdot 2\partial_t\slin.
\end{align}
The contribution of the cross terms in $\left\|(\Psilin|_{\overline{\Sigma}},\slashednabla_{n_{\overline{\Sigma}}}\Psilin|_{\overline{\Sigma}})\right\|_{\mathcal{E}^{T,RW}_{\overline{\Sigma}}}$ containing $\slin$ is
\begin{align}\label{cross term slin 1}
\begin{split}
    48M\int_{\overline{\Sigma}}dr\dw\, &\frac{3\Omega^2+1}{r^2}\,r^3\slin \cdot \partial_t r^3\slin-\frac{1}{r^2}(\mathring{\slashed{\Delta}}+4)r^3\slin\cdot \partial_tr^3\slin\\&+\Omega^2\partial_{r}r^3\slin\cdot\partial_{r}\partial_tr^3\slin+\Omega^{-2}\partial_tr^3\slin\cdot \partial_t^2r^3\slin
\end{split}
\end{align}
Note that since $\Psilin-\Psilinb=4\mathring{\fancydstar_2}\mathring{\fancydstar_1}(0,r^3\slin)$, it is clear that $\slin$ satisfies the following equation
\begin{align}\label{wave equation for slin}
    \left[\partial_t^2-\partial_{r_*}^2-\frac{\Omega^2}{r^2}\mathring{\slashed{\Delta}}-6M\frac{\Omega^2}{r^3}\right]r^3\slin=0.
\end{align}
For an asymptotically flat initial data set $\mathfrak{D}$ on $\overline{\Sigma}$, we may evaluate the contribution of \bref{cross term slin 1} integrating by parts in $r_*$ and using \bref{wave equation for slin} to get 
\begin{align}
    \int_{\overline{\Sigma}}dr\dw\,\partial_r(|\partial_tr^3\slin|^2),
\end{align}
which produces vanishing terms at $\mathcal{B}$ (since $\partial_t$ vanishes there) and in the limit towards $i^0$ since $\mathfrak{D}$ is asymptotically flat.
\end{remark}

\begin{defin}
Define the space of future scattering states $\mathcal{E}^{T}_{\mathscr{H}^+_{\geq0}}$ on $\mathscr{H}^+_{\geq0}$ to be the completion of $\Gamma_c^2(\mathscr{H}^+_{\geq0})$ under the norm
\begin{align}\label{scattering norm on H+}   
\begin{split}
    \|X\|^2_{\mathcal{E}^{T}_{\mathscr{H}^+_{\geq0}}}=&\left\|\mathcal{A}_2(\mathcal{A}_2-2)X\right\|^2_{L^2(\mathscr{H}^+_{\geq0})}+\left\|6M\partial_v X\right\|^2_{L^2(\mathscr{H}^+_{\geq0})}\\&+\int_{S^2_{\infty,0}}\sin\theta d\theta d\phi \left(\left|\mathring{\slashed{\Delta}}X\right|^2+6\left|\mathring{\slashednabla}X\right|^2+8\Big|X\Big|^2\right).
\end{split}
\end{align}
Similarly, define the space of future scattering states $\mathcal{E}^{T}_{\overline{\mathscr{H}^+}}$ via the norm
\begin{align}
     &\|X\|^2_{\mathcal{E}^{T}_{\overline{\mathscr{H}^+}}}=\left\|\mathcal{A}_2(\mathcal{A}_2-2)X\right\|^2_{L^2(\overline{\mathscr{H}^+})}+\left\|6M\partial_v X\right\|^2_{L^2(\overline{\mathscr{H}^+})}.
\end{align}
\end{defin}

\begin{defin}
Define the space of future scattering states $\mathcal{E}^{T}_{\mathscr{I}^+}$ on $\mathscr{I}^+$ to be the completion of $\Gamma_c^2(\mathscr{I}^+)$ under the norm
\begin{align}\label{scattering norm at scri+}
  \|\underline{X}\|^2_{\mathcal{E}^{T}_{\mathscr{I}^+}}=&\left\|\mathcal{A}_2(\mathcal{A}_2-2)\underline{X}\right\|^2_{L^2(\mathscr{I}^+)}+\left\|6M\partial_u\underline{X}\right\|^2_{L^2(\mathscr{I}^+)}.
\end{align}
\end{defin}

\begin{defin}
Define the space of future scattering states $\mathcal{E}^{T}_{{\overline{\mathscr{H}^-}}}$ on ${\overline{\mathscr{H}^-}}$ to be the completion of $\Gamma_c^2({\overline{\mathscr{H}^-}})$ under the norm
\begin{align}
    \|\underline{X}\|^2_{\mathcal{E}^{T}_{\overline{\mathscr{H}^-}}}=\left\|\mathcal{A}_2(\mathcal{A}_2-2)\underline{X}\right\|^2_{L^2({\overline{\mathscr{H}^-}})}+\left\|6M\partial_u\underline{X}\right\|^2_{L^2({\overline{\mathscr{H}^-}})},
\end{align}
\end{defin}

\begin{defin}
Define the space of future scattering states $\mathcal{E}^{T}_{\mathscr{I}^-}$ on $\mathscr{I}^-$ to be the completion of $\Gamma_c^2(\mathscr{I}^-)$ under the norm
\begin{align}\label{scattering norm at scri+}
\|{X}\|^2_{\mathcal{E}^{T}_{\mathscr{I}^-}}=&\left\|\mathcal{A}_2(\mathcal{A}_2-2)\partial_v{X}\right\|^2_{L^2(\mathscr{I}^-)}+\left\|6M\partial_v^2 {X}\right\|^2_{L^2(\mathscr{I}^-)}.
\end{align}
\end{defin}

\subsection{Main theorems}\label{Section 6.2: statement of results}

We now present the full statements of Theorems II, III and Corollary II. The statement of Theorem II is contained in Theorems IIA, IIB and IIC below.

\subsubsection{Theorem IIA: Forward scattering to $\mathscr{I}^+$ and $\mathscr{H}^+_{\geq0}$, $\overline{\mathscr{H}^+}$}\label{Theorem IIA}

\begin{defin}
For a solution $\mathfrak{S}$ to the system \fullsystem defined on $J^+(\Sigma^*)$ arising from asymptotically flat initial data on $\Sigma^*_+$, we denote by $\mathbb{E}_{\Sigma^*_+\rightarrow \mathscr{I}^+}[\mathfrak{S}]$ the $L^2(\Sigma^*_+)$ quantity
\begin{align}
\begin{split}
    \mathbb{E}_{{\Sigma^*_+}\rightarrow \mathscr{I}^+}[\mathfrak{S}]&:=\mathbb{F}^2_{{\Sigma^*_+}}[\Psilin]+\int_{{\Sigma^*_+}}dr\sin\theta d\theta d\phi \left[r^{\frac{13}{2}}|\Omega\plin|^2+r^{\frac{9}{2}}|\Omega^2\alin|^2\right]\\& +\int_{{\Sigma^*_+}}\Omega^2du\sin\theta d\theta d\phi\Bigg[\left|\frac{1}{\Omega}\nablagml\left(\frac{1}{\Omega}\nablagml(r^2\Omega\xlin)\right)\right|^2+\left|\frac{1}{\Omega}\nablagml(r^2\Omega\xlin)\right|^2+\frac{1}{r^2}|r^2\Omega\xlin|^2\Bigg].
\end{split}
\end{align}
\end{defin}

\begin{defin}
For a solution $\mathfrak{S}$ to the system \fullsystem defined on $J^+(\Sigma^*_+)$ arising from data on $\Sigma^*_+$ which is constructed from smooth, compactly supported seed data via \Cref{realising Sigmastar gauge future}, we denote by $\mathbb{E}_{\Sigma^*_+\rightarrow \mathscr{I}^+\cup \mathscr{H}^+_{\geq0}}[\mathfrak{S}]$ the quantity
\begin{align}
\begin{split}
    \mathbb{E}_{\Sigma^*_+\rightarrow \mathscr{I}^+\cup \mathscr{H}^+_{\geq0}}[\mathfrak{S}]:=\mathbb{E}_{{\Sigma^*_+}\rightarrow \mathscr{I}^+}[\mathfrak{S}]+&\int_{\Sigma^*_+}dr\dw\,|r^3\Omega^{-1}\pblin|^2+|r\Omega^{-1}\ablin|^2+\int_{\Sigma^*_+\cap\mathscr{H}^+_{\geq0}}\dw\,\left|\Omega^{-1}\pblin\right|^2.
\end{split}
\end{align}
\end{defin}

\begin{namedtheorem}[Theorem IIA]
\namedlabel{forward scattering full system thm}{Theorem IIA}
Let $\mathfrak{D}$ be an initial data set for the linearised Einstein equations  \fullsystem on $\Sigma^*_+$ in the sense of \Cref{def of initial data on spacelike surface}, and assume that $\mathfrak{D}$ is asymptotically flat in the sense of \Cref{def of asymptotic flatness at spacelike infinity} and that it satisfies the conditions of the $\Sigma^*_+$-gauge. Then the solution $\mathfrak{S}_{\Sigma^*_+}$ to the system \fullsystem arising from $\mathfrak{D}$ via \Cref{EinsteinWP} gives rise to smooth radiation fields $\xblins_{\mathscr{I}^+}\in \Gamma^{(2)}(\mathscr{I}^+)\cap \mathcal{E}^T_{\mathscr{I}^+}$, $\xlins_{\mathscr{H}^+}\in \Gamma^{(2)}(\mathscr{H}^+_{\geq0})\cap \mathcal{E}^T_{\mathscr{H}^+_{\geq0}}$ via
\begin{align}\label{radiation thm scri+}
    \xblins_{\mathscr{I}^+}(u,\theta^A)=\lim_{v\longrightarrow\infty}r\xblin(u,v,\theta^A),\qquad\qquad\xlins_{\mathscr{H}^+}(v,\theta^A)=\Omega\xlin|_{\mathscr{H}^+_{\geq0}},
\end{align}
and we have
\begin{align}\label{radiation thm H+}
    \|\,\mathfrak{D}\,\|^2_{\mathcal{E}^T_{{\Sigma^*_+}}}=\|\,\xblins_{\mathscr{I}^+}\,\|^2_{\mathcal{E}^T_{\mathscr{I}^+}}+\|\,\xlins_{\mathscr{H}^+}\,\|^2_{\mathcal{E}^T_{\mathscr{H}^+_{\geq0}}}.
\end{align}
Furthermore, there exists a unique pair of pure gauge solutions $(\mathfrak{G}_{\mathscr{I}^+},\underline{\mathfrak{G}}_{\mathscr{H}^+_{\geq0}})$, such that $\mathfrak{S}_{\Sigma^*_+}+\mathfrak{G}_{\mathscr{I}^+}$ is future Bondi-normalised (as in \Cref{future Bondi gauge}) and satisfies the horizon gauge conditions \eqref{initial horizon gauge condition}, $\mathfrak{S}_{\Sigma^*_+}+\underline{\mathfrak{G}}_{\mathscr{H}^+_{\geq0}}$ is in the $\mathscr{H}^+_{\geq0}$-gauge (as in \Cref{def of H+ geq 0 gauge}), and $\mathfrak{S}_{\Sigma^*_+}+\mathfrak{G}_{\mathscr{I}^+}+\underline{\mathfrak{G}}_{\mathscr{H}^+_{\geq0}}$ satisfies both the $\mathscr{I}^+$ and $\mathscr{H}^+_{\geq0}$ gauge conditions. The radiation fields $\xblins_{\mathscr{I}^+}$ and $\xlins_{\mathscr{H}^+}$ of $\mathscr{S}_{\Sigma^*_+}$ are invariant under the the addition of $\mathfrak{G}_{\mathscr{I}^+}$ and $\underline{\mathfrak{G}}_{\mathscr{H}^+_{\geq0}}$, and the gauge solutions $\mathfrak{G}_{\mathscr{I}^+}$ and $\underline{\mathfrak{G}}_{\mathscr{H}^+_{\geq0}}$ are both bounded in terms of $\mathfrak{D}$ via the estimate as follows: the generator of $\mathfrak{G}_{\mathscr{I}^+}$, constructed via \Cref{outwards gauge solutions}, is controlled by the estimate
\begin{align}
   \|\partial_u\outwardsgaugefunction\|^2_{L^2(\mathbb{R}\times S^2)}\lesssim \mathbb{E}_{\Sigma^*_+\rightarrow \mathscr{I}^+}[\mathfrak{S}].
\end{align}
The generators of the gauge solution $\underline{\mathfrak{G}}_{\mathscr{H}^+_{\geq0}}$, constructed via \Cref{inwards gauge solutions}, \Cref{residual gauge solutions}, are controlled by the estimate
\begin{align}
     \|\,\inwardsgaugefunction\,\|^2_{L^\infty_v H^2(S^2)}+\|\partial_v\inwardsgaugefunction\|^2_{L^2(\mathbb{R}\times S^2)}+\|(\partial_v\mathfrak{q}_1,\partial_v\mathfrak{q}_2)\|^2_{L^2(\mathbb{R}\times S^2)}\lesssim \mathbb{E}_{\Sigma^*_+\rightarrow \mathscr{I}^+\cup \mathscr{H}^+_{\geq0}}[\mathfrak{S}].
\end{align}

The above construction extends to a unitary map 
\begin{align}\label{forward map}
    \mathscr{F}^+:\mathcal{E}^T_{{\Sigma^*_+}}\longrightarrow \mathcal{E}^T_{\mathscr{I}^+}\oplus \mathcal{E}^T_{\mathscr{H}^+_{\geq0}}.
\end{align}
A similar statement applies to forward evolution from initial data on $\overline{\Sigma}$ and we similarly have the unitary maps
\begin{align}
    &\mathscr{F}^+:\mathcal{E}^T_{\overline{\Sigma},+}\longrightarrow \mathcal{E}^T_{\mathscr{I}^+}\oplus \mathcal{E}^T_{\overline{\mathscr{H}^+}}.
\end{align}
\end{namedtheorem}

\subsubsection{Theorem IIB: Backwards scattering from $\mathscr{I}^+$ and  $\mathscr{H}^+_{\geq0}$, $\overline{\mathscr{H}^+}$}\label{Theorem IIB}

\begin{defin}
Let $\mathfrak{S}$ be a solution to \fullsystem which is both future horizon and Bondi-normalised, such that $\mathfrak{S}$ defines smooth, compactly supported elements of $\Gamma^{(2)}(\mathbb{R}\times S^2)$ via
\begin{align}
    \xblins_{\mathscr{I}^+}:=\lim_{v\longrightarrow\infty}r\xblin(u,v,\theta^A),\qquad\qquad \xlins_{\mathscr{H}^+}:=\Omega\xlin|_{\mathscr{H}^+_{\geq0}}.
\end{align}
We denote by $\mathbb{E}_{\mathscr{H}^+_{\geq0}\cup\mathscr{I}^+\rightarrow \Sigma^*_+}[\mathfrak{S}]$ the quantity given by
\begin{align}
    \begin{split}
\mathbb{E}_{\mathscr{H}^+_{\geq0}\cup\mathscr{I}^+\rightarrow \Sigma^*_+}[\mathfrak{S}]&:=e^{\frac{4}{M}v_+}\sum_{i=0,1}\left[\left\|\partial_v^i\xlins_{\mathscr{H}^+}\right\|^2_{\mathcal{E}^T_{\mathscr{H}^+_{\geq0}}}+\sum_{|\gamma|\leq 2}\left\|\mathring{\slashednabla}^\gamma \partial_u^i \xblins_{\mathscr{I}^+}\right\|^2_{\mathcal{E}^T_{\mathscr{I}^+}}\right]\\
        &+e^{\frac{5}{2M}v_+}\int_{\mathscr{H}^+_{\geq0}}d\bar{v}\dw\,|\mathring{\slashednabla}\ablins_{\mathscr{H}^+}|^2+|\ablins_{\mathscr{H}^+}|^2.
    \end{split}
\end{align}
where $v_+$ denotes the end point of the support of $\glinhs_{\mathscr{H}^+}$ at $\mathscr{H}^+_{\geq0}$. 
If $\mathfrak{S}$ is defined over $D^+(\overline{\Sigma})$ we denote by $\mathbb{E}_{\overline{\mathscr{H}^+}\cup\mathscr{I}^+\rightarrow \overline{\Sigma}}[\mathfrak{S}]$ the quantity
\begin{align}
\begin{split}
    \mathbb{E}_{\overline{\mathscr{H}^+}\cup\mathscr{I}^+\rightarrow \overline{\Sigma}}[\mathfrak{S}]:=&\mathbb{E}_{\mathscr{H}^+_{\geq0}\cup\mathscr{I}^+\rightarrow \Sigma^*_+}[\mathfrak{S}]+\int_{\overline{\mathscr{H}^+}\cap\{V\leq1\}}d\bar{v}\left[|V^{-1}\xlins_{\mathscr{H}^+}|^2+|\slashednabla_VV^{-1}\xlins_{\mathscr{H}^+}|^2+|\slashednabla_V^2V^{-1}\xlins_{\mathscr{H}^+}|^2+|\mathring{\slashednabla}\Psilinb|^2\right].
\end{split}
\end{align}
See \Cref{labels of radiation fields at scri} and \Cref{other radiation fields at infinity} for the definitions of the quantities mentioned above.
\end{defin}

\begin{namedtheorem}[Theorem IIB]
\namedlabel{backwards scattering full system thm}{Theorem IIB}
Let $\xblins_{\mathscr{I}^+}\in\Gamma^{(2)}_c(\mathscr{I}^+)$, $\xlins_{\mathscr{H}^+}\in\Gamma^{(2)}_c(\mathscr{H}^+_{\geq0})$. There exists a unique solution $\mathfrak{S}_{\mathscr{H}^+_{\geq0}\cup\mathscr{I}^+}$ to the linearised Einstein equations \fullsystem which realises $(\xlins_{\mathscr{I}^+}$, $\xlins_{\mathscr{H}^+})$ as its radiation fields on $\mathscr{I}^+$, $\mathscr{H}^+_{\geq0}$ respectively,
\begin{align}
    \xblins_{\mathscr{I}^+}(u,\theta^A)=\lim_{v\longrightarrow\infty}r\xblin(u,v,\theta^A),\qquad\qquad\xlins_{\mathscr{H}^+}(v,\theta^A)=\Omega\xlin|_{\mathscr{H}^+_{\geq0}}.
\end{align}
The solution $\mathfrak{S}_{\mathscr{H}^+_{\geq0}\cup\mathscr{I}^+}$ induces asymptotically flat initial data on $\Sigma^*_+$ and satisfies both the $\mathscr{I}^+$ and the $\mathscr{H}^+_{\geq0}$ gauge conditions. Furthermore, there exists a unique pure gauge solution $\mathfrak{G}_{\Sigma^*_+}$ such that the data induced by $\mathfrak{S}_{\mathscr{H}^+_{\geq0}\cup\mathscr{I}^+}+\mathfrak{G}_{\Sigma^*_+}$ on $\Sigma^*_+$ is asymptotically flat to order $(1,\infty)$ and satisfies the conditions of the $\Sigma^*_+$ gauge. The generators of the gauge solution $\mathfrak{G}_{\Sigma^*_+}$ are bounded in terms of the scattering data $\xblins_{\mathscr{I}^+}$, $\xlins_{\mathscr{H}^+}$ via the estimate
\begin{align}
    \|\mathfrak{f}_{\Sigma^*_+}\|^2_{H^2_rL^2(S^2)}+ \|\underline{\mathfrak{f}}_{\Sigma^*_+}\|^2_{L^2(\Sigma^*_+)}+\|(\mathfrak{q}_{\Sigma^*_+,1},\mathfrak{q}_{\Sigma^*_+,2})\|^2_{L^2(\Sigma^*_+)}\lesssim \mathbb{E}_{\mathscr{H}^+_{\geq0}\cup\mathscr{I}^+\rightarrow \Sigma^*_+}[\mathfrak{S}_{\mathscr{H}^+_{\geq0}\cup\mathscr{I}^+}].
\end{align}

Finally, the initial data $\mathfrak{D}$ on ${\Sigma^*_+}$ induced by $\mathfrak{S}_{\mathscr{H}^+_{\geq0}\cup\mathscr{I}^+}+\mathfrak{G}_{\Sigma^*_+}$ is in $\mathcal{E}^T_{{\Sigma^*_+}}$ and we have
\begin{align}
     \|\,\mathfrak{D}[\mathfrak{S}_{\mathscr{H}^+_{\geq0}\cup\mathscr{I}^+}+\mathfrak{G}_{\Sigma^*_+}]\,\|^2_{\mathcal{E}^T_{{\Sigma^*_+}}}=\|\,\xblins_{\mathscr{I}^+}\,\|^2_{\mathcal{E}^T_{\mathscr{I}^+}}+\|\,\xlins_{\mathscr{H}^+}\,\|^2_{\mathcal{E}^T_{\mathscr{H}^+_{\geq0}}}.
\end{align}
The above defines a map
\begin{align}
    \mathscr{B}^-:\mathcal{E}^T_{\mathscr{I}^-}\oplus \mathcal{E}^T_{\mathscr{H}^+_{\geq0}}\longrightarrow\mathcal{E}^T_{{\Sigma^*_+}}
\end{align}
which inverts the map $\mathscr{F}^+$ of \ref{forward scattering full system thm}:
\begin{align}
\mathscr{B}^-\circ\mathscr{F}^+=Id_{\mathcal{E}^T_{{\Sigma^*_+}},\qquad} \mathscr{F}^+\circ\mathscr{B}^-=Id_{\mathcal{E}^T_{\mathscr{I}^+}\oplus\mathcal{E}^T_{\mathscr{H}^+_{\geq0}}}.
\end{align}
The same conclusions apply replacing $\mathscr{H}^+_{\geq0}$ with  $\overline{\mathscr{H}^+}$, ${\Sigma^*_+}$ with $\overline{\Sigma}$, and the $\Sigma^*_+$-gauge with the $\overline{\Sigma}_+$-gauge.
\end{namedtheorem}

\subsubsection{Theorem IIC: Scattering on $J^-(\overline{\Sigma})$}\label{Theorem IIC}

\begin{defin}
Let $\mathfrak{S}$ be a solution to \fullsystem which is both past horizon and Bondi-normalised, such that $\mathfrak{S}$ defines smooth, compactly supported elements of $\Gamma^{(2)}(\mathbb{R}\times S^2)$ via
\begin{align}
    \xlins_{\mathscr{I}^-}:=\lim_{u\longrightarrow-\infty}r\xlin(u,v,\theta^A),\qquad\qquad \xblins_{\mathscr{H}^-}:=\Omega\xblin|_{\overline{\mathscr{H}^-}}.
\end{align}
We denote by $\mathbb{E}_{\overline{\mathscr{H}^-}\cup\mathscr{I}^-\rightarrow \overline{\Sigma}}[\mathfrak{S}]$ the quantity defined analogously to $\mathbb{E}_{\overline{\mathscr{H}^+}\cup\mathscr{I}^+\rightarrow \overline{\Sigma}}[\mathfrak{S}]$ exchanging $\mathscr{I}^+$ with $\mathscr{I}^-$, $\overline{\mathscr{H}^+}$ with $\overline{\mathscr{H}^-}$, $u$ with $v$, and barred quantities with unbarred ones.
\end{defin}

\begin{namedtheorem}[Theorem IIC]
\namedlabel{thm for past scattering}{Theorem IIC}
Identical statements to \ref{forward scattering full system thm} and \ref{backwards scattering full system thm} apply to evolution on $D^-(\overline{\Sigma})$, replacing $\Sigma^*_+$ with $\Sigma^*_-$, $\mathscr{H}^+_{\geq0}$ with $\mathscr{H}^-_{\leq0}$, $\overline{\mathscr{H}^+}$ with $\overline{\mathscr{H}^-}$, $\mathscr{I}^+$ with $\mathscr{I}^-$, and $\xlin$ with $\xblin$, $\xlins_{\mathscr{I}^+}$ with $\xblins_{\mathscr{I}^-}$, and $\overline{\Sigma}_+$ with $\overline{\Sigma}_-$ everywhere. In particular, we obtain the Hilbert space isomorphisms
\begin{align}
    &\mathscr{F}^-:\mathcal{E}^T_{\Sigma^*_-}\longrightarrow \mathcal{E}^T_{\mathscr{I}^-}\oplus \mathcal{E}^T_{\mathscr{H}^+_{\leq0}},\\
    &\mathscr{F}^-:\mathcal{E}^T_{\overline{\Sigma},-}\longrightarrow \mathcal{E}^T_{\mathscr{I}^-}\oplus \mathcal{E}^T_{\overline{\mathscr{H}^-}}.
\end{align}
\begin{align}
    &\mathscr{B}^+:\mathcal{E}^T_{\mathscr{I}^-}\oplus \mathcal{E}^T_{\mathscr{H}^+_{\leq0}}\longrightarrow\mathcal{E}^T_{\Sigma^*_-},\\
    &\mathscr{B}^+:\mathcal{E}^T_{\mathscr{I}^-}\oplus \mathcal{E}^T_{\overline{\mathscr{H}^-}}\longrightarrow\mathcal{E}^T_{\overline{\Sigma},-}.
\end{align}
with 
\begin{align}
&\mathscr{B}^+\circ\mathscr{F}^-=Id_{\mathcal{E}^T_{{\Sigma^*_-}},\qquad} \mathscr{F}^-\circ\mathscr{B}^+=Id_{\mathcal{E}^T_{\mathscr{I}^-}\oplus\mathcal{E}^T_{\mathscr{H}^-_{\leq0}}},\\
&\mathscr{B}^+\circ\mathscr{F}^-=Id_{\mathcal{E}^T_{{\overline{\Sigma}},-},\qquad} \mathscr{F}^-\circ\mathscr{B}^+=Id_{\mathcal{E}^T_{\mathscr{I}^-}\oplus\mathcal{E}^T_{\overline{\mathscr{H}^-}}}.
\end{align}
\end{namedtheorem}

\subsubsection{Theorem III: Scattering from $\overline{\mathscr{H}^-}$, $\mathscr{I}^-$ to $\overline{\mathscr{H}^+}$, $\mathscr{I}^+$}\label{statement of corollary ii}

\begin{defin}\label{control of global gauge}
    Let $\mathfrak{S}$ be a solution to \fullsystemK which is $\overline{\mathscr{H}^-}$ and $\mathscr{I}^-$-normalised, such that $\xlins_{\mathscr{I}^-}\in \Gamma^{(2)}_c(\mathscr{I}^-)$ and  $U^{-1}\xblins|_{\overline{\mathscr{H}^+}}\in\Gamma^{(2)}_c(\overline{\mathscr{H}^-})$. We denote by $\mathbb{E}_{\mathscr{I}^-\longrightarrow\mathscr{I}^+}[\mathfrak{S}]$ the quantity given by
    \begin{align}\label{E scri- to scri+}
        \begin{split}
            \mathbb{E}_{\mathscr{I}^-\longrightarrow\mathscr{I}^+}[\mathfrak{S}]:=& \int_{\overline{\mathscr{H}^-}\cap\{|U|\leq 1\}}d|U|\dw\,|U\xlins_{\mathscr{H}^-}|^2+|\slashednabla_U U\xlins_{\mathscr{H}^-}|^2+|\slashednabla_U^2 U\xlins_{\mathscr{H}^-}|^2\\
            &+\int_{\overline{\mathscr{H}^-}\cap\{|U|\leq 1\}}d|U|\dw\,|U^2\alins|^2+|U^{2}\alin|^2+|\slashednabla_U U^{2}\alin|^2+|\mathring{\slashednabla}\upPsilinb_{\mathscr{H}^+}|^2
            \\&+e^{-\frac{4}{M}u_-}\int_{\overline{\mathscr{H}^-}\cap\{|U|\geq 1\}}du\dw\,|\mathring{\slashednabla}\plins_{\mathscr{H}^-}|^2+|\plins_{\mathscr{H}^-}|^2+\sum_{|\gamma|\leq 3}|\mathring{\slashednabla}^\gamma \alins_{\mathscr{H}^-}|^2\\&+e^{-\frac{1}{2M}u_-}\int_{\overline{\mathscr{H}^-}\cap\{|U|\geq 1\}}du\dw\,|\mathring{\slashednabla}\upPsilin_{\mathscr{H}^-}|^2+|\upPsilin_{\mathscr{H}^-}|^2\\&+\sum_{i=0,1}\left\|\partial_u^i\xblins_{\mathscr{H}^-}\right\|^2_{\mathcal{E}^T_{\overline{\mathscr{H}^-}}}+\sum_{|\gamma|\leq3, i=0,1}\left\|\partial_v^i\mathring{\slashednabla}^\gamma\xlins_{\mathscr{I}^-}\right\|^2_{\mathcal{E}^T_{\mathscr{I}^-}}\\&+\int_{\mathscr{I}^-}d\bar{u}\dw\, \frac{1}{1+|v|^{\frac{3}{2}}}\left[|\mathring{\slashednabla}\upPsilin_{\mathscr{I}^+}|^2+|\upPsilin_{\mathscr{I}^+}|^2\right].
        \end{split}
    \end{align}
    Similarly, for a solution $\mathfrak{S}$ attaining a smooth, compactly supported $\xblins_{\mathscr{I}^+}$ and smooth $\xlins_{\mathscr{H}^+}$ such that $V^{-1}\xlins_{\mathscr{H}^+}\in \Gamma^{(2)}_{\overline{\mathscr{H}^+}}$, define $\mathbb{E}_{\mathscr{I}^+\longrightarrow\mathscr{I}^-}[\mathfrak{S}]$ analogously to $\mathbb{E}_{\mathscr{I}^-\longrightarrow\mathscr{I}^+}[\mathfrak{S}]$ by exchanging barred quantities with unbarred ones, $u$ with $-v$, $U$ with $-V$ and $\overline{\mathscr{H}^-}$, $\mathscr{I}^-$ with $\overline{\mathscr{H}^+}$, $\mathscr{I}^+$ respectively. 
\end{defin}

\begin{defin}
     Let $\mathfrak{S}$ be a solution to \fullsystemK which is $\overline{\mathscr{H}^-}$ and $\mathscr{I}^-$-normalised, such that $\xlins_{\mathscr{I}^-}\in \Gamma^{(2)}_c(\mathscr{I}^-)$ and  $U^{-1}\xblins|_{\overline{\mathscr{H}^+}}\in\Gamma^{(2)}_c(\overline{\mathscr{H}^-})$. We denote by $\mathbb{E}_{\mathscr{I}^-\cup\overline{\mathscr{H}^-}\longrightarrow\mathscr{I}^+\cup\overline{\mathscr{H}^+}}[\mathfrak{S}]$ the quantity given by
     \begin{align}
         \begin{split}
             \mathbb{E}_{\mathscr{I}^-\cup\overline{\mathscr{H}^-}\longrightarrow\mathscr{I}^+\cup\overline{\mathscr{H}^+}}[\mathfrak{S}]:=&\mathbb{E}_{\mathscr{I}^-\longrightarrow\mathscr{I}^+}[\mathfrak{S}]+\int_{\overline{\mathscr{H}^-}\cap\{|U|\leq 1\}}d|U|\dw\, |\slashednabla_U U^{-2}\ablins_{\mathscr{H}^+}|^2+|U^{-2}\ablins_{\mathscr{H}^+}|^2\\&+e^{-\frac{1}{2M}u_-}\sum_{|\gamma|\leq2}\left[\left\|\mathring{\slashednabla}\xlins_{\mathscr{I}^-}\right\|^2_{\mathcal{E}^T_{\mathscr{I}^-}}+\left\|\mathring{\slashednabla}\xblins_{\mathscr{H}^-}\right\|^2_{\mathcal{E}^T_{\overline{\mathscr{H}^-}}}\right].
         \end{split}
     \end{align}
    Similarly, for a solution $\mathfrak{S}$ attaining a smooth, compactly supported $\xblins_{\mathscr{I}^+}$ and smooth $\xlins_{\mathscr{H}^+}$ such that $V^{-1}\xlins_{\mathscr{H}^+}\in \Gamma^{(2)}_{\overline{\mathscr{H}^+}}$, define $\mathbb{E}_{\mathscr{I}^+\cup\overline{\mathscr{H}^+}\longrightarrow\mathscr{I}^-\cup\overline{\mathscr{H}^-}}[\mathfrak{S}]$ analogously to $\mathbb{E}_{\mathscr{I}^-\cup\overline{\mathscr{H}^-}\longrightarrow\mathscr{I}^+\cup\overline{\mathscr{H}^+}}[\mathfrak{S}]$ by exchanging barred quantities with unbarred ones, $u$ with $-v$, $U$ with $-V$ and $\overline{\mathscr{H}^-}$, $\mathscr{I}^-$ with $\overline{\mathscr{H}^+}$, $\mathscr{I}^+$ respectively. 
\end{defin}

\begin{namedtheorem}[Theorem III]
\namedlabel{global scattering corollary}{Theorem III}
Given $\xlins_{\mathscr{I}^-}\in\Gamma^{(2)}_c(\mathscr{I}^-)$, $\xblins_{\mathscr{H}^-}\in\Gamma^{(2)}_c(\overline{\mathscr{H}^-})$, let $\mathfrak{S}_{\overline{\mathscr{H}^-}\cup\mathscr{I}^-}$ be the solution arising from $\xlins_{\mathscr{I}^-}$, $\xblins_{\mathscr{H}^-}$ via \ref{thm for past scattering}. Then $\mathfrak{S}_{\overline{\mathscr{H}^-}\cup\mathscr{I}^-}$ is strongly asymptotically flat at $\mathscr{I}^+$ in the sense of \Cref{defin of DHR flatness at scri+} and it satisfies the conditions of the past scattering gauge. 

Furthermore, there exists a unique pure gauge solution $\mathfrak{G}_{\mathscr{I}^+}$, which can be estimated in terms of the initial data of $\mathfrak{S}_{\overline{\mathscr{H}^-}\cup\mathscr{I}^-}$, such that $\mathfrak{S}_{\overline{\mathscr{H}^-}\cup\mathscr{I}^-}+\mathfrak{G}_{\mathscr{I}^+}$ is in the global scattering gauge, and there exists a unique pure gauge solution $\mathfrak{G}_{\overline{\mathscr{H}^+}}$ such that $\mathfrak{S}_{\overline{\mathscr{H}^-}\cup\mathscr{I}^-}+\mathfrak{G}_{\mathscr{I}^+}+\mathfrak{G}_{\overline{\mathscr{H}^+}}$ is in the future scattering gauge. The solution $\mathfrak{G}_{\overline{\mathscr{H}^+}\cup\mathscr{I}^+}:=\mathfrak{G}_{\mathscr{I}^+}+\mathfrak{G}_{\overline{\mathscr{H}^+}}$ is the unique pure gauge solution such that $\mathfrak{S}_{\overline{\mathscr{H}^-}\cup\mathscr{I}^-}+\mathfrak{G}_{\mathscr{I}^+}+\mathfrak{G}_{\overline{\mathscr{H}^+}}$ is in the future scattering gauge. The scalar function $\underline{\mathfrak{f}}_{\mathscr{I}^+}$ generating $\mathfrak{G}_{\mathscr{I}^+}$ via \Cref{inwards gauge solutions}, the scalar function ${\mathfrak{f}}_{\overline{\mathscr{H}^+}}$, and the scalar pair $({\mathfrak{q}_{\overline{\mathscr{H}^+},1}}, {\mathfrak{q}_{\overline{\mathscr{H}^+},2}})$ generating $\mathfrak{G}_{\overline{\mathscr{H}^+}}$ via \Cref{outwards gauge solutions} and \Cref{residual gauge solutions} respectively, are controlled via
\begin{align}\label{global gauge estimate Theorem III}
\begin{split}
&\|\partial_u\outwardsgaugefunction\|^2_{L^2(\mathbb{R}\times S^2)}+\chi_{v\in[0,\infty)}\|{\mathfrak{f}}_{\overline{\mathscr{H}^+}}\|^2_{H^2(S^2_{\infty,v})}+\|(\partial_v\mathfrak{q}_{\overline{\mathscr{H}^+},1},\partial_v\mathfrak{q}_{\overline{\mathscr{H}^+},2})\|^2_{L^2(\{v\geq0\})H^2(S^2_{\infty,v})}
\\&+\chi_{V\in[0,1]}\|V{\mathfrak{f}}_{\overline{\mathscr{H}^+}}\|^2_{H^2(S^2_{\infty,2M\log(|v|)})}+\|(\partial_V\mathfrak{q}_{\overline{\mathscr{H}^+},1},\partial_V\mathfrak{q}_{\overline{\mathscr{H}^+},2})\|^2_{L^2_V(V\in[0,1])H^2(S^2_{\infty,2M\log(|v|)})}
\\&\lesssim  \mathbb{E}_{\mathscr{I}^-\cup\overline{\mathscr{H}^-}\longrightarrow\mathscr{I}^+\cup\overline{\mathscr{H}^+}}[\mathfrak{S}_{\overline{\mathscr{H}^-}\cup\mathscr{I}^-}]
\end{split}
\end{align}
\begin{align}
     \|\mathfrak{f}_{\overline{\mathscr{H}^+}}\|^2_{H^2(S^2_{\infty,v})}+\|\partial_u\outwardsgaugefunction\|^2_{L^2(\mathbb{R}\times S^2)}\lesssim \mathbb{E}_{\overline{\mathscr{H}^-}\cup\mathscr{I}^-\rightarrow \overline{\Sigma}}[\mathfrak{S}_{\overline{\mathscr{H}^-}\cup\mathscr{I}^-}].
\end{align}
Finally, the radiation fields $\xblins_{\mathscr{I}^+}$, $\xlins_{\mathscr{H}^+}$ of $\Sscriscri$ satisfy
\begin{align}\label{global unitarity Theorem 3}
    \|\xblins_{\mathscr{I}^+}\|^2_{\mathcal{E}^T_{\mathscr{I}^+}}+\|\xlins_{\mathscr{H}^+}\|^2_{\mathcal{E}^T_{\overline{\mathscr{H}^+}}}=\|\xlins_{\mathscr{I}^-}\|^2_{\mathcal{E}^T_{\mathscr{I}^-}}+\|\xblins_{\mathscr{H}^-}\|^2_{\mathcal{E}^T_{\overline{\mathscr{H}^-}}}.
\end{align}
The relation \eqref{global unitarity Theorem 3} applies to $\mathfrak{S}_{\overline{\mathscr{H}^-}\cup\mathscr{I}^-}+\Gscriscri$ and is unaffected by the gauge transformations $\mathfrak{G}_{\mathscr{I}^+}$ and $\mathfrak{G}_{\overline{\mathscr{H}^+}}$. The above defines a map which extends to a Hilbert space-isomorphism \begin{align}
    \mathscr{S}:\mathcal{E}^T_{\mathscr{I}^-}\oplus\mathcal{E}^T_{\overline{\mathscr{H}^-}}\longrightarrow \mathcal{E}^T_{\mathscr{I}^+}\oplus\mathcal{E}^T_{\overline{\mathscr{H}^+}}.
\end{align}
The map $\mathscr{S}$ is related to $\mathscr{B}^\pm$, $\mathscr{F}^\pm$ by
\begin{align}
    \mathscr{S}=\mathscr{B}^+\circ\mathscr{F}^+,\qquad \mathscr{S}^{-1}=\mathscr{F}^-\circ\mathscr{B}^-.
\end{align}
\end{namedtheorem}

\subsubsection{Corollary II: matching the past and future linear memory effects}\label{statement of corollary ii 2}

\begin{namedcorollary}[Corollary II]
\namedlabel{matching proposition}{Corollary II}
    The radiation field $\xlins_{\mathscr{I}^+}$ belonging to $\mathfrak{S}_{\overline{\mathscr{H}^-}\cup\mathscr{I}^-}+\mathfrak{G}_{\mathscr{I}^+}$ of \ref{global scattering corollary} decays as $u\longrightarrow\infty$. Let $\Sigma^\pm_{\mathscr{I}^+}$, $\Sigma^\pm_{\mathscr{I}^-}$ be defined by
        \begin{align}\label{def of Sigma scri pm}
            \Sigma_{\mathscr{I}^+}^\pm:=\lim_{u\longrightarrow\pm\infty}\xlins_{\mathscr{I}^+},\qquad \Sigma_{\mathscr{I}^-}^\pm:=\lim_{v\longrightarrow\pm\infty}\xblins_{\mathscr{I}^-}.
        \end{align}
        Then $\left(\Sigma^-_{\mathscr{I}^+}-\Sigma^+_{\mathscr{I}^+}\right)$, $\left(\Sigma^+_{\mathscr{I}^-}-\Sigma^-_{\mathscr{I}^-}\right)$ are related via
        \begin{align}
            \left(\Sigma^-_{\mathscr{I}^+}-\Sigma^+_{\mathscr{I}^+}\right)=\overline{\left(\Sigma^+_{\mathscr{I}^-}-\Sigma^-_{\mathscr{I}^-}\right)\circ\mathscr{A}},
        \end{align}
        where the conjugate $\overline{\Xi}$ of a symmetric traceless $\mathcal{S}^2_{u,v}$ 2-tensor $\Xi$ was defined in \Cref{definition of conjugates} and $\mathscr{A}$ is the antipodal map on $S^2$.
\end{namedcorollary}

\section{Forward scattering}\label{Section 7 forward scattering}

In this section we construct radiation fields arising from asymptotically flat initial data on $\Sigma^*_+$ satisfying the horizon gauge conditions \bref{initial horizon gauge condition}. We organise this section as follows: in \Cref{Section 9.1 forward scattering estimates} we gather preliminary boundedness and decay estimates on the Regge--Wheeler and Teukolsky equations and on the full system \fullsystem which were obtained in \cite{DHR16} and \cite{Mas20}. In \Cref{Section 9.2 radiation fields at scri+} we construct radiation fields at $\mathscr{I}^+$ for solutions arising from asymptotically flat initial data on $\Sigma^*_+$ satisfying \bref{initial horizon gauge condition}, and in \Cref{Section 9.3 radiation at H+} we do the same on $\mathscr{H}^+_{\geq0}$. 

\subsection{Preliminary estimates on forwards evolution}\label{Section 9.1 forward scattering estimates}

\subsubsection[Estimates on $\protect\Psilin$, $\protect\Psilinb$ from the Regge--Wheeler equation]{Estimates on $\Psilin, \Psilinb$ from the Regge--Wheeler equation}

Define the following energies;
\begin{align}
    F_{u}^T[\Psi](v_1,v_2)&:=\int_{\mathscr{C}_u\cap\{v\in[v_1,v_2]\}}\sin\theta d\theta d\phi dv\left[|\Omega\slashed\nabla_4\Psi|^2+\Omega^2|\slashed\nabla\Psi|^2+\frac{\Omega^2}{r^2}(3\Omega^2+1)|\Psi|^2\right].
\end{align}
\begin{align}
    \underline{F}_{v}^T[\Psi](u_1,u_2)&:=\int_{\underline{\mathscr{C}}_v\cap\{u\in[u_1,u_2]\}}\sin\theta d\theta d\phi du\left[|\Omega\slashed\nabla_3\Psi|^2+\Omega^2|\slashed\nabla\Psi|^2+\frac{\Omega^2}{r^2}(3\Omega^2+1)|\Psi|^2\right]
\end{align}
\begin{align}
   \mathbb{F}^T_{\mathcal{U}}[\Psi]&:= \int_{\mathcal{U}}dr\sin\theta d\theta d\phi\; (2-\Omega^2)|\slashednabla_{t^*_+}\Psi|^2+\Omega^2|\slashednabla_r\Psi|^2+|\slashednabla\Psi|^2+(3\Omega^2+1)\frac{|\Psi|^2}{r^2}\quad \text{ for }\;\mathcal{U}\in{\Sigma^*_+},
\end{align}
\begin{align}
    \mathbb{F}^T_{\mathcal{U}}[\Psi]=\int_{\mathcal{U}} \sin\theta dr d\theta d\phi \;\frac{1}{\Omega^2}|\slashednabla_t\Psi|^2+\Omega^2|\slashednabla_r \Psi|^2+|\slashednabla\Psi|^2+(3\Omega^2+1)\frac{|\Psi|^2}{r^2}\quad \text{ for }\; \mathcal{U}\in\overline{\Sigma},
\end{align}
\begin{align}
    F_{u}[\Psi](v_1,v_2):=\int_{\mathscr{C}_u\cap\{v\in[v_1,v_2]\}}\sin\theta d\theta d\phi dv \left[|\Omega\slashed\nabla_4\Psi|^2+|\slashed\nabla\Psi|^2+\frac{1}{r^2}|\Psi|^2\right],
\end{align}

\begin{align}
     \underline{F}_{v}[\Psi](v_1,v_2):=\int_{\underline{\mathscr{C}}_v\cap\{u\in[u_1,u_2]\}}\sin\theta d\theta d\phi du\Omega^2 \left[|\Omega^{-1}\slashed\nabla_3\Psi|^2+|\slashed\nabla\Psi|^2+\frac{1}{r^2}|\Psi|^2\right],
\end{align}
\begin{align}
    \mathbb{F}_{\mathcal{U}}[\Psi]=\int_{\mathcal{U}} \sin\theta dr d\theta d\phi\left[|\slashednabla_{t^*_+}\Psi|^2+|\slashednabla_r\Psi|^2+\frac{1}{r^2}|\Psi|^2+|\slashednabla\Psi|^2\right]\quad\text{ for }\mathcal{U}\in{\Sigma^*_+},
\end{align}
\begin{align}
     \mathbb{F}^{n}_{{\Sigma^*_+}}[\Psi]:=\sum_{i_1+i_2+|\alpha|\leq n}\mathbb{F}_{{\Sigma^*_+}}\left[\slashednabla_r^{i_1}\left(\Omega^{-1}\slashednabla_3\right)^{i_2}(r\slashednabla)^\alpha\Psi\right].
\end{align}

\begin{proposition}\label{RW T-energy}
Let $\Psi$ be a solution to (\ref{RW}) arising from data as in \Cref{RWwpCauchy}, then we have
\begin{align}
    F^T_u[\Psi](v_0,v)+\underline{F}^T_v[\Psi](u_0,u)= F^T_{u_0}[\Psi](v_0,v)+\underline{F}^T_{v_0}[\Psi](u_0,u),
\end{align}
\begin{align}\label{this 07 06 2021 2}
     F^T_u[\Psi](v_0,v)+\underline{F}^T_v[\Psi](u_0,u)=\mathbb{F}^T_{{\Sigma^*_+}\cap J^-(\mathscr{C}_u)\cap J^-(\underline{\mathscr{C}}_v)}[\Psi],
\end{align}
for any $u,v$ such that $(u,v,\theta^A)\in J^+({\Sigma^*_+})$.
\end{proposition}
\begin{proof}
We prove \bref{this 07 06 2021 2}. Let $X=\Omega e_3+\Omega e_4$, multiply (\ref{RW}) by $\slashed{\nabla}_X$ and integrate by parts over $S^2$ to obtain
\begin{align}\label{T derivative identity}
    \nablau\left[|\nablav\Psi|^2+\Omega^2|\slashed\nabla \Psi|^2+V|\Psi|^2\right]+\nablav\left[|\nablau\Psi|^2+\Omega^2|\slashed\nabla \Psi|^2+V|\Psi|^2\right]\stackrel{S^2}{\equiv}0.
\end{align}    
Integrating \bref{T derivative identity} over $J^+({\Sigma^*_+})\cap J^-(\mathscr{C}_u)\cap J^-(\underline{\mathscr{C}}_v)$ yields the result.
\end{proof}
To obtain a boundedness estimate for an energy which does not degenerate near $\mathscr{H}^+_{\geq0}$, we repeat the proof of \Cref{RW T-energy} modifying $X$ with a multiple of $\frac{1}{\Omega}\nablagml$, following the recipe detailed in \cite{DR08} to obtain:
\begin{proposition}\label{RWredshift}
Let $\Psi$ be a solution to (\ref{RW}) arising from data as in \Cref{RWwpCauchy}, then we have
\begin{align}
    F_u[\Psi](v_0,\infty)+\underline{F}_v[\Psi](u_0,\infty)\lesssim \mathbb{F}_{{\Sigma^*_+}}[\Psi],
\end{align}
for any $u,v$ such that $(u,v,\theta^A)\in J^+({\Sigma^*_+})$.
\end{proposition}

With \Cref{RWredshift} the following decay estimates apply \cite{DHR16}:

\begin{proposition}\label{RWILED}
Let $\Psi$ be a solution to (\ref{RW}) arising from data as in \Cref{RWwpCauchy}, $\mathscr{D}_{u,v}=J^+(\mathscr{C}_u)\cap J^+(\underline{\mathscr{C}}_v)$, and define
\begin{align}\label{RWILEDestimate}
\begin{split}
    \mathbb{I}_{deg}^{u,v}[\Psi]= \int_{\mathscr{D}_{u,v}}d\bar{u}d\bar{v} \sin\theta &d\theta d\phi \Omega^2 \Bigg[\frac{1}{r^2}|\slashed{\nabla}_{r^*}\Psi|^2+\frac{1}{r^3}|\Psi|^2+\frac{1}{r}\left(1-\frac{3M}{r}\right)^2\left(|\slashed{\nabla}\Psi|^2+\frac{1}{r^2}|\slashednabla_t\Psi|^2\right)\Bigg].
\end{split}
\end{align}
\begin{align}\label{RWILEDestimate degenerate}
\begin{split}
    \mathbb{I}^{u,v}[\Psi]= \int_{\mathscr{D}_{u,v}}d\bar{u}d\bar{v} \sin\theta &d\theta d\phi \Omega^2 \Bigg[\frac{1}{r^2}|\slashed{\nabla}_{r^*}\Psi|^2+\frac{1}{r^3}|\Psi|^2\\
    &+\frac{1}{r}\left(1-\frac{3M}{r}\right)^2\left(|\slashed{\nabla}\Psi|^2+\frac{1}{r^2}|\nablav\Psi|^2+\frac{\Omega^2}{r^2}|\Omega^{-1}\nablagml\Psi|^2\right)\Bigg].
\end{split}
\end{align}
then we have
\begin{align}\label{ILED estimate RW degenerate}
    \begin{split}
        \mathbb{I}_{deg}^{u,v}[\Psi]\lesssim F^T_u[\Psi](v,\infty)+\underline{F}^T_v[\Psi](u,\infty),
    \end{split}
\end{align}
\begin{align}
    \begin{split}
        \mathbb{I}^{u,v}[\Psi]\lesssim F_u[\Psi](v,\infty)+\underline{F}_v[\Psi](u,\infty)
    \end{split}
\end{align}
A similar statement holds for
\begin{align}
\mathbb{I}_{deg}^{u,v,n,T,\slashed{\nabla}}[\Psi]:=\sum_{i+|\alpha|\leq n} \mathbb{I}_{deg}^{u,v}[\slashednabla_t^i(r\slashed{\nabla})^\alpha \Psi], \qquad\mathbb{I}^{u,v,n,T,\slashed{\nabla}}[\Psi]:=\sum_{i+|\alpha|\leq n} \mathbb{I}^{u,v}[\slashednabla_t^i(r\slashed{\nabla})^\alpha \Psi]
\end{align}
and 
\begin{align}
\begin{split}
&\mathbb{I}_{deg}^{u,v,n}[\Psi]:=\sum_{i+j+|\alpha|\leq n}\mathbb{I}_{deg}^{u,v,n}[(\Omega^{-1}\nablagml)^i(r\nablav)^j(r\slashed{\nabla})^\alpha\Psi],\\& \mathbb{I}^{u,v,n}[\Psi]:=\sum_{i+j+|\alpha|\leq n}\mathbb{I}^{u,v,n}[(\Omega^{-1}\nablagml)^i(r\nablav)^j(r\slashed{\nabla})^\alpha\Psi].
\end{split}
\end{align}
\end{proposition}

\begin{corollary}\label{RWrp}
Let $\Psi$ be a solution to (\ref{RW}) arising from data as in \Cref{RWwpCauchy}, and define
\begin{align}
    {\mathbb{I}_p}_{u_0,v_0}^{u,v}[\Psi]=\int_{\mathscr{D}_{{\Sigma^*_+}}^{u,v}\cap\{r>R\}} dudv\sin\theta d\theta d\phi\; r^{p-1}\left[p|\nablav\Psi|^2+(2-p)|\slashednabla\Psi|^2+{r^{-2}}|\Psi|^2\right],
\end{align}
then we have for $p\in [0,2]$
\begin{align}\label{RW rp estimate}
\begin{split}
 \int_{\mathscr{C}_u\cap\{r>R\}}  dv \sin\theta d\theta d\phi \;r^p |\nablav \Psi|^2+ {\mathbb{I}_p}_{u_0,v_0}^{u,v}[\Psi]\lesssim \mathbb{F}_{{\Sigma^*_+}}[\Psi]+\int_{{\Sigma^*_+}\cap\{r>R\}} r^p|\nablav\Psi|^2 dr\sin\theta d\theta d\phi.
\end{split}
\end{align}

Commuting $T^i (r\slashed{\nabla})^\alpha$, a similar estimate holds for 
\begin{align}
{{\mathbb{I}_p}_{u_0,v_0}^{u,v}}^{n,T,\slashed{\nabla}}[\Psi]=\sum_{i+|\alpha|\leq n}{\mathbb{I}_p}_{u_0,v_0}^{u,v}[\partial_t^i (r\slashed{\nabla})^\alpha\Psi]
\end{align}
and for  $(\Omega^{-1}\nablagml)^i(r^h\nablav)^j(r\slashed{\nabla})^\alpha\Psi$ with
\begin{align}
{{\mathbb{I}_p}_{u_0,v_0}^{u,v}}^{n,h}[\Psi]=\sum_{i+j+|\alpha|\leq n}{\mathbb{I}_p}_{u_0,v_0}^{u,v}[(\Omega^{-1}\nablagml)^i(r^h\nablav)^j(r\slashed{\nabla})^\alpha\Psi]
\end{align}
if $0\leq h\leq1$.\\

Define $\left(\frac{r^2}{\Omega^2}\nablav\right)^n\Psi:=\Phi^{(n)}$. Then we have
\begin{align}\label{RWrp k=1}
\begin{split}
     \int_{{\mathscr{C}}_u\cap\{r>R\}}d\bar{v}d\omega\; r^p|\nablav\Phi^{(1)}&|^2+\int_{\mathscr{D}^{u,v}_{{\Sigma^*_+}}\cap\{r>R\}} d\bar{u}d\bar{v}d\omega\;r^{p-1}\left[(p+4)|\nablav\Phi^{(1)}|^2+(2-p)|\slashednabla\Phi^{(1)}|^2+r^{-2}|\Phi^{(1)}|^2\right]
    \\ &\lesssim \int_{{\Sigma^*_+}\cap\{r>R\}}d\bar{v}d\omega\; r^p\left[|\nablav\Phi^{(1)}|^2+|\nablav\Psi|^2\right]+\mathbb{F}^1_{{\Sigma^*_+}}[\Psi].
\end{split}
\end{align}
\begin{align}\label{RWrp k=2}
    \begin{split}
        \int_{{\mathscr{C}}_u\cap\{r>R\}} d\bar{v}d\omega\;& r^p|\nablav\Phi^{(2)}|^2+\int_{\mathscr{D}^{u,v}_{{\Sigma^*_+}}\cap\{r>R\}} d\bar{u}d\bar{v}d\omega\; r^{p-1}\left[(p+8)|\nablav\Phi^{(2)}|^2+(2-p)|\slashednabla\Phi^{(2)}|^2+r^{-2}|\Phi^{(2)}|^2\right]
    \\ &\lesssim \int_{{\Sigma^*_+}\cap\{r>R\}}d\bar{v}d\omega\; r^p\left[|\nablav\Phi^{(2)}|^2+|\nablav\Phi^{(1)}|^2+|\nablav\Psi|^2\right]+\mathbb{F}^2_{{\Sigma^*_+}}[\Psi].
    \end{split}
\end{align}
\end{corollary}
\begin{corollary}\label{Polynomial decay of energy of RW towards future}
Let $\Psi$ be a solution to (\ref{RW}) arising from data as in \Cref{RWwpCauchy} such that
\begin{align}
    \mathbb{F}_{{\Sigma^*_+}}[\Psi]+\int_{{\Sigma^*_+}\cap\{r>R\}} r^p|\nablav\Psi|^2 dr\sin\theta d\theta d\phi<\infty,
\end{align}
with $p=1$ or $2$. Fix $r_0$. For any $u,v$ such that $r(u,v)\leq r_0$, we have
\begin{align}\label{actual polynomial decay of energy of RW towards future}
    F_u[\Psi](v,\infty)+\underline{F}_v[\Psi](u,\infty)\lesssim \frac{1}{v^p}\mathbb{F}_{{\Sigma^*_+}}^{2}[\Psi].
\end{align}
For $(u,v)$ such that $r(u,v)\geq r_0$, the above holds replacing $v^{-2}$ by $u^{-2}$.
\end{corollary}

\subsubsection[Estimates on $\protect\alin$, $\protect\ablin$ from the Teukolsky equations]{Estimates on $\alin, \ablin$ from the Teukolsky equations}\label{forward scattering prelim estimate on alin alinb}

The estimates listed here have already been presented in \cite{Mas20}, adapted from \cite{DHR16} where they occur in the context of a characteristic value problem. We state them again here to conclude that we may apply the polynomial decay estimates on $\alin,\,\plin,\,\ablin,\,\pblin$ of Theorem 2 of \cite{DHR16}.

\begin{proposition}\label{psiILED}
Let $\alpha, \psi, \Psi$ be as in  \bref{hier} and \Cref{+-2 implies RW}. Then for any $u$ and any $v>0$ such that $(u,v,\theta^A)\in J^+({\Sigma^*_+})$, the following estimates hold for sufficiently small $\epsilon>0$ :
\begin{align}\label{locallabel1}
\begin{split}
        &\int_{\mathscr{C}_{u}\cap J^+({\Sigma^*_+})\cap J^-(\underline{\mathscr{C}}_v)}d\bar{v}d\omega\; r^{8-\epsilon}\Omega^2|\psi|^2 + \int_{J^-(\mathscr{C}_{u})\cap J^+({\Sigma^*_+})\cap J^-(\underline{\mathscr{C}}_v)}d\bar{u}d\bar{v}d\omega\; r^{7-\epsilon}\Omega^2|\psi|^2\\[10pt]&\lesssim \mathbb{F}_{{\Sigma^*_+}\cap J^-(\mathscr{C}_u)}[\Psi]+\int_{{\Sigma^*_+}\cap J^-(\mathscr{C}_u)\cap J^-(\underline{\mathscr{C}}_v)}drd\omega\; r^{8-\epsilon}\Omega^2|\psi|^2.
\end{split}
\end{align}
\begin{align}\label{locallabel2}
\begin{split}
        &\int_{\mathscr{C}_{u}\cap J^+({\Sigma^*_+})\cap J^-(\underline{\mathscr{C}}_v)}d\bar{v}d\omega\; r^{6-\epsilon}\Omega^4|\alpha|^2+\int_{J^-(\mathscr{C}_{u})\cap J^+({\Sigma^*_+})\cap J^-(\underline{\mathscr{C}}_v)}d\bar{u}d\bar{v}d\omega\; r^{5-\epsilon}\Omega^4|\alpha|^2\\[10pt]&\lesssim \mathbb{F}_{{\Sigma^*_+}\cap J^-(\mathscr{C}_u)}[\Psi]+\int_{{\Sigma^*_+}\cap J^-(\mathscr{C}_u)\cap J^-(\underline{\mathscr{C}}_v)}drd\omega\; r^{8-\epsilon}\Omega^2|\psi|^2+r^{6-\epsilon}\Omega^4|\alpha|^2.
\end{split}
\end{align}
    \begin{align}
\begin{split}
    &\int_{\mathscr{C}_{u}\cap J^+({\Sigma^*_+})\cap J^-(\underline{\mathscr{C}}_v)} d\bar{v}d\omega\;r^{8-\epsilon} \left[|\nablav(\Omega\psi)|^2+ |-2r^2\fancydstar_2\fancyd_2(r^3\Omega\psi)|^2\right]\\[10pt]&+\int_{J^-(\mathscr{C}_{u})\cap J^+({\Sigma^*_+})\cap J^-(\underline{\mathscr{C}}_v)} d\bar{u}d\bar{v}d\omega\;r^{7-\epsilon} \left[|\nablav(\Omega\psi)|^2+ |-2r^2\fancydstar_2\fancyd_2(r^3\Omega\psi)|^2\right]\\[10pt]&\lesssim \mathbb{F}_{{\Sigma^*_+}\cap J^-(\mathscr{C}_u)}[\Psi]+\int_{{\Sigma^*_+}\cap\J^-(\mathscr{C}_u)\cap J^-(\underline{\mathscr{C}}_v)}dr d\omega\;r^{8-\epsilon}\left[|\Omega\psi|^2+|\Omega^{-1}\nablagml\psi|^2+|r\nablav\psi|^2\right]
\end{split}
\end{align}
\begin{align}
       \begin{split}
        &\int_{\mathscr{C}_{u}\cap J^+({\Sigma^*_+})\cap J^-(\underline{\mathscr{C}}_v)}d\bar{v}d\omega\;r^{6-\epsilon}\left[|r\fancyd_2\Omega^2\alpha|^2+|\Omega^{-1}\nablagml\Omega^2\alpha|^2+|r\nablav\Omega^2\alpha|^2\right]\\[10pt]&+\int_{J^-(\mathscr{C}_{u})\cap J^+({\Sigma^*_+})\cap J^-(\underline{\mathscr{C}}_v)}d\bar{u}d\bar{v}d\omega\;r^{5-\epsilon}\left[|r\fancyd_2\Omega^2\alpha|^2+|\Omega^{-1}\nablagml\Omega^2\alpha|^2+|r\nablav\Omega^2\alpha|^2\right]\\[10pt] &\lesssim \mathbb{F}_{{\Sigma^*_+}\cap J^-(\mathscr{C}_u)}[\Psi]+\int_{{\Sigma^*_+}\cap\J^-(\mathscr{C}_u)\cap J^-(\underline{\mathscr{C}}_v)}drd\omega\;\Bigg\{r^{8-\epsilon}\left[|r\fancyd_2\Omega\psi|^2+|\Omega^{-1}\nablagml\Omega\psi|^2+|r\nablav\Omega\psi|^2\right]\\&\qquad\qquad\qquad\qquad\qquad\qquad\qquad\qquad+ r^{6-\epsilon}\left[|r\fancyd_2\Omega^2\alpha|^2+|\Omega^{-1}\nablagml\Omega^2\alpha|^2+|r\nablav\Omega^2\alpha|^2\right]\Bigg\},
        \end{split}
    \end{align}
\end{proposition}
\begin{corollary}\label{quadratic decay alin plin}
Let $\alpha, \psi, \Psi$ be as in  \bref{hier} and \Cref{+-2 implies RW}, Then for any $u$ and any $v>0$ such that $(u,v,\theta^A)\in J^+({\Sigma^*_+})$, the following estimate holds:
\begin{align}
    |r^3\Omega\psi|^2\lesssim \frac{1}{v^2}\left[\mathbb{F}^2_{{\Sigma^*_+}}[\Psi]+\sup_{{\Sigma^*_+}}\;|r^4\Omega\psi|^2\right],\qquad\qquad|r^2\Omega^2\alpha|^2\lesssim\frac{1}{v^2}\left[\mathbb{F}^2_{{\Sigma^*_+}}[\Psi]+\sup_{{\Sigma^*_+}}\;|r^3\Omega^2\alpha|^2\right].
\end{align}
\end{corollary}

\begin{proposition}\label{L2 estimates of ablin pblin}
Let $\underline\alpha,\,\underline\psi,\,\underline\Psi$ be as in \bref{hier} and \Cref{+-2 implies RW}. Then for any $(u,v)$ such that $(u,v,\theta^A)\in J^+({\Sigma^*_+})$, the following estimates hold:
\begin{align}\label{Flux and ILED for pblin}
\begin{split}
    &\int_{\underline{\mathscr{C}}_v\cap J^+({\Sigma^*_+})\cap J^-(\mathscr{C}_u)}\Omega^2 d\bar{u}d\omega\;r^6\Omega^{-2}|\underline\psi|^2+\int_{J^+(\Sigma^*_+)\cap J^-(\underline{\mathscr{C}}_v)}d\bar{u}d\bar{v}\dw\,r^{5-\epsilon}|\pblin|^2\\&\lesssim \mathbb{F}_{{\Sigma^*_+}\cap J^-(\mathscr{C}_u)}[\underline\Psi]+\int_{{\Sigma^*_+}\cap J^-(\mathscr{C}_u)\cap J^-(\underline{\mathscr{C}}_v)}drd\omega\; r^6\Omega^{-2}|\underline\psi|^2.
\end{split}
\end{align}
\begin{align}\label{Flux and ILED for ablin}
\begin{split}
   &\int_{\underline{\mathscr{C}}_v\cap J^+({\Sigma^*_+})\cap J^-(\mathscr{C}_u)}\Omega^2 d\bar{u}d\omega\;r^2\Omega^{-4}|\underline\alpha|^2+\int_{J^+(\Sigma^*_+)\cap J^-(\underline{\mathscr{C}}_v)}d\bar{u}d\bar{v}\dw\,r^{1-\epsilon}\Omega^{-2}|\ablin|^2\\&\lesssim \mathbb{F}_{{\Sigma^*_+}\cap J^-(\mathscr{C}_u)}[\underline\Psi]+\int_{{\Sigma^*_+}\cap J^-(\mathscr{C}_u\cap J^-(\underline{\mathscr{C}}_v)}drd\omega\;r^6\Omega^{-2}|\underline\psi|^2+ r^2\Omega^{-4}|\underline\alpha|^2.
\end{split}
\end{align}
\begin{align}\label{2ndderivativeofpsibar}
 \begin{split}
        \int_{\underline{\mathscr{C}}_v\cap J^+({\Sigma^*_+})\cap J^-(\mathscr{C}_u)}\Omega^2 d\bar{u}d\omega\;\left|-2r^2\fancydstar_2\fancyd_2(r^3\Omega^{-1}\underline\psi)\right|^2\lesssim \mathbb{F}_{{{\Sigma^*_+}}\cap J^-(\mathscr{C}_u)}[\underline\Psi]+\int_{{{\Sigma^*_+}}}drd\omega\;r^6\Omega^{-2}\left|\underline\psi\right|^2+r^2\Omega^{-4}\left|\underline\alpha\right|^2.
        \end{split}
    \end{align}
    \begin{align}\label{-2 d3 psibar ILED estimate}
\begin{split}
    &\int_{\underline{\mathscr{C}}_v\cap J^+({\Sigma^*_+})\cap J^-(\mathscr{C}_u)}\Omega^2 d\bar{u}d\omega\;r^6|\Omega^{-1}\nablagml(\Omega^{-1}\underline\psi)|^2 \\[10pt]&\lesssim \mathbb{F}_{{{\Sigma^*_+}}\cap J^-(\mathscr{C}_u)}[\underline\Psi]+\int_{{{\Sigma^*_+}}}drd\omega\;r^4\Omega^{-2}\left[|\underline\psi|^2+|r\fancyd_2\underline\psi|^2+|\Omega^{-1}\nablagml(\Omega^{-1}\underline\psi)|^2+|r\nablav(\Omega^{-1}\underline\psi)|^2\right].
\end{split}
\end{align}
\begin{align}
\begin{split}
    &\int_{\underline{\mathscr{C}}_v\cap J^+({\Sigma^*_+})\cap J^-(\mathscr{C}_u)}\Omega^2 d\bar{u}d\omega\;|r^2\fancydstar_2\fancyd_2 r\Omega^{-2}\underline\alpha|^2
     \\&\;\lesssim\; \mathbb{F}_{{{\Sigma^*_+}}\cap J^-(\mathscr{C}_u)}[\underline\Psi]+\int_{{{\Sigma^*_+}}}drd\omega\;r^4\Omega^{-2}\Bigg[|\underline\psi|^2+|r\fancyd_2\underline\psi|^2+|\Omega^{-1}\nablagml(\Omega^{-1}\underline\psi)|^2+|r\nablav(\Omega^{-1}\underline\psi)|^2+|r\Omega^{-2}\underline\alpha|^2\Bigg].
\end{split}
\end{align}
\begin{align}\label{-2 d3 alphabar ILED estimate}
\begin{split}
    &\int_{\underline{\mathscr{C}}_v\cap J^+({\Sigma^*_+})\cap J^-(\mathscr{C}_u)}\Omega^2 d\bar{u}d\omega\; \left[|r\Omega^{-2}\underline\alpha|^2+|r\fancyd_2r\Omega^{-2}\underline\alpha|^2+|\Omega^{-1}\nablagml r\Omega^{-2}\underline\alpha|^2\right]\\[10pt]& \lesssim \mathbb{F}_{{{\Sigma^*_+}}\cap J^-(\mathscr{C}_u)}[\underline\Psi]+\int_{{{\Sigma^*_+}}}drd\omega\;r^6\left[|\Omega^{-1}\underline\psi|^2+|r\fancyd_2\Omega^{-1}\underline\psi|^2+|\Omega^{-1}\nablagml(\Omega^{-1}\underline\psi)|^2\right]\\[10pt]& +\int_{{{\Sigma^*_+}}}drd\omega\; r^2\left[|\Omega^{-2}\underline\alpha|^2+|r\fancyd_2\Omega^{-2}\underline\alpha|^2+|\Omega^{-1}\nablagml\Omega^{-2}\underline\alpha|^2\right].
\end{split}
\end{align}
\end{proposition}

\begin{corollary}\label{polynomial decay of pblin and ablin}
Let $\underline\alpha, \underline\psi, \underline\Psi$ be as in \bref{hier} and \Cref{+-2 implies RW}. Fix $r_0$ and let $(u,v)$ be such that $r(u,v)=r_0$. Then for any $u$ and any $v>0$ such that $(u,v,\theta^A)\in J^+({\Sigma^*_+})$. We have
\begin{align}\label{L2 polynomial decay estimate pblin ablin}
    \int_{u:\;r(u,v)=r_0}^\infty d\bar{u}\Omega^2\,\int_{S^2}d\omega\, \left(\frac{1}{r^\epsilon}|r^3\Omega^{-1}\underline\psi|^2+|r\Omega^{-2}\underline\alpha|^2\right)\lesssim \frac{1}{v^2}\left[\mathbb{F}_{{\Sigma^*_+}}^2[\Psi]+\int_{{\Sigma^*_+}}drd\omega\, \left(r^6\Omega^{-2}|\underline\psi|^2+r^2\Omega^{-4}|\underline\alpha|^2\right)\right].
\end{align}
In addition, if $(u,v)$ are such that $r(u,v)\leq r_0$ we have
\begin{align}\label{pointwise polynomial decay estimate pblin ablin}
\begin{split}
    &|r^3\Omega^{-1}\underline\psi(u,v,\theta^A)|^2\lesssim C(r_0)\frac{1}{v^2}\left[\mathbb{F}^2_{{\Sigma^*_+}\cap J^-(\mathscr{C}_u)}[\underline\Psi]+\sup_{{\Sigma^*_+}}\;|r^3\Omega^{-1}\underline\psi|^2\right],\\&|r\Omega^{-2}\underline\alpha|^2\lesssim C(r_0)\frac{1}{v^2}\left[\mathbb{F}^2_{{\Sigma^*_+}}[\underline\Psi]+\sup_{{\Sigma^*_+}}\;|r\Omega^{-2}\underline\alpha|^2\right].
\end{split}
\end{align}
For $(u,v)$ such that $r(u,v)\geq r_0$ the above holds replacing $v^{-2}$ by $u^{-2}$.
\end{corollary}

We will need the following $L^1$ estimate:
\begin{corollary}\label{L1 estimate}
Let $\underline\alpha, \underline\psi, \underline\Psi$ be as in  \bref{hier} and \Cref{+-2 implies RW}. Then for any $u$ and any $v>0$ such that $(u,v,\theta^A)\in J^+({\Sigma^*_+})$,  we have 
\begin{align}\label{L1 estimate estimate}
\begin{split}
    &\left\{\int_{\underline{\mathscr{C}}_v\cap J^+({\Sigma^*_+})\cap J^-(\mathscr{C}_u)} \Omega^2d\bar{u}\sin\theta d\theta d\phi \left[\left|\mathcal{A}_2r^3\Omega^{-1}\underline{\psi}\right|+\left|\mathcal{A}_2 r\Omega^{-2}\underline\alpha\right|\right]\right\}^2\lesssim\mathbb{F}^2_{{\Sigma^*_+}\cap J^-(\mathscr{C}_u)}[\underline\Psi]\;+\mathbb{E}_1,
\end{split}
\end{align}
where
\begin{align}\label{def of E1}
    \mathbb{E}_1:=\int_{{\Sigma^*_+}\cap J^-(\mathscr{C}_u)}dr\sin\theta d\theta d\phi \left\{|r^3\Omega^{-1}\underline\psi|^2+|\mathring{\slashednabla}r^3\Omega^{-1}\underline\psi|^2+|\Omega^{-1}\nablagml r^3\Omega^{-1}\underline\psi|^2+|r\nablav r^3\Omega^{-1}\underline\psi|^2+|r\Omega^{-2}\underline\alpha|^2\right\}.
\end{align}
\end{corollary}

We complement the estimates of \Cref{L1 estimate} with estimates on the region $u\leq u_0$ of $J^+({\Sigma^*_+})$ to obtain the following $L^1$ estimate:
\begin{lemma}\label{L1 estimate near i0}
Let $\underline\alpha$ be a solution to \bref{T-2} and let $\underline\psi, \underline\Psi$ be as in  \bref{hier}. Then for any $u$ and any $v>0$ such that $(u,v,\theta^A)\in J^+({\Sigma^*_+})$, we have
\begin{align}\label{this 10 06 2021 first estimate}
\begin{split}
\int_{S^2}d\omega\int_{u_{{\Sigma^*_+},v}}^{u}d\bar{u}\,|r^3\Omega\underline\psi|\lesssim& \int_{u_{{\Sigma^*_+},v}}^ud\bar{u} \frac{1}{\sqrt{r(\bar{u},v_{{\Sigma^*_+},\bar{u}})}}\sqrt{\mathbb{F}^T_{{\Sigma^*_+}\cap\{r\geq r(\bar{u},v_{{\Sigma^*_+},\bar{u}})\}}[\underline\Psi]}\\&+\int_{S^2}d\omega\int_{\usigmap{v}}^{u}d\bar{u}|r^3\Omega\underline\psi|_{(\vsigmap{\bar{u}})},
\end{split}
\end{align}
\begin{align}\label{this 10 06 2021 second estimate}
    \begin{split}
         \int_{S^2}d\omega\int_{\usigmap{v}}^{u}d\bar{u}\,|r\Omega^2\underline\alpha|\lesssim &\frac{1}{r(u,\vsigmap{u})}\Bigg\{\int_{u_{{\Sigma^*_+},v}}^ud\bar{u} \frac{1}{\sqrt{r(\bar{u},v_{{\Sigma^*_+},\bar{u}})}}\sqrt{\mathbb{F}^T_{{\Sigma^*_+}\cap\{r\geq r(\usigmap{v},v)\}}[\underline\Psi]}\\&\qquad\qquad\qquad+\int_{S^2}d\omega\int_{\usigmap{v}}^{u}d\bar{u}|r^3\Omega\underline\psi|_{(\bar{u},\vsigmap{\bar{u}})}\Bigg\}\\&+\int_{S^2}d\omega\int_{\usigmap(v)}^{u}d\bar{u}|r\Omega^2\underline\alpha|_{(\bar{u},\vsigmap{\bar{u}})}.
    \end{split}
\end{align}
\end{lemma}

\begin{proof}
We will only prove the first estimate \bref{this 10 06 2021 first estimate} as the same method leads to the second estimate \bref{this 10 06 2021 second estimate}. In the following, the first step is to integrate between $(\bar{u},v)$ and $(\bar{u},\vsigmap{\bar{u}})$ for $\bar{u}$ such that $\usigmap{v}\leq \bar{u}\leq u$, and then integrate in $\bar{u}$ between $u$ and $\Sigma^*_-$ to find
\begin{align}
\begin{split}
   \int_{S^2}d\omega\int_{u_{{\Sigma^*_+},v}}^{u}d\bar{u}\,|r^3\Omega\underline\psi|\leq& \int_{S^2}d\omega\int_{u_{{\Sigma^*_+},v}}^{u}d\bar{u}\,\int_{v_{{\Sigma^*_+},\bar{u}}}^v d\bar{v}\,|\nablav r^3\Omega\underline\psi|+\int_{S^2}d\omega\int_{u_{{\Sigma^*_+},v}}^{u}d\bar{u}|r^3\Omega\underline\psi|_{(\bar{u},\vsigmap{\bar{u}})}\\
   &\leq\int_{u_{{\Sigma^*_+},v}}^ud\bar{u}\,\sqrt{\int_{S^2}d\omega\int_{v_{{\Sigma^*_+},\bar{u}}}^v d\bar{v}\frac{\Omega^2}{r^2}}\,\sqrt{\int_{S^2}d\omega\int_{v_{{\Sigma^*_+},\bar{u}}}^v d\bar{v}\frac{\Omega^2}{r^2}|\underline\Psi|^2}\\&\qquad\qquad+\int_{S^2}d\omega\int_{u_{{\Sigma^*_+},v}}^{u}d\bar{u}|r^3\Omega\underline\psi|_{(\bar{u},v_{{\Sigma^*_+},\bar{u}})}\\
   &\leq \int_{\usigmap{v}}^ud\bar{u} \sqrt{4\pi\left(\frac{1}{r(\bar{u},v_{{\Sigma^*_+},\bar{u}})}-\frac{1}{r(\bar{u},v)}\right)} \sqrt{\int_{S^2}d\omega\int_{v_{{\Sigma^*_+},\bar{u}}}^v\frac{\Omega^2}{r^2}|\underline\Psi|^2}\\&\qquad+\int_{S^2}d\omega\int_{u_{{\Sigma^*_+},v}}^{u}d\bar{u}|r^3\Omega\underline\psi|_{(\bar{u},v_{{\Sigma^*_+},\bar{u}})}\\
   &\lesssim \int_{\usigmap{v}}^ud\bar{u} \frac{1}{\sqrt{r(\bar{u},v_{{\Sigma^*_+},\bar{u}})}}\sqrt{\mathbb{F}^T_{{\Sigma^*_+}\cap\{r\geq r(\bar{u},v_{{\Sigma^*_+},\bar{u}})\}}[\underline\Psi]}\\&\qquad+\int_{S^2}d\omega\int_{\usigmap{v}}^{u}d\bar{u}|r^3\Omega\underline\psi|_{(\bar{u},v_{{\Sigma^*_+},\bar{u}})}.
\end{split}
\end{align}
In the final step above we used the  $\partial_t$-energy conservation of \Cref{RW T-energy}.
\end{proof}

\subsubsection[Estimates on $\protect\xlin$, $\protect\xblin$]{Estimates on $\xlin$, $\xblin$}

In the following we consider a solution to the full system \fullsystem, so $\alin, \ablin$ satisfy the Teukolsky equations and the corresponding $\plin, \Psilin, \pblin, \Psilinb$ defined according to \Cref{linearised hier} satisfy the estimates of \Cref{forward scattering prelim estimate on alin alinb}. 

\begin{proposition}\label{decay for xlin on H+}
The quantities $\xlin, \blin, \elin$ belonging to a solution to the system \fullsystem arising from initial data constructed according to \Cref{constructing compactly supported data} satisfy the following estimates on $\mathscr{H}^+_{\geq0}$:
\begin{align}\label{L2 in v polynomial decay estimate on H+}
\begin{split}
    \int_v^\infty d\bar{v} \int_{S^2_{\infty,v}}d\omega&\,\Big[|\mathcal{A}_2^2\Omega\xlin|^2+|\divo\mathcal{A}_2\Omega^{-1}\nablagml \Omega\xlin|^2+|\divo\mathcal{A}_2\fancydstar_2\elin|^2+|\mathcal{A}_2\fancydstar_2\Omega\blin|^2\Big]\\&\lesssim \frac{1}{v^2}\left[\mathbb{F}^2_{{\Sigma^*_+}}[\Psilin]+\int_{{\Sigma^*_+}}dr\sin\theta d\theta d\phi \left(r^{8-\epsilon}|\Omega\plin|^2+r^{6-\epsilon}|\Omega^2\alin|^2\right)\right]
\end{split}
\end{align}
\begin{align}\label{L2 polynomial decay estimate on H+}
\begin{split}
    \int_{S^2_{\infty,v}}\sin\theta d\theta d\phi&\left[|\mathcal{A}_2\Omega\xlin|^2+|\mathcal{A}_2\Omega^{-1}\nablagml \Omega\xlin|^2+|\mathcal{A}_2\fancydstar_2\elin|^2+|\mathcal{A}_2\fancydstar_2\Omega\blin|^2\right]\\&\lesssim \frac{1}{v^2}\left[\mathbb{F}^2_{{\Sigma^*_+}}[\Psilin]+\int_{{\Sigma^*_+}}dr\sin\theta d\theta d\phi \left(r^{8-\epsilon}|\Omega\plin|^2+r^{6-\epsilon}|\Omega^2\alin|^2\right)+\sup_{{\Sigma^*_+}}|r^4\Omega\plin|^2\right]
\end{split}
\end{align}
\end{proposition}
\begin{proof}
See Propositions 13.2.4, 13.2.5 of \cite{DHR16}. Note that it is crucial here that $\otx|_{\mathscr{H}^+_{\geq0}\cap{\Sigma^*_+}}=0$,\\  $\left[\rlin-\rlin_{\ell=0}+\divr\elin\right]|_{\mathscr{H}^+_{\geq0}\cap{\Sigma^*_+}}=0$.
\end{proof}

\begin{proposition}\label{ILED on xlin}
Let $\xlin$ belong to a solution to the system \fullsystem arising from initial data constructed according to \Cref{constructing compactly supported data}. We then have for any $\epsilon>0$ and any $v\geq0$,
\begin{align}\label{ILED estimate on xlin}
\begin{split}
    &\int_{J^+({\Sigma^*_+})\cap J^-(\underline{\mathscr{C}}_{v})}\Omega^2 dudv\sin\theta d\theta d\phi \frac{\Omega^2}{r^{1+\epsilon}}\Bigg[\left|\frac{1}{\Omega}\nablagml\left(\frac{1}{\Omega}\nablagml(r^2\Omega\xlin)\right)\right|^2+\left|\frac{1}{\Omega}\nablagml(r^2\Omega\xlin)\right|^2+|r^2\Omega\xlin|^2\Bigg]
    \\&+\int_{\underline{\mathscr{C}}_{v}\cap J^+({\Sigma^*_+})}\Omega d\bar{u}\dw\,\frac{\Omega^2}{r^{\epsilon}}\Bigg[\left|\frac{1}{\Omega}\nablagml\left(\frac{1}{\Omega}\nablagml(r^2\Omega\xlin)\right)\right|^2+\left|\frac{1}{\Omega}\nablagml(r^2\Omega\xlin)\right|^2+\frac{1}{r}|r^2\Omega\xlin|^2\Bigg]
    \\&\lesssim\int_{{\Sigma^*_+}}\Omega^2du\sin\theta d\theta d\phi\Bigg[\left|\frac{1}{\Omega}\nablagml\left(\frac{1}{\Omega}\nablagml(r^2\Omega\xlin)\right)\right|^2+\left|\frac{1}{\Omega}\nablagml(r^2\Omega\xlin)\right|^2+\frac{1}{r^{1+\epsilon}}|r^2\Omega\xlin|^2\Bigg]\\&\qquad\qquad\qquad+\mathbb{F}_{{\Sigma^*_+}}[\Psilin]+\int_{{\Sigma^*_+}}dr\sin\theta d\theta d\phi \left[r^{7-\epsilon}|\Omega\plin|^2+r^{5-\epsilon}|\Omega^2\alin|^2\right].
\end{split}
\end{align}
\end{proposition}
\begin{proof}
We simply repeat the work of Section 13.3 of \cite{DHR16} integrating between $\mathscr{H}^+_{\geq0}$ and ${\Sigma^*_+}$. Note that the proof of this estimate requires $L^2(\mathscr{H}^+_{\geq0})$ estimates on $\Omega\xlin$ and $\Omega^{-1}\nablagml\Omega\xlin$ contained in Proposition 13.2.1 and Proposition 13.2.2 of \cite{DHR16}, which in turn make crucial use of the initial gauge conditions \bref{initial horizon gauge condition}.
\end{proof}

\begin{remark}
The estimate quoted in \cite{DHR16} in fact has a weight of $r^{-\epsilon}$ multiplying $|r^2\Omega\xlin|^2$ on the left hand side and higher $r$-weights multiplying $\alin$, $\pblin$ on the right hand side. The weights used in \Cref{ILED on xlin} will suffice for our purposes.
\end{remark}

To obtain boundedness for $\xblin$, we make use of the $L^1$ estimate of \Cref{L1 estimate} on $\ablin, \pblin$ and the quantity $\Ylin$ as was done in \cite{DHR16}. We define $\Ylin$ by\footnote{This definition differs from that in \cite{DHR16} by a factor of $-\frac{1}{2}$.}
\begin{align}\label{ylin}
    \Ylin:=\mathcal{A}_2\frac{r^2\xblin}{\Omega}-r^4\Omega^{-1}\pblin.
\end{align}
Note that
\begin{align}\label{du ylin}
    \nablau\Ylin=r^3\Omega\pblin-6Mr\ablin.
\end{align}
Knowing the estimate of \Cref{L1 estimate}, we know that 
\begin{align}
\begin{split}
    |\Ylin(u,v,\theta^A)|^2\lesssim |\Ylin(u_0,v,\theta^A)|^2+\mathbb{F}_{{\Sigma^*_+}}^2[\Psilinb]+\int_{{\Sigma^*_+}}dr\sin\theta d\theta d\phi \Big\{&|r^3\Omega^{-1}\underline\psi|^2+|\mathring{\slashednabla}r^3\Omega^{-1}\underline\psi|^2+|\Omega^{-1}\nablagml r^3\Omega^{-1}\underline\psi|^2\\&+|r\nablav r^3\Omega^{-1}\underline\psi|^2+|r\Omega^{-2}\underline\alpha|^2\Big\}.
\end{split}
\end{align}
Combining this with the polynomial decay of $\pblin$ of \Cref{polynomial decay of pblin and ablin} leads to boundedness for $\xblin$ as follows:
\begin{proposition}
Let $\xblin$ belong to a solution to the system \fullsystem arising from initial data constructed according to \Cref{constructing compactly supported data}. We then have
\begin{align}
\begin{split}
    \left|\mathcal{A}_2\left(\frac{r^2\xblin}{\Omega}\right)(u,v,\theta^A)\right|^2\lesssim&\; \sup_{{\Sigma^*_+}}|\Ylin|^2+\sup_{{\Sigma^*_+}}|r^3\Omega^{-1}\pblin|^2+\mathbb{F}_{{\Sigma^*_+}}^2[\Psilinb]\\&+\int_{{\Sigma^*_+}}dr\sin\theta d\theta d\phi\left[r^{8-\epsilon}|\Omega\plin|^2+r^{6-\epsilon}|\Omega^2\alin|^2\right]
\end{split}
\end{align}
\end{proposition}
We also have
\begin{proposition}
Let $\xblin$ belong to a solution $\mathfrak{S}$ to the system \fullsystem arising from initial data constructed according to \Cref{constructing compactly supported data}. We then have
\begin{align}\label{L2 polynomial decay of dv xblin at H+}
    \left\|\mathcal{A}_2\nablav(r\Omega^{-1}\xblin)(\infty,v,\theta^A)\right\|^2_{L^2(S^2_{\infty,v})}\lesssim \frac{1}{v^2}\mathbb{E}_{{\Sigma^*_+}}[\mathfrak{S}]
\end{align}
where 
\begin{align}
    \mathbb{E}_{{\Sigma^*_+}}[\mathfrak{S}]:=\sup_{{\Sigma^*_+}}|r^4\Omega\pblin|^2+\sup_{{\Sigma^*_+}}|r^3\ablin|^2+\mathbb{F}^2_{{\Sigma^*_+}}[\Psilinb]+\mathbb{E}_1.
\end{align}
Here $\mathbb{E}_1$ was defined in \bref{def of E1}.
\end{proposition}

\subsection{Radiation fields at $\mathscr{I}^+$}\label{Section 9.2 radiation fields at scri+}

The main result of this section is the following:

\begin{proposition}\label{proposition of asymptotic flatness at null infinity}
Assume $\mathfrak{S}$ is a solution to the system \fullsystem arising from asymptotically flat initial data $\mathfrak{D}$ on $\Sigma^*_+$. Then $\mathfrak{S}$ is strongly asymptotically flat at null infinity in the sense of \Cref{defin of DHR flatness at scri+}.
\end{proposition}

\subsubsection[Radiation fields of $\protect\Psilin$, $\protect\Psilinb$, $\protect\plin$, $\protect\pblin$, $\protect\alin$, $\protect\ablin$]{Radiation fields of $\Psilin$, $\Psilinb$, $\plin$, $\pblin$, $\alin$, $\ablin$}

We begin with constructing the radiation fields at $\mathscr{I}^+$ for $\Psilin$, $\Psilinb$:

\begin{proposition}\label{forwards scattering for Psilin Psilinb at scri+}
Assume $\Psilin$, $\Psilinb$ arise from the system \fullsystem and \Cref{linearised hier} from asymptotically flat initial data on $\Sigma^*_+$. We have that $\mathbb{F}_{{\Sigma^*_+}}^T[\Psilin]$, $\mathbb{F}_{{\Sigma^*_+}}^T[\Psilinb]$ are finite and $\Psilin$, $\Psilinb$ define smooth pointwise limits at $\mathscr{I}^+$,
\begin{align}
    \upPsilin_{\mathscr{I}^+}=\lim_{v\longrightarrow\infty}\Psilin,\qquad\qquad\upPsilinb_{\mathscr{I}^+}=\lim_{v\longrightarrow\infty}\Psilinb,
\end{align}
Define also $\upPsilin_{\mathscr{H}^+}:=\Psilin|_{\mathscr{H}^+_{\geq0}}$, $\upPsilinb_{\mathscr{H}^+}:=\Psilinb|_{\mathscr{H}^+_{\geq0}}$. We have the following for $\upPsilin_{\mathscr{I}^+}$,  $\upPsilin_{\mathscr{H}^+}$:
\begin{align}
    \lim_{u\longrightarrow\infty}\upPsilin_{\mathscr{I}^+}=0,\qquad\qquad \lim_{v\longrightarrow\infty} \upPsilin_{\mathscr{H}^+}=0,
\end{align}
and the energy identity
\begin{align}\label{this energy identity 09 06 2021}
    \|\upPsilin_{\mathscr{H}^+}\|^2_{\mathcal{E}^{T,RW}_{\mathscr{H}^+_{\geq0}}}+ \|\upPsilin_{\mathscr{I}^+}\|^2_{\mathcal{E}^{T,RW}_{\mathscr{I}^+}}=\left\|(\Psilin|_{{\Sigma^*_+}},\slashednabla_{n_{{\Sigma^*_+}}}\Psilin|_{{\Sigma^*_+}})\right\|^2_{\mathcal{E}^{T,RW}_{{\Sigma^*_+}}},
\end{align}
where
\begin{align}
    \left\|(\Psilin|_{{\Sigma^*_+}},\slashednabla_{n_{{\Sigma^*_+}}}\Psilin|_{{\Sigma^*_+}})\right\|^2_{\mathcal{E}^{T,RW}_{{\Sigma^*_+}}}=\mathbb{F}_{{\Sigma^*_+}}^T[\Psilin],
\end{align}
\begin{align}
    \|\upPsilin_{\mathscr{H}^+}\|^2_{\mathcal{E}^{T,RW}_{\mathscr{H}^+_{\geq0}}}=\int_{\mathscr{H}^+_{\geq0}}\sin\theta d\theta d\phi \,|\partial_v\upPsilin_{\mathscr{H}^+}|^2,\qquad\qquad\|\upPsilin_{\mathscr{I}^+}\|^2_{\mathcal{E}^{T,RW}_{\mathscr{I}^+}}=\int_{\mathscr{I}^+}\sin\theta d\theta d\phi \,|\partial_u\upPsilin_{\mathscr{I}^+}|^2.
\end{align}
A similar statement applies for $\upPsilinb_{\mathscr{I}^+}$, $\upPsilinb_{\mathscr{H}^+}$.
\end{proposition}
\begin{proof}
The assumption of asymptotic flatness on initial data for \fullsystem implies that $\Psilin|_{\Sigma^*_+}, \Psilinb|_{\Sigma^*_+}\sim O_\infty\left(\frac{1}{r^{\gamma-1}}\right)$ for $\gamma>1$, thus the $r^p$-estimate \bref{RW rp estimate} applies with $p=\gamma$. The existence of the radiation field $\Psilin$, $\Psilinb$ now follows from the argument of Proposition 5.2.3 of \cite{Mas20}, which we reproduce here (see also \cite{DRSR14}). For a fixed $u$ and for $v_2>v_1$, we have
\begin{align}\label{argument for radiation field}
\begin{split}
|\Psilin(u,v_2,\theta^A)-\Psilin(u,v_1,\theta^A)|^2\lesssim& \left[\sum_{|\gamma|\leq 3} \int_{S^2}d\omega\; |\slashed{\mathcal{L}}^\gamma_{S^2} (\Psi(u,v_2,\theta,\phi)-\Psi(u,v_1,\theta,\phi))|\right]^2\\
&=  \left[\sum_{|\gamma|\leq 3} \int_{S^2}d\omega\int_{v_1}^{v_2}dv |\slashed{\mathcal{L}}^\gamma_{S^2} \Omega\slashed\nabla_4\Psi|\right]^2\\
&\leq \frac{1}{r(u,v_1)^{\gamma-1}}\Bigg[\sum_{|\gamma|\leq 3} \int_{\mathscr{C}_u\cap\{v>v_1\}}dvd\omega\; r^\gamma|\slashed{\mathcal{L}}^\gamma_{S^2}\Omega\slashed{\nabla}_4\Psi|^2dv\sin \theta d\theta d\phi\Bigg],
\end{split}
\end{align}
where we used Cauchy--Schwarz to obtain the last estimate and $\slashed{\mathcal{L}}^\gamma_{S^2}=\mathcal{L}_{\Omega_1}^{\gamma_1}\mathcal{L}_{\Omega_2}^{\gamma_2}\mathcal{L}_{\Omega_3}^{\gamma_3}$ denotes Lie differentiation on $S^2$ with respect to its $so(3)$ algebra of Killing fields. The argument above implies that for any sequence ${v_n}$ with $v_n\longrightarrow\infty$ as $n\longrightarrow\infty$, $\Psilin(u,v_n,\theta^A)$ is Cauchy in $L^\infty([u_1,u_2]\times S^2)$, hence a limit $\Psi_{\mathscr{I}^+}$ exists in  $L^\infty([u_1,u_2]\times S^2)$. Commuting with $\mathring{\slashednabla}^k$, the argument above goes through to give us smoothness in the angular direction. Commuting with $\slashednabla_t^i$, $(r\nablav)^i$, the argument above can be repeated given the higher order estimates of \Cref{RWrp} to show that $\nablau^i\Psi\longrightarrow \partial_u^i\Psi_{\mathscr{I}^+}$. All of the above leads to the same conclusions for $\Psilinb$ which hence induces a smooth radiation field $\Psilinb_{\mathscr{I}^+}$ on $\mathscr{I}^+$.\\

To show that $\upPsilin_{\mathscr{I}^+}$ decays towards the future end of $\mathscr{I}^+$, we may apply the argument of Proposition 5.2.4 of \cite{Mas20}, which we reproduce here: integrating $\nablav |\Psi|^2$ in $v$ between $r=R$ and $\mathscr{I}^+$, using Cauchy--Schwarz and a Hardy estimate gives us:
\begin{align}
\begin{split}
     \int_{S^2_{u,\infty}}d\omega\,|\upPsilin_{\mathscr{I}^+}|^2&= \int_{S^2}d\omega\,|\Psilin_{r=R}|^2+\int_{v(R,u)}^\infty dv\,\int_{S^2}d\omega\; 2\Psilin\times\nablav\Psilin\\
     &\leq \int_{S^2}d\omega\,|\Psilin_{r=R}|^2 + \sqrt{\int_{S^2}d\omega\,\int_{v(R,u)}^\infty dv\, \frac{1}{r^2}|\Psilin|^2}\times\sqrt{\int_{S^2}d\omega\,\int_{v(R,u)}^\infty dv\, r^2|\nablav\Psilin|^2}.
\end{split}
\end{align}

\Cref{RWrp} commuted with $\slashednabla_t$ implies by \Cref{technical lemma} that the $L^2$ integral of $r^{-\frac{\gamma}{2}}\Psilin$ above decays as $u\longrightarrow\infty$, while the $r$-weighted estimate \bref{RW rp estimate} says that the $L^2$ integral of $r^\gamma|\nablav\Psi|^2$ is bounded. Finally, using a higher order version of \Cref{Polynomial decay of energy of RW towards future} and a Sobolev estimate, we get that $\Psilin|_{r=R}$ decays as $t\longrightarrow\infty$ (this also applies to $\Psilin|_{r=2M}$).\\

Now fix $r_0$ and take $u_+,v_+$ such that $r(u_+,v_+)=r_0$ and let $u_-<u_+$, $v_\infty>v_+$. We apply the energy identity in the region $\mathscr{D}$ bounded by ${\Sigma^*_+}\cap\{u\geq u_-\}$,  $\mathscr{H}^+_{\geq0}\cap\{v\leq v_+\}$,  $\underline{\mathscr{C}}_{v_+}\cap\{u\geq u_+\}$,  $\mathscr{C}_{u_+}\cap\{v_+\leq v \leq v_\infty\}$, $\underline{\mathscr{C}}_{v_\infty}\cap\{u_-\leq u \leq u_+\}$, $\mathscr{C}_{u_-}\cap J^+({\Sigma^*_+})$ (i.e.~integrate \bref{T derivative identity} over $\mathscr{D}$):
\begin{align}
\begin{split}
    \int_{\mathscr{H}^+_{\geq0}\cap\{v\leq v_+\}} |\partial_v \Psilin|^2\,+\,&\underline{F}_{v_+}^T[\Psilin](u_+,\infty)\,+\,{F}_{u_+}[\Psilin](v_+,v_\infty)+\underline{F}_{v_\infty}[\Psilin](u_-,u_+)\\&=\mathbb{F}^T_{{\Sigma^*_+}\cap\{u\geq u_-\}}+ {F}^T_{u_-}[\Psilin](\vsigmap{u_-},v_\infty).
\end{split}
\end{align}
As $\Psilin(u,v,\theta^A)$ converges uniformly in $C^\infty(\mathscr{I}^+\cap\{u\in[u_-,u_+]\})$ as $v\longrightarrow\infty$, we have
\begin{align}
    \underline{F}_{v_\infty}[\Psilin](u_-,u_+)\longrightarrow \int_{\mathscr{I}^+\cap\{u\in[u_-,u_+]\}}\sin\theta d\theta d\phi du |\partial_u\upPsilin_{\mathscr{I}^+}|^2.
\end{align}
Next, we take the limit of ${F}^T_{u_-}[\Psilin](\vsigmap{u_-},\infty)$ as $u_-\longrightarrow-\infty$. We know that $\Psilin=O\left(\frac{1}{r^{\gamma-1}}\right)$, $\nablau\Psilin, \nablav\Psilin=O\left(\frac{1}{r^\gamma}\right)$, so $\mathbb{F}^T_{{\Sigma^*_+}}[\Psilin]<\infty$. Knowing that 
\begin{align}
    {F}^T_{u_-}[\Psilin](\vsigmap{u_-},\infty)\leq \mathbb{F}_{{\Sigma^*_+}\cap\{u\leq u_-\}}[\Psilin],
\end{align}
it is clear that ${F}^T_{u_-}[\Psilin](\vsigmap{u_-},\infty)\longrightarrow 0$ as $u_-\longrightarrow-\infty$. Finally, the polynomial decay estimates of \Cref{Polynomial decay of energy of RW towards future} give us the energy identity \bref{this energy identity 09 06 2021}  by taking $u_+, v_+\longrightarrow\infty$.
\end{proof}

\begin{corollary}\label{finite energy unitarity RW}
Assume $\Psilin$, $\Psilinb$ arise from initial data on $\Sigma^*_+$ such that $\mathbb{F}_{\Sigma^*_+}[\Psilin]$, $\mathbb{F}_{\Sigma^*_+}[\Psilinb]$ are finite. For any $r_0>2M$, we have
\begin{align}
    \lim_{v\longrightarrow\infty} \underline{F}_v[\Psilin](u,\infty)+F_u[\Psilin](v,\infty)=0,
\end{align}
where $u$ is such that $r(u,v)=r_0$. Thus \bref{this energy identity 09 06 2021} holds.
\end{corollary}
\begin{proof}
    We have proven the claim for asymptotically flat initial data in \Cref{forwards scattering for Psilin Psilinb at scri+}. The claim then follows by the density of initial data of decay $O_\infty(r^{-\gamma})$ in the space $\mathcal{E}^{T,RW}_{\Sigma^*_+}$ and the linearity of \bref{RW} for $\gamma>-\frac{1}{2}$.
\end{proof}

For weakly asymptotically flat initial data, we have the following:

\begin{corollary}\label{weak boundedness of Psilin weak a f}
Assume $\Psilin$, $\Psilinb$ arise from the system \fullsystem and \Cref{linearised hier} from weakly asymptotically flat initial data on $\Sigma^*_+$. We have that $\mathbb{F}_{{\Sigma^*_+}}^T[\Psilin]$, $\mathbb{F}_{{\Sigma^*_+}}^T[\Psilinb]$ are finite and $\nablau\Psilin$ define smooth pointwise limits at $\mathscr{I}^+$. Furthermore, for any $\delta>0$ we have
\begin{align}
    \lim_{v\longrightarrow\infty}\frac{\Psilin}{r^\delta}(u,v,\theta^A)=0.
\end{align}
The same conclusions hold for $\Psilinb$.
\end{corollary}
\begin{proof}
    For $\delta, \delta'>0$ we have
    \begin{align}
        r^{1+\delta'}\left|\nablav \frac{\Psilin}{r^\delta}\right|^2\lesssim |\nablav\Psilin|^2 r^{1+\delta'-2\delta}+\delta^2\frac{\Omega^4}{r^{1+2\delta-\delta'}}|\Psilin|^2.
    \end{align}
    For $0<\delta, \delta'<1$ with $\delta'<2\delta$, we may integrate the above and apply a Hardy inequality to estimate the last term by the first term on the right hand side. An $r$-weighted estimate \bref{RW rp estimate} with $p=\delta'$ can be combined with an argument repeating the steps of \bref{argument for radiation field} to prove the claim.
\end{proof}

We now turn to the radiation fields of $\pblin, \ablin$. 

\begin{proposition}\label{forwards scattering for ablin pblin at scri+}
Assume $\ablin$, $\pblin$ arise from the system \fullsystem and the relations \bref{hier} via \Cref{EinsteinWP} from weakly asymptotically flat initial data  $\mathfrak{D}$ on $\Sigma^*_+$. We have that $\ablin$ $\pblin$ define smooth pointwise limits towards $\mathscr{I}^+$ via the limits,
\begin{align}
    \ablins_{\mathscr{I}^+}=\lim_{v\longrightarrow\infty}r\ablin,\qquad\qquad\pblins_{\mathscr{I}^+}=\lim_{v\longrightarrow\infty}r^3\pblin.
\end{align}
Furthermore, the field $\ablins_{\mathscr{I}^+}$ satisfies
\begin{align}\label{2 integrals of ablins vanish at scri+}
    \int_{-\infty}^\infty du_1\, \ablins_{\mathscr{I}^+}=0.
\end{align}
If $\mathfrak{D}$ is asymptotically flat to order $(1+\epsilon,\infty)$ for some $\epsilon>0$, we moreover have
\begin{align}
    \int_{-\infty}^\infty du_1 \int_{-\infty}^{u_1} du_2\, \ablins_{\mathscr{I}^+}=0.
\end{align}
\end{proposition}

\begin{proof}
Energy boundedness for $\Psilinb$ is sufficient to obtain a radiation field for $\pblin$:
\begin{align}\label{argument for radiation field of pblin}
\begin{split}
|r^3\Omega\pblin(u,v_2,\theta^A)-r^3\Omega\pblin(u,v_1,\theta^A)|^2\lesssim& \left[\sum_{|\gamma|\leq 3} \int_{S^2}d\omega\; |\slashed{\mathcal{L}}^\gamma_{S^2} (r^3\Omega\pblin(u,v_2,\theta,\phi)-r^3\Omega\pblin(u,v_1,\theta,\phi))|\right]^2\\
&=  \left[\sum_{|\gamma|\leq 3} \int_{S^2}d\omega\int_{v_1}^{v_2}dv |\slashed{\mathcal{L}}^\gamma_{S^2} \Omega\slashed\nabla_4r^3\Omega\pblin|\right]^2\\
&\leq \frac{1}{r(u,v_1)}\Bigg[\sum_{|\gamma|\leq 3} \int_{\mathscr{C}_u\cap\{v>v_1\}}dvd\omega\; \frac{r^2}{\Omega^2}|\slashed{\mathcal{L}}^\gamma_{S^2}\Omega\slashed{\nabla}_4r^3\Omega\pblin|^2dv\sin \theta d\theta d\phi\Bigg],\\
&= \frac{1}{r(u,v_1)}\Bigg[\sum_{|\gamma|\leq 3} \int_{\mathscr{C}_u\cap\{v>v_1\}}dvd\omega\; \frac{\Omega^2}{r^2}|\slashed{\mathcal{L}}^\gamma_{S^2}\Psilinb|^2dv\sin \theta d\theta d\phi\Bigg],\\
&= \frac{1}{r(u,v_1)}\sum_{|\gamma|\leq 3} F_u[\slashed{\mathcal{L}}^\gamma_{S^2}\Psilinb](v_1,\infty)
\end{split}
\end{align}
Thus $r^3\Omega\pblin\,(u,v_n,\theta^A)$ is Cauchy along any sequence $\{v_n\}_n$ with $v_n\longrightarrow\infty$. Similarly, since we have
\begin{align}\label{argument for radiation field of ablin}
\begin{split}
    \int_{S^2}d\omega\,|r\Omega^2\ablin(u,v_2,\theta^A)-r\Omega^2\ablin(u,v_1,\theta^A)|^2&\leq\int_{S^2}d\omega\left|\int_{v_1}^{v_2} d\overline{v} \,\nablav r\Omega^2\ablin(u,\overline{v},\theta^A)\right|^2\\&\leq \frac{1}{r(u,v)}\int_{\mathscr{C}_u\cap\{\overline{v}\geq v\}}d\omega d\overline{v} \frac{r^2}{\Omega^2}|\nablav r\Omega^2\ablins|^2
    \\&\leq\frac{1}{r(u,v_1)}\int_{\mathscr{C}_u\cap\{\bar{v}\geq v\}}d\omega d\bar{v} \frac{\Omega^2}{r^2}|r^3\Omega\pblin|^2
    \\&\leq \frac{1}{r(u,v_1)}\int_{\mathscr{C}_u\cap\{\overline{v}\geq v\}}d\omega\, d\bar{v} \frac{\Omega^2}{r(u,v)^2}|r^3\Omega\pblin-\pblins_{\mathscr{I}^+}|^2\\&\;\;+\frac{1}{r(u,v_1)^2}\int_{S^2}d\omega\,|\pblins_{\mathscr{I}^+}|^2
    \\&\leq \frac{1}{r(u,v_1)}\int_{\mathscr{C}_u\cap\{\overline{v}\geq v\}}d\omega\, d\bar{v} \frac{\Omega^2(u,\bar{v})}{r(u,\bar{v})^2}\Bigg|\int_{\bar{v}}^\infty d\bar{\bar{v}} \nablav r^3\Omega\pblin\,\Bigg|^2\\&\;\;+\frac{1}{r(u,v_1)^2}\int_{S^2}d\omega\,|\pblins_{\mathscr{I}^+}|^2
    \\&\leq \frac{1}{r(u,v_1)}\int_{\mathscr{C}_u\cap\{\overline{v}\geq v\}}d\omega\, d\bar{v}\left[ \frac{\Omega^2(u,\bar{v})}{r(u,\bar{v})^3}\int_{\bar{v}}^\infty d\bar{\bar{v}} \frac{\Omega^2}{r^2} |\Psilinb|^2\right]\\&\;\;+\frac{1}{r(u,v_1)^2}\int_{S^2}d\omega\,|\pblins_{\mathscr{I}^+}|^2,
\end{split}
\end{align}
we conclude that
\begin{align}\label{argument for radiation field of ablin appendix}
    \int_{S^2}d\omega\,|r\Omega^2\ablin(u,v_2,\theta^A)-r\Omega^2\ablin(u,v_1,\theta^A)|^2\lesssim \frac{1}{r(u,v_1)^3}F_u^T[\Psilinb](v,\infty)+\frac{1}{r(u,v_1)^2}\int_{S^2}d\omega\,|\pblins_{\mathscr{I}^+}|^2.
\end{align}
Commuting the above twice with Lie derivatives along the generators of the $SO(3)$, we conclude that $r\Omega^2\ablin$ converges. Further commutations with powers of $\mathring{\slashednabla}$ leads to smoothness of $\ablins_{\mathscr{I}^+}$, $\pblins_{\mathscr{I}^+}$ in the angular directions. The convergence of $r^3\Omega\pblin$ towards $\mathscr{I}^+$ implies that $\nablau r\Omega^2\ablin\longrightarrow \partial_u \ablins_{\mathscr{I}^+}$ in $L^{\infty}([u_1,u_2]\times S^2)$ as $v\longrightarrow\infty$ for any interval $[u_1,u_2]$ in $u$.  Further commutations with $\slashednabla_t$ and the convergence of $(r\nablav)^k\Psilinb$ imply smoothness of $\ablins_{\mathscr{I}^+}$ in the $u$-direction and that $\nablau^k r\Omega^2\ablin\longrightarrow \partial_u^k\ablins_{\mathscr{I}^+}$.\\

To show that $\ablins_{\mathscr{I}^+}$, $\pblins_{\mathscr{I}^+}$ decay as $u\longrightarrow\infty$, fix $r_0$ and for any $u$ repeat the arguments \bref{argument for radiation field of pblin}, \bref{argument for radiation field of ablin} taking $v_2\longrightarrow\infty$ and $v_1$ such that $r(u,v_1)=r_0$; the decay of $\ablin|_{r=r_0}$ and $\pblin|_{r=r_0}$ as $u\longrightarrow\infty$ follows from \Cref{polynomial decay of pblin and ablin}, and the decay of the right hand side at the end of the arguments \bref{argument for radiation field of pblin}, \bref{argument for radiation field of ablin appendix} simply follows from the energy decay given by \Cref{finite energy unitarity RW}. Turning to the limits of $\ablins_{\mathscr{I}^+}$, $\pblins_{\mathscr{I}^+}$ for $u\longrightarrow-\infty$, the fact that $r\ablin|_{{\Sigma^*_+}}$ and $r^3\pblin|_{{\Sigma^*_+}}$ decay as $r\longrightarrow\infty$, together with the energy inequality \bref{RW T-energy} and the fact that the energy $\|(\Psilinb|_{{\Sigma^*_+}},\slashednabla_{n_{{\Sigma^*_+}}}\Psilinb|_{{{\Sigma^*_+}}})\|_{\mathcal{E}^{T,RW}_{\Sigma^*_+}}$ is finite by \Cref{forwards scattering for Psilin Psilinb at scri+}, implies that we may apply the arguments \bref{argument for radiation field of pblin}, \bref{argument for radiation field of ablin appendix} integrating along a constant $u$ surface between ${\Sigma^*_+}$ and $\mathscr{I}^+$ to obtain decay for $\ablins_{\mathscr{I}^+}$, $\pblins_{\mathscr{I}^+}$ as $u\longrightarrow-\infty$.\\

Finally, for fixed $u_1$ and $u_2>u_1$, we have that $\int_{u_1}^{u_2} du\, r\Omega^2\ablin\longrightarrow \int_{u_1}^{u_2}du \ablins_{\mathscr{I}^+}$ since after a sufficiently large value of $v_\infty$, $r\Omega^2\ablin(u,v,\theta^A)$ is uniformly bounded on $\cup_{v\geq v_{\infty}}\underline{\mathscr{C}}_{v}\cap\{u\in[u_1,u_2]\}$, so the convergence of $r\Omega^2\ablin$ towards a limit $\ablins_{\mathscr{I}^+}$ implies the convergence of the integrals by Lebesgue's boundedness convergence theorem. Therefore, we have from the $-2$ Teukolsky equation \bref{T-2} that
\begin{align}\label{pblins at scri+ in terms of integral of ablins}
    \pblins_{\mathscr{I}^+}(u_2,\theta^A)-\pblins_{\mathscr{I}^+}(u_1,\theta^A)=\int_{u_1}^{u_2}du \,\ablins_{\mathscr{I}^+},
\end{align}
and since $\pblins_{\mathscr{I}^+}(u,\theta^A)\longrightarrow0$ as $u\longrightarrow\pm\infty$, we have that
\begin{align}\label{integral of ablins along scri+ vanishes}
\int_{-\infty}^\infty du\; \ablins_{\mathscr{I}^+}=0.
\end{align}
If $\mathfrak{D}$ has decay of order $(1+\epsilon,\infty)$ for $\epsilon>0$, another iteration of this argument together with the decay of $\Psilinb_{\mathscr{I}^+}$ as $u\longrightarrow\pm\infty$ (using the results of \Cref{forwards scattering for Psilin Psilinb at scri+}) implies that
\begin{align}\label{integral of pblins along scri+ vanishes}
    \int_{-\infty}^\infty d\bar{u}\;\pblins_{\mathscr{I}^+}= \int_{-\infty}^\infty du_1 \int_{-\infty}^{u_1} du_2\, \ablins_{\mathscr{I}^+}=0
\end{align}
\end{proof}

\begin{remark}
Had we had that the data for $\alpha$ and its derivatives on ${\Sigma^*_+}$ decay like $O_\infty(r^{-11/2-\delta})$ for $\delta>0$, we may apply the $r$-weighted estimate \bref{RWrp k=1} to conclude that $\Phi^{(1)}$ defines a radiation field $\upphi^{(1)}_{\mathscr{I}^+}$ via an identical argument to \bref{argument for radiation field}. Having yet an additional power of $r$ in the decay rate, we may reach the same conclusion for $\Phi^{(2)}$ while also concluding that $\Phi^{(1)}_{\mathscr{I}^+}\longrightarrow0$ as $u\longrightarrow\infty$. In \cite{Mas20} we work with data of compact support and by iterating the process just described in this remark and in \Cref{forwards scattering for ablin pblin at scri+}, we conclude that $\int_{-\infty}^\infty du\, \upPsilinb_{\mathscr{I}^+}=0$ $\int_{-\infty}^\infty du_1 \int_{-\infty}^{u_1} du_2\, \upPsilinb_{\mathscr{I}^+}=0$, whence we may in a similar manner to the argument in \Cref{forwards scattering for Psilin Psilinb at scri+} conclude that 
\begin{align}
    \int_{-\infty}^\infty du_1 \int_{-\infty}^{u_1} du_2\int_{-\infty}^{u_2}du_3\, \ablins_{\mathscr{I}^+}=\int_{-\infty}^\infty du_1 \int_{-\infty}^{u_1} du_2\int_{-\infty}^{u_2}du_3\int_{-\infty}^{u_3} du_4\, \ablins_{\mathscr{I}^+}=0
\end{align}
\end{remark}

In fact, we have

\begin{lemma}\label{integral of ablin converges towards scri+}
The field $\ablin$ of \Cref{forwards scattering for ablin pblin at scri+} satisfy
\begin{align}
  \int_{\usigmap{v}}^u d\bar{u}\, r^3\pblin \longrightarrow \int_{-\infty}^u d\bar{u}\, \pblins_{\mathscr{I}^+}, 
\end{align}
as $v\longrightarrow\infty$. If the initial data has decay of order $(\gamma,4)$ for $\gamma>1$, we also have
\begin{align}
  \int_{\usigmap{v}}^u d\bar{u}\, r\ablin \longrightarrow \int_{-\infty}^u d\bar{u}\, \ablins_{\mathscr{I}^+}.
\end{align}
\end{lemma}

\begin{proof}
The decay rates as $r\longrightarrow\infty$ on $\pblin|_{{\Sigma^*_+}}$, $\Psilinb|_{{\Sigma^*_+}}$ afforded by the assumption of $(1+\delta,4)$-asymptotic flatness imply that
\begin{align}
    \mathbb{F}^T_{\Sigma^*_+\cap B_{[2M,r]}}[\Psilinb]=O\left(\frac{1}{r^{1+2\delta}}\right),
\end{align}
where $B_{[2M,R]}=\{(u,v,\theta^A): r(u,v)\in[2M,R]\}$. Together with the estimates of \Cref{L1 estimate near i0}, this implies that
\begin{align}
    \int_{S^2}d\omega\int_{\usigmap{v}}^{u}d\bar{u}\,|r^3\Omega\pblin|\leq \frac{C}{r(u,\vsigmap{u})^\delta},\qquad \int_{S^2}d\omega\int_{\usigmap{v}}^{u}d\bar{u}\,|r\Omega^2\ablin|\leq \frac{C}{r(u,\vsigmap{u})^{1+\delta}}.
\end{align}
A Sobolev embedding on the sphere gives us
\begin{align}\label{integrals of pblins ablins near i0}
   \int_{\usigmap{v}}^{u}d\bar{u}\,|r^3\Omega\pblin|\leq \frac{C}{r(u,\vsigmap{u})^\delta},\qquad \int_{\usigmap{v}}^{u}d\bar{u}\,|r\Omega^2\ablin|\leq \frac{C}{r(u,\vsigmap{u})^{(1+\delta)}},
\end{align}
for a constant $C$ that depends only on the size of initial data on ${\Sigma^*_+}$.\\

Fix an arbitrary $\epsilon>0$. As we have \bref{2 integrals of ablins vanish at scri+}, we may choose $u_-$ small enough such that
\begin{align}
    \left|\int_{-\infty}^{u_-} d\bar{u} \,\ablins_{\mathscr{I}^+}\right|\leq \frac{\epsilon}{3},\qquad \frac{C}{r(u_-,\vsigmap{u_-})^{1+\delta}} <\frac{\epsilon}{3}.
\end{align}
For $[u_-,u]$, we may find $v$ large enough such that
\begin{align}
    \left|\int_{u_-}^u d\bar{u}\, r\ablin-\int_{u_-}^u d\bar{u}\, \ablins_{\mathscr{I}^+}\right|<\frac{\epsilon}{3}
\end{align}
Thus
\begin{align}
    \left|\int_{\usigmap{v}}^u d\bar{u}\, r\ablin-\int_{-\infty}^u d\bar{u}\, \ablins_{\mathscr{I}^+}\right|<\left|\int_{u_-}^u d\bar{u}\, r\ablin-\int_{u_-}^u d\bar{u}\, \ablins_{\mathscr{I}^+}\right|+\left|\int_{-\infty}^{u_-}d\bar{u}\,
    \ablins_{\mathscr{I}^+}\right|+\left|\int_{\usigmap{v}}^{u_-} d\bar{u} \,r\ablin\,\right|<\epsilon.
\end{align}
An identical argument works for $r^3\Omega\pblin$.
\end{proof}

\subsubsection[Radiation fields of $\protect\xlin$, $\protect\xblin$, $\protect\bblin$, $\protect\elin$, $\protect\Olino$ and $\protect\glinh$]{Radiation fields of  $\protect\xlin$, $\protect\xblin$, $\protect\bblin$, $\protect\elin$, $\protect\Olino$ and $\protect\glinh$}

We now study the radiation field of $\xblin$:

\begin{proposition}\label{forwards scattering for xblin at scri+}
Assume $\xblin$ arises from a solution to the system \fullsystem via \Cref{EinsteinWP} from weakly asymptotically flat initial data on ${\Sigma^*_+}$. We have that $\xblin$ defines smooth pointwise limit towards $\mathscr{I}^+$,
\begin{align}
    \xblins_{\mathscr{I}^+}(u,\theta^A)=\lim_{v\longrightarrow\infty}r\xblin(u,v,\theta^A).
\end{align}
Thus $\glinh(u,v,\theta^A)\longrightarrow0$ for any finite $u$. 
\end{proposition}

\begin{proof}
    If initial data is assumed to be only weakly asymptotically flat, we will not in general be able to define a pointwise limit of $\Psilinb$ towards $\mathscr{I}^+$. We can however still repeat the argument of \Cref{forwards scattering for Psilin Psilinb at scri+} to obtain a pointwise limit for $\nablau\Psilin$ at $\mathscr{I}^+$. For any fixed $v$, we can now integrate in $u$ from $\Sigma^*_+$ to any finite value of $u$ the relation  \bref{further constraint +}, which reads
    \begin{align}
        \nablau^2\Psilinb-\frac{3\Omega^2-1}{r}\nablau\Psilinb=\mathcal{A}_2(\mathcal{A}_2-2)r\Omega^2\ablin+6M\nablau r\Omega^2\ablin+6M\frac{\Omega^2}{r^2}r^3\Omega\pblin.
    \end{align}
    Note that
    \begin{align}
    \left|\int_{\usigmap{v}}^ud\bar{u}\frac{\Omega^2}{r^2}\Psilinb\right|^2\lesssim \frac{1}{r(\usigmap{v},v)}\mathbb{F}_{\Sigma^*_+}^2,
    \end{align}
    \begin{align}
        \left|\int_{\usigmap{v}}^ud\bar{u}\frac{\Omega^2}{r^2}r^3\Omega\plin\right|^2\lesssim \frac{1}{r(\usigmap{v},v)}\left[|r^3\Omega\plin|^2_{\Sigma^*_+}+\mathbb{F}_{\Sigma^*_+}^2\right],
    \end{align}
    thus
    \begin{align}
        \int_{\usigmap{v}}^ud\bar{u}\frac{\Omega^2}{r}\Psilinb,\quad\int_{\usigmap{v}}^ud\bar{u}\frac{\Omega^2}{r^2}r^3\Omega\pblin\longrightarrow0
    \end{align}
    as $v\longrightarrow\infty$. Since $\xblin|_{\Sigma^*_+}$ decays at order $O_\infty(r^{-\gamma-1})$ for $\gamma\geq1$, we conclude that $r\xblin(u,v,\theta^A)$ has a pointwise limit for any $u$ as $v\longrightarrow\infty$. Commuting with powers of $\partial_t$ and Lie derivatives along the generators of the $SO(3)$ symmetry algebra shows that the limits is smooth. We can similarly deduce the decay of $\int_{\usigmap{v}}^ud\bar{u}\xblin(\bar{u},v,\theta^A)$ as $v\longrightarrow\infty$ which gives the decay of $\glinh(u,v,\theta^A)$ towards $\mathscr{I}^+$.
\end{proof}

\begin{corollary}
Assume that the initial data on $\Sigma^*_+$ for \fullsystem in \Cref{forwards scattering for xblin at scri+} is asymptotically flat to order $(1+\delta,\infty)$ for $\delta>0$. Then we have for $u,v$ such that $r(u,v)>4M$ that
\begin{align}
    |r\xblin(u,v,\theta^A)|\sim O_{\infty}(r(u,\vsigmap{u})^{-\delta}).
\end{align}
\end{corollary}

\begin{proof}
    If the initial data has decay of order $(1+\delta,2)$ for $\delta>0$, we can use \Cref{L1 estimate near i0} to derive a quantitative estimate on the $r\xblin$ near $i^0$: fix $u_-$, then given that $\ablin|_{{\Sigma^*_+}}=O(r^{-3-\delta}), \pblin|_{{\Sigma^*_+}}=O(r^{-4-\delta})$|, we have for $u\leq u_-$ that
\begin{align}\label{xblins decays pointwise near scri+-}
   \int_{\usigmap{v}}^{u}d\bar{u}\,|r\Omega^2\ablin|\leq \frac{C}{r(u,\vsigmap{u})^{\delta}},
\end{align}
for a constant $C$ that is implicit in the assumption of asymptotic flatness as in \Cref{def of asymptotic flatness at spacelike infinity}. Therefore, on ${r\geq R, u\leq u_0}$ we have
\begin{align}\label{pointwise estimate on xblin near i0}
    |r\xblin(u,v,\theta^A)|\leq \frac{C'}{r(u,\vsigmap{u})^\delta}+|r\Omega\xblin(\usigmap{v},v,\theta^A)|\lesssim\frac{1}{r(u,\vsigmap{u})^\delta},
\end{align}
where $C'$ is a universal constant that results from the application of a Sobolev estimate on the sphere.\\
\end{proof}

\begin{corollary}\label{radiation field for glinh at scri+ forward scattering}
Assume that the initial data on $\Sigma^*_+$ for \fullsystem in \Cref{forwards scattering for xblin at scri+} is asymptotically flat to order $(1+\delta,\infty)$ for $\delta>0$. Then we have for $u,v$ such that $r(u,v)>4M$ that
Moreover, we have that
\begin{align}
    \int_{\usigmap{v}}^u d\bar{u}\, r\xblin
\end{align}
converges as $v\longrightarrow\infty$. In particular, the radiation field defined by the limit 
\begin{align}
    \glinhs_{\mathscr{I}^+}(u,\theta^A)=\lim_{v\longrightarrow\infty}r\glinh(u,v,\theta^A).
\end{align}
exists and defines an element of of $\Gamma^{(2)}(\mathscr{I}^+)$.
\end{corollary}

\begin{proof}
Observe that
\begin{align}\label{this 11 06 2021}
   \int_{\usigmap{v}}^u d\bar{u}\, r^3\Omega\pblin= \int_{\usigmap{v}}^u d\bar{u}\, (r^3\Omega\pblin)|_{(\usigmap{v},v)}+ \int_{\usigmap{v}}^u d\bar{u} \int_{\usigmap{v}}^{\bar{u}}d\bar{\bar{u}}\,\nablau r^3\Omega\pblin
\end{align}
The second term above in \bref{this 11 06 2021}  evaluates to
\begin{align}
    \int_{\usigmap{v}}^u d\bar{u} \int_{\usigmap{v}}^{\bar{u}}d\bar{\bar{u}}\,\nablau r^3\Omega\pblin= \int_{\usigmap{v}}^u d\bar{u} \int_{\usigmap{v}}^{\bar{u}}d\bar{\bar{u}}\,\left[\left(\mathcal{A}_2-\frac{6M}{r}\right)r\Omega^2\ablin+\frac{3\Omega^2-1}{r}r^3\Omega\pblin\right].
\end{align}
We will now show that
\begin{align}\label{laborious decay argument}
    \int_{\usigmap{v}}^u d\bar{u} \int_{\usigmap{v}}^{\bar{u}}d\bar{\bar{u}}\,\frac{3\Omega^2-1}{r}|r^3\Omega\pblin|\longrightarrow 0 \text{ as } v\longrightarrow \infty.
\end{align}
Split the integral above into
\begin{align}\label{split integral}
    \int_{\usigmap{v}}^{\usigmapt{v}} d\bar{u} \int_{\usigmap{v}}^{\bar{u}}d\bar{\bar{u}}\,\frac{3\Omega^2-1}{r}|r^3\Omega\pblin|+\int_{\usigmapt{v}}^{u} d\bar{u} \int_{\usigmap{v}}^{\bar{u}}d\bar{\bar{u}}\,\frac{3\Omega^2-1}{r}|r^3\Omega\pblin|.
\end{align}
We estimate the first integral in \bref{split integral} by
\begin{align}
\begin{split}
    \int_{\usigmap{v}}^{\usigmapt{v}} d\bar{u} \int_{\usigmap{v}}^{\bar{u}}d\bar{\bar{u}}\,\frac{3\Omega^2-1}{r}|r^3\Omega\pblin|&\leq \int_{\usigmap{v}}^{\usigmapt{v}} d\bar{u}\frac{3\Omega^2-1}{r}\times \int_{\usigmap{v}}^{\usigmapt{v}} d\bar{u}|r^3\Omega\pblin|.
    \\&\lesssim \log\left(\frac{r(\usigmap{v},v)^2}{r(\usigmapt{v},v)^2}\right)\times \int_{\usigmap{v}}^{\usigmapt{v}} d\bar{u}|r^3\Omega\pblin|.
\end{split}
\end{align}
We further have
\begin{align}
    \begin{split}
        \int_{\usigmap{v}}^{\usigmapt{v}} d\bar{u}|r^3\Omega\pblin(\bar{u},v,\theta^A)|\lesssim &\int_{\usigmap{v}}^{\usigmapt{v}} d\bar{u}|r^3\Omega\pblin(\bar{u},\vsigmap{\bar{u}},\theta^A)|\\&+\int_{J^+(\Sigma^*_+)\cap\{\bar{u}\leq u, \bar{v}\leq v\}} d\bar{u}d\bar{v}\frac{\Omega^2}{r^2}|\Psilinb(\bar{u},\bar{v},\theta^A)|.
    \end{split}
\end{align}
Since initial data on $\Sigma^*_+$ has decay of order $(1+\delta,\infty)$ with $\delta>0$, we can apply \Cref{RWrp} to get
\begin{align}
    \begin{split}
        \left[\int_{J^+(\Sigma^*_+)\cap\{\bar{u}\leq u, \bar{v}\leq v\}} d\bar{u}d\bar{v}\dw\,\frac{\Omega^2}{r^2}|\Psilinb(\bar{u},\bar{v},\theta^A)|\right]^2\lesssim \frac{1}{r^\delta}\times \int_{\Sigma^*_+}dr&\dw\,\Big[r^{1+\delta}|\nablav\Psilinb|^2\\&+\frac{1}{r^{1-\delta}}\left(|\mathring{\slashednabla}\Psilinb|^2+4|\Psilinb|^2\right)\Big],
    \end{split}
\end{align}
and in addition 
\begin{align}
    \int_{\usigmap{v}}^{\usigmapt{v}} d\bar{u}|r^3\Omega\pblin(\bar{u},\vsigmap{\bar{u}},\theta^A)|\lesssim \frac{1}{r(\usigmapt{v},\vsigmap{\log\left(\frac{v}{2M}-1\right)})^\delta}\lesssim \frac{1}{\log\left(\frac{v}{2M}-1\right)^\delta}.
\end{align}
Thus the first term in \bref{split integral} decays as $v\longrightarrow\infty$. The second term in \bref{split integral} also decays since
\begin{align}
    \begin{split}
        \int_{\usigmapt{v}}^{u} d\bar{u} \int_{\usigmap{v}}^{\bar{u}}d\bar{\bar{u}}\,\frac{3\Omega^2-1}{r}|r^3\Omega\pblin|\lesssim \log\left(\frac{r(\usigmapt{v},v)^2}{r(u,v)^2}\right)\times \int_{\usigmap{v}}^{u} d\bar{u}|r^3\Omega\pblin|,
    \end{split}
\end{align}
and since
\begin{align}
    \log\left(\frac{r(\usigmap{v},v)^2}{r(\usigmapt{v},v)^2}\right)=2\log\left(\frac{v}{v-u}\right)\longrightarrow0 \text{ as } v\longrightarrow\infty,
\end{align}
we have that the second term in \bref{split integral} also decays as $v\longrightarrow\infty$.\\

An identical argument gives us
\begin{align}
     \int_{\usigmap{v}}^u d\bar{u} \int_{\usigmap{v}}^{\bar{u}}d\bar{\bar{u}}\,\frac{6M}{r}|r\Omega^2\ablin|\;\;\longrightarrow\;\;0 \text{ as } v\longrightarrow \infty.
\end{align}
The first term in \bref{this 11 06 2021} gives
\begin{align}
    \left|\int_{\usigmap{v}}^u d\bar{u}\, (r^3\Omega\pblin)|_{(\usigmap{v},v)}\right|\leq |r^4\Omega\pblin|_{\usigmap{v}}\;\longrightarrow 0 \text{ as } v\longrightarrow\infty
\end{align}
Therefore, $\int_{\usigmap{v}}^u d\bar{u}\int_{\usigmap{v}}^{\bar{u}} d\bar{\bar{u}}\, r\ablins$ converges and we must have
\begin{align}
    \int_{\usigmap{v}}^u d\bar{u}\int_{\usigmap{v}}^{\bar{u}} d\bar{\bar{u}}\, r\ablin\,\longrightarrow\,\int_{-\infty}^ud\bar{u} \int_{-\infty}^{\bar{u}} d\bar{\bar{u}}\, \ablins_{\mathscr{I}^+}.
\end{align}
We have
\begin{align}\label{take limit for integral of xblin}
    \int_{\usigmap{v}}^u d\bar{u}\int_{\usigmap{v}}^{\bar{u}} d\bar{\bar{u}}\, r\Omega^2\ablin=\int_{\usigmap{v}}^u d\bar{u}\int_{\usigmap{v}}^{\bar{u}} d\bar{\bar{u}}\,\left[\nablau r\Omega\xblin -(2\Omega^2-1)\Omega\xblin\right].
\end{align}
Let $\underline{\mathcal{X}}(u,v,\theta^A):= \int_{\usigmap{v}}^u d\bar{u}\; |r\Omega\xblin|$. We then have
\begin{align}
    \underline{\mathcal{X}}(u,v,\theta^A)\leq \int_{\usigmap{v}}^u d\bar{u} \frac{2\Omega^2-1}{r}\underline{\mathcal{X}}(\bar{u},v,\theta^A)+\left|\int_{\usigmap{v}}^u d\bar{u}\int_{\usigmap{v}}^{\bar{u}} d\bar{\bar{u}}\;r\Omega^2\ablin\,\right|
\end{align}
Gr\"onwall's inequality implies
\begin{align}\label{this 11 06 2021 Gronwall}
     \underline{\mathcal{X}}(u,v,\theta^A) \lesssim \frac{r(\usigmap{v},v)}{r(u,v)}\times\left|\int_{\usigmap{v}}^u d\bar{u}\int_{\usigmap{v}}^{\bar{u}} d\bar{\bar{u}}\;r\Omega^2\ablin\,\right|
\end{align}
The double integral on the right hand side above in \bref{this 11 06 2021 Gronwall} converges as $v\longrightarrow\infty$, so $\underline{\mathcal{X}}$ is bounded uniformly in $v$ for any fixed $u$. We now deduce by a similar argument to the one above leading to \bref{laborious decay argument}
\begin{align}
\begin{split}
    \left|\int_{\usigmap{v}}^u d\bar{u}\int_{\usigmap{v}}^{\bar{u}} d\bar{\bar{u}}\; \frac{2\Omega^2-1}{r}r\Omega\xblin\;\right|\;\longrightarrow 0 \text{ as } v\longrightarrow\infty.
\end{split}
\end{align}
Since $\xblin|_{{\Sigma^*_+}}$ decays faster than $r^{-2}$, the expression \bref{take limit for integral of xblin} tells us that $\int_{\usigmap{v}}^u d\bar{u} \,r\Omega\xblin$ converges as $v\longrightarrow\infty$.

Now since
\begin{align}
    \int_{u(({\Sigma^*_+},v)}^u d\bar{u}\, \glinh=\int_{\usigmap{v}}^u d\bar{u}\int_{\usigmap{v}}^{\bar{u}} d\bar{\bar{u}}\; \Omega\xblin \longrightarrow 0 \text{ as } v\longrightarrow\infty, 
\end{align}
We know from \bref{metric transport in 3 direction traceless},
\begin{align}
    \nablau r\glinh=-\Omega^2\glinh+2r\Omega\xblin,
\end{align}
that the limit $\glinhs_{\mathscr{I}^+}$ of $r\glinh(u,v,\theta^A)$ exists as $v\longrightarrow\infty$, satisfying $\partial_u\glinhs_{\mathscr{I}^+}=2\xblins_{\mathscr{I}^+}$. The smoothness of $\glinhs_{\mathscr{I}^+}$ in $u$-direction follows from the smoothness of $\xblins_{\mathscr{I}^+}$, and smoothness in the angular direction follows by commuting the argument of this proposition by Lie derivatives along the generators of $SO(3)$.
\end{proof}

\begin{proposition}\label{forwards scattering for bblin at scri+}
Assume $\bblin$ arises from a solution to the system \fullsystem via \Cref{EinsteinWP} from weakly asymptotically flat initial data on ${\Sigma^*_+}$. We have that $\xblin$ defines smooth pointwise limit towards $\mathscr{I}^+$,
\begin{align}
    \bblins_{\mathscr{I}^+}(u,\theta^A)=\lim_{v\longrightarrow\infty}r^2\bblin_{\ell\geq2}(u,v,\theta^A).
\end{align}
If initial data is asymptotically flat to order $O_2(r^{-1-\delta})$, we further have that
\begin{align}
    \int_{\usigmap{v}}^u d\bar{u}\, r^2\bblin_{\ell\geq 2}
\end{align}
converges as $v\longrightarrow\infty$.
\end{proposition}

\begin{proof}
    We may repeat the arguments of \Cref{forwards scattering for xblin at scri+} and \Cref{radiation field for glinh at scri+ forward scattering} replacing $r\xblin$ with $r^2\bblin_{\ell\geq2}$ and using \bref{Bianchi-1a} instead of \bref{D3Chihatbar}.
\end{proof}

\begin{proposition}\label{forward scattering xlin weak a f}
Assume $\xblin$ arises from a solution to the system \fullsystem via \Cref{EinsteinWP} from weakly asymptotically flat initial data on $\Sigma^*_+$. Then for any finite $u$ and any $\delta>0$ we have
\begin{align}
    \lim_{v\longrightarrow\infty}r^{1+\delta}\xlin(u,v,\theta^A).
\end{align}
\end{proposition}

\begin{proof}
    From \bref{D4Chihat} we derive for any $\delta>0$
    \begin{align}
        \left|\nablav \frac{r^{1+\delta}\xlin}{\Omega}\right|^2+(1-\delta)^2\frac{\Omega^4}{r^2}\left|\frac{r^{1+\delta}\xlin}{\Omega}\right|^2+(1-\delta)\frac{\Omega^2}{r}\partial_v\left|\frac{r^{1+\delta}\xlin}{\Omega}\right|^2=|r^{1+\delta}\alin|^2.
    \end{align}
    Thus we have for $p>1$
    \begin{align}\label{06 02 2022}
        \frac{r^p}{\Omega^2} \left|\nablav \frac{r^{1+\delta}\xlin}{\Omega}\right|^2+\left(1-\delta\right)\left(2-\delta-p\right){\Omega^2}r^{p+2\delta}\left|\frac{r^{1+\delta}\xlin}{\Omega}\right|^2+(1-\delta)\partial_v\left[\frac{r^{p+1+2\delta}|\xlin|^2}{\Omega^2}\right]=r^{p+2+2\delta}|\alin|^2.
    \end{align}
    For $0<\delta<1$, choose $p=2-2\delta$, then \bref{06 02 2022} shows that $r^{1+\delta}\xlin(u,v,\theta^A)$ has a pointwise limit as $v\longrightarrow\infty$ if
    \begin{align}\label{07 02 2022}
        r^2|\alin|\in L^2\left(\mathscr{C}_u\cap J^+(\Sigma^*_+)\right).
    \end{align}
    It can be shown, repeating through the argument leading to \bref{locallabel2} in \cite{DHR16}, that \bref{07 02 2022} holds using the $r$-weighted estimate \bref{RW rp estimate} with $p=\delta$, which in turn holds provided the initial data has decay of order $O(r^{-\frac{1}{2}-\delta})$.
\end{proof}

\begin{corollary}\label{r eblins decays at scri+ weakly a f}
Assume $\eblin$ arises from a solution to the system \fullsystem via \Cref{EinsteinWP} from weakly asymptotically flat initial data on $\Sigma^*_+$. Then for any finite $u$ we have
\begin{align}
    \lim_{v\longrightarrow\infty}r\eblin(u,v,\theta^A)=0.
\end{align}
\end{corollary}
\begin{proof}
    This follows from applying \Cref{forwards scattering for xblin at scri+}, \Cref{forward scattering xlin weak a f} to equation \bref{D4Chihatbar}.
\end{proof}
\begin{proposition}\label{forwards scattering of xlin at scri+}
Assume $\mathfrak{S}$ is a solution to the system \fullsystem arising via \Cref{EinsteinWP} from asymptotically flat initial data on ${\Sigma^*_+}$. We have that the component $\xlin$ defines smooth pointwise limit towards $\mathscr{I}^+$ via
\begin{align}
    \xlins_{\mathscr{I}^+}(u,\theta^A)=\lim_{v\longrightarrow\infty}r^2\xlin(u,v,\theta^A).
\end{align}
\end{proposition}

\begin{proof}
For fixed $u$, equation \bref{D4Chihat} tells us that for sufficiently small $\epsilon$ we have
\begin{align}\label{this 11 06 2021 2}
    \int_{\mathscr{C}_u\cap\{\bar{v}\geq v\}} d\bar{v} \sin\theta d\theta d\phi\, r^{2-\epsilon}\left|\nablav \frac{r^2\xlin}{\Omega}\right|^2= \int_{\mathscr{C}_u\cap\{\bar{v}\geq v\}} d\bar{v} \sin\theta d\theta d\phi\, r^{6-\epsilon}|\alin|^2.
\end{align}
As the right hand side of \bref{this 11 06 2021 2} is bounded by \bref{locallabel2} due to the decay rates given by the asymptotic flatness of $\mathfrak{D}$, we may conclude by repeating the argument \bref{argument for radiation field} with the left hand side of \bref{this 11 06 2021 2}. We may repeat the same argument commuting arbitrary powers of $\slashednabla_t$, $r\nablav$ and $\mathring{\slashednabla}$ to conclude that $\xlins_{\mathscr{I}^+}$ is smooth and that $\nablau^k\mathring{\slashednabla}^\gamma r^2\xlin\longrightarrow \partial_u^k\mathring{\slashednabla}^\gamma \xlins_{\mathscr{I}^+}$ for any integer $k$ and index $\gamma$.
\end{proof}

\begin{corollary}\label{forward scattering nablau xlin weakly a f}
Assume $\xlin$ arises from a solution to the system \fullsystem via \Cref{EinsteinWP} from weakly asymptotically flat initial data on $\Sigma^*_+$. Then $\nablau r^2\xlin(u,v,\theta^A)$ has a pointwise limit as $v\longrightarrow\infty$ that defines an element of $\Gamma^{(2)}(\mathscr{I}^+)$.
\end{corollary}
\begin{proof}
    Under the assumption of weak asymptotic flatness, we may commute the argument of \Cref{forwards scattering of xlin at scri+} by $\partial_t$ and use the fact that $\partial_t\alin|_{\Sigma^*_+}$ has the appropriate decay rate.
\end{proof}

\begin{corollary}\label{forward scattering Olino elin at scri+ weak a f}
For a solution $\mathfrak{S}$ to \fullsystem arising from weakly asymptotically flat initial data on $\Sigma^*_+$, we have that $\elin(u,v,\theta^A)$ and $\Olino(u,v,\theta^A)$ define smooth radiation fields at $\mathscr{I}^+$ via
\begin{align}
    \elins_{\mathscr{I}^+}(u,\theta^A):=\lim_{v\longrightarrow\infty} r\elin_{\ell\geq2}(u,v,\theta^A),\qquad \Olinos(u,\theta^A):=\lim_{v\longrightarrow\infty}\Olino_{\ell\geq2}(u,v,\theta^A).
\end{align}
\end{corollary}
\begin{proof}
    This limit can be obtained by applying \Cref{forwards scattering for xblin at scri+} and \Cref{forward scattering nablau xlin weakly a f}. The second limit follows from the existence of $\elins_{\mathscr{I}^+}$ and applying \Cref{r eblins decays at scri+ weakly a f} to \bref{elin eblin Olin}.
\end{proof}

\begin{corollary}\label{forwards scattering of elin eblin at scri+}
Assume $\mathfrak{S}$ is a solution to the system \fullsystem arising via \Cref{EinsteinWP} from asymptotically flat initial data on ${\Sigma^*_+}$. We have that $\elin_{\ell\geq 2}$, $\eblin_{\ell\geq 2}$ define smooth pointwise limit towards $\mathscr{I}^+$ via
\begin{align}
    \eblins_{\mathscr{I}^+}(u,\theta^A)=\lim_{v\longrightarrow\infty}r^2\eblin_{\ell\geq 2}(u,v,\theta^A).
\end{align}
Moreover, we have that
\begin{align}
    \int_{\usigmap{v}}^u d\bar{u}\, r\elin_{\ell\geq 2}
\end{align}
converges as $v\longrightarrow\infty$.
\end{corollary}

\begin{proof}
The convergence of $r\elin_{\ell\geq2}$ and its integral in the $u$-direction follows from applying \Cref{forwards scattering for xblin at scri+,,forwards scattering of xlin at scri+} to equation \bref{D3Chihat}. We then get the convergence of $r^2\eblin$ by integrating \bref{D3etabar} in the $u$-direction.
\end{proof}

\subsubsection{Radiation properties of the remaining components}

\begin{corollary}\label{forwards scattering of rlin slin at scri+}
Assume $\mathfrak{S}$ is a solution to the system \fullsystem arising via \Cref{EinsteinWP} from weakly asymptotically flat initial data on ${\Sigma^*_+}$. We have that for any $u$ and any $\delta>0$, $r^{2+\delta}\rlin_{\ell\geq 2}(u,v,\theta^A)$ and  $r^{2+\delta}\slin_{\ell\geq 2(u,v,\theta^A)}$ are decaying as $v\longrightarrow\infty$. If the initial data is asymptotically flat then $\rlin$, $\slin$ define smooth pointwise limit towards $\mathscr{I}^+$ via
\begin{align}
    \rlins_{\mathscr{I}^+}(u,\theta^A)=\lim_{v\longrightarrow\infty}r^3\rlin_{\ell\geq 2}(u,v,\theta^A),\qquad\qquad \slins_{\mathscr{I}^+}(u,\theta^A)=\lim_{v\longrightarrow\infty}r^3\slin_{\ell\geq 2}(u,v,\theta^A)
\end{align}
\end{corollary}

\begin{proof}
This follows from the properties of $\Psilinb$, $\xblin$ and $\xlin$ and their radiation fields given by \Cref{weak boundedness of Psilin weak a f},  \Cref{forwards scattering for xblin at scri+} and \Cref{forward scattering xlin weak a f}.
\end{proof}

\begin{corollary}\label{forwards scattering of olinos bmlins at scri+}
Assume $\mathfrak{S}$ is a solution to the system \fullsystem arising via \Cref{EinsteinWP} from asymptotically flat initial data on ${\Sigma^*_+}$. We have that $\bmlin_{\ell\geq 2}$ defines a smooth pointwise limit towards $\mathscr{I}^+$ via
\begin{align}
    \bmlins_{\mathscr{I}^+}(u,\theta^A):=\lim_{v\longrightarrow\infty}r\bmlin_{\ell\geq 2}(u,v,\theta^A).
\end{align}
\end{corollary}
\begin{proof}
This is an immediate consequence of applying \Cref{forwards scattering of xlin at scri+} and \Cref{radiation field for glinh at scri+ forward scattering} to equations \bref{partial_u b} and \bref{D3Chihat}.

\end{proof}

\begin{corollary}\label{forwards scattering of otx otxb at scri+}
If $\mathfrak{S}$ is a solution to the system \fullsystem arising via \Cref{EinsteinWP} from weakly asymptotically flat initial data on ${\Sigma^*_+}$, We have that $r^{1+\delta}\otx_{\ell\geq2}(u,v,\theta^A)\longrightarrow0$ as $v\longrightarrow\infty$ for any finite $u$ and any $\delta\in[0,1)$. Furthermore, $\otxb_{\ell\geq 2}$ defines a smooth pointwise limit towards $\mathscr{I}^+$ via
\begin{align}\label{defin of radiation field of otx otxb}
     \otxbs_{\mathscr{I}^+}(u,\theta^A):=\lim_{v\longrightarrow\infty}r\otxb_{\ell\geq 2}(u,v,\theta^A),
\end{align}
thus $\tr\glin(u,v,\theta^A)\longrightarrow0$ as $v\longrightarrow\infty$ for any finite $u$. If the initial data of $\mathfrak{S}$ is assumed to be asymptotically flat, we can define a radiation field for $\otx_{\ell\geq2}$ via
\begin{align}
    \otxs_{\mathscr{I}^+}(u,\theta^A):=\lim_{v\longrightarrow\infty}r^2\otx_{\ell\geq 2}(u,v,\theta^A),
\end{align}
 and moreover we have that
\begin{align}
    \int_{\usigmap{v}}^u d\bar{u}\, r\otxb_{\ell\geq 2}
\end{align}
converges as $v\longrightarrow\infty$. In particular, the component $\tr\glin$ defines a smooth radiation field $\tr\glins_{\mathscr{I}^+}$ via
\begin{align}
    \tr\glins_{\mathscr{I}^+}(u,\theta^A)=\lim_{v\longrightarrow\infty}r\tr\glin(u,v,\theta^A).
\end{align}
\end{corollary}

\begin{proof}
The convergence of $r\otxb$ and its integral in the $u$-direction follows from applying \Cref{forwards scattering for xblin at scri+} and \Cref{forwards scattering for bblin at scri+}, \Cref{forwards scattering of elin eblin at scri+}. When data is asymptotically flat, we have that $r^3\rlin$ is bounded uniformly on $\mathscr{D}=J^+({\Sigma^*_+})\cap r>R\cap\{u\geq u_+\}$ for any fixed $R, u_+$, thus we may apply the argument leading to \bref{laborious decay argument} to deduce
\begin{align}\label{model argument for unsaturated r weights}
    \int_{\usigmap{v}}^u d\bar{u} \,|r^2\rlin|\,\longrightarrow 0 \text{ as } v\longrightarrow\infty.
\end{align}
The same applies to the $u$ integrals of $r\rlin$, $\Olino$ and $\otxb$ from ${\Sigma^*_+}$ to any fixed $u$. Integrating \bref{Bianchi-1b} in $u$, applying the above and using the fact that $\big|\blin|_{{\Sigma^*_+}}\big|=O\left(\frac{1}{r^{3+\delta}}\right)$ with $\delta>0$ gives us that $r^2\blin$ decays towards $\mathscr{I}^+$ along $\mathscr{C}_u$ for any fixed $u$. We can then conclude from the Codazzi equation \bref{elliptic equation 2} that
\begin{align}
    \int_{\usigmap{v}}^u d\bar{u}\, r\otx_{\ell\geq 2}\longrightarrow 0 \text{ as } v\longrightarrow\infty. 
\end{align}
The equation \bref{D3TrChi} gives the expression
\begin{align}
    \nablau r^2\otx=2\Omega^2\left(\divo r\elin+r^2\rlin-\frac{4M}{r}\Olin\right)-r\Omega^2(\otx+\otxb),
\end{align}
which we integrate in $u$ to conclude that $r^2\otx_{\ell\geq2}$ converges as $v\longrightarrow\infty$. The existence of $\tr\glins_{\mathscr{I}^+}$ follows from integrating \bref{metric transport in 3 direction trace} in $u$ and applying the above.
\end{proof}

Applying the above results to the linearised Gauss equation \bref{Gauss}, we arrive at the following corollary:

\begin{corollary}
For a solution $\mathfrak{S}$ to the system \fullsystem arising from weakly asymptotically flat initial data on $\Sigma^*_+$, we have that the limit
\begin{align}
    \lim_{v\longrightarrow\infty}r^2\Klin(u,v,\theta^A)
\end{align}
exists and is constant in $u$. If initial data is asymptotically flat then the pointwise limit of $r^3\Klin(u,v,\theta^A)$ as $v\longrightarrow\infty$ exists and defines an element of $\Gamma^{(0)}(\mathscr{I}^+)$.
\end{corollary}

\subsection{Radiation fields on $\mathscr{H}^+_{\geq0}$}\label{Section 9.3 radiation at H+}

We now define the radiation fields and compute the radiation flux at $\mathscr{H}^+$.

\begin{corollary}
Let $\mathfrak{S}$ be a solution to the system \fullsystem arising from smooth initial data $\mathfrak{D}$ on ${\Sigma^*_+}$ via \Cref{EinsteinWP}. The following quantities define smooth fields on $\mathscr{H}^+_{\geq0}$
\begin{align*}
\begin{split}
    &\glinh,\;\bmlin,\;\Olin,\;\tr\glin,\\
        \Omega\xlin,\;\Omega^{-1}\xblin,\;&\elin,\;\eblin,\;\otx,\;\otxb,\;\olin,\;\olinb.\\
        \Omega^2\alin,\;&\Omega^{-2}\ablin,\;\Omega\blin,\;\Omega^{-1}\bblin,\;\rlin,\;\slin
\end{split}
\end{align*}
\end{corollary}

\begin{defin}\label{definition of radiation fields at H+}
Let $\mathfrak{S}$ be a solution to the system \fullsystem arising from compactly supported initial data in the sense of \Cref{EinsteinWP}. Define
\begin{alignat}{2}
    &\alins_{\mathscr{H}^+}:=\Omega^2\alin\Big|_{\overline{\mathscr{H}^+}}\;\;,\qquad\qquad\qquad&&\ablins_{\mathscr{H}^+}:=\Omega^{-2}\ablin\Big|_{\overline{\mathscr{H}^+}}\;\;,\\
    &\blins_{{\mathscr{H}^+}}:=\Omega\blin\Big|_{\overline{\mathscr{H}^+}}\;\;,\qquad\qquad\qquad&&\bblins_{\mathscr{H}^+}:=\Omega^{-1}\bblin\Big|_{\overline{\mathscr{H}^+}}\;\;,\\
    &\rlins_{{\mathscr{H}^+}}:=\rlin_{\ell\geq2}\Big|_{\overline{\mathscr{H}^+}}\;\;,\qquad\qquad\qquad&&\slins_{\mathscr{H}^+}:=\slin_{\ell\geq2}\Big|_{\overline{\mathscr{H}^+}}\;\;,\\
   &\xlins_{\mathscr{H}^+}:=\Omega\xlin\Big|_{\overline{\mathscr{H}^+}}\;\;,\qquad\qquad\qquad&&\xblins_{\mathscr{H}^+}:=\Omega^{-1}\xblin\Big|_{\overline{\mathscr{H}^+}}\;\;,\\
    &\elins_{\mathscr{H}^+}:=\elin_{\ell\geq2}\Big|_{\overline{\mathscr{H}^+}}\;\;,\qquad\qquad\qquad&&\eblins_{\mathscr{H}^+}:=\eblin_{\ell\geq2}\Big|_{\overline{\mathscr{H}^+}}\;\;,\\
    &\otxs_{\mathscr{H}^+}:=\otx_{\ell\geq2}\Big|_{\overline{\mathscr{H}^+}}\;\;,\qquad\qquad\qquad&&\otxbs_{\mathscr{H}^+}:=\otxb_{\ell\geq2}\Big|_{\overline{\mathscr{H}^+}}\;\;,\\
    &\olins_{\mathscr{H}^+}:=\olin_{\ell\geq2}\Big|_{\overline{\mathscr{H}^+}}\;\;,\qquad\qquad\qquad&&\olinbs_{\mathscr{H}^+}:=\olinb_{\ell\geq2}\Big|_{\overline{\mathscr{H}^+}}\;\;,\\
    &\glinhs_{\mathscr{H}^+}:=\glinh\Big|_{\overline{\mathscr{H}^+}}\;\;,\qquad\qquad\qquad&&\bmlins_{\mathscr{H}^+}:=\bmlin_{\ell\geq2}\Big|_{\overline{\mathscr{H}^+}}\;\;,\\
    &\tr\glins_{\mathscr{H}^+}:=\tr\glin_{\ell\geq2}\Big|_{\overline{\mathscr{H}^+}}\;\;,\qquad\qquad\qquad&&\Olinos_{\mathscr{H}^+}:=\Olin_{\ell\geq2}\Big|_{\overline{\mathscr{H}^+}}\;\;.
\end{alignat}
Additionally, we define
\begin{align}
    &\upPsilin_{\mathscr{H}^+}:=\Psilin|_{\mathscr{H}^+_{\geq0}},\qquad\qquad\qquad \plins_{\mathscr{H}^+}:=\Omega\plin|_{\mathscr{H}^+_{\geq0}},\\
    &\upPsilinb_{\mathscr{H}^+}:=\Psilinb|_{\mathscr{H}^+_{\geq0}},\qquad\qquad\qquad \pblins_{\mathscr{H}^+}:=\Omega^{-1}\pblin|_{\mathscr{H}^+_{\geq0}}.
\end{align}
\end{defin}

\begin{proposition}
Let $\mathfrak{S}$, $\xlins_{\mathscr{H}^+}$, $\alins_{\mathscr{H}^+}$ be as in \Cref{definition of radiation fields at H+} and assume $\mathfrak{S}$ arises from asymptotically flat initial data on ${\Sigma^*_+}$ with the gauge conditions \bref{initial horizon gauge condition} satisfied. Then $\xlins_{\mathscr{H}^+}\in \mathcal{E}^{T}_{\mathscr{H}^+_{\geq0}}$ and 
\begin{align}
    \|\xlins_{\mathscr{H}^+}\|^2_{\mathcal{E}^T_{\mathscr{H}^+_{\geq0}}}=\|\alins_{\mathscr{H}^+}\|^2_{\mathcal{E}^{T,+2}_{\mathscr{H}^+_{\geq0}}}.
\end{align}
\end{proposition}

\begin{proof}
Integrating \bref{D4Chihat} restricted to $\overline{\mathscr{H}^+}\cap\{v\in[v_1,v_2]\}$ gives
\begin{align}\label{Sweet Dreams}
    \frac{1}{2M}\int_{v_1}^{v_2}dv\,e^{-\frac{1}{2M}(v_1-\bar{v})}\alins_{\mathscr{H}^+}=-e^{\frac{1}{2M}v_1}\int_{v_1}^{v_2}dv\,\partial_v\left(e^{-\frac{1}{2M}\bar{v}}\xlins_{\mathscr{H}^+}\right)=\xlins_{\mathscr{H}^+}(v_1,\theta^A)-e^{-\frac{1}{2M}(v_2-v_1)}\xlins_{\mathscr{H}^+}(v_2,\theta^A).
\end{align}
The estimate \bref{L2 in v polynomial decay estimate on H+} commuted with $\partial_v$ and $\mathring{\slashednabla}^\gamma$ show via a Sobolev estimate on $S^2$ that $\xlins_{\mathscr{H}^+}=O(|v|^{-1})$ as $v\longrightarrow\infty$. Thus taking the limit $v_2\longrightarrow\infty$ in \bref{Sweet Dreams} gives
\begin{align}
    \frac{1}{2M}\int_{v_1}^{v_2}dv\,e^{-\frac{1}{2M}(v_1-\bar{v})}\alins_{\mathscr{H}^+}=\xlins_{\mathscr{H}^+}.
\end{align}
\end{proof}

\subsubsection[Boundedness estimates on $\protect\bmlin|_{\mathscr{H}^+_{\geq0}}$]{Estimates on $\Olin|_{\mathscr{H}^+_{\geq0}}$, $\bmlin|_{\mathscr{H}^+_{\geq0}}$}

We now derive estimates on the lapse and shift at $\mathscr{H}^+_{\geq0}$ assuming \underline{only} the horizon gauge conditions \bref{initial horizon gauge condition} and asymptotic flatness to order $(2,\infty)$. This will be important later when deriving estimates on gauge solutions used to pass to a future horizon-normalised gauge. 

Note that, having shown that $r\xblin$ converges at $\mathscr{I}^+$ in \Cref{forwards scattering for xblin at scri+} for solutions arising from asymptotically flat initial data, we can in fact prove an integrated decay estimate for $\Omega\xblin$:
\begin{proposition}\label{spacetime integral of xblin proposition}
Let $\mathfrak{S}$ be a solution to \fullsystem arising from asymptotically flat initial data on $\Sigma^*_+$. Then we have
 \begin{align}\label{spacetime integral of xblin estimate}
 \begin{split}
        \int_{J^+(\Sigma^*_+)}d\bar{u}d\bar{v}\dw\,&\Omega^2|\xblin|^2\\&\lesssim \mathbb{F}^2_{\Sigma^*}[\mathfrak{S}]+\int_{\Sigma^*_+}dr\dw\,r^2|\nablav\Psilinb|^2+|r^3\Omega\mathcal{A}_2\pblin|^2+|r\Omega^2\mathcal{A}_2\ablin|^2+|r^2\mathcal{A}_2\xblin|^2.
    \end{split}
    \end{align}
\end{proposition}
\begin{proof}
    Recall that the quantity $\Ylin$ satisfies 
    \begin{align}
        \nablau\nablav\Ylin=\frac{\Omega^2}{r^2}\left[\Psilinb-\frac{6M}{r}r^3\Omega\mathcal{A}_2\pblin+\frac{12M^2\Omega^2}{r^2}r\Omega^2\mathcal{A}_2\ablin\right],
    \end{align}
    which implies
    \begin{align}
        \nablau |\nablav\Ylin|^2=2\nablav\Ylin\cdot\frac{\Omega^2}{r^2}\left[\Psilinb-\frac{6M}{r}r^3\Omega\mathcal{A}_2\pblin+\frac{12M^2\Omega^2}{r^2}r\Omega^2\mathcal{A}_2\ablin\right],
    \end{align}
    \begin{align}
        \nablau r\Omega^{-2}|\nablav\Ylin|^2+(2-\Omega^{-2})|\nablav\Ylin|^2=2\nablav\Ylin\cdot\frac{\Omega^2}{r}\left[\Psilinb-\frac{6M}{r}r^3\Omega\mathcal{A}_2\pblin+\frac{12M^2}{r^2}r\Omega^2\mathcal{A}_2\ablin\right],
    \end{align}
    Integrate the above over $r\geq 4M$ and apply Cauchy--Schwartz to find
    \begin{align}\label{29 01 2022}
    \begin{split}
        &\int_{r=4M\cap{J^+(\Sigma^*_+)}}d\bar{t}\dw\,|\nablav\Ylin|^2+\int_{J^+(\Sigma^*_+)}d\bar{u}d\bar{v}\dw\,\Omega^2|\nablav\Ylin|^2\\&\lesssim \int_{J^+(\Sigma^*_+)\cap \{r\geq 4M\}}d\bar{u}d\bar{v}\dw\,\frac{\Omega^2}{r^2}\left[|\Psilinb|^2+|r^3\Omega\mathcal{A}_2\pblin|^2+|r\Omega^2\mathcal{A}_2\ablin\right]+\int_{\Sigma^*_+}dr\dw\,|\nablav\Ylin|^2.
        \\&\lesssim \mathbb{F}^2_{\Sigma^*}[\mathfrak{S}]+\int_{\Sigma^*_+}r^2|\nablav\Psilinb|^2+|r^3\Omega\mathcal{A}_2\pblin|^2+|r\Omega^2\mathcal{A}_2\ablin|^2+|r^2\mathcal{A}_2\xblin|^2.
    \end{split}
    \end{align}
    A Hardy estimate together with \Cref{L2 estimates of ablin pblin} imply
    \begin{align}
        \int_{J^+(\Sigma^*)\cap \{r\geq 4M\}}d\bar{u}d\bar{v}\dw\,|\xblin|^2\lesssim \text{ Right hand side of }\bref{29 01 2022}.
    \end{align}
    In the region $r\leq 4M$ we integrate the equation
    \begin{align}
        \nablav r^2\Omega\xblin+\frac{4M}{r^2}|r^2\Omega\xblin|^2=-2r^2\ablin\cdot r^2\Omega\xblin,
    \end{align}
    and apply Cauchy--Schwartz and \bref{29 01 2022} to obtain
    \begin{align}
        \int_{J^+(\Sigma^*_+)\cap\{r\leq 4M\}}d\bar{u}d\bar{v}\dw\,\Omega^2|\xblin|^2\lesssim \text{ Right hand side of }\bref{29 01 2022}.
    \end{align}
\end{proof}
Combining \bref{spacetime integral of xblin estimate} and \bref{29 01 2022} immediately leads to 
\begin{corollary}\label{spacetime integral of dv xblin proposition}
Under the assumptions of \Cref{spacetime integral of xblin proposition}, we have
 \begin{align}\label{spacetime integral of dv xblin estimate}
        \int_{J^+(\Sigma^*_+)}d\bar{u}d\bar{v}\dw\,|\nablav r\Omega\xblin|^2\lesssim \mathbb{F}^2_{\Sigma^*}[\mathfrak{S}]+\int_{\Sigma^*_+}r^2|\nablav\Psilinb|^2+|r^3\Omega\mathcal{A}_2\pblin|^2+|r\Omega^2\mathcal{A}_2\ablin|^2+|r^2\mathcal{A}_2\xblin|^2.
    \end{align}
\end{corollary}
\begin{corollary}
For a solution $\mathfrak{S}$ arising from asymptotically flat initial data on $\Sigma^*_+$, the quantity $\bmlin$ satisfies
\begin{align}
    &\int_{\mathscr{H}^+_{\geq0}}d\bar{v}\dw\,|\bmlins_{\mathscr{H}^+}|^2+\int_{J^+(\Sigma^*_+)}d\bar{u}d\bar{v}\dw\,\frac{\Omega^2}{r^2}|\bmlin|^2\\&\lesssim \mathbb{F}^2_{\Sigma^*}[\mathfrak{S}]+\int_{\Sigma^*_+}r^2|\nablav\Psilinb|^2+|r^3\Omega\mathcal{A}_2\pblin|^2+|r\Omega^2\mathcal{A}_2\ablin|^2+|r^2\mathcal{A}_2\xblin|^2+r^{8-\epsilon}|\Omega\plin|^2+r^{6-\epsilon}|\Omega^2\alin|^2\\&+\int_{{\Sigma^*_+}}\Omega^2du\sin\theta d\theta d\phi\Bigg[\left|\frac{1}{\Omega}\nablagml\left(\frac{1}{\Omega}\nablagml(r^2\Omega\xlin)\right)\right|^2+\left|\frac{1}{\Omega}\nablagml(r^2\Omega\xlin)\right|^2+\frac{1}{r^\epsilon}|r^2\Omega\xlin|^2\Bigg]
\end{align}
\end{corollary}
\begin{proof}
    Equation \bref{partial_u b} implies
    \begin{align}
        \nablau r^{-1}|\bmlin|^2+\frac{\Omega^2}{r^2}|\bmlin|^2=4\Omega^2(\elin-\eblin)\cdot r^{-1}\bmlin.
    \end{align}
    Integrating the above over the whole of $J^+(\Sigma^*_+)$ and using Cauchy--Schwartz gives
    \begin{align}
        \int_{\mathscr{H}^+_{\geq0}}d\bar{v}\dw\,|\bmlins_{\mathscr{H}^+}|^2+\int_{J^+(\Sigma^*_+)}d\bar{u}d\bar{v}\dw\,\frac{\Omega^2}{r^2}|\bmlin|^2\lesssim \int_{J^+(\Sigma^*_+)}d\bar{u}d\bar{v}\dw\,\Omega^2\left[|\elin|^2+|\eblin|^2\right].
    \end{align}
    We may estimate the right hand side above using equations \bref{D3Chihat}, \bref{D4Chihatbar} and the estimates of \Cref{ILED on xlin},  \Cref{spacetime integral of xblin proposition}, and \Cref{spacetime integral of dv xblin proposition}.
\end{proof}

\begin{remark}
Recall that in the $\Sigma^*_+$ and $\overline{\Sigma}^+$-gauges the ingoing shear $\xblin$ is proportional to $\slin$ and its $\nablav$-derivative, thus it can be estimated in terms of $\Psilin-\Psilinb$.
\end{remark}

\subsection{Passing to a future Bondi-normalised gauge in forward scattering}\label{Section 10 Passing to Bondi gauge}

We now construct the gauge solutions $\mathfrak{G}_{\mathscr{I}^+}$ and $\mathfrak{G}_{\mathscr{H}^+_{\geq0}}$ of \ref{forward scattering full system thm}.

We show in this section that given a solution $\mathfrak{S}$ to \fullsystem arising from asymptotically flat initial data (thus in particular satisfying the results of Sections \ref{Section 9.1 forward scattering estimates}--\ref{Section 9.3 radiation at H+}), we can find a pure gauge solution $\mathfrak{G}_{\mathscr{I}^+}$, constructed via \Cref{inwards gauge solutions}, such that $\mathfrak{S}+\mathfrak{G}_{\mathscr{I}^+}$ satisfies the conditions of the Bondi gauge specified in \Cref{future Bondi gauge}.

\subsubsection{Early and late time asymptotics for $\mathfrak{S}$ at $\mathscr{I}^+$}\label{early and late time asymptotics section}

We start with some decay estimates on the radiation fields $\xblins_{\mathscr{I}^+}$, $\xlins_{\mathscr{I}^+}$, $\Olinos_{\mathscr{I}^+}$, $\bmlins_{\mathscr{I}^+}$ on $\mathscr{I}^+$ as $u\longrightarrow\pm\infty$.

\begin{proposition}\label{ingredient for gauge asymptotic flatness}
The radiation fields of the solution $\mathfrak{S}$ to \fullsystem arising via \Cref{proposition of asymptotic flatness at null infinity} satisfy
\begin{align}\label{decay of xblin on scri+ before bonfi gaugification}
    \partial_u \xlins_{\mathscr{I}^+}, \xblins_{\mathscr{I}^+}, \bblins_{\mathscr{I}^+}=O\left(\frac{1}{|u|^{\gamma}}\right),\qquad \ablins_{\mathscr{I}^+}=O\left(\frac{1}{|u|^{\gamma+1}}\right),
\end{align}
as $u\longrightarrow-\infty$, where the initial data on $\Sigma^*_+$ is assumed to decay to order $(\gamma,\infty)$ for $\gamma>1$.
\end{proposition}
\begin{proof}
For a given $u$, we estimate
\begin{align}
     \int_{S^2}d\omega\,\left|\pblins_{\mathscr{I}^+}(u,\theta^A)-r^3\Omega\pblin(u,v_{({\Sigma^*_+},u)},\theta^A)\right|^2&\leq \frac{1}{r(u,v_{{\Sigma^*_+},u})}\int_{\mathscr{C}_u\cap J^+({\Sigma^*_+})}d\bar{v}d\omega\frac{\Omega^2}{r^2}|\Psilinb|^2\\
     &\leq\frac{1}{r(u,v_{{\Sigma^*_+},u})}\,\mathbb{F}_{{\Sigma^*_+}\cap\{v\geq r(u,v_{{\Sigma^*_+},u})\}}[\Psilinb]
     \\&\lesssim  \frac{1}{r(u,v_{{\Sigma^*_+},u})^{2\gamma}},
\end{align}
thus
\begin{align}
    \int_{S^2}d\omega\,|\pblins_{\mathscr{I}^+}(u,\theta^A)|^2\lesssim  \frac{1}{r(u,v_{{\Sigma^*_+},u})^{2\gamma}}+\int_{S^2}d\omega\,\left|r^3\Omega\pblin(u,v_{({\Sigma^*_+},u)},\theta^A)\right|^2 \lesssim  \frac{1}{r(u,v_{{\Sigma^*_+},u})^{2\gamma}}.
\end{align}
Similarly,
\begin{align}
\begin{split}
    \int_{S^2}d\omega\,\Big|\ablins_{\mathscr{I}^+}(u,\theta^A)&-r\Omega^2\ablin(u,v_{({\Sigma^*_+},u)},\theta^A))\Big|^2\leq \frac{1}{r(u,v_{{\Sigma^*_+},u})}\int_{\mathscr{C}_u\cap J^+({\Sigma^*_+})}d\bar{v}d\omega \frac{\Omega^2}{r^2}|r^3\Omega\pblin|^2,\\
    &\leq \frac{1}{r(u,v_{{\Sigma^*_+},u})}\int_{\mathscr{C}_u\cap J^+({\Sigma^*_+})}d\bar{v}d\omega \frac{\Omega^4}{r^4}|\Psilinb|^2+\frac{1}{r(u,\vsigmap{u})^2}\int_{S^2}d\omega\,|\pblins_{\mathscr{I}^+}|^2\\
    &\leq \frac{1}{r(u,v_{{\Sigma^*_+},u})^3}\,\mathbb{F}_{{\Sigma^*_+}\cap\{v\geq r(u,v_{{\Sigma^*_+},u})\}}[\Psilinb]+\frac{1}{r(u,\vsigmap{u})^2}\int_{S^2}d\omega\,|\pblins_{\mathscr{I}^+}|^2\\
    &\lesssim \frac{1}{r(u,v_{{\Sigma^*_+},u})^{2+2\gamma}},
\end{split}
\end{align}
where we used throughout the asymptotics $\partial_v^n\ablin|_{{\Sigma^*_+}}=O\left(\frac{1}{r^{\gamma+2+n}}\right)$ which follow from the assumption of asymptotic flatness on $\mathfrak{D}$. Commuting the above with up to $2$ angular derivatives and using a Sobolev embedding, we find that
\begin{align}\label{polynomial decay of ablins at scri+-}
    \ablins_{\mathscr{I}^+}(u,\theta^A)=O\left(\frac{1}{r(u,v_{{\Sigma^*_+},u})^{1+\gamma}}\right)=O\left(\frac{1}{|u|^{1+\gamma}}\right)
\end{align}
as $u\longrightarrow-\infty$. Taking the limit towards $\mathscr{I}^+$ on the left hand side of \bref{xblins decays pointwise near scri+-}, we see that $\xblins_{\mathscr{I}^+}\longrightarrow0$ as $u\longrightarrow-\infty$, thus \bref{polynomial decay of ablins at scri+-} tells us that $\xblins_{\mathscr{I}^+}=O\left(\frac{1}{|u|^{\gamma}}\right)$  as $u\longrightarrow-\infty$. An identical argument works for $\bblins_{\mathscr{I}^+}$, applying the argument above to \bref{Bianchi-1a} instead of \bref{D3Chihatbar}.\\

Finally, for $\nablau \xlins_{\mathscr{I}^+}$ we estimate using \bref{locallabel1}, \bref{locallabel2}:
\begin{align}
\begin{split}
    \int_{S^2}d\omega\,|\partial_u\xlins_{\mathscr{I}^+}(u,\theta^A)-\nablau \frac{r^2\xlin}{\Omega}(u,v_{{\Sigma^*_+},u},\theta^A)|^2&\leq\int_{S^2}d\omega\,\left|\int_{v_{{\Sigma^*_+},u}}^\infty dv\, \nablau r^2\alin \right|^2\\
    &\leq \int_{S^2}\left[\int_{v_{{\Sigma^*_+},u}}^\infty dv |r\Omega^2\alin|+r\Omega|\plin|\right]^2\\
    &\leq \frac{1}{r(u,v_{{\Sigma^*_+},u})}\left[\int_{\mathscr{C}_u\cap J^+({\Sigma^*_+})}drd\omega\,r^{4}|\Omega^2\alin|^2+r^{5}|\Omega\plin|^2\right]\\&\lesssim\frac{1}{r(u,v_{{\Sigma^*_+},u})}\Bigg[\mathbb{F}_{\Sigma^*_+\cap\{r\geq r(u,\vsigmap{u})\}}[\Psilin]\\&\qquad+\int_{{\Sigma^*_+}\cap v\geq v_{{\Sigma^*_+},u}}drd\omega\,r^{4}|\Omega^2\alin|^2+r^{5}|\Omega\plin|^2\Bigg].
\end{split}
\end{align}
Note that since $\alin|_{{\Sigma^*_+}}=O\left(r^{-(2+\gamma)}\right)$, $\plin|_{{\Sigma^*_+}}=O\left(r^{-(3+\gamma)}\right)$,
\begin{align}
    &\int_{{\Sigma^*_+}\cap v\geq v_{{\Sigma^*_+},u}}drd\omega\,r^{5}|\Omega\plin|^2\lesssim\frac{1}{r^{2\gamma}},\\
    &\int_{{\Sigma^*_+}\cap v\geq v_{{\Sigma^*_+},u}}drd\omega\,r^{4}|\Omega^2\alin|^2\lesssim\frac{1}{r^{2\gamma-1}}.
\end{align}
Thus
\begin{align}
     \int_{S^2}d\omega\,\left|\partial_u\xlins_{\mathscr{I}^+}(u,\theta^A)-\nablau \frac{r^2\xlin}{\Omega}(u,v_{{\Sigma^*_+},u},\theta^A) \right|^2= O\left(\frac{1}{r(u,\vsigmap{u})^{2\gamma}}\right).
\end{align}
Commuting with angular derivatives up to 2\textsuperscript{nd} order, a Sobolev embedding says that $\partial_u\xlins_{\mathscr{I}^+}(u,\theta^A)-\nablau \frac{r^2\xlin}{\Omega}(u,v_{{\Sigma^*_+},u},\theta^A)=O\left(|u|^{-\gamma}\right)$. The result follows knowing that\\
$\nablau \frac{r^2\xlin}{\Omega}(u,v_{{\Sigma^*_+},u},\theta^A)=O\left(r(u,v_{{\Sigma^*_+},v})^{-\gamma}\right)=O\left(|u|^{-\gamma}\right)$ as $u\longrightarrow-\infty$. 
\end{proof}

\begin{remark}\label{xblins estimate before Bondi gaugification weakly asymptotically flat data}
    If $\mathfrak{S}$ in \Cref{ingredient for gauge asymptotic flatness} is assumed to arise from weakly asymptotically flat initial data on $\Sigma^*_+$, the estimate on \bref{decay of xblin on scri+ before bonfi gaugification} still applies to $\xblins_{\mathscr{I}^+}$, $\ablins_{\mathscr{I}^+}$ and the smooth field on $\mathscr{I}^+$ defined by the pointwise limit of $\nablau r^2\xlin (u,v,\theta^A)$ as $v\longrightarrow\infty$.
\end{remark}

\begin{corollary}\label{decay of Olinos towards - infty before Bondiification}
    The radiation fields of the solution $\mathfrak{S}$ to \fullsystem arising via \Cref{proposition of asymptotic flatness at null infinity} satisfy
    \begin{align}
        \Olinos_{\mathscr{I}^+}, \elins_{\mathscr{I}^+}, \otxbs_{\mathscr{I}^+}=O\left(\frac{1}{|u|^\gamma}\right)
    \end{align}
    as $u\longrightarrow-\infty$.
\end{corollary}
\begin{proof}
The result for $\elins_{\mathscr{I}^+}$, $\Olinos_{\mathscr{I}^+}$ follows by applying \Cref{ingredient for gauge asymptotic flatness} to \bref{D3Chihat}. The decay rate of $\otxbs_{\mathscr{I}^+}$ then follows from the Codazi equation \bref{elliptic equation 1}.
\end{proof}

\begin{corollary}
    Assume $\mathfrak{S}$ arises from weakly asymptotically flat initial data on $\overline{\Sigma}$. Then we have
    \begin{align}
        \otxbs_{\mathscr{I}^+}+4\Olinos_{\mathscr{I}^+}=0,\qquad\qquad \divo\xblins_{\mathscr{I}^+}=\bblins_{\mathscr{I}^+}.
    \end{align}
\end{corollary}

We can show that $\Olinos_{\mathscr{I}^+}$, $\partial_u\bmlins_{\mathscr{I}^+}$ decay as $u\longrightarrow\infty$ when data on $\Sigma^*_+$ is asymptotically flat. We can derive that from the following:

\begin{lemma}\label{decay of xlin towards future of scri+}
    Assume that $\mathfrak{S}$ is a solution to \fullsystem arising from asymptotically flat initial data on ${\Sigma^*_+}$ which satisfies the horizon gauge conditions \bref{initial horizon gauge condition}. The radiation field $\xlins_{\mathscr{I}^+}$ defined in \Cref{forwards scattering of xlin at scri+} and its $\partial_u$, $\mathring{\slashednabla}$ derivatives have vanishing limits as $u\longrightarrow\pm\infty$.
\end{lemma}

\begin{proof}
    We start with the limit as $u\longrightarrow\infty$. For $r_0>2M$ we estimate
    \begin{align}\label{this 21 06 2021}
    \begin{split}
        \int_{S^2}d\omega\, |\xlins_{\mathscr{I}^+}|^2&=\int_{S^2}d\omega\,\left|\frac{r^2\xlin}{\Omega}\Big|_{r=r_0}\right|^2 +2\int_{S^2}d\omega\,\int_{\frac{1}{2}(r_0^*+t)}^\infty dv \frac{r^2\xlin}{\Omega}\times r^2\alin\\
        &\leq \int_{S^2}d\omega\,\left|\frac{r^2\xlin}{\Omega}\Big|_{r=r_0}\right|^2 + \sqrt{\int_{\underline{\mathscr{C}}_{\frac{1}{2}(r_0^*+t)\cap\{r\geq r_0\}}}dv d\omega\, r^{6-\epsilon}|\alin|^2}\,\times\\&\qquad\qquad\qquad \sqrt{\int_{\underline{\mathscr{C}}_{\frac{1}{2}(r_0^*+t)\cap\{r\geq r_0\}}}dvd\omega\,\frac{1}{r^{2-\epsilon}}\left|\frac{r^2\xlin}{\Omega}\right|^2}.
    \end{split}
    \end{align}
With the data on $\Sigma^*_+$ decaying to order $(1+\delta,\infty)$, the flux term for $\alin$ in \bref{this 21 06 2021} above is bounded via \bref{locallabel2} for
\begin{align}\label{choice of epsilon}
    \epsilon>1-2\delta.
\end{align}

As for the flux term of $\xlin$, we note that commuting the spacetime estimate \bref{ILED estimate on xlin} with $\slashednabla_t$ and using \Cref{technical lemma} gives 
\begin{align}\label{decay of xlin flux near scri+}
    \lim_{u\longrightarrow\infty}\int_{{\mathscr{C}}_u\cap\{r\geq r_0\}}d\bar{v}d\omega\,\frac{1}{r^{1+\epsilon}}\left|\frac{r^2\xlin}{\Omega}\right|^2=0,
\end{align}    
\begin{align}\label{decay of xlin flux near H+}
    \lim_{v\longrightarrow\infty}\int_{\underline{\mathscr{C}}_v\cap\{r\leq r_0\}}{\Omega^2}d\bar{u}d\omega\,\left[\left|\left(\frac{1}{\Omega}\nablagml\right)^2 \Omega\xlin\right|^2+\left|\frac{1}{\Omega}\nablagml \Omega\xlin\right|^2+\left|\Omega\xlin\right|^2\right]=0.
\end{align}
The decay of the $r=r_0$ term in \bref{this 21 06 2021} follows from \bref{decay of xlin flux near H+} and a Sobolev embedding. The decay of $\partial_u \xlins_{\mathscr{I}^+}$ follows by commuting the argument above with $\partial_t$ and using the fact that $r^2\alin$ has a vanishing limit towards $\mathscr{I}^+$. The vanishing of the limit as $u\longrightarrow-\infty$ can be established by a similar estimate to \bref{this 21 06 2021} integrating between $\Sigma^*_+$ and $\mathscr{I}^+$ and noting that \bref{locallabel2} implies 
\begin{align}
    \lim_{R\longrightarrow\infty} \int_{\Sigma^*_+\cap\{r\geq R\}}d{r}\dw\, {r}^{6-\epsilon}|\alin|^2=0.
\end{align}
\end{proof}

\begin{remark}\label{asymptotic flatness implies trivial memory}
    In particular, note that solutions to \fullsystem arising from asymptotically flat initial data (that is, data that decays to order $(1+\delta,\infty)$ for $\delta>0$ in the sense of \Cref{def of asymptotic flatness at spacelike infinity}) give rise to radiation fields at $\mathscr{I}^+$ exhibiting \underline{trivial memory}.
\end{remark}

\begin{corollary}\label{olinos bmlins decay towards future of scri+}
    The radiation fields $\partial_u\bmlins_{\mathscr{I}^+}$, $\Olinos_{\mathscr{I}^+}$ of \Cref{forwards scattering of olinos bmlins at scri+} have vanishing limits as $u\longrightarrow\infty$.
\end{corollary}

\begin{proof}
    This follows from applying \Cref{decay of xlin towards future of scri+} (taking into account \Cref{xblins estimate before Bondi gaugification weakly asymptotically flat data}) to
    \begin{align}\label{28 11 2021}
        2\mathring{\fancydstar_2}\elins_{\mathscr{I}^+}=-\xblins_{\mathscr{I}^+}-\lim_{v\longrightarrow\infty}\nablau r^2\xlin,\qquad \elins_{\mathscr{I}^+}=2\mathring{\slashednabla}\Olinos_{\mathscr{I}^+}=\frac{1}{2}\partial_u\bmlins_{\mathscr{I}^+}.
    \end{align}
\end{proof}

\begin{corollary}\label{asymptotic flatness implies enhanced decay 1}
    Assuming $\mathfrak{S}$ arises from asymptotically flat initial data to order $(1+\delta,\infty)$ with $\delta>0$ in {\Cref{proposition of asymptotic flatness at null infinity}}, we have in addition that $\Olinos_{\mathscr{I}^+}=O_{\infty}(|u|^{-1-\delta})$, $\bmlins_{\mathscr{I}^+}=O_{\infty}(|u|^{-\delta})$ and $\glinhs_{\mathscr{I}^+}=O_\infty(|u|^{-\delta})$ as $u\longrightarrow-\infty$.
\end{corollary}

\begin{proof}
    The asymptotics for $\Olinos_{\mathscr{I}^+}$ follows from \bref{28 11 2021} and \Cref{ingredient for gauge asymptotic flatness}, while those of $\bmlins_{\mathscr{I}^+}$ follow by integrating \bref{28 11 2021} and noting that $r\bmlin|_{\Sigma^*_+}\longrightarrow0$ as $r\longrightarrow\infty$. 

    As for $\glinhs_{\mathscr{I}^+}$, note that the convergence of $r\glinh$ towards $\mathscr{I}^+$ implies that $r\nablav r\glinh(u,v,\theta^A)\longrightarrow0$ as $v\longrightarrow\infty$. Thus we have by \bref{metric transport in 4 direction traceless} 
    \begin{align}\label{relation between glinhs and xlins before Bondi gaugification}
        \glinhs_{\mathscr{I}^+}=-2\,\xlins_{\mathscr{I}^+}-2\fancydstarring_2\bmlins_{\mathscr{I}^+}.
    \end{align}
\end{proof}
\begin{corollary}\label{asymptotic flatness implies enhanced decay 2}
    Under the assumptions of \Cref{asymptotic flatness implies enhanced decay 1}, we have that  $\Olinos_{\mathscr{I}^+}$, $\bmlins_{\mathscr{I}^+}$ and $\glinhs_{\mathscr{I}^+}$ decay as $u\longrightarrow\infty$.
\end{corollary}
\begin{proof}
    Note that \bref{integral of pblins along scri+ vanishes} implies that $\lim_{u\longrightarrow\infty}\glinhs_{\mathscr{I}^+}=0$, which implies that $\lim_{u\longrightarrow\infty}\bmlins_{\mathscr{I}^+}=0$ by \Cref{relation between glinhs and xlins before Bondi gaugification}.
\end{proof}

We will now derive an $L^2(\mathscr{I}^+)$ estimate of $\partial_u\xlins_{\mathscr{I}^+}$ in the case of solutions to\fullsystem arising from asymptotically flat initial data: commuting $\nablau\Omega^2$ past  \bref{D3Chihatbar} gives
\begin{align}
    \nablau\nablav r^2\Omega\xlin-\frac{2M}{r^2}\nablau r^2\Omega\xlin-\frac{4M\Omega^2}{r^3}r^2\Omega\xlin=-r\Omega^4\alin-r^2\Omega^3\plin.
\end{align}
Squaring both sides above, integrating over $J^+(\Sigma^*_+)$ and using Cauchy--Schwarz and the estimates \bref{locallabel1}, \bref{ILED estimate on xlin} we get
\begin{proposition}\label{primitive estimate on xlin at scri+}
Under the assumptions of \Cref{decay of xlin towards future of scri+}, we have
\begin{align}
\begin{split}
    &\int_{J^+(\Sigma^*_+)}d\bar{v}\dw\,r^{1+\epsilon}\left|\nablav\nablau \frac{r^2\xlin}{\Omega}\right|^2\\[10pt]&\lesssim \mathbb{F}^T_{\Sigma^*_+}[\Psilin]+ \int_{\Sigma^*_+}dr\dw\,\left[r^{4+\epsilon}|\alin|^2+r^{6+\epsilon}|\plin|^2+\left|\frac{1}{\Omega}\nablagml\left(\frac{1}{\Omega}\nablagml(r^2\Omega\xlin)\right)\right|^2+\left|\frac{1}{\Omega}\nablagml(r^2\Omega\xlin)\right|^2+\frac{1}{r^{2}}|r^2\Omega\xlin|^2\right],
\end{split}
\end{align}
for any $\epsilon>0$ such that the right hand side above is finite.
\end{proposition}

We can estimate $\glinhs_{\mathscr{I}^+}$ in $L^2(\mathscr{I}^+)$ directly using the $r^p$-estimates of  \Cref{RWrp}.
\begin{proposition}\label{estimate on glinhs at scri+}
The radiation field $\glinhs_{\mathscr{I}^+}$ of \Cref{radiation field for glinh at scri+ forward scattering} satisfies
\begin{align}\label{scri+ estimate on glinh}
\begin{split}
    &\int_{\mathscr{I}^+\cap\{\bar{u}\leq u\}}d\bar{u}\dw\,|\mathcal{A}_2(\mathcal{A}_2-2)\glinhs_{\mathscr{I}^+}|^2\\&\lesssim \mathbb{F}_{{\Sigma^*_+}\cap\{r\geq r(u,\vsigmap{u}\}}[\Psilin]+\int_{{\Sigma^*_+}\cap\{r\geq r(u,\vsigmap{u}\}}dr\dw\, r^2|\nablav\Psilin|^2.
\end{split}
\end{align}
\end{proposition}

We may estimate $\partial_u\glinhs_{\mathscr{I}^+}$ in $L^2(\mathscr{I}^+)$ using energy conservation alone:
\begin{proposition}\label{estimate on partial u glinhs at scri+}
The radiation field $\glinhs_{\mathscr{I}^+}$ of \Cref{forwards scattering of olinos bmlins at scri+} satisfies
\begin{align}
    \int_{\mathscr{I}^+\cap\{\bar{u}\leq u\}}d\bar{u}\dw\,|\mathcal{A}_2(\mathcal{A}_2-2)\partial_u\glinhs_{\mathscr{I}^+}|^2\lesssim \mathbb{F}_{{\Sigma^*_+}\cap\{r\geq r(u,\vsigmap{u}\}}[\Psilin].
\end{align}
\end{proposition}

We now arrive at

\begin{corollary}
The radiation field $\Olinos_{\mathscr{I}^+}$ of \Cref{forwards scattering of olinos bmlins at scri+} satisfies
\begin{align}\label{master gauge estimate forwards scattering}
\begin{split}
    &\int_{\mathscr{I}^+\cap\{\bar{u}\leq u\}}d\bar{u}\dw\,|\Olinos_{\mathscr{I}^+}|^2\\[10pt]&\;\lesssim\; \mathbb{F}_{{\Sigma^*_+}\cap\{r\geq r(u,\vsigmap{u}\}}[\Psilin]+\int_{{\Sigma^*_+}\cap\{r\geq r(u,\vsigmap{u}\}}dr\sin\theta d\theta d\phi \left[r^{\frac{7}{2}}|\Omega\plin|^2+r^{\frac{5}{2}}|\Omega^2\alin|^2\right]\\& +\int_{{\Sigma^*_+}\cap\{r\geq r(u,\vsigmap{u}\}}\Omega^2du\sin\theta d\theta d\phi\Bigg[\left|\frac{1}{\Omega}\nablagml\left(\frac{1}{\Omega}\nablagml(r^2\Omega\xlin)\right)\right|^2+\left|\frac{1}{\Omega}\nablagml(r^2\Omega\xlin)\right|^2+\frac{1}{r^{2}}|r^2\Omega\xlin|^2\Bigg].
\end{split}
\end{align}
\end{corollary}

\begin{remark}\label{superpolynomial decay is transported to scri+}
Note that the estimates \bref{scri+ estimate on glinh}, \bref{master gauge estimate forwards scattering} allow us to transport the decay rates of the initial data on $\Sigma^*_+$ in $r$ as $r\longrightarrow\infty$ to corresponding decay rates towards $u\longrightarrow-\infty$ along $\mathscr{I}^+$. In particular, If the initial data decays superpolynomially (as in the case of data constructed via \Cref{realising Sigmastar gauge future} from \underline{compactly supported} seed data), we have that the radiation fields on $\mathscr{I}^+$ also decay superpolynomially as $u\longrightarrow-\infty$.
\end{remark}

\subsubsection{The definition of the pure gauge solution $\mathfrak{G}_{\mathscr{I}^+}$}

\begin{defin}\label{transformation to Bondi gauge}
Let $\Olinos_{\mathscr{I}^+}$ be as in \Cref{forward scattering Olino elin at scri+ weak a f}. Define
\begin{align}
   \partial_u\outwardsgaugefunction(u,\theta^A)=-\Olinos_{\mathscr{I}^+},\qquad \outwardsgaugefunction|_{\ell=0,1}=0,\qquad \lim_{u\longrightarrow\infty}{\outwardsgaugefunction}_{,\ell\geq2}(u,\theta^A)=0.
\end{align}
Define $\mathfrak{G}_{\mathscr{I}^+}$ to be the gauge solution generated by $\outwardsgaugefunction$ via \Cref{outwards gauge solutions}.
\end{defin}

\begin{remark}
As $\mathfrak{G}_{\mathscr{I}^+}$ arises from a function that depends only on $u,\theta^A$ via \Cref{outwards gauge solutions}, it is easy to see that the radiation fields associated to $\mathfrak{G}_{\mathscr{I}^+}$ are all well-defined and that $\mathfrak{G}_{\mathscr{I}^+}$ is in fact asymptotically flat at $\mathscr{I}^+$ according to \Cref{defin of DHR flatness at scri+}.
\end{remark}

The estimates of the previous section lead to the following:

\begin{corollary}\label{otxbs of gauge solution decays forward scattering}
    The quantities
    \begin{align}
        \Olinos_{\mathscr{I}^+}[\mathfrak{G}_{\mathscr{I}^+}], \otxbs_{\mathscr{I}^+}[\mathfrak{G}_{\mathscr{I}^+}]
    \end{align}
    arising from $\mathfrak{G}_{\mathscr{I}^+}$ satisfy 
    \begin{align}
        \Olinos_{\mathscr{I}^+}[\mathfrak{G}_{\mathscr{I}^+}], \otxbs_{\mathscr{I}^+}[\mathfrak{G}_{\mathscr{I}^+}]=O\left(\frac{1}{|u|^\gamma}\right)
    \end{align}
    as $u\longrightarrow-\infty$.
\end{corollary}
\begin{proof}
    This follows from \Cref{decay of Olinos towards - infty before Bondiification} and the fact that
    \begin{align}
        \Olinos_{\mathscr{I}^+}[\mathfrak{G}_{\mathscr{I}^+}]=\Olinos_{\mathscr{I}^+}[\mathfrak{S}],
    \end{align}
    \begin{align}
        \otxbs_{\mathscr{I}^+}[\mathfrak{G}_{\mathscr{I}^+}]=-2\partial_u\outwardsgaugefunction.
    \end{align}
\end{proof}

We are now in a position to assert that $\mathfrak{S}+\mathfrak{G}_{\mathscr{I}^+}$ is Bondi-normalised at $\mathscr{I}^+$: it is clear from \Cref{transformation to Bondi gauge} that $\Olinos_{\mathscr{I}^+}$ arising from $\mathfrak{S}+\mathfrak{G}_{\mathscr{I}^+}$ vanishes. We also have

\begin{corollary}
The solution $\mathfrak{S}+\mathfrak{G}_{\mathscr{I}^+}$ satisfies
\begin{align}
    \lim_{v\longrightarrow\infty}r^2\Klin(u,v,\theta^A)[\mathfrak{S}+\mathfrak{G}_{\mathscr{I}^+}]=0.
\end{align}
\end{corollary}

\begin{proof}
    It is clear that for $\mathfrak{S}+\mathfrak{G}_{\mathscr{I}^+}$ we have
    \begin{align}
        \lim_{v\longrightarrow\infty}r^2\Klin(u,v,\theta^A)[\mathfrak{S}+\mathfrak{G}_{\mathscr{I}^+}]=-\frac{1}{2}\otxbs_{\mathscr{I}^+}[\mathfrak{S}+\mathfrak{G}_{\mathscr{I}^+}].
    \end{align}
    Given that $\Olinos_{\mathscr{I}^+}[\mathfrak{S}+\mathfrak{G}_{\mathscr{I}^+}]=0$, equation \bref{D3TrChibar} implies
    \begin{align}
        \partial_u \otxbs_{\mathscr{I}^+}[\mathfrak{S}+\mathfrak{G}_{\mathscr{I}^+}]=0.
    \end{align}
    Note that by \Cref{decay of Olinos towards - infty before Bondiification} and \Cref{otxbs of gauge solution decays forward scattering} we have that 
    \begin{align}
        \lim_{u\longrightarrow-\infty}\otxbs_{\mathscr{I}^+}[\mathfrak{S}+\mathfrak{G}_{\mathscr{I}^+}]=0.
    \end{align}
    Thus $\otxbs_{\mathscr{I}^+}[\mathfrak{S}+\mathfrak{G}_{\mathscr{I}^+}]=0$ identically on $\mathscr{I}^+$. The vanishing of $r^2\Klin$ towards $\mathscr{I}^+$ now follows via the Gauss equation \bref{Gauss}.
\end{proof}

\begin{corollary}\label{scri+ gauge solution is also asymptotically flat at spacelike infinity}
    The gauge solution $\mathfrak{G}_{\mathscr{I}^+}$ of \Cref{transformation to Bondi gauge} is asymptotically flat at spacelike infinity and the same is true of $\mathfrak{S}+\mathfrak{G}_{\mathscr{I}^+}$. Moreover, the $\ell\geq2$ component of $\mathfrak{S}+\mathfrak{G}_{\mathscr{I}^+}$, viewed as a solution to \fullsystem with vanishing $\ell=0,1$ modes, is future Bondi-normalised in the sense of \Cref{future Bondi gauge}.
\end{corollary}

\begin{corollary}   
    If $\mathfrak{S}$ is assumed to arise from asymptotically flat initial data, we have that $\outwardsgaugefunction\longrightarrow0$ as $u\longrightarrow\infty$, and thus the gauge conditions \bref{initial horizon gauge condition} are satisfied by $\mathfrak{G}_{\mathscr{I}^+}$. Thus the arguments of \Cref{Section 9.2 radiation fields at scri+} as well as \Cref{ingredient for gauge asymptotic flatness}, \Cref{decay of xlin towards future of scri+} and \Cref{olinos bmlins decay towards future of scri+} all apply to $\mathfrak{S}+\mathfrak{G}_{\mathscr{I}^+}$, and moreover, the solution $\mathfrak{G}_{\mathscr{I}^+}$ does not radiate to $\mathscr{H}^+_{\geq0}$, i.e.~the radiation fields at $\mathscr{H}^+_{\geq0}$ defined for $\mathfrak{S}$ via \Cref{definition of radiation fields at H+} remain unmodified after the application of $\mathfrak{G}_{\mathscr{I}^+}$. Finally, the $\ell\geq2$ component of $\mathfrak{S}+\mathfrak{G}_{\mathscr{I}^+}$, viewed as a solution to \fullsystem with vanishing $\ell=0,1$ modes, is future Bondi-normalised in the sense of \Cref{future Bondi gauge}.
\end{corollary}

\begin{remark}
    While it is possible to achieve a future Bondi-normalisation for solutions arising from weakly asymptotically flat initial data, it is not true in general that $\mathfrak{G}_{\mathscr{I}^+}$ does not radiate to $\mathscr{H}^+_{\geq0}$. 
\end{remark}
\begin{corollary}
Assume that $\mathfrak{S}$ is a solution to \fullsystem arising from initial data on $\Sigma^*_+$ which is asymptotically flat to order $(1,\infty)$. Then the radiation fields of $\Olin$ exists as defined in \Cref{forwards scattering for Psilin Psilinb at scri+}, and for any $u\in\mathbb{R}$ we have the estimate.
\begin{align}\label{master gauge estimate forwards scattering (1,infty)}
\begin{split}
    &\int_{\mathscr{I}^+\cap\{\bar{u}\leq u\}}d\bar{u}\dw\,|\Olinos_{\mathscr{I}^+}|^2\\[10pt]&\;\lesssim\; \mathbb{F}_{{\Sigma^*_+}\cap\{r\geq r(u,\vsigmap{u}\}}[\Psilin]+\int_{{\Sigma^*_+}\cap\{r\geq r(u,\vsigmap{u}\}}dr\sin\theta d\theta d\phi \left[r^{6+\epsilon}|\Omega\plin|^2+r^{4+\epsilon}|\Omega^2\alin|^2\right]\\& +\int_{{\Sigma^*_+}\cap\{r\geq r(u,\vsigmap{u}\}}dr\sin\theta d\theta d\phi\Bigg[\left|\frac{1}{\Omega}\nablagml\left(\frac{1}{\Omega}\nablagml(r\Omega\xlin)\right)\right|^2+\left|\frac{1}{\Omega}\nablagml(r\Omega\xlin)\right|^2+\frac{1}{r^{1+\epsilon}}|r\Omega\xlin|^2\Bigg].
\end{split}
\end{align}
\end{corollary}

\begin{remark}
    When initial data is asymptotically flat to order $(1,\infty)$, it is not the case in general that we can bound $\outwardsgaugefunction$ in terms of initial data. While solutions arising from past scattering data on $\mathscr{I}^-$, $\overline{\mathscr{H}^-}$ with nontrivial memory at $\mathscr{I}^-$ give rise to data on $\overline{\Sigma}$ decaying to order $(1,\infty)$ only, we will be able to derive boundedness estimates on $\outwardsgaugefunction$ thanks to the tame asymptotics of the solution towards $i^0$. See already \Cref{Section 14: global scattering problem}.
\end{remark}

\begin{remark}\label{remark on regularity of Bondi gauge at H+}
    Note that in the context of \ref{forward scattering full system thm}, solutions arising from asymptotically flat Cauchy data on $\Sigma^*_+$ or $\overline{\Sigma}$ have that $\Olinos_{\mathscr{I}^+}$ is integrable on $\mathscr{I}^+$, thus the generating function $\outwardsgaugefunction$ of $\mathfrak{G}_{\mathscr{I}^+}$ can be shown to decay polynomially as $u\longrightarrow\infty$. In particular, since
    \begin{align}
        \Omega^{-2}\partial_u \Omega^2 \outwardsgaugefunction=\partial_u\outwardsgaugefunction+\frac{2M}{r^2}\outwardsgaugefunction,
    \end{align}
    we have that $\Omega^{-1}\nablagml \Omega\xlin[\mathfrak{G}_{\mathscr{I}^+}]$ is well defined on $\mathscr{H}^+_{\geq0}$, thus $\mathfrak{G}_{\mathscr{I}^+}$ is $C^1$. In order to guarantee the existence of $(\Omega^{-1}\nablagml)^n\Omega\xlin[\mathfrak{G}_{\mathscr{I}^+}]$ for $n>1$ while simultaneously ensuring a Bondi-normalisation at $\mathscr{I}^+$, we need that $\outwardsgaugefunction$ and its $u$ derivatives decay exponentially, which is generically not the case even for solutions to \fullsystem arising from compactly supported Cauchy data. 
\end{remark}

\subsection{Passing to a future horizon-normalised gauge in forward scattering}\label{Section 11 Horizon gauge}

Having obtained the future Bondi-normalised solution $\mathfrak{S}+\mathfrak{G}_{\mathscr{I}^+}$ in \Cref{Section 10 Passing to Bondi gauge}, we now construct the pure gauge solution $\mathfrak{G}_{\mathscr{H}^+_{\geq0}}$ of \ref{forward scattering full system thm}.

The construction of $\mathfrak{G}_{\mathscr{H}^+_{\geq0}}$ will proceed in two steps: First, a pure gauge solution $\mathfrak{G}_{\mathscr{H}^+_{\geq0},I}$ is found via \Cref{inwards gauge solutions} such that $\mathfrak{S}+\mathfrak{G}_{\mathscr{I}^+}+\mathfrak{G}_{\mathscr{H}^+_{\geq0},I}$ satisfies $\Olinos_{\mathscr{H}^+_{\geq0}}=0$. We will then construct another pure gauge solution $\mathfrak{G}_{\mathscr{H}^+_{\geq0},II}$ via \Cref{residual gauge solutions} to achieve, in addition to having $\Olinos_{\mathscr{H}^+}=0$, that $\bmlins_{\mathscr{H}^+_{\geq0}}=0$ for $\mathfrak{S}+\mathfrak{G}_{\mathscr{I}^+}+\mathfrak{G}_{\mathscr{H}^+_{\geq0},I}-\mathfrak{G}_{\mathscr{H}^+_{\geq0},II}$. We take $\mathfrak{G}_{\mathscr{H}^+_{\geq0}}=\mathfrak{G}_{\mathscr{H}^+_{\geq0},I}+\mathfrak{G}_{\mathscr{H}^+_{\geq0},II}$.  We will show that the addition of $\mathfrak{G}_{\mathscr{H}^+_{\geq0}}$ preserves Bondi-normalisation at $\mathscr{I}^+$. 

\subsubsection[Preliminary boundedness estimates on radiation fields at $\protect\mathscr{H}^+_{\geq0}$]{Further boundedness estimates on radiation fields at $\protect\mathscr{H}^+_{\geq0}$}\label{Section 11.1 preliminary estimates on xblin at H+}

In this section we collect estimate on the radiation fields of the solution $\mathfrak{S}$ which will allow us later to estimate the gauge solution $\mathfrak{G}_{\mathscr{H}^+_{\geq0},I}$. 

\begin{proposition}
    The radiation fields of $\mathfrak{S}$ on $\mathscr{H}^+_{\geq0}$ satisfy the following relations:
    \begin{alignat*}{2}
    \refstepcounter{equation}\latexlabel{horizon equation 1}
    \refstepcounter{equation}\latexlabel{horizon equation 2}
     \refstepcounter{equation}\latexlabel{horizon equation 3}
        &\left(\mathring{\slashed{\Delta}}-\frac{3}{2}\right)\bblins_{\mathscr{H}^+}=\frac{1}{(2M)^2}\fancydstarring_1\left(\frac{1}{2\Omega^2}\partial_ur^3\rlin\Big|_{\mathscr{H}^+_{\geq0}},\frac{1}{2\Omega^2}\partial_ur^3\slin\Big|_{\mathscr{H}^+_{\geq0}}\right)&&-M\divo\pblins_{\mathscr{H}^+}-\frac{3}{4M}\elins_{\mathscr{H}^+},\tag{\ref{horizon equation 1}}\\
        &\left(\partial_v+\frac{1}{2M}\right)\frac{1}{\Omega^2}\partial_u\Psilinb\Big|_{\mathscr{H}^+}=\frac{1}{(2M)^2}\left(\mathring{\slashed{\Delta}}-1\right)\upPsilinb_{\mathscr{H}^+},\qquad
        &&\frac{1}{(2M)^2}\upPsilinb_{\mathscr{H}^+}=\left(\partial_v+\frac{1}{2M}\right)\pblins_{\mathscr{H}^+}.\tag{\ref{horizon equation 2},\,\ref{horizon equation 3}}
    \end{alignat*}
\end{proposition}
\begin{proof}
    Equation \bref{horizon equation 1} is attained by combining equations \bref{Bianchi-2}, \bref{Bianchi-0} and \bref{elliptic equation 1}. Equations \bref{horizon equation 2} and \bref{horizon equation 3} follow easily by restricting \bref{RW}, \bref{hier'} respectively to $\mathscr{H}^+_{\geq0}$.
\end{proof}

\begin{proposition}\label{estimate on xblins at H+ proposition}
    For any solution $\mathfrak{S}$ to \fullsystem the following estimate applies to $\xblins_{\mathscr{H}^+}$,
    \begin{align}\label{estimate on xblins at H+}
        \|\xblins_{\mathscr{H}^+}\|_{L^2(S^2_{\infty,v})}^2\lesssim \left\|\pblins_{\mathscr{H}^+}\right\|_{L^2(\Sigma^*_+\cap \mathscr{H}^+_{\geq0})}^2+\sum_{|\gamma|\leq 2} \mathbb{F}_{\Sigma^*_+}[\mathring{\slashednabla}^\gamma\Psilinb],
    \end{align}
    provided the right hand side is finite.
\end{proposition}

\begin{proof}
    We will find an $L^2(S^2_{\infty,v})$ estimate on $\pblins_{\mathscr{H}^+}$ via \bref{horizon equation 3} and then estimate $\|\bblins_{\mathscr{H}^+}\|_{L^2(S^2_{\infty,v})}$ via \bref{horizon equation 1}. Squaring \bref{horizon equation 3} and integrating by parts immediately gives
    \begin{align}
        \begin{split}
&\left\|\partial_v\pblins_{\mathscr{H}^+}\right\|^2_{L^2([0,v]\times S^2)}+\frac{1}{(2M)^2}\|\pblins_{\mathscr{H}^+}\|^2_{L^2([0,v]\times S^2)}\|^2_{L^2([0,v]\times S^2)}+\frac{1}{2M}\|\pblins_{\mathscr{H}^+}\|^2_{L^2(S^2_{\infty,v})}\\&\lesssim \frac{1}{(2M)^4}\|\upPsilinb_{\mathscr{H}^+}\|^2_{L^2([0,v]\times S^2)}+\frac{1}{2M}\|\pblins_{\mathscr{H}^+}\|^2_{L^2(\Sigma^*_+\cap\mathscr{H}^+_{\geq0})}
\\&\lesssim \frac{1}{(2M)^4}\mathbb{F}_{\Sigma^*_+}[\Psilinb]+\frac{1}{2M}\|\pblins_{\mathscr{H}^+}\|^2_{L^2(\Sigma^*_+\cap\mathscr{H}^+_{\geq0})}.
        \end{split}
    \end{align}
The same can be done with \bref{horizon equation 2} to get the following bound using a redshift estimate from \Cref{RWredshift}:
\begin{align}
    \begin{split}
        &\left\|\partial_v\upPsilinb_{\mathscr{H}^+}\right\|^2_{L^2([0,v]\times S^2)}+\frac{1}{(2M)^2}\|\upPsilinb_{\mathscr{H}^+}\|^2_{L^2([0,v]\times S^2)}\|^2_{L^2([0,v]\times S^2)}+\frac{1}{2M}\|\upPsilinb_{\mathscr{H}^+}\|^2_{L^2(S^2_{\infty,v})}\,\lesssim \sum_{|\gamma|\leq 2} \mathbb{F}_{\Sigma^*_+}[\mathring{\slashednabla}^\gamma\Psilinb],
    \end{split}
\end{align}
Using the horizon gauge condition \eqref{initial horizon gauge condition}, it is easy to estimate $\elins_{\mathscr{H}^+}$ and $\Omega^{-1}\nablagml\xlin|_{\mathscr{H}^+_{\geq0}}$ as was done for Theorem 3 in \cite{DHR16}:
\begin{align}\label{12 10 2022}
\begin{split}
    \left\|\fancydstarring_2\fancydstarring_1\fancydring_1\elins_{\mathscr{H}^+}\right\|^2_{L^2(S^2_{\infty,v})}&\lesssim \left\|\fancydstarring_2\fancydstarring_1(\rlins_{\mathscr{H}^+},\slins_{\mathscr{H}^+})\right\|^2_{L^2(S^2_{\infty,v})}\lesssim \|\upPsilinb_{\mathscr{H}^+}\|^2_{L^2(S^2_{\infty,v})}+\|\xlins_{\mathscr{H}^+}\|^2_{L^2(S^2_{\infty,v})}\\
    &\lesssim \mathbb{F}_{\Sigma^*_+}[\Psilinb].
    \end{split}
\end{align}
Thus an estimate on $\Omega^{-1}\nablagml \xlin$ on $\mathscr{H}^+_{\geq0}$ follows now using \bref{D3Chihat} and \bref{12 10 2022}:
\begin{align}
    \left\|\Omega^{-1}\nablagml\Omega\xlin\Big|_{\mathscr{H}^+}\right\|^2_{L^2(S^2_{\infty,v})}\lesssim\mathbb{F}_{\Sigma^*_+}[\Psilinb].
\end{align}
Applying all of the above to \bref{horizon equation 1}, we conclude the following estimate on $\bblins_{\mathscr{H}^+}$:
\begin{align}\label{12 10 2022 2}
    \|\bblins_{\mathscr{H}^+}\|_{H^2(S^2_{\infty,v})}^2\lesssim \sum_{|\gamma|\leq 2} \mathbb{F}_{\Sigma^*_+}[\mathring{\slashednabla}^\gamma\Psilinb]+\frac{1}{2M}\|\pblins_{\mathscr{H}^+}\|^2_{L^2(\mathscr{H}^+_{\geq0}\cap\Sigma^*_+)}.
\end{align}
The estimate \bref{estimate on xblins at H+} now follows from applying \bref{12 10 2022 2} to equation \bref{Bianchi-2}.
\end{proof}

\subsubsection[Setting $\protect\Olinos_{\mathscr{H}^+}=0$: the definition of $\mathfrak{G}_{\mathscr{H}^+_{\geq0},I}$]{Setting $\protect\Olinos_{\mathscr{H}^+}=0$: the definition of $\mathfrak{G}_{\mathscr{H}^+_{\geq0},I}$}\label{Killing Olinos at H+ in forwards scattering}

We now construct the pure gauge solution $\mathfrak{G}_{\mathscr{H}^+_{\geq0},I}$ and use the estimates of \Cref{Section 11.1 preliminary estimates on xblin at H+} to derive an estimate on $\mathfrak{G}_{\mathscr{H}^+_{\geq0},I}$. 

\begin{defin}\label{def of inwards gauge solution}
Let $\Olinos_{\mathscr{H}^+}$ be as in \Cref{definition of radiation fields at H+}. Define $\inwardsgaugefunction(v,\theta^A)$ with domain $(v,\theta^A)\in[0,\infty)\times S^2$ by
\begin{align}
    2{(\Olinos_{\mathscr{H}^+}[\mathfrak{S}_{\Sigma^*_+}])}_{\ell\geq2}=-\partial_v \inwardsgaugefunction-\frac{1}{2M}\inwardsgaugefunction.
\end{align}
with the condition that $\inwardsgaugefunction|_{v=0}=0$. Define $\underline{\mathfrak{G}}_{\mathscr{H}^+_{\geq0},I}$ to be the gauge solution generated by $\inwardsgaugefunction$ via \Cref{inwards gauge solutions}.
\end{defin}
\begin{remark}
Note that the above definition can be used to set $\Olin|_{\mathscr{H}^+_{\geq0}}=0$ for either $\mathfrak{S}_{\Sigma^*_+}$ satisfying \Cref{proposition of asymptotic flatness at null infinity} or $\mathfrak{S}_{\Sigma^*_+}+\mathfrak{G}_{\mathscr{I}^+}$ with $\mathfrak{G}_{\mathscr{I}^+}$ defined in \Cref{transformation to Bondi gauge}, since the gauge solution $\mathfrak{G}_{\mathscr{I}^+}$ does not radiate to $\mathscr{H}^+$ according to \Cref{scri+ gauge solution is also asymptotically flat at spacelike infinity}. 
\end{remark}

Using that $\Olinos_{\mathscr{H}^+}=0$ for $\mathfrak{S}+\mathfrak{G}_{\mathscr{H}^+_{\geq0},I}$, we can use \bref{12 10 2022} to estimate $\xblins_{\mathscr{H}^+}[\mathfrak{S}+\mathfrak{G}_{\mathscr{H}^+_{\geq0},I}]$ utilising \bref{elin eblin Olin}. Note that according to the definition of $\mathfrak{G}_{\mathscr{H}^+_{\geq0},I}$ via \Cref{inwards gauge solutions}, we have that $\elin\left[\mathfrak{S}+\mathfrak{G}_{\mathscr{H}^+_{\geq0},I}\right]=\elin[\mathfrak{S}]$. We can thus conclude the following

\begin{proposition}
    For the solution $\mathfrak{S}+\mathfrak{G}_{\mathscr{H}^+_{\geq0},I}$ is a solution to \fullsystem on $J^+(\Sigma^*_+)$, we have that $\xblins_{\mathscr{H}^+}$ is bounded in $L^2(S^2_{\infty,v})$ via
    \begin{align}
        \begin{split}
            &\left\|\partial_v\xblins_{\mathscr{H}^+}\right\|^2_{L^2([0,v]\times S^2)}+\frac{1}{(2M)^2}\|\xblins_{\mathscr{H}^+}\|^2_{L^2([0,v]\times S^2)}\|^2_{L^2([0,v]\times S^2)}+\frac{1}{2M}\|\xblins_{\mathscr{H}^+}\|^2_{L^2(S^2_{\infty,v})}\\
            &\lesssim \mathbb{F}_{\Sigma^*_+}[\Psilinb].
        \end{split}
    \end{align}
\end{proposition}
\begin{proof}
    This follows by restricting \bref{D4Chihatbar} to $\mathscr{H}^+_{\geq0}$ and using $\Olinos_{\mathscr{H}^+}=0$ to estimate $\eblins_{\mathscr{H}^+}$ via \bref{12 10 2022}. Note that the $\Sigma^*_+$ gauge conditions and the fact that $\inwardsgaugefunction$ vanishes at $v=0$ implies that $\divo^2\xblins_{\mathscr{H}^+}\Big|_{\Sigma^*_+\cap\mathscr{H}^+_{\geq0}}=0$, while $\curlo\divo\xblins_{\mathscr{H}^+}\Big|_{\Sigma^*_+\cap\mathscr{H}^+_{\geq0}}$ is proportional to $\slin|_{\Sigma^*_+\cap\mathscr{H}^+_{\geq0}}$ and $\partial_t\slin|_{\Sigma^*_+\cap\mathscr{H}^+_{\geq0}}$. Thus for the solution $\mathfrak{S}+\mathfrak{G}_{\mathscr{H}^+_{\geq0},I}$, we can bound $\|\xblins_{\mathscr{H}^+}\|_{L^2(\Sigma^*_+\cap\mathscr{H}^+_{\geq0})}$ by $\mathbb{F}_{\Sigma^*_+}[\Psilinb]$.
\end{proof}
\begin{corollary}
    The scalar function $\inwardsgaugefunction$ of \Cref{def of inwards gauge solution} generating $\mathfrak{G}_{\mathscr{H}^+_{\geq0},I}$ is bounded in terms of the initial data giving rise to $\mathfrak{S}$ via
    \begin{align}
       \|\inwardsgaugefunction\|^2_{H^2(S^2_{\infty,v})}\lesssim \left\|\pblins_{\mathscr{H}^+}\right\|_{L^2(\Sigma^*_+\cap \mathscr{H}^+_{\geq0})}^2+\mathbb{F}^2_{\Sigma^*_+}[\Psilinb].
    \end{align}
\end{corollary}


\begin{remark}
    In order to show that $\mathfrak{G}_{\mathscr{H}^+_{\geq0},I}$ is weakly asymptotically flat at spacelike infinity, we need to show that $\partial_v\inwardsgaugefunction=O(v^{-1})$ as $v\longrightarrow\infty$. This is however not necessary to construct and estimate the solution $\mathfrak{G}_{\mathscr{H}^+_{\geq0},II}$.
\end{remark}

\subsubsection[Setting $\protect\bmlins_{\mathscr{H}^+}=0$: the definition of $\mathfrak{G}_{\mathscr{H}^+_{\geq0},II}$]{Setting $\protect\bmlins_{\mathscr{H}^+}=0$: the definition of $\mathfrak{G}_{\mathscr{H}^+_{\geq0},II}$}

With $\Olinos_{\mathscr{H}^+}=0$ we can make use of the estimates of Theorem 4 in \cite{DHR16} to prove integrated decay estimates on $\xblin$. We recall these estimates in this section and use them in the following section to estimate the pure gauge solution $\mathfrak{G}_{\mathscr{H}^+_{\geq0},II}$.

\begin{proposition}\label{another DHR estimate}
    For the solution $\mathfrak{S}+\mathfrak{G}_{\mathscr{H}^+_{\geq0},I}$, we have for any $\epsilon\in(0,1)$,
    \begin{align}
        \int_{J^+(\Sigma^*_+)}\dw d\bar{v}d\bar{u}\,{\Omega^2}\left[{r^{1-\epsilon}}\left|\xblin\right|^2+r^{3-\epsilon}|\nablav\Omega^{-1}\xblin|^2\right]\lesssim \mathbb{F}_{\Sigma^*_+}[\Psilinb]+\int_{\Sigma^*_+}dr\dw\, r^6\Omega^{-2}|\pblin|^2+r^2\Omega^{-4}|\ablin|^2
    \end{align}
\end{proposition}
\begin{proof}
    See Proposition 14.2.1 and Proposition 14.2.2 of \cite{DHR16}.
\end{proof}

\begin{corollary}
     For the solution $\mathfrak{S}+\mathfrak{G}_{\mathscr{H}^+_{\geq0},I}$, we have
     \begin{align}
         \int_{\mathscr{H}^+_{\geq0}}d\bar{v}\dw\,\left[\left|\glinhs_{\mathscr{H}^+}\right|^2+\left|\fancydstarring_2\bmlins_{\mathscr{H}^+}\right|^2\right]\lesssim \mathbb{F}_{\Sigma^*_+}[\Psilinb]+\int_{\Sigma^*_+}dr\dw\, r^6\Omega^{-2}|\pblin|^2+r^2\Omega^{-4}|\ablin|^2,
     \end{align}
     for $\epsilon\in(0,1)$.
\end{corollary}
\begin{proof}
    We derive from \bref{metric transport in 3 direction traceless},
    \begin{align}
        \begin{split}
            \partial_u r^\epsilon|\glinh|^2+\frac{\Omega^2}{r^{1-\epsilon}}|\glinh|^2=4r^\epsilon\cdot\Omega\xblin.
        \end{split}
    \end{align}
    We integrate over $J^+(\Sigma^*_+)$ and use Cauchy--Schwartz and \Cref{another DHR estimate} to obtain an  $L^2(\mathscr{H}^+_{\geq0})$ estimate on $\glinhs_{\mathscr{H}^+}$.\\
    
    We similarly derive from \bref{metric transport in 3 direction traceless}. Note that $\glinh|_{\Sigma^*_+}=0$ in the $\Sigma^*_+$ gauge.
    
    \begin{align}
        \partial_u\left(r|\nablav\glinh|^2\right)+\Omega^2|\nablav\glinh|^2=4\nablav\glinh\cdot\left(\Omega^2\nablav r\Omega^{-1}\xblin-\frac{\Omega^2}{r}(2\Omega^2-1)r\Omega^{-1}\xblin\right).
    \end{align}
    We integrate again over $J^+(\Sigma^*_+)$ and use Cauchy--Schwartz and \Cref{another DHR estimate} to obtain an  $L^2(\mathscr{H}^+_{\geq0})$ estimate on $\partial_v\glinhs_{\mathscr{H}^+}$. We can then use energy conservation for \bref{RW} to estimate $\bmlins_{\mathscr{H}^+}$ in $L^2(\mathscr{H}^+_{\geq0})$ using \bref{metric transport in 4 direction traceless}.
\end{proof}
\begin{defin}\label{defin of q_1 q_2}
For the solution $\mathfrak{S}_{\Sigma^*_+}+\underline{\mathfrak{G}}_{\mathscr{H}^+_{\geq0},I}$, define the scalars $\mathfrak{q}_1(v,\theta^A),\mathfrak{q}_2(v,\theta^A)$ by 
\begin{align}
    \bmlins_{\mathscr{H}^+}[\mathfrak{S}_{\Sigma^*_+}+\underline{\mathfrak{G}}_{\mathscr{H}^+_{\geq0},I}]=:(2M)^2\fancydstarring_1(\partial_v\mathfrak{q}_1,\partial_v\mathfrak{q}_2)
\end{align}
with the condition that $\mathfrak{q}_1,\mathfrak{q}_2$ decay as $v\longrightarrow\infty$ and have no $\ell=0$ components. Define the pure gauge solution $\underline{\mathfrak{G}}_{\mathscr{H}^+_{\geq0},II}$ via \Cref{residual gauge solutions} with $(\mathfrak{q}_1,\mathfrak{q}_2)$ as generators.
\end{defin}

\begin{defin}
Define the pure gauge solution $\underline{\mathfrak{G}}_{\mathscr{H}^+_{\geq0}}$ by
\begin{align}
    \underline{\mathfrak{G}}_{\mathscr{H}^+_{\geq0}}:=\underline{\mathfrak{G}}_{\mathscr{H}^+_{\geq0},I}+\underline{\mathfrak{G}}_{\mathscr{H}^+_{\geq0},II}.
\end{align}
\end{defin}

\begin{corollary}
    The solution
    \begin{align}
        \mathfrak{S}_{\Sigma^*_+}+\mathfrak{G}_{\mathscr{I}^+}+\mathfrak{G}_{\mathscr{H}^+_{\geq0}}+\mathfrak{G}_{out}\left[\frac{1}{2}\mathfrak{m}v\right]+\mathfrak{G}_{in}\left[\frac{1}{2}\mathfrak{m}u\right]
    \end{align}
    is both future horizon and Bondi-normalised.
\end{corollary}

\subsection{Scattering from $\overline{\Sigma}$ to $\mathscr{I}^+$, $\overline{\mathscr{H}^+}$}\label{forward scattering from Sigma bar}

The results of \Cref{Section 7 forward scattering} carry over to scattering from data on $\overline{\Sigma}$ and we immediately have the following

\begin{corollary}\label{forward scattering Sigma bar}
    Assume $\mathfrak{D}_{\overline{\Sigma}}=(\,\glinhs, \tr\glins$, $\Olinos$, $\bmlins$, $\glinhs'$, $\tr\glins'$, $\Olinos'$, $\bmlins'\,)$ is a smooth, asymptotically flat initial data set for the system \fullsystem on $\overline{\Sigma}$ in the sense of \Cref{def of initial data on spacelike surface}, such that
    \begin{align}
        \divr\left(\elin-\eblin\right)|_{\mathcal{B}}=0,\qquad \left[\rlin-\rlin_{\ell=0}+\divr\elin\right]|_{\mathcal{B}}=0.
    \end{align}
    Then the solution $\mathfrak{S}_{\overline{\Sigma}}$ to \fullsystem arising from $\mathfrak{D}_{\overline{\Sigma}}$ is strongly asymptotically flat at $\mathscr{I}^+$
\end{corollary}

\begin{proof}
    The solution $\mathfrak{S}$ arising from the data on $\mathfrak{D}$ on $\overline{\Sigma}$ induces an initial data set $\mathfrak{D}_{{\Sigma^*_+}}$ which satisfies \bref{initial horizon gauge condition} (since the regularity of $\mathfrak{S}$ as a solution to \fullsystemK implies that $\otxs|_{\mathscr{H}^+}=0$). Note that Cauchy stability implies that $\mathfrak{D}_{{\Sigma^*_+}}$ is also an asymptotically flat initial data set.
\end{proof}

We may also repeat the procedure carried out in \Cref{Section 10 Passing to Bondi gauge} to construct a pure gauge solution $\mathfrak{G}_{\overline{\Sigma}\rightarrow\mathscr{I}^+}$ such that $\mathfrak{S}_{\overline{\Sigma}}+\mathfrak{G}_{\overline{\Sigma}\rightarrow\mathscr{I}^+}$ is $\mathscr{I}^+$-normalised and estimate $\mathfrak{G}_{\overline{\Sigma}\rightarrow\mathscr{I}^+}$ in terms of $\mathfrak{D}_{\overline{\Sigma}}$:

\begin{corollary}\label{Bondi normalisation from Sigma bar}
    For the solution $\mathfrak{S}_{\overline{\Sigma}}$ constructed in \Cref{forward scattering Sigma bar}, there exists a unique pure gauge solution $\mathfrak{G}_{\overline{\Sigma}\rightarrow\mathscr{I}^+}$ such that $\mathfrak{S}_{\overline{\Sigma}}+\mathfrak{G}_{\overline{\Sigma}\rightarrow\mathscr{I}^+}$ is $\mathscr{I}^+$-normalised as in \Cref{future Bondi gauge}. Moreover, we have the estimate
    \begin{align}\label{Bondi normalisation from Sigma bar estimate}
        \begin{split}
            &\int_{\mathscr{I}^+}d\bar{u}\dw\,|\partial_u\mathfrak{f}_{\overline{\Sigma}\rightarrow\mathscr{I}^+}|^2\\&\lesssim\mathbb{F}_{\overline{\Sigma}}[\Psilin]+\int_{\overline{\Sigma}\cap\{V\geq1\}}dr\sin\theta d\theta d\phi \left[r^{6+\epsilon}|\Omega\plin|^2+r^{4+\epsilon}|\Omega^2\alin|^2+\left|\frac{1}{\Omega}\nablagml\left(\frac{1}{\Omega}\nablagml(r^2\Omega\xlin)\right)\right|^2+\left|\frac{1}{\Omega}\nablagml(r^2\Omega\xlin)\right|^2+\frac{1}{r^{1+\epsilon}}|r^2\Omega\xlin|^2\right]\\&+
        \int_{\overline{\Sigma}\cap\{V\leq1\}}dr\dw\,\left[\left|\slashednabla_U \alinK\right|^2+|\alinK|^2+\left|\slashednabla_U^2\xlinK\right|^2+\left|\slashednabla_U\xlinK\right|^2+|\xlinK|^2\right],
        \end{split}
    \end{align}
     where $\mathfrak{f}_{\overline{\Sigma}\rightarrow\mathscr{I}^+}$ is the scalar function generating $\mathfrak{G}_{\overline{\Sigma}\rightarrow\mathscr{I}^+}$ via \Cref{outwards gauge solutions}.
\end{corollary}

\begin{proof}
    Since the conditions \eqref{initial horizon gauge condition} are satisfied everywhere on $\overline{\mathscr{H}^+}$, it suffices to estimate the right hand side of \eqref{master gauge estimate forwards scattering} in terms of $\mathfrak{D}_{\overline{\Sigma}}$. By repeating the proof of \Cref{ILED estimate on xlin} and using the equations \eqref{D4Chihat Kruskal} and \eqref{Kruskal hier+} in the region $V\leq1$, we easily deduce the estimate
    \begin{align}
    \begin{split}
        &\mathbb{F}_{{\Sigma^*_+}}[\Psilin]+\int_{{\Sigma^*_+}}dr\sin\theta d\theta d\phi \left[r^{6+\epsilon}|\Omega\plin|^2+r^{4+\epsilon}|\Omega^2\alin|^2+\left|\frac{1}{\Omega}\nablagml\left(\frac{1}{\Omega}\nablagml(r^2\Omega\xlin)\right)\right|^2+\left|\frac{1}{\Omega}\nablagml(r^2\Omega\xlin)\right|^2+\frac{1}{r^{1+\epsilon}}|r^2\Omega\xlin|^2\right]\\
        &\lesssim\mathbb{F}_{\overline{\Sigma}}[\Psilin]+\int_{\overline{\Sigma}\cap\{V\geq1\}}dr\sin\theta d\theta d\phi \left[r^{6+\epsilon}|\Omega\plin|^2+r^{4+\epsilon}|\Omega^2\alin|^2+\left|\frac{1}{\Omega}\nablagml\left(\frac{1}{\Omega}\nablagml(r^2\Omega\xlin)\right)\right|^2+\left|\frac{1}{\Omega}\nablagml(r^2\Omega\xlin)\right|^2+\frac{1}{r^{1+\epsilon}}|r^2\Omega\xlin|^2\right]\\&+
        \int_{\overline{\Sigma}\cap\{V\leq1\}}dr\dw\,\left[\left|\slashednabla_U \alinK\right|^2+|\alinK|^2+\left|\slashednabla_U^2\xlinK\right|^2+\left|\slashednabla_U\xlinK\right|^2+|\xlinK|^2\right].
    \end{split}
    \end{align}
\end{proof}

In a similar manner, we may conclude
\begin{corollary}\label{transformation to H+ bar gauge forward scattering}
    For the solution $\mathfrak{S}_{\overline{\Sigma}}+\mathfrak{G}_{\overline{\Sigma}\rightarrow\mathscr{I}^+}$ constructed in \Cref{forward scattering Sigma bar} and \Cref{Bondi normalisation from Sigma bar}, there exists a unique pure gauge solution $\mathfrak{G}_{\overline{\Sigma}\rightarrow \overline{\mathscr{H}^+}\cup\mathscr{I}^+}$ such that $\mathfrak{S}_{\overline{\Sigma}}+\mathfrak{G}_{\overline{\Sigma}\rightarrow\mathscr{I}^+}+\mathfrak{G}_{\overline{\Sigma}\rightarrow \overline{\mathscr{H}^+}\cup\mathscr{I}^+}$ is both $\mathscr{I}^+$ and $\overline{\mathscr{H}^+}$-normalised. Moreover, we have the estimate 
    \begin{align}\label{transformation to H+ bar gauge forward scattering estimate}
        \begin{split}
            &\chi_{v\in[0,\infty)}\|\mathfrak{f}_{\overline{\Sigma}\rightarrow\overline{\mathscr{H}^+}\cup\mathscr{I}^+}\|^2_{H^2(S^2_{\infty,v})}+\|(\partial_v{\mathfrak{q}_1}_{\overline{\Sigma}\rightarrow\overline{\mathscr{H}^+}\cup\mathscr{I}^+},\partial_v{\mathfrak{q}_2}_{\overline{\Sigma}\rightarrow\overline{\mathscr{H}^+}\cup\mathscr{I}^+})\|^2_{L^2(\{v\geq0\})H^2(S^2_{\infty,v})}\\
            &+\chi_{V\in[0,1]}\|V\mathfrak{f}_{\overline{\Sigma}\rightarrow\overline{\mathscr{H}^+}\cup\mathscr{I}^+}\|^2_{H^2(S^2_{\infty,2M\log(|v|)})}+\|(\partial_V{\mathfrak{q}_1}_{\overline{\Sigma}\rightarrow\overline{\mathscr{H}^+}\cup\mathscr{I}^+},\partial_V{\mathfrak{q}_2}_{\overline{\Sigma}\rightarrow\overline{\mathscr{H}^+}\cup\mathscr{I}^+})\|^2_{L^2_V(V\in[0,1])H^2(S^2_{\infty,2M\log(|v|)})}
            \\&\lesssim \sum_{|\gamma|\leq2}\mathbb{F}_{\Sigma^*_+}[\mathring{\slashednabla}^\gamma\Psilinb]+\|\slashednabla_V\pblinK\|_{L^2(\mathcal{B})}^2+\int_{\overline{\Sigma}\cap\{V\geq1\}}dr\dw\,r^6\Omega^{-2}|\pblin|^2+r^4\Omega^{-4}|\ablin|^2\\&+\int_{\overline{\Sigma}\cap\{V\leq1\}}dr\dw\,|\slashednabla_V\ablinK|^2+|\ablinK|^2,
        \end{split}
    \end{align}
    where $\chi_{\mathcal{S}}$ is the characteristic function over the set $\mathcal{S}$, $\mathfrak{f}_{\overline{\Sigma}\rightarrow\overline{\mathscr{H}^+}}$ generates a pure gauge solution $\mathfrak{G}_{\overline{\Sigma}\rightarrow\overline{\mathscr{H}^+}\cup\mathscr{I}^+,I}$ via \Cref{outwards gauge solutions}, $({\mathfrak{q}_1}_{\overline{\Sigma}\rightarrow\overline{\mathscr{H}^+}\cup\mathscr{I}^+},{\mathfrak{q}_2}_{\overline{\Sigma}\rightarrow\overline{\mathscr{H}^+}\cup\mathscr{I}^+})$ generates a pure gauge solution $\mathfrak{G}_{\overline{\Sigma}\rightarrow\overline{\mathscr{H}^+}\cup\mathscr{I}^+,II}$ via \Cref{residual gauge solutions}, and $\mathfrak{G}_{\overline{\Sigma}\rightarrow\overline{\mathscr{H}^+}\cup\mathscr{I}^+}=\mathfrak{G}_{\overline{\Sigma}\rightarrow\overline{\mathscr{H}^+}\cup\mathscr{I}^+,I}+\mathfrak{G}_{\overline{\Sigma}\rightarrow\overline{\mathscr{H}^+}\cup\mathscr{I}^+,II}$.
\end{corollary}

\begin{proof}
    The construction of \Cref{Section 11 Horizon gauge} implies the result, modifying the estimate on the gauge transformations by taking into account the Kruskal-normalised equations \fullsystemK in the region $V\leq 1$.
\end{proof}

\subsection{Scattering from $\overline{\Sigma}$ to $\mathscr{I}^-$, $\overline{\mathscr{H}^-}$}\label{Section 8.7: Scattering from Sigmabar to H- and I-}

For $\alin$ satisfying to the $+2$ Teukolsky equations \bref{T+2} on $J^-(\overline{\Sigma})$, the quantity $\overone{\invertedalpha}(u,v):=\alin(-u,-v)$ defines a smooth solution to the $-2$ Teukolsky equation \bref{T-2} on $J^+(\overline{\Sigma})$. Therefore, if $\alin$ arises from asymptotically flat initial data on $\overline{\Sigma}$, we may deduce the behaviour of $\alin$ towards $\mathscr{I}^-$ from the estimates and asymptotics proven for solutions to the $-2$ Teukolsky equation \bref{T-2} arising from asymptotically flat initial data. An analogous statement applies to solutions of \bref{T-2} on $J^-(\overline{\Sigma})$. Consequently, the estimates proven on $\xlin$, $\blin$, $\elin$, $\otx$ on $J^+(\overline{\Sigma})$ apply to $\xblin$,  $\bblin$, $\eblin$, $\otxb$ on $J^-(\overline{\Sigma})$ replacing $t$ with $-t$ everywhere in the estimates (in particular exchanging $u$ and $v$), and vice versa.
Using the above we can state the following:

\begin{proposition}\label{forward scattering past}
    Assume $\mathfrak{D}_{\overline{\Sigma},-}=(\,\glinhs, \tr\glins$, $\Olinos$, $\bmlins$, $\glinhs'$, $\tr\glins'$, $\Olinos'$, $\bmlins'\,)$ is a smooth, asymptotically flat initial data set on ${\overline{\Sigma}}$ for the system \fullsystem in the sense of \Cref{def of initial data on spacelike surface}. The solution $\mathfrak{S}_{\overline{\Sigma},-}$ to \fullsystem arising from $\mathfrak{D}_{\overline{\Sigma},-}$ via \Cref{EinsteinWP Sigmabar past} gives rise to smooth fields at $\mathscr{I}^-$ via the following pointwise limits:
   
    \begin{alignat}{4}
       &\xlins_{\mathscr{I}^-}:={\lim_{u\longrightarrow-\infty}}r\xlin,\qquad &&\xblins_{\mathscr{I}^-}:=\lim_{u\longrightarrow-\infty}r^2\xblin,\qquad &&\elins_{\mathscr{I}^-}:=\lim_{u\longrightarrow-\infty}r^2\elin,\qquad && \eblins_{\mathscr{I}^-}:=\lim_{u\longrightarrow-\infty}r\eblin,
    \end{alignat}
    \begin{alignat}{4}
    & \qquad &&\otxs_{\mathscr{I}^-}:=\lim_{u\longrightarrow-\infty} r\otx,\qquad && \otxbs_{\mathscr{I}^-}:=\lim_{u\longrightarrow-\infty} r^2\otxb, \qquad &&  \qquad
    \end{alignat}
    \begin{alignat}{4}
        &\alins_{\mathscr{I}^-}:=\lim_{u\longrightarrow-\infty}r\alin,\qquad && \blins_{\mathscr{I}^-}:=\lim_{u\longrightarrow-\infty} r^2\blin,\qquad && \rlins_{\mathscr{I}^-}:=\lim_{u\longrightarrow-\infty} r^3\rlin,\qquad && \slins_{\mathscr{I}^-}:=\lim_{u\longrightarrow-\infty} r^3\slin,
    \end{alignat}
    \begin{alignat}{1}
        &\Olinos_{\mathscr{I}^-}=\lim_{u\longrightarrow-\infty} \Olin_{\ell\geq2}.
    \end{alignat}
\end{proposition}
\begin{remark}
    Note that \Cref{forward scattering past} does not require $\mathfrak{D}_{\overline{\Sigma},-}$ to be in the $\overline{\Sigma}_-$ gauge.
\end{remark}
\begin{proposition}
The radiation fields defined in \Cref{forward scattering past} satisfy 
\begin{align}\label{global energy conservation Sigmastar past}
    \|\mathfrak{D}_{\overline{\Sigma},-}\|_{\mathcal{E}^{T}_{\Sigma^*_-}}^2=\|\xlins_{\mathscr{I}^-}\|_{\mathcal{E}^{T}_{\mathscr{I}^-}}^2+\|\xblins_{\mathscr{H}^-}\|_{\mathcal{E}^{T}_{\mathscr{H}^-_{\leq0}}}^2.
\end{align}
\end{proposition}

When it comes to the metric components, we will only be able to define radiation fields on $\mathscr{I}^-$ by passing to a Bondi-normalised gauge at $\mathscr{I}^-$. We first establish preliminary decay estiamtes on $\bmlin$, $\glinh$ near $\mathscr{I}^-$

\begin{proposition}
For $\mathfrak{S}$ of \Cref{forward scattering past}, we have for any $\delta>0$
\begin{align}
    \lim_{u\longrightarrow-\infty}r^{-\delta}\bmlin=0.
\end{align}
\end{proposition}
\begin{proof}
    Rewrite \bref{partial_u b} as
    \begin{align}
        \nablau r^{-\delta}\bmlin+(1-\delta)\frac{\Omega^2}{r}r^{-\delta}\bmlin=\frac{1}{2r^{\delta}\Omega^2}(\elin-\eblin)
    \end{align}
    we integrate the in u from $\Sigma^*_-$ towards $\mathscr{I}^-$ to deduce
    \begin{align}
        |\,r^{-\delta}\bmlin\,(u,v)\,|\leq |\,r^{-\delta}\bmlin\,(\usigmam{v},v)\,|+\int_{\usigmam{v}}^u d\bar{u}\,\left[(1-\delta)\frac{\Omega^2}{r} r^{-\delta}|\,\bmlin\,(\bar{u},v)\,|+\int_{\usigmam{v}}^u d\bar{u}\,\frac{1}{2r^{\delta}\Omega^2}|\,\elin-\eblin\,|\right].
    \end{align}
    Gr\"onwall's inequality implies
    \begin{align}
         |\,r^{-\delta}\bmlin\,(u,v)\,|\,\leq \log(r^{1-\delta})\left[|\,r^{-\delta}\bmlin\,(\usigmam{v},v)\,|+\frac{1}{2r^{\delta}\Omega^2}|\,\elin-\eblin\,|\right],
    \end{align}
    which in turn implies for $0<\delta<1/2$,
    \begin{align}
         |\,r^{-\delta}\bmlin\,(u,v)\,|\,\leq r^{-\delta}\log(r^{1-2\delta})\left[|\,r^{-2\delta}\bmlin\,(\usigmam{v},v)\,|+\frac{1}{2r^{2\delta}\Omega^2}|\,\elin-\eblin\,|\right].
    \end{align}
    The integral of $|\elin-\eblin|$ converges for any $\delta>0$ since \Cref{forward scattering past} gives that $r^2\elin$, $r\eblin$ have smooth limits at $\mathscr{I}^-$.
\end{proof}
Applying the argument above to \bref{metric transport in 3 direction traceless}, using the fact that $r^2\xblin$ has a smooth limit at $\mathscr{I}^-$,  gives
\begin{proposition}
For $\mathfrak{S}$ of \Cref{forward scattering past}, we have for any $\delta>0$
\begin{alignat}{1}
    \lim_{u\longrightarrow-\infty}r^{\delta}\glinh=0,\qquad\qquad\lim_{u\longrightarrow-\infty}r^\delta \tr\glin=0.
\end{alignat}
\end{proposition}

We now show how to construct radiation fields for the metric components in the past Bondi-normalised gauge:

\begin{proposition}\label{forward scattering past metric}
For a solution $\mathfrak{S}$ arising from asymptotically flat initial data on $\Sigma^*_-$ as in \Cref{forward scattering past}, there exists a pure gauge solution $\mathfrak{G}_{\mathscr{I}^-}$ such that the radiation fields given in \Cref{grand proposition backwards scattering from past} exist and in addition the metric components of $\mathfrak{S}+\mathfrak{G}$ define radiation fields on $\mathscr{I}^-$ via the limits
 \begin{alignat}{4}
        &\glinhs_{\mathscr{I}^-}=\lim_{u\longrightarrow-\infty}r\glinh,\qquad && \tr\glins_{\mathscr{I}^-}=\lim_{u\longrightarrow-\infty} \tr\glin_{\ell\geq2},\qquad &&  \bmlins_{\mathscr{I}^-}=\lim_{u\longrightarrow-\infty} \bmlin_{\ell\geq2}.
    \end{alignat}
\end{proposition}
We now pass to a $\mathscr{I}^-$-normalised gauge
\begin{defin}\label{Bondi normalisation forward scattering past}
    Assume $\mathfrak{D}_{\overline{\Sigma},-}$ is a Cauchy data set on $\overline{\Sigma}$ which is normalised to the $\overline{\Sigma}_-$ gauge, and let $\mathfrak{S}_{\overline{\Sigma}_-}$ be the corresponding solution to \fullsystem according to \Cref{EinsteinWP Sigmabar past}. Define $\mathfrak{G}_{\overline{\Sigma}_-\rightarrow\mathscr{I}^-}$ to be the pure gauge solution generated according to \Cref{outwards gauge solutions} via ${\mathfrak{f}}_{\overline{\Sigma}_-\rightarrow\mathscr{I}^-}$ given by
    \begin{align}
    \partial_v{\mathfrak{f}}_{\overline{\Sigma}_-\rightarrow\mathscr{I}^-}=2\Olinos_{\mathscr{I}^-},\qquad\lim_{v\longrightarrow\infty}{\mathfrak{f}}_{\overline{\Sigma}_-\rightarrow\mathscr{I}^-}=0.
\end{align}
\end{defin}

We will now  estimate ${\mathfrak{f}}_{\overline{\Sigma}_-\rightarrow\mathscr{I}^-}$. Equation \bref{D4Chihatbar} implies in the limit towards $\mathscr{I}^-$,
\begin{align}\label{22 11 2021}
    2\mathring{\fancydstar_2}\eblins_{\mathscr{I}^-}=2\mathring{\fancydstar_2}\mathring{\slashednabla}\Olinos_{\mathscr{I}^-}=\xlins_{\mathscr{I}^-}-\partial_v \xblins_{\mathscr{I}^-}
\end{align}
We estimate $\partial_v\xblins_{\mathscr{I}^-}$ analogously to the estimate \bref{primitive estimate on xlin at scri+}:
\begin{align}
\begin{split}
    &\int_{J^-(\Sigma^*_-)}d\bar{u}d\bar{v}\dw\,r^{1+\epsilon}\left|\nablau {r^2\Omega\xblin}\right|^2\\[10pt]&\lesssim \mathbb{F}^T_{\Sigma^*_-}[\Psilinb]+ \int_{\Sigma^*_-}dr\dw\,\left[r^{5+\epsilon}|\ablin|^2+r^{7+\epsilon}|\pblin|^2+\left|\frac{1}{\Omega}\nabladlt\left(\frac{1}{\Omega}\nabladlt(r^2\Omega\xblin)\right)\right|^2+\left|\frac{1}{\Omega}\nabladlt(r^2\Omega\xblin)\right|^2+\frac{1}{r^\epsilon}|r^2\Omega\xblin|^2\right].
\end{split}
\end{align}
Combining the above with energy conservation and a Poincar\'e's estimate gives
\begin{proposition}
The gauge solution $\mathfrak{G}_{\mathscr{I}^-}$ satisfies 
\begin{align}
\begin{split}
    &\int_{-\infty}^\infty\int_{S^2} d\bar{v}\dw\, |\partial_v\mathfrak{f}_{\overline{\Sigma}_-\rightarrow\mathscr{I}^-}|^2\;\lesssim\; \mathbb{F}_{\overline{\Sigma}}[\Psilinb]\\[10pt]& +\int_{\overline{\Sigma}\cap\{V\geq1\}}dr\dw\,\left[r^{4+\epsilon}|\ablin|^2+r^{6+\epsilon}|\pblin|^2+\left|\frac{1}{\Omega}\nabladlt\left(\frac{1}{\Omega}\nabladlt(r^2\Omega\xblin)\right)\right|^2+\left|\frac{1}{\Omega}\nabladlt(r^2\Omega\xblin)\right|^2+\frac{1}{r^{1+\epsilon}}|r^2\Omega\xblin|^2\right]\\
    &+\int_{\overline{\Sigma}\cap\{V\leq1\}}dr\dw\,\left[|\ablinK|^2+|\slashednabla\ablinK|^2+|\slashednabla_V^2\xblinK|^2+|\slashednabla_V\xblinK|^2+|\xblinK|^2\right].
\end{split}
\end{align}
\end{proposition}

Now that $\Olinos_{\mathscr{I}^-}=0$, we can show that $r\bmlin_A$ attains a limit at $\mathscr{I}^-$:
\begin{proposition}
The shift vector $\bmlin$ of $\mathfrak{S}_{\overline{\Sigma}_-}+\mathfrak{G}_{\overline{\Sigma}_-\rightarrow\mathscr{I}^-}$ satisfies
\begin{align}
    \lim_{u\longrightarrow-\infty}r\bmlin(u,v,\theta^A)=-\elins_{\mathscr{I}^-}.
\end{align}
\end{proposition}

\begin{proof}
    We integrate \bref{D3etabar} from $\mathscr{I}^-$, noting that now $\eblins_{\mathscr{I}^-}=0$:
    \begin{align}
        r^2\eblin(u,v,\theta^A)=-r(u,v)\int_{-\infty}^u d\bar{u} \left[r\Omega\bblin-\Omega^2\elin\right]\Big|_{(\bar{u},v,\theta^A)}.
    \end{align}
    Since $r^3\bblin\longrightarrow0$, $r^2\elin\longrightarrow \elins_{\mathscr{I}^-}$, we have
    \begin{align}
        r^2\eblin(u,v,\theta^A)-\elins_{\mathscr{I}^-}(v,\theta^A)
        =-r(u,v)\int_{-\infty}^u d\bar{u}\frac{1}{r^2}\left[r^3\Omega\bblin(\bar{u},v,\theta^A)-r^2\Omega^2\elin(\bar{u},v,\theta^A)+\Omega^2\elins_{\mathscr{I}^-}(v,\theta^A)\right]=
    \end{align}
    Given $\epsilon>0$, there exists $u_-$ such that for $u<u_-$ we have $\left|r^2\elin(u,v,\theta^A)-\elins_{\mathscr{I}^-}(v,\theta^A)\right|<\frac{\epsilon}{2}$ and $|r^3\Omega\bblins(u,v,\theta^A)|<\frac{\epsilon}{2}$, thus
    \begin{align}
        \left|r^2\eblin(u,v,\theta^A)-\elins_{\mathscr{I}^-}(v,\theta^A)\right|<r(u,v)\times\frac{1}{r(u,v)}\times \epsilon.
    \end{align}
    An identical argument to the above applied to \bref{partial_u b} using that $r^2\elin$, $r^2\eblin$ converge yields the result for $\bmlin$.
\end{proof}

\begin{corollary}\label{convergence of bmlin forward scattering past}
For the solution $\mathfrak{S}_{\overline{\Sigma}_-}+\mathfrak{G}_{\overline{\Sigma}_-\rightarrow\mathscr{I}^-}$, the limit
\begin{align}
    \lim_{u\longrightarrow-\infty}{r\glinh}(u,v,\theta^A)
\end{align}
exists and defines an element of $\Gamma^{(2)}({\mathscr{I}^-})$.
\end{corollary}

\begin{proof}
    We can deduce following a similar argument to that leading to \bref{model argument for unsaturated r weights} that as $u\longrightarrow-\infty$,
    \begin{align}\label{integral of bmlin decays forward scattering past}
        \int_{v}^{\vsigmap{u}}d\bar{v}\;\bmlin\longrightarrow0.
    \end{align}
    The argument of \Cref{radiation field for glinh at scri+ forward scattering} applies to $\xlin$ near $\mathscr{I}^-$ and we deduce that $\int_{v}^{\vsigmap{u}}d\bar{v}\;r\xlin$ converges to $\int^\infty_{v}d\bar{v}\;\xlins_{\mathscr{I}^-}$. Integrating \bref{metric transport in 4 direction traceless} and taking the limit as $u\longrightarrow-\infty$ gives the result.
\end{proof}

\begin{corollary}
For the solution $\mathfrak{S}_{\overline{\Sigma}_-}+\mathfrak{G}_{\overline{\Sigma}_-\rightarrow\mathscr{I}^-}$, the limit
\begin{align}
    \lim_{u\longrightarrow-\infty}r\tr\glin(u,v,\theta^A)
\end{align}
vanishes at $\mathscr{I}^-$.
\end{corollary}

\begin{proof}
    The convergence of $\int_{v}^{\vsigmap{u}}d\bar{v}\;\divo r\xlin$, $\int_{v}^{\vsigmap{u}}d\bar{v}\;r^2\blin$ to $\int^\infty_{v}d\bar{v}\;\divo\xlins_{\mathscr{I}^-}$, $\int^\infty_{v}d\bar{v}\;\blins_{\mathscr{I}^-}$ respectively implies, using the Codazzi equation \bref{elliptic equation 2}, that $\int_{v}^{\vsigmap{u}}d\bar{v}\;r^2\otx_{\ell\geq2}$ converges to $\int^\infty_{v}d\bar{v}\;\otxs_{\mathscr{I}^-}$. Together with \bref{integral of bmlin decays forward scattering past}, we get using \bref{metric transport in 4 direction trace} that $r\tr\glin_{\ell\geq2}(u,v,\theta^A)$ converges as $u\longrightarrow-\infty$ to a limit $\tr\glins_{\mathscr{I}^-}(v,\theta^A)$. Now the linearised Gauss equation \bref{Gauss} implies $\otxs_{\mathscr{I}^-}=0$, thus $\tr\glins_{\mathscr{I}^-}$ is constant. Integrating \bref{metric transport in 4 direction trace} from $\Sigma^*_-$ in $v$ and using \bref{integral of bmlin decays forward scattering past} gives that $\tr\glins_{\mathscr{I}^-}=0$.
\end{proof}

In light of the asymptotics given above, we have the following relations between the radiation fields at $\mathscr{I}^-$:

\begin{corollary}\label{scattering data at scri- forward scattering}
The radiation fields of $\mathfrak{S}_{\overline{\Sigma}_-}+\mathfrak{G}_{\overline{\Sigma}_-\rightarrow\mathscr{I}^-}$ defined as in \Cref{grand proposition backwards scattering from past}, \Cref{forward scattering past metric} satisfy 
\begin{alignat}{2}
    & \xblins_{\mathscr{I}^-}=\frac{1}{2}\glinhs_{\mathscr{I}^-}, \qquad && \qquad \elins_{\mathscr{I}^-}=\frac{1}{2}\divo\glinhs_{\mathscr{I}^-}\\
    & \xlins_{\mathscr{I}^-}=\frac{1}{2}\partial_v \glinhs_{\mathscr{I}^-}, \qquad && \blins_{\mathscr{I}^-}=-\divo \xlins_{\mathscr{I}^-}=-\frac{1}{2}\divo\partial_v\glinhs_{\mathscr{I}^-} 
\end{alignat}
\begin{alignat}{3}
    & \alins_{\mathscr{I}^-}=-\frac{1}{2}\partial_v^2 \divo^2 \glinhs_{\mathscr{I}^-}, \qquad && \rlins_{\mathscr{I}^-}=-\frac{1}{2}\divo^2\glinhs_{\mathscr{I}^-},\qquad && \slins_{\mathscr{I}^-}=-\frac{1}{2}\curlo\divo \glinhs_{\mathscr{I}^-}.
\end{alignat}
\begin{align}
    \Olinos_{\mathscr{I}^-}=\tr\glins_{\mathscr{I}^-}=\otxs_{\mathscr{I}^-}=\otxbs_{\mathscr{I}^-}=0.
\end{align}
\end{corollary}

We now turn to $\overline{\mathscr{H}^-}$.

\begin{defin}\label{definition of radiation fields at H-}
Let $\mathfrak{S}$ be a solution to the system \fullsystem arising via \Cref{EinsteinWP Sigmabar past} from smooth data on $\overline{\Sigma}$. Define
\begin{alignat}{2}
    &\alins_{\mathscr{H}^-}:=\Omega^{-2}\alin\Big|_{\overline{\mathscr{H}^-}}\;\;,\qquad\qquad\qquad&&\ablins_{\mathscr{H}^-}:=\Omega^{2}\ablin\Big|_{\overline{\mathscr{H}^-}}\;\;,\\
    &\blins_{{\mathscr{H}^-}}:=\Omega^{-1}\blin\Big|_{\overline{\mathscr{H}^-}}\;\;,\qquad\qquad\qquad&&\bblins_{\mathscr{H}^-}:=\Omega\bblin\Big|_{\overline{\mathscr{H}^-}}\;\;,\\
    &\rlins_{{\mathscr{H}^-}}:=\rlin_{\ell\geq2}\Big|_{\overline{\mathscr{H}^-}}\;\;,\qquad\qquad\qquad&&\slins_{\mathscr{H}^-}:=\slin_{\ell\geq2}\Big|_{\overline{\mathscr{H}^-}}\;\;,\\
   &\xlins_{\mathscr{H}^-}:=\Omega^{-1}\xlin\Big|_{\overline{\mathscr{H}^-}}\;\;,\qquad\qquad\qquad&&\xblins_{\mathscr{H}^-}:=\Omega\xblin\Big|_{\overline{\mathscr{H}^-}}\;\;,\\
    &\elins_{\mathscr{H}^-}:=\elin_{\ell\geq2}\Big|_{\overline{\mathscr{H}^-}}\;\;,\qquad\qquad\qquad&&\eblins_{\mathscr{H}^-}:=\eblin_{\ell\geq2}\Big|_{\overline{\mathscr{H}^-}}\;\;,\\
    &\otxs_{\mathscr{H}^-}:=\Omega^{-2}\otx_{\ell\geq2}\Big|_{\overline{\mathscr{H}^-}}\;\;,\qquad\qquad\qquad&&\otxbs_{\mathscr{H}^-}:=\otxb_{\ell\geq2}\Big|_{\overline{\mathscr{H}^-}}\;\;,\\
    &\olins_{\mathscr{H}^-}:=\Omega^{-2}\olin_{\ell\geq2}\Big|_{\overline{\mathscr{H}^-}}\;\;,\qquad\qquad\qquad&&\olinbs_{\mathscr{H}^-}:=\olinb_{\ell\geq2}\Big|_{\overline{\mathscr{H}^-}}\;\;,\\
    &\glinhs_{\mathscr{H}^-}:=\glinh\Big|_{\overline{\mathscr{H}^-}}\;\;,\qquad\qquad\qquad&&\bmlins_{\mathscr{H}^-}:=\Omega^{-2}\bmlin_{\ell\geq2}\Big|_{\overline{\mathscr{H}^-}}\;\;,\\
    &\tr\glins_{\mathscr{H}^-}:=\tr\glin_{\ell\geq2}\Big|_{\overline{\mathscr{H}^-}}\;\;,\qquad\qquad\qquad&&\Olinos_{\mathscr{H}^-}:=\Olin_{\ell\geq2}\Big|_{\overline{\mathscr{H}^-}}\;\;.
\end{alignat}
Additionally, we define
\begin{align}
    &\upPsilin_{\mathscr{H}^-}:=\Psilin|_{\overline{\mathscr{H}^-}},\qquad\qquad\qquad \plins_{\mathscr{H}^+}:=\Omega^{-1}\plin|_{{\mathscr{H}^-}},\\
    &\upPsilinb_{\mathscr{H}^-}:=\Psilinb|_{\overline{\mathscr{H}^-}},\qquad\qquad\qquad \pblins_{\mathscr{H}^-}:=\Omega\pblin|_{\overline{\mathscr{H}^-}}.
\end{align}
\end{defin}

\begin{corollary}\label{scattering data at H- forward scattering}
The radiation fields of $\mathfrak{S}$ on $\mathscr{H}^-$, defined as in \Cref{grand proposition backwards scattering from past}, \Cref{forward scattering past metric} satisfy the following equations
\begin{alignat}{2}
    &(2M)^{-1}\partial_U\glinhs_{\mathscr{H}^-}=-2U^{-1}\xblins_{\mathscr{H}^-},\qquad &&\tr\glin_{\mathscr{H}^-}=0,\\[5pt]
    &(2M)^{-1}\partial_U U\bmlins_{\mathscr{H}^-}=-2\left(\elins_{\mathscr{H}^-}-\eblins_{\mathscr{H}^-}\right),\qquad && \elins_{\mathscr{H}^-}+\eblins_{\mathscr{H}^-}=2{\slashednabla}_A \Olinos_{\mathscr{H}^-},\\[5pt]
    &(2M)^{-1}\partial_U \elins_{\mathscr{H}^-}=-U^{-1}\bblins_{\mathscr{H}^-},\qquad && (2M)^{-1}\partial_U\left(\rlins_{\mathscr{H}^-},\slins_{\mathscr{H}^-}\right)={\fancyd_1}U^{-1}\bblins_{\mathscr{H}^-},
\end{alignat}
\begin{align}
    & (2M)^{-1}\partial_U U\otxs_{\mathscr{H}^-}=-\left(2\divr\elins_{\mathscr{H}^-}+2\rlins_{\mathscr{H}^-}-\frac{1}{M}\Olins_{\mathscr{H}^-}\right), 
    \\[5pt]& (2M)^{-1}\partial_U U^{-1}\xblins_{\mathscr{H}^-}= U^{-2}\ablins_{\mathscr{H}^-},
    \\[5pt]&(2M)^{-1}\partial_U\xlins_{\mathscr{H}^-}=2{\fancydstar_2}\elins_{\mathscr{H}^-}+\xblins_{\mathscr{H}^-},
    \\[5pt]&(2M)^{-1}\partial_U U^2\alins_{\mathscr{H}^-}=2{\fancydstar_2}U\blins_{\mathscr{H}^-}-\frac{3}{2M^2}U\xlins_{\mathscr{H}^-},\\[5pt] & (2M)^{-1}\partial_U U^{-1}\bblins_{\mathscr{H}^-}=-\divr\, U^{-2}\ablins_{\mathscr{H}^-},
    \\[5pt] & (2M)^{-1}\partial_U \blins_{\mathscr{H}^-}=\fancydstar_1\left(\rlins_{\mathscr{H}^-},-\slins_{\mathscr{H}^-}\right)+3\elins_{\mathscr{H}^-},
    \\[5pt]&(2M)^{-1}\partial_U\olins_{\mathscr{H}^-}=\rlins_{\mathscr{H}^-}+\frac{1}{2M^3}\Olinos_{\mathscr{H}^-},
\end{align}
\begin{align}
    &\divr\,\xblins_{\mathscr{H}^-}=\bblins_{\mathscr{H}^-},\qquad \curlr\,\elins_{\mathscr{H}^-}=\slins_{\mathscr{H}^-},\qquad
    \frac{1}{2}\divr^2\glinhs_{\mathscr{H}^-}=-\rlins_{\mathscr{H}^-}.
\end{align}
\begin{align}
    \divr \,\xlins_{\mathscr{H}^-}=-\frac{1}{2M}\eblins_{\mathscr{H}^-}-\blins_{\mathscr{H}^-}+\frac{1}{2}\slashednabla \otxs_{\mathscr{H}^-}
\end{align}
\end{corollary}

Passing to a $\overline{\mathscr{H}^-}$-normalised gauge proceeds analogously to the procedure followed in \Cref{Killing Olinos at H+ in forwards scattering}. We state the resulting definition of the gauge transformation the requisite estimate:
\begin{corollary}
    For the solution $\mathfrak{S}_{\overline{\Sigma}_-}+\mathfrak{G}_{\overline{\Sigma}_-\rightarrow\mathscr{I}^-}$ constructed in \Cref{forward scattering past} and \Cref{Bondi normalisation forward scattering past}, there exists a unique pure gauge solution $\mathfrak{G}_{\overline{\Sigma}_-\rightarrow \overline{\mathscr{H}^-}\cup\mathscr{I}^-}$ such that $\mathfrak{S}_{\overline{\Sigma}_-}+\mathfrak{G}_{\overline{\Sigma}_-\rightarrow\mathscr{I}^-}+\mathfrak{G}_{\overline{\Sigma}_-\rightarrow \overline{\mathscr{H}^-}\cup\mathscr{I}^-}$ is both $\mathscr{I}^-$ and $\overline{\mathscr{H}^-}$-normalised. Moreover, we have the estimate 
    \begin{align}
        \begin{split}
            &\chi_{u\in(-\infty,0]}\|\underline{\mathfrak{f}}_{\overline{\Sigma}_-\rightarrow\overline{\mathscr{H}^-}\cup\mathscr{I}^-}\|^2_{H^2(S^2_{u,-\infty})}+\chi_{U\in[-1,0]}\|U\underline{\mathfrak{f}}_{\overline{\Sigma}_-\rightarrow\overline{\mathscr{H}^-}\cup\mathscr{I}^-}\|^2_{H^2(S^2_{2M\log(|u|),-\infty})}
            \\&\lesssim \mathbb{F}^2_{\overline{\Sigma}}[\Psilinb]+\|\slashednabla_U\alinK\|_{L^2(\mathcal{B})}^2+\int_{\overline{\Sigma}\cap\{|U|\geq1\}}dr\dw\,r^6|\plin|^2+r^2|\alin|^2+\int_{\overline{\Sigma}\cap\{|U|\leq1\}}dr\dw\,|\slashednabla_U\alinK|^2+|\alinK|^2,
        \end{split}
    \end{align}
    where $\chi_{\mathcal{S}}$ is the characteristic function over the set $\mathcal{S}$, and $\underline{\mathfrak{f}}_{\overline{\Sigma}_-\rightarrow\overline{\mathscr{H}^-}\cup\mathscr{I}^-}$ generates a pure gauge solution $\mathfrak{G}_{\overline{\Sigma}_-\rightarrow\overline{\mathscr{H}^-}\cup\mathscr{I}^-}$ via \Cref{inwards gauge solutions}.
\end{corollary}

\section{Backwards scattering}\label{Section 6 Backwards scattering}

In this section we will prove the following:

\begin{proposition}\label{grand proposition backwards scattering}
Given $\xblins_{\mathscr{I}^+}\in\Gamma_c^{(2)}(\mathscr{I}^+)$, $\xlins_{\mathscr{H}^+}\in\Gamma_c^{(2)}(\mathscr{H}^+_{\geq0})$ and $(\mathfrak{m},\mathfrak{a})\in \mathbb{R}\times \mathbb{R}^3$, there exists a unique solution $\mathfrak{S}$ to the system \fullsystem on $D^+({\Sigma^*_+})$ that satisfies both the $\mathscr{H}^+_{\geq0}$ and the $\mathscr{I}^+$ gauge conditions and which attains $\xblins_{\mathscr{I}^+}$, $\xlins_{\mathscr{H}^+}$ via
\begin{align}\label{radiation fields attained main proposition backwards scattering}
    \lim_{v\longrightarrow\infty}  r\xblin (u,v,\theta^A)=\xblins_{\mathscr{I}^+}(u,\theta^A),\qquad\qquad \Omega\xlin|_{\mathscr{H}^+_{\geq0}}=\xlins_{\mathscr{H}^+}.
\end{align}
The solution $\mathfrak{S}$ defines data on ${\Sigma^*_+}$ which satisfies the gauge conditions \bref{initial horizon gauge condition} and which is asymptotically flat at spacelike infinity in the sense of \cref{def of asymptotic flatness at spacelike infinity} to order $(1,\infty)$. 
\end{proposition}

We will also show that
\begin{proposition}\label{grand proposition backwards scattering addendum}
If, in addition to the assumptions of \Cref{grand proposition backwards scattering}, we have that $\xblins_{\mathscr{I}^+}$ satisfies
\begin{align}
    \int_{-\infty}^\infty d\bar{u}\,\xblins_{\mathscr{I}^+}(\bar{u},\theta^A)=0,
\end{align}
then the data induced by $\mathfrak{S}$ on $\Sigma^*_+$ decays to order $(2,\infty)$ towards $i^0$.
\end{proposition}

\subsection{Constructing a solution from scattering data on $\mathscr{H}^+$, $\mathscr{I}^+$}\label{Section 9.1: Constructing a solution from scattering data on H+, I+}

Recall the following result from \cite{Mas20}:

\begin{proposition}\label{Mas20 corollary}
Let $\underline\upalpha_{\mathscr{I}^+}\in\Gamma^2_c(\mathscr{I}^+)$, $\upalpha_{\mathscr{H}^+}\in\Gamma^2_c(\mathscr{H}^+_{\geq0})$. Then there exists a unique pair $\alpha, \underline\alpha$ which are smooth on $D^+({\Sigma^*_+})$, such that $\alpha$ satisfies the $+2$ Teukolsky equation \bref{T+2}, and $\underline\alpha$ satisfies the $-2$ Teukolsky equation \bref{T-2}, with
\begin{align}
    \lim_{v\longrightarrow\infty} \mathring{\slashednabla}^\gamma \slashednabla_t^k r\underline\alpha(u,v,\theta^A)=\mathring{\slashednabla}^\gamma\partial_u^k\underline\upalpha_{\mathscr{I}^+},\qquad\qquad \mathring{\slashednabla}^\gamma \slashednabla_t^k\Omega^2\alpha|_{\mathscr{H}^+_{\geq0}}=\mathring{\slashednabla}^\gamma\partial_v^k\alpha_{\mathscr{H}^+},
\end{align}
for any index $\gamma$ or integer $k\in\mathbb{N}$.
\end{proposition}

\Cref{Mas20 corollary} says that if a solution $\mathfrak{S}$ to \fullsystem is such that $\alins_{\mathscr{H}^+}$ and $\ablins_{\mathscr{I}^+}$ are vanishing, the corresponding pair $\alin, \ablin$ are also vanishing on $D^+({\Sigma^*_+})$, which implies that $\mathfrak{S}$ is a pure gauge solution.\\

We begin with the following corollary to \Cref{Horizon gauge fixes gauge}:

\begin{proposition}\label{gauge solution ner i+}
For a given $u_+,v_+\in\mathbb{R}$, assume $\mathscr{S}$ is a solution to \fullsystem on $J^+(\Sigma^*_+)$ which is future Bondi-normalised and satisfies on $J^+(\underline{\mathscr{C}}_{v_+})\cap J^+(\mathscr{C}_{u_+})$ the conditions
\begin{align}
    \lim_{v\longrightarrow\infty}r\xblin(u,v,\theta^A)=0,\qquad \Omega\xlin|_{\mathscr{H}^+}(v,\theta^A)=0,
\end{align}
as well as 
    \begin{alignat}{2}
        &\Olin_{\ell\geq2}|_{\mathscr{H}^+_{\geq0}}=0,\qquad &&\left[\rlin-\rlin_{\ell=0}+\divr\elin\right]|_{\mathscr{H}^+_{\geq0},v=v_+}=0,\qquad \otx|_{\mathscr{H}^+_{\geq0},v=v_+}=0,\\
        &\divo^2\Omega^{-1}\xblin|_{\mathscr{H}^+_{\geq0},v=v_+}=0,\qquad &&\bmlin_{\ell\geq2}|_{\mathscr{H}^+_{\geq0}}=0.
    \end{alignat}
Then on $J^+(\underline{\mathscr{C}}_{v_+})\cap J^+(\mathscr{C}_{u_+})$ the $\ell\geq2$ component of $\mathfrak{S}$ is trivial.
\end{proposition}
\begin{proof}
    The vanishing of $\bmlin_{\ell\geq2}$, $\Omega\xlin$ and $\otx_{\ell\geq2}$ on $\mathscr{H}^+_{\geq0}$ implies that $\glinh$ and $\tr\glin_{\ell\geq2}$ are constant in $v$ on $\mathscr{H}^+_{\geq0}$. The linearised Gauss equation \bref{Gauss} then says that $\rlin_{\ell\geq2}|_{\mathscr{H}^+_{\geq0}}$ is constant in $v$. Similarly, the Codazzi equation \bref{elliptic equation 1} implies that $\Omega\blin_{\ell\geq2}|_{\mathscr{H}^+_{\geq0}}$, which by \bref{Bianchi+1b} says that $\slin_{\ell\geq2}|_{\mathscr{H}^+_{\geq0}}$ is constant in $v$. Thus $\Psilin$, $\Psilinb$ are constant in $v$ on $\mathscr{H}^+_{\geq0}$. The pointwise decay of $r\xblin$ towards $\mathscr{I}^+$ implies that $\int_{u_1}^{u_2}\int_{S^2}d\bar{u}\dw\,|\partial_u\Psilinb|^2$ decays towards $\mathscr{I}^+$ (see \Cref{missing note from Part I}). Thus by the uniqueness clause of Theorem 1 in \cite{Mas20} we have that $\Psilinb=0$ and $\ablin=0$ on $J^+(\underline{\mathscr{C}}_{v_+})\cap J^+(\mathscr{C}_{u_+})$. Thus $\bblin_{\ell\geq2}=0$ and $\xblin=0$ on $J^+(\underline{\mathscr{C}}_{v_+})\cap J^+(\mathscr{C}_{u_+})$. Equation \bref{Bianchi-1b} implies that $\slin_{\ell\geq2}$ is constant in $u,v$ on $\overline{\mathscr{M}}\cap J^+(\underline{\mathscr{C}}_{v_+})\cap J^+(\mathscr{C}_{u_+})$, which then implies using the Regge--Wheeler equation \bref{RW} that $\slin_{\ell\geq2}$ vanishes identically on $J^+(\underline{\mathscr{C}}_{v_+})\cap J^+(\mathscr{C}_{u_+})$. Therefore, $\Psilin=0$ and integrating \bref{hier} we get that $\alin=0$ on $J^+(\underline{\mathscr{C}}_{v_+})\cap J^+(\mathscr{C}_{u_+})$. Thus the solution to \fullsystem on $J^+(\underline{\mathscr{C}}_{v_+})\cap J^+(\mathscr{C}_{u_+})$ is a pure gauge solution by Theorem B.1 of \cite{DHR16}. We can now deduce that $\mathfrak{G}$ is trivial following a similar argument to that of \Cref{Horizon gauge fixes gauge}.
\end{proof}

As the equations \fullsystem are linear, we can apply Theorem I to construct $\alin, \ablin$ and use them to source transport equations for the remaining quantities. We will detail this construction in the following:
\begin{proof}[Proof of \Cref{grand proposition backwards scattering}]
We divide the construction of $\mathfrak{S}$ into steps:\\
\textbf{Constructing data at $\mathscr{H}^+_{\geq0}$:} We are given $\xlins_{\mathscr{H}^+}\in\Gamma^2_c(\mathscr{H}^+_{\geq0})$. Let $v_+\in\mathbb{R}_+$ lies beyond the support of $\xlins_{\mathscr{H}^+}$.  Define
\begin{alignat}{2}
    &\glinhs_{\mathscr{H}^+}=-2\int_v^{v_+}d\bar{v} \,\xlins_{\mathscr{H}^+},\qquad\qquad\quad &&V^{-2}\alins_{\mathscr{H}^+}=-\frac{1}{2M}\partial_V V^{-1}\xlins_{\mathscr{H}^+},\label{backwards radiation fields at H+ 1}\\
    &\blins_{\mathscr{H}^+}:=-\frac{1}{2M}\divo \xlins_{\mathscr{H}^+},\qquad\qquad\quad
    &&\tr\glins_{\mathscr{H}^+}=0,\label{backwards radiation fields at H+ 2}\\ &\rlins_{\mathscr{H}^+}:=-\frac{1}{8M^2}\divo^2\glinhs_{\mathscr{H}^+},\qquad\qquad\quad &&\elins_{\mathscr{H}^+}:=\frac{1}{4M}\divo\glinhs_{\mathscr{H}^+},\label{backwards radiation fields at H+ 3}\\
    &\slins_{\mathscr{H}^+}:=\frac{1}{2M}\curlo\, \elins_{\mathscr{H}^+}=\frac{1}{8M^2}\curlo\divo\glinhs_{\mathscr{H}^+}.\qquad&&\label{backwards radiation fields at H+ 4}
\end{alignat}
We define $\olinbs_{\mathscr{H}^+}$ to be the solution to 
\begin{align}
    \partial_V V\olinbs_{\mathscr{H}^+}=-2M\rlins_{\mathscr{H}^+},\qquad\qquad \olinbs_{\mathscr{H}^+}|_{v=v_+}=0,
\end{align}
which gives
\begin{align}
    \olinbs_{\mathscr{H}^+}=-\frac{1}{8M^2}\int_v^{\infty}d\bar{v}\,e^{-\frac{1}{2M}(v-\bar{v})}\divo^2\glinhs_{\mathscr{H}^+}
\end{align}
Similarly, we define $\xblins_{\mathscr{H}^+}$ via
\begin{align}\label{this 07 07 2021 2}
    \partial_V V\xblins_{\mathscr{H}^+}=\left[2\mathring{\fancydstar_2}\elins_{\mathscr{H}^+}+\xlins_{\mathscr{H}^+}\right],\qquad\qquad \xblins_{\mathscr{H}^+}|_{v=v_+}=0,
\end{align}
which gives
\begin{align}
    \xblins_{\mathscr{H}^+}=-\frac{1}{2M}\int_v^{\infty}d\bar{v}\,e^{-\frac{1}{2M}(v-\bar{v})}\left[2\mathring{\fancydstar_2}\elins_{\mathscr{H}^+}+\xlins_{\mathscr{H}^+}\right],
\end{align}
and we define $\otxbs_{\mathscr{H}^+}$ by
\begin{align}
    \partial_V V\otxbs_{\mathscr{H}^+}+\frac{1}{2M}\otxbs_{\mathscr{H}^+}=-4\divo \elins_{\mathscr{H}^+},\qquad\qquad \otxbs_{\mathscr{H}^+}\Big|_{v=v_+}=0,
\end{align}
which gives
\begin{align}
    \otxbs_{\mathscr{H}^+}=-\frac{2}{M}\int_v^{\infty}d\bar{v}\,e^{-\frac{1}{2M}(v-\bar{v})}\divo\elins_{\mathscr{H}^+}.
\end{align}
Define
\begin{align}
    \bblins_{\mathscr{H}^+}:=\frac{1}{2M}\left[\divo\xblins_{\mathscr{H}^+}-\elins_{\mathscr{H}^+}-\frac{1}{2}\mathring{\slashednabla}\otxbs_{\mathscr{H}^+}\right]
\end{align}
which implies
\begin{align}
    \partial_V V\bblins_{\mathscr{H}^+}=-\frac{3}{2M}\elins_{\mathscr{H}^+}+2M\mathring{\fancydstar_1}(\rlins_{\mathscr{H}^+},\slins_{\mathscr{H}^+}).
\end{align}
Finally, define $\ablins_{\mathscr{I}^+}$ to be the solution to
\begin{align}
    \partial_V V^2\ablins_{\mathscr{H}^+}=2\mathring{\fancydstar_2}V\bblins_{\mathscr{H}^+}+\frac{3}{2M}V\xblins_{\mathscr{H}^+},\qquad\qquad \ablins_{\mathscr{H}^+}|_{v=v_+}=0.
\end{align}
\textbf{Constructing data at $\mathscr{I}^+$}: We are given data $\xblins_{\mathscr{I}^+}\in\Gamma^2_c(\mathscr{I}^+)$ and vanishing $\Olinos_{\mathscr{I}^+}$, $\bmlins_{\mathscr{I}^+}$. Define
\begin{alignat}{2}
    &\glinhs_{\mathscr{I}^+}=-2\int_{u}^{u_+}d\bar{u}\,\xblins_{\mathscr{I}^+},\qquad\qquad &&\ablins_{\mathscr{I}^+}=\partial_u\xblins_{\mathscr{I}^+}\\
    &\bblins_{\mathscr{I}^+}=-\frac{1}{2}\divo\xblins_{\mathscr{I}^+},\qquad\qquad &&\slins_{\mathscr{I}^+}=-\frac{1}{2}\curlo\divo\glinhs_{\mathscr{I}^+}.
\end{alignat}
Define
\begin{align}
    \xlins_{\mathscr{I}^+}=-\frac{1}{2}\glinhs_{\mathscr{I}^+},
\end{align}
so that
\begin{align}
    \partial_u\xlins_{\mathscr{I}^+}=-\xblins_{\mathscr{I}^+},\qquad\qquad\xlins_{\mathscr{I}^+}|_{u=u_+}=0,\label{this 07 07 2021 3}.
\end{align}
i.e.~so that $\xlins_{\mathscr{I}^+}=0$ on $u\geq u_+$. Similarly, we make the following definitions for the remaining radiation fields on $\mathscr{I^+}$ so that they vanish on $u\geq u_+$ and are related by the relations
\begin{alignat}{2}
    &\rlins_{\mathscr{I}^+}=-\frac{1}{2}\divo^2\glinhs_{\mathscr{I}^+}\qquad\qquad&&\eblins_{\mathscr{I}^+}=-\frac{1}{2}\divo\glins_{\mathscr{I}^+},\qquad\\&\blins_{\mathscr{I}^+}=\int^{\infty}_ud\bar{u}\,\mathring{\fancydstar_1}\mathring{\fancyd_1}\mathring{\fancyd_2}\glinhs_{\mathscr{I}^+},\qquad\qquad\qquad\olins_{\mathscr{I}^+}=&&\int^\infty_u d\bar{u}\,\rlins_{\mathscr{I}^+}=-\frac{1}{2}\int^\infty_ud\bar{u}\,\divo^2\glinhs_{\mathscr{I}^+}.
\end{alignat}
We define $\alins_{\mathscr{I}^+}$ to be the solution to
\begin{align}
    \partial_u\alins_{\mathscr{I}^+}=-2\mathring{\fancydstar_2}\blins_{\mathscr{I}^+}+6M\xlins_{\mathscr{I}^+},\qquad\qquad \alins_{\mathscr{I}^+}|_{u=u_+}=0,
\end{align}
which gives
\begin{align}\label{alins at scri+ backwards scattering}
     \alins_{\mathscr{I}^+}=\int^{\infty}_u d\bar{u}\int^{\infty}_{\bar{u}}d\bar{\bar{u}}\,\left[{2\mathring{\fancydstar_2}\mathring{\fancydstar_1}\mathring{\fancyd_1}\mathring{\fancyd_2}}\,\glinhs_{\mathscr{I}^+}6M\partial_u\glinhs_{\mathscr{I}^+}\right].
\end{align}
The remaining quantities, $\tr\glins_{\mathscr{I}^+}$, $\otxbs_{\mathscr{I}^+}$, $\otxs_{\mathscr{I}^+}$, $\olinbs_{\mathscr{I}^+}$ are all taken to vanish.\\

\textbf{Constructing $\alin$, $\ablin$ $\xlin$, $\xblin$, $\blin$, $\bblin$, $\elin$ and $\eblin$}: With data $\alins_{\mathscr{H}^+}\in\Gamma^{(2)}_c(\mathscr{H}^+_{\geq0})$, $\ablins_{\mathscr{I}^+}\in\Gamma^{(2)}_c(\mathscr{I}^+)$, we may apply \Cref{Mas20 corollary} to obtain $\alin$ that satisfies the $+2$ Teukolsky equation \bref{T+2} and $\ablin$ that satisfies the $-2$ Teukolsky equation \bref{T-2}, such that $(\alin,\ablin)$ satisfy the Teukolsky--Starobinsky identities \bref{eq:TS-}, \bref{eq:TS+}, each of $\ablin$ and $\alin$ realise their radiation fields on $\mathscr{I}^+$, $\mathscr{H}^+_{\geq0}$ respectively, and
\begin{align}\label{this 07 07 2021}
\alin=\ablin=0
\end{align}
on $J^+(\mathscr{C}_{u_+})\cap J^+(\underline{\mathscr{C}}_{v_+})$.\\

Having constructed $\alin, \ablin$, we may solve for $\xlin, \xblin$ via \bref{D4Chihat},
\begin{align}
    \nablav \frac{r^2\xlin}{\Omega}=-r^2\alin,\qquad\qquad \nablau \frac{r^2\xblin}{\Omega}=-r^2\ablin,
\end{align}
with data $\xlins_{\mathscr{I}^+}$ and $\xblins_{\mathscr{H}^+}$ respectively. We may similarly construct $\blin, \bblin$ using \bref{Bianchi+1a}, \bref{Bianchi-1a},
\begin{align}\label{bianchi +1a'}
    \nablav \frac{r^4\blin}{\Omega}=\divo r^3\alin,\qquad\qquad \nablau \frac{r^4\bblin}{\Omega}=-\divo r^3\ablin,
\end{align}
with data $\blins_{\mathscr{I}^+}$ and $\bblins_{\mathscr{H}^+}$.\\

Note that our construction of data is such that  $\xblins_{\mathscr{H}^+}$, $\bblins_{\mathscr{H}^+}$ vanish on $v\geq v_+$ which, given \bref{this 07 07 2021}, implies that 
\begin{align}
    \xlin=\xblin=0,\qquad \blin=\bblin=0
\end{align}
on $J^+(\mathscr{C}_{u_+})\cap J^+(\underline{\mathscr{C}}_{v_+})$. 
We can trace this gauge choice to the initial conditions used to solve for $\xlins_{\mathscr{I}^+}$, $\xblins_{\mathscr{H}^+}$, $\blins_{\mathscr{I}^+}$, $\bblins_{\mathscr{H}^+}$ (e.g.~the initial condition on $\xblins_{\mathscr{H}^+}$ in \bref{this 07 07 2021 2}).\\

It is clear that $\lim_{v\longrightarrow\infty}r^2\xlin(u,v,\theta^A)=\xlins_{\mathscr{I}^+}(u,\theta^A)$ and $\lim_{u\longrightarrow\infty}\Omega^{-1}\xblin(u,v,\theta^A)=\xblins_{\mathscr{H}^+}(v,\theta^A)$. To see that $\lim_{u\longrightarrow\infty} \Omega\xlin(u,v,\theta^A)=\xlins_{\mathscr{H}^+}(v,\theta^A)$, note that
\begin{itemize}
    \item $\Omega^2\alin(u,v,\theta^A)$ converges uniformly as $u\longrightarrow\infty$ by \Cref{Mas20 corollary}, and so we have $\slashednabla_V \frac{r^3}{2M}e^{\frac{r\Omega^2}{2M}}V^{-1}\Omega\xlin$ also converges uniformly as $u\longrightarrow\infty$,
    \item both $\xlins_{\mathscr{H}^+}=0$ on $v\geq v_+$ and $\Omega\xlin$ vanishes on $J^+(\mathscr{C}_{u_+})\cap J^+(\underline{\mathscr{C}}_{v_+})$.
\end{itemize}
A similar argument applies to $\xblin$ to show that $\lim_{v\longrightarrow\infty} r\xblin(u,v,\theta^A)=\xblins_{\mathscr{I}^+}(u,v,\theta^A)$. We also see in an identical manner that $\blin$, $\bblin$ realise their radiation data on both $\mathscr{H}^+_{\geq0}$, $\mathscr{I}^+$. Commuting $\xlin, \xblin, \blin, \bblin$ with $\mathring{\slashednabla}^\gamma \slashednabla_t^k$ for any index $\gamma$, $k\in\mathbb{N}$ and using Theorem I from \cite{Mas20} in the same manner as above tells us that
\begin{alignat}{2}
    &\lim_{v\longrightarrow\infty} \mathring{\slashednabla}^\gamma \slashednabla_t^k r\xblin(u,v,\theta^A)=\mathring{\slashednabla}^\gamma \partial_u^k \xblins_{\mathscr{I}^+}(u,\theta^A),\qquad\qquad && \lim_{v\longrightarrow\infty} \mathring{\slashednabla}^\gamma \slashednabla_t^k r^2\bblin(u,v,\theta^A)=\mathring{\slashednabla}^\gamma \partial_u^k \bblins_{\mathscr{I}^+}(u,\theta^A),\\
    &\lim_{u\longrightarrow\infty} \mathring{\slashednabla}^\gamma \slashednabla_t^k \Omega\xlin(u,v,\theta^A)=\mathring{\slashednabla}^\gamma \partial_v^k \xlins_{\mathscr{I}^+}(v,\theta^A),\qquad\qquad&&\lim_{u\longrightarrow\infty} \mathring{\slashednabla}^\gamma \slashednabla_t^k \Omega\blin(u,v,\theta^A)=\mathring{\slashednabla}^\gamma \partial_v^k \blins_{\mathscr{I}^+}(v,\theta^A).
\end{alignat}

To show that $\lim_{u\longrightarrow\infty}\nablav \Omega^{-1}\xblin=\partial_v \xblins_{\mathscr{H}^+}$, note that since $\nablau\nablav\frac{r^2\xblin}{\Omega}=-\nablav r^2\ablin$ and since $\Omega^{-2}\ablin, \nablav\ r^2\Omega^{-2}\ablin$ converge uniformly towards $\mathscr{H}^+_{\geq0}$, we see by integrating in $u$ that $\nablav \Omega^{-1}\xblin(u_n,v,\theta^A)$ converges uniformly for any $(u_n)_n$ with $u_n\longrightarrow\infty$. The result now follows since both $\partial_v\xblins_{\mathscr{H}^+}$ and $\lim_{u\longrightarrow\infty}\nablav \Omega^{-1}\xblin(u,v,\theta^A)$ vanish on $v\geq v_+$. We can repeat the argument above commuting with powers of $\nablav$, $\nablau$ and $\mathring{\slashednabla}$ to obtain higher order statements showing that $\xlin, \xblin$ are smooth. We can similarly show that $\lim_{v\longrightarrow\infty} \mathring{\slashednabla}^\gamma\nablau^k \frac{r^2\xlin}{\Omega}=\mathring{\slashednabla}^\gamma\partial_u^k\xlins_{\mathscr{I}^+}$ for any index $\gamma$, integer $k\in\mathbb{N}$. Applying the above to \bref{bianchi +1a'}, we obtain the relevant statements for $\blin, \bblin$.\\

We may now check that the equations \bref{Bianchi+2}, \bref{Bianchi-2} are satisfied. Note that the $+2$ Teukolsky equation \bref{T+2} satisfied by $\alin$ implies, together with \bref{D4Chihat}, \bref{Bianchi+1a}, that
\begin{align}
    \nablav\left[\frac{r^4}{\Omega^4}\nablau r\Omega^2\alin+2\mathring{\fancydstar_2}\frac{r^4\blin}{\Omega}-6M\frac{r^2\xlin}{\Omega}\right]=0.
\end{align}
and so $\alin, \xlin, \blin$ satisfy \bref{Bianchi+2} since
\begin{align}
    \lim_{v\longrightarrow\infty}\left[\frac{r^4}{\Omega^4}\nablau r\Omega^2\alin+2\mathring{\fancydstar_2}\frac{r^4\blin}{\Omega}-6M\frac{r^2\xlin}{\Omega}\right]=\partial_u \alins_{\mathscr{I}^+}+2\mathring{\fancydstar_2}\blins_{\mathscr{I}^+}-6M\xlins_{\mathscr{I}^+}=0.
\end{align}
A similar argument shows that $\ablin, \xblin, \bblin$ satisfy \bref{Bianchi-2}.\\

Now we define $\elin,\eblin$ up to $\ell\geq2$ from $\xlin,\xblin$ via \bref{D3Chihat}, \bref{D4Chihatbar}. Using \bref{D3Chihat}, \bref{D4Chihatbar}, \bref{D4Chihat} and \bref{Bianchi+2} we can check that equations \bref{D3etabar}, \bref{D4eta} are satisfied.  Taking the limit of \bref{D3Chihat} towards $\mathscr{I}^+$, it is clear that $r\elin\longrightarrow0$ as $v\longrightarrow\infty$. Similarly, equation \bref{D4Chihatbar} on $\mathscr{H}^+$ tells us that $\eblin_{\ell\geq2}|_{\mathscr{H}^+}=\eblins|_{\mathscr{H}^+}$. To see that $\elin_{\ell\geq2}|_{\mathscr{H}^+}=\elins_{\mathscr{H}^+}$, note that by taking the limit of \bref{D4eta} towards $\mathscr{H}^+_{\geq0}$, we find that $\nablav \elin_{\ell\geq2}\longrightarrow \partial_v\elins_{\mathscr{H}^+}$, thus $\elin_{\ell\geq2}\longrightarrow\elins_{\mathscr{H}^+}$ since both $\elins_{\mathscr{H}^+}$ and $\lim_{u\longrightarrow\infty}\elin_{\ell\geq2}$ vanish on $v\geq v_+$. A similar argument shows that $r^2\eblin_{\ell\geq2}\longrightarrow \eblins_{\mathscr{I}^+}$ as $v\longrightarrow\infty$. Given that $\xlin$, $\xblin$, $\blin$, $\bblin$ are smooth, commuting the above argument with powers of $\nablau, \nablav, \mathring{\slashednabla}$ shows that $\elin$, $\eblin$ are smooth and that
\begin{align}\label{23 10 2022}
     &\lim_{v\longrightarrow\infty} \mathring{\slashednabla}^\gamma \slashednabla_t^k r^2\eblin(u,v,\theta^A)=\mathring{\slashednabla}^\gamma \partial_u^k \eblins_{\mathscr{I}^+}(u,\theta^A),\qquad\qquad && \lim_{u\longrightarrow\infty} \mathring{\slashednabla}^\gamma \slashednabla_t^k \elin(u,v,\theta^A)=\mathring{\slashednabla}^\gamma \partial_v^k \elins_{\mathscr{I}^+}(v,\theta^A).
\end{align}

\textbf{Propagating the Codazzi equations}: We may now modify the argument presented in the proof of \Cref{EinsteinWP} to conclude that equations \bref{equation of C}, \bref{equation of Codazzi+}, \bref{equation of Codazzi-} are satisfied as follows: Given that data for the system \bref{equation of C}, \bref{equation of Codazzi+}, \bref{equation of Codazzi-} is trivial on $\mathscr{H}^+_{\geq0}$, $\underline{\mathscr{C}}_{v_+}\cap\{u\geq u_+\}$, the system \bref{equation of C}, \bref{equation of Codazzi+}, \bref{equation of Codazzi-} holds on $D^-(\mathscr{H}^+_{\geq0}\cup \underline{\mathscr{C}}_{v_+}\cap\{u\geq u_+\})$. To propagate the constraints near $\mathscr{I}^+$ we prove the following lemma:

\begin{lemma}
    There exists $v_\infty\in\mathbb{R}$ such that the system \bref{equation of C}, \bref{equation of Codazzi+}, \bref{equation of Codazzi-} admits only the trivial solution on $J^+(\Sigma^*_+)\cap \{v\geq v_\infty\}$.
\end{lemma}
\begin{proof}
    We can use equation \ref{Codazzi+} to deduce that $r^3\partial_v r^2\text{C}(u,v,\theta^A)$ as $v\longrightarrow\infty$, since $r^3\nablav r^2\xlin$, $r^2\nablav r^4\blin$ all converge towards $\mathscr{I}^+$, and by \bref{23 10 2022} we know that $r^2\nablav\eblin_{\ell\geq2}$ also converges towards $\mathscr{I}^+$. Integrating \bref{Codazzi+} we get
    \begin{align}
        r^2\text{C}(u,v,\theta^A)=-\frac{r^3}{\Omega}\text{Codazzi}^+(u,v,\theta^A)+\int_{v}^\infty d\bar{v} \,r^2\Omega\text{Codazzi}^+(u,\bar{v},\theta^A),
    \end{align}
    which implies
    \begin{align}
        \left|r^2\text{C}(u,v,\theta^A)\right|\leq \left|\frac{r^3}{\Omega}\text{Codazzi}^+(u,v,\theta^A)\right|+\frac{1}{r(u,v)}\sup_{\bar{v}\in[v,\infty)}\left|\frac{r^4}{\Omega}\text{Codazzi}^+(u,\bar{v},\theta^A)\right|.
    \end{align}
    We then have that $r^3\text{C}(u,v,\theta^A)\longrightarrow0$ as $v\longrightarrow\infty$. Thus $r^2\nablav r^3 \text{C}$ decays, and we can commute the above argument by $\partial_t$ to find that $\partial_u^i r^n \text{C}\sim \frac{1}{r^{4-n}}$ towards $\mathscr{I}^+$ for $n\in\{0,1,2,3,4\}$ and $i=0,1$.

    Define
    \begin{align}
        \text{F}^+:=\frac{r^4}{\Omega}\text{Codazzi}^++r^3\text{C},\qquad \text{F}^-:=\frac{r^4}{\Omega}\text{Codazzi}^--r^3\text{C}.
    \end{align}
    The system \bref{equation of C}, \bref{equation of Codazzi+}, \bref{equation of Codazzi-} implies
    \begin{align}\label{equation of r3C}
        \partial_u\partial_v r^3\text{C}-\frac{\Omega^2}{r^2}\mathring{\slashed{\Delta}} r^3\text{C}-\frac{\Omega^2}{r^2}(\Omega^2+1)r^3\text{C}=\frac{1}{r}\left(\partial_u \Omega^2\text{F}^++\partial_v\Omega^2\text{F}^-\right),
    \end{align}
    \begin{align}
        \refstepcounter{equation}\latexlabel{equation of F+}
        \refstepcounter{equation}\latexlabel{equation of F-}
        \partial_v \text{F}^+=r^2\Omega^2\text{C},\qquad\qquad \partial_u \text{F}^-=r^2\Omega^2\text{C}\tag{\ref{equation of F+},\;\ref{equation of F-}}
    \end{align}
    For given $u,v$, we multiply \bref{equation of r3C} by $2r^2\partial_tr^3\text{C}$ and integrate by parts in the region $\mathscr{D}_{u,v}^{u_+,\infty}$ to get
    \begin{align}\label{T energy for constraints}
        \begin{split}
            &\int_{\mathscr{C}_{u}\cap\{\bar{v}\geq v\}}d\bar{v}\dw\,r^2|\partial_v r^3\text{C}|^2+\Omega^2|\mathring{\slashednabla} r^3\text{C}|^2+\int_{\mathscr{C}_v\cap\{\bar{u}\in[u,u_+]\}}d\bar{u}\dw\,r^2|\partial_u r^3\text{C}|^2+\Omega^2|\mathring{\slashednabla} r^3\text{C}|^2\\
            &+\int_{\mathscr{D}_{u,v}^{u_+,\infty}}d\bar{u}d\bar{v}\dw\,2r\Omega^2|\partial_u r^3\text{C}|^2\\&=\int_{\mathscr{D}_{u,v}^{u_+,\infty}}d\bar{u}d\bar{v}\dw\,2r\Omega^2|\partial_v r^3\text{C}|^2+\int_{\mathscr{D}_{u,v}^{u_+,\infty}}d\bar{u}d\bar{v}\dw\,2r\partial_t r^3\text{C}\times \left[\partial_u \left(\Omega^2\text{F}^+\right)+\partial_v \left(\Omega^2\text{F}^-\right)\right]
        \end{split}
    \end{align}
    We estimate the last term on the right hand side above as follows: the term
    \begin{align}
        \begin{split}
            \int_{\mathscr{D}_{u,v}^{u_+,\infty}}d\bar{u}d\bar{v}\dw\,r\partial_u r^3\text{C}\times&\partial_u\Omega^2\text{F}^+
         \end{split}
    \end{align}
    evaluates to
    \begin{align}\label{temp label 1}
    \begin{split}         
            \int_{\mathscr{D}_{u,v}^{u_+,\infty}}d\bar{u}d\bar{v}\dw\,r\partial_u r^3\text{C}\times \left[-\Omega^2\int_{\bar{v}}^{\infty}d\bar{\bar{v}}\frac{\Omega^2}{r}\partial_ur^3\text{C}+r\Omega^2(2\Omega^2-1)\text{C}+\frac{2M\Omega^2}{r^2}\int_{\bar{v}}^{\infty}d\bar{\bar{v}}r^2\Omega^2\text{C}\right].
        \end{split}
    \end{align} 
    The first term in \bref{temp label 1} is bounded via
    \begin{align}
        \begin{split}
             \int_v^\infty d\bar{v}\int_u^{u_+}d\bar{u}\dw\, r\Omega^2\partial_u r^3\text{C}\int_{\bar{v}}^{\infty}d\bar{\bar{u}}\frac{\Omega^2}{r}\partial_u r^3\text{C}\leq&\; \epsilon_1\int_{\mathscr{D}_{u,v}^{u_+,\infty}}d\bar{u}d\bar{v}\dw\,r|\partial_u r^3\text{C}|^2\\&+\frac{1}{4\epsilon_1}\int_v^\infty d\bar{v}\int_u^{u_+}d\bar{u}\dw\,r\left(\int_{\bar{v}}^\infty \frac{\Omega^2}{r}\partial_u r^3\text{C}\right)^2
        \end{split}
    \end{align}
    \begin{align}\label{temp label eins}
       \int_v^\infty d\bar{v}\int_u^{u_+}d\bar{u}\dw\, r\Omega^2\partial_u r^3\text{C}\int_{\bar{v}}^{\infty}d\bar{\bar{u}}\frac{\Omega^2}{r}\partial_u r^3\text{C} \lesssim\left(\epsilon_1+\frac{1}{4\epsilon_1 r(u_+,v)}\right)\int_{\mathscr{D}_{u,v}^{u_+,\infty}}d\bar{u}d\bar{v}\dw\,r|\partial_u r^3\text{C}|^2.
    \end{align}
    The second term in \bref{temp label 1} is bounded via
    \begin{align}\label{temp label zwei}
        \begin{split}
            &\int_{\mathscr{D}_{u,v}^{u_+,\infty}}d\bar{u}d\bar{v}\dw\,r\partial_u r^3\text{C}\times\Omega^2\int_{\bar{v}}^\infty d\bar{\bar{v}} r\Omega^2(2\Omega^2-1)\text{C}\\&\leq \epsilon_2 \int_{\mathscr{D}_{u,v}^{u_+,\infty}}d\bar{u}d\bar{v}\dw\,r|\partial_u r^3\text{C}|^2+\frac{1}{4\epsilon_2}\int_{\mathscr{D}_{u,v}^{u_+,\infty}}d\bar{u}d\bar{v}\dw\,r\Omega^2\left(\int_{\bar{v}}^\infty d\bar{\bar{v}}2\frac{\Omega^2}{r^2}|r^3\text{C}|\right)^2
            \\&\lesssim\epsilon_2 \int_{\mathscr{D}_{u,v}^{u_+,\infty}}d\bar{u}d\bar{v}\dw\,r|\partial_u r^3\text{C}|^2+\frac{1}{4\epsilon_2}\int_u^{u_+}\int_v^\infty d\bar{u}d\bar{v}\dw \frac{1}{r^3}\int_{\bar{v}}^\infty d\bar{\bar{v}}|r^3\text{C}|^2\\
            &\lesssim \epsilon_2 \int_{\mathscr{D}_{u,v}^{u_+,\infty}}d\bar{u}d\bar{v}\dw\,r|\partial_u r^3\text{C}|^2+\frac{1}{4\epsilon_2\times r(u_+,v)^2}\int_{\mathscr{D}_{u,v}^{u_+,\infty}}d\bar{u}d\bar{v}\dw\,r^2|\partial_v r^3\text{C}|^2.
        \end{split}
    \end{align}
    Note that in the last step above we applied a Hardy estimate knowing that $r|r^3\text{C}|^2\longrightarrow0$ as $v\longrightarrow \infty$. A similar procedure gives a bound on the third term in \bref{temp label 1} via
    \begin{align}\label{temp label drei}
        \begin{split}
            &\int_{\mathscr{D}_{u,v}^{u_+,\infty}}d\bar{u}d\bar{v}\dw\,r\partial_u r^3\text{C}\times\frac{2M\Omega^2}{r^2}\int_{\bar{v}}^\infty d\bar{\bar{v}}r^2\Omega^2\text{C}\\
            &\lesssim \epsilon_3 \int_{\mathscr{D}_{u,v}^{u_+,\infty}}d\bar{u}d\bar{v}\dw\,r|\partial_u r^3\text{C}|^2+\frac{1}{4\epsilon_3\times r(u_+,v)^3}\int_{\mathscr{D}_{u,v}^{u_+,\infty}}d\bar{u}d\bar{v}\dw\,r^2|\partial_v r^3\text{C}|^2.
        \end{split}
    \end{align}
    An identical procedure gives
    \begin{align}\label{temp label vier}
    \begin{split}
        &\int_{\mathscr{D}_{u,v}^{u_+,\infty}}d\bar{u}d\bar{v}\dw\,r\partial_ur^3\text{C}\times \partial_v \Omega^2 \text{F}^-\\&=-\int_{\mathscr{D}_{u,v}^{u_+,\infty}}d\bar{u}d\bar{v}\dw\, r\partial_u r^3\text{C}\times\Omega^2\left[\frac{2M}{r^2}\int_{\bar{u}}^{u_+}d\bar{\bar{u}}\frac{\Omega^2}{r}r^3\text{C}+\int_{\bar{u}}^{u_+}d\bar{\bar{u}} \frac{\Omega^2}{r}\partial_v r^3\text{C}-\int_{\bar{u}}^{u_+}d\bar{\bar{u}}\frac{\Omega^2}{r^2}(2\Omega^2-1)r^3\text{C}\right]\\
        &\lesssim \epsilon_4\int_{\mathscr{D}_{u,v}^{u_+,\infty}}d\bar{u}d\bar{v}\dw\, r|\partial_u r^3\text{C}|^2+\frac{1}{4\epsilon_4 r(u_+,v)^3}\int_{\mathscr{D}_{u,v}^{u_+,\infty}}d\bar{u}d\bar{v}\dw\,r^2|\partial_v r^3\text{C}|^2
    \end{split}
    \end{align}
We now turn to the integral of $r\partial_v r^3\text{C}\times \partial_u \Omega^2 \text{F}^+$:
\begin{align}\label{temp label 2}
    \begin{split}
        \int_{\mathscr{D}_{u,v}^{u_+,\infty}}d\bar{u}d\bar{v}\dw\,r\partial_vr^3\text{C}\times \left[-\Omega^2\int_{\bar{v}}^{\infty}d\bar{\bar{v}}\frac{\Omega^2}{r}\partial_ur^3\text{C}+r\Omega^2(2\Omega^2-1)\text{C}+\frac{2M\Omega^2}{r^2}\int_{\bar{v}}^{\infty}d\bar{\bar{v}}r^2\Omega^2\text{C}\right].
    \end{split}
\end{align}
We estimate the first term in \bref{temp label 2} via
\begin{align}\label{temp label funf}
    \begin{split}
    &\int_{\mathscr{D}_{u,v}^{u_+,\infty}}d\bar{u}d\bar{v}\dw\,r\partial_vr^3\text{C}\times\Omega^2\int_{\bar{v}}^{\infty}d\bar{\bar{v}}\frac{\Omega^2}{r}\partial_ur^3\text{C}\\
    &\lesssim \epsilon_5 \int_{\mathscr{D}_{u,v}^{u_+,\infty}}d\bar{u}d\bar{v}\dw\,\left(\int_{\bar{v}}^\infty d\bar{\bar{v}}\frac{\Omega^2}{r}|\partial_u r^3\text{C}|\right)^2+\frac{1}{4\epsilon_5}\int_{\mathscr{D}_{u,v}^{u_+,\infty}}d\bar{u}d\bar{v}\dw\, r^2|\partial_v r^3\text{C}|^2\\
    &\lesssim \frac{\epsilon_5}{r(u_+,v)^2} \int_{\mathscr{D}_{u,v}^{u_+,\infty}}d\bar{u}d\bar{v}\dw\,r|\partial_u r^3\text{C}|^2+\frac{1}{4\epsilon_5}\int_{\mathscr{D}_{u,v}^{u_+,\infty}}d\bar{u}d\bar{v}\dw\,r^2|\partial_vr^3\text{C}|^2.
    \end{split}
\end{align}
The second term and third terms in \bref{temp label 2} can be bounded by applying Young's and Hardy's inequalities to obtain the bound
\begin{align}
    \begin{split}
        &\int_{\mathscr{D}_{u,v}^{u_+,\infty}}d\bar{u}d\bar{v}\dw\,r\partial_vr^3\text{C} \int_{\bar{v}}^\infty d\bar{\bar{v}}\frac{\Omega^2}{r^2}r^3\text{C}+\int_{\mathscr{D}_{u,v}^{u_+,\infty}}d\bar{u}d\bar{v}\dw\,\frac{2M\Omega^2}{r}\partial_vr^3\text{C} \int_{\bar{v}}^\infty  d\bar{\bar{v}}r^2\Omega^2\text{C}\\&\lesssim  \int_{\mathscr{D}_{u,v}^{u_+,\infty}}d\bar{u}d\bar{v}\dw\,r^2|\partial_vr^3\text{C}|^2.
    \end{split}
\end{align}
Similarly, we can show
\begin{align}
    \begin{split}
        &\int_{\mathscr{D}_{u,v}^{u_+,\infty}}d\bar{u}d\bar{v}\dw\,r \partial_v r^3\text{C}\times\Omega^2\left[\frac{2M}{r^2}\int_{\bar{u}}^{u_+}d\bar{\bar{u}}\frac{\Omega^2}{r}r^3\text{C}+\int_{\bar{u}}^{u_+}d\bar{\bar{u}} \frac{\Omega^2}{r}\partial_v r^3\text{C}-\int_{\bar{u}}^{u_+}d\bar{\bar{u}}\frac{\Omega^2}{r^2}(2\Omega^2-1)r^3\text{C}\right]\\ 
        &\lesssim \int_{\mathscr{D}_{u,v}^{u_+,\infty}}d\bar{u}d\bar{v}\dw\,r^2|\partial_vr^3\text{C}|^2.
    \end{split}
\end{align}
    Choosing $\epsilon_1, \epsilon_2, \epsilon_3, \epsilon_4, \epsilon_5$ small enough and $v$ large enough, we can absorb the spacetime integrals of $r|\partial_u r^3\text{C}|^2$ in \bref{temp label eins}, \bref{temp label zwei}, \bref{temp label drei}, \bref{temp label vier} and \bref{temp label funf} by the left hand side of \bref{temp label 1}. We may now conclude with a Gr\"onwall inequality applied to the resulting estimate,
    \begin{align}
        \begin{split}
            \int_{\mathscr{C}_u}d\bar{v}\dw\, r^2|\partial_v r^3\text{C}|^2\lesssim \int_u^{u_+} d\bar{u}\int_{\mathscr{C}_{\bar{u}}} d\bar{v}\dw\, r^2|\partial_v r^3\text{C}|^2.
        \end{split}
    \end{align}
\end{proof}

We can now conclude

\begin{corollary}
The equations \bref{equation of C}, \bref{equation of Codazzi+}, \bref{equation of Codazzi-} are satisfied throughout $J^+(\Sigma^*_+)$.
\end{corollary}

\textbf{Constructing the remaining components}: Now we can define $\Olin_{\ell\geq2}$ via \bref{elin eblin Olin}. It is easy to check that $\lim_{u\longrightarrow\infty}\Olin=0$,\, $\lim_{v\longrightarrow\infty}\Olino=0$. Define $\glinh$ by solving \bref{metric transport in 3 direction traceless} with data $\glinhs_{\mathscr{H}^+}$, and define $\bmlin$ up to $\ell\geq2$ via \bref{metric transport in 4 direction traceless}. Taking the $\partial_u$-derivative of \bref{metric transport in 4 direction traceless} and using \bref{D3Chihat} and  \bref{D4Chihatbar}, we see that $\bmlin_{\ell\geq2}$ satisfies \bref{partial_u b}. Applying the arguments above to the limit of \bref{metric transport in 3 direction traceless} towards $\mathscr{I}^+$, we obtain that $r\glinh(u,v,\theta^A)\longrightarrow\glinhs_{\mathscr{I}^+}(u,\theta^A)$ and we can easily deduce that $\glinh$ is smooth from the fact that $\xblin$ is smooth.\\

We now define $\otxb$, $\otx$ via \bref{elliptic equation 1}, \bref{elliptic equation 2},  respectively. It is clear that $\otxb$, $\otx$ are smooth and each of them realises its radiation data on each of $\mathscr{H}^+$ and $\mathscr{I}^+$. It is now easy to check that equations \bref{D4TrChi} and \bref{D3TrChi} are satisfied.\\

Now define $\rlin, \slin$ up to $\ell\geq2$ by \bref{Bianchi-1b}. It is easy to check that $\rlin_{\ell\geq2}, \slin_{\ell\geq2}$ realise their radiation data at $\mathscr{I}^+$. As for radiation on $\mathscr{H}^+_{\geq0}$, we compute
\begin{align}
    \nablav \frac{r^2}{\Omega^2}\nablau r^2\Omega\blin=-\mathring{\fancydstar_2}\mathring{\fancyd_1}r^2\Omega\blin+12M r\Omega\blin+6M\divo\Omega\xlin,
\end{align}
which, using the Codazzi equation \bref{elliptic equation 2}, gives us \bref{Bianchi+0}, \bref{Bianchi+0*} for the $\partial_v$-derivatives of $\rlin, \slin$. Thus $\partial_v r^3\rlin_{\ell\geq2}, \partial_v r^3\slin_{\ell\geq2}$ converge uniformly as $u\longrightarrow\infty$, which implies the same for $\rlin_{\ell\geq2}, \slin_{\ell\geq2}$ given that both vanish on $\mathscr{H}^+_{\geq0}\cap\{v\geq v_+\}$.\\

Equation \bref{Bianchi-1b} implies that $\Psilin$ is given by \bref{expression for Psilin}. To see that \bref{expression for Psilinb} is also satisfied, note that $\Psilin-\Psilinb+4\mathring{\fancydstar_2}\mathring{\fancydstar_1}(0,r^3\slin)$ vanishes in the limit towards both $\mathscr{I}^+$ and $\mathscr{H}^+$. Since this expression also satisfies the Regge--Wheeler equation, Theorem 4.1.2 of \cite{Mas20} implies that $\Psilin-\Psilinb+4\mathring{\fancydstar_2}\mathring{\fancydstar_1}(0,r^3\slin)=0$ everywhere on $J^+({\Sigma^*_+})$. We then have that \bref{Bianchi+1b}, and subsequently \bref{Bianchi-0}, \bref{Bianchi-0*}, as well as \bref{D4TrChi} are satisfied.\\

Now define $\tr\glin$ up to $\ell\geq2$ via \bref{Gauss}. It is easy to check that \bref{metric transport in 3 direction trace}, \bref{metric transport in 4 direction trace} are satisfied and that $r\tr\glin\longrightarrow 0$ towards $\mathscr{H}^+_{\geq0}$, $\mathscr{I}^+$.\\

Finally, we now perform a gauge transformation to ensure that $\mathfrak{S}$ satisfies the $\mathscr{H}^+_{\geq0}$ gauge conditions. Let $C$ be the quantity
\begin{align}
    C:=-\frac{1}{2M}\int_{v=0}^{v_+} d{v}\,e^{-\frac{1}{2M}(v_+-{v})}\left[2\,\divo^2\mathring{\fancydstar_2}\elins_{\mathscr{H}^+}+\divo^2\xlins_{\mathscr{H}^+}\right].
\end{align}
Now define the pure gauge solution $\mathfrak{G}_i$ via \Cref{outwards gauge solutions} by
\begin{align}\label{13 11 2022}
    f(v)=-Ce^{-\frac{1}{2M}v},
\end{align}
It is easy to check that the solution $\mathfrak{S}+\mathfrak{G}_i$ is both future Bondi and future horizon normalised.
\end{proof}

\subsection{Boundedness estimates and asymptotic flatness near $i^0$}\label{Section asymptotic flatness near i0 forwards scattering}

In what follows, we derive boundedness estimates on the solution $\mathfrak{S}$ constructed in \Cref{grand proposition backwards scattering} in terms of scattering data:

\subsubsection[Boundedness on $\protect\Psilin$, $\protect\Psilinb$]{Boundedness on $\Psilin$, $\Psilinb$}\label{Boundedness of Psilin Psilinb backwards scattering}

Our strategy here is as follows: Near $\mathscr{I}^+$ we will use a backwards $r^p$-estimate to obtain boundedness on $\Psilin$.

Near the event horizon we will employ a \textit{blueshift} estimate: an estimate on the transverse null derivatives, $\Omega^{-1}\nablagml\Psilin, \Omega^{-1}\nablagml\Psilinb$, with a constant that diverges exponentially in $v_+$, the future cutoff on scattering data on $\mathscr{H}^+_{\geq0}$. 

Finally, $\partial_t$-energy conservation will suffice to estimate $\Psilin, \Psilinb$ in any compact region that is bounded away from $\mathscr{H}^+_{\geq0}$.

Outside of these two regions, energy conservation will be sufficient to control all derivatives of $\Psilin$.

\subsubsection*{Estimates near $\mathscr{I}^+$}

We start by deriving the necessary backwards $r^p$-estimates near $\mathscr{I}^+$:

\begin{proposition}\label{backwards rp RW}
Let $\mathcal{R}_{\mathscr{I}^+}>3M$, $v_\infty$ such that $r(u_+,v_\infty)=\mathcal{R}_{\mathscr{I}^+}$. The quantities $\Psilin$, $\Psilinb$ belonging to $\mathfrak{S}$ of \Cref{grand proposition backwards scattering} satisfy the following estimate for any $u\leq u_+$:
\begin{align}\label{backwards rp estimate RW}
    \int_{\mathscr{C}_u\cap\{\bar{v}\geq v_\infty\}}d\bar{v}\sin\theta d\theta d\phi\,\frac{r^2}{\Omega^2}|\nablav\Psilin|^2\leq \left(\frac{r^2\Omega^{-2}(u_+,v_\infty)}{r^2\Omega^{-2}(u,v_\infty)}\right) \int_{\mathscr{I}^+\cap\{\bar{u}\in[u,u_+]\}}d\bar{u}\sin\theta d\theta d\phi\, \left[|\mathring{\slashednabla}\upPsilin_{\mathscr{I}^+}|^2+4|\upPsilin_{\mathscr{I}^+}|^2\right].
\end{align}
\begin{align}\label{backwards integrated rp RW}
\begin{split}
    &\int_{\bar{u}\in[u,u_+],\bar{v}\geq v_\infty}d\bar{u}d\bar{v}\sin\theta d\theta d\phi\,\frac{3\Omega^2-1}{r}\frac{r^2}{\Omega^2}|\nablav\Psilin|^2+\int_{\underline{\mathscr{C}}_{v_\infty}\cap\{\bar{u}\in [u,u_+]\}}d\bar{u}\sin\theta d\theta d\phi\,\left[|\mathring{\slashednabla}\Psilin|^2+(3\Omega^2+1)|\Psilin|^2\right]\\&+\int_{\bar{v}\geq v_\infty,\bar{u}\in[u,u_+]}d\bar{u}d\bar{v}\sin\theta d\theta d\phi\,\frac{\Omega^2}{r^2}|\Psilin|^2\leq\left(\frac{r^2\Omega^{-2}(u,v_\infty)}{r^2\Omega^{-2}(u_+,v_\infty)}\right) \int_{\mathscr{I}^+\cap\{\bar{u}\in[u,u_+]\}}d\bar{u}\sin\theta d\theta d\phi\, \left[|\mathring{\slashednabla}\upPsilin_{\mathscr{I}^+}|^2+4|\upPsilin_{\mathscr{I}^+}|^2\right].
\end{split}
\end{align}
The same estimate applies with $\Psilinb$ instead of $\Psilin$, $\upPsilinb_{\mathscr{I}^+}$ instead of $\upPsilin_{\mathscr{I}^+}$.
\begin{proof}
The Regge--Wheeler equation \bref{RW} implies
\begin{align}\label{this 11 07 2021}
    \nablau \frac{r^2}{\Omega^2}|\nablav\Psilin|^2+\frac{3\Omega^2-1}{r}\frac{r^2}{\Omega^2}|\nablav\Psilin|^2+\nablav \left[|\mathring{\slashednabla}\Psilin|^2+(3\Omega^2+1)|\Psilin|^2\right]-6M\frac{\Omega^2}{r^2}|\Psilin|^2=0.
\end{align}
Integrate \bref{this 11 07 2021} over the region $\mathscr{D}$ bounded by $\mathscr{C}_{u_+}$, $\mathscr{C}_u$, $\underline{\mathscr{C}}_{v_\infty}$ and $\mathscr{I}^+$ to get:
\begin{align}
\begin{split}
    &\int_{\mathscr{C}_{u}\cap\{v\geq v_\infty\}}du \sin\theta d\theta d\phi\,\frac{r^2}{\Omega^2}|\nablav\Psi|^2+6M\int_{\mathscr{D}}d\bar{u}d\bar{v}\sin\theta d\theta d\phi\,\frac{\Omega^2}{r^2}|\Psilin|^2\\&+\int_{\underline{\mathscr{C}}_{v_\infty}\cap\{\bar{u}\in[u,u_+]\}}d\bar{u}\sin\theta d\theta d\phi\, \left[|\mathring{\slashednabla}\Psilin|^2 +(3\Omega^2+1)|\Psilin|^2\right]\\&=\int_{\mathscr{D}}d\bar{u}d\bar{v}\sin\theta d\theta d\phi\,\frac{3\Omega^2-1}{r}\frac{r^2}{\Omega^2}|\nablav\Psilin|^2+\int_{\mathscr{I}^+\cap\{\bar{u}\in[u,u_+]\}}d\bar{u}\sin\theta d\theta d\phi\,\left[|\mathring{\slashednabla}\upPsilin_{\mathscr{I}^+}|^2+4|\upPsilin_{\mathscr{I}^+}|^2\right].
\end{split}
\end{align}
Thus we have
\begin{align}\label{this 11 07 2021 2}
    \begin{split}
         \int_{\mathscr{C}_{u}\cap\{v\geq v_\infty\}}dv \sin\theta d\theta d\phi\,\frac{r^2}{\Omega^2}|\nablav\Psi|^2&\leq \int_u^{u_+}d\bar{u}\frac{3\Omega(\bar{u},v_\infty)^2-1}{r(\bar{u},v_\infty)}\int_{\mathscr{C}_{\bar{u}}}\sin\theta d\theta d\phi\, \frac{r^2}{\Omega^2}|\nablav\Psilin|^2\\
         &+\int_{\mathscr{I}^+\cap\{\bar{u}\in[u,u_+]\}}d\bar{u}\sin\theta d\theta d\phi\,\left[|\mathring{\slashednabla}\upPsilin_{\mathscr{I}^+}|^2+4|\upPsilin_{\mathscr{I}^+}|^2\right].
    \end{split}
\end{align}
The result follows by applying Gr\"onwall's inequality (\Cref{Gronwall inequality}) to \bref{this 11 07 2021 2}.
\end{proof}
\end{proposition}

\begin{remark}\label{ratio of rs is bounded}
Note that the ratio $\frac{r(u,v)}{r(u_+,v)}$ is uniformly bounded on $J^+(\overline{\Sigma})$ if $u\leq u_+$ and $v\geq v_\infty$, where $v_\infty$ is such that $r(u_+,v_\infty)\geq 3M$.
\end{remark}


Similar estimates to \bref{backwards rp estimate RW} can be derived for $\left(\frac{r}{\Omega^2}\nablav\right)^k\Psilin$, $\left(\frac{r}{\Omega^2}\nablav\right)^k\Psilinb$ for any integer $k\in\mathbb{N}$. We now show how to do this for $k=1$:
\begin{proposition}\label{rdv Psi near scri+}
    Let $\upphi^{(1)}:=\left(\frac{r}{\Omega^2}\nablav\right)\Psilin$, where $\Psilin$ belongs to the solution $\mathfrak{S}$ of \fullsystem constructed in \Cref{grand proposition backwards scattering}. Then for $\mathcal{R}_{\mathscr{I}^+}, v_\infty$ as in \bref{backwards rp estimate RW} and for $v\geq v_\infty$, the quantity $\upphi^{(1)}$ satisfies
    \begin{align}\label{this 16 07 2021}
    \begin{split}
        \int_{S^2}\sin\theta d\theta d\phi\, |\upphi^{(1)}(u,v,\theta^A)|^2\lesssim\int_{\mathscr{I}^+\cap\{\bar{u}\in[u,u_+]\}}d\bar{u}\sin\theta d\theta d\phi\,\left[\sum_{|\gamma|\leq3}|\mathring{\slashednabla}^\gamma\upPsilin_{\mathscr{I}^+}|^2\right],
    \end{split}
    \end{align}
    \begin{align}
        \int_u^{u_+}\int_{S^2}d\bar{u}\sin\theta d\theta d\phi\,\frac{\Omega^2}{r}|\upphi^{(1)}|^2\lesssim\int_{\mathscr{I}^+\cap\{\bar{u}\in[u,u_+]\}}d\bar{u}\sin\theta d\theta d\phi\,\left[\sum_{|\gamma|\leq3}|\mathring{\slashednabla}^\gamma\upPsilin_{\mathscr{I}^+}|^2\right].
    \end{align}
\end{proposition}
\begin{proof}
Equation \bref{RW} implies
\begin{align}\label{this 14 07 2021}
    \nablau \upphi^{(1)}+\frac{2\Omega^2-1}{r}\upphi^{(1)}-\frac{1}{r}\mathring{\slashed{\Delta}}\Psilin+\frac{1}{r}(3\Omega^2+1)\Psilin=0.
\end{align}
From \bref{this 14 07 2021} we can derive
\begin{align}
\begin{split}\label{this 14 07 2021 2}
    \partial_u |\upphi^{(1)}|^2 +2\frac{2\Omega^2-1}{r}|\upphi^{(1)}|^2+\upphi^{(1)}\cdot\left(\frac{2}{r}(3\Omega^2+1)\Psilin-\frac{2}{r}\mathring{\slashed{\Delta}}\Psilin\right)=0.
\end{split}
\end{align}
Integrating \bref{this 14 07 2021} over $\underline{\mathscr{C}}_{v_\infty}$ between $u$ and $u_+$ and using Cauchy--Schwarz, we get
\begin{align}\label{this 15 07 2021}
    \begin{split}
        \int_{S^2}\sin\theta d\theta d\phi\,|\upphi^{(1)}|^2&\leq\int_u^{u_+}\int_{S^2}d\bar{u}\sin\theta d\theta d\phi\,\left[4\frac{\Omega^2+1}{r}|\upphi^{(1)}|^2+\frac{1}{r}|\mathring{\slashed{\Delta}}\Psilin|^2+\frac{4}{r}|\Psilin|^2\right]\\
        &\leq \int_u^{u_+}\int_{S^2}d\bar{u}\sin\theta d\theta d\phi\,\frac{16\Omega^2}{r}|\upphi^{(1)}|^2\\&\qquad\qquad\qquad\qquad+\int_{u}^{u_+}\int_{S^2}d\bar{u}\sin\theta d\theta d\phi\,\frac{1}{r}\left(|\mathring{\slashed{\Delta}}\Psilin|^2+{4}|\Psilin|^2\right).
    \end{split}
\end{align}
In the above, note that $u_+,v_+$ are chosen so that on $u=u_+,v\geq v_+$ we have $r\geq 3M$ which implies $\Omega^{-2}\leq 3$. Gr\"onwall's inequality (\Cref{Gronwall inequality}) applied to \bref{this 15 07 2021} implies
\begin{align}
     \int_{S^2}\sin\theta d\theta d\phi\,|\upphi^{(1)}|^2\leq \frac{r(u,v_\infty)^{16}}{r(u_+,v_\infty)^{16}}\int_{u}^{u_+}\int_{S^2}d\bar{u}\sin\theta d\theta d\phi\,\frac{1}{r}\left(|\mathring{\slashed{\Delta}}\Psilin|^2+{4}|\Psilin|^2\right),
\end{align}
\begin{align}
    \int_u^{u_+}\int_{S^2}d\bar{u}\sin\theta d\theta d\phi\,\frac{16\Omega^2}{r}|\upphi^{(1)}|^2\leq \frac{r(u,v_\infty)^{16}}{r(u_+,v_\infty)^{16}}\int_{u}^{u_+}\int_{S^2}d\bar{u}\sin\theta d\theta d\phi\,\frac{1}{r}\left(|\mathring{\slashed{\Delta}}\Psilin|^2+{4}|\Psilin|^2\right)
\end{align}
As $\mathring{\slashednabla}^\gamma\Psi(u,v,\theta^A)$ is uniformly bounded in $v$ for any index $\gamma$, we estimate
\begin{align}
\begin{split}
    \int_u^{u_+}\int_{S^2}d\bar{u}\sin\theta d\theta d\phi\,\frac{1}{r}|\Psilin|^2&=\int_u^{u_+}\int_{v_\infty}^\infty\int_{S^2}d\bar{u}d\bar{v}\sin\theta d\theta d\phi\,\left[\frac{2\Psilin\times\nablav\Psilin}{r}-\frac{|\Psilin|^2}{r^2}\right].\\&
    \leq \int_u^{u_+}\int_{v_\infty}^\infty\int_{S^2}d\bar{u}d\bar{v}\sin\theta d\theta d\phi\,|\nablav\Psilin|^2
\end{split}
\end{align}
\begin{align}
    \int_u^{u_+}\int_{S^2}d\bar{u}\sin\theta d\theta d\phi\,\frac{1}{r}|\mathring{\slashed{\Delta}}\Psilin|^2 \leq \int_u^{u_+}\int_{v_\infty}^\infty\int_{S^2}d\bar{u}d\bar{v}\sin\theta d\theta d\phi\,|\nablav\mathring{\slashed{\Delta}}\Psilin|^2.
\end{align}
Using \bref{backwards integrated rp RW} commuted with $\mathring{\slashednabla}^\gamma$ for $|\gamma|\leq2$, we get
\begin{align}
\begin{split}
     \int_{S^2}\sin\theta d\theta d\phi\,|\upphi^{(1)}|^2&\leq 12\frac{r(u,v_\infty)^{18}}{r(u_+,v_\infty)^{18}} \int_{\mathscr{I}^+\cap\{\bar{u}\in[u,u_+]\}}d\bar{u}\sin\theta d\theta d\phi\, \left[\sum_{|\gamma|\leq3}|\mathring{\slashednabla}^\gamma\upPsilin_{\mathscr{I}^+}|^2\right].\\
     &\leq 12 \int_{\mathscr{I}^+\cap\{\bar{u}\in[u,u_+]\}}d\bar{u}\sin\theta d\theta d\phi\, \left[\sum_{|\gamma|\leq3}|\mathring{\slashednabla}^\gamma\upPsilin_{\mathscr{I}^+}|^2\right].
\end{split}
\end{align}
And similarly, 
\begin{align}
     \int_u^{u_+}\int_{S^2}d\bar{u}\sin\theta d\theta d\phi\,\frac{\Omega^2}{r}|\upphi^{(1)}|^2\,\leq 12 \int_{\mathscr{I}^+\cap\{\bar{u}\in[u,u_+]\}}d\bar{u}\sin\theta d\theta d\phi\, \left[\sum_{|\gamma|\leq3}|\mathring{\slashednabla}^\gamma\upPsilin_{\mathscr{I}^+}|^2\right].
\end{align}
\end{proof}

We can find similar bounds to \bref{this 16 07 2021} for $\upphi^{(k)}$ by induction on the equation
\begin{align}
    \begin{split}
        &\nablau \upphi^{(n+1)}+(n+1)\left(\frac{1}{r}-\frac{4M}{r^2}\right)\upphi^{(n+1)}+(-1)^n\left[\upphi^{(1)}\left(\frac{1}{r}-\frac{4M}{r^2}2^n\right)+2\Psilin\left(\frac{1}{r}-\frac{3M}{r^2}2^n\right)+\frac{1}{r}\mathring{\slashed{\Delta}}\Psilin\right]\\
        &+\sum_{j=1}^{n-1}(-1)^{n-j}\upphi^{(j+1)}{n \choose j}\left[\left(\frac{n+1}{n-j+1}-\frac{2(n-j)}{j+1}\right)\frac{1}{r}+\left(\frac{3(n-j)}{j+1}-\frac{2(n+1)}{n-j+1}\right)\frac{2M}{r^2}2^{n-j}\right]\\
        &+\sum_{j=1}^n(-1)^{n-j}\frac{1}{r}\mathring{\slashed{\Delta}}\upphi^{(j)}=0.
    \end{split}
\end{align}
\begin{corollary}\label{rdv Psi near scri+ k}
 For $\upphi^{(k)}=\left(\frac{r}{\Omega^2}\nablav\right)^k\Psilin$, where $\Psilin$ belongs to the solution $\mathfrak{S}$ of \fullsystem constructed in \Cref{grand proposition backwards scattering}, we have the following estimate on $v\geq v_\infty$ for $\mathcal{R}_{\mathscr{I}^+}, v_\infty$ as in \Cref{backwards rp RW}:
   \begin{align}\label{this 16 07 2021 2} 
    \begin{split}
        \int_{S^2}\sin\theta d\theta d\phi\, |\upphi^{(k)}(u,v,\theta^A)|^2\leq C(k)\int_{\mathscr{I}^+\cap\{\bar{u}\in[u,u_+]\}}d\bar{u}\sin\theta d\theta d\phi\,\left[\sum_{|\gamma|\leq2k+1}|\mathring{\slashednabla}^\gamma\upPsilin_{\mathscr{I}^+}|^2\right],
    \end{split}
    \end{align}
    \begin{align}
        \int_u^{u_+}\int_{S^2}d\bar{u}\sin\theta d\theta d\phi\,\frac{\Omega^2}{r}|\upphi^{(k)}|^2\leq C(k)\int_{\mathscr{I}^+\cap\{\bar{u}\in[u,u_+]\}}d\bar{u}\sin\theta d\theta d\phi\,\left[\sum_{|\gamma|\leq2k+1}|\mathring{\slashednabla}^\gamma\upPsilin_{\mathscr{I}^+}|^2\right].
    \end{align}
\end{corollary}

\begin{corollary}
For $\Psilin$ belonging to the solution $\mathfrak{S}$ of \fullsystem constructed in \Cref{grand proposition backwards scattering}, we have 
\begin{align}
    \lim_{v\longrightarrow\infty} \partial_t^k \Psilin=\partial_u^k \upPsilin_{\mathscr{I}^+},\qquad\qquad\lim_{v\longrightarrow\infty} \partial_t^k \Psilinb=\partial_u^k \upPsilinb_{\mathscr{I}^+}
\end{align}
\end{corollary}

In \cite{Mas20} we also proved the following $r$-weighted estimates on $\upPhi^{(1)}:=\left(\frac{r^2}{\Omega^2}\nablav\right)\Psilin$. Note that the limit of $\Phi^{(1)}(u,v,\theta^A)$ as $v\longrightarrow\infty$ exists and is equal to
\begin{align}
    \Phi^{(1)}_{\mathscr{I}^+}:=\lim_{v\longrightarrow\infty}\Phi^{(1)}(u,v,\theta^A)=2\mathring{\fancydstar_2}\mathring{\fancydstar_1}\mathring{\fancyd}_1\blins_{\mathscr{I}^+}+\mathring{\fancydstar_2}\mathring{\fancydstar_1}(6M\otxs_{\mathscr{I}^+},0)+12M\mathring{\fancydstar_2}\eblins_{\mathscr{I}^+}.
\end{align}
Taking $\underline{\Phi}^{(k)}:=\left(\frac{r^2}{\Omega^2}\nablav\right)^k\Psilinb$, we have a similar expression for the limit of $\underline{\Phi}^{(1)}$ towards $\mathscr{I}^+$:
\begin{align}
    \underline{\Phi}^{(1)}_{\mathscr{I}^+}:=\lim_{v\longrightarrow\infty}\underline{\Phi}^{(1)}(u,v,\theta^A)=2\mathring{\fancydstar_2}\mathring{\fancydstar_1}\mathring{\overline{\fancyd}_1}\blins_{\mathscr{I}^+}+\mathring{\fancydstar_2}\mathring{\fancydstar_1}(6M\otxs_{\mathscr{I}^+},0)+12M\mathring{\fancydstar_2}\eblins_{\mathscr{I}^+}.
\end{align}

\begin{proposition}\label{backwards rp Phi 1}
For $\upPhi^{(k)}=\left(\frac{r^2}{\Omega^2}\nablav\right)^k\Psilin$, where $\Psilin$ belongs to the solution $\mathfrak{S}$ of \fullsystem constructed in \Cref{grand proposition backwards scattering}, we have the following estimate on $v\geq v_\infty$ for $\mathcal{R}_{\mathscr{I}^+}, v_\infty$ as in \Cref{backwards rp RW}:
\begin{align}
\begin{split}
    &\int_{\mathscr{C}_u\cap\{\bar{v}\geq v_\infty\}}d\bar{v}\sin\theta d\theta d\phi\,\frac{r^2}{\Omega^2}|\nablav\Phi^{(1)}|^2+
    \int_u^{u_+}\int_{\mathscr{C}_{\bar{u}}\cap\{\bar{v}\geq v_\infty\}}d\bar{u}d\bar{v}\sin\theta d\theta d\phi\,\left[\frac{r^2}{\Omega^2}|\nablav\Phi^{(1)}|^2+\frac{6M\Omega^2}{r^2}|\Phi^{(1)}|^2\right]\\&+\int_{\underline{\mathscr{C}}_{v_\infty}\cap\{\bar{u}\geq u\}}d\bar{u}\sin\theta d\theta d\phi\,|\Phi^{(1)}|^2\,\leq \frac{r(u,v)^6}{r(u_+,v)^6}\int_{\mathscr{I}^+\cap\{\bar{u}\in[u,u_+]\}}d\bar{u}\sin\theta d\theta d\phi\,[|\Phi^{(1)}_{\mathscr{I^+}}|^2+|\upPsilinb_{\mathscr{I}^+}|^2],
\end{split}
\end{align}
where the limit The same estimate applies replacing $\Psilin$ with $\Psilinb$
\end{proposition}

\begin{proof}
    The Regge--Wheeler equation \bref{RW} implies that $\Phi^{(1)}$ satisfies
    \begin{align}
        \nablau\nablav \Phi^{(1)}+\frac{3\Omega^2-1}{r}\nablav\Phi^{(1)}+\frac{\Omega^2}{r^2}(5-3\Omega^2)\Phi^{(1)}-\Omega^2\slashed\Delta \Phi^{(1)}=-6M\frac{\Omega^2}{r^2}\Psilin,
    \end{align}
    which implies
    \begin{align}
    \begin{split}
        &\partial_u \left[\frac{r^2}{\Omega^2}|\nablav \Phi^{(1)}|^2\right]+3\frac{3\Omega^2-1}{r}\frac{r^2}{\Omega^2}|\nablav\Phi^{(1)}|^2+\partial_v\Big[(5-3\Omega^2)|\Phi^{(1)}|^2+|\mathring{\slashednabla}\Phi^{(1)}|^2+12M\Psilin\cdot\Phi^{(1)}\Big]\\&-6M\frac{\Omega^2}{r^2}|\Phi^{(1)}|^2\,=0.
    \end{split}
    \end{align}
    Taking $v_\infty$ such that $r(u_+,v_\infty)>3M$, we integrate the above over the region $\mathscr{D}=[u,u_+]\times [v_\infty,\infty)\times S^2$ gives
    \begin{align}
    \begin{split}
        &-\int_{\mathscr{C}_u\cap\{\bar{v}\geq v_\infty\}}d\bar{v}\sin\theta d\theta d\phi\,\frac{r^2}{\Omega^2}|\nablav\Phi^{(1)}|^2+\int_{\mathscr{D}}d\bar{u}d\bar{v}\sin\theta d\theta d\phi\,3\frac{3\Omega^2-1}{r}\frac{r^2}{\Omega^2}|\nablav\Phi^{(1)}|^2\\&+\int_{\mathscr{I}^+\cap\{\bar{u}\in[u,u_+]\}}d\bar{u}\sin\theta d\theta d\phi\,\left[12M\upPsilin_{\mathscr{I}^+}\cdot \Phi^{(1)}_{\mathscr{I}^+}+|\mathring{\slashednabla}\Phi^{(1)}_{\mathscr{I}^+}|^2+2|\Phi^{(1)}_{\mathscr{I}^+}|^2\right]-6M\int_{\mathscr{D}}d\bar{u}d\bar{v}\sin\theta d\theta d\phi\,\frac{\Omega^2}{r^2}|\Phi^{(1)}|^2\\&-\int_{\underline{\mathscr{C}}_{v_\infty}\cap\{\bar{u}\in[u,u_+]\}}d\bar{u}\sin\theta d\theta d\phi\,\left[(5-3\Omega^2)|\Phi^{(1)}|^2+|\mathring{\slashednabla}\Phi^{(1)}|^2+12M\Psilin\cdot\Phi^{(1)}\right]=0.
    \end{split}
    \end{align}
    Thus
    \begin{align}
        \begin{split}
            &\int_{\mathscr{C}_u\cap\{\bar{v}\geq v_\infty\}}d\bar{v}\sin\theta d\theta d\phi\,\frac{r^2}{\Omega^2}|\nablav\Phi^{(1)}|^2+6M\int_{\mathscr{D}}d\bar{u}d\bar{v}\sin\theta d\theta d\phi\,\frac{\Omega^2}{r^2}|\Phi^{(1)}|^2\\&\leq \int_{u}^{u_+}d\bar{u}\,3\frac{3\Omega^2-1}{r}\int_{\mathscr{C}_{\bar{u}}}d\bar{v}\sin\theta d\theta d\phi\,\frac{r^2}{\Omega^2}|\nablav\Phi^{(1)}|^2+2(12M)^2\int_{\underline{\mathscr{C}}_{v_\infty}\cap\{\bar{u}\in[u,u_+]\}}d\bar{u}\sin\theta d\theta d\phi\,|\Psilin|^2\\&+\int_{\mathscr{I}^+\cap\{\bar{u}\in[u,u_+]\}}d\bar{u}\sin\theta d\theta d\phi\,\left[2(12M)^2|\upPsilin_{\mathscr{I}^+}|^2+|\mathring{\slashednabla}\Phi^{(1)}_{\mathscr{I}^+}|^2+3|\Phi^{(1)}_{\mathscr{I}^+}|^2\right].
        \end{split}
    \end{align}
    Gr\"onwall's inequality (\Cref{Gronwall inequality}), together with \Cref{backwards rp RW}, imply the result.
\end{proof}

We can prove similar estimates on $\Phi^{(k)}$, $\underline{\Phi}^{(k)}$ by iterating the above argument using the equations
\begin{align}\label{transport equation for nth r^2 dv derivative of Psi}
\begin{split}
    &\nablau\Phi^{(n)}+n\left(\frac{2}{r}-\frac{6M}{r^2}\right)\Phi^{(n)}+\left[\left(4-\frac{6M}{r}\right)-n(n-1)\left(1-\frac{6M}{r}\right)\right]\Phi^{(n-1)}\\&-2M(n^2-1)(n-3)\Phi^{(n-2)}-\mathring{\slashed{\Delta}}\Phi^{(n-1)}=0,
\end{split}
\end{align}

\begin{align}\label{wave equation for nth r^2 dv derivative of Psi}
    \begin{split}
        &\nablau\nablav \Phi^{(n)}-\frac{\Omega^2}{r^2}\mathring{\slashed{\Delta}}\Phi^{(n)}+n\left(\frac{2}{r}-\frac{6M}{r^2}\right)\nablav \Phi^{(n)}+\left[4-\frac{6M}{r}-n(n+1)\left(1-\frac{6M}{r}\right)\right]\frac{\Omega^2}{r^2}\Phi^{(n)}\\
        &-2Mn(n-2)(n+2)\frac{\Omega^2}{r^2}\Phi^{(n-1)}=0.
    \end{split}
\end{align}

The first step is to show that $\Phi^{(n)}$ tends to a limit $\Phi^{(n)}_{\mathscr{I}^+}$ which defines a smooth field on $\mathscr{I}^+$:
\begin{proposition}\label{backwards transport estimate of Phi n}
    Let $\Phi^{(n)}$ be as in \Cref{backwards rp Phi 1}. The limit
    \begin{align}\label{limit of Phi n}
        \Phi^{(n)}_{\mathscr{I}^+}(u,v,\theta^A)=\lim_{v\longrightarrow\infty}\Phi^{(n)}(u,\theta^A)
    \end{align}
    exists and defines an element of $\Gamma(\mathscr{I}^+)$. The same applies to angular derivatives of $\Phi^{(n)}$ and 
    \begin{align}\label{this 16 08 2021}
        \lim_{v\longrightarrow\infty}\mathring{\slashednabla}^\gamma\Phi^{(n)}_{\mathscr{I}^+}=\mathring{\slashednabla}^\gamma \Phi^{(n)}_{\mathscr{I}^+}
    \end{align}
for any index $\gamma$.
\end{proposition}
\begin{proof}
    We will prove this result by induction on $n$. Assume the result holds for $\Phi^{(n-1)}$ and $\Phi^{(n-2)}$. Let $v_\infty$ be such that $r(u_+,v_\infty)>3M$.
    Integrate equation \bref{transport equation for nth r^2 dv derivative of Psi} from $(u_+,v,\theta^A)$ to $(u,v,\theta^A)$ to obtain:
    \begin{align}
        \begin{split}
            |\Phi^{(n)}|\leq\int_u^{u_+}d\bar{u}\, n\left(\frac{2}{r}-\frac{6M}{r^2}\right)|\Phi^{(n)}|+\int_u^{u_+}d\bar{u}\,\Big[&(7+4n(n-1)|\times|\Phi^{(n-1)}|\\&+|\mathring{\slashed{\Delta}}\Phi^{(n-1)}|+2M|(n^2-1)(n-3)|\times|\Phi^{(n-2)}|\Big].
        \end{split}
    \end{align}
    Gr\"onwall's inequality (\Cref{Gronwall inequality}) implies
    \begin{align}
    \begin{split}
        |\Phi^{(n)}|\leq\left(\frac{r^2\Omega^{-2}(u,v)}{r^2\Omega^{-2}(u_+,v)}\right)^n\times |u_+-u|\Big[|\Phi^{(n-2)}|+(&7+4n(n-1))|\Phi^{(n-1)}|+|\mathring{\slashed{\Delta}}\Phi^{(n-1)}|\\&+2M|(n^2-1)(n-3)|\times|\Phi^{(n-2)}|\Big].
    \end{split}
    \end{align}
    Equation \bref{transport equation for nth r^2 dv derivative of Psi} is solved by
    \begin{align}
    \begin{split}
        \Phi^{(n)}(u,v,\theta^A)=\int_u^{u_+}d\bar{u} \left(\frac{r^2\Omega^{-2}(u,v,\theta^A)}{r^2\Omega^{-2}(\bar{u},v,\theta^A)}\right)^n\Bigg\{&\left[4-\frac{6M}{r}-n(n+1)\left(1-\frac{6M}{r}\right)+\mathring{\slashed{\Delta}}\right]\Phi^{(n-1)}\\&\quad-2Mn(n^2-1)(n-3)\Phi^{(n-2)}\Bigg\}.
    \end{split}
    \end{align}
    \Cref{Lebesgue's BCT} implies that the limit \bref{limit of Phi n} exists and is equal to
    \begin{align}\label{formula for radiation field of Phi n}
        \Phi^{(n)}_{\mathscr{I}^+}=\int_u^{u_+}d\bar{u}\left\{\left[\mathring{\slashed{\Delta}}+2-n(n+1)\right]\Phi^{(n-1)}_{\mathscr{I}^+}-2M n(n^2-1)(n-3)\Phi^{(n-2)}_{\mathscr{I}^+}\right\}.
    \end{align}
    Commuting the above argument with ${\mathring{\slashednabla}^\gamma}$ produces \bref{this 16 08 2021}.
\end{proof}

\begin{remark}\label{asymptotics of Phi n scri}
    We can use the formula \bref{formula for radiation field of Phi n} and apply an inductive argument to conclude that $\mathring{\slashednabla}^\gamma\partial_u^i\Phi^{(n)}_{\mathscr{I}^+}=O(|u|^{n-i-1})$ as $u\longrightarrow-\infty$ for any $i\in\mathbb{N}$ and any index $\gamma$ if $\upPsilin_{\mathscr{I}^+}$ is compactly supported. If $\upPsilin_{\mathscr{I}^+}$ is not compactly supported, we have $\mathring{\slashednabla}^\gamma\partial_u^i\Phi^{(n)}_{\mathscr{I}^+}=O(|u|^{n-i})$.
\end{remark}

We can now state
\begin{proposition}\label{backwards rp RW Phi n}
Let $\Psilin$ be as in \Cref{grand proposition backwards scattering} and let $\Phi^{(n)}$ as defined in \Cref{backwards rp Phi 1}. For $v_\infty$ be such that $r(u_+,v_\infty)>3M$, we have
\begin{align}\label{backwards rp estimate RW Phi n}
\begin{split}
    &\int_{\mathscr{C}_u\cap\{\bar{v}\in[v_\infty,\infty)\}}d\bar{v}\sin\theta d\theta d\phi\,\frac{r^2}{\Omega^2}|\nablav\Phi^{(n)}|^2+\int_{\bar{u}\in[u,u_+],\bar{v}\geq v_\infty}d\bar{u}d\bar{v}\sin\theta d\theta d\phi\,\left[\frac{\Omega^2}{r^2}|\Phi^{(n)}|^2+\frac{r}{\Omega^2}|\nablav\Phi^{(n)}|^2\right]\\
    &+\int_{\underline{\mathscr{C}}_{v_\infty}}d\bar{u}\sin\theta d\theta d\phi\,|\Phi^{(n)}|^2\,\leq\, C(n,M,u_+)\int_{\mathscr{I}^+\cap\{\bar{u}\in[u,u_+]\}}d\bar{u}\sin\theta d\theta d\phi\,\left[|\mathring{\slashednabla}\Phi^{(n)}_{\mathscr{I}^+}|^2+|\Phi^{(n)}_{\mathscr{I}^+}|^2+|\Phi^{(n-1)}_{\mathscr{I}^+}|^2\right].
\end{split}
\end{align}
\end{proposition}

\begin{proof}
   We apply the same procedure followed in the proof of \Cref{backwards rp Phi 1}, multiplying equation \bref{wave equation for nth r^2 dv derivative of Psi} by  $2r^2{\Omega^{-2}}\nablav\Phi^{(n)}$ and integrating by parts over $\mathscr{D}=[u,u_+]\times(v_\infty)\times S^2$, where $v_\infty$ is such that $r(u_+,v_\infty)>3M$,  to get
   \begin{align}\label{this 16 08 2021 2}
       \begin{split}
           &-\int_{\mathscr{C}_u\cap\{v\geq v_\infty\}}d\bar{v}\sin\theta d\theta d\phi\, \frac{r^2}{\Omega^2}|\nablav\Phi^{(n)}|^2+\int_{\mathscr{D}}d\bar{u}d\bar{v}\sin\theta d\theta d\phi\,(2n+1)\left(\frac{2}{r}-\frac{6M}{r^2}\right)\frac{r^2}{\Omega^2}|\nablav\Phi^{(n)}|^2\\
           &-\int_{\mathscr{D}}d\bar{u}d\bar{v}\sin\theta d\theta d\phi\,\frac{2M\Omega^2}{r^2}\left[3(1-n(n+1))|\Phi^{(n)}|^+2n(n^2-4)\Phi^{(n-1)}\cdot\nablav\Phi^{(n)}\right]
           \\&-\int_{\underline{\mathscr{C}}_{v_\infty}}d\bar{u}\sin\theta d\theta d\phi\,\left\{\left[4-n(n+1)-\frac{6M}{r}(1-n(n+1))\right]|\Phi^{(n)}|^2+|\mathring{\slashednabla}\Phi^{(n)}|^2\right\}\\
           &+\int_{\mathscr{I}^+}d\bar{u}\sin\theta d\theta d\phi\,\left[|\mathring{\slashednabla}\Phi^{(n)}_{\mathscr{I}^+}|^2+(4-n(n+1))|\Phi^{(n)}_{\mathscr{I}^+}|^2\right]=0.
       \end{split}
   \end{align}
    We estimate $\int_{\underline{\mathscr{C}}_{v_\infty}}|\Phi^{(n)}|^2$ as follows: the fundamental theorem of calculus and Cauchy--Schwarz imply
    \begin{align}\label{model estimate near scri+}
        \begin{split}
            \int_{\underline{\mathscr{C}}_{v_\infty}}d\bar{u}\sin\theta d\theta d\phi\,|\Phi^{(n)}|^2&\leq2\int_{\mathscr{I}^+\cap\{\bar{u}\in[u,u_+]\}}d\bar{u}\sin\theta d\theta d\phi\,|\Phi^{(n)}_{\mathscr{I}^+}|^2+\int_{\underline{\mathscr{C}}_{v_\infty}}d\bar{u}\sin\theta d\theta d\phi\,\left|\int_{v_\infty}^\infty d\bar{v}\nablav\Phi^{(n)}\right|\\
            \leq&2\int_{\mathscr{I}^+\cap\{\bar{u}\in[u,u_+]\}}d\bar{u}\sin\theta d\theta d\phi\,|\Phi^{(n)}_{\mathscr{I}^+}|^2\\&+\int_u^{u_+}d\bar{u}\,\frac{2}{r(\bar{u},v)}\int_{\underline{\mathscr{C}}_{v_\infty}}d\bar{v}\sin\theta d\theta d\phi\,\frac{r^2}{\Omega^2}|\nablav\Phi^{(n)}|^2.
        \end{split}
    \end{align}
    Following similar steps, we estimate $\int_{\mathscr{D}}\frac{\Omega^2}{r^2}|\Phi^{(n)}|^2$ via
    \begin{align}
        \begin{split}
            \int_{\mathscr{D}}d\bar{u}d\bar{v}\sin\theta d\theta d\phi\,\frac{\Omega^2}{r^2}|\Phi^{(n)}|^2\leq &\frac{1}{M}\int_{\mathscr{I}^+\cap\{\bar{u}\in[u,u_+]\}}d\bar{u}\sin\theta d\theta d\phi\,|\Phi^{(n)}_{\mathscr{I}^+}|^2\\&+\int_u^{u_+}d\bar{u}\,\frac{2}{r(\bar{u},v)^2}\int_{\underline{\mathscr{C}}_{v_\infty}}d\bar{v}\sin\theta d\theta d\phi\,\frac{r^2}{\Omega^2}|\nablav\Phi^{(n)}|^2.
        \end{split}
    \end{align}
    Applying the same steps to estimate $\int_{\mathscr{D}}\frac{\Omega^2}{r^2}|\Phi^{(n-1)}|^2$, we get
    \begin{align}
        \begin{split}
            \int_{\mathscr{D}}d\bar{u}d\bar{v}\sin\theta d\theta d\phi\,\frac{\Omega^2}{r^2}|\Phi^{(n-1)}|^2\,\leq& \,\frac{1}{(3M)^2}\int_{\mathscr{D}}d\bar{u}d\bar{v}\sin\theta d\theta d\phi\,\frac{\Omega^2}{r^2}|\Phi^{(n)}|^2\\&+\frac{2}{3M}\int_{\mathscr{I}^+\cap\{\bar{u}\in[u,u_+]\}}d\bar{u}\sin\theta d\theta d\phi\,|\Phi^{(n-1)}_{\mathscr{I}^+}|^2
            \\\leq\,&\frac{1}{(3M)^2}\int_u^{u_+}d\bar{u}\frac{2}{r^2}\int_{\underline{\mathscr{C}}_{v_\infty}}d\bar{v}\sin\theta d\theta d\phi\,\frac{r^2}{\Omega^2}|\nablav\Phi^{(n)}|^2\\&+\int_{\mathscr{I}^+\cap\{\bar{u}\in[u,u_+]\}}d\bar{u}\sin\theta d\theta d\phi\,\left[\frac{2}{3M}|\Phi^{(n-1)}_{\mathscr{I}^+}|^2+\frac{2}{(3M)^3}|\Phi^{(n)}_{\mathscr{I}^+}|^2\right].
        \end{split}
    \end{align}
Going back to \bref{this 16 08 2021 2}, we now can derive
\begin{align}
    \begin{split}
        &\int_{\mathscr{C}_u\cap\{v\geq v_\infty\}}d\bar{v}\sin\theta d\theta d\phi\, \frac{r^2}{\Omega^2}|\nablav\Phi^{(n)}|^2\leq \int_u^{u_+}d\bar{u}\,\Bigg\{(2n+1)\left(\frac{2}{r}-\frac{6M}{r^2}\right)+4Mn(n^2-4)\frac{\Omega^2}{r^4}\\&+\frac{2}{r}(n(n+1)+2)+\frac{12M}{r^2}[n(n+1)-1]+n|n^2-4|\frac{32}{9Mr^2}\Bigg\}
        \int_{\mathscr{C}_{\bar{u}}\cap\{\bar{v}\geq v_\infty\}}d\bar{v}\sin\theta d\theta d\phi\,\frac{r^2}{\Omega^2}|\nablav\Phi^{(n)}|^2\\&+\int_{\mathscr{I}^+\cap\{\bar{u}\in[u,u_+]\}}d\bar{u}\sin\theta d\theta d\phi\,4Mn|n^2-4|\times\\&\qquad\qquad\qquad\qquad\left[|\mathring{\slashednabla}\Phi^{(n)}_{\mathscr{I}^+}|^2+\left(|4-n(n+1)|+\frac{8n|n^2-4|}{27M^2}\right)|\Phi^{(n)}_{\mathscr{I}^+}|^2+\frac{8n|n^2-4|}{27M^2}|\Phi^{(n-1)}_{\mathscr{I}^+}|^2\right].
    \end{split}
\end{align}
Gr\"onwall's inequality (\Cref{Gronwall inequality}) implies
\begin{align}
\begin{split}
    &\int_{\mathscr{C}_u\cap\{v\geq v_\infty\}}d\bar{v}\sin\theta d\theta d\phi\, \frac{r^2}{\Omega^2}|\nablav\Phi^{(n)}|^2+\int_{\bar{u}\in[u,u_+],\bar{v}\geq v_\infty}d\bar{u}d\bar{v}\sin\theta d\theta d\phi\,\frac{r}{\Omega^2}|\nablav\Phi^{(n)}|^2\,\leq\\&  \left(\frac{r^2\Omega^{-2}(u,v,\theta^A)}{r^2\Omega^{-2}(\bar{u},v,\theta^A)}\right)^{2n+1}\times\left(\frac{r\Omega^2(u,v)}{r\Omega^2(\bar{u},v)}\right)^{2n(n+1)+4}\times\left(\frac{\Omega^2(u,v)}{\Omega^2(u_+,v)}\right)^{6(n(n+1)-1)+\frac{4n|n^2-4|}{9M^2}} \times I,
\end{split}
\end{align}
where 
\begin{align}
\begin{split}
    I=\int_{\mathscr{I}^+\cap\{\bar{u}\in[u,u_+]\}}d\bar{u}\sin\theta d\theta d\phi\,&4Mn|n^2-4|\times\\&\left[|\mathring{\slashednabla}\Phi^{(n)}_{\mathscr{I}^+}|^2+\left(|4-n(n+1)|+\frac{8n|n^2-4|}{27M^2}\right)|\Phi^{(n)}_{\mathscr{I}^+}|^2+\frac{8n|n^2-4|}{27M^2}|\Phi^{(n-1)}_{\mathscr{I}^+}|^2\right]
\end{split}
\end{align}
\end{proof}

\paragraph[Boundedness on $\protect\Psilin$, $\protect\Psilinb$ near $\mathscr{H}^+$]{Boundedness on $\Psilin$, $\Psilinb$ near $\mathscr{H}^+$}\label{Boundedness of Psilin Psilinb backwards scattering near H+}

We now study the boundedness of $\frac{1}{\Omega^2}\nablau$-derivatives of $\Psilin, \Psilinb$ near $\mathscr{H}^+$. An analogous procedure to that leading to \Cref{rdv Psi near scri+} and \Cref{rdv Psi near scri+ k} applies to derive boundedness for $\left(\frac{1}{\Omega^2}\nablau\right)^k\Psilin$ near $\mathscr{H}^+_{\geq0}$, by considering the equation

\begin{align}\label{equation for transverse null derivative near H+}
    \begin{split}
        &\nablav \left(\frac{1}{\Omega^2}\nablau\right)^{n+1}\Psilin+(n+1)\frac{2M}{r^2}\left(\frac{1}{\Omega^2}\nablau\right)^{n+1}\Psilin+\frac{(n+1)!}{r^{n+2}}\left(4\Psilin-\frac{3M}{r}(n+2)\Psilin-\mathring{\slashed{\Delta}}\Psilin\right)\\
        &+\sum_{k=1}^n\frac{n!}{k!}\frac{1}{r^{n-k+2}}\left(4(n-k+1)-\frac{M}{r}k(n-k+2)(3(n-k)+1)\right)\left(\frac{1}{\Omega^2}\nablau\right)^{k}\Psilin\\
        &-\sum_{k=1}^n\frac{n!(n-k+1)}{k!}\frac{1}{r^{n-k+2}}\left(\frac{1}{\Omega^2}\nablau\right)^{k}\mathring{\slashed{\Delta}}\Psilin=0.
    \end{split}
\end{align}

In Proposition 5.4.1 of \cite{Mas20}, a similar procedure to the proof of \Cref{backwards rp RW} was applied to \bref{equation for transverse null derivative near H+} with $n=1$ to show the following estimates using the Regge--Wheeler equation \bref{RW}:

\begin{align}\label{RW exponential backwards near H+}
\begin{split}
     \int_{\underline{\mathscr{C}}_v\cap\{\bar{u}\in[u_\infty,\infty]\}}d\bar{u}d\omega\;\frac{1}{\Omega^2}&|\nablau\Psilin|^2\\\lesssim  e^{\frac{1}{2M}(v_+-v)}\times&\left\{\int_{\mathscr{H}^+\cap\{\bar{v}\in[v,v_+]\}}d\bar{v}\sin\theta d\theta d\phi\, \left[|\mathring{\slashednabla}{\upPsilin}_{\mathscr{H}^+}|^2+|\bm{\uppsi}_{\mathscr{H}^+}|^2\right]\right\}.
\end{split}
\end{align}


\begin{align}\label{RW integrated exponential backwards near H+}
\begin{split}
     \int_{\bar{u}\geq u_\infty, \bar{v}\in[v,v_+]\}}d\bar{u}d\bar{v}d\omega\;\frac{1}{\Omega^2}&|\nablau\Psilin|^2\\\lesssim  e^{\frac{1}{2M}(v_+-v)}\times&\left\{\int_{\mathscr{H}^+\cap\{\bar{v}\in[v,v_+]\}}d\bar{v}\sin\theta d\theta d\phi\, \left[|\mathring{\slashednabla}{\upPsilin}_{\mathscr{H}^+}|^2+|\bm{\uppsi}_{\mathscr{H}^+}|^2\right]\right\}.
\end{split}
\end{align}

Using an entirely analogous procedure to the proof of \Cref{rdv Psi near scri+}, we derive from \bref{equation for transverse null derivative near H+} with $n=0$ the following estimate

\begin{align}\label{this 17 07 2021}
    \begin{split}
        \int_{S^2}\sin\theta d\theta d\phi\,\left|\frac{1}{\Omega^2}\nablau\Psilin(u,v,\theta^A)\right|^2&\leq \int_{\mathscr{C}_{u}\cap\{\bar{v}\in[v,v_+]\}}d\bar{v}\sin\theta d\theta d\phi\,\frac{8M}{r^2}\left|\frac{1}{\Omega^2}\nablau\Psilin(\bar{u},\bar{v},\theta^A)\right|^2\\
        &+\int_{\mathscr{C}_{u}\cap\{\bar{v}\in[v,v_+]\}}d\bar{v}\sin\theta d\theta d\phi\,\frac{2}{Mr^2} \left[\left(4-\frac{6M}{r}\right)^2|\Psilin|^2+|\mathring{\slashed{\Delta}}\Psilin|^2\right].
    \end{split}
\end{align}

The last term on the right hand side in \bref{this 17 07 2021} can be estimated via \bref{RW integrated exponential backwards near H+} from
\begin{align}
\begin{split}
    \int_{\mathscr{C}_{u}\cap\{\bar{v}\in[v,v_+]\}}d\bar{v}\sin\theta d\theta d\phi\,|\Psilin|^2&\leq  \int_{\mathscr{H}_{\geq0}^+\cap\{\bar{v}\in[v,v_+]\}}d\bar{v}\sin\theta d\theta d\phi\,|\upPsilin_{\mathscr{H}^+}|^2\\
    &+\int_{\bar{u}\geq u, \bar{v}\in [v,v_+]}d\bar{u}d\bar{v}\sin\theta d\theta d\phi\,\left|\frac{1}{\Omega^2}\nablau\Psilin(u,v,\theta^A)\right|^2\\
    &\lesssim  e^{\frac{1}{2M}(v_+-v)}\times\left\{\int_{\mathscr{H}^+\cap\{\bar{v}\in[v,v_+]\}}d\bar{v}\sin\theta d\theta d\phi\, \left[|\mathring{\slashednabla}{\upPsilin}_{\mathscr{H}^+}|^2+|\bm{\uppsi}_{\mathscr{H}^+}|^2\right]\right\}.
\end{split}
\end{align}
\begin{align}
    \int_{\mathscr{C}_{u}\cap\{\bar{v}\in[v,v_+]\}}d\bar{v}\sin\theta d\theta d\phi\,|\mathring{\slashed{\Delta}}\Psilin|^2&\lesssim  e^{\frac{1}{2M}(v_+-v)}\times\left\{\int_{\mathscr{H}^+\cap\{\bar{v}\in[v,v_+]\}}d\bar{v}\sin\theta d\theta d\phi\, \left[\sum_{|\gamma|\leq 3}|\mathring{\slashednabla}^\gamma{\upPsilin}_{\mathscr{H}^+}|^2\right]\right\}.
\end{align}

Gr\"onwall's inequality (\Cref{Gronwall inequality}) on \bref{this 17 07 2021} gives

\begin{proposition}
    For $\Psilin$ as in \Cref{grand proposition backwards scattering} and for $u\geq u_+$, the following estimates apply:
    \begin{align}
         \int_{S^2}\sin\theta d\theta d\phi\,\left|\frac{1}{\Omega^2}\nablau\Psilin(u,v,\theta^A)\right|^2\lesssim e^{\frac{5}{2M}(v_+-v)}\left\{\int_{\mathscr{H}^+\cap\{\bar{v}\in[v,v_+]\}}d\bar{v}\sin\theta d\theta d\phi\, \left[\sum_{|\gamma|\leq 3}|\mathring{\slashednabla}^\gamma{\upPsilin}_{\mathscr{H}^+}|^2\right]\right\}
    \end{align}
    \begin{align}
        \int_v^{v_+}\int_{S^2}d\bar{v}\sin\theta d\theta d\phi\,\left|\frac{1}{\Omega^2}\nablau\Psilin({u},\bar{v},\theta^A)\right|^2\lesssim e^{\frac{5}{2M}(v_+-v)}\left\{\int_{\mathscr{H}^+\cap\{\bar{v}\in[v,v_+]\}}d\bar{v}\sin\theta d\theta d\phi\, \left[\sum_{|\gamma|\leq 3}|\mathring{\slashednabla}^\gamma{\upPsilin}_{\mathscr{H}^+}|^2\right]\right\}
    \end{align}
\end{proposition}

We can prove boundedness estimates on  $\left(\frac{1}{\Omega^2}\nablau\right)^k\Psilin$ near $\mathscr{H}^+_{\geq0}$ by induction on the equation \bref{equation for transverse null derivative near H+} to find

\begin{proposition}
    For $\Psilin$ as in \Cref{grand proposition backwards scattering} and for $u\geq u_+$, the following estimates apply:
    \begin{align}\label{blueshift estimate on Psilin n backwards scattering 1}
         \int_{S^2}\sin\theta d\theta d\phi\,\left|\left(\frac{1}{\Omega^2}\nablau\right)^n\Psilin(u,v,\theta^A)\right|^2\lesssim& e^{\frac{n^2+4}{2M}(v_+-v)}\times\\&\left\{\int_{\mathscr{H}^+\cap\{\bar{v}\in[v,v_+]\}}d\bar{v}\sin\theta d\theta d\phi\, \left[\sum_{|\gamma|\leq 2n+1}|\mathring{\slashednabla}^\gamma{\upPsilin}_{\mathscr{H}^+}|^2\right]\right\}
    \end{align}
    \begin{align}\label{blueshift estimate on Psilin n backwards scattering 2}
    \begin{split}
        \int_v^{v_+}\int_{S^2}d\bar{v}\sin\theta d\theta d\phi\,\left|\left(\frac{1}{\Omega^2}\nablau\right)^n\Psilin({u},\bar{v},\theta^A)\right|^2\lesssim& e^{\frac{n^2+4}{2M}(v_+-v)}\times\\&\left\{\int_{\mathscr{H}^+\cap\{\bar{v}\in[v,v_+]\}}d\bar{v}\sin\theta d\theta d\phi\, \left[\sum_{|\gamma|\leq 2n+1}|\mathring{\slashednabla}^\gamma{\upPsilin}_{\mathscr{H}^+}|^2\right]\right\}
    \end{split}
    \end{align}
\end{proposition}

As a corollary of  \Cref{backwards rp RW} and the estimate \bref{RW exponential backwards near H+} we have
\begin{corollary}\label{complicated backwards blueshift RW}
For $\Psilin$ arising from $\mathfrak{S}$ of \Cref{grand proposition backwards scattering} and for all $u\leq u_+$, $0\leq v\leq v_+$ we have
    \begin{align}\label{complicated backwards blue shift estimate Rw}
    \begin{split}
         F_u[\Psilin](v,\infty)+\underline{F}_v[\Psilin](u,\infty)&\lesssim e^{\frac{1}{2M}(v_+-v)}\times\left\{\int_{\mathscr{H}^+\cap\{\bar{v}\in[v,v_+]\}}d\bar{v}\sin\theta d\theta d\phi\, \left[|\mathring{\slashednabla}{\upPsilin}_{\mathscr{H}^+}|^2+|\bm{\uppsi}_{\mathscr{H}^+}|^2\right]\right\}\\&+\int_{\mathscr{H}^+\cap\{\bar{v}\in[v,v_+]\}}d\bar{v}\sin\theta d\theta d\phi\,|\partial_v \upPsilin_{\mathscr{H}^+}|^2+\int_{\mathscr{I}^+\cap\{\bar{u}\in[u,u_+]\}}d\bar{u}\sin\theta d\theta d\phi\,|\partial_u \upPsilin_{\mathscr{I}^+}|^2\\
         &+ \left(\frac{r^2\Omega^{-2}(u_+,v_\infty)}{r^2\Omega^{-2}(u,v_\infty)}\right) \int_{\mathscr{I}^+\cap\{\bar{u}\in[u,u_+]\}}d\bar{u}\sin\theta d\theta d\phi\, \left[|\mathring{\slashednabla}\upPsilin_{\mathscr{I}^+}|^2+4|\upPsilin_{\mathscr{I}^+}|^2\right].
    \end{split}
    \end{align}
\end{corollary}

\paragraph{Integrated local energy decay estimates}

Finally, we note the following $L^2$ spacetime estimates, which immediately follow from  \Cref{RWILED,,RW T-energy} and \Cref{complicated backwards blueshift RW}.

\begin{proposition}\label{backwards ILED RW}
    Let $\Psilin$ be as in \Cref{grand proposition backwards scattering}. Then we have
    \begin{align}\label{backwards degenerate ILED RW}
         \mathbb{I}_{deg}^{u,v}[\Psi]\lesssim \int_{\mathscr{H}^+\cap\{\bar{v}\geq v\}}d\bar{v}\sin\theta d\theta d\phi\,|\partial_v \upPsilin_{\mathscr{H}^+}|^2+\int_{\mathscr{I}^+\cap\{\bar{u}\geq u\}}d\bar{u}\sin\theta d\theta d\phi\,|\partial_u \upPsilin_{\mathscr{I}^+}|^2,
    \end{align}
    \begin{align}\label{backwards nondegenerate ILED RW}
        \mathbb{I}^{u,v}[\Psi]\lesssim \text{ Right hand side of } \bref{complicated backwards blue shift estimate Rw}
    \end{align}
    The same applies to $\Psilinb$.
\end{proposition}

The estimate \bref{backwards integrated rp RW} allows us to improve on the $r$-weights of $r$-weights of \Cref{backwards ILED RW} near $\mathscr{I}^+$:
\begin{proposition}\label{backwards ILED RW better r weight Gronwall}
 Let $\Psilin$ be as in \Cref{grand proposition backwards scattering} and let $u,v$ be such that $S^2_{u,v}\in J^+({\Sigma^*_+})$. Then we have
\begin{align}
\begin{split}
     \int_{\mathscr{D}_{u,v}\cap J^+({\Sigma^*_+})}d\bar{u}d\bar{v}\sin\theta d\theta d\phi\,\frac{\Omega^2}{r^2}|\Psilin|^2&\lesssim \int_{\mathscr{I}^+\cap\{\bar{u}\in[u,u_+]\}}d\bar{u}\sin\theta d\theta d\phi\, \left[|\mathring{\slashednabla}\upPsilin_{\mathscr{I}^+}|^2+4|\upPsilin_{\mathscr{I}^+}|^2+|\partial_u \upPsilin_{\mathscr{I}^+}|^2\right]\\
     &+\int_{\mathscr{H}^+\cap\{\bar{v}\geq v\}}d\bar{v}\sin\theta d\theta d\phi\,|\partial_v \upPsilin_{\mathscr{H}^+}|^2
    \end{split}
\end{align}
The same estimate applies with $\Psilinb$, $\upPsilinb_{\mathscr{I}^+}$, $\upPsilinb_{\mathscr{H}^+}$ replacing $\Psilin$, $\upPsilin_{\mathscr{I}^+}$, $\upPsilin_{\mathscr{H}^+}$.
\end{proposition}
We can also improve on the weights by combining \Cref{backwards ILED RW} with the backwards $r^p$-estimates of Section 5.5.2 of \cite{Mas20}, adapted from \cite{AAG16a}. For given $u,v$, denote $\mathscr{D}_{u,v}=J^+(\mathscr{C}_u)\cup J^+(\underline{\mathscr{C}}_v)$.

\begin{proposition}\label{backwards ILED RW better r weight energy identity at infinity}
    Let $\Psilin$ be as in \Cref{grand proposition backwards scattering} and let $u,v$ be such that $S^2_{u,v}\in J^+({\Sigma^*_+})$. Then we have
    \begin{align}\label{backwards ILED estimate RW better r weight}
    \begin{split}
        \int_{\mathscr{D}_{u,v}\cap J^+({\Sigma^*_+})}d\bar{u}d\bar{v}\sin\theta d\theta d\phi\,\frac{\Omega^2}{r^2}|\Psilin|^2&\lesssim \int_{\mathscr{I}^+\cap\{\bar{u}\geq u\}}d\bar{u}\sin\theta d\theta d\phi\,(1+|u_+-u|)|\partial_u \upPsilin_{\mathscr{I}^+}|^2\\&+\int_{\mathscr{H}^+\cap\{\bar{v}\geq v\}}d\bar{v}\sin\theta d\theta d\phi\,|\partial_v \upPsilin_{\mathscr{H}^+}|^2.
    \end{split}
    \end{align}
     The same estimate applies with $\Psilinb$, $\upPsilinb_{\mathscr{I}^+}$, $\upPsilinb_{\mathscr{H}^+}$ replacing $\Psilin$, $\upPsilin_{\mathscr{I}^+}$, $\upPsilin_{\mathscr{H}^+}$.
\end{proposition}

\paragraph[Asymptotics of $\protect\Psilin$, $\protect\Psilinb$ near $i^0$]{Asymptotics of $\Psilin$, $\Psilinb$ near $i^0$}

The estimates of \Cref{backwards rp RW Phi n} allows us to study the asymptotics of $\Psilin, \Psilinb$ as $r\longrightarrow\infty$ on ${\Sigma^*_+}$:

\begin{proposition}\label{asymptotic flatness Psilin}
Under the assumptions of \Cref{grand proposition backwards scattering}, $\Psilin$, $\Psilinb$ satisfy
\begin{align}
    \Psilin|_{{\Sigma^*_+}}, \Psilinb|_{{\Sigma^*_+}} =O\left(1\right).
\end{align}
as $r\longrightarrow\infty$. If the assumptions of \Cref{grand proposition backwards scattering addendum} apply then we have
\begin{align}
    \Psilin|_{{\Sigma^*_+}}, \Psilinb|_{{\Sigma^*_+}} =O\left(\frac{1}{r}\right).
\end{align}
\end{proposition}

\begin{proof}
   We compare $\Psilin$ to $\upPsilin_{\mathscr{I}^+}$ to get
    \begin{align}\label{this 22 08 2021 2}
    \begin{split}
        \int_{S^2}\sin\theta d\theta d\phi\,\left|\Psilin|_{{\Sigma^*_+}}(u,\theta^A)-\upPsilin_{\mathscr{I}^+}(u,\theta^A)\right|^2&=\int_{S^2}\sin\theta d\theta d\phi\,\left|\int_{v}^\infty d\bar{v} \nablav \Psilin\right|^2\\&\leq\frac{1}{r(u,v_{{\Sigma^*_+},u})}\int_{\mathscr{C}_u\{\cap\{\bar{v}\geq v_{{\Sigma^*_+}}(u)\}\}}d\bar{v}\sin\theta d\theta d\phi\,\frac{r^2}{\Omega^2}|\nablav\Psilin(u,\bar{v},\theta^A)|^2\\
        &\lesssim \frac{1}{r(u,\vsigmap{{u}})}\int_{\mathscr{I}^+\cap\{\bar{u}\geq u\}}d\bar{u}\sin\theta d\theta d\phi\,|\mathring{\slashednabla}\upPsilin_{\mathscr{I}^+}|^2+|\upPsilin_{\mathscr{I}^+}|^2.
    \end{split}
    \end{align}
    If $\int_{u_-}^{u_+}d\bar{u}\,\xblins_{\mathscr{I}^+}\neq0$, commuting the above with the generators of $so(3)$ and applying a Sobolev estimate shows that $\Psilin|_{\Sigma^*_+}=O(1)$ as $r\longrightarrow\infty$ (since in Bondi-normalised coordinates $\upPsilin_{\mathscr{I}^+}=\mathring{\slashed{\Delta}}(\mathring{\slashed{\Delta}}+2)\xlins_{\mathscr{I}^+}$). If we have that $\int_{u_-}^{u_+}d\bar{u}\,\xblins_{\mathscr{I}^+}=0$ then the above gives $\Psilin|_{\Sigma^*_+}$ a decay of $O(r^{-\frac{1}{2}})$. We can improve the decay rate as follows: for $u\leq u_-$ we can estimate
    \begin{align}\label{decay of Psilin towards i0 backwards scattering}
        \begin{split}
             \int_{S^2}\sin\theta d\theta d\phi\,\left|\Psilin|_{{\Sigma^*_+}}(u,\theta^A)\right|^2\leq&\frac{1}{r(u,v_{{\Sigma^*_+},u})}\int_{\mathscr{C}_u\{\cap\{\bar{v}\geq v_{{\Sigma^*_+},u}\}\}}d\bar{v}\sin\theta d\theta d\phi\,\frac{r^2}{\Omega^2}|\nablav\Psilin(u,\bar{v},\theta^A)|^2
            \\&\lesssim \frac{1}{r(u,v_{{\Sigma^*_+},u})^2}\times\sup_{\bar{v}\geq v}|\Phi^{(1)}|^2
            \\&\lesssim \frac{1}{r(u,v_{{\Sigma^*_+},u})^2}\Bigg[|\upPhi^{(1)}_{\mathscr{I}^+}(u,\theta^A)|^2\\&+\frac{1}{r(u,\vsigmap{u})}\int_{\mathscr{C}_{u}\cap\{\bar{v}\geq \vsigmap{u}\}} d\bar{v}\dw\,\frac{r^2}{\Omega^2}|\nablav\Phi^{(1)}|^2\Bigg]
            \\& \lesssim \frac{1}{r(u,v_{{\Sigma^*_+},u})^2}\Bigg[|\upPhi^{(1)}_{\mathscr{I}^+}(u,\theta^A)|^2\\&+\frac{1}{r(u,\vsigmap{u})}\int_{\mathscr{I}^+\cap\{\bar{u}\in[u,u_+]\}}d\bar{u}\dw\,|\mathring{\slashednabla}\upPhi_{\mathscr{I}^+}^{(1)}|^2+|\upPhi^{(1)}_{\mathscr{I}^+}|^2+|\upPsilin_{\mathscr{I}^+}|^2\Bigg].
        \end{split}
    \end{align}
    The claim follows knowing that \bref{formula for radiation field of Phi n} implies that $\mathring{\slashednabla}^{\gamma}\partial_u^i\Phi^{(n)}_{\mathscr{I}^+}=O\left(|u|^{n-i-1}\right)$ for any index $\gamma$ (see \Cref{asymptotics of Phi n scri}).
\end{proof}

\begin{proposition}\label{asymptotic flatness Psilin n}
    In addition to \Cref{asymptotic flatness Psilin}, we have
    \begin{align}\label{asymptotic flatness Psilin statement}
        \nablav^n\Psilin|_{{\Sigma^*_+}}, \nablav^n\Psilinb|_{{\Sigma^*_+}}, \nablau^n\Psilin|_{{\Sigma^*_+}}, \nablau^n\Psilinb|_{{\Sigma^*_+}} =O\left(\frac{1}{r^{n}}\right).
    \end{align}
    If the assumptions of \Cref{grand proposition backwards scattering addendum} apply then we can improve \bref{asymptotic flatness Psilin statement} by one power of $\frac{1}{r}$.
\end{proposition}
\begin{proof}
    Since
    \begin{align}
        \nablav^n \Psilin=\sum_{k=1}^nO\left(\frac{1}{r^{n+k}}\right){\Phi}^{(k)}
    \end{align}
    as $r\longrightarrow\infty$, the statement concerning $\nablav^n\Psilin$ follows if
    \begin{align}\label{this 23 08 2021}
        \Phi^{(n)}|_{\Sigma^*_+}=O\left(r^{n-1}\right).
    \end{align}
    Since $\mathring{\slashednabla}^{\gamma}\partial_u^i\Phi^{(n)}_{\mathscr{I}^+}=O\left(|u|^{n-i}\right)$, the estimate \bref{this 23 08 2021} holds since \bref{backwards rp estimate RW Phi n} implies for sufficiently large $r$ that
    \begin{align}\label{this 23 08 2021 2}
        \begin{split}
            |\Phi^{(n)}|_{{\Sigma^*_+}}(r,\theta^A)|^2&\lesssim |\Phi^{(n)}_{\mathscr{I}^+}(u)|^2+\frac{C(u_+,M,n)}{r}\int_{\mathscr{I}^+\cap\{\bar{u}\in[u,u_+]\}}d\bar{u}\sin\theta d\theta d\phi\,\left[|\mathring{\slashednabla}\Phi^{(n)}_{\mathscr{I}^+}|^2+|\Phi^{(n)}_{\mathscr{I}^+}|^2+|\Phi^{(n-1)}_{\mathscr{I}^+}|^2\right],
        \end{split}
    \end{align}
    where $u$ above in \bref{this 23 08 2021 2} is such that $r|_{\mathscr{C}_u\cap {\Sigma^*_+}}=r$. We will now show that
    \begin{align}\label{16 10 2021}
        \partial_t^n\Psilin=O\left(\frac{1}{r^{n}}\right).
    \end{align}
    Let $u_-$ be such that $\xblins_{\mathscr{I}^+}$ is not supported on $u\leq u_-$ and take $r$ such that $r|_{\mathscr{C}_u\cap{\Sigma^*_+}}=r$, $v$ such that $r(u,v)=r$. We estimate starting from \bref{decay of Psilin towards i0 backwards scattering}:
    \begin{align}\label{16 10 2021 1}
    \begin{split}
        &\int_{S^2}\dw\,\left|\Psilin|_{{\Sigma^*_+}}(r,\theta^A)-\upPsilin_{\mathscr{I}^+}(u,\theta^A)\right|^2\\&\lesssim
         \frac{1}{r(u,v_{{\Sigma^*_+},u})^2}\Bigg[\int_{S^2}\dw\,|\upPhi^{(1)}_{\mathscr{I}^+}(u,\theta^A)|^2+\frac{1}{r(u,\vsigmap{u})}\int_{\mathscr{C}_{u}\cap\{\bar{v}\geq \vsigmap{u}\}} d\bar{v}\dw\,\frac{r^2}{\Omega^2}|\nablav\Phi^{(1)}|^2\Bigg]\\
         &\lesssim \frac{1}{r(u,v_{{\Sigma^*_+},u})^2}\Bigg[\int_{S^2}\dw\,|\upPhi^{(1)}_{\mathscr{I}^+}(u,\theta^A)|^2+\frac{1}{r(u,\vsigmap{u})}\int_{\mathscr{C}_{u}\cap\{\bar{v}\geq \vsigmap{u}\}} d\bar{v}\dw\,\frac{\Omega^2}{r^2}|\Phi^{(2)}|^2\Bigg]\\
         &\lesssim  \frac{1}{r(u,v_{{\Sigma^*_+},u})^2}\int_{S^2}\dw\,|\upPhi^{(1)}_{\mathscr{I}^+}(u,\theta^A)|^2+\frac{1}{r(u,\vsigmap{u})^4}\int_{S^2}\dw\,|\upPhi^{(2)}_{\mathscr{I}^+}(u,\theta^A)|^2\\&+\frac{1}{r(u,\vsigmap{u})^5}\int_{\mathscr{C}_{u}\cap\{\bar{v}\geq \vsigmap{u}\}} d\bar{v}\dw\,\frac{r^2}{\Omega^2}|\nablav\Phi^{(2)}|^2\\
         &\lesssim \frac{1}{r(u,v_{{\Sigma^*_+},u})^2}\int_{S^2}\dw\,|\upPhi^{(1)}_{\mathscr{I}^+}(u,\theta^A)|^2+\frac{1}{r(u,\vsigmap{u})^4}\int_{S^2}\dw\,|\upPhi^{(2)}_{\mathscr{I}^+}(u,\theta^A)|^2\\&+\frac{1}{r(u,\vsigmap{u})^5}\int_{\mathscr{I}^+\cap\{\bar{u}\in[u,u_+]\}}d\bar{u}\dw\,|\mathring{\slashednabla}\upPhi_{\mathscr{I}^+}^{(2)}|^2+|\upPhi^{(2)}_{\mathscr{I}^+}|^2+|\upPhi^{(1)}_{\mathscr{I}^+}|^2.
    \end{split}
    \end{align}
    Commuting the above with $\partial_t$, we get on $u\leq u_-$ (where $\partial_u\upPhi^{(1)}_{\mathscr{I}^+}=(\mathring{\slashed{\Delta}}-4)\upPsilin_{\mathscr{I}^+}$):
    \begin{align}\label{16 10 2021 2}
    \begin{split}
         \int_{S^2}\dw\,|\partial_t\Psilin|_{{\Sigma^*_+}}(r,\theta^A)|^2&\lesssim \frac{1}{r(u,v_{{\Sigma^*_+},u})^2}\int_{S^2}\dw\,|(\mathring{\slashed{\Delta}}-4)\upPsilin_{\mathscr{I}^+}(u,\theta^A)|^2\\&+\frac{1}{r(u,\vsigmap{u})^4}\int_{S^2}\dw\,|\partial_u\upPhi^{(2)}_{\mathscr{I}^+}(u,\theta^A)|^2\\&+\frac{1}{r(u,\vsigmap{u})^5}\int_{\mathscr{I}^+\cap\{\bar{u}\in[u,u_+]\}}d\bar{u}\dw\,|\mathring{\slashednabla}\partial_u\upPhi_{\mathscr{I}^+}^{(2)}|^2+|\partial_u\upPhi^{(2)}_{\mathscr{I}^+}|^2+|\partial_u\upPhi^{(1)}_{\mathscr{I}^+}|^2.
    \end{split}
    \end{align}
    Knowing $\mathring{\slashednabla}^{\gamma}\partial_u^i\Phi^{(n)}_{\mathscr{I}^+}=O\left(|u|^{n-i}\right)$, commuting the above with the generators of $so(3)$ and applying a Sobolev estimate, we obtain \bref{16 10 2021} for $n=1$. The statement \bref{16 10 2021} can also be obtained for all $n$ by iterating \bref{16 10 2021 1} and \bref{16 10 2021 2} to get:
    \begin{align}\label{07 09 2022}
    \begin{split}
        \int_{S^2}\dw\,|\Psilin|_{{\Sigma^*_+}}(r,\theta^A)-\upPsilin_{\mathscr{I}^+}(u,\theta^A)|^2&\leq \frac{1}{r}\int_{\mathscr{C}_u\cap\{\bar{v}\geq v\}}d\bar{v}\sin\theta d\theta d\phi\,\frac{\Omega^2}{r^2}|\Phi^{(1)}|^2\\
        &\lesssim \frac{\sup_{S^2_{u,\infty}}|\Phi^{(1)}_{\mathscr{I}^+}|^2}{r^2}+\frac{1}{r^4}\,\sup_{\bar{v}\geq v}|\Phi^{(2)}(u,\bar{v},\theta^A)|^2\\
        &\lesssim_{k}\sum_{j=1}^k\frac{1}{r^{2j}}\sup_{S^2_{u,\infty}}|\Phi^{(j)}_{\mathscr{I}^+}|^2+\frac{1}{r^{2k+1}}\,\int_{\mathscr{C}_u\cap\{\bar{v}\geq v \}}d\bar{v}\dw\,\frac{r^2}{\Omega^2}|\nablav\Phii{k}|^2
        \\&\lesssim_{k} \sum_{j=1}^k\frac{1}{r^{2j}}\sup_{S^2_{u,\infty}}|\Phi^{(j)}_{\mathscr{I}^+}|^2\\&+\frac{1}{r^{2k+1}}\int_{\mathscr{I}^+\cap\{\bar{u}\geq u\}}d\bar{u}\sin\theta d\theta d\phi\,|\mathring{\slashednabla}\Phi^{(k+1)}_{\mathscr{I}^+}|^2+|\Phi^{(k+1)}_{\mathscr{I}^+}|^2+M|\Phi^{(k)}_{\mathscr{I}^+}|^2.
    \end{split}
    \end{align}
    Taking $k=n$ above and commuting the above with $\partial^n_t$, we get
    \begin{align}\label{16 10 2021 3}
    \begin{split}
        |\partial_t^n\Psilin|_{{\Sigma^*_+}}(r,\theta^A)|^2&\lesssim_n \frac{1}{r^{2n}}\sup_{S^2_{u,\infty}}|\partial_u^n\Phi^{(n)}_{\mathscr{I}^+}|^2+\frac{1}{r^{2n+1}}\int_{\mathscr{I}^+\cap\{\bar{u}\geq u\}}d\bar{u}\sin\theta d\theta d\phi\,|\mathring{\slashednabla}\partial_u^n\Phi^{(n)}_{\mathscr{I}^+}|^2+|\partial_u^n\Phi^{(n)}_{\mathscr{I}^+}|^2+M|\partial_u^n\Phi^{(n-1)}_{\mathscr{I}^+}|^2\\
        &= O\left(\frac{1}{r^{n}}\right),
    \end{split}
    \end{align}
    since $\mathring{\slashednabla}^{\gamma}\partial_u^i\Phi^{(n)}_{\mathscr{I}^+}=O\left(|u|^{n-i}\right)$. In the case that the assumptions of \Cref{grand proposition backwards scattering addendum} are satisfied, the first term in \bref{16 10 2021 3} vanishes, thus we may take $k=n+1$ in \bref{07 09 2022} and use the fact that $\mathring{\slashednabla}^{\gamma}\partial_u^i\Phi^{(n)}_{\mathscr{I}^+}=O\left(|u|^{n-1-i}\right)$, so we may obtain an additional power of decay in $r$ using the above argument.
\end{proof}

\subsubsection[Boundedness on $\protect\alin$, $\protect\ablin$]{Boundedness on $\alin$, $\ablin$}

We now estimate $\alin$, $\ablin$ using the estimates on $\Psilin$, $\Psilinb$ of the previous section and the transport relations \bref{hier}.

\paragraph[Boundedness on $\protect\ablin$]{Boundedness on $\protect\ablin$}

Near $\mathscr{H}^+$ we will need a blueshift estimate to obtain boundedness on $\Omega^{-1}\nablagml\Omega^{-2}\ablin$. Away from $\mathscr{H}^+_{\geq0}$, energy boundedness and \bref{hier} will be sufficient to estimate $\pblin$ $\ablin$ all the way to $\mathscr{I}^+$.

The following estimates were proven in \cite{Mas20}:

\begin{proposition}
    Let $\Psilinb$, $\pblin$, $\ablin$ be as in \Cref{grand proposition backwards scattering}. Then we have
    \begin{align}
        &\int_{{\mathscr{C}}_u} \frac{r^2}{\Omega^2}|\nablav r^3\Omega\underline\psi|^2\leq {F}^T_u[\underline\Psi](v,\infty),\label{backwards rp pblin}\\
        &\int_{{\mathscr{C}}_u} \frac{r^2}{\Omega^2}|\nablav r\Omega^2\underline\alpha|^2\lesssim \frac{1}{r^2}{F}^T_u[\underline\Psi](v,\infty),\label{backwards rp ablin}
    \end{align}
    for any $u,v$. Moreover, we have
    \begin{align}
    \begin{split}
        \int_{\underline{\mathscr{C}}_v\cap\{r\leq3M\}}d\bar{u}\sin\theta d\theta d\phi\,\frac{r^2}{\Omega^2}|\nablau r^5\Omega^{-2}\ablin|^2+&\int_{v}^{v_+}\int_{\underline{\mathscr{C}}_{\bar{v}}\cap\{r\leq3M\}}d\bar{v} d\bar{u}\sin\theta d\theta d\phi\,\frac{r^2}{\Omega^2}|\nablau r^5\Omega^{-2}\ablin|^2\\+\int_{v}^{v_+}\int_{\underline{\mathscr{C}}_{\bar{v}}\cap\{r\leq3M\}}d\bar{v} d\bar{u}\sin\theta d\theta d\phi\,\frac{\Omega^2}{r^2}|r^5\Omega^{-2}&\ablin|^2\\&\lesssim e^{\frac{5}{2M}(v_+-v)}\int_{\mathscr{H}^+}d\bar{v}\sin\theta d\theta d\phi\,\left[|\mathring{\slashednabla}\ablins_{\mathscr{H}^+}|^2+|\ablins_{\mathscr{H}^+}|^2\right],\label{backwards blueshift ablin}
    \end{split}
    \end{align}
    \begin{align}
        \begin{split}
            &\int_{\underline{\mathscr{C}}_v\cap\{r\leq3M\}}d\bar{u}\sin\theta d\theta d\phi\,\frac{r^2}{\Omega^2}|\nablau r^5\Omega^{-1}\pblin|^2+\int_{v}^{v_+}\int_{\underline{\mathscr{C}}_{\bar{v}}\cap\{r\leq3M\}}d\bar{v} d\bar{u}\sin\theta d\theta d\phi\,\frac{r^2}{\Omega^2}|\nablau r^5\Omega^{-1}\pblin|^2\\&+\int_{v}^{v_+}\int_{\underline{\mathscr{C}}_{\bar{v}}\cap\{r\leq3M\}}d\bar{v} d\bar{u}\sin\theta d\theta d\phi\,\frac{\Omega^2}{r^2}|r^5\Omega^{-1}\pblin|^2\\&\qquad\qquad\qquad\lesssim_{M} e^{\frac{4}{M}(v_+-v)}\int_{\mathscr{H}^+_{\geq0}}d\bar{v}\sin\theta d\theta d\phi\,\Bigg[|\mathring{\slashednabla}\pblins_{\mathscr{H}^+}|^2+|\pblins_{\mathscr{H}^+}|^2+\sum_{|\gamma|\leq3}|\mathring{\slashednabla}^\gamma \ablins_{\mathscr{H}^+}|^2\Bigg].\label{backwards blueshift pblin}
        \end{split}
    \end{align}
\end{proposition}

\begin{proof}
    The estimates \bref{backwards rp pblin},  \bref{backwards rp ablin} follow immediately from \bref{hier}. We derive \bref{backwards blueshift ablin} from the $-2$ Teukolsky equation \bref{T-2} as follows: write \bref{T-2} as
    \begin{align}\label{T-2 normalised to H+}
        \nablau\nablav r^5\Omega^{-2}\ablin-2\frac{3\Omega^2-1}{r}\nablau r^5\Omega^{-2}\ablin-\frac{\Omega^2}{r^2}(3\Omega^2-2)r^5\Omega^{-2}\ablin=\frac{\Omega^2}{r^2}\left(\mathcal{A}_2-\frac{6M}{r}\right)r^5\Omega^{-2}\ablin,
    \end{align}
    which leads to 
    \begin{align}
    \begin{split}
        &\partial_v \left(\frac{r^2}{\Omega^2}|\nablau r^5\Omega^{-2}\ablin|^2\right)-5\frac{3\Omega^2-1}{r}\frac{r^2}{\Omega^2}|\nablau r^5\Omega^{-2}\ablin|^2+\partial_u\left((13-15\Omega^2)|r^5\Omega^{-2}\ablin|^2+|\mathring{\slashednabla}r^5\Omega^{-2}\ablin|^2\right)\\&-30M\frac{\Omega^2}{r^2}|r^5\Omega^{-2}\ablin|^2=0.
    \end{split}
    \end{align}
    Integrate in the region $\mathscr{D}=J^+({\Sigma^*_+})\cap J^+(\underline{\mathscr{C}}_v)\cap\{r\leq3M\}$ to get
    \begin{align}\label{this 06 08 2021}
    \begin{split}
        &\int_{\underline{\mathscr{C}}_{v}\cap\{r\leq 3M\}}d\bar{u}\sin\theta d\theta d\phi\,\frac{r^2}{\Omega^2}|\nablau r^5\Omega^{-2}\ablin|^2+\int_{r=3M\cap\{J^+(\underline{\mathscr{C}}_v)\}}dt\sin\theta d\theta d\phi\left[|\mathring{\slashednabla}r^5\Omega^{-2}\ablin|^2+(13-15\Omega^2)|r^5\Omega^{-2}\ablin|^2\right]\\&+\int_{\mathscr{D}}d\bar{u}d\bar{v}\sin\theta d\theta d\phi\,\left[5\frac{3\Omega^2-1}{r}\frac{r^2}{\Omega^2}|\nablau r^5\Omega^{-2}\ablin|^2+30M\frac{\Omega^2}{r^2}|r^5\Omega^{-2}\ablin|^2\right]\\&-\int_{\mathscr{H}^+_{\geq0}\cap J^+(\underline{\mathscr{C}}_v)}d\bar{v}\sin\theta d\theta d\phi\,|\mathring{\slashednabla}r^5\Omega^{-2}\ablin|^2+13|r^5\Omega^{-2}\ablin|^2=0.
    \end{split}
    \end{align}
    Thus
    \begin{align}\label{this 06 08 2021 2}
    \begin{split}
        \int_{\underline{\mathscr{C}}_{v}\cap\{r\leq 3M\}}d\bar{u}\sin\theta d\theta d\phi\,\frac{r^2}{\Omega^2}|\nablau r^5\Omega^{-2}\ablin|^2&\leq \frac{5}{2M}\int_v^{v_+}d\bar{v}\int_{\underline{\mathscr{C}}_{\bar{v}}\cap\{r\leq 3M\}}d\bar{u}\sin\theta d\theta d\phi\,\left[\frac{r^2}{\Omega^2}|\nablau r^5\Omega^{-2}\ablin|^2\right]\\&+\int_{\mathscr{H}^+_{\geq0}\cap J^+(\underline{\mathscr{C}}_v)}d\bar{v}\sin\theta d\theta d\phi\,|\mathring{\slashednabla}r^5\Omega^{-2}\ablin|^2+13|r^5\Omega^{-2}\ablin|^2
    \end{split}
    \end{align}
    Gr\"onwall's inequality (\Cref{Gronwall inequality}) applied to \bref{this 06 08 2021 2} leads to \bref{backwards blueshift ablin}. The estimate \bref{backwards blueshift pblin} follows from a similar argument applied to the equation
    \begin{align}
        \nablau\nablav r^5\Omega^{-1}\pblin-\frac{3\Omega^2-1}{r}\nablau r^5\Omega^{-1}\pblin-\frac{\Omega^2}{r^2}\left(\mathcal{A}_2-\frac{6M}{r}\right)r^5\Omega^{-1}\pblin=6Mr\Omega^2\ablin
    \end{align}
    and using the estimate \bref{backwards blueshift ablin}.
\end{proof}

We can derive higher order versions of the estimates above using
\begin{align}
\begin{split}
    &\nablav \left(\frac{r^2}{\Omega^2}\nablau\right)^nr^5\Omega^{-1}\pblin-(n+1)\left(\frac{2}{r}-\frac{6M}{r^2}\right)\left(\frac{r^2}{\Omega^2}\nablau\right)^nr^5\Omega^{-1}\pblin\\&-\left(n^2+n-4-(n^2+n)\frac{6M}{r}\right) \left(\frac{r^2}{\Omega^2}\nablau\right)^{n-1}r^5\Omega^{-1}\pblin\\&+2Mn(n^2-1) \left(\frac{r^2}{\Omega^2}\nablau\right)^{n-2}r^5\Omega^{-1}\pblin= \left(\frac{r^2}{\Omega^2}\nablau\right)^n\Psilinb,
\end{split}
\end{align}

\begin{align}
\begin{split}
    &\nablav \left(\frac{r^2}{\Omega^2}\nablau\right)^nr^5\Omega^{-2}\ablin-(n+2)\left(\frac{2}{r}-\frac{6M}{r^2}\right)\left(\frac{r^2}{\Omega^2}\nablau\right)^nr^5\Omega^{-2}\ablin\\&-\left(n^2+3n-6-(n^2+3n+2)\frac{4M}{r}\right) \left(\frac{r^2}{\Omega^2}\nablau\right)^{n-1}r^5\Omega^{-2}\ablin\\&+2M(n-1)(n+2)(n+3) \left(\frac{r^2}{\Omega^2}\nablau\right)^{n-2}r^5\Omega^{-2}\ablin= \left(\frac{r^2}{\Omega^2}\nablau\right)^nr^5\Omega^{-1}\pblin.
\end{split}
\end{align}

\begin{corollary}\label{blueshift on ablin pblin n commuted}
Let $\Psilinb$, $\pblin$, $\ablin$ be as in \Cref{grand proposition backwards scattering}. Then for $\tilde{u}$ such that $r(\tilde{u},v_+)>3M$ and for any $u\geq \tilde{u}$, we have
\begin{align}\label{blueshift estimate pblin near H+ n times}
\begin{split}
    &\int_{S^2}\dw\left|\,\left(\frac{r^2}{\Omega^2}\nablau\right)^nr^5\Omega^{-1}\pblin\,\right|^2+\int_v^{v_+}d\bar{v}\int_{S^2}\dw \left|\,\left(\frac{r^2}{\Omega^2}\nablau\right)^nr^5\Omega^{-1}\pblin\,\right|^2\\
    &\lesssim C(n,M)e^{\frac{1}{2M}(n^2+2n+3)\times(v_+-v)}\Bigg\{\int_{\mathscr{H}^+_{\geq0}}d\bar{v}\dw\sum_{|\gamma|\leq 2n+1}\left[|\mathring{\slashednabla}^\gamma\pblins_{\mathscr{H}^+}|^2+|\mathring{\slashednabla}^\gamma\partial_v\upPsilinb_{\mathscr{H}^+}|^2+|\mathring{\slashednabla}^\gamma\upPsilinb_{\mathscr{H}^+}|^2\right]\\&\qquad\qquad\qquad\qquad\qquad\qquad\qquad\qquad\qquad\qquad\qquad+\sum_{|\gamma|\leq2n+3}|\mathring{\slashednabla}^\gamma \ablins_{\mathscr{H}^+}|^2\Bigg\},
\end{split}
\end{align}
\begin{align}\label{blueshift estimate ablin near H+ n times}
\begin{split}
    &\int_{S^2}\dw\left|\,\left(\frac{r^2}{\Omega^2}\nablau\right)^nr^5\Omega^{-2}\ablin\,\right|^2+\int_v^{v_+}d\bar{v}\int_{S^2}\dw \left|\,\left(\frac{r^2}{\Omega^2}\nablau\right)^nr^5\Omega^{-2}\ablin\,\right|^2\\
    &\lesssim C(n,M)e^{\frac{1}{2M}(n^2+4n+5)\times(v_+-v)}\left\{\int_{\mathscr{H}^+_{\geq0}}d\bar{v}\dw\sum_{|\gamma|\leq 2n+1}\left[|\mathring{\slashednabla}^\gamma\pblins_{\mathscr{H}^+}|^2+|\mathring{\slashednabla}^\gamma\partial_v\upPsilinb_{\mathscr{H}^+}|^2+|\mathring{\slashednabla}^\gamma\upPsilinb_{\mathscr{H}^+}|^2+|\mathring{\slashednabla}^\gamma\ablins_{\mathscr{H}^+}|^2\right]\right\}.
\end{split}
\end{align}
\end{corollary}

The following estimates will be helpful in later sections when estimating the connection and curvature components:

\begin{proposition}\label{backwards estimate on ablin ingoing direction beyond H+}
Let $\Psilinb$, $\pblin$, $\ablin$ be as in \Cref{grand proposition backwards scattering}. Then we have for any $u,v$ such that $S^2_{u,v}\in J^+({\Sigma^*_+})$,
\begin{align}\label{backwards overall ingoing estimate pblin better}
\begin{split}
    \int_{\underline{\mathscr{C}}_v}d\bar{u}\sin\theta d\theta d\phi\,|r^3\Omega\pblin|^2+&\int_u^{\infty}\int_{\underline{\mathscr{C}}_{\bar{v}}} d\bar{u}d\bar{v}\sin\theta d\theta d\phi\,\frac{\Omega^2}{r^2}|r^3\Omega\pblin|^2\\&\lesssim \int_{\mathscr{I}^+\cap\{\bar{u}\geq u\}}d\bar{u}\sin\theta d\theta d\phi\,\left[|\pblins_{\mathscr{I}^+}|^2+|\upPsilinb_{\mathscr{I}^+}|^2+|\mathring{\slashednabla}\upPsilinb_{\mathscr{I}^+}|^2+|\partial_u\upPsilinb_{\mathscr{I}^+}|^2\right],\\
    &+\int_{\mathscr{H}^+_{\geq0}\cap\{\bar{v}\geq v\}}d\bar{v}\dw\,|\partial_v\upPsilinb_{\mathscr{H}^+}|^2.
\end{split}
\end{align}
\begin{align}\label{backwards overall ingoing estimate ablin better}
\begin{split}
    \int_{\underline{\mathscr{C}}_{v}}d\bar{u}\sin\theta d\theta d\phi\,|r\Omega^2\ablin|^2+&\int_u^{\infty}\int_{\underline{\mathscr{C}}_{\bar{v}}} d\bar{u}d\bar{v}\sin\theta d\theta d\phi\,\frac{\Omega^2}{r^2}|r\Omega^2\ablin|^2\\&\lesssim
    \int_{\mathscr{I}^+\cap\{\bar{u}\geq u\}}d\bar{u}\sin\theta d\theta d\phi\,\left[|\ablins_{\mathscr{I}^+}|^2+|\pblins_{\mathscr{I}^+}|^2+|\upPsilinb_{\mathscr{I}^+}|^2+|\mathring{\slashednabla}\upPsilinb_{\mathscr{I}^+}|^2+|\partial_u\upPsilinb_{\mathscr{I}^+}|^2\right]
    \\&+\int_{\mathscr{H}^+_{\geq0}\cap\{\bar{v}\geq v\}}d\bar{v}\dw\,|\partial_v\upPsilinb_{\mathscr{H}^+}|^2.
\end{split}
\end{align}
\end{proposition}
\begin{proof}
Using the fundamental theorem of calculus and the Cauchy--Schwarz inequality, we can show that
    \begin{align}
    \begin{split}
        \int_{\underline{\mathscr{C}}_v} d\bar{u}\sin\theta d\theta d\phi\,|r^3\Omega\pblin|^2&\leq 2\int_{u}^\infty \int_{S^2} d\bar{u}\sin\theta d\theta d\phi\,|\pblins_{\mathscr{I}^+}|^2+2\int_{\underline{\mathscr{C}}_v}d\bar{u}\sin\theta d\theta d\phi\, |r^3\Omega\pblin-\pblins_{\mathscr{I}^+}|^2\\
        &\leq 2\int_{u}^\infty \int_{S^2} d\bar{u}\sin\theta d\theta d\phi\,|\pblins_{\mathscr{I}^+}|^2+2\int_{\underline{\mathscr{C}}_v}d\bar{u}\sin\theta d\theta d\phi\, \frac{1}{r(\bar{u},v)} \int_{\bar{v}}^\infty d\bar{v}\frac{\Omega^2}{r^2}|\Psilinb|^2\\
        &\leq 2\int_{u}^\infty \int_{S^2} d\bar{u}\sin\theta d\theta d\phi\,|\pblins_{\mathscr{I}^+}|^2+\frac{1}{M}\int_{u}^\infty \int_{\mathscr{C}_{\bar{u}}\cap\{\bar{v}\geq v\}}d\bar{u}d\bar{v}\sin\theta d\theta d\phi\,\frac{\Omega^2}{r^2}|\Psilinb|^2.
    \end{split}
    \end{align}
    \begin{align}
    \begin{split}
        \int_u^\infty \int_{\mathscr{C}_{\bar{u}}}d\bar{u}\sin\theta d\theta d\phi\,\frac{\Omega^2}{r^2}|r^3\Omega\pblin|^2&\leq 2\int_u^\infty \int_v^\infty \int_{S^2}d\bar{u}d\bar{v}\sin\theta d\theta d\phi\,\frac{\Omega^2}{r^2}|\pblins_{\mathscr{I}^+}|^2\\&+\int_u^\infty \int_v^\infty \int_{S^2}d\bar{u}d\bar{v}\sin\theta d\theta d\phi\,\frac{\Omega^2}{r^2}\int_{\bar{v}}^\infty d\bar{\bar{v}}\frac{\Omega^2}{r^2}|\Psilinb|^2\\
        &\leq \frac{1}{M}\int_u^\infty  \int_{S^2}d\bar{u}d\bar{v}\sin\theta d\theta d\phi\,\frac{\Omega^2}{r^2}|\pblins_{\mathscr{I}^+}|^2+\frac{1}{M}\int_{u}^\infty \int_{v}^\infty d\bar{u}d\bar{v}\sin\theta d\theta d\phi\,\frac{\Omega^2}{r^2}|\Psilinb|^2
    \end{split}
    \end{align}
The estimate \bref{backwards overall ingoing estimate pblin better} now follows from \Cref{backwards ILED RW better r weight Gronwall}. An identical argument using \bref{backwards overall ingoing estimate pblin better} goes through for $\ablin$, leading to \bref{backwards overall ingoing estimate ablin better}.
\end{proof}

\begin{remark}\label{backwards integral of ablin pblin on Sigma*}
An identical argument to the proof of \Cref{backwards estimate on ablin ingoing direction beyond H+} can be used to show
\begin{align}\label{backwards overall ingoing estimate pblin better Sigma^*}
\begin{split}
    \int_{{\Sigma^*_+}}dr\sin\theta d\theta d\phi\,|r^3\Omega\pblin|^2+&\int_{J^+({\Sigma^*_+})} d\bar{u}d\bar{v}\sin\theta d\theta d\phi\,\frac{\Omega^2}{r^2}|r^3\Omega\pblin|^2\\&\lesssim \int_{\mathscr{I}^+}d\bar{u}\sin\theta d\theta d\phi\,\left[|\pblins_{\mathscr{I}^+}|^2+|\upPsilinb_{\mathscr{I}^+}|^2+|\mathring{\slashednabla}\upPsilinb_{\mathscr{I}^+}|^2+|\partial_u\upPsilinb_{\mathscr{I}^+}|^2\right],\\
    &+\int_{\mathscr{H}^+_{\geq0}}d\bar{v}\dw\,|\partial_v\upPsilinb_{\mathscr{H}^+}|^2.
\end{split}
\end{align}
\begin{align}\label{backwards overall ingoing estimate ablin better Sigma*}
\begin{split}
    \int_{{\Sigma^*_+}}d\bar{u}\sin\theta d\theta d\phi\,|r\Omega^2\ablin|^2+&\int_{J^+({\Sigma^*_+})} d\bar{u}d\bar{v}\sin\theta d\theta d\phi\,\frac{\Omega^2}{r^2}|r\Omega^2\ablin|^2\\&\lesssim \int_{\mathscr{I}^+}d\bar{u}\sin\theta d\theta d\phi\,\left[|\ablins_{\mathscr{I}^+}|^2+|\pblins_{\mathscr{I}^+}|^2+|\upPsilinb_{\mathscr{I}^+}|^2+|\mathring{\slashednabla}\upPsilinb_{\mathscr{I}^+}|^2+|\partial_u\upPsilinb_{\mathscr{I}^+}|^2\right]
    \\&+\int_{\mathscr{H}^+_{\geq0}}d\bar{v}\dw\,|\partial_v\upPsilinb_{\mathscr{H}^+}|^2.
\end{split}
\end{align}
\end{remark}

\begin{corollary}\label{backwards ingoing overall on ablin}
Let $\Psilinb$, $\pblin$, $\ablin$ be as in \Cref{grand proposition backwards scattering}. Then we have for any $u,v$ such that $S^2_{u,v}\in J^+({\Sigma^*_+})$,
\begin{align}\label{backwards ingoing overall estimate on ablin}
\begin{split}
    \int_{\underline{\mathscr{C}}_v\cap\{\bar{u}\geq u\}}d\bar{u}\sin\theta d\theta d\phi\,\frac{1}{\Omega^2}|\ablin(\bar{u},v,\theta^A)|^2&\lesssim_M \int_{S^2}\sin\theta d\theta d\phi\,|\ablins_{\mathscr{H}^+}(v,\theta^A)|^2\\&+e^{\frac{5}{2M}(v_+-v)}\int_{\mathscr{H}^+_{\bar{v}\geq v}}d\bar{v}\sin\theta d\theta d\phi\,\left[|\mathring{\slashednabla}(2M)^5\ablins_{\mathscr{H}^+}|^2+|(2M)^5\ablins_{\mathscr{H}^+}|^2\right]\\ &+\int_{\mathscr{I}^+\cap\{\bar{u}\geq u\}}d\bar{u}\sin\theta d\theta d\phi\left[|\ablins_{\mathscr{I}^+}|^2+|\pblins_{\mathscr{I}^+}|^2+|\partial_u \upPsilinb_{\mathscr{I}^+}|^2\right]\\&+\int_{\mathscr{H}^+\cap\{\bar{v}\geq v\}}d\bar{v}\sin\theta  d\theta d\phi\,|\partial_v\upPsilinb_{\mathscr{H}^+}|^2.
\end{split}
\end{align}
\end{corollary}

\begin{proof}
    Given \Cref{backwards estimate on ablin ingoing direction beyond H+}, it suffices to handle the integral on the left hand side in the region $r\leq 3M$, which we estimate using \bref{backwards blueshift ablin}, since
    \begin{align}
    \begin{split}
        \int_{\underline{\mathscr{C}}_v\cap\{r\leq 3M\}}d\bar{u}\sin\theta d\theta d\phi\,\frac{1}{\Omega^2}|\ablin(\bar{u},v,\theta^A)|^2&\leq \frac{2}{(2M)^{10}}\int_{\underline{\mathscr{C}}_v\cap\{r\leq 3M\}}d\bar{u}\sin\theta d\theta d\phi\,r^{10}\Omega^2|\ablins_{\mathscr{H}^+}|^2\\&+\int_{\underline{\mathscr{C}}_v\cap\{r\leq 3M\}}d\bar{u}\sin\theta d\theta d\phi\,\Omega^2\left|\int_{\bar{u}}^\infty d\bar{\bar{u}}\nablau r^5\Omega^{-2}\ablin\right|^2\\
        &\leq \frac{6M}{11}\left(\frac{3}
        {2}\right)^{10}\int_{S^2}\sin\theta d\theta d\phi\,|\ablins_{\mathscr{H}^+}(v,\theta^A)|^2\\&+\frac{1}{2M^{9}}\int_u^\infty\int_{S^2}d\bar{u}\sin\theta d\theta d\phi\,\frac{r^2}{\Omega^2}|\nablau r^5\Omega^{-2}\ablin|^2.
    \end{split}
    \end{align}
\end{proof}

An identical argument to the above yields the following when applied to $\pblin$:

\begin{corollary}\label{backwards ingoing overall on pblin}
Let $\Psilinb$, $\pblin$, $\ablin$ be as in \Cref{grand proposition backwards scattering}. Then we have for any $u,v$ such that $S^2_{u,v}\in J^+({\Sigma^*_+})$,
\begin{align}\label{backwards ingoing overall estimate on pblin}
        \begin{split}
             \int_{\underline{\mathscr{C}}_v\cap\{\bar{u}\geq u\}}d\bar{u}\sin\theta d\theta &d\phi\,|\pblin(\bar{u},v,\theta^A)|^2\lesssim_M\int_{S^2}\sin\theta d\theta d\phi\,|\pblins_{\mathscr{H}^+}(v,\theta^A)|^2+\int_{\mathscr{H}^+\cap\{\bar{v}\geq v\}}d\bar{v}\sin\theta  d\theta d\phi\,|\partial_v\upPsilinb_{\mathscr{H}^+}|^2
             \\&+\int_{\mathscr{I}^+\cap\{\bar{u}\geq u\}}d\bar{u}\sin\theta d\theta d\phi\left[|\pblins_{\mathscr{I}^+}|^2+|\partial_u \upPsilinb_{\mathscr{I}^+}|^2\right]
             \\&+(2M)^{10}e^{\frac{4}{M}(v_+-v)}\int_{\mathscr{H}^+_{\geq0}}d\bar{v}\sin\theta d\theta d\phi\,\Bigg[|\mathring{\slashednabla}\pblins_{\mathscr{H}^+}|^2+|\pblins_{\mathscr{H}^+}|^2+\sum_{|\gamma|\leq3}|\mathring{\slashednabla}^\gamma \ablins_{\mathscr{H}^+}|^2\Bigg].
        \end{split}
    \end{align}
\end{corollary}

\paragraph[Asymptotics of $\protect\ablin$ towards $i^0$]{Asymptotics of $\ablin$ towards $i^0$}
\begin{corollary}\label{asymptotic flatness at spacelike infinity for ablin}
    The quantity $\ablin$ constructed in \Cref{grand proposition backwards scattering} satisfies
    \begin{align}
        \partial_t^n\ablin|_{{\Sigma^*_+}}=O_\infty\left(\frac{1}{r^{n+3}}\right).
    \end{align}
    If the assumptions of \Cref{grand proposition backwards scattering addendum} are satisfied then we have
    \begin{align}
        \partial_t^n\ablin|_{{\Sigma^*_+}}=O_\infty\left(\frac{1}{r^{n+4}}\right).
    \end{align}
\end{corollary}
\begin{proof}
    Given that $\ablins_{\mathscr{I}^+}$, $\pblins_{\mathscr{I}^+}$ vanish for $u\leq u_-$, we may follow the steps leading to \bref{16 10 2021} using \bref{hier} to get
    \begin{align}
         \int_{S^2}\dw\,\left|r^3\Omega\pblin|_{{\Sigma^*_+}}(r,\theta^A)\right|^2&\lesssim \frac{1}{r(u,\vsigmap{u})^2}\Bigg[ \int_{S^2}\dw\,|\upPsilinb_{\mathscr{I}^+}(u,\theta^A)|^2\\&\;\;+\frac{1}{r(u,\vsigmap{u})}\int_{\mathscr{I}^+\cap\{\bar{u}\in[u,u_+]\}}d\bar{u}\dw\,|\mathring{\slashednabla}\upPsilinb_{\mathscr{I}^+}|^2+|\upPsilinb_{\mathscr{I}^+}|^2\Bigg],
    \end{align}
    \begin{align}\label{17 10 2021}
    \begin{split}
         \int_{S^2}\dw\,\left|r\Omega^2\ablin|_{{\Sigma^*_+}}(r,\theta^A)\right|^2&\lesssim \frac{1}{r(u,\vsigmap{u})^4}\Bigg[ \int_{S^2}\dw\,|\upPsilinb_{\mathscr{I}^+}(u,\theta^A)|^2\\&\;\;+\frac{1}{r(u,\vsigmap{u})}\int_{\mathscr{I}^+\cap\{\bar{u}\in[u,u_+]\}}d\bar{u}\dw\,|\mathring{\slashednabla}\upPsilinb_{\mathscr{I}^+}|^2+|\upPsilinb_{\mathscr{I}^+}|^2\Bigg].
    \end{split}
    \end{align}
    which, knowing that $\mathring{\slashednabla}^{\gamma}\partial_u^i\underline{\Phi}^{(n)}_{\mathscr{I}^+}=O\left(|u|^{n-i}\right)$, implies $\nablav^i\ablin|_{{\Sigma^*_+}}=O(r^{-3-i})$ for $i=0$, $1$. Combining this with \Cref{asymptotic flatness Psilin n} implies $\nablav^i\ablin|_{{\Sigma^*_+}}=O(r^{-4-i})$ for all $i\in\mathbb{N}$. The statement $\partial_t\ablin|_{{\Sigma^*_+}}=O(r^{-4})$ follows by applying the argument of \Cref{asymptotic flatness Psilin n} leading to \bref{16 10 2021 2} to \bref{17 10 2021} to get for $u\leq u_-$:
   \begin{align}\label{17 10 2021 1}
    \begin{split}
         \int_{S^2}\dw\,\left|\partial_t r\Omega^2\ablin|_{{\Sigma^*_+}}(r,\theta^A)\right|^2&\lesssim \frac{1}{r(u,\vsigmap{u})^6}\int_{S^2}\dw\,|\partial_u\underline{\upPhi}^{(1)}_{\mathscr{I}^+}(u,\theta^A)|^2\\\;\;&+\frac{1}{r(u,\vsigmap{u})^7}\int_{\mathscr{I}^+\cap\{\bar{u}\in[u,u_+]\}}d\bar{u}\dw\,|\mathring{\slashednabla}\partial_u\underline{\upPhi}_{\mathscr{I}^+}^{(1)}|^2+|\partial_u\underline{\upPhi}_{\mathscr{I}^+}^{(1)}|^2+M|\partial_u\upPsilinb_{\mathscr{I}^+}|^2.
    \end{split}
    \end{align}
    The statement for higher order derivatives $\partial_t$ of $\ablin$ can be obtained by iterating the estimate above to get
    \begin{align}\label{17 10 2021 11}
    \begin{split}
        \int_{S^2}\dw\,\left|\partial_t^n r\Omega^2\ablin|_{{\Sigma^*_+}}(r,\theta^A)\right|^2&\lesssim \frac{1}{r(u,\vsigmap{u})^{4+2n}} \int_{S^2}\dw\,|\partial_u^n\underline{\upPhi}^{(n)}_{\mathscr{I}^+}(u,\theta^A)|^2\\&\;\;+\frac{1}{r(u,\vsigmap{u})^{5+2n}}\int_{\mathscr{I}^+\cap\{\bar{u}\in[u,u_+]\}}d\bar{u}\dw\,\Bigg[|\mathring{\slashednabla}\partial_u^n\underline{\upPhi}_{\mathscr{I}^+}^{(n)}|^2+|\partial_u^n\underline{\upPhi}_{\mathscr{I}^+}^{(n)}|^2\\&\qquad\qquad\qquad\qquad\qquad\qquad\qquad\qquad\qquad\qquad+M|\partial_u^n\underline{\upPhi}_{\mathscr{I}^+}^{(n-1)}|^2\Bigg].
    \end{split}
    \end{align}
    When the assumptions of \Cref{grand proposition backwards scattering addendum} are satisfied, we have $\mathring{\slashednabla}^{\gamma}\partial_u^i\underline{\Phi}^{(n)}_{\mathscr{I}^+}=O\left(|u|^{n-1-i}\right)$, so we may use the above argument to estimate $\partial_t^n r\Omega^2\ablin|_{\Sigma^*_+}$ in terms of $\partial_u^{n+1}\underline{\upPhi}^{(n+1)}_{\mathscr{I}^+}$ and obtain an additional power of decay in $r$.
\end{proof}

\paragraph[Boundedness on $\protect\alin$]{Boundedness on $\alin$}

Near $\mathscr{H}^+_{\geq0}$, energy boundedness and the transport relations \bref{hier} are sufficient to obtain an estimate on $\Omega^{-1}\nablagml\Omega^2\alin$ and subsequently $\alin$. Away from $\mathscr{H}^+_{\geq0}$ we will derive backwards $r^p$-estimates using an identical argument to the one applied to the Regge--Wheeler equation \bref{RW}, now applied ton \bref{T+2}. 

Note that the transport relations \bref{hier} allow us to obtain a better blueshift factor near $\mathscr{H}^+_{\geq0}$ for $(\Omega^{-1}\nablagml)^k\Omega^2\alin$ than for $\Psilin$.

\begin{proposition}\label{backwards ingoing overall on alin}
Let $\Psilin$, $\plin$, $\alin$ be as in \Cref{grand proposition backwards scattering}. Then we have
\begin{align}\label{backwards ingoing overall estimate on alin}
    \int_{\underline{\mathscr{C}}_v\cap\{\bar{u}\geq u\}}d\bar{u}\sin\theta d\theta d\phi\,\frac{r^2}{\Omega^2}|\nablau r^3\Omega\plin|^2+\frac{r^2}{\Omega^2}|\nablau r\Omega^2\alin|^2\leq \int_{\underline{\mathscr{C}}_v\cap\{\bar{u}\geq u\}}d\bar{u}\sin\theta d\theta d\phi\,\frac{2\Omega^2}{r^2}|\Psilin|^2,
\end{align}
\end{proposition}

To estimate $\xlin$ we will also need energy boundedness along $\mathscr{C}_u$:

\begin{proposition}\label{backwards ILED plin}
Let $\Psilin$, $\plin$, $\alin$ be as in \Cref{grand proposition backwards scattering}. Let $\mathscr{D}_{u,v}=J^+(\mathscr{C}_u)\cap J^+(\underline{\mathscr{C}}_v)$. Then we have
\begin{align}\label{backwards ILED estimate plin}
\begin{split}
     \int_{{\mathscr{C}}_u\cap\{\bar{v}\geq v\}}&d\bar{v}\sin\theta d\theta d\phi\,r^{6-\epsilon} \Omega^2|\plin|^2+\int_{\mathscr{\mathscr{D}}_{u,v}}d\bar{u}d\bar{v}\sin\theta d\theta d\phi\,r^{5-\epsilon}\Omega^4|\plin|^2
    \,\\&\lesssim_{\epsilon,M} \int_{\mathscr{H}^+_{\geq0}\cap\{\bar{v}\geq v\}}d\bar{v}\sin\theta d\theta d\phi\,\left[|\partial_v\upPsilin_{\mathscr{H}^+}|^2+|\plins_{\mathscr{H}^+}|^2\right]+\int_{\mathscr{I}^+\cap\{\bar{u}\geq u\}}d\bar{u}\sin\theta d\theta d\phi\,|\partial_u\upPsilin_{\mathscr{I}^+}|^2.
\end{split}
\end{align}
\end{proposition}

\begin{proof}
    We derive from \bref{hier}
    \begin{align}
        \nablau r^{n+6}\Omega^2|\plin|^2+nr^{n+5}\Omega^4|\plin|^2=2\Omega^3 r^{n+1} \Psilin\cdot\plin.
    \end{align}
    Integrate the above over $\mathscr{D}$ to find
    \begin{align}
    \begin{split}
        &\int_{\mathscr{H}^+_{\geq0}}d\bar{u}\sin\theta d\theta d\phi\,(2M)^{6+n}|\plins_{\mathscr{H}^+}|^2-\int_{\mathscr{C}_u\cap\{\bar{v}\geq v\} }d\bar{v}\sin\theta d\theta d\phi\,r^{n+6}\Omega^2|\plin|^2\\&+\int_{\mathscr{D}} d\bar{v} \dw nr^{n+5}\Omega^2|\plin|^2+2\Omega^3 r^{n+1}\Psilin\cdot \plin=0.
    \end{split}
    \end{align}
    Cauchy--Schwarz gives us
    \begin{align}
    \begin{split}
        \int_{{\mathscr{C}}_u\cap\{\bar{v}\geq v\}}&d\bar{v}\sin\theta d\theta d\phi\,r^{6-\epsilon} \Omega^2|\plin|^2+\frac{\epsilon}{2}\int_{\mathscr{\mathscr{D}}_{u,v}}d\bar{u}d\bar{v}\sin\theta d\theta d\phi\,r^{5-\epsilon}\Omega^4|\plin|^2
    \,\\&\leq \int_{\mathscr{H}^+_{\geq0}\cap\{\bar{v}\geq v\}}d\bar{v}\sin\theta d\theta d\phi\,(2M)^{6-\epsilon}|\plins_{\mathscr{H}^+}|^2+\int_{\mathscr{D}_{u,v}}d\bar{u}\sin\theta d\theta d\phi\,\frac{8}{\epsilon}\frac{\Omega^2}{r^{3+\epsilon}}|\Psilin|^2.
    \end{split}
    \end{align}
    We now use the degenerate ILED estimates of \Cref{backwards ILED RW}.
\end{proof}

We may derive from \bref{hier},
    \begin{align}\label{formula for alins plins}
        \plins_{\mathscr{I}^+}=-\int_u^{u_+}d\bar{u}\;\upPsilin_{\mathscr{I}^+},\qquad \alins_{\mathscr{I}^+}=\int_u^{u_+}d\bar{u}\int_{\bar{u}}^{u_+}d\bar{\bar{u}}\;\upPsilin_{\mathscr{I}^+}.
    \end{align}
Thus
\begin{align}\label{asymptotics of alins plins}
    \plins_{\mathscr{I}^+}=O(|u|),\qquad \alins_{\mathscr{I}^+}=O(|u|^2),
\end{align}
as $u\longrightarrow-\infty$.

An identical argument to that of \Cref{backwards ILED plin} above, when applied to $\alin$, gives
\begin{proposition}\label{backwards ILED alin}
Let $\Psilin$, $\plin$, $\alin$ be as in \Cref{grand proposition backwards scattering}. Let $\mathscr{D}=J^+(\mathscr{C}_u)\cap J^+(\underline{\mathscr{C}}_v)$. Then we have
\begin{align}\label{backwards ILED estimate alin}
\begin{split}
     &\int_{{\mathscr{C}}_u\cap\{\bar{v}\geq v\}}d\bar{v}\sin\theta d\theta d\phi\,r^{2-\epsilon} \Omega^4|\alin|^2+\int_{\mathscr{\mathscr{D}}_{u,v}}d\bar{u}d\bar{v}\sin\theta d\theta d\phi\,r^{1-\epsilon}\Omega^6|\alin|^2
    \,\\&\lesssim_{\epsilon,M} \int_{\mathscr{H}^+_{\geq0}\cap\{\bar{v}\geq v\}}d\bar{v}\sin\theta d\theta d\phi\,\left[|\partial_v\upPsilin_{\mathscr{H}^+}|^2+|\plins_{\mathscr{H}^+}|^2+|\alins_{\mathscr{H}^+}|^2\right]+\int_{\mathscr{I}^+\cap\{\bar{u}\geq u\}}d\bar{u}\sin\theta d\theta d\phi\,|\partial_u\upPsilin_{\mathscr{I}^+}|^2.
\end{split}
\end{align}
\end{proposition}

We may write the $+2$ Teukolsky equation \bref{T+2} as
\begin{align}\label{this 21 08 2021}
    \begin{split}
        \nablav\nablau r^5\Omega^{-2}\alin+2\frac{3\Omega^2-1}{r}\nablav r^5\Omega^{-2}\alin-\frac{4\Omega^2}{r^2}(3\Omega^2-2)r^5\Omega^{-2}\alin=\frac{\Omega^2}{r^2}\left(\mathcal{A}_2-\frac{6M}{r}\right)r^5\Omega^{-2}\alin.
    \end{split}
\end{align}

Note that \bref{this 21 08 2021} is identical to the equation governing $\Phi^{(2)}, \underline{\Phi}^{(2)}$ (we may also see this from the Teukolsky--Starobinsky identity \bref{eq:TS+}). 

\begin{defin}\label{definition of Alin n}
Consider the quantity $\alin$ belonging to a solution $\mathfrak{S}$ of \fullsystem arising from scattering data as in \Cref{grand proposition backwards scattering}. For $n\geq0$ define 
\begin{align}
    \mathtt{A}^{(n+2)}:=\left(\frac{r^2}{\Omega^2}\nablav\right)^n r^5\Omega^{-2}\alin.
\end{align}
We also take $\mathtt{A}^{(1)}:=0$.
\end{defin}

We immediately infer that the following applies to $\alin$ and its $\nablav$-derivatives, by repeating the same steps leading to \Cref{backwards transport estimate of Phi n}, \Cref{rdv Psi near scri+ k} and \Cref{backwards rp RW Phi n}, replacing $\Phi^{(n)}$ by $\mathtt{A}^{(n)}$: 

\begin{proposition}\label{backwards rp alin n}
Let $\alin, \mathtt{A}^{(n)}$ be as in \Cref{definition of Alin n}. Then the limit
\begin{align}
    \mathtt{A}_{\mathscr{I}^+}^{(n)}(u,v,\theta^A)=\lim_{v\longrightarrow\infty}\mathtt{A}^{(n)}(u,v,\theta^A).
\end{align}
exists and defines an element of $\Gamma(\mathscr{I}^+)$. Moreover, the estimates of \Cref{backwards transport estimate of Phi n}, \Cref{rdv Psi near scri+ k} and \Cref{backwards rp RW Phi n} apply with $\Phi^{(n)}$ replaced by $\mathtt{A}^{(n)}$, $\upphi^{(n)}$ replaced by $r^{-1}\mathtt{A}^{(n)}$, and $\Phi^{(n)}_{\mathscr{I}^+}$ replaced by $\mathtt{A}^{(n)}_{\mathscr{I}^+}$ for $n\geq 2$. We explicitly state the analogous estimate to \bref{backwards rp estimate RW Phi n}:
\begin{align}\label{backwards rp estimate Alin n}
\begin{split}
    &\int_{\mathscr{C}_u\cap\{\bar{v}\in[v_\infty,\infty)\}}d\bar{v}\sin\theta d\theta d\phi\,\frac{r^2}{\Omega^2}|\nablav\mathtt{A}^{(n)}|^2+\int_{\bar{u}\in[u,u_+],\bar{v}\geq v_\infty}d\bar{u}d\bar{v}\sin\theta d\theta d\phi\,\left[\frac{\Omega^2}{r^2}|\mathtt{A}^{(n)}|^2+\frac{r}{\Omega^2}|\nablav\mathtt{A}^{(n)}|^2\right]\\
    &+\int_{\underline{\mathscr{C}}_{v_\infty}\cap\{\bar{u}\geq u\}}d\bar{u}\sin\theta d\theta d\phi\,|\mathtt{A}^{(n)}|^2\,\leq\, C(n,M,u_+)\int_{\mathscr{I}^+\cap\{\bar{u}\in[u,u_+]\}}d\bar{u}\sin\theta d\theta d\phi\,\left[|\mathring{\slashednabla}\mathtt{A}^{(n)}_{\mathscr{I}^+}|^2+|\mathtt{A}^{(n)}_{\mathscr{I}^+}|^2+M|\mathtt{A}^{(n-1)}_{\mathscr{I}^+}|^2\right].
\end{split}
\end{align}
\end{proposition}

The statement of \Cref{backwards rp alin n} was proved in \cite{Mas20} for $n=1$. In particular, following the steps leading to \bref{model estimate near scri+}, we have
\begin{proposition}\label{backwards estimate on alin away from scri+}
Let $\alin$ be as in \Cref{grand proposition backwards scattering} and $\mathtt{A}^{(n)}$ be as in \Cref{definition of Alin n}. Then we have
\begin{align}
\begin{split}
    \int_{{\mathscr{C}}_u\cap\{\bar{v}\geq v\}}d\bar{v}\dw\,\frac{\Omega^2}{r^{1+\epsilon}}|r^5\Omega^{2}\alin|^2\lesssim C(M,\epsilon)\Bigg\{&\int_{S^2}\dw|\alins_{\mathscr{I}^+}(u,\theta^A)|^2\\&+ \text{ Right hand side of }\bref{backwards rp estimate Alin n}\text{ with }n=2\Bigg\}.
\end{split}
\end{align}
\end{proposition}

\subsubsection*{Asymptotics of $\protect\alin$ near $i^0$}

\begin{corollary}\label{asymptotic flatness at spacelike infinity for alin}
 Let $\alin$, be as in \Cref{grand proposition backwards scattering}. Then
 \begin{align}
     \partial_t^n\alin|_{{\Sigma^*_+}}=O_{\infty}(r^{-n-3}).
 \end{align}
With the assumptions of \Cref{grand proposition backwards scattering addendum}, we have
\begin{align}
    \partial_t^n \alin|_{\Sigma^*_+}=O_{\infty}(r^{-n-4}).
\end{align}
\end{corollary}

\begin{proof}
    Let $u_+$, $u_-$ be the future and past cutoffs of $\glinh$. The formula \bref{alins at scri+ backwards scattering} implies $\plins_{\mathscr{I}^+}$ is constant and $\alins_{\mathscr{I}^+}$ is linear in $u$ on $u\leq u_-$, while \Cref{backwards rp alin n} implies $\mathring{\slashednabla}^\gamma\partial_u^i\mathtt{A}^{(n)}_{\mathscr{I}^+}=O(|u|^{n-1-i})$ for all multiindices $\gamma$ and all $i\in\mathbb{N}$. We estimate on $r>r(u_-,\vsigmap{u_-})$:
    \begin{align}
    \begin{split}
        \int_{S^2}\dw\,\left|r^5\Omega^{-1}\plin|_{{\Sigma^*_+}}(r,\theta^A)\right|^2&\leq \int_{S^2}\dw\,\left|\plins_{\mathscr{I}^+}(u,\theta^A)\right|^2+\frac{1}{r}\int_{\mathscr{C}_u\cap\{\bar{v}\geq v\}}d\bar{v}\dw\,\frac{r^2}{\Omega^2}\left|\nablav r^5\Omega^{-1}\plin\right|^2\\
        &=\int_{S^2}\dw\,\left|\plins_{\mathscr{I}^+}(u,\theta^A)\right|^2\\&+\frac{1}{r}\int_{\mathscr{C}_u\cap\{\bar{v}\geq v\}}d\bar{v}\dw\,\frac{\Omega^2}{r^2}\left|\left(\mathcal{A}_2-\frac{6M}{r}\right)\nablav r^5\Omega^{-1}\plin\right|^2\\
        &\leq \int_{S^2}\dw\,\left|\plins_{\mathscr{I}^+}(u,\theta^A)\right|^2+\frac{1}{r^2}\sup_{\bar{v}\geq v}\left|\left(\mathcal{A}_2-\frac{6M}{r}\right)r^5\Omega^{-2}\alin\right|^2\\&\leq\int_{S^2}\dw\,\left|\plins_{\mathscr{I}^+}(u,\theta^A)\right|^2\\&\qquad+\frac{1}{r(u,\vsigmap{u})^2}\left[\int_{S^2}\dw\,\sum_{|\gamma|\leq2}|\mathring{\slashednabla}^\gamma\alins_{\mathscr{I}^+}(u,\theta^A)|^2\right]\\&\qquad+\frac{1}{r(u,\vsigmap{u})^3}\int_{\mathscr{C}_u\cap\{\bar{v}\geq v\}}d\bar{v}\dw\,\frac{r^2}{\Omega^2}\sum_{|\gamma|\leq2}\left|\nablav\mathring{\slashednabla}^\gamma r^5\Omega^{-2}\alin\right|^2
        \\&\leq \int_{S^2}\dw\,\left|\plins_{\mathscr{I}^+}(u,\theta^A)\right|^2\\&\qquad+\frac{1}{r(u,\vsigmap{u})^2}\left[\int_{S^2}\dw\,\sum_{|\gamma|\leq2}|\mathring{\slashednabla}^\gamma\alins_{\mathscr{I}^+}(u,\theta^A)|^2\right]\\&\qquad+\frac{1}{r(u,\vsigmap{u})^3}\int_{\mathscr{I}^+\cap\{\bar{u}\geq u\}}d\bar{u}\dw\,\sum_{|\gamma|\leq3}|\mathring{\slashednabla}^\gamma\alins_{\mathscr{I}^+}(\bar{u},\theta^A)|^2.
    \end{split}
    \end{align}
    Commuting the above with the $\mathcal{L}^\gamma$ for multiindices $|\gamma|\leq2$ and $\mathcal{L}$ ranging over the generators of $SO(3)$, we get that $\plin|_{{\Sigma^*_+}}=O(r^{-4})$ as $r\longrightarrow\infty$ given \bref{formula for alins plins} and \bref{asymptotics of alins plins}. An identical argument shows that $\alin|_{{\Sigma^*_+}}=O(r^{-3})$ as $r\longrightarrow\infty$. \Cref{asymptotic flatness Psilin n} implies that 
    $\nablau^n\alin|_{{\Sigma^*_+}}=O(r^{-3-n})$ as $r\longrightarrow\infty$. We can deduce $\partial_t^n \alin|_{{\Sigma^*_+}}=O(r^{{-3-n}})$ as $r\longrightarrow\infty$ using the arguments above in an identical manner to that leading to \bref{16 10 2021 3}. When the assumptions of \Cref{grand proposition backwards scattering addendum} are satisfied, we have $\mathring{\slashednabla}^{\gamma}\partial_u^i\Phi^{(n)}_{\mathscr{I}^+}=O\left(|u|^{n-1-i}\right)$, so we may use the above argument to obtain an additional power of decay in $r$.
\end{proof}

\subsubsection[Boundedness on $\protect\xlin$, $\protect\xblin$]{Boundedness on $\xlin$, $\xblin$}\label{section 6.2.1.1 backwards boundedness on xlin xblin}


\subsubsection*{Boundedness of $\xblin$}

\begin{proposition}\label{backwards ingoing overall  on xblin}
For the solution $\mathfrak{S}$ constructed in \Cref{grand proposition backwards scattering}, we have for any $u,v$ the estimate
\begin{align}\label{backwards ingoing overall estimate on xblin}
\begin{split}
    \int_{\underline{\mathscr{C}}_v\cap\{\bar{u}\geq u\}}d\bar{u}\sin\theta d\theta d\phi\,& \left[\frac{1}{\Omega^2}|\nablau \Omega^{-1}\xblin(\bar{u},v,\theta^A)|^2+\frac{6}{r^2}|\xblin(\bar{u},v,\theta^A)|^2\right]+\int_{S^2_{u,v}}\dw\,\frac{1}{r(u,v)}|\Omega^{-1}\xblin(u,v,\theta^A)|^2\\
    &\lesssim \frac{1}{2M}\int_{S^2}\sin\theta d\theta d\phi\,|\xblins_{\mathscr{H}^+}(u,\theta^A)|^2+\text{Right hand side of }\bref{backwards ingoing overall estimate on ablin}.
\end{split}
\end{align}

\end{proposition}
\begin{proof}
    Write \bref{D3Chihatbar} as
    \begin{align}
        \nablau \Omega^{-1}\xblin-\frac{2\Omega^2}{r}\Omega^{-1}\xblin=-\ablin,
    \end{align}
    which implies
    \begin{align}
        \frac{1}{\Omega^2}|\nablau\Omega^{-1}\xblin|^2+\frac{6\Omega^2}{r^2}|\Omega^{-1}\xblin|^2-2\partial_u \left(\frac{1}{r}|\Omega^{-1}\xblin|^2\right)=\frac{1}{\Omega^2}|\ablin|^2.
    \end{align}
    Integrate over $\underline{\mathscr{C}}_v$ from $u$ to $\mathscr{H}^+$ to obtain 
        \begin{align}
    \begin{split}
        \int_{\underline{\mathscr{C}}_v\cap\{\bar{u}\geq u\}}d\bar{u}\sin\theta d\theta d\phi\,& \left[\frac{1}{\Omega^2}|\nablau \Omega^{-1}\xblin(\bar{u},v,\theta^A)|^2+\frac{6}{r^2}|\xblin(\bar{u},v,\theta^A)|^2\right]+\frac{1}{r(u,v)}\int_{S^2}\dw\,|\Omega^{-1}\xblin(u,v,\theta^A)|^2\\&\leq\frac{1}{2M}\int_{S^2}\sin\theta d\theta d\phi\,|\xblins_{\mathscr{H}^+}(v,\theta^A)|^2+\int_{\underline{\mathscr{C}}_v}d\bar{u}\sin\theta d\theta d\phi\,\frac{1}{\Omega^2}|\ablin(\bar{u},v,\theta^A)|^2.
    \end{split}
    \end{align}
    The result now follows by applying \Cref{backwards ingoing overall on ablin} to the above.
\end{proof}

Applying the above argument to the equation
\begin{align}
    \nablau\nablav\Omega^{-1}\xblin-\frac{2\Omega^2}{r}\nablav \Omega^{-1}\xblin+\frac{2\Omega^2}{r^2}(2\Omega^2-1)\Omega^{-1}\xblin=\frac{1}{r}\Omega^{-2}\ablin-r^2\Omega^{-1}\pblin
\end{align}
leads to an analogous result to \Cref{backwards ingoing overall  on xblin}:

\begin{corollary}\label{backwards ingoing nablav commuted overall xblin}
For the solution $\mathfrak{S}$ constructed in \Cref{grand proposition backwards scattering}, we have 
\begin{align}
    \begin{split}
        &\int_{\underline{\mathscr{C}}_v\cap\{\bar{u}\geq u\}}d\bar{u}\sin\theta d\theta d\phi\, \Bigg[\frac{1}{\Omega^2}|\nablau\nablav \Omega^{-1}\xblin(\bar{u},v,\theta^A)|^2+\frac{\Omega^2}{r^2}|\nablav \Omega^{-1}\xblin|^2\Bigg]\\&+\int_{S^2}\sin\theta d\theta d\phi\,\frac{2}{r(u,v)}|\nablav\Omega^{-1}\xblin|^2\lesssim
        \text{ Right hand sides of }\bref{backwards ingoing overall estimate on ablin}, \bref{backwards ingoing overall estimate on pblin} \text{ and }\bref{backwards ingoing overall estimate on xblin}.
    \end{split}
\end{align}
\end{corollary}
The following estimates will be used to show asymptotic flatness near $i^0$:
\begin{proposition}\label{backwards rp xblin}
For the solution $\mathfrak{S}$ constructed in \Cref{grand proposition backwards scattering}, let $v_\infty$ be such that $r(u_+,v_\infty)\geq 16M$. We have for any $u\geq u_+$
\begin{align}\label{backwards rp estimate xblin}
\begin{split}
   &\int_{\mathscr{C}_u\cap\{\bar{v}\geq v_\infty\}\cap J^+({\Sigma^*_+})}d\bar{v}\sin\theta d\theta d\phi\,\frac{r^2}{\Omega^2}|\nablav r\Omega\xblin|^2+\int_{\mathscr{D}}d\bar{u}d\bar{v}\sin\theta d\theta d\phi\,4M\Omega^4|\xblin|^2\\&\qquad\leq C(u_+) \int_{\mathscr{I}^+\cap\{\bar{u}\in[u,u_+]\}}d\bar{u}\sin\theta d\theta d\phi\,\left[|\xblins_{\mathscr{I}^+}|^2+|\pblins_{\mathscr{I}^+}|^2+|\upPsilinb_{\mathscr{I}^+}|^2+|\partial_u\upPsilinb_{\mathscr{I}^+}|^2+|\mathring{\slashednabla}\upPsilinb_{\mathscr{I}^+}|^2\right],
\end{split}
\end{align}
where $\mathscr{D}=[u,u_+]\times[v_\infty,\infty)\times S^2$
\end{proposition}
\begin{proof}
    Commuting $\nablav$ past \bref{D3Chihat} gives
    \begin{align}
        \nablau \nablav r\Omega\xblin-\frac{2\Omega^2-1}{r}\nablav r\Omega\xblin+\frac{\Omega^2}{r^2}(4\Omega^2-3)r\Omega\xblin=-\frac{\Omega^2}{r^2}r^3\Omega\pblin.
    \end{align}
    which implies
    \begin{align}
    \begin{split}
        \partial_u \frac{r^2}{\Omega^2}|\nablav r\Omega\xblin|^2+\frac{4M}{r^2}\frac{r^2}{\Omega^2}|\nablav r\Omega\xblin|^2+\partial_v\Big[(4\Omega^2-3)|r\Omega\xblin|^2\Big]&-\frac{8M\Omega^2}{r^2}|r\Omega\xblin|^2\\&=-2\nablav\left[r^3\Omega\pblin\cdot r\Omega\xblin\right]+\frac{2\Omega^2}{r^2}\Psilinb\cdot r\Omega\xblin.
    \end{split}
    \end{align}
    Integrate the above in the spacetime region $\mathscr{D}=[u,u_+]\times [v_\infty,\infty)\times S^2$ to get
    \begin{align}
    \begin{split}
        &-\int_{\mathscr{C}_u\cap\{\bar{v}\geq v_\infty\}\cap J^+({\Sigma^*_+})}d\bar{v}\sin\theta d\theta d\phi\,\frac{r^2}{\Omega^2}|\nablav r\Omega\xblin|^2+\int_{\mathscr{D}}d\bar{u}d\bar{v}\sin\theta d\theta d\phi\,\frac{4M}{r^2}\frac{r^2}{\Omega^2}|\nablav r\Omega\xblin|^2\\&+\int_{\mathscr{I}^+\cap\{\bar{u}\in[u,u_+]\}}d\bar{u}\sin\theta d\theta d\phi\,|\xblins_{\mathscr{I}^+}|^2-\int_{\underline{\mathscr{C}}_{v_\infty}\cap\{\bar{u}\in[u,u_+]\}}d\bar{u}\sin\theta d\theta d\phi\, (4\Omega^2-3)|r\Omega\xblin|^2\\&-\int_{\mathscr{D}}d\bar{u}d\bar{v}\sin\theta d\theta d\phi\,8M\frac{\Omega^2}{r^2}|r\Omega\xblin|^2=-2\int_{\underline{\mathscr{C}}_{v_\infty}\cap\{\bar{u}\in[u,u_+]\}}d\bar{u}\sin\theta d\theta d\phi\, r^3\Omega\pblin\cdot r\Omega\xblin\\&\qquad\qquad\qquad\qquad\qquad\qquad\qquad\qquad+2\int_{\mathscr{I}^+\cap\{\bar{u}\in[u,u_+]\}}d\bar{u}\sin\theta d\theta d\phi\,\pblins_{\mathscr{I}^+}\cdot \xblins_{\mathscr{I}^+}.
    \end{split}
    \end{align}
    We can now estimate
    \begin{align}
    \begin{split}
        &\int_{\mathscr{C}_u\cap\{\bar{v}\geq v_\infty\}\cap J^+({\Sigma^*_+})}d\bar{v}\sin\theta d\theta d\phi\,\frac{r^2}{\Omega^2}|\nablav r\Omega\xblin|^2+\int_{\underline{\mathscr{C}}_{v_\infty}\cap\{\bar{u}\in[u,u_+]\}}d\bar{u}\sin\theta d\theta d\phi\, \frac{8\Omega^2-7}{2}|r\Omega\xblin|^2\\
        &+\int_{\mathscr{D}}d\bar{u}d\bar{v}\sin\theta d\theta d\phi\,\frac{4M\Omega^2}{r^2}|r\Omega\xblin|^2\,\leq\, \int_{\mathscr{I}^+\cap\{\bar{u}\in[u,u_+]\}}d\bar{u}\sin\theta d\theta d\phi\,\left[2|\xblins_{\mathscr{I}^+}|^2+|\pblins_{\mathscr{I}^+}|^2\right]\\
        &\qquad\qquad+2\int_{\underline{\mathscr{C}}_{v_\infty}\cap\{\bar{u}\in[u,u_+]\}}d\bar{u}\sin\theta d\theta d\phi\,|r^3\Omega\pblin|^2+\frac{1}{4M}\int_{\mathscr{D}}d\bar{u}d\bar{v}\sin\theta d\theta d\phi\,\frac{\Omega^2}{r^2}|\Psilinb|^2.
    \end{split}
    \end{align}
The result now follows by applying Gr\"onwall's inequality, using \bref{backwards overall ingoing estimate pblin better} to estimate $\pblin$ on $\underline{\mathscr{C}}_{v_\infty}$ and \Cref{backwards ILED RW better r weight Gronwall} to estimate the spacetime integral of $\Psilinb$ above.
\end{proof}

We can prove a generalisation of \Cref{backwards rp xblin} analogous to \Cref{backwards rp RW Phi n} by applying the arguments leading to \Cref{backwards rp RW Phi n} to \bref{D3Chihatbar}:

\begin{defin}\label{definition of Xblin n}
Let $\xblin$ arise from a solution $\mathfrak{S}$ to the system \fullsystem in the sense of \Cref{EinsteinWP} or \Cref{grand proposition backwards scattering}. Define
\begin{align}
    \underline{\mathcal{X}}^{(n)}:=\left(\frac{r^2}{\Omega^2}\nablav\right)^nr\Omega\xblin.
\end{align}
\end{defin}

\begin{lemma}\label{transport on xblin n}
    The quantity $ \underline{\mathcal{X}}^{(n)}$ defined in \Cref{definition of Xblin n} satisfies the transport equation
    \begin{align}\label{transport equation Xblin n}
    \begin{split}
        &\nablau  \underline{\mathcal{X}}^{(n)}-\left(\frac{2n-1}{r}-\frac{2M}{r^2}(3n-2)\right) \underline{\mathcal{X}}^{(n)}-n\left[n-2-\frac{2M}{r}(3n-7)\right] \underline{\mathcal{X}}^{(n-1)}\\&-2Mn(n-1)(n-4) \underline{\mathcal{X}}^{(n-2)}=-\underline{\Phi}^{(n-2)},
    \end{split}
    \end{align}
    where we now take $\underline{\Phi}^{(-2)}:=r\Omega^2\ablin,\,\underline{\Phi}^{(-1)}:=r^3\Omega\pblin$. Moreover, $ \underline{\mathcal{X}}^{(n)}$ satisfies the equation
    \begin{align}\label{wave equation Xblin n}
    \begin{split}
        &\nablau\nablav  \underline{\mathcal{X}}^{(n)}+\left[\frac{2n-1}{r}-\frac{2M}{r^2}(3n-2)\right]\nablav \underline{\mathcal{X}}^{(n)}-n\left[n-2-\frac{2M}{r}(3n-7)\right] \underline{\mathcal{X}}^{(n-1)}\\
        &-2Mn(n-1)(n-4) \underline{\mathcal{X}}^{(n-2)}=-\underline{\Phi}^{(n-1)}.
    \end{split}
    \end{align}
\end{lemma}

We can show the following in an identical manner to how \Cref{backwards transport estimate of Phi n} was proven:

\begin{proposition}
 Let $\underline{\mathcal{X}}^{(n)}$ be as in \Cref{definition of Xblin n}. The limit
    \begin{align}\label{limit of Xlin n}
       \underline{\mathcal{X}}^{(n)}_{\mathscr{I}^+}(u,v,\theta^A)=\lim_{v\longrightarrow\infty}\underline{\mathcal{X}}^{(n)}(u,\theta^A)
    \end{align}
    exists and defines an element of $\Gamma(\mathscr{I}^+)$. The same applies to angular derivatives of $\underline{\mathcal{X}}^{(n)}$ and 
    \begin{align}\label{this 16 08 2021 3}
        \lim_{v\longrightarrow\infty}\mathring{\slashednabla}^\gamma\underline{\mathcal{X}}^{(n)}_{\mathscr{I}^+}=\mathring{\slashednabla}^\gamma \underline{\mathcal{X}}^{(n)}_{\mathscr{I}^+}
    \end{align}
for any index $\gamma$.
\end{proposition}

\begin{proposition}\label{backwards rp Xblin n}
For the solution $\mathfrak{S}$ constructed in \Cref{grand proposition backwards scattering} and for $n\geq2$, $v_\infty$ such that $r(u_+,v_\infty)>3M$, we have the following on $\mathscr{D}=[u,u_+]\times [v_\infty,\infty)\times S^2$:
\begin{align}\label{backwards rp estimates Xblin n}
    \begin{split}
        &\int_{\mathscr{C}_u\cap\{\bar{v}\geq v_\infty\}}d\bar{v}\sin\theta d\theta d\phi\,\frac{r^2}{\Omega^2}|\nablav\underline{\mathcal{X}}^{(n)}|^2+\int_{\mathscr{D}}d\bar{u}d\bar{v}\sin\theta d\theta d\phi\,\frac{\Omega^2}{r^2}|\underline{\mathcal{X}}^{(n)}|^2+\frac{r}{\Omega^2}|\nablav \underline{\mathcal{X}}^{(n)}|^2\\&
        +\int_{\underline{\mathscr{C}}_{v_\infty}}d\bar{u}\sin\theta d\theta d\phi\,|\underline{\mathcal{X}}^{(n)}|^2\leq C(n,M,u_+)\int_{\mathscr{I}^+\cap\{\bar{u}\in[u,u_+]\}}d\bar{u}\sin\theta d\theta d\phi\,\Bigg[|\underline{\mathcal{X}}^{(n)}_{\mathscr{I}^+}|^2+|\underline{\mathcal{X}}^{(n-1)}_{\mathscr{I}^+}|^2+|\mathring{\slashednabla}\Phi^{(n-1)}_{\mathscr{I}^+}|^2\\&\qquad\qquad\qquad\qquad\qquad\qquad\qquad\qquad\qquad\qquad\qquad\qquad\qquad\qquad\qquad+|\mathring{\slashednabla}\underline{\Phi}^{(n-2)}_{\mathscr{I}^+}|^2+|\underline{\Phi}^{(n-3)}_{\mathscr{I}^+}|^2\Bigg]
    \end{split}
\end{align}
\end{proposition}
\begin{proof}
    Applying the steps leading to the proof of \Cref{backwards rp RW Phi n} to the equation \bref{wave equation Xblin n}, we integrate the following equation over $\mathscr{D}=[u,u_+]\times[v_\infty,\infty)\times S^2$ 
    \begin{align}\label{this 19 08 2021}
        \begin{split}
           & -\int_{\mathscr{C}_u\cap\{\bar{v}\geq v_\infty\}}d\bar{v}\sin\theta d\theta d\phi\, \frac{r^2}{\Omega^2}|\nablav\underline{\mathcal{X}}^{(n)}|^2+\int_{\mathscr{D}}d\bar{u}d\bar{v}\sin\theta d\theta d\phi\,\left(\frac{4n}{r}-\frac{2M}{r^2}(6n-1)\right)\frac{r^2}{\Omega^2}|\nablav\underline{\mathcal{X}}^{(n)}|^2\\
           &-\int_{\mathscr{I}^+\cap\{\bar{u}\in[u,u_+]\}}d\bar{u}\sin\theta d\theta d\phi\,(n^2-1)|\underline{\mathcal{X}}^{(n)}_{\mathscr{I}^+}|^2\\&+\int_{\underline{\mathscr{C}}_{v_\infty}\cap\{\bar{u}\in[u,u_+]\}}d\bar{u}\sin\theta d\theta d\phi\,\left[n^2-1-\frac{2M}{r}(n+1)(3n-4)\right]|\underline{\mathcal{X}}^{(n)}|^2\\&      +4Mn(n^2-2n-2)\Big\{\int_{\mathscr{I}^+\cap\{\bar{u}\in[u,u_+]\}}d\bar{u}\sin\theta d\theta d\phi\,\underline{\mathcal{X}}^{(n-1)}_{\mathscr{I}^+}\cdot \underline{\mathcal{X}}^{(n)}_{\mathscr{I}^+}\\&-\int_{\underline{\mathscr{C}}_{v_\infty}\cap\{\bar{u}\in[u,u_+]}d\bar{u}\sin\theta d\theta d\phi\,\underline{\mathcal{X}}^{(n-1)}\cdot \underline{\mathcal{X}}^{(n)}\Big\}
           \\&-2M[2n^3-5n^2-n+4]\int_{\mathscr{D}}d\bar{u}d\bar{v}\sin\theta d\theta d\phi\,\frac{\Omega^2}{r^2}|\underline{\mathcal{X}}^{(n)}|^2=-\int_{\mathscr{I}^+\cap\{\bar{u}\in[u,u_+]\}}d\bar{u}\sin\theta d\theta d\phi\,\underline{\Phi}^{(n-2)}_{\mathscr{I}^+}\cdot\underline{\mathcal{X}}^{(n)}_{\mathscr{I}^+}\\&\qquad\qquad+\int_{\underline{\mathscr{C}}_{v_\infty}\cap\{\bar{u}\in[u,u_+]\}}d\bar{u}\sin\theta d\theta d\phi\,\underline{\Phi}^{(n-2)}\cdot\underline{\mathcal{X}}^{(n)}+\int_{\mathscr{D}}d\bar{u}d\bar{v}\sin\theta d\theta d\phi\,\frac{\Omega^2}{r^2}\underline{\Phi}^{(n-1)}\cdot\underline{\mathcal{X}}^{(n)}.
        \end{split}
    \end{align}
    As in the proof of \Cref{backwards rp RW Phi n}, we estimate
    \begin{align}
        \begin{split}
            \int_{\mathscr{D}}d\bar{u}d\bar{v}\sin\theta d\theta d\phi\,\frac{\Omega^2}{r^2}|\underline{\mathcal{X}}^{(n)}|^2\,\leq\,& \frac{2}{3M}\int_{\mathscr{I}^+\cap\{\bar{u}\in[u,u_+]\}}d\bar{u}\sin\theta d\theta d\phi\,|\underline{\mathcal{X}}^{(n-1)}_{\mathscr{I}^+}|^2\\&+\int_u^{u_+}d\bar{u}\frac{2}{r^2}\int_{\mathscr{C}_{\bar{u}}\cap\{\bar{v}\geq v_\infty\}}d\bar{v}\sin\theta d\theta d\phi\frac{r^2}{\Omega^2}|\nablav\underline{\mathcal{X}}^{(n)}|^2,
        \end{split}
    \end{align}
    \begin{align}
        \begin{split}
            \int_{\mathscr{D}}d\bar{u}d\bar{v}\sin\theta d\theta d\phi\,\frac{\Omega^2}{r^2}|\underline{\mathcal{X}}^{(n-1)}|^2\,\leq\,& \frac{2}{3M}\int_{\mathscr{I}^+\cap\{\bar{u}\in[u,u_+]\}}d\bar{u}\sin\theta d\theta d\phi\,|\underline{\mathcal{X}}^{(n)}_{\mathscr{I}^+}|^2\\&+\int_u^{u_+}d\bar{u}\frac{5}{2r^2}\int_{\mathscr{C}_{\bar{u}}\cap\{\bar{v}\geq v_\infty\}}d\bar{v}\sin\theta d\theta d\phi\frac{r^2}{\Omega^2}|\nablav\underline{\mathcal{X}}^{(n)}|^2,
        \end{split}
    \end{align}
    \begin{align}
        \begin{split}
            \int_{\underline{\mathscr{C}}_{v_\infty}\cap\{\bar{u}\in[u,u_+]\}}d\bar{u}\sin\theta d\theta d\phi\,|\underline{\mathcal{X}}^{(n)}|^2\leq\,& 2\int_{\mathscr{I}^+\cap\{\bar{u}\in[u,u_+\}}d\bar{u}\sin\theta d\theta d\phi\,|\underline{\mathcal{X}}^{(n)}_{\mathscr{I}^+}|^2\\&+2\int_u^{u_+}d\bar{u}\frac{2}{r}\int_{\mathscr{C}_{\bar{u}}\cap\{\bar{v}\geq v_\infty\}}d\bar{v}\sin\theta d\theta d\theta d\phi\,\frac{r^2}{\Omega^2}|\nablav\underline{\mathcal{X}}^{(n)}|^2,
        \end{split}
    \end{align}
     \begin{align}
        \begin{split}
            \int_{\underline{\mathscr{C}}_{v_\infty}\cap\{\bar{u}\in[u,u_+]\}}d\bar{u}\sin\theta d\theta d\phi\,|\underline{\mathcal{X}}^{(n-1)}|^2\leq\,& 2\int_{\mathscr{I}^+\cap\{\bar{u}\in[u,u_+\}}d\bar{u}\sin\theta d\theta d\phi\,|\underline{\mathcal{X}}^{(n-1)}_{\mathscr{I}^+}|^2+|\underline{\mathcal{X}}^{(n)}_{\mathscr{I}^+}|^2\\&+\int_u^{u_+}d\bar{u}\frac{2}{r^2}\int_{\mathscr{C}_{\bar{u}}\cap\{\bar{v}\geq v_\infty\}}d\bar{v}\sin\theta d\theta d\theta d\phi\,\frac{r^2}{\Omega^2}|\nablav\underline{\mathcal{X}}^{(n)}|^2.
        \end{split}
    \end{align}
    Thus \bref{this 19 08 2021} implies
    \begin{align}\label{this 20 08 2021}
        \begin{split}
            &\int_{\mathscr{C}_u\cap\{\bar{v}\geq v_\infty\}}d\bar{v}\sin\theta d\theta d\phi\, \frac{r^2}{\Omega^2}|\nablav\underline{\mathcal{X}}^{(n)}|^2\leq \int_{\mathscr{I}^+\cap\{\bar{u}\in[u,u_+]\}}d\bar{u}(1+n^2)|\underline{\mathcal{X}}^{(n)}_{\mathscr{I}^+}|^2
            \\&+\int_{\mathscr{I}^+\cap\{\bar{u}\in[u,u_+]\}}d\bar{u}\sin\theta d\theta d\phi\,\left(\frac{125}{9}n^2+4n+\frac{106}{3}\right)|\underline{\mathcal{X}}^{(n)}_{\mathscr{I}^+}|^2
            \\&+\int_u^{u_+}d\bar{u}\left(\frac{2n^2+16n+16}{3r}+\frac{2M(6n^2+n+2)}{r^2}\right)\int_{\mathscr{C}_{\bar{u}}\cap\{\bar{v}\geq v_\infty\}}d\bar{v}\sin\theta d\theta d\phi\,\frac{r^2}{\Omega^2}|\nablav\underline{\mathcal{X}}^{(n)}|^2\\
            &+M^2\int_{\mathscr{D}}d\bar{u}d\bar{v}\sin\theta d\theta d\phi\,\frac{\Omega^2}{r^2}|\underline{\Phi}^{(n-1)}|^2+\frac{1}{4}\int_{\mathscr{I}^+\cap\{\bar{u}\in[u,u_+]\}}d\bar{u}\sin\theta d\theta d\phi\,|\underline{\Phi}^{(n-2)}_{\mathscr{I}^+}|^2\\
            &+\frac{1}{4}\int_{\underline{\mathscr{C}}_{v_\infty}\cap\{\bar{u}\in[u,u_+]\}}d\bar{u}\sin\theta d\theta d\phi\,|\underline{\Phi}^{(n-2)}|^2.
        \end{split}
    \end{align}
    \Cref{backwards rp RW Phi n} and Gr\"onwall's inequality (\Cref{Gronwall inequality}) can now be applied to \bref{this 20 08 2021} to obtain the result.
\end{proof}

\begin{corollary}\label{asymptotic flatness at spacelike infinity for xblin}
 The quantity $\xblin$ of \Cref{grand proposition backwards scattering} satisfies
 \begin{align}\label{decay of xblin on Sigma star}
     \partial_t^n \xblin|_{{\Sigma^*_+}}=O_{\infty}(r^{-2-n}).
 \end{align}
 When the assumptions of \Cref{grand proposition backwards scattering addendum} are satisfied, we have
 \begin{align}
     \partial_t^n \xblin|_{\Sigma^*_+}=O_\infty(r^{-3-n}).
 \end{align}
\end{corollary}
\begin{proof}
    Taking the limit of \bref{transport equation Xblin n} as $v\longrightarrow\infty$ for $n=1$, $n=2$ gives
    \begin{align}
        \partial_u \underline{\mathcal{X}}^{(2)}_{\mathscr{I}^+}=-6M\xblins_{\mathscr{I}^+}-\upPsilinb_{\mathscr{I}^+},\qquad \qquad \partial_u \underline{\mathcal{X}}^{(1)}_{\mathscr{I}^+}=(\mathring{\slashed{\Delta}}-3)\xblins_{\mathscr{I}^+}.
    \end{align}
    So we have
    \begin{align}\label{06 11 2022 1}
        \underline{\mathcal{X}}^{(1)}_{\mathscr{I}^+}(u,\theta^A)=-(\mathring{\slashed{\Delta}}-3)\xlins_{\mathscr{I}^+},
    \end{align}
    \begin{align}\label{06 11 2022 2}
        \underline{\mathcal{X}}^{(2)}_{\mathscr{I}^+}(u,\theta^A)=6M\xlins_{\mathscr{I}^+}-(\mathring{\slashed{\Delta}}-2)(\mathring{\slashed{\Delta}}-4)\int^\infty_ud\bar{u}\;\xlins_{\mathscr{I}^+}(\bar{u},\theta^A).
    \end{align}
    Thus $\underline{\mathcal{X}}^{(1)}_{\mathscr{I}^+}(u,\theta^A)$ vanishes and $\underline{\mathcal{X}}^{(2)}_{\mathscr{I}^+}(u,\theta^A)$ becomes constant in $u$ for $u< u_-$. Following the argument made in \bref{16 10 2021}, we can estimate for $u< u_-$
    \begin{align}
    \begin{split}
        \int_{S^2}\dw\,\left|r\Omega\xblin|_{{\Sigma^*_+}}(r,\theta^A)\right|^2&\leq\frac{1}{r(u,\vsigmap{u})^4}\Bigg[\int_{S^2}\dw\,|\underline{\mathcal{X}}^{(2)}_{\mathscr{I}^+}|^2\\&\qquad\qquad\qquad+\frac{1}{r}\int_{\mathscr{C}_u\cap\{\bar{v}\geq v\}}d\bar{v}\dw\,\frac{r^2}{\Omega^2}\left|\nablav\underline{\mathcal{X}}^{(2)}\right|^2\Bigg]
        \\&\lesssim \frac{1}{r(u,\vsigmap{u})^4}\Bigg[\int_{S^2}\dw\,|\underline{\mathcal{X}}^{(2)}_{\mathscr{I}^+}|^2\\&+\frac{1}{r}\int_{\mathscr{I}^+\cap\{\bar{u}\geq u\}}d\bar{u}\dw\,\Big(|\underline{\mathcal{X}}^{(2)}_{\mathscr{I}^+}|^2+|\underline{\mathcal{X}}^{(1)}_{\mathscr{I}^+}|^2+|\mathring{\slashednabla}\underline{\upPhi}_{\mathscr{I}^+}^{(1)}|^2\\&\qquad\qquad\qquad\qquad\qquad\qquad+|\mathring{\slashednabla}\upPsilinb_{\mathscr{I}^+}|^2+|\underline{\upPhi}_{\mathscr{I}^+}^{(1)}|^2+|\upPsilinb_{\mathscr{I}^+}|^2\Big)\Bigg].
    \end{split}
    \end{align}
    Given that $\mathring{\slashednabla}^\gamma\partial_u^i\underline{\Phi}^{(n)}_{\mathscr{I}^+}=O(|u|^{n-i})$ for all multiindices $\gamma$ and $i\in\mathbb{N}$, commuting the above with up to to orders of derivatives along the generators of $SO(3)$ and using a Sobolev estimate we get that $\xblin|_{{\Sigma^*_+}}=O(r^{-2})$. We may similarly argue for $\partial_t\xblin|_{{\Sigma^*_+}}$ to get 
    \begin{align}
    \begin{split}
        \int_{S^2}\dw\,\left|r\Omega\partial_t\xblin|_{{\Sigma^*_+}}(r,\theta^A)\right|^2&\leq\frac{1}{r(u,\vsigmap{u})^6}\Bigg[\int_{S^2}\dw\,|\partial_u\underline{\mathcal{X}}^{(3)}_{\mathscr{I}^+}|^2\\&\qquad\qquad\qquad+\frac{1}{r}\int_{\mathscr{C}_u\cap\{\bar{v}\geq v\}}d\bar{v}\dw\,\frac{r^2}{\Omega^2}\left|\partial_t\nablav\underline{\mathcal{X}}^{(3)}\right|^2\Bigg]
        \\&\lesssim \frac{1}{r(u,\vsigmap{u})^6}\Bigg[\int_{S^2}\dw\,|\partial_u\underline{\mathcal{X}}^{(3)}_{\mathscr{I}^+}|^2\\&+\frac{1}{r}\int_{\mathscr{I}^+\cap\{\bar{u}\geq u\}}d\bar{u}\dw\,\sum_{i=0}^1\Big(|\partial_u\underline{\mathcal{X}}^{(2+i)}_{\mathscr{I}^+}|^2+|\partial_u\underline{\mathcal{X}}^{(1+i)}_{\mathscr{I}^+}|^2+|\mathring{\slashednabla}\partial_u\underline{\upPhi}_{\mathscr{I}^+}^{(2+i)}|^2\\&\qquad\qquad\qquad\qquad\qquad\qquad\qquad+|\partial_u\mathring{\slashednabla}\underline{\upPhi}^{(i)}_{\mathscr{I}^+}|^2+|\partial_u\underline{\upPhi}_{\mathscr{I}^+}^{(2+i)}|^2+|\partial_u\underline{\upPhi}_{\mathscr{I}^+}^{(1+i)}|^2\\&\qquad\qquad\qquad\qquad\qquad\qquad\qquad+|\partial_u\underline{\upPhi}_{\mathscr{I}^+}^{(i)}|^2+|\partial_u\upPhi^{(i-1)}_{\mathscr{I}^+}|^2\Big)\Bigg].
    \end{split}
    \end{align}
    Using that 
    \begin{align}
        \underline{\mathcal{X}}_{\mathscr{I}^+}^{(3)}=-\int^{\infty}_ud\bar{u}\;6\underline{\mathcal{X}}_{\mathscr{I}^+}^{(2)}-12M\underline{\mathcal{X}}_{\mathscr{I}^+}^{(1)}-\underline{\upPhi}^{(1)}_{\mathscr{I}^+}
    \end{align}
    we get that $\underline{\mathcal{X}}_{\mathscr{I}^+}^{(3)}$ is linear in $u$ and $\partial_u\underline{\mathcal{X}}_{\mathscr{I}^+}^{(3)}$ is constant for $u<u_-$. Thus, as for $\xblin$, we conclude that $\partial_t\xblin|_{{\Sigma^*_+}}=O(r^{-4})$ as $r\longrightarrow\infty$. We finally arrive at \bref{decay of xblin on Sigma star} iterating the above argument using 
    \begin{align}
        \partial_u\underline{\mathcal{X}}^{(n)}_{\mathscr{I}^+}=n(n-2)\underline{\mathcal{X}}^{(n-1)}_{\mathscr{I}^+}+2Mn(n-1)(n-4)\underline{\mathcal{X}}^{(n-2)}_{\mathscr{I}^+}-\underline{\upPhi}^{(n-2)}_{\mathscr{I}^+}
    \end{align}
    and
     \begin{align}
    \begin{split}
        \int_{S^2}\dw\,\left|r\Omega\partial_t^n\xblin|_{{\Sigma^*_+}}(r,\theta^A)\right|^2&\lesssim \frac{1}{r(u,\vsigmap{u})^{4+2n}}\Bigg[\int_{S^2}\dw\,|\partial_u^n\underline{\mathcal{X}}^{(2)}_{\mathscr{I}^+}|^2\\&+\frac{1}{r}\int_{\mathscr{I}^+\cap\{\bar{u}\geq u\}}d\bar{u}\dw\,\sum_{i=0}^n\Big(|\partial_u\underline{\mathcal{X}}^{(2+i)}_{\mathscr{I}^+}|^2+|\partial_u\underline{\mathcal{X}}^{(1+i)}_{\mathscr{I}^+}|^2+|\mathring{\slashednabla}\partial_u\underline{\upPhi}_{\mathscr{I}^+}^{(2+i)}|^2\\&\qquad\qquad\qquad\qquad\qquad\qquad\qquad+|\partial_u\mathring{\slashednabla}\underline{\upPhi}^{(i)}_{\mathscr{I}^+}|^2+|\partial_u\underline{\upPhi}_{\mathscr{I}^+}^{(2+i)}|^2+|\partial_u\underline{\upPhi}_{\mathscr{I}^+}^{(1+i)}|^2\\&\qquad\qquad\qquad\qquad\qquad\qquad\qquad+|\partial_u\underline{\upPhi}_{\mathscr{I}^+}^{(i)}|^2+|\partial_u\upPhi^{(i-1)}_{\mathscr{I}^+}|^2\Big)\Bigg].
    \end{split}
    \end{align}
    When the assumptions of \Cref{grand proposition backwards scattering addendum} are satisfied, we have $\mathring{\slashednabla}^{\gamma}\partial_u^i\Phi^{(n)}_{\mathscr{I}^+}=O\left(|u|^{n-1-i}\right)$, so we may use the above argument to obtain an additional power of decay in $r$.
\end{proof}

\subsubsection*{Estimates on $\xlin$}

Turning to $\xlin$, we may use \Cref{backwards rp alin n} to derive $r$-weighted estimates on $\nablav\xlin$ from \bref{D4Chihat} as follows:

\begin{defin}\label{definition of Xlin n}
Consider the quantity $\xlin$ belonging to a solution $\mathfrak{S}$ of \fullsystem arising from scattering data as in \Cref{grand proposition backwards scattering}. For $n\geq0$ define 
\begin{align}
    \mathcal{X}^{(n)}:=\left(\frac{r^2}{\Omega^2}\nablav\right)^n\frac{r^2\xlin}{\Omega}.
\end{align}
\end{defin}

\begin{proposition}
    Let $\xlin$, $\mathcal{X}^{(n)}$ be as in \Cref{definition of Xlin n}. Let $v_\infty$ be such that $r(u_+,v_\infty)>3M$. Then we have for $n\geq1$
    \begin{align}
        \begin{split}
            \int_{\mathscr{C}_u\cap\{\bar{v}\geq v_\infty\}}d\bar{v}\sin\theta d\theta d\phi\,\frac{\Omega^2}{r^2}|\mathcal{X}^{(n)}|^2\leq \int_{\mathscr{C}_u\cap\{\bar{v}\geq v_\infty\}}d\bar{v}\sin\theta d\theta d\phi\,\frac{\Omega^2}{r^2}|\mathtt{A}^{(n+1)}|^2.
        \end{split}
    \end{align}
\end{proposition}
\begin{proof}
    Straightforward computation using \bref{D4Chihat}.
\end{proof}

\begin{proposition}
Take $\mathscr{D}_{u,v}=J^+(\mathscr{C}_u)\cap J^+(\underline{\mathscr{C}}_v)$. For the solution $\mathfrak{S}$ constructed in \Cref{grand proposition backwards scattering}, we have
\begin{align}\label{this 30 08 2021}
    \begin{split}
        \int_{\mathscr{D}_{u,v}}d\bar{u}d\bar{v}&\dw\,\left[r^2|\nablav r^2\Omega\xlin|^2+(2M)^2r^2\Omega^2|\xlin|^2\right]+2M\int_{\underline{\mathscr{C}}_{v}\cap\{\bar{u}\geq u\}}d\bar{u}\dw\,r^4\Omega^4|\xlin|^2\\
        &\leq C(M,u_+)\int_{\mathscr{I}+\cap\{\bar{u}\geq u\}}d\bar{u}\dw\,\left[|\xlins_{\mathscr{I}^+}(\bar{u},\theta^A)|^2+|\mathring{\slashednabla}{\mathtt{A}^{(3)}}_{\mathscr{I}^+}|^2+|{\mathtt{A}^{(3)}}_{\mathscr{I}^+}|^2+|\alins_{\mathscr{I}^+}|^2\right],
    \end{split}
\end{align}
\begin{align}\label{this 30 08 2021 2}
    \begin{split}
        &\int_{\mathscr{C}_{u}\cap\{\bar{v}\geq v\}}d\bar{u}d\bar{v}\dw\,\left[r^2|\nablav r^2\Omega\xlin|^2+(2M)^2r^2\Omega^2|\xlin|^2\right]+2M\int_{S^2}\dw\,r^4\Omega^4|\xlin(u,\theta^A)|^2\\
            &\leq C(M,u_+)\Big\{\int_{S^2}\dw\,|\xlins_{\mathscr{I}^+}(u,\theta^A)|^2\\&\qquad\qquad\qquad+\int_{\mathscr{I}+\cap\{\bar{u}\geq u\}}d\bar{u}\dw\,\left[|\xlins_{\mathscr{I}^+}(\bar{u},\theta^A)|^2+|\mathring{\slashednabla}{\mathtt{A}^{(3)}}_{\mathscr{I}^+}|^2+|{\mathtt{A}^{(3)}}_{\mathscr{I}^+}|^2+|\alins_{\mathscr{I}^+}|^2\right]\Big\}.
    \end{split}
\end{align}
\end{proposition}
\begin{proof}
    We will only show \bref{this 30 08 2021} as \bref{this 30 08 2021 2} is similar. We square both sides of
    \begin{align}
        \nablav r^2\Omega\xlin -\frac{2M}{r^2} r^2\Omega\xlin=-2r^2\Omega^2\alin,
    \end{align}
    multiply by $r^2$ and integrate over $\mathscr{D}_{u,v}$ to get
    \begin{align}
        \begin{split}
            \int_{\mathscr{D}_{u,v}}d\bar{u}d\bar{v}\dw\,&\left[r^2|\nablav r^2\Omega\xlin|^2+(2M)^2r^2\Omega^2|\xlin|^2\right]+2M\int_{\underline{\mathscr{C}}_{v}\cap\{\bar{u}\geq u\}}d\bar{u}\dw\,r^4\Omega^4|\xlin|^2\\
            &=2M\int_{\mathscr{I}+\cap\{\bar{u}\geq u\}}d\bar{u}\dw\,|\xlins_{\mathscr{I}^+}(\bar{u},\theta^A)|^2+\int_{\mathscr{D}_{u,v}}d\bar{u}d\bar{v}\dw\,r^6\Omega^6|\alin|^2.
        \end{split}
    \end{align}
    We estimate the last term above using \Cref{backwards rp alin n,,backwards estimate on alin away from scri+,,backwards ILED alin}.
\end{proof}

\begin{proposition}\label{bounded r estimate on xlin backwards scattering}
Take $\mathscr{D}_{u,v}=J^+(\mathscr{C}_u)\cap J^+(\underline{\mathscr{C}}_v)$. For the solution $\mathfrak{S}$ constructed in \Cref{grand proposition backwards scattering}, we have
\begin{align}
    \begin{split}
        \int_{\underline{\mathscr{C}}_v\cap\{\bar{u}\geq u\}}d\bar{u}&\dw\,r^4\Omega^4|\xlin|^2+\int_{\mathscr{D}_{u,v}}d\bar{u}d\bar{v}\dw\,r^2\Omega^4|\xlin|^2\\
        &\leq C(M) \int_{\mathscr{I}+\cap\{\bar{u}\geq u\}}d\bar{u}\dw\,\left[|\xlins_{\mathscr{I}^+}(\bar{u},\theta^A)|^2+|\mathring{\slashednabla}{\mathtt{A}^{(3)}}_{\mathscr{I}^+}|^2+|{\mathtt{A}^{(3)}}_{\mathscr{I}^+}|^2+|\alins_{\mathscr{I}^+}|^2\right].
    \end{split}
\end{align}
\end{proposition}

\begin{proof}
    We derive from \bref{D4Chihat}
    \begin{align}
        \nablav \Omega^2|r^2\Omega\xlin|^2-\frac{6M\Omega^2}{r^2}|r^2\Omega\xlin|^2=-2r^4\Omega^5 \alin\cdot \xlin.
    \end{align}
    Integrate over $\mathscr{D}_{u,v}$ and apply Cauchy--Schwarz to get
    \begin{align}
        \begin{split}
            \int_{\underline{\mathscr{C}}_v\cap\{\bar{u}\geq u\}}d\bar{u}&\dw\,r^4\Omega^4|\xlin|^2+3M\int_{\mathscr{D}_{u,v}}d\bar{u}d\bar{v}\dw\,r^2\Omega^4|\xlin|^2\\
        &\leq \int_{\mathscr{I}+\cap\{\bar{u}\geq u\}}d\bar{u}\dw\,|\xlins_{\mathscr{I}^+}(\bar{u},\theta^A)|^2+\frac{1}{3M}\int_{\mathscr{D}_{u,v}}d\bar{u}d\bar{v}\dw\,r^4\Omega^6|\alin|^2.
        \end{split}
    \end{align}
    The last term on the right hand side above can be estimated using \Cref{backwards rp alin n,,backwards estimate on alin away from scri+,,backwards ILED alin}. 
\end{proof}

\begin{proposition}\label{redshift estimate on xlin backwards scattering}
Take $\mathscr{D}_{u,v}=J^+(\mathscr{C}_u)\cap J^+(\underline{\mathscr{C}}_v)$. For the solution $\mathfrak{S}$ constructed in \Cref{grand proposition backwards scattering}, we have
\begin{align}
    \begin{split}
        &\int_{\underline{\mathscr{C}}_v\cap\{\bar{u}\geq u\}}d\bar{u}\dw\,\frac{1}{\Omega^2}|\nablau r^2\Omega^2\xlin|^2+\int_{\mathscr{D}_{u,v}}d\bar{u}d\bar{v}\dw\,\frac{1}{r^2\Omega^2}|\nablau r^2\Omega\xlin|^2\\&\qquad\leq\; C(M,u_+)\Big\{\int_{\mathscr{H}^+\cap\{\bar{v}\geq v\}}d\bar{v}\dw\,|\partial_v \upPsilin_{\mathscr{H}^+}|^2+|\plins_{\mathscr{H}^+}|^2+|\alins_{\mathscr{H}^+}|^2+|\xlins_{\mathscr{H}^+}|^2\\&\qquad\qquad\qquad+\int_{\mathscr{I}^+\cap\{\bar{u}\geq u\}}d\bar{u}\dw\,|\partial_u \upPsilin_{\mathscr{I}^+}|^2+|\mathring{\slashednabla}\mathtt{A}^{(3)}_{\mathscr{I}^+}|^2+|\mathtt{A}^{(2)}_{\mathscr{I}^+}|^2+|\alins_{\mathscr{I}^+}|^2+|\partial_u\xlins_{\mathscr{I}^+}|^2\Big\}.
    \end{split}
\end{align}
\end{proposition}
\begin{proof}
    Commuting \bref{D4Chihat} with $\nablau$ gives
    \begin{align}
        \nablau\nablav r^2\Omega\xlin-\frac{2M}{r^2}\nablau r^2\Omega\xlin+\frac{4M\Omega^2}{r^3}r^2\Omega\xlin=-r\Omega^4\alin-r^2\Omega^3\plin.
    \end{align}
    Multiply by $\Omega^{-1}\nablagml r^2\Omega\xlin$ and integrate over $\mathscr{D}_{u,v}$ to get
    \begin{align}
    \begin{split}
        &\int_{\underline{\mathscr{C}}_v\cap\{\bar{v}\geq v\}}d\bar{v}\dw\,\frac{1}{\Omega^2}|\nablau r^2\Omega^2\xlin|^2+\int_{\mathscr{D}_{u,v}}d\bar{u}d\bar{v}\dw\,\left[\frac{2M}{r^2\Omega^2}|\nablau r^2\Omega\xlin|^2+12M^2\frac{\Omega^4}{r^2}|\xlin|^2\right]\\
        &+\int_{\mathscr{C}_u\cap\{\bar{v}\geq v\}}d\bar{v}\dw\,r\Omega^2|\xlin|^2=\int_{\mathscr{H}^+\cap\{\bar{v}\geq v\}}d\bar{v}\dw\,8M^2|\xlins_{\mathscr{H}^+}|^2\\&+\int_{\mathscr{I}^+\cap\{\bar{u}\geq u\}}d\bar{u}\dw\,|\partial_u\xlins_{\mathscr{I}^+}|^2
        +2\int_{\mathscr{D}_{u,v}}d\bar{u}d\bar{v}\dw\,\nablau r^2\Omega\xlin\cdot\left[r\Omega^2\alin+r^2\Omega\plin\right].
    \end{split}
\end{align}
Cauchy--Schwarz gives
  \begin{align}
    \begin{split}
        &\int_{\underline{\mathscr{C}}_v\cap\{\bar{v}\geq v\}}d\bar{v}\dw\,\frac{1}{\Omega^2}|\nablau r^2\Omega^2\xlin|^2+\int_{\mathscr{D}_{u,v}}d\bar{u}d\bar{v}\dw\,\left[\frac{M}{r^2\Omega^2}|\nablau r^2\Omega\xlin|^2\right]
        \\&\leq\int_{\mathscr{H}^+\cap\{\bar{v}\geq v\}}d\bar{v}\dw\,8M^2|\xlins_{\mathscr{H}^+}|^2+\int_{\mathscr{I}^+\cap\{\bar{u}\geq u\}}d\bar{u}\dw\,|\partial_u\xlins_{\mathscr{I}^+}|^2
        \\&+\int_{\mathscr{D}_{u,v}}d\bar{u}d\bar{v}\dw\,\frac{1}{M}\Omega^2\left[|r\Omega^2\alin|^2+|r^2\Omega\plin|^2\right].
    \end{split}
\end{align}
We estimate the bulk integral of $\plin$ on the right hand side above using \Cref{backwards ILED plin}:
\begin{align}
    \begin{split}
        \int_{\mathscr{D}_{u,v}}d\bar{u}d\bar{v}\dw\,r^4\Omega^4|\plin|^2\lesssim& \int_{\mathscr{H}^+\cap\{\bar{v}\geq v\}}d\bar{v}\dw\,|\partial_v \upPsilin_{\mathscr{H}^+}|^2+|\plins_{\mathscr{H}^+}|^2\\&+\int_{\mathscr{I}^+\cap\{\bar{u}\geq u\}}d\bar{u}\dw\,|\partial_u \upPsilin_{\mathscr{I}^+}|^2,
    \end{split}
\end{align}
while the bulk integral of $\alin$ is estimated using \Cref{backwards ILED alin} and \Cref{backwards rp alin n}:
\begin{align}
    \begin{split}
        \int_{\mathscr{D}_{u,v}}d\bar{u}d\bar{v}\dw\,r^2\Omega^6|\alin|^2\lesssim& \int_{\mathscr{H}^+\cap\{\bar{v}\geq v\}}d\bar{v}\dw\,|\partial_v \upPsilin_{\mathscr{H}^+}|^2+|\plins_{\mathscr{H}^+}|^2+|\alins_{\mathscr{H}^+}|^2\\&+\int_{\mathscr{I}^+\cap\{\bar{u}\geq u\}}d\bar{u}\dw\,|\partial_u \upPsilin_{\mathscr{I}^+}|^2+|\mathring{\slashednabla}\mathtt{A}^{(3)}_{\mathscr{I}^+}|^2+|\mathtt{A}^{(2)}_{\mathscr{I}^+}|^2+|\alins_{\mathscr{I}^+}|^2.
    \end{split}
\end{align}
\end{proof}

Finally, we can estimate higher $\Omega^{-1}\nablagml$-derivatives of $\xlin$ starting with the equation
\begin{align}\label{formula for transverse derivative of xlin near H+}
\begin{split}
    &\nablav \left(\frac{r^2}{\Omega^2}\nablau\right)^nr^2\Omega\xlin -\left(\frac{2n}{r}-\frac{2M}{r^2}(3n-2)\right)\left(\frac{r^2}{\Omega^2}\nablau\right)^nr^2\Omega\xlin\\&+\left(n(n-1)-\frac{2M}{r}(3n^2-7n+6)\right)\left(\frac{r^2}{\Omega^2}\nablau\right)^{n-1}r^2\Omega\xlin-(n-1)(n^2+3n-6)2M\left(\frac{r^2}{\Omega^2}\nablau\right)^{n-2}r^2\Omega\xlin\\&=-\left(\frac{r^2}{\Omega^2}\nablau\right)^nr^2\Omega^2\alin.
\end{split}
\end{align}

Note that $\left(\frac{r^2}{\Omega^2}\nablau\right)^nr^2\Omega\xlin$ suffers from a blueshift only for $n\geq2$, where we have
  \begin{align}\label{blueshift estimate xlin backwards scattering 1}
  \begin{split}
         &\int_{S^2}\sin\theta d\theta d\phi\,\left|\left(\frac{r^2}{\Omega^2}\nablau\right)^nr^2\Omega\xlin(u,v,\theta^A)\right|^2\\&\lesssim e^{\frac{n^2+4}{2M}(v_+-v)}\times\Bigg\{\int_{\mathscr{H}^+\cap\{\bar{v}\in[v,v_+]\}}d\bar{v}\sin\theta d\theta d\phi\, \left[|\alins_{\mathscr{H}^+}|^2+|\plins_{\mathscr{H}^+}|^2+|\partial_v\upPsilin_{\mathscr{H}^+}|^2+\sum_{|\gamma|\leq 2n+1}|\mathring{\slashednabla}^\gamma{\upPsilin}_{\mathscr{H}^+}|^2\right]\\&\qquad\qquad\qquad\qquad+\int_{\mathscr{I}^+}d\bar{u}\dw\,|\partial_u\upPsilin_{\mathscr{I}^+}|^2\Bigg\},
    \end{split}
    \end{align}
    \begin{align}\label{blueshift estimate xlin backwards scattering 2}
    \begin{split}
        &\int_v^{v_+}\int_{S^2}d\bar{v}\sin\theta d\theta d\phi\,\left|\left(\frac{r^2}{\Omega^2}\nablau\right)^nr^2\Omega\xlin(u,v,\theta^A)\right|^2\\&\lesssim e^{\frac{n^2+4}{2M}(v_+-v)}\times\Bigg\{\int_{\mathscr{H}^+\cap\{\bar{v}\in[v,v_+]\}}d\bar{v}\sin\theta d\theta d\phi\, \left[|\alins_{\mathscr{H}^+}|^2+|\plins_{\mathscr{H}^+}|^2+|\partial_v\upPsilin_{\mathscr{H}^+}|^2+\sum_{|\gamma|\leq 2n+1}|\mathring{\slashednabla}^\gamma{\upPsilin}_{\mathscr{H}^+}|^2\right]\\&\qquad\qquad\qquad\qquad+\int_{\mathscr{I}^+}d\bar{u}\dw\,|\partial_u\upPsilin_{\mathscr{I}^+}|^2\Bigg\}
    \end{split}.
    \end{align}

We now turn to the asymptotic behaviour of $\xlin$ near $i^0$.

\begin{corollary}\label{asymptotic flatness at spacelike infinity for xlin}
Let $\xlin$ be as in \Cref{grand proposition backwards scattering}. Then we have
\begin{align}\label{17 10 2021 33}
    \partial_t^n\xlin|_{{\Sigma^*_+}}=O_\infty(r^{-2-n}).
\end{align}
\end{corollary}
\begin{proof}
    We estimate $\xlin|_{{\Sigma^*_+}}$ on $r>r(u_-,\vsigmap{u_-})$ using \Cref{backwards rp alin n} and  \bref{D4Chihat} to get:
    \begin{align}\label{17 10 2021 44}
    \begin{split}
        \int_{S^2}\dw\,\left|\frac{r^2\xlin}{\Omega}-\xlins_{\mathscr{I}^+}\right|^2\lesssim \frac{1}{r(u,\vsigmap{u})^4}&\Bigg\{ \int_{S^2}\dw\,|\alins_{\mathscr{I}^+}|^2\\&+\frac{1}{r(u,\vsigmap{u})}\int_{\mathscr{I}^+\cap\{\bar{u}\geq u\}}d\bar{u}\dw\,|\mathring{\slashednabla}\alins_{\mathscr{I}^+}|^2+|\alins_{\mathscr{I}^+}|^2\Bigg\}.
    \end{split}
    \end{align}
    Commuting with the generators of $SO(3)$ to order $2$ gives the result for $n=0$. We have that $\nablav^n\xlin|_{{\Sigma^*_+}}=O_{\infty}(r^{-2-n})$ using \Cref{asymptotic flatness at spacelike infinity for alin}. The argument for $n\geq1$ in \bref{17 10 2021 33} follows by iterating \bref{17 10 2021 44} above to get
    \begin{align}\label{17 10 2021 nn}
    \begin{split}
        \int_{S^2}\dw\,\left|\partial_t^n\frac{r^2\xlin}{\Omega}\right|^2\lesssim \frac{1}{r(u,\vsigmap{u})^{2+2n}}&\Bigg\{ \int_{S^2}\dw\,|\partial_u^n\mathtt{A}^{(n+1)}_{\mathscr{I}^+}|^2\\&+\frac{1}{r(u,\vsigmap{u})}\int_{\mathscr{I}^+\cap\{\bar{u}\geq u\}}d\bar{u}\dw\sum_{i=0}^n\,|\mathring{\slashednabla}\partial_u^n\mathtt{A}^{(n+1)}_{\mathscr{I}^+}|^2+|\partial_u^n\mathtt{A}^{(n+1)}_{\mathscr{I}^+}|^2\Bigg\}.
    \end{split}
    \end{align}
    When the assumptions of \Cref{grand proposition backwards scattering addendum} are satisfied, we have $\mathtt{A}^{n+2}$ grows at one less power of $|u|$, so we may use the above argument to obtain an additional power of decay in $r$.
\end{proof}
\subsubsection[Boundedness on $\protect\blin$,  $\protect\bblin$]{Boundedness on $\blin$, $\bblin$}\label{section 6.2.1.2 backwards boundedness on blin bblin}

As the equations relating $\blin$, $\bblin$ to $\alin$, $\ablin$ respectively are identical in structure to those relating $\xlin$ to $\alin$, $\xblin$ to $\ablin$, the estimates of \Cref{section 6.2.1.1 backwards boundedness on xlin xblin} also apply to $\blin$, $\bblin$. We list the needed estimates in this section.

\subsubsection*{Boundedness on $\bblin$}

\begin{proposition}\label{backwards ingoing overall  on bblin}
For the solution $\mathfrak{S}$ constructed in \Cref{grand proposition backwards scattering}, we have 
\begin{align}\label{backwards ingoing overall estimate on bblin}
\begin{split}
    \int_{\underline{\mathscr{C}}_v}d\bar{u}\sin\theta d\theta d\phi\,& \left[\frac{1}{\Omega^2}|\nablau \Omega^{-1}\bblin(\bar{u},v,\theta^A)|^2+\frac{6}{r^2}|\bblin(\bar{u},v,\theta^A)|^2\right]+\int_{S^2}\frac{1}{r(u,v)}|\Omega^{-1}\bblin(u,v,\theta^A)|^2\\
    &\lesssim_M \frac{1}{2M}\int_{S^2}\sin\theta d\theta d\phi\,|\bblins_{\mathscr{H}^+}(u,\theta^A)|^2\\&+\text{Right hand side of }\bref{backwards ingoing overall estimate on ablin}\text{ commuted with }\divo.
\end{split}
\end{align}
\end{proposition}

\begin{corollary}
For the solution $\mathfrak{S}$ constructed in \Cref{grand proposition backwards scattering}, we have 
\begin{align}
    \begin{split}
        &\int_{\underline{\mathscr{C}}_v\cap\{\bar{u}\geq u\}}d\bar{u}\sin\theta d\theta d\phi\, \Bigg[\frac{1}{\Omega^2}|\nablau\nablav \Omega^{-1}\bblin(\bar{u},v,\theta^A)|^2+\frac{\Omega^2}{r^2}|\nablav \Omega^{-1}\bblin|^2\Bigg]\\&+\int_{S^2}\sin\theta d\theta d\phi\,\frac{2}{r(u,v)}|\nablav\Omega^{-1}\bblin|^2\;\lesssim\;
        \frac{1}{2M}\int_{S^2}\sin\theta d\theta d\phi\,|\bblins_{\mathscr{H}^+}(u,\theta^A)|^2\\&+\text{ Right hand sides of }\bref{backwards ingoing overall estimate on ablin}\text{ and }\bref{backwards ingoing overall estimate on pblin}\text{ commuted with }\divo.
    \end{split}
\end{align}
\end{corollary}

\begin{proposition}\label{backwards rp bblin}
Let $v_\infty$ be such that $r(u_+,v_\infty)\geq 16M$. For the solution $\mathfrak{S}$ constructed in \Cref{grand proposition backwards scattering}, we have for any $u\geq u_+$,
\begin{align}\label{backwards rp estimate bblin}
\begin{split}
   &\int_{\mathscr{C}_u\cap\{\bar{v}\geq v_\infty\}\cap J^+({\Sigma^*_+})}d\bar{v}\sin\theta d\theta d\phi\,\frac{r^2}{\Omega^2}|\nablav r\Omega\bblin|^2+\int_{\mathscr{D}}d\bar{u}d\bar{v}\sin\theta d\theta d\phi\,4M\Omega^4|\bblin|^2\\&\qquad\leq C(u_+) \int_{\mathscr{I}^+\cap\{\bar{u}\in[u,u_+]\}}d\bar{u}\sin\theta d\theta d\phi\,\Big[|\divo\bblins_{\mathscr{I}^+}|^2+|\divo\pblins_{\mathscr{I}^+}|^2+|\divo\pblins_{\mathscr{I}^+}|^2+|\mathring{\slashednabla}\divo\upPsilinb_{\mathscr{I}^+}|^2\\&\qquad\qquad\qquad\qquad\qquad\qquad\qquad\qquad+|\divo\upPsilinb_{\mathscr{I}^+}|^2+|\mathring{\slashednabla}\divo\upPsilinb_{\mathscr{I}^+}|^2\Big]
\end{split}
\end{align}
\end{proposition}

\begin{defin}\label{definition of Bblin n}
Let $\bblin$ arise from a solution $\mathfrak{S}$ to the system \fullsystem in the sense of \Cref{EinsteinWP} or \Cref{grand proposition backwards scattering}. Define
\begin{align}
    \underline{\mathcal{B}}^{(n)}:=\left(\frac{r^2}{\Omega^2}\nablav\right)^nr^2\Omega\bblin.
\end{align}
\end{defin}

\begin{proposition}
 Let $\underline{\mathcal{B}}^{(n)}$ be as in \Cref{definition of Bblin n}. The limit
    \begin{align}\label{limit of Bblin n}
       \underline{\mathcal{B}}^{(n)}_{\mathscr{I}^+}(u,v,\theta^A)=\lim_{v\longrightarrow\infty}\underline{\mathcal{B}}^{(n)}(u,\theta^A)
    \end{align}
    exists and defines an element of $\Gamma(\mathscr{I}^+)$. The same applies to angular derivatives of $\underline{\mathcal{B}}^{(n)}$ and 
    \begin{align}\label{this 16 08 2021 4}
        \lim_{v\longrightarrow\infty}\mathring{\slashednabla}^\gamma\underline{\mathcal{B}}^{(n)}_{\mathscr{I}^+}=\mathring{\slashednabla}^\gamma \underline{\mathcal{B}}^{(n)}_{\mathscr{I}^+}
    \end{align}
for any index $\gamma$.
\end{proposition}

\begin{proposition}\label{backwards rp Bblin n}
Take $n\geq2$, $v_\infty$ such that $r(u_+,v_\infty)>3M$, and let $\mathscr{D}$ be the region $[u,u_+]\times [v_\infty,\infty)\times S^2$. For the solution $\mathfrak{S}$ constructed in \Cref{grand proposition backwards scattering} we have
\begin{align}\label{backwards rp estimates Bblin n}
    \begin{split}
        &\int_{\mathscr{C}_u\cap\{\bar{v}\geq v_\infty\}}d\bar{v}\sin\theta d\theta d\phi\,\frac{r^2}{\Omega^2}|\nablav\underline{\mathcal{B}}^{(n)}|^2+\int_{\mathscr{D}}d\bar{u}d\bar{v}\sin\theta d\theta d\phi\,\frac{\Omega^2}{r^2}|\underline{\mathcal{B}}^{(n)}|^2+\frac{r}{\Omega^2}|\nablav \underline{\mathcal{B}}^{(n)}|^2\\&
        +\int_{\underline{\mathscr{C}}_{v_\infty}}d\bar{u}\sin\theta d\theta d\phi\,|\underline{\mathcal{B}}^{(n)}|^2\\&\leq C(n,M,u_+)\int_{\mathscr{I}^+\cap\{\bar{u}\in[u,u_+]\}}d\bar{u}\sin\theta d\theta d\phi\,\Bigg[|\underline{\mathcal{B}}^{(n)}_{\mathscr{I}^+}|^2+|\underline{\mathcal{B}}^{(n-1)}_{\mathscr{I}^+}|^2+|\mathring{\slashednabla}\divo\Phi^{(n)}_{\mathscr{I}^+}|^2\\&\qquad\qquad\qquad\qquad\qquad+|\mathring{\slashednabla}\divo\Phi^{(n-2)}_{\mathscr{I}^+}|^2+|\Phi^{(n)}_{\mathscr{I}^+}|^2+|\divo\Phi^{(n-1)}_{\mathscr{I}^+}|^2+|\divo\Phi^{(n-2)}_{\mathscr{I}^+}|^2+|\divo\Phi^{(n-3)}_{\mathscr{I}^+}|^2\Bigg]
    \end{split}
\end{align}
\end{proposition}

Analogously to \Cref{asymptotic flatness at spacelike infinity for xblin}, we have
\begin{corollary}\label{asymptotic flatness at spacelike infinity for bblin}
 The quantity $\bblin$ of \Cref{grand proposition backwards scattering} satisfies
 \begin{align}\label{decay of bblin on Sigma star}
     \partial_t^n \bblin|_{{\Sigma^*_+}}=O_{\infty}(r^{-3-n}).
 \end{align}
 When the assumptions of \Cref{grand proposition backwards scattering addendum} are satisfied, we have that
 \begin{align}\label{decay of bblin on Sigma star}
     \partial_t^n \bblin|_{{\Sigma^*_+}}=O_{\infty}(r^{-4-n}).
 \end{align}
\end{corollary}

\subsubsection*{Boundedness on $\protect\blin$}

\begin{defin}\label{definition of Blin n}
Consider the quantity $\blin$ belonging to a solution $\mathfrak{S}$ of \fullsystem arising from scattering data as in \Cref{grand proposition backwards scattering}. For $n\geq0$ define 
\begin{align}
    \mathcal{B}^{(n)}:=\left(\frac{r^2}{\Omega^2}\nablav\right)^n\frac{r^4\blin}{\Omega}.
\end{align}
\end{defin}

\begin{proposition}
    Let $\blin$, $\mathcal{B}^{(n)}$ be as in \Cref{definition of Blin n}. Let $v_\infty$ be such that $r(u_+,v_\infty)>3M$. Then for the solution $\mathfrak{S}$ constructed in \Cref{grand proposition backwards scattering} we have for $n\geq1$
    \begin{align}
        \begin{split}
            \int_{\mathscr{C}_u\cap\{\bar{v}\geq v_\infty\}}d\bar{v}\sin\theta d\theta d\phi\,\frac{\Omega^2}{r^2}|\mathcal{B}^{(n)}|^2\leq \int_{\mathscr{C}_u\cap\{\bar{v}\geq v_\infty\}}d\bar{v}\sin\theta d\theta d\phi\,\frac{\Omega^2}{r^2}|\divo\mathtt{A}^{(n+1)}|^2.
        \end{split}
    \end{align}
\end{proposition}

\begin{proposition}
Take $\mathscr{D}_{u,v}=J^+(\mathscr{C}_u)\cap J^+(\underline{\mathscr{C}}_v)$. For the solution $\mathfrak{S}$ constructed in \Cref{grand proposition backwards scattering} we have
\begin{align}\label{this 30 08 2021 blin}
    \begin{split}
        &\int_{\mathscr{D}_{u,v}}d\bar{u}d\bar{v}\dw\,\left[r^2|\nablav r^4\Omega\blin|^2+(2M)^2r^6\Omega^2|\blin|^2\right]+2M\int_{\underline{\mathscr{C}}_{v}\cap\{\bar{u}\geq u\}}d\bar{u}\dw\,r^8\Omega^4|\blin|^2\\
        &\leq C(M,u_+)\int_{\mathscr{I}+\cap\{\bar{u}\geq u\}}d\bar{u}\dw\,\left[|\blins_{\mathscr{I}^+}(\bar{u},\theta^A)|^2+|\mathring{\slashednabla}\divo\mathtt{A}^{(3)}_{\mathscr{I}^+}|^2+|\divo\mathtt{A}^{(3)}_{\mathscr{I}^+}|^2+|\divo\alins_{\mathscr{I}^+}|^2\right],
    \end{split}
\end{align}
\begin{align}\label{this 30 08 2021 blin 2}
    \begin{split}
        &\int_{\mathscr{C}_{u}\cap\{\bar{v}\geq v\}}d\bar{u}d\bar{v}\dw\,\left[r^2|\nablav r^4\Omega\blin|^2+(2M)^2r^6\Omega^2|\blin|^2\right]+2M\int_{S^2}\dw\,r^8\Omega^4|\blin(u,v,\theta^A)|^2\\
            &\leq C(M,u_+)\Big\{\int_{S^2}\dw\,|\blins_{\mathscr{I}^+}(u,\theta^A)|^2\\&\qquad\qquad\qquad+\int_{\mathscr{I}+\cap\{\bar{u}\geq u\}}d\bar{u}\dw\,\left[|\divo\blins_{\mathscr{I}^+}(\bar{u},\theta^A)|^2+|\mathring{\slashednabla}\divo\mathtt{A}^{(3)}_{\mathscr{I}^+}|^2+|\divo\mathtt{A}^{(3)}_{\mathscr{I}^+}|^2+|\divo\alins_{\mathscr{I}^+}|^2\right]\Big\}.
    \end{split}
\end{align}
\end{proposition}

\begin{proposition}
Take $\mathscr{D}_{u,v}=J^+(\mathscr{C}_u)\cap J^+(\underline{\mathscr{C}}_v)$. For the solution $\mathfrak{S}$ constructed in \Cref{grand proposition backwards scattering} we have
\begin{align}
    \begin{split}
        \int_{\underline{\mathscr{C}}_v\cap\{\bar{u}\geq u\}}d\bar{u}&\dw\,r^8\Omega^4|\blin|^2+\int_{\mathscr{D}_{u,v}}d\bar{u}d\bar{v}\dw\,r^6\Omega^4|\blin|^2\\
        &\leq C(M) \int_{\mathscr{I}+\cap\{\bar{u}\geq u\}}d\bar{u}\dw\,\left[|\blins_{\mathscr{I}^+}(\bar{u},\theta^A)|^2+|\mathring{\slashednabla}\divo\mathtt{A}^{(3)}_{\mathscr{I}^+}|^2+|\divo\mathtt{A}^{(3)}_{\mathscr{I}^+}|^2+|\divo\alins_{\mathscr{I}^+}|^2\right].
    \end{split}
\end{align}
\end{proposition}

\begin{proposition}
Take $\mathscr{D}_{u,v}=J^+(\mathscr{C}_u)\cap J^+(\underline{\mathscr{C}}_v)$. For the solution $\mathfrak{S}$ constructed in \Cref{grand proposition backwards scattering} we have
\begin{align}
    \begin{split}
       &\int_{\underline{\mathscr{C}}_v\cap\{\bar{v}\geq v\}}d\bar{v}\dw\,\frac{1}{\Omega^2}|\nablau r^4\Omega\blin|^2+\int_{\mathscr{D}_{u,v}}d\bar{u}d\bar{v}\dw\,\frac{1}{r^2\Omega^2}|\nablau r^4\Omega\blin|^2\\&\leq\; C(M,u_+)\Big\{\int_{\mathscr{H}^+\cap\{\bar{v}\geq v\}}d\bar{v}\dw\,|\partial_v \divo\upPsilin_{\mathscr{H}^+}|^2+|\divo\plins_{\mathscr{H}^+}|^2+|\divo\alins_{\mathscr{H}^+}|^2+|\blins_{\mathscr{H}^+}|^2\\&\qquad\qquad+\int_{\mathscr{I}^+\cap\{\bar{u}\geq u\}}d\bar{u}\dw\,|\partial_u \divo\upPsilin_{\mathscr{I}^+}|^2+|\mathring{\slashednabla}\divo\mathtt{A}^{(3)}_{\mathscr{I}^+}|^2+|\divo\mathtt{A}^{(2)}_{\mathscr{I}^+}|^2+|\divo\alins_{\mathscr{I}^+}|^2+|\partial_u\blins_{\mathscr{I}^+}|^2\Big\}.
    \end{split}
\end{align}
\end{proposition}
Similarly to \bref{blueshift estimate xlin backwards scattering 1} and \bref{blueshift estimate xlin backwards scattering 2} for $\xlin$, we have
  \begin{align}\label{blueshift estimate blin backwards scattering 1}
  \begin{split}
         &\int_{S^2}\sin\theta d\theta d\phi\,\left|\left(\frac{r^2}{\Omega^2}\nablau\right)^nr^2\Omega\blin(u,v,\theta^A)\right|^2\\&\lesssim e^{\frac{n^2+4}{2M}(v_+-v)}\times\Bigg\{\int_{\mathscr{H}^+\cap\{\bar{v}\in[v,v_+]\}}d\bar{v}\sin\theta d\theta d\phi\, \Bigg[|\divo\alins_{\mathscr{H}^+}|^2+|\divo\plins_{\mathscr{H}^+}|^2+|\partial_v\divo\upPsilin_{\mathscr{H}^+}|^2\\&\qquad\qquad\qquad\qquad+\sum_{|\gamma|\leq 2n+1}|\mathring{\slashednabla}^\gamma\divo{\upPsilin}_{\mathscr{H}^+}|^2\Bigg]+\int_{\mathscr{I}^+}d\bar{u}\dw\,|\partial_u\divo\upPsilin_{\mathscr{I}^+}|^2\Bigg\},
    \end{split}
    \end{align}
    \begin{align}\label{blueshift estimate blin backwards scattering 2}
    \begin{split}
        &\int_v^{v_+}\int_{S^2}d\bar{v}\sin\theta d\theta d\phi\,\left|\left(\frac{r^2}{\Omega^2}\nablau\right)^nr^2\Omega\blin(u,v,\theta^A)\right|^2\\&\lesssim e^{\frac{n^2+4}{2M}(v_+-v)}\times\Bigg\{\int_{\mathscr{H}^+\cap\{\bar{v}\in[v,v_+]\}}d\bar{v}\sin\theta d\theta d\phi\, \Bigg[|\divo\alins_{\mathscr{H}^+}|^2+|\divo\plins_{\mathscr{H}^+}|^2+|\partial_v\divo\upPsilin_{\mathscr{H}^+}|^2\\&\qquad\qquad\qquad\qquad+\sum_{|\gamma|\leq 2n+1}|\mathring{\slashednabla}^\gamma\divo{\upPsilin}_{\mathscr{H}^+}|^2\Bigg]+\int_{\mathscr{I}^+}d\bar{u}\dw\,|\partial_u\divo\upPsilin_{\mathscr{I}^+}|^2\Bigg\}
    \end{split}.
    \end{align}
Analogously to \Cref{asymptotic flatness at spacelike infinity for xlin}, we have
\begin{corollary}\label{asymptotic flatness at spacelike infinity for blin}
Let $\blin$ be as in \Cref{grand proposition backwards scattering}. Then we have
\begin{align}\label{17 10 2021 66 blin}
    \partial_t^n\blin|_{{\Sigma^*_+}}=O_\infty(r^{-3-n}).
\end{align}
When the assumptions of \Cref{grand proposition backwards scattering addendum} are satisfied, we have
\begin{align}\label{17 10 2021 666 blin}
    \partial_t^n\blin|_{{\Sigma^*_+}}=O_\infty(r^{-4-n}).
\end{align}
\end{corollary}

\subsubsection[Boundedness on $\protect\glinh$]{Boundedness on $\glinh$}

Given \bref{backwards ingoing overall estimate on xblin} and applying to \bref{metric transport in 3 direction traceless} a similar argument to that leading to \bref{backwards ingoing overall  on xblin}, we can show the following:

\begin{proposition}
For the solution $\mathfrak{S}$ constructed in \Cref{grand proposition backwards scattering} we have
\begin{align}
    \int_{\underline{\mathscr{C}}_v}d\bar{u}\sin\theta d\theta d\phi\,\frac{1}{r\Omega^2}|\nablau\glinh|^2\lesssim_M \text{Right hand side of }\bref{backwards ingoing overall estimate on xblin}
\end{align}
\end{proposition}

\begin{defin}
For $\glinh$ belonging to a solution $\mathfrak{S}$ of \fullsystem, define
\begin{align}
    \hat{\mathtt{g}}^{(n)}:=\left(\frac{r^2}{\Omega^2}\nablav\right)^n r\glinh
\end{align}
\end{defin}

\begin{lemma}
Take $\mathscr{D}_{u,v}=J^+(\mathscr{C}_u)\cap J^+(\underline{\mathscr{C}}_v)$. For the solution $\mathfrak{S}$ constructed in \Cref{grand proposition backwards scattering} we have
    \begin{align}
    \begin{split}
        &\nablau \glinhn{n}+\left(\frac{2n+1}{r}-\frac{2M}{r^2}(3n+1)\right)\glinhn{n}-\frac{\Omega^2}{r^2}\left(n^2-\frac{2M}{r}(3n^2-n)\right)\glinhn{n-1}\\&-2M(n-1)^2n\glinhn{n-2}=2\underline{\mathcal{X}}^{(n)}.
    \end{split}
    \end{align}
    \begin{align}\label{this 22 08 2021}
        \begin{split}
            &\nablau\nablav \glinhn{n}+\left(\frac{2n+1}{r}-\frac{2M}{r^2}(3n+1)\right)\nablav\glinhn{n}-\frac{\Omega^2}{r^2}\left((n+1)^2-\frac{2M}{r}(3n^2+5n+2)\right)\glinhn{n}\\&-2M\frac{\Omega^2}{r^2}n^2(n+1)\glinhn{n-1}=2\frac{\Omega^2}{r^2}\underline{\mathcal{X}}^{(n+1)}
        \end{split}
    \end{align}
\end{lemma}

By applying the argument of the backwards $r^p$ estimates of \Cref{backwards rp Xblin n} to \bref{this 22 08 2021} and using \bref{backwards rp estimate RW Phi n} and the results of \Cref{backwards rp Xblin n}, we obtain

\begin{proposition}\label{rp glinh n backwards scattering}
Take $v_\infty$ such that $r(u_+,v_\infty)>3M$ and let $\mathscr{D}$ be the region $[u,u_+]\times [v_\infty,\infty)\times S^2$. Take $\mathscr{D}_{u,v}=J^+(\mathscr{C}_u)\cap J^+(\underline{\mathscr{C}}_v)$. For the solution $\mathfrak{S}$ constructed in \Cref{grand proposition backwards scattering} we have 
\begin{align}
    \begin{split}
        &\int_{\mathscr{C}_u\cap\{\bar{v}\geq v_\infty\}}d\bar{v}\sin\theta d\theta d\phi\,\frac{r^2}{\Omega^2}|\nablav\glinhn{n}|^2+\int_{\mathscr{D}}d\bar{u}d\bar{v}\sin\theta d\theta d\phi\,\frac{\Omega^2}{r^2}|\glinhn{n}|^2\\&+\int_{\underline{\mathscr{C}}_{v_\infty}}d\bar{u}\sin\theta d\theta d\phi\,|\glinhn{n}|^2\leq C(n,M,u_+)\int_{\mathscr{I}^+\cap\{\bar{u}\in[u,u_+]\}}d\bar{u}\sin\theta d\theta d\phi\,\Bigg[|\glinhnscr{n}_{\mathscr{I}^+}|^2+|\glinhnscr{n-1}_{\mathscr{I}^+}|^2\\&\qquad\qquad\qquad\qquad\qquad\qquad\qquad\qquad+\sum_{k=0}^1|\underline{\mathcal{X}}^{(n+k)}_{\mathscr{I}^+}|^2+\sum_{k=0}^1|\mathring{\slashednabla}\underline{\Phi}^{(n+k)}_{\mathscr{I}^+}|^2+|\underline{\Phi}^{(n-2)}_{\mathscr{I}^+}|^2\Bigg].
    \end{split}
\end{align}
In the above, we take $\glinhn{n}=0$ for $n<0$.
\end{proposition}

We can prove the following by applying the arguments of \Cref{asymptotic flatness at spacelike infinity for xblin} and \Cref{rp glinh n backwards scattering} to \bref{metric transport in 3 direction traceless}:
\begin{corollary}\label{asymptotic flatness at spacelike infinity for glinh}
 The quantity $\xblin$ of \Cref{grand proposition backwards scattering} satisfies
 \begin{align}\label{decay of glinh on Sigma star}
     \partial_t^n \glinh|_{{\Sigma^*_+}}=O_{\infty}(r^{-1-n}).
 \end{align}
 When the assumptions of \Cref{grand proposition backwards scattering addendum} are satisfied we have
 \begin{align}\label{decay of glinh on Sigma star}
     \partial_t^n \glinh|_{{\Sigma^*_+}}=O_{\infty}(r^{-2-n}).
 \end{align}
\end{corollary}

\begin{remark}\label{remark on glinh}
We can estimate $\glinh$ everywhere using \Cref{backwards ingoing overall  on xblin}, \Cref{backwards ingoing overall on ablin} and the energy estimates on $\Psilinb$, $\Psilin$ commuted with $\partial_t$ to arbitrary higher orders. Near $\mathscr{H}^+_{\geq0}$, we can estimate arbitrary orders of $\Omega^{-1}\nablagml$-derivatives using \Cref{blueshift on ablin pblin n commuted}.
\end{remark}
\subsubsection{Boundedness on the remaining components}\label{section 6.2.1.3 backwards boundedness on glinh and rest}

We are now in a position to estimate  from scattering data all components of the solution and all their derivatives.

\subsubsection*{Boundedness on $\rlin$, $\slin$}

\begin{corollary}\label{rlin estimate near H+ backwards scattering}
    Let $\mathfrak{S}$, $u_+$, $v_+$ be as in \Cref{grand proposition backwards scattering}. For $\tilde{u}$ such that $r(\tilde{u},v_+)<3M$ and for $u\geq\tilde{u}$ we have for the $\ell\geq2$ of $\rlin$
    \begin{align}
    \begin{split}
        \int_{S^2}\dw\left|\left(\frac{1}{\Omega^2}\nablau\right)^n\rlin_{\ell\geq2}(u,v,\theta^A)\right|^2\lesssim& \text{ Right hand side of } \bref{blueshift estimate on Psilin n backwards scattering 1} \text{ for both } \Psilin, \Psilinb \\&+ \text{Right hand side of } \bref{blueshift estimate xlin backwards scattering 1}, \bref{backwards ingoing overall estimate on xblin}, \bref{blueshift estimate ablin near H+ n times}.
    \end{split}
    \end{align}
\end{corollary}

\begin{proof}
    Use \Cref{formula for Psilin Psilinb in terms of linearised system} and the estimates mentioned above.
\end{proof}
\begin{corollary}
 Let $\mathfrak{S}$, $u_+$, $v_+$ be as in \Cref{grand proposition backwards scattering}, and let $\rhoup^{(n)}:=\left(\frac{r^2}{\Omega^2}\nablav\right)^nr^3\rlin_{\ell\geq2}$. For $v_\infty$ such that $r(u_+,v_\infty)>3M$ we have the following:
 \begin{align}
 \begin{split}
     \int_{\mathscr{C}_u\cap\{\bar{v}\geq v_\infty\}}d\bar{v}\dw\,\frac{r^2}{\Omega^2}|\nablav \rhoup^{(n)}|^2\lesssim &\text{ Right hand side of }\bref{backwards rp estimate RW Phi n} \text{ applied to both }\Psilin, \Psilinb\\& +\text{ Right hand side of } \bref{backwards rp estimate Alin n} \text{ and }\bref{backwards rp estimates Xblin n}.
\end{split}
 \end{align}
\end{corollary}

\begin{corollary}\label{work for slin already done backwards scattering}
The results of \Cref{Boundedness of Psilin Psilinb backwards scattering} apply to
the $\ell\geq2$ component of $\slin$ replacing $\Psilin$ with $\slin$, $\upPsilin_{\mathscr{I}^+}$ with $\slins_{\mathscr{I}^+}$ and $\upPsilin_{\mathscr{H}^+}$ with $\slins_{\mathscr{H}^+}$ on both sides everywhere.
\end{corollary}
\begin{proof}
    Use $\Psilin-\Psilinb=4\mathring{\fancydstar_2}\mathring{\fancydstar_1}(0,\slin)$.
\end{proof}

\subsubsection*{Boundedness on $\elin$, $\eblin$, $\Olino$}

\begin{corollary}\label{elin estimate near H+ backwards scattering}
    Let $\mathfrak{S}$, $u_+$, $v_+$ be as in \Cref{grand proposition backwards scattering}. For $\tilde{u}$ such that $r(\tilde{u},v_+)<3M$ and for $u\geq\tilde{u}$ we have for the $\ell\geq2$ of $\elin$ and $\eblin$
    \begin{align}
    \begin{split}
        \int_{S^2}\dw\left|\left(\frac{1}{\Omega^2}\nablau\right)^n\elin_{\ell\geq2}(u,v,\theta^A)\right|^2\lesssim\text{ Right hand side of } \bref{blueshift estimate xlin backwards scattering 1}, \bref{backwards ingoing overall estimate on xblin}, \bref{blueshift estimate ablin near H+ n times},
    \end{split}
    \end{align}
      \begin{align}
    \begin{split}
        \int_{S^2}\dw\left|\left(\frac{1}{\Omega^2}\nablau\right)^n\eblin_{\ell\geq2}(u,v,\theta^A)\right|^2\lesssim\text{ Right hand side of } \bref{blueshift estimate xlin backwards scattering 1}, \bref{backwards ingoing overall estimate on xblin}, \bref{blueshift estimate ablin near H+ n times}.
    \end{split}
    \end{align}u
    We may replace $\elin$, $\eblin$ above with $\Olino$.
\end{corollary}
\begin{proof}
    Use the equations \bref{D3Chihat}, \bref{D4Chihatbar}, \bref{elin eblin Olin} and the estimates \bref{blueshift estimate xlin backwards scattering 1}, \bref{backwards ingoing overall estimate on xblin}, \bref{blueshift estimate ablin near H+ n times} mentioned above.
\end{proof}

\begin{corollary}\label{rp elin eblin backwards scattering}
 Let $\mathfrak{S}$, $u_+$, $v_+$ be as in \Cref{grand proposition backwards scattering}, and let $\etaup^{(n)}:=\left(\frac{r^2}{\Omega^2}\nablav\right)^nr\elin_{\ell\geq2}$ $\underline{\etaup}^{(n)}:=\left(\frac{r^2}{\Omega^2}\nablav\right)^nr^2\eblin_{\ell\geq2}$. For $v_\infty$ such that $r(u_+,v_\infty)>3M$ we have the following:
 \begin{align}\label{rp estimate elin backwards scattering}
 \begin{split}
     \int_{\mathscr{C}_u\cap\{\bar{v}\geq v_\infty\}}d\bar{v}\dw\,\frac{r^2}{\Omega^2}|\nablav \etaup^{(n)}|^2\lesssim \text{ Right hand side of } \bref{backwards rp estimate Alin n} \text{ and }\bref{backwards rp estimates Xblin n},
\end{split}
 \end{align}
  \begin{align}\label{rp estimate eblin backwards scattering}
 \begin{split}
     \int_{\mathscr{C}_u\cap\{\bar{v}\geq v_\infty\}}d\bar{v}\dw\,\frac{r^2}{\Omega^2}|\nablav \underline{\etaup}^{(n)}|^2\lesssim \text{ Right hand side of } \bref{backwards rp estimate Alin n} \text{ and }\bref{backwards rp estimates Xblin n}
\end{split}
 \end{align}
for all $n\in\mathbb{N}$. We may replace either of $\etaup^{(n)}$ or $\underline{\etaup}^{(n)}$ with $\left(\frac{r^2}{\Omega^2}\nablav\right)^n\Olino$ above.
\end{corollary}

\subsubsection*{Boundedness on $\otx$, $\otxb$, $\tr\glin$, $\bmlin$}

\begin{corollary}
 The results of \Cref{rp elin eblin backwards scattering} apply to the $\ell\geq2$ components of $\otx$ and $\otxb$ replacing $r\elin$ or $r^2\eblin$ with $r^2\otx$ or $r\otxb$. 
\end{corollary}
\begin{proof}
    Use the linearised Codazzi equations \bref{elliptic equation 1}, \bref{elliptic equation 2} and apply \Cref{rp elin eblin backwards scattering}, \Cref{backwards rp Bblin n} and \Cref{backwards rp alin n}.
\end{proof}

Similarly,

\begin{corollary}
 The results of \Cref{elin estimate near H+ backwards scattering} apply to the $\ell\geq2$ components of $\otx$ and $\otxb$ replacing $\elin$ or $\eblin$ with $\otx$ or $\Omega^{-1}\otxb$. 
\end{corollary}

As for $\tr\glin$, applying the above results and using the linearised Gauss equation \bref{Gauss} we get

\begin{corollary}
 The results of \Cref{elin estimate near H+ backwards scattering} apply to the $\ell\geq2$ component of $\tr\glin$ replacing $\elin$ or $\eblin$ with $\tr\glin$.
\end{corollary}

\begin{corollary}
 The results of \Cref{elin estimate near H+ backwards scattering} apply to the $\ell\geq2$ component of $\tr\glin$ replacing $\elin$ or $\eblin$ with $r\tr\glin$. 
\end{corollary}

\subsubsection{Asymptotic flatness of induced data on ${\Sigma^*_+}$}

With the results of this section we can finally conclude

\begin{corollary}\label{backwards scattering asymptotic flatness of induced data on Sigma*}
 The solution $\mathfrak{S}$ constructed in \Cref{grand proposition backwards scattering} induced an initial data set on ${\Sigma^*_+}$ which is asymptotically flat to order $(1,\infty)$ in the sense of \Cref{def of asymptotic flatness at spacelike infinity}. When the assumptions of \Cref{grand proposition backwards scattering addendum} are satisfied, the data set induced on $\Sigma^*_+$ decays to order $(2,\infty)$.
\end{corollary}
\begin{proof}
    \Cref{asymptotic flatness Psilin n,,asymptotic flatness at spacelike infinity for ablin,,asymptotic flatness at spacelike infinity for alin,,asymptotic flatness at spacelike infinity for xblin,,asymptotic flatness at spacelike infinity for xlin,,asymptotic flatness at spacelike infinity for bblin,,asymptotic flatness at spacelike infinity for blin,,asymptotic flatness at spacelike infinity for glinh} give the result for $\alin$, $\ablin$, $\blin$, $\bblin$, $\xlin$, $\xblin$, $\glinh$. The result for the remaining components now follow immediately from the equations \fullsystem.
\end{proof}

We can now complete the proof of \ref{forward scattering full system thm} and \ref{backwards scattering full system thm}:
\begin{proof}[Proof of \ref{forward scattering full system thm} and \ref{backwards scattering full system thm}]
    The ${\Sigma}^*_+$ gauge condition ensures that $\mathcal{E}^{T,RW}_{{\Sigma^*_+}}$ and $\mathcal{E}^{T}_{{\Sigma^*_+}}$ are isomorphic, while the gauge used to construct solutions in \Cref{grand proposition backwards scattering} ensures that $\mathcal{E}^{T,RW}_{\mathscr{H}^+_{\geq0}}\oplus\mathcal{E}^{T,RW}_{\mathscr{I}^+}$ and $\mathcal{E}^{T}_{\mathscr{H}^+_{\geq0}}\oplus\mathcal{E}^{T}_{\mathscr{I}^+}$ are isomorphic. 
\end{proof}


\subsection{Scattering from data on $\overline{\mathscr{H}^+}$ and $\mathscr{I}^+$}\label{Section 13 11 2022}

We collect here the corresponding results for backwards scattering from smooth, compactly supported data on $\overline{\mathscr{H}^+}$, $\mathscr{I}^+$, all of which follow immediately from the results of \Cref{Section 6 Backwards scattering} obtained so far.

\begin{proposition}
    Let $\xblins_{\mathscr{I}^+}\in\Gamma^{(2)}_c(\mathscr{I}^+)$ and $\xlins_{\mathscr{H}^+}$ be such that $V^{-1}\xlins_{\mathscr{H}^+}\in\Gamma^{(2)}_c(\overline{\mathscr{H}^+})$. There exists a unique solution $\SscriHf$ to \fullsystemK which is $\mathscr{I}^+$ and $\overline{\mathscr{H}^+}$-normalised such that
    \begin{align}
        \lim_{v\longrightarrow\infty}r\xblin(u,v,\theta^A)=\xblins_{\mathscr{I}^+}(u,\theta^A),\qquad \Omega\xlin|_{\overline{\mathscr{H}^+}}=\xlins_{\mathscr{H}^+}.
    \end{align}
    The solution $\SscriHf$ is smooth and it induces Cauchy data on $\overline{\Sigma}$ which is weakly asymptotically flat towards $i^0$.
\end{proposition}

\begin{proof}
    Given $\xlins_{\mathscr{H}^+}$ and $\xblins_{\mathscr{I}^+}$, we may construct data for the full system \fullsystemK on $\overline{\mathscr{H}^+}$ using the same steps followed in the proof of \Cref{grand proposition backwards scattering}. We then apply \Cref{grand proposition backwards scattering} to the data $(\xblins_{\mathscr{I}^+},\xlins_{\mathscr{H}^+}|_{\mathscr{H}^+_{\geq0}})$ to obtain $\SscriHf$ on $J^+(\Sigma^*_+)$. We now apply \Cref{EinsteinWP Sigmastar H+ past} to construct $\SscriHf$ on all of $D^+(\overline{\Sigma})$. It is clear that $\SscriHf$ satisfies the conclusions of \Cref{backwards scattering asymptotic flatness of induced data on Sigma*}. Note that we may set $\divo^2\xblinK=0$ at $\mathcal{B}$ in an identical manner as in the proof of \Cref{grand proposition backwards scattering} (setting the integral to start from $-\infty$ in \eqref{13 11 2022}).
\end{proof}

\subsection{An estimate on the $\Sigma^*_+$, $\Sigma^*_-$ and $\overline{\Sigma}^\pm$ gauges}\label{an estimate on initial gauge from backwards scattering}

Recall that we have shown in \Cref{realising Sigmastar gauge future} how to pass to the $\Sigma^*_+$ for a smooth initial data set. We now show in this section how to estimate the gauge transformation of \Cref{realising Sigmastar gauge future}.

\subsubsection[Estimating $\protect\xlin$ and $\protect\alin$ on $\protect\Sigma^*_+$, $\protect\overline{\Sigma}$]{Estimating $\protect\xlin$ and  $\protect\alin$ on $\Sigma^*_+$, $\overline{\Sigma}$}

Let $\fbar(v,\theta^A)$ be the gauge generator appearing in \Cref{realising Sigmastar gauge future}. Using the coordinate system $(r,\theta^A)$ with $\partial_r=-\frac{2-\Omega^2}{2\Omega^2}\partial_{u^*}$, integrating the squares of both sides of \bref{instinctive dismissal} and using a Poincar\'e inequality, we get
\begin{align}\label{23 11 2021}
\begin{split}
    &\int_{\Sigma^*_+}dr\dw\,|\partial_r^2 \Omega^2\fbar|^2+5|\partial_r\Omega^2\fbar|^2+4|\Omega^2\fbar|^2\lesssim \int_{\Sigma^*_+\cap\mathscr{H}^+}\dw\,5|\partial_r\Omega^2\fbar|^2+8|\Omega^2\fbar|^2\\[10pt]&+\int_{\Sigma^*_+}dr\dw\,|\partial_r^2 \divo^2\Omega\xlin|^2+5|\partial_r\divo^2\Omega\xlin|^2+4|\divo^2\Omega\xlin|^2
   +\int_{\Sigma^*_+\cap\mathscr{H}^+}\dw\,5|\partial_r\divo^2\Omega\xlin|^2+8|\divo^2\Omega\xlin|^2\\[10pt]
   &\lesssim \int_{\Sigma^*_+\cap\mathscr{H}^+}\dw\,5|\partial_r\Omega^2\fbar|^2+8|\Omega^2\fbar|^2+\int_{\Sigma^*_+}dr\dw\,|\partial_r^2 \divo^2\Omega\xlin|^2+5|\partial_r\divo^2\Omega\xlin|^2+4|\divo^2\Omega\xlin|^2,
\end{split}
\end{align}
where the final estimate is derived using Morrey's inequality. Note that in the case of the solution $\mathfrak{S}$ of \Cref{grand proposition backwards scattering}, the conditions \eqref{initial horizon gauge condition} are already satisfied, thus $\Omega^2\underline{f}|_{\Sigma^*_+\cap\mathscr{H}^+_{\geq0}}$, $\partial_r\Omega^2\underline{f}|_{\Sigma^*_+\cap\mathscr{H}^+_{\geq0}}$ vanish. An $H^2_rL^2_{S^2}(\Sigma^*_+)$ estimate of $\Omega\xlin|_{\Sigma^*_+}$ thus implies an estimate on $\Omega^2\fbar$ via \bref{23 11 2021} above. We will estimate $r^{\frac{3}{2}-\epsilon}\Omega\xlin$ in $H^2_r(\Sigma^*_+)$, which will lead to the the required estimate on $\Omega^2\fbar$ as well as give an estimate on the quantity that appears on the right hand side of \bref{ILED estimate on xlin} in terms of scattering data on $\mathscr{I}^+$ and $\mathscr{H}^+_{\geq0}$ or $\overline{\mathscr{H}^+}$. This will be an ingredient for an estimate on gauge solutions that appear in scattering between $\overline{\mathscr{H}^-}$, $\mathscr{I}^-$ and $\overline{\mathscr{H}^+}$, $\mathscr{I}^+$.

Let $u^+$, $v^+$ be the future cutoffs of $\xlins_{\mathscr{H}^+}$, $\xblins_{\mathscr{I}^+}$ respectively, and let $u_-$ be the past cutoff on $\xblins_{\mathscr{I}^+}$. Assume without loss of generality that $r(u_+,v_+)=3M$, and take $r_-:=r(u_+,\vsigmap{u_+})$,  $r_+:=r(\usigmap{v_+},v_+), r(u_-,v_+)$ so that $2M<r_-<3M$, $r_+>3M$ and $\usigmap{r_+}\leq u_-$. The estimate is obtained  on each component of the partition of $\Sigma^*_+$ as follows:

\begin{itemize}
    \item For $2M\leq r\leq r_-$ we can repeat the arguments leading to \Cref{bounded r estimate on xlin backwards scattering}, \Cref{redshift estimate on xlin backwards scattering}, and the estimate \bref{blueshift estimate xlin backwards scattering 1} so that they apply to $\Sigma^*_+$ instead of an ingoing null hypersurface.
    
    \item For $r_- \leq r \leq r_+$ it clearly suffices to estimate $\xlin$ using the argument leading to \Cref{bounded r estimate on xlin backwards scattering}, $\Omega^{-1}\nablagml r^2\Omega\xlin$ using \Cref{redshift estimate on xlin backwards scattering}, $\alin$ using the argument leading to the estimate \bref{backwards ILED estimate alin}, and $\plin$ using the argument leading to the estimate \bref{backwards ILED estimate plin}, since the estimate for $\Omega^{-1}\nablagml r^2\Omega\xlin$ can be commuted by $\partial_t$ to recover the required derivative.
    
    \item For $r\geq r_+$ we apply \Cref{asymptotic flatness at spacelike infinity for xlin}.
\end{itemize}

We begin with the estimate in the region $\Sigma^*_+\cap\{r\geq r_+\}$:

\begin{proposition}\label{gauge estimate xlin near i0}
    For the solution $\mathfrak{S}$ of \Cref{grand proposition backwards scattering}, we have
    \begin{align}
    \begin{split}
        \int_{\Sigma^*_+\cap\{r\geq r_+\}}dr\dw\,|r\xlin|^2\lesssim\sum_{|\gamma|\leq 1}\left\|\mathring{\slashednabla}^\gamma\xblins_{\mathscr{I}^+}\right\|^2_{\mathcal{E}^T_{\mathscr{I}^+}}.
    \end{split}
    \end{align}
\end{proposition}

\begin{proof}
    For $u,v$ such that $(u,v,\theta^A)\in \Sigma^*_+$ and $r(u,v)\geq r_+$, we apply \Cref{asymptotic flatness at spacelike infinity for xlin} and estimate
    \begin{align}
        \begin{split}
            \int_{S^2}\dw\, |r\xlin(u,v)|^2\lesssim& \frac{1}{|u|^2}\int_{S^2}\dw\,|\xlins_{\mathscr{I}^+}|^2+\frac{1}{|u|^6}\int_{S^2}\dw\,|\alins_{\mathscr{I}^+}|^2\\&+\frac{1}{|u|^7}\int_{\mathscr{I}^+\cap\{\bar{u}\geq u\}}d\bar{u}\dw\,|\mathring{\slashednabla}\alins_{\mathscr{I}^+}|^2+|\alins_{\mathscr{I}^+}|^2.
        \end{split}
    \end{align}
    We integrate the above in $u$ towards $-\infty$, and apply Hardy's inequality to the right hand side noting that
    \begin{align}
        \partial_u\plins_{\mathscr{I}^+}=\partial_u^2\alins_{\mathscr{I}^+}=\upPsilin_{\mathscr{I}^+}. 
    \end{align}
\end{proof}

\begin{proposition}\label{gauge estimate alin plin near i0}
For the solution $\mathfrak{S}$ of \Cref{grand proposition backwards scattering}, we have
\begin{align}
    \begin{split}
        \int_{\overline{\Sigma}\cap\{r\geq r_+\}}dr\dw\, r^{6}|\plin|^2+r^{4}|\alin|^2\lesssim \sum_{|\gamma|\leq 1}\left\|\mathring{\slashednabla}^\gamma\xblins_{\mathscr{I}^+}\right\|^2_{\mathcal{E}^T_{\mathscr{I}^+}}.
    \end{split}
\end{align}
\end{proposition}
\begin{proof}
The argument of \Cref{gauge estimate xlin near i0} can be applied to $\alin$ and $\plin$ using \Cref{asymptotic flatness at spacelike infinity for alin} to get
    \begin{align}
    \begin{split}
        \int_{\overline{\Sigma}\cap\{r\geq r_+\}}dr\dw\, r^{6}|\plin|^2\lesssim\int^{\usigmap{r_+}}_{-\infty}du \int_{S^2} \dw\,\frac{1}{|u|^{4}}|\plins_{\mathscr{I}^+}|^2+\frac{1}{|u|^{6}}\sum_{|\gamma|\leq3}|\mathring{\slashednabla}^\gamma\alins_{\mathscr{I}^+}|^2.
    \end{split}
\end{align}
    \begin{align}
    \begin{split}
        \int_{\overline{\Sigma}\cap\{r\geq r_+\}}dr\dw\, r^{4}|\alin|^2\lesssim \int^{\usigmap{r_+}}_{-\infty}du \int_{S^2}\dw\,\frac{1}{|u|^{6}}\left[|\mathring{\slashednabla}\alins_{\mathscr{I}^+}|^2+|\alins_{\mathscr{I}^+}|^2+|\partial_u\alins_{\mathscr{I}^+}|^2\right].
    \end{split}
\end{align}
We now apply Hardy's inequality to the right hand side.
\end{proof}

\begin{corollary}\label{gauge estimate dr xlin near i0}
    For the solution $\mathfrak{S}$ of \Cref{grand proposition backwards scattering}, we have
    \begin{align}
        \begin{split}
              \int_{\Sigma^*_+\cap\{r\geq r_+\}}dr\dw\,|\slashednabla_r r^2\xlin|^2\lesssim\sum_{|\gamma|\leq 1}\left\|\mathring{\slashednabla}^\gamma\xblins_{\mathscr{I}^+}\right\|^2_{\mathcal{E}^T_{\mathscr{I}^+}}.
        \end{split}
    \end{align}
\end{corollary}
\begin{proof}
    The estimate follows from \eqref{gauge estimate ablin pblin near i0} and the argument of \Cref{gauge estimate xlin near i0} applied to $\partial_t r^2\xlin$.
\end{proof}

\begin{corollary}\label{gauge estimate dr dr xlin near i0}
    For the solution $\mathfrak{S}$ of \Cref{grand proposition backwards scattering}, we have
    \begin{align}
        \begin{split}
              \int_{\Sigma^*_+\cap\{r\geq r_+\}}dr\dw\,|\slashednabla_r^2 r^2\xlin|^2\lesssim\sum_{|\gamma|\leq 3}\left\|\mathring{\slashednabla}^\gamma\xblins_{\mathscr{I}^+}\right\|^2_{\mathcal{E}^T_{\mathscr{I}^+}}.
        \end{split}
    \end{align}
\end{corollary}

We now turn to the estimate near $\mathscr{H}^+_{\geq0}$:

\begin{proposition}\label{gauge estimate xlin near H+}
    For the solution $\mathfrak{S}$ of \Cref{grand proposition backwards scattering}, we have
    \begin{align}
        \int_{\Sigma^*_+\cap\{r\leq r_-\}}dr\dw\, |\Omega\xlin|^2\lesssim \sum_{|\gamma|\leq 2}\left\|\mathring{\slashednabla}^\gamma \xblins_{\mathscr{I}^+}\right\|^2_{\mathcal{E}^T_{\mathscr{I}^+}}.
    \end{align}
\end{proposition}
\begin{proof}
    Using the argument of \Cref{bounded r estimate on xlin backwards scattering}, we derive
    \begin{align}
        \int_{\Sigma^*_+\cap\{r\leq r_-\}}dr\dw\, |\Omega\xlin|^2\lesssim \int_{\mathscr{I}+\cap\{\bar{u}\geq u_-\}}d\bar{u}\dw\,\left[|\xlins_{\mathscr{I}^+}(\bar{u},\theta^A)|^2+|\mathring{\slashednabla}{\mathtt{A}^{(3)}}_{\mathscr{I}^+}|^2+|{\mathtt{A}^{(3)}}_{\mathscr{I}^+}|^2+|\alins_{\mathscr{I}^+}|^2\right],
    \end{align}
    where $u$ above is such that $r(u,v)=r_-$. We conclude by applying Hardy's inequality to the right hand side.
\end{proof}

\begin{proposition}
     For the solution $\mathfrak{S}$ of \Cref{grand proposition backwards scattering}, we have
     \begin{align}
         \begin{split}
             \int_{\Sigma^*_+\cap\{r\leq r_-\}}dr\dw\, |\slashednabla_r\Omega\xlin|^2\lesssim \left\|\xlins_{\mathscr{H}^+}\right\|^2_{\mathcal{E}^T_{\mathscr{H}^+_{\geq0}}}+\sum_{|\gamma|\leq 2, i\in\{0,1\}}\left\|\partial_u^i\mathring{\slashednabla}^\gamma \xblins_{\mathscr{I}^+}\right\|^2_{\mathcal{E}^T_{\mathscr{I}^+}}.
         \end{split}
     \end{align}
\end{proposition}
\begin{proof}
    We derive using the argument of \Cref{redshift estimate on xlin backwards scattering} the following estimate
    \begin{align}
        \begin{split}
            \int_{\Sigma^*_+\cap\{r\leq r_-\}}dr\dw\, \left|\frac{1}{\Omega}\nablagml\Omega\xlin\right|^2\lesssim& \Big\{\int_{\mathscr{H}^+_{\geq0}}d\bar{v}\dw\,|\partial_v \upPsilin_{\mathscr{H}^+}|^2+|\plins_{\mathscr{H}^+}|^2+|\alins_{\mathscr{H}^+}|^2+|\xlins_{\mathscr{H}^+}|^2\\&+\int_{\mathscr{I}^+\cap\{\bar{u}\geq u_-\}}d\bar{u}\dw\,|\partial_u \upPsilin_{\mathscr{I}^+}|^2+|\mathring{\slashednabla}\mathtt{A}^{(3)}_{\mathscr{I}^+}|^2+|\mathtt{A}^{(2)}_{\mathscr{I}^+}|^2+|\alins_{\mathscr{I}^+}|^2+|\partial_u\xlins_{\mathscr{I}^+}|^2\Big\},
        \end{split}
    \end{align}
    where $u,v$ are such that $r(u,v)=r_-$. Similarly, we get from \Cref{backwards ingoing overall on alin},
    \begin{align}
        \begin{split}
             \int_{\Sigma^*_+\cap\{r\leq r_-\}}dr\dw\,|\Omega^2\alin|^2\lesssim \left\|\xlins_{\mathscr{H}^+}\right\|^2_{\mathcal{E}^T_{\mathscr{H}^+_{\geq0}}}+\left\|\xblins_{\mathscr{I}^+}\right\|^2_{\mathcal{E}^T_{\mathscr{I}^+}}.
        \end{split}
    \end{align}
\end{proof}

Estimating $\slashednabla_r \Omega\xlin$ near $\mathscr{H}^+_{\geq0}$ incurs an exponentially growing factor in $v_+$n due to the blueshift suffered by $(\Omega^{-1}\nablagml)^2\Omega\xlin$ in \eqref{formula for transverse derivative of xlin near H+} and \eqref{blueshift estimate xlin backwards scattering 1}:

\begin{proposition}
    For the solution $\mathfrak{S}$ constructed in \Cref{grand proposition backwards scattering}, we have
    \begin{align}
        \begin{split}
            \int_{\Sigma^*_+\cap\{r\leq r_-\}}dr\dw\, |\slashednabla_r^2 \Omega\xlin|^2\lesssim e^{\frac{4}{M}v_+}\times \Bigg\{\sum_{i=0,1}\left\|\partial_v^i\xlins_{\mathscr{H}^+}\right\|^2_{\mathcal{E}^T_{\mathscr{H}^+_{\geq0}}}+\sum_{|\gamma|\leq2, i\in\{0,1\}}\left\|\mathring{\slashednabla}^\gamma \partial_u^i\xblins_{\mathscr{I}^+}\right\|^2_{\mathcal{E}^T_{\mathscr{I}^+}}\Bigg\}.
        \end{split}
    \end{align}
\end{proposition}
\begin{proof}
    We apply the estimates of \Cref{gauge estimate xlin near H+}, \Cref{gauge estimate alin plin near i0}, \Cref{gauge estimate dr xlin near i0} and \eqref{blueshift estimate xlin backwards scattering 1} with $n=2$.
\end{proof}


Applying the steps above leads to the following estimate:

\begin{proposition}\label{grand estimate on Sigmastar gauge backwards scattering}
For the solution $\mathfrak{S}$ of \Cref{grand proposition backwards scattering}, the right hand side of \eqref{23 11 2021} is bounded via
\begin{align}
    \begin{split}
   &\int_{\Sigma^*_+}dr\dw\,|\partial_r^2 \divo^2\Omega\xlin|^2+5|\partial_r\divo^2\Omega\xlin|^2+4|\divo^2\Omega\xlin|^2
        \\&\lesssim e^{\frac{4}{M}v_+}\times \Bigg\{\sum_{i=0,1}\left\|\partial_v^i\xlins_{\mathscr{H}^+}\right\|^2_{\mathcal{E}^T_{\mathscr{H}^+_{\geq0}}}+\sum_{|\gamma|\leq2, i\in\{0,1\}}\left\|\mathring{\slashednabla}^\gamma \partial_u^i\xblins_{\mathscr{I}^+}\right\|^2_{\mathcal{E}^T_{\mathscr{I}^+}}\Bigg\}.
    \end{split}
\end{align}
\end{proposition}

The above estimate applies to induced data on $\overline{\Sigma}$, where in applying the argument above we must treat a neighborhood of $\mathcal{B}$ of bounded Kruskal coordinates using the system \fullsystemK. The resulting estimate is as follows:

\begin{proposition}\label{grand estimate on Sigmabar gauge backwards scattering}
For the solution $\mathfrak{S}$ of \Cref{grand proposition backwards scattering}, the right hand side of \bref{ILED estimate on xlin} satisfies
\begin{align}
\begin{split}
    &\int_{{\overline{\Sigma}}\cap\{V>1\}}dr\sin\theta d\theta d\phi|\partial_r^2 \divo^2\Omega\xlin|^2+5|\partial_r\divo^2\Omega\xlin|^2+4|\divo^2\Omega\xlin|^2
    \\[10pt]&+\int_{{\overline{\Sigma}}\cap\{V\leq1\}}dr\sin\theta d\theta d\phi\Bigg[\left|\slashednabla_r\left(\slashednabla_r(U\xlin)\right)\right|^2+\left|\slashednabla_r(U\xlin)\right|^2+|U\xlin|^2\Bigg]
    \\&\lesssim   e^{\frac{4}{M}v_+}\times \Bigg\{\sum_{i=0,1}\left\|\partial_v^i\xlins_{\mathscr{H}^+}\right\|^2_{\mathcal{E}^T_{\mathscr{H}^+_{\geq0}}}+\sum_{|\gamma|\leq2, i\in\{0,1\}}\left\|\mathring{\slashednabla}^\gamma \partial_u^i\xblins_{\mathscr{I}^+}\right\|^2_{\mathcal{E}^T_{\mathscr{I}^+}}
    \\&+\int_{\overline{\mathscr{H}^+}\cap\{V\leq 1\}}d\bar{v}\sin\theta d\theta d\phi\, \left[|V^{-1}\xlins_{\mathscr{H}^+}|^2+|V^{-2}\alins_{\mathscr{H}^+}|^2+|V^{-1}\plins_{\mathscr{H}^+}|^2+|\partial_v\upPsilin_{\mathscr{H}^+}|^2+\sum_{|\gamma|\leq 5}|\mathring{\slashednabla}^\gamma{\upPsilin}_{\mathscr{H}^+}|^2\right]\Bigg\}.
\end{split}
\end{align}
\end{proposition}

Finally, We turn to $\alin$ and $\plin$ near $\mathcal{B}$.

\begin{corollary}
    Let $\mathfrak{S}$ be a solution to the system \fullsystemK arising from scattering data $\xblins_{\mathscr{I}^+}$, $\xlins_{\mathscr{H}^+}$ such that $\xblins_{\mathscr{I}^+}\in \Gamma_c^{(2)}(\mathscr{I}^+)$ and $V^{-1}\xlins_{\mathscr{H}^+}\in \Gamma_c^{(2)}(\overline{\mathscr{H}^+})$. Then we have
    \begin{align}
        \begin{split}
            \int_{\overline{\Sigma}\cap\{V\leq1\}}dr\dw\, |\slashednabla_U V^{-2}\Omega^2\alin|^2+|V^{-2}\Omega^2\alin|^2\lesssim&  \int_{\overline{\mathscr{H}^+}\cap\{V\leq 1\}}|\slashednabla_U V^{-2}\alins_{\mathscr{H}^+}|^2+|V^{-2}\alins_{\mathscr{H}^+}|^2+|\upPsilin_{\mathscr{H}^+}|^2+|\mathring{\slashednabla}\upPsilin_{\mathscr{H}^+}|^2\\&+e^{\frac{1}{2M}v_+}\sum_{|\gamma|\leq 1}\left[\left\|\mathring{\slashednabla}^\gamma\xblins_{\mathscr{I}^+}\right\|^2_{\mathcal{E}^T_{\mathscr{I}^+}}+\left\|\mathring{\slashednabla}^\gamma\xlins_{\mathscr{H}^+}\right\|^2_{\mathcal{E}^T_{\overline{\mathscr{H}^+}}}\right].
        \end{split}
    \end{align}
\end{corollary}
\begin{proof}
    Using the relations \eqref{Kruskal hier+}, we may derive the following estimate in the region $J^+(\overline{\Sigma})\cap\{V\leq 1\}$ using Gr\"onwall's inequality:
    \begin{align}
    \begin{split}
        \int_{\overline{\Sigma}\cap\{V\leq1\}}dr\dw\, |\slashednabla_U V^{-2}\Omega^2\alin|^2+|V^{-2}\Omega^2\alin|^2\lesssim& \int_{\overline{\mathscr{H}^+}\cap\{V\leq 1\}}|\slashednabla_U V^{-2}\alins_{\mathscr{H}^+}|^2+|V^{-2}\alins_{\mathscr{H}^+}|^2+|\upPsilin_{\mathscr{H}^+}|^2+|\mathring{\slashednabla}\upPsilin_{\mathscr{H}^+}|^2\\
        &+\int_{\{V=1\}\cap J^+(\overline{\Sigma})}dr\dw\, |\Omega^{-1}\nablagml\Psilin|^2.
    \end{split}
    \end{align}
    We now conclude by appealing to \eqref{RW exponential backwards near H+}.
\end{proof}

\subsubsection[Estimating $\protect\xblin$ and $\protect\ablin$ on $\protect\Sigma^*_+$, $\protect\overline{\Sigma}$]{Estimating $\protect\xblin$ and $\protect\ablin$ on $\Sigma^*_+$, $\overline{\Sigma}$}

In an entirely analogous manner, we can establish a similar estimate on $\xblin$ at $\overline{\Sigma}$ as follows:
\begin{itemize}
    \item for $2M\leq r\leq r_-$ We can make use of the second equation of \bref{D4Chihat Kruskal} and its $\slashednabla_V$-commuted versions and apply Gr\"onwall's inequality. We can then use \Cref{backwards ingoing overall  on xblin} and the estimate \bref{blueshift estimate ablin near H+ n times},
    
    \item for $r_- \leq r \leq r_+$ it clearly suffices to estimate $\xblin$ using the argument leading to \Cref{backwards ingoing overall  on xblin}, $\nabladlt\xblin$ and $\partial_t\nabladlt\xblin$ using the estimates leading to \Cref{backwards ingoing nablav commuted overall xblin}, $\ablin$ and $\pblin$ using the estimates listed in  \Cref{backwards integral of ablin pblin on Sigma*},
    
    \item and finally, for $r\geq r_+$ we apply \Cref{asymptotic flatness at spacelike infinity for xblin}.
\end{itemize}

We will prove the estimate on $2M\leq r\leq r_-$ in detail:

\begin{proposition}\label{xblin gauge on Sigmastar near H+}
For the solution $\mathfrak{S}$ constructed in \Cref{grand proposition backwards scattering}, we have
    \begin{align}
        \begin{split}
            \int_{\Sigma^*_+\cap\{r\leq r_-\}}dr\dw\,|\Omega^{-1}\xblin|^2\lesssim &\left\|\xblins_{\mathscr{I}^+}\right\|^2_{\mathcal{E}^T_{\mathscr{I}^+}}+\left\|\xlins_{\mathscr{H}^+}\right\|^2_{\mathcal{E}^T_{\mathscr{H}^+_{\geq0}}}+\int_{\mathscr{H}^+_{\geq0}}dv\dw\, |\xblins_{\mathscr{H}^+}|^2\\&+e^{\frac{5}{2M}v_+}\int_{\mathscr{H}^+_{\geq 0}}d\bar{v}\sin\theta d\theta d\phi\,\left[|\mathring{\slashednabla}(2M)^5\ablins_{\mathscr{H}^+}|^2+|(2M)^5\ablins_{\mathscr{H}^+}|^2\right].
        \end{split}
    \end{align}
\end{proposition}
\begin{proof}
We may apply the argument of \Cref{backwards ingoing overall  on xblin} integrating towards $\Sigma^*_+$ instead of $\underline{\mathscr{C}}_v$.
\end{proof}

\begin{proposition}\label{xblin gauge on Sigmabar}
For the solution $\mathfrak{S}$ of \Cref{grand proposition backwards scattering}, we have the following estimate on $V\Omega^{-1}\xblin$ on $\overline{\Sigma}\cap \{V\leq 1\}$:
\begin{align}\label{estimate on xblin on Sigmabar}
\begin{split}
    &\int_{\overline{\Sigma}\cap\{V<1\}}dr\dw\,\left|V\Omega^{-1}\xblin\right|^2+\left|\slashednabla_VV\Omega^{-1}\xblin\right|^2+\left|\slashednabla_V^2V\Omega^{-1}\xblin\right|^2+|V^{-2}\ablin|^2+|\slashednabla_V V^{-2}\ablin|^2\\&
    \lesssim \int_{\overline{\mathscr{H}^+}\cap\{V\leq1\}}d\bar{v}\left[|V\xblins_{\mathscr{H}^+}|^2+\left|\slashednabla_VV\xblins_{\mathscr{H}^+}\right|^2+\left|\slashednabla_V^2V\xblins_{\mathscr{H}^+}\right|^2+|\mathring{\slashednabla}\upPsilinb_{\mathscr{H}^+}|^2+|V^{-2}\ablin|^2+|\slashednabla_V V^{-2}\ablin|^2\right]\\&+\int_{\{V=1,U\leq 0\}\cap J^+(\overline{\Sigma})}dr\dw\,\left[\Omega^{-4}|\ablin|^2+\Omega^{-2}|\pblin|^2+|\Omega^{-1}\nablagml \Psilinb|^2\right].
\end{split}
\end{align}
\end{proposition}
\begin{proof}
    The second equation of \bref{D4Chihat Kruskal} implies
    \begin{align}
        \slashednabla_U \left|\frac{r^2}{2M}V\Omega^{-1}\xblin\right|^2=2\frac{r^2}{\fr}V^2\Omega^{-2}\ablin \times \frac{r^2}{2M}V{\Omega^{-1}}\xblin.
    \end{align}
    For $0 \leq V\leq1$ Integrate the above on $[U,0]\times S^2$ for $S^2_{U,V}\in J^+(\overline{\Sigma})$ to get
    \begin{align}
    \begin{split}    
        &\int_{S^2_{U,V}}\dw\,\left|\frac{r^2}{2M}V\Omega^{-1}\xblin\right|^2\lesssim \int_{S^2_{0,V}}\dw\,\left|(2M)V\xblins_{\mathscr{H}^+}\right|^2
        \\&+\int_{U}^0\int_{S^2}d\overline{U}\dw\,\left[\left|\frac{r^2}{2M}V\Omega^{-1}\xblin(\overline{U},V,\theta^A)\right|^2+\left|\frac{r^2}{\fr}V^2\Omega^{-2}\ablin(\overline{U},V,\theta^A)|^2\right|^2\right].
    \end{split}
    \end{align}
    Gr\"onwall's inequality implies (note that $0\leq |U|\leq, 1\leq V\leq1$):
    \begin{align}
    \begin{split}
        \int_{S^2_{U,V}}\dw\,\left|\frac{r^2}{2M}V\Omega^{-1}\xblin\right|^2\lesssim& \int_{S^2_{0,V}}\dw\,\left|(2M)V\xblins_{\mathscr{H}^+}\right|^2\\&+\int_{U}^0\int_{S^2}d\overline{U}\dw\,\left|\frac{r^2}{2M}V\Omega^{-1}\xblin(\overline{U},V,\theta^A)\right|^2.
    \end{split}
    \end{align}    
    For each $V\in[0,1]$, take $U=-V$ in the above and integrate over $V\in[0,1]$ to get
    \begin{align}\label{3 12 2021}
    \begin{split}
        \int_{\overline{\Sigma}\cap\{V\leq1\}}dr\dw\,\left|\frac{r^2}{2M}V\Omega^{-1}\xblin\right|^2\lesssim& \int_{\overline{\mathscr{H}^+}\cap\{V\leq 1\}}dV\dw\,\left|(2M)V\xblins_{\mathscr{H}^+}\right|^2\\&+\int_{J^+(\overline{\Sigma})\cap\{V\leq1\}}d{U}dV\dw\,\left|\frac{r^2}{\fr}V^2\Omega^{-2}\ablin({U},V,\theta^A)|^2\right|^2.
    \end{split}
    \end{align}   
    We now estimate the bulk term on the right hand side in \bref{3 12 2021}. The equation relating $\ablin$ and $\pblin$ in \bref{hier} can be written in terms of the quantity $V\Omega^{-1}\pblin$, which is regular on $\overline{\mathscr{H}^+}$, as
    \begin{align}\label{4 12 2021}
    V\slashednabla_V rV^2\Omega^{-2}\ablin+2\left(1-\frac{4M^2}{r^2}\right)V^2r\Omega^{-2}\ablin=4MV rV\Omega^{-1}\pblin.
    \end{align}
    Multiply by $\slashednabla_V rV^2\Omega^{-2}\ablin$ and derive
    \begin{align}
        \slashednabla_V V|rV^2\Omega^{-2}\ablin|^2+2\left(\frac{8M^2}{r^2}-3\right)|V^2r\Omega^{-2}\ablin|^2=8MV rV\Omega^{-1}\pblin\times V^2r\Omega^{-2}\ablin.
    \end{align}
    Integrate over $\bigtriangleup=J^+(\overline{\Sigma})\cap\{V\leq1\}$ and use Cauchy--Schwarz to get
    \begin{align}
    \begin{split}
        &\int_{\overline{\Sigma}\cap\{V\leq 1\}}dr\dw\,r^2V^5\Omega^{-4}|\ablin|^2+\int_{\bigtriangleup}dUdV\dw\,|V^2r\Omega^{-2}\ablin|^2\\&\lesssim_M \int_{\{V=1,U\leq0\}\cap J^+(\overline{\Sigma})}dr\dw\, \Omega^{-4}|\ablin|^2+\int_{\bigtriangleup}dUdV\dw\, V^4\Omega^{-2}|\pblin|^2.
    \end{split}
    \end{align}
    We can also estimate the integral of $|\slashednabla_V V^2\Omega^{-2}\ablin|^2$ in $\bigtriangleup$ by squaring \bref{4 12 2021} and integrating over $\mathscr{D}$ to get
    \begin{align}
    \begin{split}
        &\int_{\bigtriangleup}dUdV\dw\,|\slashednabla_VV^2r\Omega^{-2}\ablin|^2+\frac{2U^2}{\fr^2}\left(\frac{4M}{r}+1\right)|V^2r\Omega^{-2}\ablin|^2\\[4pt]
        &+\int_{\{V=1\}\cap J^+(\overline{\Sigma})}dU\dw\, \frac{2U}{\fr}\left(\frac{2M}{r}+1\right)|V^2r\Omega^{-2}\ablin|^2-\int_{\overline{\Sigma}\cap\{V\leq1\}}dV\dw\,\frac{2U}{\fr}\left(\frac{2M}{r}+1\right)|V^2r\Omega^{-2}\ablin|^2\\[4pt]&=\int_{\bigtriangleup}dUdV\dw\,\frac{4M^2}{r^4}|r^3V\Omega^{-1}\pblin|^2.
    \end{split}
    \end{align}
    Remembering that $V\geq |U|$ on $J^+(\overline{\Sigma})$, we obtain
    \begin{align}\label{4 12 2021 33}
        \begin{split}
            &\int_{\bigtriangleup}dUdV\dw\,|\slashednabla_VV^2r\Omega^{-2}\ablin|^2+\int_{\overline{\Sigma}\cap\{V\leq 1\}}dr\dw\,\frac{2V}{\fr}\left(\frac{2M}{r}+1\right)|V^2r\Omega^{-2}\ablin|^2\\[4pt]&\lesssim \int_{\bigtriangleup}dUdV\dw\,\frac{2V^2}{f^2}\left(\frac{4M}{r}+1\right)|V^2r\Omega^{-2}\ablin|^2+\frac{4M^2}{r^4}|r^3V\Omega^{-1}\pblin|^2\\[6pt]&+\int_{\{V=1\}\cap J^+(\overline{\Sigma})}dr\dw\,|\Omega^{-2}\ablin|^2.
        \end{split}
    \end{align}
    In turn, we estimate $V\Omega^{-1}\pblin$ using \bref{hier} treating $\Psilinb$ as a source:
    \begin{align}
    \begin{split}
        &\int_{\overline{\Sigma}\cap\{V\leq 1\}}dr\dw\,r^6V^5\Omega^{-1}|\pblin|^2+4\int_{\mathscr{D}}dUdV\dw\,V^4r^6|\Omega^{-1}\pblin|^2\\[4pt]&\lesssim_M \int_{\{V=1,U\leq0\}\cap J^+(\overline{\Sigma})}dr\dw\, \Omega^{-2}|\pblin|^2+\int_{\mathscr{D}}dUdV \frac{4M}{r^2}V^4|\Psilinb|^2.
    \end{split}
    \end{align}
    We can estimate the $L^2$ integral of $\slashednabla_V V\Omega^{-1}\pblin$ on the $\bigtriangleup$ applying the same steps leading to \bref{4 12 2021 33} to get
       \begin{align}\label{4 12 2021 333}
        \begin{split}
            &\int_{\bigtriangleup}dUdV\dw\,|\slashednabla_VVr^3\Omega^{-1}\pblin|^2+\int_{\overline{\Sigma}\cap\{V\leq 1\}}dr\dw\,\frac{V}{\fr}\left(\frac{2M}{r}+1\right)|Vr^3\Omega^{-1}\pblin|^2\\[4pt]&\lesssim \int_{\bigtriangleup}dUdV\dw\,\frac{V^2}{\fr^2}\frac{4M}{r^2}|Vr\Omega^{-1}\pblin|^2+\frac{4M^2}{r^4}|\Psilinb|^2+\int_{\{V=1\}\cap J^+(\overline{\Sigma})}dr\dw\,|\Omega^{-1}\pblin|^2.
        \end{split}
    \end{align}
    To estimate the bulk term in $\Psilinb$ above, we turn to the Regge--Wheeler equation in Kruskal coordinates \bref{RW in Kruskal}: Multiply \bref{RW in Kruskal} by $\slashednabla_U\Psilinb$ and integrate by parts over $\bigtriangleup$ to get
    \begin{align}\label{RW in Kruskal blueshift}
    \begin{split}
        &\int_{\bigtriangleup}dUdV\dw\, \frac{V}{(rf)^2}\left[\left(\frac{6M}{r}+1\right)|\mathring{\slashednabla}\Psilinb|^2+\left((\frac{6M}{r}+1)(3\Omega^2+1)-\frac{12M^2}{r^2}\right)|\Psilinb|^2\right]\\[4pt]&+\int_{\overline{\Sigma}\cap\{V\leq 1\}}dV\dw\, |\slashednabla_U\Psilinb|^2+\frac{1}{r^2f}\left[|\mathring{\slashednabla}\Psilinb|^2+(3\Omega^2+1)|\Psilinb|^2\right]
        \\[4pt]&\leq\int_{\overline{\mathscr{H}^+}\cap\{V\leq 1\}}dV\dw\,\left[|\mathring{\slashednabla}\Psilinb|^2+|\Psilinb|^2\right]+\int_{\{V=1\}\cap J^+(\overline{\Sigma})}dU\dw\,|\slashednabla_U\Psilinb|^2.
    \end{split}
    \end{align}
    To conclude, the procedure above applies to the second equation of \bref{D4Chihat Kruskal} commuted with $\slashednabla_V$, $\slashednabla_V^2$ since the estimates above cover all the new $\bigtriangleup$-integrals of $\ablin$, $\pblin$, $\Psilinb$ that arise.
\end{proof}

Applying the estimates \bref{RW exponential backwards near H+}, \bref{backwards blueshift ablin} and \bref{backwards blueshift pblin} to the right hand side of \bref{estimate on xblin on Sigmabar} of \Cref{xblin gauge on Sigmabar} above, we can now conclude

\begin{corollary}\label{estimate on xblin on Sigmabar proposition}
For the solution $\mathfrak{S}$ of \Cref{grand proposition backwards scattering}, the right hand side of \bref{ILED estimate on xlin} satisfies
\begin{align}\label{estimate on xblin on Sigmabar}
\begin{split}
    &\int_{\overline{\Sigma}\cap\{V<1\}}dr\dw\,\left|V\Omega^{-1}\xblin\right|^2+\left|\slashednabla_VV\Omega^{-1}\xblin\right|^2+\left|\slashednabla_V^2V\Omega^{-1}\xblin\right|^2+|V^{-2}\ablin|^2+|\slashednabla_V V^{-2}\ablin|^2\\&
    \lesssim \int_{\overline{\mathscr{H}^+}\cap\{V\leq1\}}d\bar{v}\left[|V\xblins_{\mathscr{H}^+}|^2+\left|\slashednabla_VV\xblins_{\mathscr{H}^+}\right|^2+\left|\slashednabla_V^2V\xblins_{\mathscr{H}^+}\right|^2+|\mathring{\slashednabla}\upPsilinb_{\mathscr{H}^+}|^2+|V^{2}\ablin|^2+|\slashednabla_V V^{2}\ablin|^2\right]\\&+e^{\frac{1}{2M}v_+}\int_{\overline{\mathscr{H}^+}\cap\{V\geq 1\}}|\mathring{\slashednabla}\upPsilinb_{\mathscr{H}^+}|^2+|\upPsilinb_{\mathscr{H}^+}|^2\,+e^{\frac{4}{M}v_+}\int_{\overline{\mathscr{H}^+}\cap\{V\geq 1\}}|\mathring{\slashednabla}\pblins_{\mathscr{H}^+}|^2+|\pblins_{\mathscr{H}^+}|^2+\sum_{|\gamma|\leq 3}|\mathring{\slashednabla}^\gamma\ablins_{\mathscr{H}^+}|^2.
\end{split}
\end{align}
\end{corollary}

\begin{proposition}\label{gauge estimate xblin in intermediate region}
For any $r_-,r_+$ with $r_->2M$, the left hand side of \bref{ILED estimate on xlin} satisfies
\begin{align}
    \begin{split}
        &\int_{\overline{\Sigma}\cap\{r\in[r_-,r_+]\}}\Omega^2dr\sin\theta d\theta d\phi\Bigg[\left|\frac{1}{\Omega}\nabladlt\left(\frac{1}{\Omega}\nabladlt(r^2\Omega\xblin)\right)\right|^2+\left|\frac{1}{\Omega}\nabladlt(r^2\Omega\xblin)\right|^2+\frac{1}{r^{2}}|r^2\Omega\xblin|^2\Bigg]
        \\[10pt]&\lesssim_{r_-}\int_{\mathscr{H}^+_{\geq0}}d\bar{v}\dw\,|\xblins_{\mathscr{H}^+}|^2+\left\|\xlins_{\mathscr{H}^+}\right\|^2_{\mathcal{E}^T_{\overline{\mathscr{H}^+}}}+\sum_{|\gamma|\leq 2}\left\|\mathring{\slashednabla}^\gamma\xblins_{\mathscr{I}^+}\right\|^2_{\mathcal{E}^T_{\mathscr{I}^+}}\\
        &+\int_{\mathscr{H}^+_{\geq0}}d\bar{v}\dw\,\sum_{|\gamma|\leq3}|\mathring{\slashednabla}^\gamma\ablins_{\mathscr{H}^+}|^2+\sum_{|\gamma|\leq1}|\mathring{\slashednabla}^\gamma\pblins_{\mathscr{H}^+}|^2.
    \end{split}
\end{align}
\end{proposition}

\begin{proposition}\label{07 11 2022 A}
Let $r_+\geq r(u_-,v_+)$, where $u_-,v_+$ are as in \Cref{grand proposition backwards scattering}. The solution $\mathfrak{S}$ of \Cref{grand proposition backwards scattering} satisfies
\begin{align}
    \int_{\overline{\Sigma}\cap\{r\geq r_+\}}dr\dw\, r^2|\xblin|^2\lesssim \sum_{|\gamma|\leq 3}\left\|\mathring{\slashednabla}^\gamma \xblins_{\mathscr{I}^+}\right\|^2_{\mathcal{E}^T_{\mathscr{I}^+}}.
\end{align}
\end{proposition}

\begin{proof}
    We begin by estimating the integral of $|r\xblin|^2$ on ${\Sigma}^*_+\cap\{r\geq r_+\}$. Using \Cref{asymptotic flatness at spacelike infinity for xblin}, we have 
    \begin{align}
    \begin{split}
        \int_{\overline{\Sigma}\cap\{r\geq r_+\}}dr\dw\,|r\Omega^{-1}\xblin|^2\lesssim &\int_{-\infty}^{u_-}du\frac{1}{|u|^{4}}\int_{S^2}\left|\underline{\mathcal{X}}^{(2)}_{\mathscr{I}^+}\right|^2\\
        &+\int_{-\infty}^{u_-}du\frac{1}{|u|^{5}}\Bigg[\int_{\mathscr{I}^+\cap\{\bar{u}\geq u\}}d\bar{u}\dw\,\Big(|\underline{\mathcal{X}}^{(2)}_{\mathscr{I}^+}|^2+|\underline{\mathcal{X}}^{(1)}_{\mathscr{I}^+}|^2+|\mathring{\slashednabla}\underline{\upPhi}_{\mathscr{I}^+}^{(1)}|^2\\&\qquad\qquad\qquad\qquad\qquad+|\mathring{\slashednabla}\upPsilinb_{\mathscr{I}^+}|^2+|\underline{\upPhi}_{\mathscr{I}^+}^{(1)}|^2+|\upPsilinb_{\mathscr{I}^+}|^2\Big)\Bigg].
    \end{split}
    \end{align}
    The right hand side above is integrable since $\partial_u^i\underline{\mathcal{X}}^{(n)}=O(|u|^{n-1-i})$, $\partial_u^i\underline{\upPhi}^{(n)}=O(|u|^{n-i})$ when $\int_{-\infty}^{\infty}d\bar{u}\xblins_{\mathscr{I}^+}\neq0$. Using Hardy's inequality we estimate the right hand side in terms of $\xblins_{\mathscr{I}^+}$: using \eqref{06 11 2022 1}, \eqref{06 11 2022 2} we have
    \begin{align}\label{07 11 2022 Hardy}
         \int_{\mathscr{I}^+}du\dw\,\frac{1}{1+|u|^4}\left|\underline{\mathcal{X}}^{(2)}_{\mathscr{I}^+}\right|^2\lesssim \int_{\mathscr{I}^+}du\dw\,\left|\partial_u^2\underline{\mathcal{X}}^{(2)}_{\mathscr{I}^+}\right|^2=\left\|\xblins_{\mathscr{I}^+}\right\|^2_{\mathcal{E}^T_{\mathscr{I}^+}}.
    \end{align}
    Integration by parts yields
    \begin{align}
        \int_{-\infty}^{u_-}d{u}\frac{1}{|u|^5}\int_{u}^{u_-}d\bar{u}\dw\,\left|\mathcal{X}^{(2)}_{\mathscr{I}^+}\right|^2=\frac{1}{4}\int_{-\infty}^{u_-}du\dw\, \frac{1}{|u|^4}\left|\underline{\mathcal{X}}^{(2)}_{\mathscr{I}^+}\right|^2,
    \end{align}
    as we have that $\underline{\mathcal{X}}^{(2)}_{\mathscr{I}^+}$. Since
    \begin{align}
\partial_u\underline{\Phi}^{(1)}_{\mathscr{I}^+}=\left(\mathcal{A}_2-2\right)\upPsilinb_{\mathscr{I}^+},
    \end{align}
    we conclude by applying the computation \eqref{07 11 2022 Hardy} to the remaining terms.
    \end{proof}
\begin{proposition}\label{07 11 2022 B}
    Let $r_+\geq r(u_-,v_+)$, where $u_-,v_+$ are as in \Cref{grand proposition backwards scattering}. The solution $\mathfrak{S}$ of \Cref{grand proposition backwards scattering} satisfies
    \begin{align}
         \int_{\overline{\Sigma}\cap\{r\geq r_+\}}dr\dw\,|\nablav r^2\Omega^{-1}\xblin|^2\lesssim \| \sum_{|\gamma\leq 3}\left\|\xblins_{\mathscr{I}^+}\right\|^2_{\mathcal{E}^T_{\mathscr{I}^+}}.
    \end{align}
\end{proposition}
    
    \begin{proof}
    To estimate the integral of $|\nablav r^2\xblin|^2$, we estimate $r^2\ablin$ and  $\partial_t r^2\xblin$ in $L^2(\Sigma^*_+\cap\{r\geq r_+\})$, using \Cref{asymptotic flatness at spacelike infinity for ablin} and \Cref{asymptotic flatness at spacelike infinity for xblin} respectively. We then obtain using \eqref{17 10 2021}
    \begin{align}
        \int_{\overline{\Sigma}\cap\{r\geq r_+\}}dr\dw\,|r^2\ablin|^2\lesssim \int_{-\infty}^{u_-}du\dw\, \frac{1}{|u|^4}\left[|\mathring{\slashednabla}\upPsilinb_{\mathscr{I}^+}|^2+|\upPsilinb_{\mathscr{I}^+}|^2\right],
    \end{align}
    while for $\partial_t^2r^2\Omega^{-1}\xblin$ we get using
    \begin{align}
    \begin{split}
        \int_{\Sigma^*_+\cap\{r\geq r_+\}}dr\dw\,|\partial_t r^2\Omega\xblin|^2\lesssim \int_{\mathscr{I}^+}du\dw\,\frac{1}{|u|^2}\Big[&\left|\partial_u\underline{\mathcal{X}}^{(2)}_{\mathscr{I}^+}\right|^2+\left|\partial_u\underline{\mathcal{X}}^{(1)}_{\mathscr{I}^+}\right|^2
\\&+\left|\mathring{\slashednabla}\partial_u\underline{\Phi}^{(1)}\right|^2+\left|\mathring{\slashednabla}\upPsilinb_{\mathscr{I}^+}\right|^2\Big].
    \end{split}
    \end{align}
Applying Hardy's inequality yields the result.
\end{proof}
\begin{proposition}\label{07 11 2022 C}
    Let $r_+\geq r(u_-,v_+)$, where $u_-,v_+$ are as in \Cref{grand proposition backwards scattering}. The solution $\mathfrak{S}$ of \Cref{grand proposition backwards scattering} satisfies
    \begin{align}
         \int_{\overline{\Sigma}\cap\{r\geq r_+\}}dr\dw\,|(\nablav)^2 r^2\Omega^{-1}\xblin|^2\lesssim \sum_{|\gamma\leq 3}\left\|\xblins_{\mathscr{I}^+}\right\|^2_{\mathcal{E}^T_{\mathscr{I}^+}}.
    \end{align}
\end{proposition}

\begin{proof}
    To estimate $(\nablav)^2r^2\Omega^{-1}\xblin$ in $L^2(\Sigma^*_+\cap\{r\geq r_+\})$, we use that
    \begin{align}
        (\nablav)^2r^2\Omega\xblin=2M\frac{\Omega^2}{r^2} r\Omega\xblin+\frac{2M\Omega^2}{r^3}\underline{\mathcal{X}}^{(1)}+\frac{\Omega^4}{r^3}\underline{\mathcal{X}}^{(2)}
    \end{align}
    thus we obtain from \Cref{asymptotic flatness at spacelike infinity for xblin}
    \begin{align}
        \int_{\overline{\Sigma}\cap\{r\geq r_+\}}dr\dw\,\frac{1}{r^6}\left|\underline{\mathcal{X}}^{(1)}\right|^2\lesssim \int_{-\infty}^{u_-}\int_{S^2}du\dw\,\frac{1}{|u|^6}\Big[\left|\underline{\mathcal{X}}^{(1)}_{\mathscr{I}^+}\right|^2+\left|\mathring{\slashednabla}\upPsilinb_{\mathscr{I}^+}\right|^2\Big],
    \end{align}
    \begin{align}
    \begin{split}
        \int_{\overline{\Sigma}\cap\{r\geq r_+\}}dr\dw\,\frac{1}{r^6}\left|\underline{\mathcal{X}}^{(2)}\right|^2\lesssim \int_{-\infty}^{u_-}\int_{S^2}du\dw\,\frac{1}{|u|^6}\Bigg[&\left|\underline{\mathcal{X}}^{(2)}_{\mathscr{I}^+}\right|^2+\left|\underline{\mathcal{X}}^{(1)}_{\mathscr{I}^+}\right|^2+\left|\mathring{\slashednabla}\underline{\Phi}^{(1)}_{\mathscr{I}^+}\right|^2        \\&+\left|\mathring{\slashednabla}\upPsilinb_{\mathscr{I}^+}\right|^2\Bigg].
    \end{split}
    \end{align}
   We conclude by applying Hardy's inequality to the above estimates. 
\end{proof}

\begin{corollary}\label{gauge estimate on xblin ablin pblin near i0}
Let $r_+>r(u_-,v_+)$ where $u_-$, $v_+$ are as in \Cref{grand proposition backwards scattering}. For the solution $\mathfrak{S}$ of \Cref{grand proposition backwards scattering}, the left hand side of \cref{ILED estimate on xlin} satisfies
\begin{align}
    \begin{split}
        &\int_{\overline{\Sigma}\cap\{r\geq r_+\}}dr\sin\theta d\theta d\phi\Bigg[\left|\frac{1}{\Omega}\nabladlt\left(\frac{1}{\Omega}\nabladlt(r^2\Omega\xblin)\right)\right|^2+\left|\frac{1}{\Omega}\nabladlt(r^2\Omega\xblin)\right|^2+\frac{1}{r^{2}}|r^2\Omega\xblin|^2\Bigg]
       \lesssim \sum_{|\gamma|\leq 3}\left\|\mathring{\slashednabla}^\gamma\xblins_{\mathscr{I}^+}\right\|^2_{\mathcal{E}^T_{\mathscr{I}^+}},
    \end{split}
\end{align}

\begin{align}\label{gauge estimate ablin pblin near i0}
    \begin{split}
        \int_{\overline{\Sigma}\cap\{r\geq r_+\}}dr\dw\, r^{4+\epsilon}|\ablin|^2+r^{6+\epsilon}|\pblin|^2\lesssim \int_{\mathscr{I}^+}du\dw\,\frac{1}{1+|u|^{2-\epsilon}}\left[|\mathring{\slashednabla}\upPsilinb_{\mathscr{I}^+}|^2+|\upPsilinb_{\mathscr{I}^+}|^2\right],
    \end{split}
\end{align}
for any $\epsilon>0$. An identical statement holds for $\overline{\Sigma}$.
\end{corollary}
\begin{proof}
    The statement concerning $\xblin$ follows from \Cref{07 11 2022 A,,07 11 2022 B,,07 11 2022 C}. The statment concerning $\ablin$ and $\pblin$ follows from \Cref{asymptotic flatness at spacelike infinity for ablin}.
\end{proof}

\begin{corollary}\label{gauge estimate on xblin ablin pblin near i0 degenerate}
    In addition to \Cref{gauge estimate on xblin ablin pblin near i0}, we have
\begin{align}
    \begin{split}
        &\int_{\overline{\Sigma}\cap\{r\geq r_+\}}dr\sin\theta d\theta d\phi\Bigg[\left|\frac{1}{\Omega}\nabladlt\left(\frac{1}{\Omega}\nabladlt(r^2\Omega\slashednabla_t\xblin)\right)\right|^2+\left|\frac{1}{\Omega}\nabladlt(r^2\Omega\slashednabla_t\xblin)\right|^2+\frac{1}{r^{2}}|r^2\Omega\slashednabla_t\xblin|^2\Bigg]
      \\&+ \int_{\overline{\Sigma}\cap\{r\geq r_+\}}dr\dw\, r^{4+\epsilon}|\slashednabla_t\ablin|^2+r^{6+\epsilon}|\slashednabla_t\pblin|^2\lesssim \sum_{|\gamma|\leq 3, i\in\{0,1\}}\left\|\partial_u^i\mathring{\slashednabla}^\gamma\xblins_{\mathscr{I}^+}\right\|^2_{\mathcal{E}^T_{\mathscr{I}^+}}.
    \end{split}
\end{align}
\end{corollary}
\begin{proof}
    This follows from \Cref{gauge estimate on xblin ablin pblin near i0} commuting both sides with $\partial_t$ and applying Hardy's inequality to the $\partial_t$-commuted right hand side of \eqref{gauge estimate ablin pblin near i0}.
\end{proof}

\subsubsection[Estimating $\protect\glinh$ on $\protect\Sigma^*_+$, $\protect\overline{\Sigma}$]{Estimating $\glinh$ on $\Sigma^*_+$, $\overline{\Sigma}$}

As with $\xlin$, $\xblin$ before, for large $r$ we can estimate $\glinh$ using \Cref{rp glinh n backwards scattering}, while for bounded $r$ we can estimate $\glinh$ using \bref{metric transport in 3 direction traceless},
\begin{align}
    \nablau \frac{1}{r}|\glinh|^2-\frac{\Omega^2}{r^2}|\glinh|^2=4r^{-1}\glinh\cdot\Omega\xblin,
\end{align}
and applying \Cref{backwards ingoing overall  on xblin}. We finally obtain
\begin{proposition}
The quantity $\glinh$ belonging to $\mathfrak{S}_{\mathscr{I}^+\cup\mathscr{H}^+_{\geq0}}$ satisfies
\begin{align}\label{gauge estimate glinh on Sigma star backwards scattering}
    \begin{split}
        \int_{\Sigma^*_+}dr\dw\,|\glinh|^2\lesssim &\sum_{|\gamma|\leq 1}\left\|\mathring{\slashednabla}^\gamma\xblins_{\mathscr{I}^+}\right\|^2_{\mathcal{E}^T_{\mathscr{I}^+}}+\left\|\xlins_{\mathscr{H}^+}\right\|^2_{\mathcal{E}^T_{\mathscr{H}^+_{\geq0}}}+\int_{\mathscr{H}^+_{\geq0}}dv\dw\,|\glinhs_{\mathscr{H}^+}|^2\\
        &+e^{\frac{5}{2M}v_+}\int_{\mathscr{H}^+_{\geq0}}d\bar{v}\sin\theta d\theta d\phi\,\left[|\mathring{\slashednabla}(2M)^5\ablins_{\mathscr{H}^+}|^2+|(2M)^5\ablins_{\mathscr{H}^+}|^2\right].
    \end{split}
\end{align}
\end{proposition}

A similar estimate to \bref{gauge estimate glinh on Sigma star backwards scattering} applies on $\overline{\Sigma}$.

\subsection{Scattering from data on $\overline{\mathscr{H}^-}$, $\mathscr{I}^-$}\label{Section 13.4 scattering from scri-}

We now show how solutions to \fullsystem are constructed from past scattering data representing the asymptotics obtained in \Cref{Section 8.7: Scattering from Sigmabar to H- and I-} in the past Bondi-normalised gauge. In this section we prove the following claim:

\begin{proposition}\label{grand proposition backwards scattering from past}
Given $\xlins_{\mathscr{I}^-}\in\Gamma_c^2(\mathscr{I}^-)$, $\xblins_{\mathscr{H}^-}\in\Gamma_c^2(\overline{\mathscr{H}^-})$, there exists a unique solution $\mathfrak{S}$ to the system \fullsystem on $J^-(\overline{\Sigma})$ which is both future Bondi and future horizon normalised, such that 
\begin{alignat}{2}\label{data attainement at past}
    &\lim_{u\longrightarrow-\infty}{r\xlin}(u,v,\theta^A)=\xlins_{\mathscr{I}^-}(v,\theta^A),\qquad\qquad \Omega\xblin|_{\overline{\mathscr{H}^-}}=\xblins_{\mathscr{H}^-}. 
\end{alignat}
The solution $\mathfrak{S}$ defines data on $\overline{\Sigma}$ which satisfies the gauge conditions of \bref{initial horizon gauge condition} and which is asymptotically flat at spacelike infinity in the sense of \Cref{def of asymptotic flatness at spacelike infinity} to order $(1,\infty)$. If in addition we have
\begin{align}\label{trivial memory near scri-}
    \int_{-\infty}^\infty dv \;\xlins_{\mathscr{I}^-}=0.
\end{align}

\end{proposition}
\begin{proposition}\label{gauge solutions near i-}
    Let $\xlins_{\mathscr{I}^-}\in\Gamma^2_c(\mathscr{I}^-)$, $\xblins_{\mathscr{H}^-}\in\Gamma^2_c(\overline{\mathscr{H}^-})$, and assume $\mathfrak{S}$ is a solution to \fullsystem which is past Bondi-normalised and which attains the data $\xlins_{\mathscr{I}^-}$, $\xblins_{\mathscr{H}^-}$ as in \Cref{data attainement at past}. Let $u_-$, $v_-$ such that $\xlins_{\mathscr{I}^-}=0$ on $v\leq v_-$, $\xblins_{\mathscr{H}^-}=0$ on $u\leq u_-$ and $r(u_-,v_-)>3M$.
    Assume that $\mathfrak{S}$ satisfies
    \begin{alignat}{2}
        &\Olin_{\ell\geq2}|_{\overline{\mathscr{H}^-}}=0,\qquad &&\left[\rlin-\rlin_{\ell=0}+\divr\elin\right]|_{\overline{\mathscr{H}^-}}=0,\\
        &\divo^2 U\Omega^{-1}\xlin|_{\mathscr{H}^-,u=u_-}=0.&&
    \end{alignat}
Then on  $J^-(\underline{\mathscr{C}}_{v_-})\cap J^-(\mathscr{C}_{u_-})$ the $\ell\geq2$ component of $\mathfrak{S}$ is trivial.
\end{proposition}

\begin{proof}[Proof of \cref{grand proposition backwards scattering from past}]
    As with the proof of \Cref{grand proposition backwards scattering}, we begin by constructing scattering data on $\overline{\mathscr{H}^-}$, $\mathscr{I}^-$ for all the quantities defining a solution to \fullsystem. To find the scattering data on $\overline{\mathscr{H}^-}$ we use the equations of \Cref{scattering data at H- forward scattering} with data $\xblins_{\mathscr{H}^-}$ and with $\Olinos_{\mathscr{H}^-}=\tr\glins_{\mathscr{H}^-}=0$, solving the transport equations with vanishing data at $u=u_-$ in view of \Cref{gauge solutions near i-}. Similarly, scattering data on $\mathscr{I}^-$ is found using the equations of \Cref{scattering data at scri- forward scattering}. The construction of the solution $\mathfrak{S}$ then can be done analogously to the proof of \Cref{grand proposition backwards scattering}, and similarly to the proof of \Cref{grand proposition backwards scattering} we can perform a gauge transformation using \Cref{inwards gauge solutions} to pass to the gauge conditions of the $\mathscr{I}^-$ and $\overline{\mathscr{H}^-}$ gauges. The asymptotic flatness of $\mathfrak{S}$ on $\overline{\Sigma}$ near $i^0$ can be established using the arguments made in \Cref{Section asymptotic flatness near i0 forwards scattering} for all linearised curvature and connection components, as well as for linearised lapse $\Olino$. 
    
    We now show that
    \begin{align}
        \glinh|_{\overline{\Sigma}}\;\sim \frac{1}{r}, \partial_t\glinh|_{\overline{\Sigma}}\;\sim \frac{1}{r^2}
    \end{align}
    as $r\longrightarrow\infty$. We estimate $\glinh|_{\overline{\Sigma}}$ via  $\ablins_{\mathscr{I}^-}$. Suppressing the angular directions, we have:
\begin{align}\label{10 12 2021}
\begin{split}
    |\glinh(u,v)|_{\overline{\Sigma}}|&\lesssim\frac{1}{r}\left|\xblins_{\mathscr{I}^-}(v)\right|+ \frac{1}{r^4}|\ablins_{\mathscr{I}^-}|+\frac{1}{r(u,v)^{\frac{5}{2}}}\sqrt{\int_{\underline{\mathscr{C}}_v\cap\{\bar{u}\leq u\}}d\bar{u}\dw\,\frac{r^2}{\Omega^2}|\nablau r^5\ablin|^2},\\
    &\lesssim\frac{1}{r}\left|\xblins_{\mathscr{I}^-}(v)\right|+ \frac{1}{r^4}|\ablins_{\mathscr{I}^-}|+\frac{1}{r(u,v)^{\frac{5}{2}}}\sqrt{\int_{\mathscr{I}^-\cap\{\bar{v}\leq v_+\}}d\bar{v}\dw\,\left|\underline{\mathtt{A}}^{(3)}_{\mathscr{I}^-}(\bar{v})\right|^2},
\end{split}
\end{align}    
where $\underline{\mathtt{A}}^{(n+2)}_{\mathscr{I}^-}(v,\theta^A)=\lim_{u\longrightarrow-\infty} (r^2\nablau)^n r^5\ablin(u,v,\theta^A)$ exists and satisfies $\underline{\mathtt{A}}^{(n+2)}_{\mathscr{I}^-}(v,\theta^A)\;\sim |v|^{n+2}$ for $v\geq v_+$ (see the analogous \Cref{backwards rp alin n} near $\mathscr{I}^+$). The asymptotics for higher order $\partial_t$ derivatives proceeds analogously. When \bref{trivial memory near scri-} is satisfied, $\xblins_{\mathscr{I}^-}=0$ on $v\geq v_+$ and $\underline{\mathtt{A}}^{(n+2)}_{\mathscr{I}^-}(v,\theta^A)\;\sim |v|^{n+1}$, so we may extract an additional power of decay in $r$.
\end{proof}

\section{The global scattering problem: proof of Theorem III}\label{Section 14: global scattering problem}

We now combine the results of the previous sections to prove \ref{global scattering corollary}.\\

We may already  apply Theorem II to construct a solution to the linearised equations on $\overline{\mathscr{M}}$ from smooth, compactly supported past scattering data.

\begin{proposition}\label{grand proposition global scattering}
    For past scattering data $\xlins_{\mathscr{I}^-}\in\Gamma^{(2)}_c(\mathscr{I}^-)$, $\xblins_{\mathscr{H}^-}$ such that $U^{-1}\xblins_{\mathscr{H}^-}\in\Gamma^{(2)}_c(\overline{\mathscr{H}^-})$, the solution $\mathfrak{S}_{\mathscr{I}^-\cup\overline{\mathscr{H}^-}}$ to \fullsystemK arising from $(\xlins_{\mathscr{I}^-},\xblins_{\mathscr{H}^-})$ according to \ref{thm for past scattering} extends uniquely to a solution of \fullsystemK on the whole of $\overline{\mathscr{M}}$. Furthermore, $\mathfrak{S}_{\mathscr{I}^-\cup\overline{\mathscr{H}^-}}$ is asymptotically flat at $\mathscr{I}^+$.
\end{proposition}

\begin{proof}
    Given the data assumed in the hypothesis of \ref{global scattering corollary}, a solution $\SscriHp$ arising from smooth, compactly supported data on $\overline{\mathscr{H}^-}$, $\mathscr{I}^-$ induces weakly asymptotically flat initial data on $\overline{\Sigma}$ by \Cref{grand proposition backwards scattering from past}, and the results of \Cref{Section 7 forward scattering} enable the construction of $\ablins_{\mathscr{I}^+}$, $\bblins_{\mathscr{I}^+}$, $\xblins_{\mathscr{I}^+}$ and $\Olinos_{\mathscr{I}^+}$.
\end{proof}

The remainder of this section is devoted to showing that $\SscriHp$ is strongly asymptotically flat at $\mathscr{I}^+$ and to show how to pass to a gauge that is normalised at $\mathscr{I}^+$ and $\overline{\mathscr{H}^+}$. The argument is organised as follows: in \Cref{Section 11.1 uniform bounded of Psilin Psilinb} we show the uniform boundedness of $\Psilin, \Psilinb$. We study the asymptotics of the spherical harmonic modes of $\Psilin$, $\Psilinb$ in \Cref{Section 11.2 spherical harmonic decomposition of RW}, and we normalise the solution $\SscriHp$ to the future scattering gauge in \Cref{Section 14.3 global scattering Bondi}.


\subsection[The uniform boundedness of $\protect\Psilin$, $\protect\Psilinb$ on $\overline{\mathscr{M}}$]{The uniform boundedness of $\protect\Psilin$, $\protect\Psilinb$ on $\overline{\mathscr{M}}$}\label{Section 11.1 uniform bounded of Psilin Psilinb}
    
 We begin by showing that $\Psilin$, $\Psilinb$ are uniformly bounded using an $r^p$ estiamte with $p=1$ near $i^0$:

\begin{defin}\label{defin of global solution}
For compactly supported $\xlins_{\mathscr{I}^-}\in\Gamma^{(2)}_c(\mathscr{I}^-)$,  $\xblins_{\mathscr{H}^-}\in\Gamma^{(2)}_c(\overline{\mathscr{H}^-})$, we denote by
\begin{align}
    \SscriHp[\xlins_{\mathscr{I}^-},\xblins_{\mathscr{H}^-}]
\end{align}
the unique solution arising from $(\xlins_{\mathscr{I}^-},\xblins_{\mathscr{H}^-})$ via \Cref{grand proposition backwards scattering from past}. 
\end{defin}



\begin{defin}
    Denote 
    \begin{align}
        \pastmemoryPsilin(\theta^A):=\upPsilin_{\mathscr{I}^-}(v_+,\theta^A).
    \end{align}
    A similar definition applies to $\Psilinb$. 
\end{defin}

\begin{proposition}\label{r dv Psilin squared bounded global scattering}
    Fix $R>4M$. For the solution $\SscriHp$ and any $u\in\mathbb{R}$, we have that
    \begin{align}
        \begin{split}
            \int_{\mathscr{C}_u\cap\{r>R,v\geq v_+\}}dv\dw\, r|\nablav\Psilin|^2
        \end{split}
    \end{align}
    is bounded. The same applies replacing $\Psilin$ with $\Psilinb$.
\end{proposition}

\begin{proof}
    We first perform the estimate for $u=-v_+$. Let $u_\infty<u$, multiply \bref{RW} by $r\Omega^{-2}\nablav\Psilin$ and integrate over the region bounded by $\mathscr{C}_u$, $\mathscr{C}_{u_\infty}$, $\underline{\mathscr{C}}_{v_+}$, $\underline{\mathscr{C}}_v$. 
    \begin{align}\label{prelim estimate of r dv Psilin}
        \begin{split}
            &\int_{\mathscr{C}_u\cap\{\bar{v}\in[v_+,v]\}}d\bar{v}\dw\,\frac{r}{\Omega^2}|\nablav\Psilin|^2 +\int_{\bar{u}\in[u_\infty,u],\bar{v}\in[v_+,v]}d\bar{u}d\bar{v}\dw\,\left(2-\frac{1}{\Omega^2}\right)|\nablav\Psilin|^2\\
            &=-\int_{\underline{\mathscr{C}}_{v}\cap\{\bar{u}\in[u_\infty,u]\}}d\bar{u}\dw\,\frac{1}{r}\left[|\mathring{\slashednabla}\Psilin|^2+\left(4-\frac{6M}{r}\right)|\Psilin|^2\right]\\\;\;&+\int_{\underline{\mathscr{C}}_{v_+}\cap\{\bar{u}\in[u_\infty,u]\}}d\bar{u}\dw\,\frac{1}{r}\left[|\mathring{\slashednabla}\Psilin|^2+\left(4-\frac{6M}{r}\right)|\Psilin|^2\right]\\
            \;\;&+\int_{\bar{u}\in[u_\infty,u],\bar{v}\in[v_+,v]}d\bar{u}d\bar{v}\dw\,\frac{\Omega^2}{r^2}\left[|\mathring{\slashednabla}\Psilin|^2+4\left(1-\frac{3M}{r}\right)|\Psilin|^2\right]+\int_{\mathscr{C}_{u_\infty}\cap\{\bar{v}\in[v_+,v]\}}d\bar{v}\dw\,\frac{r}{\Omega^2}|\nablav\Psilin|^2.
        \end{split}
    \end{align}
    We estimate
    \begin{align}
    \begin{split}
        &\int_{\underline{\mathscr{C}}_{v}\cap\{\bar{u}\in[u_\infty,u]\}\cap J^-(\overline{\Sigma})}d\bar{u}\dw\,\frac{1}{r}\left[|\mathring{\slashednabla}\Psilin|^2-|\mathring{\slashednabla}\upPsilin_{\mathscr{I}^-}|^2\right]=\int_{S^2}\dw\,\int_{u_\infty}^{-v}d\bar{u}\frac{1}{r}\int_{u_\infty}^{\bar{u}}d\bar{{\bar{u}}}2\Psilin\cdot \nablau\Psilin.
    \end{split}
    \end{align}
    Note that on $J^-(\overline{\Sigma})$ we have
    \begin{align}\label{sup r2 du Psilin constant}
        r^2|\nablau\Psilin|\lesssim 2(v-v_-)\times \sup_{J^-(\overline{\Sigma})\cap\{v\geq v_+\}}\left[|\mathring{\slashed{\Delta}}\Psilin|+4|\Psilin|^2\right],
    \end{align}
    \begin{align}\label{sup Psilin constant}
        \sup_{J^-(\overline{\Sigma})\cap\{v\geq v_+\}}|\Psilin|^2\lesssim \sum_{|\gamma|\leq4}\mathcal{E}^T_{\mathscr{I}^-}[\slashed{\mathcal{L}}_{so(3)}^\gamma\xlins_{\mathscr{I}^-}]+\mathcal{E}^T_{\overline{\mathscr{H}^-}}[\slashed{\mathcal{L}}_{so(3)}^\gamma\xblins_{\mathscr{H}^-}]+\sum_{|\gamma|\leq3}\int_{S^2}|\slashed{\mathcal{L}}_{so(3)}^\gamma\pastmemoryPsilin|^2.
    \end{align}
    Therefore,
    \begin{align}\label{overall cancellation}
        \begin{split}
            \left|\int_{S^2}\dw\,\int_{u_\infty}^{-v}d\bar{u}\frac{1}{r}\int_{u_\infty}^{\bar{u}}d\bar{{\bar{u}}}\;2\Psilin\cdot \nablau\Psilin\right|\lesssim \sum_{|\gamma|\leq2}&\Bigg[\text{ Product of right hand sides of }\bref{sup r2 du Psilin constant}\text{ and }\bref{sup Psilin constant}\\&\;\;\; \text{ commuted by }\slashed{\mathcal{L}}_{so(3)}^\gamma\Bigg].
        \end{split}
    \end{align}
    We now apply the estimate \bref{overall cancellation} to \bref{prelim estimate of r dv Psilin}, adding and subtracting $\int_{u_\infty}^{-v}\int_{S^2}4|\pastmemoryPsilin|^2+|\mathring{\slashednabla}\pastmemoryPsilin|^2$ to get
      \begin{align}\label{prelim estimate of r dv Psilin 2}
        \begin{split}
            &\int_{\mathscr{C}_u\cap\{\bar{v}\in[v_+,v]\}}d\bar{v}\dw\,\frac{r}{\Omega^2}|\nablav\Psilin|^2 +\int_{\bar{u}\in[u_\infty,u],\bar{v}\in[v_+,v]}d\bar{u}d\bar{v}\dw\,\left(2-\frac{1}{\Omega^2}\right)|\nablav\Psilin|^2\\
            &+\int_{\underline{\mathscr{C}}_{v}\cap\{\bar{u}\in[-v,u]\}}d\bar{u}\dw\,\frac{1}{r}\left[|\mathring{\slashednabla}\Psilin|^2+\left(4-\frac{6M}{r}\right)|\Psilin|^2\right]\\
            &=-\int_{\underline{\mathscr{C}}_{v}\cap\{\bar{u}\in[u_\infty,-v]\}}d\bar{u}\dw\,\frac{1}{r}\left[|\mathring{\slashednabla}\Psilin|^2-|\mathring{\slashednabla}\pastmemoryPsilin|^2+\left(4-\frac{6M}{r}\right)\left(|\Psilin|^2-|\pastmemoryPsilin|^2\right)\right]\\\;\;&+\int_{\underline{\mathscr{C}}_{v_+}\cap\{\bar{u}\in[u_\infty,u]\}}d\bar{u}\dw\,\frac{1}{r}\left[|\mathring{\slashednabla}\Psilin|^2-|\mathring{\slashednabla}\pastmemoryPsilin|^2+\left(4-\frac{6M}{r}\right)\left(|\Psilin|^2-|\pastmemoryPsilin|^2\right)\right]\\
            \;\;&+\int_{\bar{u}\in[u_\infty,u],\bar{v}\in[v_+,v]}d\bar{u}d\bar{v}\dw\,\frac{\Omega^2}{r^2}\left[|\mathring{\slashednabla}\Psilin|^2-|\mathring{\slashednabla}\pastmemoryPsilin|^2+4\left(1-\frac{3M}{r}\right)\left(|\Psilin|^2-|\pastmemoryPsilin|^2\right)\right]\\&+\int_{\mathscr{C}_{u_\infty}\cap\{\bar{v}\in[v_+,v]\}}d\bar{v}\dw\,\frac{r}{\Omega^2}|\nablav\Psilin|^2.
        \end{split}
    \end{align}
    Taking the limit $u_\infty\longrightarrow-\infty$ in \bref{prelim estimate of r dv Psilin 2}, we make use of the estimates \bref{sup r2 du Psilin constant}, \bref{sup Psilin constant} to obtain
    \begin{align}
        \begin{split}
            &\int_{\mathscr{C}_u\cap\{\bar{v}\in[v_+,v]\}}d\bar{v}\dw\,\frac{r}{\Omega^2}|\nablav\Psilin|^2 +\int_{\bar{u}\in[u_\infty,u],\bar{v}\in[v_+,v]}d\bar{u}d\bar{v}\dw\,\left(2-\frac{1}{\Omega^2}\right)|\nablav\Psilin|^2\\
            &+\int_{\underline{\mathscr{C}}_{v}\cap\{\bar{u}\in[-v,u]\}}d\bar{u}\dw\,\frac{1}{r}\left[|\mathring{\slashednabla}\Psilin|^2+\left(4-\frac{6M}{r}\right)|\Psilin|^2\right]\\
            &\lesssim \sum_{|\gamma|\leq3}\Bigg[\text{ Product of right hand sides of }\bref{sup r2 du Psilin constant}\text{ and }\bref{sup Psilin constant}\text{ commuted by }\slashed{\mathcal{L}}_{so(3)}^\gamma\Bigg].
        \end{split}
    \end{align}
    The above estimate remains valid when taking $v\longrightarrow\infty$.
\end{proof}

\begin{corollary}\label{uniform boundedness of Psilin}
For $\SscriHp$, we have that $\Psilin$, $\Psilinb$ are uniformly bounded on $\overline{\mathscr{M}}$ and we have for $R>2M$
\begin{align}
    \sup_{\overline{\mathscr{M}}\cap\{r>R\}}|\Psilin|^2\lesssim \sum_{|\gamma|\leq5}\Bigg[\text{ Product of right hand sides of }\bref{sup r2 du Psilin constant}\text{ and }\bref{sup Psilin constant}\text{ commuted by }\slashed{\mathcal{L}}_{so(3)}^\gamma\Bigg].
\end{align}
\end{corollary}

\begin{proof}
    It suffices to derive the bound above near $\mathscr{I}^+$. Noe that \Cref{weak boundedness of Psilin weak a f} allows us to estimate for any $\delta>0$
    \begin{align}
        \begin{split}
            &\int_{S^2}\dw\,\frac{|\Psilin(u,v,\theta^A)|^2}{r^{2\delta}}\leq \frac{1}{r^{2\delta}}\left[\int_{\mathscr{C}_u\cap\{r>R\}}d\bar{v}\dw\, r|\nablav\Psilin|^2+\frac{|\Psilin|^2}{r}\right]\\&\lesssim \frac{1}{r^{2\delta}}\left[\int_{\mathscr{C}_u\cap\{r>R\}}d\bar{v}\dw\, r|\nablav\Psilin|^2+\sum_{|\gamma|\leq2}\mathcal{E}^T_{\mathscr{I}^-}[\slashed{\mathcal{L}}_{so(3)}^\gamma \xlins_{\mathscr{I}^-}]+\mathcal{E}^T_{\overline{\mathscr{H}^-}}[\slashed{\mathcal{L}}_{so(3)}^\gamma \xblins_{\mathscr{H}^-}]\right],
        \end{split}
    \end{align}
    where the final line above was obtained by applying a Hardy estimate to bound $\|r^{-1}\Psilin\|_{L^2(\mathscr{C}_u\cap\{r>R\})}$ in terms of $\|r\nablav\Psilin\|_{L^2(\mathscr{C}_u\cap\{r>R\})}$. The result now follows from \Cref{r dv Psilin squared bounded global scattering}.
\end{proof}

\begin{corollary}
    For the solution $\Sscriscri$ and any $R>2M$ we have 
    \begin{align}
        \sup_{\overline{\mathscr{M}}\cap\{r>R\}} r|\nablav\Psilin|\lesssim \sum_{|\gamma|\leq7}\Bigg[\text{ Product of right hand sides of }\bref{sup r2 du Psilin constant}\text{ and }\bref{sup Psilin constant}\text{ commuted by }\slashed{\mathcal{L}}_{so(3)}^\gamma\Bigg].
    \end{align}
\end{corollary}
\begin{proof}
    Integrate \bref{RW} in $u$ from $-\infty$ to $u$ to get
    \begin{align}
        r|\nablav\Psilin(u,v,\theta^A)|\lesssim \sup_{\bar{u}\in(-\infty,u]}\left[|\mathring{\slashed{\Delta}}\Psilin(\bar{u},v,\theta^A)|+4|\Psilin(\bar{u},v,\theta^A)|\right].
    \end{align}
\end{proof} 

\subsection[Approximate conservation laws for the Regge--Wheeler equation and constructing the radiation fields of $\protect\Psilin$, $\protect\Psilinb$ at $\mathscr{I}^+$]{Approximate conservation laws for the Regge--Wheeler equation and constructing the radiation fields of $\protect\Psilin$, $\protect\Psilinb$ at $\mathscr{I}^+$}\label{Section 11.2 spherical harmonic decomposition of RW}

We are now in position to study the spherical harmonic modes of $\Psilin$, $\Psilinb$. Using the methods of \cite{KehrbergerIII}, we show that each individual $(\ell,m)$-mode of the scalar components $\Psilin$ and $\Psilinb$ converges towards $\mathscr{I}^+$, leading us to conclude via \Cref{Section 11.1 uniform bounded of Psilin Psilinb} that $\upPsilin_{\mathscr{I}^+}$, $\upPsilinb_{\mathscr{I}^+}$ exist.

\begin{lemma}
    Let $\Psi$ be a solution to the Regge--Wheeler equation \bref{RW} and assume $(f,g)$ are scalar functions such that $\Psi=2\fancydstarring_2\fancydstarring_1(f,g)$ for all $u,v\in\mathbb{R}^2$. Then $f$ satisfies
    \begin{align}\label{scalar RW}
        \partial_u\partial_v f-\frac{\Omega^2}{r^2}\mathring{\slashed{\Delta}}f-\frac{6M\Omega^2}{r^3}f=0.
    \end{align}
    The same applies to $g$.
\end{lemma}
\begin{proof}
    Recall that $\mathring{\slashed{\Delta}}\fancydstarring_2\fancydstarring_1=\fancydstarring_2\fancydstarring_1\left(\mathring{\slashed{\Delta}}+4\right)$.
\end{proof}

The following is an immediate consequence of \Cref{uniform boundedness of Psilin}:

\begin{proposition}
    For the solution $\SscriHp$, let $\flin$, $\glinn$ be the scalar functions associated with $\Psilin$ via $\Psilin=2\fancydstarring_2\fancydstarring_1(\flin,\glinn)$. Then there exists sequences of smooth functions $\{\flinl\}$, $\{\glinnl\}$ such that
    \begin{align}
        &\flin(u,v,\theta^A)=\sum_{\ell=2}^\infty \sum_{m=-\ell}^\ell \flinl(u,v) Y_{\ell,m}(\theta^A),\\
        &\glinn(u,v,\theta^A)=\sum_{\ell=2}^\infty \sum_{m=-\ell}^\ell \glinnl(u,v) Y_{\ell,m}(\theta^A),
    \end{align}
    for all $u,v$, where $Y_{\ell,m}$ is the $(\ell,m)$-spherical harmonic function on the unit round sphere $S^2$. Each of $\flinl$, $\glinnl$ satisfies
    \begin{align}
        &\partial_u\partial_v \flinl+\frac{\Omega^2}{r^2}\left(\ell(\ell+1)-\frac{6M}{r}\right)\flinl=0,\label{scalar l m RW}\\
        &\partial_u\partial_v \glinnl+\frac{\Omega^2}{r^2}\left(\ell(\ell+1)-\frac{6M}{r}\right)\glinnl=0.
    \end{align}
\end{proposition}

\begin{notation*}
    We define $(\mathpzc{f}_{\mathscr{I}^-},\mathpzc{g}_{\mathscr{I}^-})$ to be the scalar functions associated with $\upPsilin_{\mathscr{I}^-}$ via
    \begin{align}
        \upPsilin_{\mathscr{I}^-}=2\fancydstarring_2\fancydstarring_1(\mathpzc{f}_{\mathscr{I}^-},\mathpzc{g}_{\mathscr{I}^-}).
    \end{align}
    We denote by $\flinslp$ the $(\ell,m)$\textsuperscript{th} coefficient in the spherical harmonic expansion of $\mathpzc{f}_{\mathscr{I}^-}$. A similar notation applies to $\glinnslp$.
\end{notation*}

\begin{lemma}
    Let $\flinslp$ be the $(\ell,m)$\textsuperscript{th} coefficient in the spherical harmonic expansion of $\mathpzc{f}_{\mathscr{I}^-}$. Then we have
    \begin{align}
        \lim_{u\longrightarrow-\infty}\flinl(u,v,\theta^A)=\flinslp(v,\theta^A).
    \end{align}
\end{lemma}

As mentioned earlier, to show that $\flinl$ converges towards $\mathscr{I}^+$ we will make use of a cancellation that arises in the equation satisfied by $(r^2\partial_v)^\ell \flinl$, which we derive from \bref{wave equation for nth r^2 dv derivative of Psi}:

\begin{notation*}
    Define 
    \begin{align}
        \Flins{\ell}{n}:=\left(\frac{r^2}{\Omega^2}\partial_v\right)^n\flinl
    \end{align}
\end{notation*}

\begin{lemma}
The quantity $\Flins{\ell}{n}$ satisfies
    \begin{align}\label{wave equation for nth r^2 dv derivative of Psi}
    \begin{split}
        &\partial_u\partial_v \Flins{\ell}{n}+\frac{\Omega^2}{r^2}\ell(\ell+1)\Flins{\ell}{n}+n\left(\frac{2}{r}-\frac{6M}{r^2}\right)\partial_v \Flins{\ell}{n}+\left[-\frac{6M}{r}-n(n+1)\left(1-\frac{6M}{r}\right)\right]\frac{\Omega^2}{r^2}\Flins{\ell}{n}\\
        &-2Mn(n-2)(n+2)\frac{\Omega^2}{r^2}\Flins{\ell}{n-1}=0.
    \end{split}
\end{align}
In particular,
\begin{align}\label{lth order wave equation for lth mode with cancellation}
    \begin{split}
        \partial_u \frac{\Omega^{2\ell}}{r^{2\ell}}\partial_v \Flins{\ell}{\ell}+6M\frac{\Omega^{2\ell+2}}{r^{2\ell+3}}(\ell(\ell+1)-1)\Flins{\ell}{\ell}-2M\ell(\ell-2)(\ell+2)\frac{\Omega^{2\ell+2}}{r^{2\ell+2}}\Flins{\ell}{\ell-1}=0.
    \end{split}
\end{align}
\end{lemma}

Using \bref{lth order wave equation for lth mode with cancellation}, we use the fact that $(r\partial_v)^n\flin$ is uniformly bounded for $n$ to show that $\flin$ converges towards $\mathscr{I}^+$:

\begin{proposition}\label{l mode convergence}
    For any $\ell$ and any fixed $u$, we have
    \begin{align}
        r^2\partial_v \flinl(u,v,\theta^A)=O(\log(v))
    \end{align}
    as $v\longrightarrow\infty$.
\end{proposition}

\begin{proof}
    Define $y:=\frac{1}{r}$. Then $\partial_y=\frac{r^2}{\Omega^2}\partial_v$. We can write \bref{lth order wave equation for lth mode with cancellation} as
    \begin{align}\label{y lth order wave equation for lth mode with cancellation}
        \partial_u \frac{\Omega^{2\ell+2}}{r^{2\ell+2}}(\partial_y)^{\ell+1} \flinl+6M\frac{\Omega^{2\ell+2}}{r^{2\ell+3}}[\ell(\ell+1)-1](\partial_y)^\ell \flinl-2M\ell(\ell-2)(\ell+2)\frac{\Omega^{2\ell+2}}{r^{2\ell+2}}(\partial_y)^{\ell-1}\flinl=0.
    \end{align}
    Since $\frac{\Omega^{2\ell+2}}{r^{2\ell+2}}(\partial_y)^{\ell+1} \flinl\longrightarrow0$ as $u\longrightarrow-\infty$, we integrate \bref{y lth order wave equation for lth mode with cancellation} from $\mathscr{I}^-$ for $v>v_+$ and use the fact that $y^\ell (\partial_y)^{\ell}\flinl$ is uniformly bounded to find
    \begin{align}
        \left|(\partial_y)^{\ell+1}\flinl\right|\lesssim y^{-\ell}.
    \end{align}
    We integrate the above $\ell$ times between $v_+$ and some finite $v>v_+$. Note that
    \begin{align}\label{ell iterated integrals}
    \begin{split}
        &\int_{y_+}^ydy_1\int_{y_+}^{y_1}dy_2\dots \int_{y_+}^{y_{\ell-1}}dy_{\ell} (\partial_y)^{\ell+1}\flinl\\&=\int_{y_+}^ydy_1\int_{y_+}^{y_1}dy_2\dots \int_{y_+}^{y_{\ell-2}}dy_{\ell-1} \left[(\partial_y)^{\ell}\flinl(u,y_{\ell})-(\partial_y)^{\ell}\flinl(u,y_{+})\right]\\
        &=\left[\int_{y_+}^ydy_1\int_{y_+}^{y_1}dy_2\dots \int_{y_+}^{y_{\ell-2}}dy_{\ell-1} (\partial_y)^{\ell}\flinl(u,y_{\ell})\right]-\frac{(y_+-y)^{\ell-1}}{(\ell-1)!}\left[(\partial_y)^{\ell}\flinl\right]\Big|_{v_+},
    \end{split}
    \end{align}
    where $y_+=(r(u,v_+))^{-1}$. Iterating the recurrence relation above we get
    \begin{align}\label{ell iterated integrals 2}
        \begin{split}
            \int_{y_+}^ydy_1\int_{y_+}^{y_1}dy_2\dots \int_{y_+}^{y_{\ell-1}}dy_{\ell} (\partial_y)^{\ell+1}\flinl=\partial_y \flinl(u,v)-\sum_{j=1}^{\ell}\frac{(y_+-y)^{j}}{j!}\left[(\partial_y)^{j}\flinl\right]\Big|_{v_+}.
        \end{split}
    \end{align}
    Since 
    \begin{align}
        y^{-\ell}=\frac{(-1)^\ell}{(\ell-1)!}\partial_y^\ell \log|y|,
    \end{align}
    we may apply the formula \bref{ell iterated integrals} to find that the $\ell$\textsuperscript{th} iterated integral of $y^\ell$ yields
    \begin{align}\label{anomaly integral}
        \begin{split}
            \int_{y_+}^ydy_1\int_{y_+}^{y_1}dy_2\dots \int_{y_+}^{y_{\ell-1}}dy_{\ell} (y_{\ell+1})^{-\ell}=\frac{(-1)^{\ell-1}}{(\ell-1)!}\log\left(\left|\frac{y}{y_+}\right|\right)-\sum_{j=1}^{\ell-1}\frac{(-1)^{\ell-j}}{{y_+}^{\ell-j+1}}{(y_+-y)^{\ell-j}},
        \end{split}
    \end{align}
    which implies the proposition.
\end{proof}

\begin{corollary}\label{radiation field of RW global scattering}
    The quantity $\flinl(u,v)$ converges as $v\longrightarrow\infty$ to a limit $\mathpzc{f}_{\mathscr{I}^+,(\ell,m)}$ that defines a smooth function of $u$. The same applies to $\glinnl$ which defines a smooth limit $\glinnslf$ at $\mathscr{I}^+$. Therefore, $\Psilin(u,v,\theta^A)$ defines a smooth radiation field $\upPsilin_{\mathscr{I}^+}$ at $\mathscr{I}^+$ via
    \begin{align}
        \Psilin(u,v,\theta^A)\longrightarrow \upPsilin_{\mathscr{I}^+}(u,\theta^A)=\sum_{\ell=2}^\infty\sum_{m=-\ell}^\ell 2\fancydstarring_2\fancydstarring_1\left(\flinslf(u),\glinnslf(u)\right)Y_{\ell,m}(\theta^A).
    \end{align}
\end{corollary}
\begin{proof}
    The existence of $\flinslf$, $\glinnslf$ is implied by \Cref{l mode convergence}. We may apply \Cref{uniform boundedness of Psilin} to each individual $(\ell,m)$ mode to get
         \begin{align}\label{sup Psilin constant l m}
        \sup_{J^-(\overline{\Sigma})\cap\{v\geq v_+\}}|\flinl|\lesssim \ell^2\left[|\flinslp(v_+)|+\int_{(-\infty,v_+]}d\bar{v}|\flinslp|+\int_{\mathbb{R}}|\mathpzc{f}_{\mathscr{H}^-}|\right].
    \end{align}
    Since the right hand side is summable over all $\ell, m$, the result for $\flins$ follows by the Weierstrass M-test. We also conclude that
    \begin{align}\label{sup Psilin scri constant ell}
         \sup_{J^-(\overline{\Sigma})\cap\{v\geq v_+\}}|\flinslf|\lesssim \ell^2\left[|\flinslp(v_+)|+\int_{(-\infty,v_+]}d\bar{v}|\flinslp|+\int_{\mathbb{R}}|\mathpzc{f}_{\mathscr{H}^-}|\right].
    \end{align}
         \begin{align}\label{sup Psilin scri constant}
        \sup_{J^-(\overline{\Sigma})\cap\{v\geq v_+\}}|\upPsilin_{\mathscr{I}^+}|^2\lesssim \sum_{|\gamma|\leq4}\mathcal{E}^T_{\mathscr{I}^-}[\slashed{\mathcal{L}}_{so(3)}^\gamma\xlins_{\mathscr{I}^-}]+\mathcal{E}^T_{\overline{\mathscr{H}^-}}[\slashed{\mathcal{L}}_{so(3)}^\gamma\xblins_{\mathscr{H}^-}]+\sum_{|\gamma|\leq3}\int_{S^2}|\slashed{\mathcal{L}}_{so(3)}^\gamma\pastmemoryPsilin|^2.
    \end{align}
\end{proof}

We can also deduce from \Cref{l mode convergence}, applying the Weierstrass M-test, the following:
\begin{corollary}\label{Psilin decays in future global scattering}
    For the solution $\SscriHp$, the radiation fields $\upPsilin_{\mathscr{I}^+}$ constructed in \Cref{radiation field of RW global scattering} decays as $u\longrightarrow\infty$. The same applies to $\upPsilinb_{\mathscr{I}^+}$.
\end{corollary}

\subsection{Passing to the future scattering gauge}\label{Section 14.3 global scattering Bondi}

We now show that $\xlins_{\mathscr{I}^+}$ for $\SscriHp$ exists. With the help of \eqref{Psilin decays in future global scattering}, we will also show that $\xlins_{\mathscr{I}^+}$. decays as $u\longrightarrow\infty$. We then construct $\mathfrak{G}_{\mathscr{I}^+}$ via \Cref{transformation to Bondi gauge} and show that it is trivial at $\overline{\mathscr{H}^+}$, thus preserving the conditions \eqref{bifurcation gauge conditions}.

\begin{proposition}
    For $\SscriHp$, $\xblins_{\mathscr{I}^+}$ is integrable in $u$ on $[u_0,\infty)$ for any $u_0$.
\end{proposition}
\begin{proof}
    As $\upPsilinb_{\mathscr{I}^+}$, $\xblins_{\mathscr{I}^+}$ decay with $u\longrightarrow\infty$ and $\mathfrak{S}$ induces weakly asymptotically flat Cauchy data on $\overline{\Sigma}$, the proposition follows from the formula \bref{further constraint ++} and \Cref{decay of xlin towards future of scri+}.
\end{proof}

\begin{proposition}\label{auxiliary 1122}
    For $\SscriHp$ and any $u\in\mathbb{R}$, the limit of $r^2\xlin(u,v,\theta^A)$ as $v\longrightarrow\infty$ exists and defines a smooth radiation field $\xlins_{\mathscr{I}^+}$ on $\mathscr{I}^+$. Moreover, $\xlins_{\mathscr{I}^+}$ decays as $u\longrightarrow\infty$. 
\end{proposition}

\begin{proof}
    We show that $r^2\alin(u,v,\theta^A)$ is integrable in $v$ for $r>4M$ using the formula \bref{further constraint -}. Note that
    \begin{align}
        \frac{\Omega^2}{r}\nablav\frac{r^2}{\Omega^2}\nablav\Psilin=\Omega^2\nablav\left[\frac{r}{\Omega^2}\nablav\Psilin+\Psilin\right],
    \end{align}
    thus we deduce from \bref{further constraint --}
    \begin{align}
        \begin{split}
            \nablav\Big[\frac{r}{\Omega^2}\nablav\Psilin+\Psilin+\left(\mathring{\slashed{\Delta}}-2\right)\left(\mathring{\slashed{\Delta}}-4\right)r^2\Omega^{-1}\xlin&+6M\partial_t r^2\Omega^{-1}\xlin\Big]=0.
        \end{split}
    \end{align}
     As $\Psilin$ and $\partial_t r^2\xlin$ converge and $r\nablav\Psilin$ decays towards $\mathscr{I}^+$, we deduce that $\left(\mathring{\slashed{\Delta}}-2\right)\left(\mathring{\slashed{\Delta}}-4\right)r^2\xlin(u,v,\theta^A)$ has a limit as $v\longrightarrow\infty$, which implies that $r^2\xlin$ also converges as $v\longrightarrow\infty$ to a limit $\xlins_{\mathscr{I}^+}$. For any fixed $R$ we have
    \begin{align}
        \begin{split}
            &\upPsilin_{\mathscr{I}^+}+\left(\mathring{\slashed{\Delta}}-2\right)\left(\mathring{\slashed{\Delta}}-4\right)\xlins_{\mathscr{I}^+}+6M\partial_u \xlins_{\mathscr{I}^+}\\
            &=\Big[R\Omega^{-2}(R)\nablav\Psilin+\Psilin+\left(\mathring{\slashed{\Delta}}-2\right)\left(\mathring{\slashed{\Delta}}-4\right)R^2\Omega^{-1}\xlin+6M\partial_t R^2\Omega^{-1}\xlin\Big]\Big|_{r=R}.
        \end{split}
    \end{align}
    We may apply \Cref{technical lemma} to \bref{RWILED} commuted with $\partial_t$ to deduce that $\Psilin|_{r=R}\longrightarrow0$ as $u\longrightarrow\infty$, and commuting with $\nablav$ implies the same for $\nablav\Psilin|_{r=R}$. A similar argument using \bref{ILED on xlin} shows that $\xlin|_{r=R}, \partial_t \xlin|_{r=R}$ also decay as $u\longrightarrow\infty$. Therefore, $\xlins_{\mathscr{I}^+}\longrightarrow0$ as $u\longrightarrow\infty$.
\end{proof}

\begin{corollary}
For $\SscriHp$, the quantity $\Olinos_{\mathscr{I}^+}$ is integrable in $u$ on $[u_0,\infty)$ for any $u_0\in\mathbb{R}$.
\end{corollary}

Given that on $\{v\leq v_-, u\leq u_-\}$ the solution $\Sscriscri$ has $\Psilin(u,v,\theta^A)=0$ and $r^2\xlin(u,v,\theta^A)\longrightarrow0$, we can repeat the argument of \Cref{auxiliary 1122} integrating between $\underline{\mathscr{C}}_{v_--1}$ and $\mathscr{I}^+$ to conclude that $\xlins_{\mathscr{I}^+}$ attains a limit as $u\longrightarrow-\infty$ with
\begin{align}\label{memory limit 1}
    \left(\mathring{\slashed{\Delta}}-2\right)\left(\mathring{\slashed{\Delta}}-4\right)\lim_{u\longrightarrow-\infty}\xlins_{\mathscr{I}^+}(u,\theta^A)=-\lim_{u\longrightarrow-\infty}\upPsilin_{\mathscr{I}^+}(u,\theta^A).
\end{align}

We are now in a position to construct the gauge solution $\mathfrak{G}_{\mathscr{I}^-\longrightarrow\mathscr{I}^+}$:

\begin{defin}\label{transformation to Bondi gauge global scattering}
Let $\Olinos_{\mathscr{I}^+}$ be as in \Cref{forward scattering Olino elin at scri+ weak a f}. Define
\begin{align}
   \partial_u\outwardsgaugefunction(u,\theta^A)=-\Olinos_{\mathscr{I}^+},\qquad \outwardsgaugefunction|_{\ell=0,1}=0, \lim_{u\longrightarrow\infty}{\outwardsgaugefunction}_{,\ell\geq2}(u,\theta^A)=0.
\end{align}
Define $\mathfrak{G}_{\mathscr{I}^+}$ to be the gauge solution generated by $\outwardsgaugefunction$ via \Cref{outwards gauge solutions}.
\end{defin}

\begin{remark}
    It is easy to read off from the solutions defined via \Cref{outwards gauge solutions} that, given 
    \begin{align}
        \lim_{u\longrightarrow\infty}{\outwardsgaugefunction}_{,\ell\geq2}(u,\theta^A)=0,
    \end{align}
    that $\mathfrak{G}_{\mathscr{I}^-\longrightarrow\mathscr{I}^+}$ does not radiate to $\overline{\mathscr{H}^+}$. Moreover, The gauge conditions \ref{bifurcation gauge conditions} are satisfied by $\SscriHp+\mathfrak{G}_{\mathscr{I}^-\longrightarrow\mathscr{I}^+}$.
\end{remark}

\begin{corollary}
    The solution $\SscriHp+\mathfrak{G}_{\mathscr{I}^+}$ is $\mathscr{I}^-$-normalised
\end{corollary}
\begin{proof}
    As $\SscriHp$ induces weakly asymptotically flat Cauchy data on $\overline{\Sigma}$ by \Cref{grand proposition backwards scattering from past}, we have that $\Olinos_{\mathscr{I}^+}$ decays as $|u|\longrightarrow\pm\infty$ by the results of \Cref{early and late time asymptotics section}. Thus  L'H\^opital's rule allows us to conclude that $|u|^{-1}\inwardsgaugefunction$ also decays as $|u|\longrightarrow\pm\infty$.
\end{proof}

\begin{corollary}
    The solution $\SscriHp+\mathfrak{G}_{\mathscr{I}^+}$ is in the global scattering gauge.
\end{corollary}

\begin{remark}
    The solution $\mathfrak{G}_{\mathscr{I}^+}$ is controlled via $\mathbb{E}_{\overline{\Sigma}\longrightarrow \mathscr{I}^+\cup\overline{\mathscr{H}^+}}$, which is in turn controlled via $\mathbb{E}_{\mathscr{I}^-\cup\overline{\mathscr{H}^-}\longrightarrow \mathscr{I}^+\cup\overline{\mathscr{H}^+}}$ according to the results of \Cref{an estimate on initial gauge from backwards scattering}.
\end{remark}

\begin{remark}
    Having constructed a $\mathscr{I}^+$-normalised solution $\mathfrak{S}_{\overline{\mathscr{H}^-}\cup\mathscr{I}^-}+\mathfrak{G}_{\mathfrak{I}^+}$, we may now apply the steps of \Cref{forward scattering from Sigma bar} to construct the unique pure gauge solution $\underline{\mathfrak{G}}_{\overline{\mathscr{H}^+}}$ such that $\mathfrak{S}_{\overline{\mathscr{H}^-}\cup\mathscr{I}^-}+\mathfrak{G}_{\mathfrak{I}^+}+\underline{\mathfrak{G}}_{\overline{\mathscr{H}^+}}$ is in the future scattering gauge.  We may then conclude
\end{remark}
\begin{proposition}
    Let $\Sscriscri:=\SscriHp+\mathfrak{G}_{\mathscr{I}^+}+\underline{\mathfrak{G}}_{\overline{\mathscr{H}^+}}$, where $\SscriHp$ is the solution constructed via \Cref{grand proposition global scattering}, $\mathfrak{G}_{\mathscr{I}^+}$ is constructed via \Cref{transformation to Bondi gauge global scattering} and $\underline{\mathfrak{G}}_{\overline{\mathscr{H}^+}}$ is constructed via \Cref{transformation to H+ bar gauge forward scattering}. Then $\Sscriscri$ satisfies the conclusions of \ref{global scattering corollary}.
\end{proposition}

The isometry $t\longrightarrow-t$ allows us to immediately use the estimates of \Cref{an estimate on initial gauge from backwards scattering} to control the gauge solutions $\mathfrak{G}_{\mathscr{I}^+}$, $\underline{\mathfrak{G}}_{\overline{\mathscr{H}^+}}$ in terms of past scattering data:

\begin{proposition}
    The pure gauge solutions $\mathfrak{G}_{\mathscr{I}^+}$, $\underline{\mathfrak{G}}_{\overline{\mathscr{H}^+}}$ satisfy the estimate \eqref{global gauge estimate Theorem III}.
\end{proposition}
\begin{proof}
    The right hand side of \eqref{Bondi normalisation from Sigma bar estimate} near $\mathcal{B}$ is estimated in terms of $\xlins_{\mathscr{I}^-}$, $\xblins_{\mathscr{H}^-}$ via \Cref{estimate on xblin on Sigmabar proposition} exchanging $\xblin$ with $\xlin$, $V$ with $-U$, $\mathscr{I}^+$ with $\mathscr{I}^-$ and $\overline{\mathscr{H}^+}$ with $\overline{\mathscr{H}^-}$ respectively. Similarly, near $i^0$, we use \Cref{gauge estimate on xblin ablin pblin near i0}, applying the same exchange of variables and barred/unbarred quantities. The remainder of $\overline{\Sigma}$ can be covered by \Cref{gauge estimate xblin in intermediate region} and \Cref{gauge estimate on xblin ablin pblin near i0}. The right hand side of \eqref{transformation to H+ bar gauge forward scattering estimate} can be similarly estimated in terms of $\xlins_{\mathscr{I}^-}$, $\xblins_{\mathscr{H}^-}$ by appealing to \eqref{RW in Kruskal blueshift}, \eqref{RW exponential backwards near H+} and \Cref{gauge estimate alin plin near i0}, again applying the isometry $t\longrightarrow-t$ and exchanging barred quantities with unbarred ones.
\end{proof}

\section{Proof of Corollary II}\label{Section 15 Proof of Theorem III}

We now collect the results accumulated in the course of proving Theorems II and III to prove Corollary II. The argument of \Cref{Section 11.2 spherical harmonic decomposition of RW} will be used \Cref{Section 12.1 matching Psilin Psilinb} to relate the limits of $\Psilin$, $\Psilinb$ near $i^0$. We then use the resulting asymptotics in \Cref{Section 12.2 memory matching} to derive the relation between $\xlins_{\mathscr{I}^+}$ and $\xblins_{\mathscr{I}^+}$ in the global scattering gauge.

\subsection[Matching $\protect\upPsilin_{\mathscr{I}^+}$ and $\protect\upPsilin_{\mathscr{I}^-}$ near $i^0$]{Matching $\protect\upPsilin_{\mathscr{I}^+}$ and $\protect\upPsilin_{\mathscr{I}^-}$ near $i^0$}\label{Section 12.1 matching Psilin Psilinb}

\begin{defin}\label{def of scriscriscri}
    For $\xlins_{\mathscr{I}^-}\in\Gamma^{(2)}_c(\mathscr{I}^-)$, $\xblins_{\mathscr{H}^-}$ such that $U^{-1}\xblins_{\mathscr{H}^-}\in\Gamma^{(2)}_c(\overline{\mathscr{H}^-})$, denote by $\Sscriscriscri$ the solution $\SscriHp+\mathfrak{G}_{\mathscr{I}^+}$ to \fullsystemK, where $\SscriHp$ is constructed as in \Cref{grand proposition global scattering} and $\mathfrak{G}_{\mathscr{I}^+}$ is constructed as in \Cref{transformation to Bondi gauge global scattering}.
\end{defin}
 
We begin by proving that $\upPsilin_{\mathscr{I}^\pm}$,  $\upPsilinb_{\mathscr{I}^\pm}$ are antipodally matched near $i^0$. First, we show that $r^{-n}\left(r^2\partial_v\right)^n\flinl$ attains a limit towards $\mathscr{I}^-$. 

\begin{proposition}
    For the solution $\Sscriscriscri$, we have that
    \begin{align}
        r^{-n}\left(r^2\partial_v\right)^n\flinl(u,v)\longrightarrow (-1)^n\frac{(\ell+n)!}{n!(\ell-n)!}\flinslp(v)
    \end{align}
    as $u\longrightarrow-\infty$ for $v\geq v_+$.
\end{proposition}
\begin{proof}
    It is easy to see that $r\nablav\Psilin(u,v,\theta^A)\longrightarrow (\mathring{\slashed{\Delta}}-4)\pastmemoryPsilin$, since using \bref{sup r2 du Psilin constant} we compute
    \begin{align}
    \begin{split}
        \nablav\Psilin(u,v,\theta^A)=&\int_{-\infty}^ud\bar{u}\,\frac{\Omega^2}{r^2}\left[\mathring{\slashed{\Delta}}-4\right]\Psilin(\bar{u},v,\theta^A)\\
        &=\int_{-\infty}^ud\bar{u}\,\frac{\Omega^2}{r^2}\left[\mathring{\slashed{\Delta}}-4\right]\pastmemoryPsilin+\int_{-\infty}^ud\bar{u}\frac{\Omega^2}{r^2}\int_{-\infty}^{\bar{u}}d\bar{\bar{u}}\left[\mathring{\slashed{\Delta}}-4\right]\nablau\Psilin(\bar{\bar{u}},v,\theta^A),
    \end{split}
    \end{align}
    thus
    \begin{align}
        \begin{split}
            \left|r\nablav\Psilin(u,v,\theta^A)-\left[\mathring{\slashed{\Delta}}-4\right]\pastmemoryPsilin\right|\lesssim \frac{1}{r}\text{ Right hand side of }\bref{sup r2 du Psilin constant}.
        \end{split}
    \end{align}
    An inductive argument using \bref{wave equation for nth r^2 dv derivative of Psi} that $r^{-n}(r^2\nablav)^n\Psilin(u,v,\theta^A)$ similarly attains a limit as $u\longrightarrow-\infty$ for $v\geq v_+$:
    \begin{align}
        \lim_{u\longrightarrow-\infty} \frac{1}{r^n}\left(\frac{r^2}{\Omega^2}\nablav\right)^n\Psilin(u,v,\theta^A)=\left[n-1+\frac{1}{n}(\mathring{\slashed{\Delta}}-4)\right]\lim_{u\longrightarrow-\infty} \frac{1}{r^{n-1}}\left(\frac{r^2}{\Omega^2}\nablav\right)^{n-1}\Psilin(u,v,\theta^A),
    \end{align}
    which leads to 
    \begin{align}
        \lim_{u\longrightarrow-\infty} \frac{1}{r^n}\left(\frac{r^2}{\Omega^2}\nablav\right)^n\Psilin(u,v,\theta^A)=\prod_{j=0}^{n-1}\left[j+\frac{1}{j+1}(\mathring{\slashed{\Delta}}-4)\right]\pastmemoryPsilin.
    \end{align}
    Turning to the $(\ell,m)$\textsuperscript{th} harmonic mode of the scalar components of $\Psilin$, the above translates to
    \begin{align}
        \begin{split}
            \Flinslp{\ell}{n}:=&\lim_{u\longrightarrow-\infty} \frac{1}{r^n}\left(\frac{r^2}{\Omega^2}\partial_v\right)^n\Flins{\ell}{n}(u,v,\theta^A)=\frac{n(n-1)-\ell(\ell+1)}{n}\Flinslp{\ell}{n-1}\\
            &=(-1)^n\frac{(\ell+n)!}{n!{(\ell-n)!}}\;\flinslp(v).
        \end{split}
    \end{align}
    \end{proof}
    
    The relations \bref{ell iterated integrals 2} and \bref{anomaly integral}, taking the limit $y\longrightarrow0$ which corresponds to $v\longrightarrow\infty$, leads to
    \begin{align}
        \begin{split}
            \flinslf(u)-\sum_{j=0}^{\ell}\frac{(-1)^j}{j!}(y_+)^j\partial_y^j\flinl\Big|_{u,v_+}.
        \end{split}
    \end{align}
    Taking the limit of the above as $u\longrightarrow-\infty$ yields
    \begin{align}
        \begin{split}
            \lim_{u\longrightarrow-\infty}\flinslf=\sum_{j=1}^\ell(-1)^j\frac{(\ell+j)!}{(j!)^2(\ell-j)!}\;\flinslp(v_+)=\sum_{j=1}^\ell(-1)^j\binom{\ell+j}{j}\binom{\ell}{j}\;\flinslp(v_+).
        \end{split}
    \end{align}
    We will use Egorychev's method to show the following
    \begin{lemma}\label{Egorychev}
        \begin{align}
            \sum_{j=1}^\ell(-1)^j\binom{\ell+j}{j}\binom{\ell}{j}=(-1)^\ell.
        \end{align}
    \end{lemma}
    \begin{proof}
        We can express the binomial coefficient in terms of a contour integral as follows: for any $\epsilon>0$ we have
        \begin{align}
            \binom{\ell}{j}=\frac{1}{2\pi i}\int_{|z|=\epsilon}\frac{1}{(1-z)^{j+1}z^{\ell-j+1}}dz.
        \end{align}
        Thus
        \begin{align}
            \begin{split}
                \sum_{j=1}^\ell(-1)^j\binom{\ell+j}{j}\binom{\ell}{j}&=\frac{1}{2\pi i}\int_{|z|=\epsilon}\sum_{j=1}^\ell(-1)^j\binom{\ell}{j}\frac{1}{(1-z)^{j+1}z^{\ell+1}}=\frac{1}{2\pi i}\int_{|z|=\epsilon}\left(1-\frac{1}{1-z}\right)^\ell\frac{1}{z^{\ell+1}(1-z)}\\
                &=\frac{(-1)^\ell}{2\pi i}\int_{|z|=\epsilon}\frac{1}{z(1-z)^{\ell+1}}=(-1)^\ell.
            \end{split}
        \end{align}     
    \end{proof}

In view of \Cref{Egorychev}, we can conclude
\begin{corollary}\label{RW antipodal map ell mode}
    The solution $\SscriHp$ satisfies
    \begin{align}
        \lim_{u\longrightarrow-\infty}\flinslf(u)=(-1)^\ell\;\flinslp(v_+).
    \end{align}
\end{corollary}
    
\begin{remark}
    Note that the spherical harmonics on $S^2$ transform under the antipodal map $(\theta,\varphi)\longrightarrow(\pi-\theta,\pi+\varphi)$ via
    \begin{align}
        Y_{\ell,m}(\pi-\theta,\pi+\varphi)=(-1)^\ell Y_{\ell,m}(\theta,\varphi).
    \end{align}
    Let $s^{(1)},s^{(2)}$ be scalar functions on the sphere whose spherical harmonic expansion coefficients satisfy
    \begin{align}
        s^{(1)}_{(\ell,m)}=(-1)^\ell s^{(2)}_{(\ell,m)},
    \end{align}
    then
    \begin{align}
        s^{(1)}(\theta,\varphi)=s^{(2)}(\pi-\theta,\pi+\varphi).
    \end{align}
\end{remark}

Given \bref{sup Psilin scri constant ell} and \Cref{RW antipodal map ell mode}, we deduce via the Weierstrass M-test that
\begin{corollary}\label{RW final matching}
    For the solution $\SscriHp$, $\upPsilin_{\mathscr{I}^+}$ attains a limit as $u\longrightarrow-\infty$ and we have
    \begin{align}
        \lim_{u\longrightarrow-\infty}\upPsilin_{\mathscr{I}^+}(u,\theta^A)=\mathscr{A}[\pastmemoryPsilin]=\lim_{v\longrightarrow\infty}[\mathscr{A}\upPsilin_{\mathscr{I}^-}](v,\theta^A),
    \end{align}
    where $\mathscr{A}$ is the antipodal map on $S^2$. The same relation applies to the pair $\upPsilinb_{\mathscr{I}^\pm}$.
\end{corollary}

\subsection{Matching past and future linear memory near $i^0$}\label{Section 12.2 memory matching}

We are now ready to prove
\begin{corollary}
    For the solution $\Sscriscriscri$ of \Cref{def of scriscriscri}, the quantities $\Sigma^\mp_{\mathscr{I}^\pm}$ defined in \eqref{def of Sigma scri pm} satisfy
    \begin{align}
       \Sigma^-_{\mathscr{I}^+}=\overline{\Sigma^+_{\mathscr{I}^-}\circ \mathscr{A}},
    \end{align}
    where $\mathscr{A}$ is the antipodal map on $S^2$ and  where the conjugate $\overline{\Xi}$ of a symmetric traceless $\mathcal{S}^2_{u,v}$ 2-tensor $\Xi$ was defined in \Cref{definition of conjugates}.
\end{corollary}

\begin{proof}

Given that $\Sscriscriscri$ is both past and future Bondi-normalised and that $\xlins_{\mathscr{I}^+},\upPsilin_{\mathscr{I}^+}\longrightarrow0$ as $u\longrightarrow\infty$, $\xblins_{\mathscr{I}^-}, \upPsilinb_{\mathscr{I}^-}\longrightarrow0$ as $v\longrightarrow-\infty$, we have from \bref{further constraint -}, \bref{further constraint +},
\begin{align}\label{memory limit 2}
    \upPsilinb_{\mathscr{I}^+}=\left(\mathring{\slashed{\Delta}}-4\right)\left(\mathring{\slashed{\Delta}}-2\right)\xlins_{\mathscr{I}^+}-6M\xblins_{\mathscr{I}^+},\qquad\qquad \upPsilin_{\mathscr{I}^-}=\left(\mathring{\slashed{\Delta}}-4\right)\left(\mathring{\slashed{\Delta}}-2\right)\xblins_{\mathscr{I}^-}+6M\xlins_{\mathscr{I}^-}.
\end{align}
 The formulae \eqref{expression for Psilin}, \eqref{expression for Psilinb} tell us that that $\rlins_{\mathscr{I}^+}$, $\slins_{\mathscr{I}^+}$ attain limits defining smooth functions on $S^2$ as $u\longrightarrow-\infty$, and decay in the limit $u\longrightarrow\infty$. Define
\begin{align}
    \Pscri_{\mathscr{I}^+}^\pm=\lim_{u\longrightarrow\pm\infty} \rlins_{\mathscr{I}^+}, \qquad \Pscri_{\mathscr{I}^-}^\pm=\lim_{v\longrightarrow\pm\infty} \rlins_{\mathscr{I}^-},
\end{align}

\begin{align}
    \Qscri_{\mathscr{I}^+}^\pm=\lim_{u\longrightarrow\pm\infty} \slins_{\mathscr{I}^+}, \qquad \Qscri_{\mathscr{I}^-}^\pm=\lim_{v\longrightarrow\pm\infty} \slins_{\mathscr{I}^-}.
\end{align}
Then we have
\begin{align}
    \lim_{u\longrightarrow-\infty}\upPsilin_{\mathscr{I}^+}=2\fancydstarring_2\fancydstarring_1\left(\Pscri_{\mathscr{I}^+}^-,-\Qscri_{\mathscr{I}^+}^-\right),\qquad  \lim_{u\longrightarrow-\infty}\upPsilinb_{\mathscr{I}^+}=2\fancydstarring_2\fancydstarring_1\left(\Pscri_{\mathscr{I}^+}^-,\Qscri_{\mathscr{I}^+}^-\right),
\end{align}
\begin{align}
    \lim_{v\longrightarrow\infty}\upPsilin_{\mathscr{I}^-}=2\fancydstarring_2\fancydstarring_1\left(\Pscri_{\mathscr{I}^-}^+,-\Qscri_{\mathscr{I}^-}^+\right),\qquad  \lim_{v\longrightarrow\infty}\upPsilinb_{\mathscr{I}^-}=2\fancydstarring_2\fancydstarring_1\left(\Pscri_{\mathscr{I}^-}^+,\Qscri_{\mathscr{I}^-}^+\right).
\end{align}
The relations \Cref{RW final matching} then imply
\begin{align}\label{PQ matching}
    \Pscri_{\mathscr{I}^+}^-=\Pscri_{\mathscr{I}^-}^+\circ\mathscr{A},\qquad \Qscri_{\mathscr{I}^+}^-=\Qscri_{\mathscr{I}^-}^+\circ\mathscr{A}.
\end{align}
The relation \eqref{memory limit 1} tells us that
\begin{align}
    \fancydring_1\fancydring_2\left[\lim_{u\longrightarrow-\infty}\xlins_{\mathscr{I}^+}\right]=(\Pscri_{\mathscr{I}^+}^-,\Qscri_{\mathscr{I}^+}^-),
\end{align}
while at $\mathscr{I}^-$ we have

\begin{align}
    \fancydring_1\fancydring_2\left[\lim_{u\longrightarrow-\infty}\xblins_{\mathscr{I}^-}\right]=(\Pscri_{\mathscr{I}^-}^+,-\Qscri_{\mathscr{I}^-}^+).
\end{align}
Let $f,g$ be the scalar components of $\lim_{u\longrightarrow-\infty}\xlins_{\mathscr{I}^+}$ according to \Cref{2-forms on S^2}. Then we have
\begin{align}
    \Pscri_{\mathscr{I}^+}^-=\left(\mathring{\slashed{\Delta}}-2\right)\left(\mathring{\slashed{\Delta}}-4\right)f,\qquad \Qscri_{\mathscr{I}^+}^-=\left(\mathring{\slashed{\Delta}}-2\right)\left(\mathring{\slashed{\Delta}}-4\right)g,
\end{align}
Let $\underline{f}$, $\underline{g}$ be the scalar components of $\lim_{v\longrightarrow\infty}\xblins_{\mathscr{I}^-}$ according to \Cref{2-forms on S^2}. Then we have
\begin{align}
    \Pscri_{\mathscr{I}^-}^+=\left(\mathring{\slashed{\Delta}}-2\right)\left(\mathring{\slashed{\Delta}}-4\right)\underline{f},\qquad \Qscri_{\mathscr{I}^-}^+=\left(\mathring{\slashed{\Delta}}-2\right)\left(\mathring{\slashed{\Delta}}-4\right)\underline{g}. 
\end{align}
Thus \eqref{PQ matching} immediately implies
\begin{align}
    f=\underline{f}\circ\mathscr{A},\qquad g=-\underline{g}\circ\mathscr{A}.
\end{align}
\end{proof}

As $\xblins_{\mathscr{I}^+}\in\mathcal{E}^T_{\mathscr{I}^+}$,  $\xlins_{\mathscr{I}^-}\in\mathcal{E}^T_{\mathscr{I}^-}$, we have that $\xblins_{\mathscr{I}^+}, \xlins_{\mathscr{I}^-}\longrightarrow0$ as $u\longrightarrow\pm\infty$ and $v\longrightarrow\pm\infty$ respectively. Thus $\Sigma^+_{\mathscr{I}^+}=\Sigma^-_{\mathscr{I}^-}=0$. This concludes the proof of \ref{matching proposition}.

\appendix 

\section[Estimates on initial data on $\Sigma^*_+$ in terms of $\protect\alin$, $\protect\ablin$]{Estimates on initial data on $\Sigma^*_+$ in terms of $\protect\alin$, $\protect\ablin$ in the $\Sigma^*_+$ gauge}

In this section we show that, given solutions $\alpha$, $\underline{\alpha}$ to the $+2$ and $-2$ Teukolsky equations (equations \bref{T+2} and \bref{T-2} respectively) on the exterior region $\overline{\mathscr{M}}$, we can construct an initial data set, unique up to linearised Kerr modes, on a given spacelike hypersurface. Without loss of generality we will work with the hypersurface $\Sigma^*_+$.

\begin{proposition}\label{Appendix A construction mode}
    Let $\alpha$, $\underline{\alpha}$ be solutions to the Teukolsky equations \eqref{T+2}, \eqref{T-2} respectively, arising from smooth data on $\Sigma^*_+$. Then there exists a unique Cauchy data set on $\Sigma^*_+$ for the system \fullsystem as in \Cref{def of initial data on spacelike surface} which satisfies the conditions of $\Sigma^*_+$ gauge.
\end{proposition}

\begin{proof}
 To start with, we define $\psi$, $\underline{\psi}$, $\Psi$, $\underline{\Psi}$ via \bref{hier}, \bref{hier'}. Then $\slin$ can be defined via
\begin{align}
&4\mathring{\fancydstar_2}\mathring{\fancydstar_1}(0,r^3\slin_{\ell\geq2}|_{{\mathscr{H}^+_{\geq0}\cap\Sigma^*_+}})=\left(\underline\Psi-{\Psi}\right)\Big|_{{\mathscr{H}^+_{\geq0}\cap\Sigma^*_+}}\label{slin in terms of Psilin Psilinb appendix}.
\end{align}
Commuting \bref{slin in terms of Psilin Psilinb appendix} with $\partial_t$, we can use \bref{elliptic equation 1} to find $\curlo\divo\,\Omega\xlin|_{\mathscr{H}^+_{\geq0}\cap\Sigma^*_+}$ and $\curlo\divo\,\Omega\xblin|_{\mathscr{H}^+_{\geq0}\cap\Sigma^*_+}$. This determines $\Omega^{-1}\xblin$ as the gauge conditions \bref{Sigma*+ gauge conditions} require $\divo^2\Omega^{-1}\xblin=0$.  We now find $\divo^2\Omega\xlin$ on all of $\Sigma^*_+$ using the first equation of \bref{Sigma*+ gauge conditions}, and this then determines $\Omega\xlin$ on $\Sigma^*_+$ as $\fancydring_1\fancydring_2\Omega\xlin$ determines $\Omega\xlin$ uniquely. To do so, we need to find $\divo^2\,\Omega\xlin$, $\divo^2\,\partial_r\Omega\xlin$ at $\mathscr{H}^+_{\geq0}\cap\Sigma^*_+$.

Since in the $\Sigma^*_+$ gauge the horizon gauge conditions \bref{initial horizon gauge condition} are satisfied, the Codazzi equation \bref{elliptic equation 2} gives
\begin{align}\label{xlin blin appendix}
    \divo^2\,\Omega\xlin|_{\Sigma^*_+\cap \mathscr{H}^+_{\geq0}}=-2M\divo\,\Omega\blin|_{\Sigma^*_+\cap \mathscr{H}^+_{\geq0}}.
\end{align}
We can then find $\divr^2\Omega\xlin$ at $\Sigma^*_+\cap \mathscr{H}^+_{\geq0}$ using \bref{Bianchi+2}, which now reads
\begin{align}\label{xlin at H+ cap Sigma+ appendix}
    \left[\divr^2\Omega^2\,\alin+2M\Omega^{-1}\nablagml \divo^2\,\Omega^2\alin\right]\Big|_{\Sigma^*_+\cap \mathscr{H}^+_{\geq0}}=-\frac{1}{2M}(\mathring{\slashed{\Delta}}-1)\divo^2\,\Omega\xlin|_{\Sigma^*_+\cap \mathscr{H}^+_{\geq0}}.
\end{align}
Equation \bref{xlin at H+ cap Sigma+ appendix} can be solved to get a unique $\Omega\xlin|_{\Sigma^*_+\cap \mathscr{H}^+_{\geq0}}$ given the left hand side. 

We now find $\partial_r\Omega\xlin|_{\Sigma^*_+\cap \mathscr{H}^+_{\geq0}}$. Knowing $\Omega\xlin|_{\Sigma^*_+\cap \mathscr{H}^+_{\geq0}}$, we can find $\rlin_{\ell\geq2}|_{\Sigma^*_+\cap \mathscr{H}^+_{\geq0}}$ in terms of $\alpha$, $\underline{\alpha}$ and $\Omega\xlin|_{\Sigma^*_+\cap \mathscr{H}^+_{\geq0}}$ using the formula for $\Psilin$ given in \bref{formula for Psilin Psilinb in terms of linearised system}: defining $\Psi$ in terms of $\alpha$ via \bref{hier}, we define $\rlin_{\ell\geq2}|_{\Sigma^*_+\cap \mathscr{H}^+_{\geq0}}$, $\slin_{\ell\geq2}|_{\Sigma^*_+\cap \mathscr{H}^+_{\geq0}}$ via 
\begin{align}\label{appendix appendix}
    2\fancydstarring_2\fancydstarring_1(\rlin_{\ell\geq2},\slin_{\ell\geq2})|_{\Sigma^*_+\cap \mathscr{H}^+_{\geq0}}:=\Psi-12M^2\Omega\xlin|_{\Sigma^*_+\cap \mathscr{H}^+_{\geq0}}.
\end{align}
We can now find $\divo\elin$ at ${\mathscr{H}^+_{\geq0}\cap\Sigma^*_+}$ using the $\Sigma^*_+$ gauge conditions, namely
\begin{align}\label{appendix appendix appendix}
    \left[\rlin-\rlin_{\ell=0}+\divr\,\elin\right]\Big|_{\mathscr{H}^+_{\geq0}\cap\Sigma^*_+}=0.
\end{align}
while $\curlo\slin$ at ${\mathscr{H}^+_{\geq0}\cap\Sigma^*_+}$ can be found using \bref{elliptic equation 3 and 4}. This gives us $\elin|_{\mathscr{H}^+_{\geq0}\cap\Sigma^*_+}$ since $\fancyd_1\elin$ determines $\elin$ uniquely. We can now find $\partial_r\Omega\xlin|_{\mathscr{H}^+_{\geq0}\cap\Sigma^*_+}$ using \bref{D3Chihat} and \bref{Bianchi+1b}.

Having found $\Omega\xlin$, $\Omega^{-1}\xblin$, we can recover $\elin_{\ell\geq2}$, $\eblin_{\ell\geq2}$ everywhere on $\Sigma^*_+$ using \bref{D3Chihat}, \bref{D4Chihatbar}, thus we can get $\Olin_{\ell\geq2}$ via \bref{elin eblin Olin}. We can extend $\rlin_{\ell\geq2}$, $\slin_{\ell\geq2}$ to the whole of $\Sigma^*_+$ via \bref{formula for Psilin Psilinb in terms of linearised system}. The quantities $\blin_{\ell\geq2}$, $\bblin_{\ell\geq2}$ can be found using the equations \bref{Bianchi+2}, \bref{Bianchi-2}. We can find $\otx_{\ell\geq2}$, $\otxb_{\ell\geq2}$ using the Codazzi equations \bref{elliptic equation 1}, \bref{elliptic equation 2}. This then determines $\tr\glin_{\ell\geq2}$ via the Gauss equation \bref{Gauss}. This procedure gives a complete seed data set for the system \fullsystem via \Cref{constructing compactly supported data}.
\end{proof}

We now estimate the components of the data set constructed above.
\begin{corollary}
    The Cauchy data set $\mathfrak{C}$ for \fullsystem constructed in \Cref{Appendix A construction mode} satisfies the following estimate: let $\overone{g}$ denote the linearised metric components, $\overone{\Gamma}$ denote the linearised connection coefficients, and $\overone{R}$ denote the linearised curvature components of the solution arising from $\mathfrak{C}$ via \Cref{EinsteinWP}, then on $\Sigma^*_+$ we have
\begin{align}\label{Appendix A estimation mode estimate}
\begin{split}
    \int_{\Sigma^*_+}dr\dw\,&r^n|\overone{g}|^2+r^{n+1}|\overone{\Gamma}|^2+|\overone{R}|^2\\
    &\lesssim \int_{\Sigma^*_+}dr\dw\,\Bigg[r^{n+2}|\mathcal{A}_2\Omega^2\alin|^2+r^{n+3}|\slashednabla_r\Omega^2\alin|^2+r^{n+4}|\slashednabla_r\slashednabla_{t^+_*}\Omega^2\alin|^2+r^{n+3}|\mathcal{A}_2\partial_{t^*_+}\Omega^2\alin|^2\\
    &\qquad\qquad\qquad\qquad +r^{n+2}|\mathcal{A}_2\Omega^{-2}\ablin|^2+r^{n+3}|\slashednabla_r\Omega^{-2}\ablin|^2+r^{n+4}|\slashednabla_r\slashednabla_{t^+_*}\Omega^{-2}\ablin|^2+r^{n+3}|\mathcal{A}_2\partial_{t^*_+}\Omega^{-2}\ablin|^2\Bigg],
\end{split}
\end{align}
\end{corollary}
\begin{proof}
To start with, note that the the $\Sigma^*_+$ gauge condition on $\divr^2\Omega\xlin$ leads gives
\begin{align}
    2\divr^2\Omega\xlin|_{\Sigma^*_+}=\left[2\divr^2\Omega\xlin|_{\Sigma^*_+\cap\mathscr{H}^+_{\geq0}}+\partial_r\divr^2\Omega\xlin\right]e^{-\left(\frac{r}{2M}-1\right)}-\left[\divr^2\Omega\xlin|_{\Sigma^*_+\cap\mathscr{H}^+_{\geq0}}+\partial_r\divr^2\Omega\xlin\right]e^{-\left(\frac{r}{M}-1\right)}.
\end{align}

Thus we find that
\begin{align}\label{gauge estimate on xlin appendix}
    \int_{\Sigma^*_+}dr\dw\,r^n|\divo^2\Omega\xlin|^2\leq C(M,n)\footnotemark \int_{\Sigma^*_+\cap\mathscr{H}^+_{\geq0}}\dw\,|\divo^2\Omega\xlin|^2+|\partial_r\divo^2\Omega\xlin|^2.
\end{align}
\footnotetext{Note that $C(n)\sim (2M)^n\frac{n!}{4^n}$}
We may estimate $\slin_{\ell\geq2}$ using \bref{slin in terms of Psilin Psilinb appendix} and Poincar\'e's inequality:
\begin{align}
\begin{split}
    \int_{\Sigma^*_+}dr\dw\,r^n|\mathring{\slashed{\Delta}}\slin_{\ell\geq2}|^2&\lesssim \int_{\Sigma^*_+}dr\dw\,r^{n-3}\left[|\Psilin|^2+|\Psilinb|^2\right]\\
    &\lesssim \int_{\Sigma^*_+}dr\dw\,\Bigg[r^n|\mathcal{A}_2\Omega^2\alin|^2+r^{n+1}|\slashednabla_r\Omega^2\alin|^2+r^{n+2}|\slashednabla_r\slashednabla_{t^+_*}\Omega^2\alin|^2+\\
    &\qquad\qquad\qquad\qquad r^n|\mathcal{A}_2\Omega^{-2}\ablin|^2+r^{n+1}|\slashednabla_r\Omega^{-2}\ablin|^2+r^{n+2}|\slashednabla_r\slashednabla_{t^+_*}\Omega^{-2}\ablin|^2\Bigg].
    \end{split}
\end{align}
An identical estimate applies to $\partial_t\slin_{\ell\geq2}$:
\begin{align}
\begin{split}
    \int_{\Sigma^*_+}dr\dw\,&r^n|\mathring{\slashed{\Delta}}\partial_t \slin_{\ell\geq2}|^2\\
    &\lesssim \int_{\Sigma^*_+}dr\dw\,\Bigg[r^n|\mathcal{A}_2\Omega^2\alin|^2+r^{n+1}|\slashednabla_r\Omega^2\alin|^2+r^{n+2}|\slashednabla_r\slashednabla_{t^+_*}\Omega^2\alin|^2+r^{n+1}|\mathcal{A}_2\partial_{t^*_+}\Omega^2\alin|^2\\
    &\qquad\qquad\qquad\qquad +r^n|\mathcal{A}_2\Omega^{-2}\ablin|^2+r^{n+1}|\slashednabla_r\Omega^{-2}\ablin|^2+r^{n+2}|\slashednabla_r\slashednabla_{t^+_*}\Omega^{-2}\ablin|^2+r^{n+1}|\mathcal{A}_2\partial_{t^*_+}\Omega^{-2}\ablin|^2\Bigg].
\end{split}
\end{align}
Combining the above with \bref{xlin at H+ cap Sigma+ appendix}, \bref{gauge estimate on xlin appendix}, \bref{appendix appendix}, \bref{appendix appendix appendix}, equation \bref{D3Chihat} and \bref{Bianchi+1a} leads to similar estimates on $\xlin$, $\xblin$:
\begin{align}
\begin{split}
    \int_{\Sigma^*_+}dr\dw\,&r^n(|\fancydring_2\fancydring_1\Omega\xlin|^2+|\fancydring_2\fancydring_1\Omega^{-1}\xblin|^2)\\
    &\lesssim \int_{\Sigma^*_+}dr\dw\,\Bigg[r^{n+1}|\mathcal{A}_2\Omega^2\alin|^2+r^{n+2}|\slashednabla_r\Omega^2\alin|^2+r^{n+3}|\slashednabla_r\slashednabla_{t^+_*}\Omega^2\alin|^2+r^{n+2}|\mathcal{A}_2\partial_{t^*_+}\Omega^2\alin|^2\\
    &\qquad\qquad\qquad\qquad +r^{n+1}|\mathcal{A}_2\Omega^{-2}\ablin|^2+r^{n+2}|\slashednabla_r\Omega^{-2}\ablin|^2+r^{n+3}|\slashednabla_r\slashednabla_{t^+_*}\Omega^{-2}\ablin|^2+r^{n+2}|\mathcal{A}_2\partial_{t^*_+}\Omega^{-2}\ablin|^2\Bigg].
\end{split}
\end{align}
  Continuing in this manner using the remainder of the equations \fullsystem, we can derive the estimate \eqref{Appendix A estimation mode estimate} from the remaining equations of the system \fullsystem.
\end{proof}
\section{The uniqueness clause in backwards scattering}\label{missing note from Part I}

For any element of $(F_{\mathscr{H}^+},F_{\mathscr{I}^+})$ $\mathcal{E}^{T,RW}_{{\mathscr{H}^+_{\geq0}}}\oplus\mathcal{E}^{T,RW}_{{\mathscr{I}^+}}$, we may use the isomorphism with $\mathcal{E}^{T,RW}_{\Sigma^*_+}$ to obtain an initial data set for the Regge--Wheeler equation with finite $\partial_t$-energy. This allows us to use the well-posedness of the Cauchy problem for \bref{RW} to construct a solution $\Psi$ to \eqref{RW} on $D^+(\Sigma^*_+)$. This solution realises $(F_{\mathscr{H}^+},F_{\mathscr{I}^+})$ in the sense that the trace of $\Psi$ on $\mathscr{H}^+_{\geq0}$ $F_{\mathscr{H}^+}$ and
\begin{align}
    \lim_{v\longrightarrow\infty}\int_{u_1}^{u_2}\int_{S^2}du \dw\,|\nablau\Psi(u,v,\theta^A)-F_{\mathscr{I}^+}(u,\theta^A)|^2\longrightarrow0
\end{align}
for any finite $u_1,u_2$. It follows immediately that a solution $\Psi$ to \bref{RW} for which $\int_{S^2}\dw\,|\nablau\Psi|^2$ decays towards $\mathscr{I}^+$ and which has a vanishing trace on $\mathscr{H}^+_{\geq0}$ must vanish. It is easy to obtain the same statement when $\nablau\Psi$ decays pointwise towards $\mathscr{I}^+$. This is the uniqueness clause in the backwards scattering constructions of Theorem I (i.e.~Theorem 1 in \cite{Mas20}). Analogous statements hold for the Teukolsky equations \bref{T+2}, \bref{T-2}.\\

In this section we present a detailed argument showing that any solution with trivial trace on $\mathscr{H}^+_{\geq0}$ such that 
\begin{align}
    \lim_{v\longrightarrow\infty}\Psi(u,v,\theta^A)=0
\end{align}
must be trivial.

\begin{lemma}\label{convergence of radiation field in finite energy}
Let $\Psi$ be a solution to the Regge--Wheeler equation \bref{RW} on $J^+(\Sigma^*_+)$, arising from finite $\partial_t$-energy initial data on $\Sigma^*_+$. Then for any finite $u_1,u_2\in\mathbb{R}$, $\partial_u\Psi(u,v,\theta^A)$  converges in $L^2([u_1,u_2]\times\mathbb{S}^2)$ as $v\longrightarrow\infty$.
\end{lemma}

\begin{proof}
Assume first that $\Psi$ is $\Gamma^{(2)}(J^+(\Sigma^*_+))$ arising from compactly supported data. Then indeed we may estimate a sequence $\partial_u\Psi(u,v_n,\theta^A)$ in $L^2([u_1,u_2]\times\mathbb{S}^2)$ using an ILED estimate: for $v_1,v_2$ sufficiently large we have
\begin{align}
\begin{split}
    \int_{u_1}^{u_2}\int_{S^2}d\bar{u}\dw|\partial_u\Psi(\bar{u},v_2,\theta^A)-\partial_u\Psi(\bar{u},v_1,\theta^A)|^2&=\int_{S^2}\dw\int_{u_1}^{u_2}d\bar{u}\left|\int_{v_1}^{v_2}\frac{1}{r^2}(\mathring{\slashed{\Delta}}\Psi+(3\Omega^2+1)\Psi\right|^2
    \\&\lesssim \frac{1}{r(u_1,v_1)}\mathbb{F}^{T,2}_{\Sigma^*_+}[\Psi].
\end{split}
\end{align}
For $\Psi$ arising from $\mathbb{F}^T_{\Sigma^*_+}[\Psi]$-finite initial data and an $\epsilon>0$, there exists $\Psi_{C^\infty}$ which solves Regge--Wheeler, is smooth and which satisfies 
\begin{align}
    \mathbb{F}^{T,2}_{\Sigma^*_+}[\Psi-\Psi_{C^\infty}]\lesssim \frac{\epsilon}{3}.
\end{align}
Then for $u_1,u_2$ finite and $v_1,v_2$ sufficiently large with $v_2>v_1$, we have
\begin{align}
    \begin{split}
        &\int_{u_1}^{u_2}\int_{S^2}d\bar{u}\dw|\partial_u\Psi(\bar{u},v_2,\theta^A)-\partial_u\Psi(\bar{u},v_1,\theta^A)|^2\\&\lesssim\int_{u_1}^{u_2}\int_{S^2}d\bar{u}\dw|\partial_u\Psi(\bar{u},v_2,\theta^A)-\partial_u\Psi_{C^\infty}(\bar{u},v_2,\theta^A)|^2\\&+\int_{u_1}^{u_2}\int_{S^2}d\bar{u}\dw|\partial_u\Psi_{C^\infty}(\bar{u},v_2,\theta^A)-\partial_u\Psi_{C^\infty}(\bar{u},v_1,\theta^A)|^2\\&+\int_{u_1}^{u_2}\int_{S^2}d\bar{u}\dw|\partial_u\Psi(\bar{u},v_1,\theta^A)-\partial_u\Psi_{C^\infty}(\bar{u},v_1,\theta^A)|^2.
    \end{split}
\end{align}
The energy identity allows to uniformly bound the first and third terms on the right hand side above by $\frac{\epsilon}{3}$. Choosing $v_1$ sufficiently large concludes the argument.
\end{proof}

\begin{corollary}\label{convergence of radiation field in L1 loc under finite energy}
Let $\Psi$ be as in \Cref{convergence of radiation field in finite energy}. Then for any $u_1,u_2$ there exists a sequence $\{v_n\}$ with $v_n\longrightarrow\infty$ as $n\longrightarrow\infty$ such that $\int_{u_1}^{u_2}d\bar{u}\;\partial_u\Psi(\bar{u},v_n,\theta^A)$ converges as $n\longrightarrow\infty$ almost everywhere in $S^2$.
\end{corollary}

\begin{corollary}\label{decay of energy towards scri+}
Let $\Psi$ be as in \Cref{convergence of radiation field in finite energy}. If for any $u\in\mathbb{R}$, $\theta^A\in \mathbb{S}^2$ there exists a sequence $\{v_n\}$ with $v_n\longrightarrow\infty$ as $n\longrightarrow\infty$ such that $\partial_u\Psi(u,v_n,\theta^A)\longrightarrow0$, then $\int_{u_1}^{u_2}d\bar{u}\;\partial_u\Psi(\bar{u},v_n,\theta^A)$ decays as $v\longrightarrow\infty$.
\end{corollary}

\begin{corollary}
Let $\Psi$ be as in \Cref{convergence of radiation field in finite energy}. Then for any $u_1,u_2$ there exists a sequence $\{v_n\}$ with $v_n\longrightarrow\infty$ as $n\longrightarrow\infty$ such that $\partial_u\Psi$ converges pointwise almost everywhere in $[u_1,u_2]\times \mathbb{S}^2$ as $n\longrightarrow\infty$.
\end{corollary}

\begin{proposition}\label{Final uniqueness clause for RW}
Assume $\Psi$ is a solution to the Regge--Wheeler equation \bref{RW} on $J^+(\Sigma^*_+)$, arising from finite $\partial_t$-energy initial data on $\Sigma^*_+$ such that
\begin{align}
    \lim_{v\longrightarrow\infty} \Psi=0,\qquad\qquad \Psi|_{\mathscr{H}^+_{\geq0}}=0.
\end{align}
Then $\Psi$ vanishes identically.
\end{proposition}

\begin{proof}
    Since $\Psi(u,v,\theta^A)$ decays for any $u$, we can immediately conclude by \Cref{convergence of radiation field in L1 loc under finite energy} that for any $u_1,u_2$ there is a sequence $v_n$ with $v_n\longrightarrow\infty$ as $n\longrightarrow\infty$, such that $\int_{u_1}^{u_2}d\bar{u}\,\partial_u\Psi(\bar{u},v_n,\theta^A)$ decays. Therefore, $\int_{u_1}^{u_2}\int_{S^2}d\bar{u}\dw|\partial_u\Psi(\bar{u},v_n,\theta^A)|^2$ decays as $n\longrightarrow\infty$. The $\partial_t$-energy identity implies that $\mathbb{F}^T_{\Sigma^*_+}[\Psi]=0$.
\end{proof}

\addcontentsline{toc}{section}{References}
\printbibliography

@inproceedings{Chandrasekhar,
      author         = "Chandrasekhar, Subrahmanyan",
      title          = "{The mathematical theory of black holes}",
      booktitle      = "{Oxford, UK: Clarendon (1992) 646 p.}",
      year           = "1985",
}

@ARTICLE{1916SPAW.......688E,
       author = {{Einstein}, Albert},
        title = "{N{\"a}herungsweise Integration der Feldgleichungen der Gravitation}",
      journal = {Sitzungsberichte der K{\"o}niglich Preu{\ss}ischen Akademie der Wissenschaften (Berlin},
         year = 1916,
        month = jan,
        pages = {688-696},
       adsurl = {https://ui.adsabs.harvard.edu/abs/1916SPAW.......688E}
}

@article{Prabhu,
    author = "Prabhu, Kartik and Shehzad, Ibrahim",
    title = "{Conservation of asymptotic charges from past to future null infinity: Lorentz charges in general relativity}",
    eprint = "2110.04900",
    archivePrefix = "arXiv",
    primaryClass = "gr-qc",
    doi = "10.1007/JHEP08(2022)029",
    journal = "JHEP",
    volume = "08",
    pages = "029",
    year = "2022"
}

@article{Capone,
    author = "Capone, Federico and Nguyen, Kevin and Parisini, Enrico",
    title = "{Charge and Antipodal Matching across Spatial Infinity}",
    eprint = "2204.06571",
    archivePrefix = "arXiv",
    primaryClass = "hep-th",
    month = "4",
    year = "2022"
}

@article{Lin-Rod,
      author         = "Lindblad, Hans and Rodnianski, Igor",
      title          = "{The Global stability of the Minkowski space-time in
                        harmonic gauge}",
      year           = "2004",
      eprint         = "math/0411109",
      archivePrefix  = "arXiv",
      primaryClass   = "math-ap",
      SLACcitation   = "%%CITATION = MATH/0411109;%%"
}

@article{Zeldovich,
    author = "Zel'dovich, Y. B. and Polnarev, A. G.",
    title = "{Radiation of gravitational waves by a cluster of superdense stars}",
    journal = "Sov. Astron.",
    volume = "18",
    pages = "17",
    year = "1974"
}

@article{BRAGINSKY,
title = {Gravitational waves and the detection of gravitational radiation},
journal = {Physics Reports},
volume = {46},
number = {5},
pages = {165-200},
year = {1978},
issn = {0370-1573},
doi = {https://doi.org/10.1016/0370-1573(78)90192-8},
url = {https://www.sciencedirect.com/science/article/pii/0370157378901928},
author = {V.B. Braginsky and V.N. Rudenko}
}

@article{KehrbergerI,
   title={The Case Against Smooth Null Infinity I: Heuristics and Counter-Examples},
   ISSN={1424-0661},
   url={http://dx.doi.org/10.1007/s00023-021-01108-2},
   DOI={10.1007/s00023-021-01108-2},
   journal={Annales Henri Poincaré},
   publisher={Springer Science and Business Media LLC},
   author={Kehrberger, Leonhard M. A.},
   year={2021},
   month={Sep}
}

@ARTICLE{KehrbergerII,
  title         = "The Case Against Smooth Null Infinity {II}: A
                   logarithmically modified Price's law",
  author        = "Kehrberger, Leonhard M A",
  month         =  may,
  year          =  2021,
  copyright     = "http://arxiv.org/licenses/nonexclusive-distrib/1.0/",
  archivePrefix = "arXiv",
  primaryClass  = "gr-qc",
  eprint        = "2105.08084"
}

@ARTICLE{KehrbergerIII,
  title         = "The case against smooth null infinity {III}: Early-time
                   asymptotics for higher $\ell$-modes of linear waves on a
                   Schwarzschild background",
  author        = "Kehrberger, Leonhard M A",
  month         =  may,
  year          =  2021,
  copyright     = "http://arxiv.org/licenses/nonexclusive-distrib/1.0/",
  archivePrefix = "arXiv",
  primaryClass  = "gr-qc",
  eprint        = "2106.00035"
}

@article{KehrbergerM,
 title="Early Time Asymptotics for Linearised Gravity Around Schwarzschild (working title)",
 author="L. M. A. Kehrberger and H. Masaood.",
 year=2022,
 note="to appear"
}

@ARTICLE{KehrbergerGajic,
  title         = "On the relation between asymptotic charges, the failure of
                   peeling and late-time tails",
  author        = "Gajic, Dejan and Kehrberger, Leonhard M A",
  month         =  feb,
  year          =  2022,
  copyright     = "http://creativecommons.org/licenses/by/4.0/",
  archivePrefix = "arXiv",
  primaryClass  = "gr-qc",
  eprint        = "2202.04093"
}

@article {ChristodoulouPRL,
    AUTHOR = {Christodoulou, Demetrios},
     TITLE = {Nonlinear nature of gravitation and gravitational-wave
              experiments},
   JOURNAL = {Phys. Rev. Lett.},
  FJOURNAL = {Physical Review Letters},
    VOLUME = {67},
      YEAR = {1991},
    NUMBER = {12},
     PAGES = {1486--1489},
      ISSN = {0031-9007},
   MRCLASS = {83B05 (83C35)},
  MRNUMBER = {1123900},
       DOI = {10.1103/PhysRevLett.67.1486},
       URL = {https://doi.org/10.1103/PhysRevLett.67.1486},
}

@INPROCEEDINGS{ChristodoulouMarcelGrossman,
       author = {{Christodoulou}, Demetrios},
        title = "{The Global Initial Value Problem in General Relativity}",
    booktitle = {The Ninth Marcel Grossmann Meeting},
         year = 2002,
       editor = {{Gurzadyan}, Vahe G. and {Jantzen}, Robert T. and {Ruffini}, Remo},
        month = dec,
        pages = {44-54},
          doi = {10.1142/9789812777386_0004},
       adsurl = {https://ui.adsabs.harvard.edu/abs/2002nmgm.meet...44C}
}

@article{Hawking:2016msc,
    author = "Hawking, Stephen W. and Perry, Malcolm J. and Strominger, Andrew",
    title = "{Soft Hair on Black Holes}",
    eprint = "1601.00921",
    archivePrefix = "arXiv",
    primaryClass = "hep-th",
    doi = "10.1103/PhysRevLett.116.231301",
    journal = "Phys. Rev. Lett.",
    volume = "116",
    number = "23",
    pages = "231301",
    year = "2016"
}

@article{PenroseZeroRestMass,
author= "Penrose, R.",
title= "Zero rest-mass fields including gravitation: asymptotic behaviour",
doi= "10.1098/rspa.1965.0058",
journal= "Proc. R. Soc. Lond. A",
volume= "284",
pages= "159–203",
}

@misc{AAG16a,
    title={Late-time asymptotics for the wave equation on spherically symmetric, stationary spacetimes},
    author={Yannis Angelopoulos and Stefanos Aretakis and Dejan Gajic},
    year={2016},
    eprint={1612.01566},
    archivePrefix={arXiv},
    primaryClass={math.AP}
}

@article {Fred,
    AUTHOR = {Alford, Fred},
     TITLE = {The scattering map on {O}ppenheimer-{S}nyder space-time},
   JOURNAL = {Ann. Henri Poincar\'{e}},
  FJOURNAL = {Annales Henri Poincar\'{e}. A Journal of Theoretical and
              Mathematical Physics},
    VOLUME = {21},
      YEAR = {2020},
    NUMBER = {6},
     PAGES = {2031--2092},
      ISSN = {1424-0637},
       DOI = {10.1007/s00023-020-00905-5},
}

@article{DRSR14,
   title={A scattering theory for the wave equation on Kerr black hole exteriors},
   volume={51},
   ISSN={1873-2151},
   %url={http://dx.doi.org/10.24033/asens.2358},
   DOI={10.24033/asens.2358},
   number={2},
   journal={Annales scientifiques de l'\'Ecole normale sup\'erieure},
   publisher={Societe Mathematique de France},
   author={Mihalis Dafermos and Igor Rodnianski and Yakov Shlapentokh-Rothman},
   year={2018},
   pages={371-486}
}

@article{Holzegel_2016,
   title={Conservation laws and flux bounds for gravitational perturbations of the Schwarzschild metric},
   volume={33},
   ISSN={1361-6382},
   %url={http://dx.doi.org/10.1088/0264-9381/33/20/205004},
   DOI={10.1088/0264-9381/33/20/205004},
   number={20},
   journal={Classical and Quantum Gravity},
   publisher={IOP Publishing},
   author={Holzegel, Gustav},
   year={2016},
   month={Sep},
   pages={205004}
}

@misc{Rita2019,
    title={Mode stability for the Teukolsky equation on extremal and subextremal Kerr spacetimes},
    author={Teixeira da Costa, Rita},
    year={2019},
    eprint={1910.02854},
    archivePrefix={arXiv},
    primaryClass={gr-qc}
}

@ARTICLE{Schwarzschild,
       author = {{Schwarzschild}, Karl},
        title = "{{\"U}ber das Gravitationsfeld eines Massenpunktes nach der Einsteinschen Theorie}",
      journal = {Sitzungsberichte der K{\"o}niglich Preu{\ss}ischen Akademie der Wissenschaften (Berlin},
         year = 1916,
        month = jan,
        pages = {189-196},
       %adsurl = {https://ui.adsabs.harvard.edu/abs/1916SPAW.......189S},
      adsnote = {Provided by the SAO/NASA Astrophysics Data System}
}

@article {Johnson,
    AUTHOR = {Johnson, Thomas William},
     TITLE = {The linear stability of the {S}chwarzschild solution to
              gravitational perturbations in the generalised wave gauge},
   JOURNAL = {Ann. PDE},
  FJOURNAL = {Annals of PDE. Journal Dedicated to the Analysis of Problems
              from Physical Sciences},
    VOLUME = {5},
      YEAR = {2019},
    NUMBER = {2},
     PAGES = {Paper No. 13, 92},
      ISSN = {2524-5317},
   MRCLASS = {58J45 (35Q75 83C57)},
  MRNUMBER = {4015163},
       DOI = {10.1007/s40818-019-0069-0},
       %URL = {https://doi.org/10.1007/s40818-019-0069-0},
}

@article{DHR16,
   title={The linear stability of the Schwarzschild solution to gravitational perturbations},
   volume={222},
   ISSN={1871-2509},
   %url={http://dx.doi.org/10.4310/ACTA.2019.v222.n1.a1},
   DOI={10.4310/acta.2019.v222.n1.a1},
   number={1},
   journal={Acta Mathematica},
   publisher={International Press of Boston},
   author={Dafermos, Mihalis and Holzegel, Gustav and Rodnianski, Igor},
   year={2019},
   pages={1-214}
}

@ARTICLE{TeukP74,
       author = {{Teukolsky}, S.~A. and {Press}, W.~H.},
        title = "{Perturbations of a rotating black hole. III. Interaction of the hole with gravitational and electromagnetic radiation.}",
      journal = {The Astrophysical Journal},
     keywords = {Black Holes (Astronomy), Electromagnetic Interactions, Gravitational Waves, Perturbation Theory, Stellar Rotation, Angular Momentum, Asymptotic Methods, Boundary Value Problems, Conservation Laws, Eigenvalues, Electromagnetic Scattering, Energy Transfer, Tables (Data), Astrophysics},
         year = "1974",
        month = "Oct",
       volume = {193},
        pages = {443-461},
          doi = {10.1086/153180},
       %adsurl = {https://ui.adsabs.harvard.edu/abs/1974ApJ...193..443T},
      adsnote = {Provided by the SAO/NASA Astrophysics Data System}
}

@misc{DR08,
    title={Lectures on black holes and linear waves},
    author={Mihalis Dafermos and Igor Rodnianski},
    year={2008},
    eprint={0811.0354},
    archivePrefix={arXiv},
    primaryClass={gr-qc}
}

@book {Ch-K,
    AUTHOR = {Christodoulou, Demetrios and Klainerman, Sergiu},
     TITLE = {The global nonlinear stability of the {M}inkowski space},
    SERIES = {Princeton Mathematical Series},
    VOLUME = {41},
 PUBLISHER = {Princeton University Press, Princeton, NJ},
      YEAR = {1993},
     PAGES = {x+514},
      ISBN = {0-691-08777-6},
   MRCLASS = {83C05 (35Q75 58G16 83C35)},
  MRNUMBER = {1316662},
MRREVIEWER = {Alan D. Rendall},
}

@misc{DHR18,
    title={A scattering theory construction of dynamical vacuum black holes},
    author={Mihalis Dafermos and Gustav Holzegel and Igor Rodnianski},
    year={2013},
    eprint={1306.5364},
    archivePrefix={arXiv},
    primaryClass={gr-qc}
}

@article{PenroseAsymptoticPropertiesOfFieldsAndSpacetimes,
  title = {Asymptotic Properties of Fields and Space-Times},
  author = {Penrose, Roger},
  journal = {Phys. Rev. Lett.},
  volume = {10},
  issue = {2},
  pages = {66--68},
  numpages = {0},
  year = {1963},
  month = {Jan},
  publisher = {American Physical Society},
  doi = {10.1103/PhysRevLett.10.66},
  url = {https://link.aps.org/doi/10.1103/PhysRevLett.10.66}
}

@article{Penrose:1964ge,
    author = "Penrose, R.",
    editor = "DeWitt, C. and DeWitt, B.",
    title = "Conformal treatment of infinity",
    doi = "10.1007s10714-010-1110-5",
    pages = "565--586",
    year = "1964"
}

@article{Mas20,
    author = "Masaood, Hamed",
    title = "{A Scattering Theory for Linearised Gravity on the Exterior of the Schwarzschild Black Hole I: The Teukolsky Equations}",
    eprint = "2007.13658",
    archivePrefix = "arXiv",
    primaryClass = "gr-qc",
    doi = "10.1007/s00220-022-04372-3",
    journal = "Commun. Math. Phys.",
    volume = "393",
    number = "1",
    pages = "477--581",
    year = "2022"
}

@ARTICLE{Teu73,
       author = {{Teukolsky}, Saul A.},
        title = "{Perturbations of a Rotating Black Hole. I. Fundamental Equations for Gravitational, Electromagnetic, and Neutrino-Field Perturbations}",
      journal = {The Astrophysical Journal},
         year = 1973,
        month = oct,
       volume = {185},
        pages = {635-648},
          doi = {10.1086/152444},
       %adsurl = {https://ui.adsabs.harvard.edu/abs/1973ApJ...185..635T},
      adsnote = {Provided by the SAO/NASA Astrophysics Data System}
}

@article{Sachs1962,
author = {Sachs, R. K. :  and Bondi, Hermann },
title = {Gravitational waves in general relativity VIII. Waves in asymptotically flat space-time},
journal = {Proceedings of the Royal Society of London. Series A. Mathematical and Physical Sciences},
volume = {270},
number = {1340},
pages = {103-126},
year = {1962}
}

@article{BMV1962,
author = {Bondi, Hermann  and Van der Burg, M. G. J.  and Metzner, A. W. K. },
title = {Gravitational waves in general relativity, VII. Waves from axi-symmetric isolated system},
journal = {Proceedings of the Royal Society of London. Series A. Mathematical and Physical Sciences},
volume = {269},
number = {1336},
pages = {21-52},
year = {1962}
}

@article{Flanagan_2020,
   title={Extensions of the asymptotic symmetry algebra of general relativity},
   volume={2020},
   ISSN={1029-8479},
   url={http://dx.doi.org/10.1007/JHEP01(2020)002},
   DOI={10.1007/jhep01(2020)002},
   number={1},
   journal={Journal of High Energy Physics},
   publisher={Springer Science and Business Media LLC},
   author={Flanagan, \'Eanna \'E. and Prabhu, Kartik and Shehzad, Ibrahim},
   year={2020},
   month={Jan}
}

@article{Campiglia_2015,
   title={New symmetries for the gravitational S-matrix},
   volume={2015},
   ISSN={1029-8479},
   url={http://dx.doi.org/10.1007/JHEP04(2015)076},
   DOI={10.1007/jhep04(2015)076},
   number={4},
   journal={Journal of High Energy Physics},
   publisher={Springer Science and Business Media LLC},
   author={Campiglia, Miguel and Laddha, Alok},
   year={2015},
   month={Apr}
}

@misc{LukOh21,
      title={Global nonlinear stability of large dispersive solutions to the Einstein equations}, 
      author={Jonathan Luk and Sung-Jin Oh},
      year={2021},
      eprint={2108.13379},
      archivePrefix={arXiv},
      primaryClass={gr-qc}
}

@misc{DHRT21,
      title={The non-linear stability of the Schwarzschild family of black holes}, 
      author={Mihalis Dafermos and Gustav Holzegel and Igor Rodnianski and Martin Taylor},
      year={2021},
      eprint={2104.08222},
      archivePrefix={arXiv},
      primaryClass={gr-qc}
}

@article {Choquet-Bruhat,
    AUTHOR = {Four\`es-Bruhat, Y.},
     TITLE = {Th\'{e}or\`eme d'existence pour certains syst\`emes d'\'{e}quations aux
              d\'{e}riv\'{e}es partielles non lin\'{e}aires},
   JOURNAL = {Acta Math.},
  FJOURNAL = {Acta Mathematica},
    VOLUME = {88},
      YEAR = {1952},
     PAGES = {141--225},
}

@article{ChristodoulouMemoryEffect,
  title = {Gravitational-wave bursts with memory: The Christodoulou effect},
  author = {Thorne, Kip S.},
  journal = {Phys. Rev. D},
  volume = {45},
  issue = {2},
  pages = {520--524},
  numpages = {0},
  year = {1992},
  month = {Jan},
  publisher = {American Physical Society},
  doi = {10.1103/PhysRevD.45.520},
  url = {https://link.aps.org/doi/10.1103/PhysRevD.45.520}
}

@book{PenroseRindler,
    author = "Penrose, R. and Rindler, W.",
    title = "{Spinors and Space-time. vol. 2: Spinor and Twistor methods in space-time geometry}",
    doi = "10.1017/CBO9780511524486",
    isbn = "978-0-521-34786-0, 978-0-511-86842-9",
    publisher = "Cambridge University Press",
    series = "Cambridge Monographs on Mathematical Physics",
    month = "4",
    year = "1988"
}
\end{document}